\documentclass[%
 reprint,
superscriptaddress,
 amsmath,amssymb,
 aps,
pra,
]{revtex4-2}

\usepackage{pauli_preamble}

\usepackage[dvipsnames,svgnames]{xcolor}

\NewDocumentCommand{\TAIL}{+m}{{\color{cyan!60!black}#1}}

\usepackage[
    colorlinks=true,
    linkcolor=black,
    filecolor=magenta,
    urlcolor=cyan,
    citecolor=green,
    pdftitle={Document Title},
    pdfpagemode=FullScreen,
]{hyperref}

\renewcommand{\thepart}{\Roman{part}}

\begin{document}

\title{From Pauli Strings to Quantum Dynamics: A Unified Characterization}

\author{Roberto Gargiulo\,\href{https://orcid.org/0009-0009-8820-6049}{\includegraphics[height=6pt]{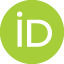}}}
\email{r.gargiulo@fz-juelich.de}
\affiliation{Forschungszentrum Jülich GmbH, Peter Grünberg Institute, Quantum Control (PGI-8),
  52425 Jülich,
  Germany}
\affiliation{%
  University of Cologne, Institute for Theoretical Physics (THP),
  50937 Köln,
  Germany
}

\author{Paul Herringer\,\href{https://orcid.org/0009-0000-1149-0998}{\includegraphics[height=6pt]{ORCID-iD_icon-64x64.png}}}
\email{paul.herringer@itp.uni-hannover.de}
\affiliation{Institut für Theoretische Physik, Leibniz Universität Hannover, Appelstra{\ss}e 2, 30167 Hannover, Germany}
\affiliation{%
  Department of Physics and Astronomy, University of British Columbia, 6224 Agricultural Road, Vancouver, British Columbia V6T1Z4, Canada
}
\affiliation{Stewart Blusson Quantum Matter Institute, University of British Columbia, Vancouver, British Columbia V6T1Z4, Canada}

\author{Robert Zeier\,\href{https://orcid.org/0000-0002-2929-612X}{\includegraphics[height=6pt]{ORCID-iD_icon-64x64.png}}}
\email{r.zeier@fz-juelich.de}
\affiliation{Forschungszentrum Jülich GmbH, Peter Grünberg Institute, Quantum Control (PGI-8),
  52425 Jülich,
  Germany}

\date{June 8, 2026}

\begin{abstract}
Understanding the dynamical properties of quantum systems is an essential task in quantum computing, quantum control, and many-body physics. Tools such as representation theory and Lie theory provide crucial information on reachability and computational power.
However, this information can be difficult to access exactly or compute
efficiently for arbitrary generating sets.
Here we focus on the setting of Pauli strings, which satisfy numerous exceptional properties that simplify the problem.
We find deep connections between Pauli Lie algebras and certain subgroups of the Clifford group generated by \emph{transvections}, through the symplectic properties of the Pauli strings.
This allows us to give an invariant-based perspective on these objects and their reachability, in the language of Pauli orbits, symmetries, and invariant subspaces.
The invariant-based approach provides efficient algorithms for identifying Lie algebras and orbits, as well as a simple framework for analyzing structured Pauli generating sets.
We also show in an elementary way that Clifford subgroups generated by transvections provide 3-designs for the corresponding Pauli Lie groups.
We illustrate the framework through structured examples from variational quantum algorithms, restricted quantum computation, many-body systems, and random circuits.
\end{abstract}

\maketitle

\section{Introduction}

Understanding dynamical properties of quantum systems is a crucial task in quantum physics,
from quantum information~\cite{LewisSwan_SafaviNaini_Kaufman_Rey_2019,Benatti_2023,Amico_Fazio_Osterloh_Vedral_2008}
to many-body physics~\cite{Girvin_Yang_2019,Ryu_Schnyder_Furusaki_Ludwig_2010,Altland_Simons_2010,Parameswaran_2018,Moudgalya_2023}.
Treating quantum circuits as dynamical systems can clarify the limitations and advantages of quantum algorithms and guide the design of new ones.
This perspective has become especially useful for variational quantum algorithms~\cite{Farhi_Goldstone_Gutmann_2014,Peruzzo_2014,Cerezo_2021,Cerezo_2025,
Kazi_Larocca_Farinati_Coles_Cerezo_Zeier_2025,singh2025groundstate},
where dynamical properties characterize trainability, expressivity and reachability.

A typical setting is digital or analog quantum computing, where a specified
gate set or set of controls generates the dynamics after choosing parameters or
pulses.
For a given set of generators, one wants not only a classification of possible dynamics,
but also methods for identifying and distinguishing them.
Typical questions include which group is generated, whether a target state or gate is reachable, and
which parameter choices realize it.
Answering these questions usually requires classical computations with matrices
or states whose size grows exponentially with the number of qubits.
In general, this complexity barrier appears unavoidable.

\PauliContentsTable[tab:contents]{Contents: summary of the parts and sections in this work}

However, one may still be able to answer these questions in restricted
settings.
We focus in this work on quantum dynamics generated by sets of Pauli strings
$P\in\{I,X,Y,Z\}^{\otimes n}$ for Pauli matrices $X$, $Y$, $Z$ and the identity $I$.
In particular, each independently controllable interaction is a single Pauli
string, rather than a sum of Pauli strings.
This setting has emerged recently for variational quantum algorithms
\cite{Kazi_Larocca_Farinati_Coles_Cerezo_Zeier_2025,Wiersema_Kokcu_Kemper_Bakalov_2024,Kokcu_Wiersema_Kemper_Bakalov_2024},
disordered many-body systems \cite{Imbrie_2016},
parity quantum computing \cite{Smith_Klaver_Nautrup_Lechner_Briegel_2025},
and other models for quantum computing \cite{Fellner_2022,Herringer_2025}.
This also connects to how Pauli strings act on each other and arbitrary operators \cite{Diaz_GarciaMartin_Kazi_Larocca_Cerezo_2023,West_Dowling_Southwell_Sevior_Usman_Modi_Quella_2025}.
More general architectures may carry additional symmetries, but the Pauli
setting provides a first simplified description of the resulting dynamical
structure
\cite{Kazi_Larocca_Farinati_Coles_Cerezo_Zeier_2025,singh2025groundstate,gargiulo2026obstructionsuniversalitygloballycontrolled}.

In this work, we classify the possible dynamics generated by Pauli strings
and give an algorithm that identifies the dynamical class of any given Pauli
generating set. The algorithm runs in time polynomial in the number of qubits
and the number of Pauli generators. We have announced
this algorithm in \cite{Natal_2026,Dresden_2026}.
We also provide a polynomial-time algorithm, in the same parameters, for separating
the Pauli orbits associated with a given Pauli generating set.
Together, these results establish a unified framework for Pauli-generated
dynamics, bringing together aspects of quantum information that were
previously treated separately.
We detail our key results in Section~\ref{sec:key:results}.

We build on a large body of mathematical results for
subgroups of the binary symplectic group generated by so-called \emph{transvections}
\cite{McLaughlin_1969,Wajnryb_1980,Humphries_1983,humphries_1985,Janssen_1983,Janssen_1985,Brown_Humphries_1986a,Brown_Humphries_1986b}.
Already at that time, Hall~\cite{Hall_1984} emphasized that the problem was
not completely solved: ``The problem remains unexhausted. Despite various present solutions,
it remains in part unanswered.''
Important developments came afterwards \cite{Cameron_Hall_1991,Shapiro_Shapiro_Vainshtein_1998,Callan_1976,Hall_1989,Sjöstrand_2025,Vorstermans},
and we particularly rely on the work of Cuypers \cite{Cuypers_2021_Whitney,Cuypers_2021_E6,Cuypers_2021}
and \cite{Seven_2005}.
More recently, Aguilar \emph{et al.} \cite{Aguilar_Cichy_Eisert_Bittel_2024} provided a novel algorithmic approach to canonical classes of generators,
resulting in different, but compatible canonical classes compared to \cite{Janssen_1983}.
Our final algorithms do not rely on transformations to canonical generators and therefore
preserve structural properties
of multiple, related problem instances which would be hidden by switching to canonical generating sets.
Very recently, Cuypers~\cite{Cuypers_2026} has applied his earlier work
to Pauli-string dynamics and also provides a classification algorithm for Pauli dynamics.
Similarities and differences to our classification algorithm are discussed in Section~\ref{sec:classification_and_algorithm_lie_algebras}.

We also develop several applications that use the structural information
provided by Pauli dynamics.
For Ising and XY interactions, the Pauli-string framework gives an elementary
description of the dynamics, simplifying the multi-angle setting of
\cite{Kazi_Larocca_Farinati_Coles_Cerezo_Zeier_2025} and naturally recovering
cases studied in
\cite{Wiersema_Kokcu_Kemper_Bakalov_2024,Kokcu_Wiersema_Kemper_Bakalov_2024}.
For parity quantum computing, we give necessary and sufficient conditions for
universality
\cite{Fellner_2022,Smith_Klaver_Nautrup_Lechner_Briegel_2025}.
We also apply the framework to Instantaneous Quantum Polynomial (IQP)
computation
\cite{Shepherd_2009,Bremner_2010,Bremner_2017,Nakata_2014,buzet2026iqpcircuits2forrelation}.
In many-body settings, we give a unified classification of free-fermionic
mappings, generalizing
\cite{Chapman_Flammia_2020,Fendley_2019,Elman_Chapman_Flammia_2021,Fendley_Pozsgay_2024}.
For Clifford groups, we simplify the proof of the 3-design property
\cite{Webb_2016,Zhu_2017,Zhu_Kueng_Grassl_Gross_2016,Bittel_Eisert_Leone_Mele_Oliviero_2025}
and extend the argument to arbitrary Clifford subgroups generated by transvections.
Finally, the resulting understanding of Pauli orbits may support improved
classical simulation methods
\cite{Goh_Larocca_Cincio_Cerezo_Sauvage_2023,Martinez_Angrisani_Pankovets_Fawzi_Franca_2025,Rudolph_Jones_Teng_Angrisani_Holmes_2025,Rall_Liang_Cook_Kretschmer_2019,Begusic_Hejazi_Chan_2025,Fontana_Rudolph_Duncan_Rungger_Cîrstoiu_2025,Ugale_Master}.

The manuscript is organized into nine parts, summarized in
Table~\ref{tab:contents}.
Part~\ref{part:introduction} gives a non-technical overview of our work,
including the main results and their relation to the literature.
Part~\ref{part:applications} then presents physical examples and applications.
Parts~\ref{part:theory}--\ref{part:final_classification} develop the Pauli
framework in detail, starting from Pauli strings and their binary
representatives and moving through graph-theoretic and invariant-theoretic
methods to Clifford groups, free-fermionic models, and the final classification
of Pauli dynamics and Pauli orbits.
Part~\ref{part:appendices} collects the methodology section, acknowledgments,
appendices, and references.

\begin{table*}
\centering
\caption{Key results as discussed in Section~\ref{sec:key:results}.}
\label{tab:main_results}
\footnotesize
\begin{tabular}{@{\hspace{1mm}}l@{\hspace{8mm}}l@{\hspace{8mm}}l@{\hspace{1mm}}}
\hline\hline
\\[-2.5mm]
 & Description & Link(s)
\\[0.5mm] \hline
\\[-2.0mm]
1 & \begin{minipage}[t]{0.66\textwidth}\raggedright Bijection between centerless Pauli Lie algebras and transvection groups\end{minipage}
& Thm.~\ref{thm:pauli:transvection}\\[1.5mm]
2 & \begin{minipage}[t]{0.66\textwidth}\raggedright Classification of invariant bilinear forms for Pauli strings and binary quadratic forms\end{minipage}
& Thm.~\ref{thm:classification_of_affine_subspaces_isomorphism_classes} and Cor.~\ref{cor:Canonical_Forms_Invariant_Bilinear_Forms}\\[2mm]
3 & \begin{minipage}[t]{0.66\textwidth}\raggedright Classification of connected Pauli Lie algebras, Clifford and binary transvection groups\end{minipage}
& Thms.~\ref{thm:full_classification_pauli_lie_algebras} and~\ref{thm:full_classification_transvection_groups}\\[2mm]
4 & \begin{minipage}[t]{0.66\textwidth}\raggedright Invariant-based algorithm for Lie-algebra and transvection-group identification\end{minipage}
& Cor.~\ref{cor:algorithm_lie_algebras}\\[1.5mm]
5 & \begin{minipage}[t]{0.66\textwidth}\raggedright General Description for Lie-algebraic dependencies\end{minipage}
& Cor.~\ref{cor:Limits_Lie_Algebraic_Dependencies_general_classification}\\[2mm]
6 & \begin{minipage}[t]{0.66\textwidth}\raggedright Complete connected Pauli orbit classification and efficient algorithm (beyond \cite{Janssen_1983,Seven_2005})\end{minipage}
& Sec.~\ref{sec:classification:orbits} and Cor.~\ref{cor:algorithm_orbits}\\[2mm]
7 & \begin{minipage}[t]{0.66\textwidth}\raggedright Invariant-based criterion for $t$-equivalent canonical frustration graphs (beyond \cite{Janssen_1983,Brown_Humphries_1986b,Aguilar_Cichy_Eisert_Bittel_2024})\end{minipage}
& Cors.~\ref{cor:distinguishing_quasi_universal_cases} and~\ref{cor:distinguishing_line_graph_cases}\\[2mm]
8 & \begin{minipage}[t]{0.66\textwidth}\raggedright Free-fermionic mappings with symmetries and algebraic dependencies (beyond \cite{Chapman_Flammia_2020})\end{minipage}
& Prop.~\ref{prop:free_fermionic_mapping_alg_dep}\\[1.5mm]
9 & \begin{minipage}[t]{0.66\textwidth}\raggedright Clifford 3-designs for Pauli Lie Groups\end{minipage}
& Thms~\ref{thm:transvection_3_design_property} and~\ref{thm:transvection_not_4_design}\\[1.5mm]
10 & \begin{minipage}[t]{0.66\textwidth}\raggedright Adjoint Commutant and Invariant Subspaces for Pauli Lie Groups\end{minipage}
& Thms~\ref{thm:Adjoint_Commutant_Paulis_Orbits} and~\ref{thm:symmetry_adapated_orbit_subspaces}\\[1.5mm]
11 & \begin{minipage}[t]{0.66\textwidth}\raggedright Symmetries for local Pauli-string Hamiltonians\end{minipage}
& Res.~\ref{result:from_local_to_global_symmetries}\\[1.5mm]
12 & \begin{minipage}[t]{0.66\textwidth}\raggedright New Proofs for 2-Local Pauli Lie Algebras and Parity Quantum Computing \end{minipage}
& Res.~\ref{result:2-local-lie_algebras}\\[1.5mm]
13 & \begin{minipage}[t]{0.66\textwidth}\raggedright Symmetries for Free-Fermions in Disguise for arbitrary boundary conditions \end{minipage}
& Section~\ref{sec:ffd_free_fermions_in_disguise}\\[1.0mm]
\hline\hline
\end{tabular}
\end{table*}

\ManuscriptPart{Overview and Summary of Results}{part:introduction}

This part serves as an overview before the more technical developments in the
later parts. It is intended as a starting point and guide to the main body of the work.
We begin in Section~\ref{sec:key:results} with a discussion of the key technical results and how they go beyond the state of the art.
We then continue Section~\ref{sec:overview_math}
with a high-level overview
of our Pauli framework and how it is connected to related work.
Section~\ref{sec:overview_phys} provides an overview of physical examples and applications. We close with a discussion and outlook in Section~\ref{sec:discussion}, after which the technical developments start.

\section{Key Results in Context}\label{sec:key:results}

We briefly highlight here our main results, indexed as in Table~\ref{tab:main_results}. We discuss their connection with previous work and also how we extend the existing literature.

(1) One of the basic ideas underlying this work is the connection between transvections and Pauli strings, here formalized as a bijection between Pauli Lie algebras and transvection groups modulo their centers.
A general connection between Lie algebras over fields $\F$ and transvection groups has been investigated in \cite{Brouwer_Cohen_Cuypers_Hall_Postma_2012}, though not at the level of Pauli strings and the binary formalism, which we precisely state here. In \cite{Cuypers_2026} the connection to Pauli Lie algebras is instead formulated in terms of \emph{cotriangular spaces}, an object closely related to transvections and transvection groups.
(2) Similarly, we take advantage of this connection to view Pauli bilinear forms as binary quadratic forms.
Classical results provide the isomorphism classes of individual bilinear forms over $\C$ and quadratic forms $\F_2$, though it was not known what the classes for \emph{sets} of such objects are.

(3) A classification of Pauli Lie algebras and binary transvection groups restricted to the spanning space was known.
We expand this classification to include the full groups as living inside the binary symplectic $\Sp(2n,\F_2)$ and Clifford $\cl_n$ groups.
Also, we provide a classification for these objects purely based on their invariant properties.

(4) We can also turn our approach into an explicit algorithm for efficient identification. A graph-theoretic approach based on graph transformations was given in \cite{Aguilar_Cichy_Eisert_Bittel_2024}, but not explicitly expanded into an algorithm.
Here we provide explicit steps based on simple linear algebra or graph-theoretic (i.e.\ root multigraph computation and cycle space) subroutines.

(5) In the context of minimal generating sets, a description of Lie-algebraic dependencies was known in the canonical case \cite{Janssen_1983,Aguilar_Cichy_Eisert_Bittel_2024}, and here it is generalized to arbitrary frustration graphs, taking advantage of the invariant description of the generating sets and specifically the \textit{orbit radical}.
This also provides an invariant-based criterion to systematically remove redundant generators from a given Pauli set.

(6) Beyond Lie algebras and transvection groups, we also discuss their action via the Pauli orbits.
We expand previous work on transvection groups \cite{Janssen_1983,Brown_Humphries_1986b,Seven_2005} to provide a full labelling of the orbits over $\Fn$ for arbitrary generating sets, as well as a description of the orbits for the exceptional free-fermionic case (Proposition~\ref{prop:orbits_in_Fn_exceptional_case_free_fermionic}).

(7) On the graph-theoretic side, we provide an invariant description of the canonical $t$-equivalent representatives not only for $\calE_6$-free graphs \cite{Janssen_1983,Brown_Humphries_1986b}, which is based on quadratic forms, but also for those $t$-equivalent to certain blown-up path graphs, using information from the root multigraph, i.e.\ the cycle space and T-joins.
(8) We also expand the work of \cite{Chapman_Flammia_2020} by providing faithful free-fermionic mappings, which reproduce all symmetries and algebraic dependencies, via line graphs of multigraphs.
This allows for a Majorana-based description of Lie algebras and Pauli orbits over the entire space.
However, the corresponding description is no longer strictly in terms of quadratic Majoranas, due to the presence of cycle symmetries, as well as the odd and exceptional cases, whose generators also include linear Majoranas and the parity.

(9) We show that arbitrary Clifford transvection groups are 3-designs for the corresponding Pauli Lie groups.
This includes the previously known cases of the full Clifford group \cite{Nebe_Rains_Sloane_2001}, real Clifford group \cite{Hashagen2018realrandomized} and matchgate Clifford group \cite{Wan_2023}, which we discuss in a unified manner.
Using the Clifford formalism, we provide a basis for the adjoint commutant for Pauli Lie algebras, which is a counterpart to the square commutant described in \cite{Diaz_GarciaMartin_Kazi_Larocca_Cerezo_2023} via the commutator graph formalism.
(10) We also provide a partial characterization of the operator invariant subspaces based on orbits and linear symmetries, which has been fully discussed in \cite{Diaz_GarciaMartin_Kazi_Larocca_Cerezo_2023} only for the even free-fermionic case.

(11) Independently of transvections, we showcase how one can exactly compute the commutant and invariant bilinear forms for certain structured families of local
Pauli-string models.
We expand and formalize the work of \cite{Wiersema_Kokcu_Kemper_Bakalov_2024}, which focused on 2-local models, to also include certain $k$-local models.
(12) We apply the tools from (11) to the case of free-fermions in disguise (FFD) with various boundary conditions to describe unitary and anti-unitary symmetries.
A full description of unitary symmetries, or the commutant algebra, was given for open boundary conditions in \cite{Vernier_Piroli_2026}.
Here we focus only on the symmetries that do not come from the `virtual' FFD modes, which still provide much information.
This also partially identifies the dynamical properties of such systems, since we also prove quasi-universality.
(13) Finally, we provide new proofs for the Lie algebras of certain Pauli generating sets, which were originally discussed in \cite{Kazi_Larocca_Farinati_Coles_Cerezo_Zeier_2025,Wiersema_Kokcu_Kemper_Bakalov_2024,Kokcu_Wiersema_Kemper_Bakalov_2024,Smith_Klaver_Nautrup_Lechner_Briegel_2025}.
This showcases the general approach to the problem of Pauli Lie algebras in structured generating sets, leading to conceptually simpler proofs.

\section{Overview of Framework and Main Results}\label{sec:overview_math}

In this section we present an overview of our mathematical framework and its main results, in the context of Pauli Lie algebras, symplectic and Clifford groups, as well its relationship to graph-theoretic properties. We touch on quantum dynamics and reachability in Section~\ref{sec:overview:quantum:dynamics}, while
covering the basic mathematical setting in Section~\ref{sec:overview:pauli}. Graph-theoretic aspects
are discussed in Section~\ref{sec:overview:math:graphs} which then to leads to Pauli orbits and transvection groups
in Section~\ref{sec:overview:orbits:transvection:groups}.
Sections~\ref{sec:overview:canonical:generators} and \ref{sec:overview:invariants} consider the perspectives
of canonical generators and invariants, respectively. The connection to Pauli Lie algebras and Clifford groups is detailed in Section~\ref{sec:overview:pauli:clifford:groups}.
We explore the quasi-universal and the free-fermionic cases in Sections~\ref{sec:quasi_universal_description}
and \ref{sec:overview:free:fermions}. Finally, Section~\ref{sec:overview:orbits} provides
an overview of our classification of the Pauli orbits.

\subsection{Quantum Dynamics, Reachability and Lie Theory}\label{sec:overview:quantum:dynamics}

We start by recalling the action resulting from a given set of infinitesimal generators $\pgens = \{H_i\}_{i=1}^L$ on a $d$-dimensional quantum system $\C^d$.
In a digital setting this corresponds to a time evolution obtained by a circuit of the form $\prod_{j=1}^L e^{\im\theta_jH_{i_j}}$ where $L$ represents the number of gates applied from the parametrized gate set $\pgens$ and $\theta_j\in\R$ are some real parameters.
In a continuous setting as for analog quantum computing or optimal control, the time evolution instead takes the form of a time-ordered exponential $\mathcal{T}\exp(\im\int_0^t\dd{\tau} \sum_i c_i(\tau)H_i)$ for some pulses represented by real functions $c_i$.
Then, from the point of view of quantum algorithms, the problem of constructing a given circuit is the choice of the angles $\theta_j$ or pulses $c_i$, given some initial state and measurement basis.

A large number of schemes exist for constructing circuits this way, depending on the specific problem and approach.
In the context of Noisy Intermediate-Scale Quantum computing (NISQ), a commonly adopted scheme is that of \emph{variational quantum algorithms} \cite{Farhi_Goldstone_Gutmann_2014,Cerezo_2021,Peruzzo_2014}.

Regardless of whether the setting is digital or analog, the long time properties are described by a subgroup of the unitary group.
This is a connected Lie group whose infinitesimal generators are the Lie algebra generated by $H_i$.
Knowledge of such an object and its algebraic properties in the context of quantum algorithms provides fundamental information about the circuit, such as: computational power and universality \cite{DiVincenzo_1995,Sleator_Weinfurter_1995,Barenco_1995,Jozsa_2009,Webb_2016}, classical simulability \cite{Goh_Larocca_Cincio_Cerezo_Sauvage_2023}, quantum resource theories \cite{Chitambar_2019,Veitch_Hamed_Mousavian_Gottesman_Emerson_2014,diaz2025unifiedapproachquantumresource}, reachability and controllability \cite{Albertini_2001,Schirmer_Solomon_Leahy_2002,dalessandro2022,Zeier_2011,Zeier_2015,ZZKS14,Zimboras_Zeier_SchulteHerbruggen_Burgarth_2015,schulte2017,Keyl_Zeier_Zimboras_Dirr_Schulte_book}, trainability \cite{Diaz_GarciaMartin_Kazi_Larocca_Cerezo_2023,Ragone_2024}.

Even beyond the context of quantum information, such tools have been used in context of many body systems, regarding symmetries, phase transitions, quantum chaos and more \cite{Girvin_Yang_2019,Ryu_Schnyder_Furusaki_Ludwig_2010,Altland_Simons_2010,Parameswaran_2018,Moudgalya_2023}.

Then, in this context fundamental questions to answer for a given set of generators include the constants of motion $f(\rho(0)) = f(\rho(t))$, operators or \emph{superoperators} which are left invariant under time evolution, as well as \emph{subspaces} of states or operators which are left invariant under time evolution.

Out of these properties, \emph{reachability} plays an especially important role, and addresses a very natural question for dynamical systems: given an initial condition, which other states can I reach?
For instance, in the context of universal quantum computation, given some initial pure state $\ket{\phi}$ in the Hilbert space, one can reach \emph{any} other pure state $\ket{\psi}$ by implementing a suitable unitary gate $U\in\lieU(d)$, viewed either as a circuit in the digital setting or as a pulse over some continuous control fields.
For arbitrary mixed states, unitary dynamics does not have arbitrary reachability, since conjugation $\rho\mapsto U\rho U^\dagger$ must conserve the eigenvalues of the density matrix.
Similarly, this holds in the Heisenberg picture for observables, which evolve as $O \mapsto U^\dagger \rho U$.
Such a perspective has recently become more relevant in the context of quantum algorithms for the computation of observables and correlation functions \cite{Abanin_Acharya_Aghababaie_Beni_Aigeldinger_Ajoy_Alcaraz_Aleiner_Andersen_Ansmann_Arute_et_al_2025,king2025simplifiedversionquantumotoc2}, as well as classical algorithms for quantum emulation \cite{Goh_Larocca_Cincio_Cerezo_Sauvage_2023,Rudolph_Jones_Teng_Angrisani_Holmes_2025,Ugale_Master,miller2025simulationfermioniccircuitsusing}, by taking advantage of certain algebraic or sparsity properties of the time evolution.

\subsection{Pauli Strings and Clifford Groups}\label{sec:overview:pauli}

We explicitly address such questions in the restricted setting where the generators are \emph{Pauli strings}, $\pgens\subseteq \PP_n = \{I,X,Y,Z\}^{\otimes n}$.
These objects satisfy various exceptional properties, which make them well suited to address many tasks in quantum information theory.
We mention a few key properties: they form an orthonormal tensor product basis for the space of operators; the set $\{\pm 1,\pm\im\}\PP_n$ is closed under multiplication and forms the \emph{Pauli group}; its normalizer group in the unitary group coincides with the \emph{Clifford group}; a commuting set of Pauli strings determines a \emph{stabilizer subspace} of $\C^{2^n}$, which reduces to a stabilizer \emph{state} if the set is generated by $n$ independent Pauli strings.

The stabilizer property is particularly relevant in the context of (stabilizer) quantum error correction, where one encodes logical qubits into special invariant subspaces with respect to commuting Pauli strings \cite{Gottesman_PhD}.
Clifford operations are often the easiest to implement in an error correction scheme, and they are maximal in the unitary group, with the Clifford+T gate set being one of the most well-known universal gate sets for logical quantum computation \cite{dawson2005solovaykitaevalgorithm}.
Moreover, by the celebrated Gottesmann-Knill theorem \cite{gottesman1998heisenbergrepresentationquantumcomputers} and related results, stabilizer states and Clifford circuits are at the center of common classical algorithms for simulation of quantum circuits \cite{Dehaene_PRA_2003,Seddon_Regula_Pashayan_Ouyang_Campbell_2021}.
The orthonormal basis property is especially important in the Heisenberg picture, since any observable may expanded into the Pauli basis.
Indeed, this has also been exploited in the context of the recent Pauli backpropagation methods, which work effectively when the observable or state is \emph{sparse} in the Pauli basis \cite{Rudolph_Jones_Teng_Angrisani_Holmes_2025}.

This shows that understanding reachability of Pauli strings is essential to probe the computational power of a given architecture.
In this work we tackle the following restricted questions: under arbitrary Pauli dynamics, which Pauli strings can be reached? Moreover, what does this imply for arbitrary reachability of observables under Pauli dynamics?
Answering this question will also provide information about \emph{symmetries} for the dynamics.

By definition, the Clifford group acts transitively on the set of Pauli strings (excluding the identity), which suggests that the question of reachability in Pauli architectures is very much related to the \emph{Clifford} properties of arbitrary Pauli Lie groups and Pauli Lie algebras.
In the universal case $\lie{\pgens} = \su(2^n)$, the answer is straightforward, given that the full Clifford group is a subgroup of the unitary group, which results in an orbit containing the identity and another containing all non-identity Pauli strings.
On the other hand, this does not guarantee that one can reach arbitrary observable from a given starting one, given that the spectral properties of the initial observables must be conserved (indeed, this is also the only constraint for hermitian operators).
Hence, beyond understanding universality, we are particularly interested in describing \emph{non-universal} cases with limited reachability.
One such non-trivial example is that of free fermions $\pgens = \{\gamma_i\gamma_j\}_{i<j=1}^{2n}$, which result in the orthogonal Lie algebra in its spinor representation $\lie{\pgens} = \so(2n)$.
In this case reachability is restricted, and each orbit consists precisely of the \emph{Majorana strings} $\gamma_i\gamma_j\gamma_k\cdots $ of a given \emph{length}, a property which is exploited in the celebrated Wick's theorem for free fermions \cite{Wick_1950,bravyi2004lagrangianrepresentationfermioniclinear,Onsager_1944,Kaufman1949}.

Our work and results make extensive use of the \emph{binary formalism} \cite{Calderbank_PRL_1997,Calderbank_IEEE_1998,Gottesman_PRA_1998,Gottesman_PhD,Nielsen_Chuang_2008,Dehaene_PRA_2003,Aaronson_Gottesmann_2004}
for Pauli strings (see Table~\ref{tab:common_symbols}).
This builds on the identification of the Pauli group as a binary symplectic space of dimension $2n$ (see Section~\ref{sec:pauli}) and the Clifford group as the symplectic group $\Sp(2n,\F_2)$ over this space (see Section~\ref{sec:clifford:transvections}).
Specifically, one can view each $X_i$ and $Z_i$ as basis vectors which generate all Pauli strings over $\F_2$.
This vector space is equipped with a symplectic product $\symp{\cdot}{\cdot}$ such that $\symp{v}{w} =0$ when the corresponding Pauli strings commute and $\symp{v}{w} =1$ when they anti-commute.

\begin{table}
    \centering
    \caption{Common symbols.}
    \label{tab:common_symbols}
    \begin{tabular}{@{\hspace{1mm}}l@{\hspace{8mm}}l@{\hspace{1mm}}}
        \hline\hline
        \\[-2.5mm]
        Symbol & Description
        \\[0.5mm] \hline
        \\[-2.0mm]
        $\iso{v}$ & Pauli string $\prod_{j=1}^n i^{v_{j}v_{n+j}} X_{j}^{v_{j}}Z_{j}^{v_{n+j}}$\\
        $\PP_n$ & Set of Pauli strings $\{ I,X,Y,Z \}^{\otimes n}$ \\
        $\pgens$ & Subset of Pauli strings\\
        $\Fn$ & Binary symplectic space of dimension $2n$\\
        $\vgens$ & Subset of binary vectors in $\Fn$\\
        $V$ & Subspace of $\Fn$\\
        $v$ & Binary vector\\
        $\symp{\cdot}{\cdot}$ & Symplectic product of two binary vectors\\
        $\matalg$ & (Pauli) matrix algebra\\
        $\rad(V)$ & Set of commuting vectors in $V$\\
        $\nullity(V)$ & Dimension of $\rad(V)$\\
        $\QQ$ & Quadratic form\\
        $\rank(\matalg)$ & Rank of a Pauli matrix algebra [Eq.~\eqref{eq:def:rank_and_nullity_Pauli_matrix_algebra}]\\
        $\lie{\pgens}$ & Lie algebra generated by $\pgens$\\
        $\Ad_M$ & Conjugation map by $M\in\GL(d)$\\
        $\ad_H$ & Commutator map by $H\in\matalg(2^n,\C)$\\
        $\graphG$ & Simple graph\\
        $\Delta$ & Multigraph\\[1.0mm]
        \hline\hline
    \end{tabular}
\end{table}

\subsection{The Graph Theoretic Formalism}\label{sec:overview:math:graphs}

\begin{figure*}
    \centering
    \includegraphics[width=0.8\textwidth]{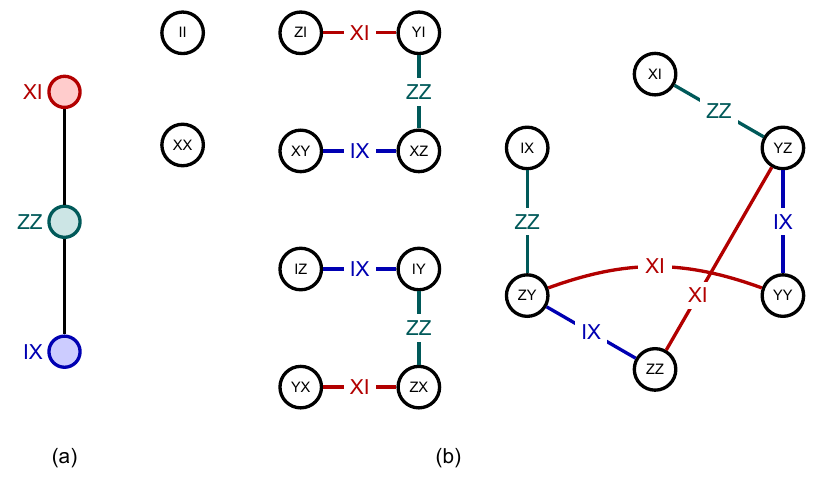}
    \caption{The graph-theoretic formalism. Example of (a) frustration graph and (b) commutator graph over two qubits, for the generating set $\pgens = \{ \text{IX, XI, ZZ} \}$.}
    \label{fig:commutator-and-frustration-graphs}
\end{figure*}

Additionally, a \emph{graph-theoretic} framework has emerged to study dynamical and spectral properties of sets of Pauli strings.
The fundamental object is the \emph{frustration graph} of a set $\pgens$ of Pauli strings, which is the graph whose vertex set is $\pgens$ and where two Pauli strings are connected if and only if they anti-commute (see Fig.~\ref{fig:commutator-and-frustration-graphs}(a) for an example).
For instance, this was used in \cite{Chapman_Flammia_2020} to understand the existence of certain \emph{free-fermionic mappings}, which guarantee the existence of exact solutions.
In \cite{de_Gois_Hansenne_Guhne_2023,xu2025simultaneousvariancespaulistrings}, it was used instead in the context of quantum correlations, entanglement and ground state energy certification.
It was also used in \cite{Aguilar_Cichy_Eisert_Bittel_2024} for a classification of \emph{Pauli} Lie algebras, which makes use of certain graph-transformations which preserve (Lie-)algebraic properties, to describe the Lie algebra generated by Pauli strings.
Specifically, they identified six distinct families of Pauli Lie algebras depending on a certain representative, up to these Lie-preserving transformations, which we shall also discuss later.

A related idea was also used in \cite{Diaz_GarciaMartin_Kazi_Larocca_Cerezo_2023,West_Dowling_Southwell_Sevior_Usman_Modi_Quella_2025} to describe dynamical properties of sets $\pgens$ of Pauli strings.
Namely, they proposed the idea of \emph{commutator graph}, where the vertices are all Pauli strings $P\in\PP_n$ and where two Pauli strings $P,Q$ are connected if and only if there is a generator $G\in\pgens$ such that $\comm{G}{P} \simeq Q$ (see Fig.~\ref{fig:commutator-and-frustration-graphs}(b) for an example).
Hence, by viewing the time evolution operator as a series of nested commutators, such an object gives information on the time evolution of Pauli strings.
It was also shown in \cite{Diaz_GarciaMartin_Kazi_Larocca_Cerezo_2023} that knowledge of the connected components of this graph provides information about the \emph{square commutant} and the operator invariant subspaces.
Hence, reachability for Pauli strings also provides information of arbitrary observables or states.
This was used to address questions of trainability in the context of barren plateaus \cite{Diaz_GarciaMartin_Kazi_Larocca_Cerezo_2023}, as well as for the computation of correlation functions and frame potentials in chaotic systems \cite{West_Dowling_Southwell_Sevior_Usman_Modi_Quella_2025}.

\subsection{Pauli Orbits and Transvection Groups}\label{sec:overview:orbits:transvection:groups}

Here we contribute to this field of knowledge by addressing the following issues in the Pauli setting:
\begin{enumerate}
    \item\label{task-a} an invariant-based classification of \emph{Pauli} Lie groups and algebras, their representations, together with an efficient algorithm to compute the Lie algebra (see Section~\ref{sec:classification_and_algorithm_lie_algebras});
    \item\label{task-b} classification of the Clifford subgroups which lie in Pauli Lie groups, for arbitrary \emph{connected} sets of Pauli strings;
    \item\label{task-c} classification of the connected components of \emph{commutator graphs}
     for sets of Pauli strings that are \emph{connected} with respect to their frustration graph, together with an efficient algorithm for checking if two Pauli strings are in the same connected component (see Section~\ref{sec:classification_and_algorithm_orbits}).
\end{enumerate}
We obtain these results by expanding upon the graph-theoretic framework developed by \cite{Chapman_Flammia_2020,Aguilar_Cichy_Eisert_Bittel_2024,Diaz_GarciaMartin_Kazi_Larocca_Cerezo_2023} and taking advantage of the well-known binary formalism for Pauli strings \cite{Aaronson_Gottesmann_2004}.
Namely, we realize this by making the basic but fundamental observation that certain special Clifford operations, known as \emph{Clifford transvections}, act as a unitary
version of commutators.
Explicitly, a Clifford transvection in the Clifford group is defined as $(I+\im P)/\sqrt{2}$, with $P\in\PP_n$ being its \emph{center}.
Then, conjugation by a Clifford transvection with center $P$ sends a Pauli $Q$ into itself if they commute, and sends $Q$ into $\im PQ$ if they anti commute.
On the other hand, the commutator $\frac{\im}{2}\comm{P}{\cdot}$ sends a Pauli $Q$ into $0$ if they commute, and sends $Q$ into $\im PQ$ if they anti commute.
Clifford transvections and their binary variant known as symplectic \emph{transvections} have already been studied in the context of generation of Clifford gates \cite{Koenig_Smolin_2014,Pllaha_Volanto_Tirkkonen_2021}
and are one of the simplest Clifford operations.

Given this connection, one finds a natural connection between Pauli Lie algebras and \emph{transvection groups} generated by a set of Pauli strings $\pgens$.
The Pauli basis for the Pauli Lie algebra, which consists of nested commutators, coincides with the \emph{orbits} containing its generators, with respect to the transvection group.
Indeed, this corresponds to the fact that, from the binary or Clifford point of view, nested commutators are simply nested products of transvections acting on the generating set itself.
We discuss precisely this connection in Section~\ref{sec:connection_Pauli_Lie_algebras_transvection_groups}.

\begin{table*}
    \centering
    \caption{The different formalisms used in our analysis: Pauli strings, the binary formalism and graphs. We highlight the connections between various objects in each formalism.}
    \label{tab:connections}
    \begin{tabular}{@{\hspace{1mm}}l@{\hspace{8mm}}l@{\hspace{8mm}}l@{\hspace{1mm}}}
        \hline\hline
        \\[-2.5mm]
        \textbf{Pauli Strings} & \textbf{Binary Vectors} & \textbf{Colorings on Graphs}
        \\[0.5mm] \hline
        \\[-2.0mm]
        Pauli Lie Algebras & Transvection Groups & Games on Graphs\\[0.5mm]
        Invariant Subspaces & Orbits & Orbits and Games on Colorings\\[0.5mm]
        Abelian Symmetries & Radical & Colorings with Even Neighborhood\\[0.5mm]
        Commutant & Orthogonal Complement & - \\[0.5mm]
        Invariant Bilinear Forms & Invariant Quadratic Forms & Euler Characteristic\\[0.5mm]
        Free fermions & Permutation Groups & Line graphs\\[1mm]
        \hline\hline
    \end{tabular}
\end{table*}

Similarly, the binary and transvection framework provides a natural view of the connected components of the commutator graph as orbits of this group on $\PP_n$.
Again, this follows from the fact that two Pauli strings lie in the same component if and only if there is a nested commutator of the generator which brings one into the other.
Hence, this corresponds to a nested application of transvections which brings a Pauli string into another.

Moreover, we can also make use the connection of the Clifford group to the \emph{binary} symplectic group $\Sp(2n,\F_2)$, under which the transvection groups are matrix groups over a $2n$-dimensional vector space.
Then, the tasks~\ref{task-a}-\ref{task-c} become equivalent to
\begin{enumerate}[label=(\arabic*)]
    \item the classification of \emph{orbits} in $\Fn$ by a transvection group, and specifically the orbit(s) containing the generating set and
    \item the classification of transvection groups for sets of binary vectors that are \emph{connected} with regard to their frustration graph.
\end{enumerate}
Under this connection, we also found that much of the graph-theoretic framework applied to Pauli strings has a natural correspondence to the rich literature of transvection groups \cite{McLaughlin_1969,Wajnryb_1980,Janssen_1983,humphries_1985,Brown_Humphries_1986a,Brown_Humphries_1986b,Seven_2005,Sjöstrand_2025,Cameron_Hall_1991,Shapiro_Shapiro_Vainshtein_1998,Brouwer_Cohen_Cuypers_Hall_Postma_2012}.
Namely, the concept of \emph{contraction} or \emph{lighting} used in \cite{Aguilar_Cichy_Eisert_Bittel_2024} corresponds to the $t$-Nielsen transformations used in \cite{humphries_1985} or basic moves as in \cite{Brown_Humphries_1986b}, which defines the concept of $t$-equivalence between graphs which are related via a series of contractions.
Such transformations conserve all algebraic properties of the original generating set.
In particular, assuming no algebraic dependencies in the generating set, one is able to completely describe the algebraic properties of these objects
(e.g., the orbits, the Lie algebra and transvection group) purely in terms of graph-theoretic properties of the frustration graphs of the generating sets.
In this context, a fundamental tool is that of a \emph{coloring} of the frustration graph, which provides a unique element in the \emph{matrix algebra} generated by $\pgens$ as a product of the colored elements.
Our work provides a accessible framework that is general enough for the intended applications
and we develop this in considerable detail to connect to the large body of known results and
extend these when required. We refer to
Table~\ref{tab:connections} to highlight how various notions in the language of Pauli strings, binary vectors and graphs are connected.

\subsection{The Canonical Generating Sets}\label{sec:overview:canonical:generators}

For the purpose of classification, a fundamental result is the fact that contractions, hence $t$-equivalence, conserve all algebraic properties of the original generating set.
This was used in \cite{Aguilar_Cichy_Eisert_Bittel_2024} for the classification of Pauli Lie algebras and in the transvection group literature for the classification of general orbits \cite{Brown_Humphries_1986b,Seven_2005,Shapiro_Shapiro_Vainshtein_1998} as well as the transvection groups themselves \cite{humphries_1985,Janssen_1983,Wajnryb_1980}.
A fundamental result in this direction is finding canonical representatives for each $t$-equivalence class.
A full set of such graphs was originally provided in \cite{Janssen_1983} and a distinct family was given in \cite{Aguilar_Cichy_Eisert_Bittel_2024}, which both consist of four distinct graph families.
We discuss these classes and the graph-theoretic formalism in Section~\ref{sec:graph_theoretic_formalism}.
One finds that such representative graphs are \emph{not} strictly in bijection with generating sets, where the gap between these objects is given by \emph{algebraic} or \emph{linear} dependencies (depending on the Pauli or binary formalism), which appear whenever products of a subset of the Pauli generating set provide the rest of the set (equivalently, a subset of the binary generating set provides the rest as linear combinations).
Under the addition of algebraic dependencies, it remains to understand what are the possible \emph{minimal} generating sets whose frustration graphs are the canonical ones.
In \cite{Janssen_1983} and \cite{Aguilar_Cichy_Eisert_Bittel_2024} it was shown that at most a \emph{single} algebraic dependency may appear in a given minimal generating set, and only for two of the possible graph families.
This results in six distinct families of Pauli Lie algebras or transvection groups, which are determined up to isomorphism (for a given system size $n$).
We discuss this background and the necessary formalism in Section~\ref{sec:graphs_to_lie_algebras}.

As only a finite number of classes are possible, one would like to distinguish between these classes for an arbitrary given Pauli generating set, as well as to characterize their properties in a way which is independent of the choice of generating set.
In \cite{Aguilar_Cichy_Eisert_Bittel_2024} and most literature of transvection groups \cite{Janssen_1983,humphries_1985,Brown_Humphries_1986a,Brown_Humphries_1986b,Seven_2005} this is viewed with respect whether a given minimal generating set is $t$-equivalent to one of six possible representatives.
Moreover, \cite{Aguilar_Cichy_Eisert_Bittel_2024} also mentions how this may be turned into a practical algorithm for classification, where one is able to both obtain a minimal generating set and a $t$-equivalent canonical representative.
Furthermore, given that Pauli Lie algebras and transvection groups are determined up to isomorphism, it suffices to find an arbitrary labelling which satisfies the same algebraic relations to determine the type of Lie algebra, as done in \cite{Aguilar_Cichy_Eisert_Bittel_2024}.

\subsection{The Invariants for Classification}\label{sec:overview:invariants}

However, we are interested here in approaching the problem in a way which makes use of \emph{invariants} of the given generating set, hence Lie algebra or transvection group, without the use of transformations.
We do this for three reasons: First, it provides an explicit algorithm for the tasks~\ref{task-a}-\ref{task-c}.
Second, it enables an invariant or \emph{symmetry}-based perspective to the problem, which naturally extends to the description of the orbits and is also suitable for \emph{proofs} of well structured (families of) generating sets. Finally but most importantly, our approach preserves structural properties
of multiple, related problem instances which would be hidden by switching to canonical generating set and graphs (as, e.g., in \cite{Aguilar_Cichy_Eisert_Bittel_2024}).
In Sections~\ref{sec:example:2-local_paulis}, \ref{sec:universality_and_computational_power}, \ref{sec:ffd_free_fermions_in_disguise}, we discuss various examples of certain Pauli generating sets which have appeared in the literature, thus highlighting the use of alternative criteria for identifying the possible classes, as well as their action on Pauli strings and arbitrary operators.

Namely, we identify the following fundamental invariant properties of Pauli generating sets:
\begin{enumerate}[label=(\roman*)]\label{enum:overview_invariants_classification}
    \item\label{overview_invariants_one} (Linear/Ordinary) Symmetries: the set of (linear) constants of motion, viewed as the set of operators $S$ which commute with the generating set, $SH_i - H_iS =0$ for all $i$ (see Sections~\ref{sec:pauli:matrix} and \ref{sec:commutant})
    \item\label{overview_invariants_two} Anti-unitary symmetries connected to invariant bilinear forms: matrices $B$ so that $BH_i^T + H_iB = 0$ for all $i$ (see Sections~\ref{sec:quadratic_bilinear_forms_intro} and \ref{sec:quadratic_bilinear_forms_for_pauli})
    \item\label{overview_invariants_three} Free-fermionicness: generating sets are of \emph{free-fermionic} type if they admit a \emph{generalized free-fermionic mapping}, which occurs if and only if the frustration graph is the line graph of a \emph{multigraph} (see Sections~\ref{sec:Line_Graphs_E6_Condition}, \ref{sec:Line_Graphs_Orbits_Lie_Algebras_Free_Fermions_Majoranas}, and \ref{sec:canonical_free_fermionic_structures})
    \item\label{overview_invariants_four} Quasi-universality: a generating set is of \emph{quasi-universal} type if it does \emph{not} admit a free fermionic mapping. For such generating sets the algebraic properties are completely determined by their (ordinary) symmetries and invariant bilinear forms, such as Pauli Lie algebras and transvection groups (see Section~\ref{sec:symmetry_classification_transvection_groups_pauli_lie_algebras})
\end{enumerate}

\subsection{Pauli Lie Algebras and Transvection Groups}\label{sec:overview:pauli:clifford:groups}

We summarize our first main result in the following, which complements the work of \cite{Aguilar_Cichy_Eisert_Bittel_2024} with an invariant-based perspective:
\begin{result}[Classification of Connected Pauli Lie Algebras, informal version; see Thm.~\ref{thm:full_classification_pauli_lie_algebras}]
Consider a set of Paulis $\pgens$ such that the frustration graph $\frustration{\pgens} = \graphG$ is connected.
Then, either $\graphG$ is the line graph of a multigraph and the Lie algebra can be mapped to a free-fermionic system, or the Lie algebra is quasi-universal and is the set of isometries for its constants of motion and invariant bilinear forms.
This can be decided in at most $\BigO(\max\{\abs{\graphG},2n\}^3)$-time.
\end{result}
We also have the corresponding statement for transvection groups, which we state here in the Pauli/Clifford setting:
\begin{result}[Transvection Groups Classification, informal version; see Thm.~\ref{thm:full_classification_transvection_groups}]
Consider a set of Pauli strings $\vgens$ such that the frustration graph $\frustration{\pgens} = \graphG$ is connected, with transvection-generated group $\cltvgroupempty\subseteq\cl_n$.
Then, either $\graphG$ is the line graph of a multigraph and $\tvgroupempty$ is an extension of a symmetric group (representation), or $\tvgroupempty$ is the group of isometries for the Pauli commutant of $\vgens$ and its invariant bilinear forms.
This can be decided in at most $\BigO(\max\{\abs{\graphG},2n\}^3)$-time.
\end{result}
In Section~\ref{sec:classification:groups_lie_algebras} we provide the precise statements of these results.
In particular, the properties \ref{overview_invariants_one}-\ref{overview_invariants_four} do \emph{not} provide a full set of invariants for a given Pauli generating set.
Specifically, we shall see in the free-fermionic case that three subcases exist, which require additional knowledge from the symmetries and root multigraph of the frustration graph (see Section~\ref{sec:Lie_Algebraic_Dependencies_in_general}).

To put these results in context, we note that computing the Lie algebra is not a simple task for arbitrary \emph{non-Pauli} generating sets.
Indeed, on the computational side, typical approaches involve linear algebra routines which scale with space dimension $d$ either as $\BigO(d^2)$ or $\BigO(d^3)$, such as computing commutators and checking for linear independence.
For qubit systems we have $d=2^n$, which makes exact computations prohibitive beyond a few qubits without
further assumptions on the Lie algebra or generators.
Even under assumption of sparse matrices, the exponential scaling cannot be avoided, and it becomes especially problematic when the dimension of the Lie algebra is exponential, which is the also most common setting.
Moreover, in both computational methods and analytical proofs, the method typically boils down to finding explicit sequences of commutators which can produce a basis for the Lie algebra, which may not be feasible.

Alternatively, symmetry-based methods have also been developed, which make use of the adjoint or square commutant
\cite{Zeier_2011,Zeier_2015,Zimboras_Zeier_SchulteHerbruggen_Burgarth_2015,Lastres_Pollmann_Moudgalya_2024}.
In this picture invariant operator subspaces take an important role as obstacle to universality \cite{Kazi_Larocca_Farinati_Coles_Cerezo_Zeier_2025,singh2025groundstate,
gargiulo2026obstructionsuniversalitygloballycontrolled}, which cannot be seen at the level of invariant subspaces for pure states.
Such an approach also address directly the action of a Lie algebra not only on itself (which is provided purely by the isomorphism class), but also on the entire operator space.
Hence, we see that restricting to the Pauli setting is crucial to be able to obtain more information about the given generating set.

Important aspects of our work and classification rely on results of Cuypers~\cite{Cuypers_2021_Whitney,Cuypers_2021_E6}.
Recently, he has applied his earlier work to the setting of Pauli strings and Pauli Lie algebras
and independently provides an algorithm for the classification of Pauli Lie algebras \cite{Cuypers_2026}.
We discuss similarities and differences to our classification algorithm in Section~\ref{sec:classification_and_algorithm_lie_algebras};
we had announced our results in~\cite{Natal_2026,Dresden_2026}.
Our results go beyond the classification of Pauli Lie algebras as we are also interested
in the full commutator graph, or set of orbits, corresponding to these algebra/groups, which we classify in all cases with connected frustration graph.
This provides essential practical information about the action of the Lie algebra on the Pauli strings and the entire operator space, which cannot be obtained solely from the Lie algebra.

\subsection{The Quasi-Universal Classes}\label{sec:quasi_universal_description}

We now discuss the implications of the given invariants \ref{overview_invariants_one}-\ref{overview_invariants_three}
and corresponding isomorphism classes, as well as how one might compute them, taking advantage of the binary formalism.
We start with \ref{overview_invariants_one} and \ref{overview_invariants_two}, which already determine the  \emph{quasi-universal} classes.

\begin{figure*}
    \centering
    \includegraphics{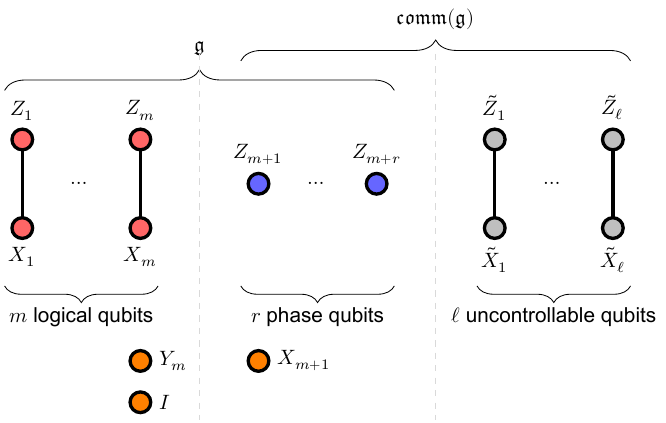}
    \caption{Up to isomorphism, the partition of the system in three parts due to symmetries: $m$ logical qubits, $r$ phase qubits, $\ell$ uncontrollable qubits.
    Also, in orange, the three possible choices of reference invariant bilinear form, $B=I$, $Y_m$, $X_{m+1}$.
    These invariant properties completely determine the \emph{quasi-universal} cases.}
    \label{fig:partition-commutant}
\end{figure*}

It is straightforward to see that the Pauli linear symmetries \ref{overview_invariants_one} correspond to the vectors which lie in the commutant of the given generating set. This corresponds to the \emph{orthogonal complement} in the binary picture, while
the so-called \emph{radical} plays an especially important role (see Section~\ref{sec:commutant}).
Explicitly, two Paulis commute $\comm{P}{Q} = 0$ iff they are orthogonal under the symplectic product $\symp{v}{w} = 0$, and anti-commute $\acomm{P}{Q} = 0$ otherwise, $\symp{v}{w} = 1$.
Hence, such spaces admit a simple classification up to isomorphism in terms of the dimension of its maximal non-degenerate subspace and the dimension of the radical.
Up to isomorphism, the symmetries determine a natural tripartition of the $n$ qubits into three sets (see Fig.~\ref{fig:partition-commutant}): $m$ \emph{logical} qubits, over which the system has a large (if not universal) control; $r$ \emph{phase} qubits, over which the system may only affect the phases in the computational basis; $\ell$ \emph{uncontrollable} qubits, which are left completely unaffected by the system.

The presence of phase qubits is due to the presence of \emph{abelian} symmetries.
Moreover, compared to \cite{Aguilar_Cichy_Eisert_Bittel_2024}, we also take into account the missing uncontrollable qubits.
These result from \emph{non-abelian} symmetries which are not generated by the given generating set.
Restricted \emph{only} by (linear) symmetries, one has full control over the logical qubits (hence acts as $\su(2^m)$ with $m$ the number of logical qubits), or equivalently can implement \emph{strictly} universal quantum computation.
Also, one has \emph{full} control over the phases of the phase qubits, which still realizes non-trivial quantum computation, and corresponds to independent copies of $\su(2^m)$ for each Z-diagonal configuration for the phase qubits, hence $\su(2^m)^{\oplus 2^r}$.

For \ref{overview_invariants_two}, we make an explicit connection between \emph{Pauli} bilinear forms, which are either symmetric or skew-symmetric, and binary \emph{quadratic} forms.
Additionally, we connect \emph{sets} of invariant $\F_2$-quadratic forms over the full non-degenerate space $\Fn$ to the \emph{unique} invariant $\F_2$-quadratic form over the $\F_2$-subspace spanned by the generating set.
Furthermore, we also provide a simple framework for identifying invariant Pauli bilinear forms, which can also be turned into an efficient algorithm.
Using a classical result of the classification of quadratic forms over $\F_2$, we show that spaces of Pauli bilinear forms come in three distinct isomorphism classes.
Equivalently, with fixed commutant in its canonical form, we can choose a \emph{reference} Pauli bilinear form among three: $I$, $Y_m$ and $X_{m+1}$.
The first two of these correspond to the well-known orthogonal $\so(2^m)$ ($B=I$) and symplectic $\usp(2^m)$ Lie algebras ($B=Y_m$), which also appear independently in each block associated to one of the $2^r$ phase qubit configurations.
The third is the unitary group $\su(2^m)$ ($B=X_{m+1}$), but with a distinct representation, which may be viewed as an intersection of the orthogonal and symplectic groups.
Indeed, a symmetric matrix $S$ in $\su(2^m)$ acts as $S\otimes Z_1$ over the first phase qubit, whereas a skew-symmetric matrix $A$ acts as $A\otimes I$.
Hence, the elements of this Lie algebra act in a correlated manner depending on whether the first phase qubit is in the $\ket{\up}$ or $\ket{\down}$ state, leading to only $2^{r-1}$ copies for the remaining phase qubits.

Under the presence of a bilinear form the system has additional constraints.
For instance, if the initial state is $B$-symmetric, $B\rho^T + \rho B = 0$, such invariance is conserved under time evolution.
Moreover, now the control of the logical and phase qubits are not independent of each other.
For instance, if $\rho = \rho_- \otimes \rho_+ \otimes \rho_0$ is $B$-skew-symmetric over the logical qubits but $B$-symmetric over the phase qubits, any transformation which makes the logical state symmetric must also make the phase state skew-symmetric, $\rho' = \rho_+'\otimes\rho_-'\otimes \rho_0$, since the global $B$-(skew-)symmetry must be maintained.
Finally, we also highlight that such objects are equivalent to anti-unitary symmetries, which have found extensive applications in many-body physics in the setting of classification of phases of matter \cite{Altland_Zirnbauer_1997,Ryu_Schnyder_Furusaki_Ludwig_2010,Uhlmann_2016} as well as for spectral constraints of Hamiltonians \cite{Wigner1932}.

\subsection{The Free-Fermionic Classes}\label{sec:overview:free:fermions}

Contrasting free-fermionicness and quasi-universality, the case~\ref{overview_invariants_three} generalizes the work of \cite{Chapman_Flammia_2020} (as well as \cite{Ruh_Elman_2025}) to take into account all symmetries in a free-fermionic mapping.
This is particularly important in a dynamical setting, since it allows us to completely characterize the Pauli Lie algebra, transvection group and their actions on the entire operator space.
In particular, we avoid projections which may lose details of the original generators, including possible algebraic dependencies.

Explicitly, in the free-fermionic cases we find instead three families: the \emph{even} type, which consists of quadratic Majoranas, up to symmetry; the \emph{odd} type, which consists of quadratic and linear Majoranas, up to symmetry; the \emph{exceptional} type, which consists of quadratic and (complement) linear Majoranas as well as the parity symmetry.
Moreover, in the Majorana description we have a related description of the modes as partitioned in three sets (see also Fig.~\ref{fig:partition-Majorana}): a set of $2m$ \emph{logical} fermionic modes, over which the system has a certain degree of control; a set of $2q$ \emph{phase} fermionic modes, over which the system may only affect the phases in the computational/occupation basis; $2\ell$ \emph{uncontrollable} modes, which are left completely unaffected by the system.

Namely, from the strictly quadratic free-fermionic point of view all cycle symmetries $\prod_{ij\in C}\gamma_i\gamma_j = I$ are trivial, whereas we allow for non-trivial cycle symmetries by adding associated \emph{phase} or occupation number operators, $\prod_{ij\in C}P_{ij} = \tilde{\gamma}_i\tilde{\gamma}_j$.
The natural framework for identifying such `free-fermionicness' in our setting is that of line-graphs of \emph{multigraphs}.
These generalize to regular line graphs and also admit a forbidden subgraph description in terms of a class of $32$ graphs \cite{Seven_2005,Cuypers_2021_E6}, which consists of the $t$-equivalent graphs to a certain $6$-vertex graph, denoted as $\graphE_6$.
We refer to this test as the $\calE_6$-criterion or condition, to distinguish between the specific forbidden subgraph $\graphE_6$ and the full class of $t$-equivalent graphs, denote as $\calE_6$.
Hence, in this setting a graph is the line-graph of a multigraph if and only if it is $\calE_6$-free. Refer to
Fig.~\ref{fig:e6-class-atlas} for the full list of six-vertex graphs contained in $\calE_6$ which includes the graph $\graphE_6$.
This provides a particularly convenient local criterion to test for free-fermionicness in structured sets, since one needs to only consider induced subgraphs of the frustration graph.
This can be often decided visually and provides a convenient avenue for proofs.
Line graphs of multigraphs (following the definition in \cite{Cuypers_2021_E6}) is in fact equivalent to being a regular line graph \emph{up to twins}, a fact which was mentioned in \cite{Chapman_Flammia_2020} and \cite{Ruh_Elman_2025}, but not fully explored in the context of line graphs of multigraphs, as well as $t$-equivalence and Lie algebras.
We also highlight that being free fermionic in our setting then does \emph{not} necessarily imply exact (in the sense of efficient) solvability and specifically does not imply \emph{polynomially} sized Lie algebras.
Indeed, the sole presence of a large amount of (abelian) symmetries may imply exponentially sized Lie algebras even for `free-fermionic' Lie algebras, which in an irreducible block may act as the usual quadratic Majoranas.
The most extreme example of these are the uniformly controlled single qubit gates, which consist of purely two Majorana modes $\gamma_1$, $\gamma_2$ together with $n{-}1$ phase operators, which control the phase in the occupation number basis.
We treat this explicitly in Section~\ref{sec:uniformly_controlled_many_symmetries}.

\subsection{The Pauli Orbits}\label{sec:overview:orbits}

In Sections~\ref{sec:classification:groups_lie_algebras} and \ref{sec:classification:orbits} we shall use these graph-theoretic (line graphs, free-fermionic mappings) and symmetry (constants of motion, bilinear/quadratic forms) tools to describe first the transvection groups and Pauli Lie algebras, and later the resulting orbits in each case.
Specifically, we find the following:

\begin{result}[Classification of Orbits, informal version; see Sec.~\ref{sec:classification:orbits} and Cor.~\ref{cor:algorithm_orbits}]
\label{thm:informal_full_classification_orbits}
Consider a set of Paulis $\pgens$ such that the frustration graph $\frustration{\pgens} = \graphG$ is connected.
There exists a compact description of the orbits in terms of properties of $\pgens$: if $\pgens$ is of quasi-universal type, this is determined by its constants of motion and invariant bilinear forms; if $\pgens$ is free-fermionic, it also requires knowledge of the length of the Majorana strings with respect to the free fermionic mapping.
This description can be computed in at most $\BigO(\max\{\abs{\graphG},2n\}^3)$-time, and for any Pauli $P\in\Pgroup_n$. Also, it takes $\BigO(\max\{\abs{\graphG},2n\}^3)$-time to decide whether two Pauli strings $P,P'$ are in the same orbit.
\end{result}

In the quasi-universal case, we find \emph{large} unitary, orthogonal, or symplectic groups with their standard (or dual) representations.
In these cases, each orbit is always exponential in the number of logical \emph{and} phase qubits $\BigO(2^{2m+r})$, with a number of orbits which scales as $\BigO(2^{2\ell+r})$ (see Section~\ref{sec:Orbits_Full_Space_Fn_Quasi_Universal_Case}).

On the other hand, for free-fermionic cases the orbit sizes can vary significantly depending on whether the given Majorana string $\majiso{v}$ acting on gates
commutes with the $r$ number or phase operators or not.
If $\majiso{v}$ commutes with the phase operators, then the orbit is `essentially' free-fermionic, since its size scales depending on its length $\BigO(2^r\binom{2m}{L})$, and there are $\BigO(2^{2\ell+r})$ such orbits.
For $L\approx m$ we also have quasi-exponential scaling in the number of logical qubits, due to the asymptotic behaviour $\binom{2m}{L} \approx 2^m/\sqrt{m}$.
If instead $\majiso{v}$ does not commute with the phase operators, then \emph{every} orbit scales as $\BigO(2^{2m+r})$, in a way which again depends purely on the symmetries and invariant bilinear forms.
This points to a surprising result on the computational power of free-fermionic circuits in the presence of abelian symmetries, beyond the parity symmetry.
Indeed, in such cases one does not need large Majorana lengths to obtain an exponential scaling in the available space of a Majorana string under time evolution.
In particular, from the symmetry point of view, these orbits and invariant subspaces are precisely the same as those of the quasi-universal case.

Viceversa, in the context of classical simulability for dynamics generated by (connected) Pauli strings, we also have a full characterization of when a Pauli string evolves in a polynomially large space.
Namely, this only happens in the free-fermionic case when the corresponding Majorana string commutes with the phase operators and has finite (or very large) Majorana length $L = \BigO(1)$ (or $2m-L = \BigO(1)$), over the first $2m$ modes.

Furthermore, the dimension of the orbits scales for all cases  exponentially in the number of abelian symmetries $\BigO(2^r)$, independently of the size of logical qubits.
Again, this points to the fact already commuting gates are capable of non-trivial quantum computation, since they are able to explore a large portion of the operator space.
Notably, in cases for which $r$ scales with system size, the Lie algebra for connected frustration graphs will scale exponentially with system size.
However, we shall see examples where in fact all non-trivial Pauli orbits are of exponential size, despite the fact that the Lie algebra itself is of polynomial size and indeed with \emph{commuting} generators (Section~\ref{sec:IQP}).
At the same time, the invariant subspaces remain trivial (one-dimensional) highlighting the difference between the behaviour of arbitrary observables and more structured observables such as Pauli strings.
We also discuss the general relation between Pauli orbits and invariant subspaces (with partial characterization) in Section~\ref{sec:invariant_subspaces}.

\section{Overview of Examples and Applications}\label{sec:overview_phys}

Complementing the overview of our framework
in Section~\ref{sec:overview_math}, we now provide an overview
tailored to applications. Section~\ref{sec:overview:ex:vqa}
focusses on dynamical systems and variational quantum algorithms
and Section~\ref{sec:overview:ex:computational:power} discusses computational
power and classical simulability. Symmetries of many-body systems
are considered in Section~\ref{sec:overview:many:body}
and Section~\ref{sec:overview:random_clifford_designs} treats
applications for random circuits and $t$-designs.

\subsection{Lie-theory for Variational Quantum Computing}\label{sec:overview:ex:vqa}

As mentioned, a Lie-theoretic approach can provide much information about the dynamical system one is considering.
In recent years, such an approach has been particularly productive in the analysis of \emph{variational quantum algorithms}, where ansatz design plays a fundamental role in terms of gate set choice and parameter initialization and update.
For instance, it has been shown that many architectures suffer from \emph{barren plateaus} \cite{Fontana_Rudolph_Duncan_Rungger_Cîrstoiu_2025,Diaz_GarciaMartin_Kazi_Larocca_Cerezo_2023,Ragone_2024}, which
result from exponentially suppressed gradients as well as local minima in the cost landscape \cite{Bittel_Kliesch_2021,Anschuetz_Kiani_2022}, which both hinder optimization.
Specifically, one common setting is that of parametrized circuits of the form
\begin{equation}
    U(\boldsymbol{\theta}) = \prod_{i=1}^L e^{\im\theta_i H_i}
\end{equation}
where $H_i$ is drawn from a certain set of generators $\pgens$.
In \cite{Ragone_2024,Fontana_Rudolph_Duncan_Rungger_Cîrstoiu_2025} it was shown that for deep circuits, the behavior of the cost function
depends on the Lie-algebraic properties, including how the cost function and initial state relate to the Lie algebra.
In particular, assuming the cost function or initial state in the Lie algebra, one consequence is that there is a barren plateau whenever the dimension of the Lie algebra scales exponentially with system size (or more precisely, when this is a \emph{simple} Lie algebra).
Specifically, this holds whenever $U(\boldsymbol{\theta})$ forms a unitary 2-design for its Lie group of transformations, i.e.,
it appears as a Haar random unitary over the Lie group generated by $\pgens$.
Such results have spurred interest in the computation of Lie algebras for many known ans\"{a}tze in common algorithms, such as: various ans\"{a}tze for QAOA \cite{Kazi_Larocca_Farinati_Coles_Cerezo_Zeier_2025,tsvelikhovskiy2025provableavoidancebarrenplateaus,Mao_Yuan_Allcock_Zhang_2025,allcock2026dynamicalliealgebrasquantum}, hardware efficient ans\"{a}tze using 2-local Pauli strings \cite{Wiersema_Kokcu_Kemper_Bakalov_2024,Kokcu_Wiersema_Kemper_Bakalov_2024}, Hamming-weight preserving \cite{monbroussou2025trainability,yan2026universal2localsymmetrypreservingquantum,stergiou2026universalityquantumgatesparticle} and XY interactions \cite{Kordonowy_Leipold_2026}, permutation-invariant circuits \cite{mansky2023permutationinvariantquantumcircuits,Kazi_2024} and many more.

Beyond variational quantum algorithms, such problems emerge also in Optimal Quantum Control \cite{kirk2004optimal,jurdjevic1972controllability,agrachev2013control,polack2009uncontrollable,dalessandro2022,pechen2011}.
Here the problem is typically formulated in a continuous setting, in terms of a drift Hamiltonian and various controls, i.e. as a bilinear control problem \cite{elliott2009bilinear}:
\begin{equation}
    H(t) = H_0 + \sum_k c_k(t) H_k
\end{equation}
where the $c_k(t)$ are control fields or pulses.
Indeed, close connections have been made between optimal quantum control problems and variational quantum algorithms \cite{Magann_Arenz_2021,meitei2021gatefreestatepreparationfast,Yang_Rahmani_2017,Tao_Wang_Wu_2026,wiedmann2025convergencevariationalquantumeigensolver}.
Here, Lie-theoretic tools have been used extensively to address questions of controllability for state transfer or gate synthesis \cite{polack2009uncontrollable,kirk2004optimal}.
Namely, for unitary dynamics, Lie-theoretic knowledge can provide necessary and sufficient conditions for the existence of pulses (or parameters) that realize specific target gates \cite{polack2009uncontrollable}, as well as convergence guarantees \cite{wiedmann2025convergencevariationalquantumeigensolver}.

Given the relevance of understanding and identifying Lie algebras in a variational setting over a variety of ans\"{a}tze, our framework
can help addressing this problem for Pauli-string generators.
In particular, by Theorem~\ref{thm:full_classification_pauli_lie_algebras}, this can be performed efficiently using linear algebra over $\F_2$ and as well as some combinatorial procedures over the frustration graph.
This finds the Lie algebra representation for all connected cases, as well as the Lie algebra up to isomorphism for all generating sets.

We showcase this in Section~\ref{sec:example:2-local_paulis} by looking at the class of 2-local Pauli strings on some connected \emph{interaction} graph, as discussed in \cite{Wiersema_Kokcu_Kemper_Bakalov_2024,Kokcu_Wiersema_Kemper_Bakalov_2024}.
These are specified by some interactions on pairs of edges as well as an interaction graph which dictates which qubits interact.
Specifically, we focus on two examples: the Ising interaction, or equivalently the multi-angle QAOA ansatz \cite{Kazi_Larocca_Farinati_Coles_Cerezo_Zeier_2025},
which consists of X-rotations on all vertices and ZZ-interactions; arbitrary anisotropic XY interactions, i.e. XX and YY interactions on the edges of some connected graph.

In both cases simple rules provide the frustration graph starting from the interaction graph, given that two 2-local Paulis may anti-commute only if they share at least one vertex.
Then, we can use the $\calE_6$-condition to determine whether a given generating set is of free-fermionic or quasi-universal type.
Thus we obtain a clear divide between free-fermionic generating sets and quasi-universal sets. The free-fermionic ones
appear only when the interaction graph is a path or a cycle (or one-dimensional with open and periodic boundary conditions),
while quasi-universal ones appear if and only if there is at least one vertex of degree three in the interaction graph.
In the free-fermionic cases we test for symmetries and algebraic dependencies, which can be done easily given that the frustration graphs are also paths and cycles. This allows to distinguish between the different free-fermionic cases.

In order to classify the quasi-universal cases instead we need the commutant and invariant bilinear forms.
We devise in Section~\ref{sec:from_local_to_global} a useful tool which allows to construct these symmetries starting purely from the local commutant and invariant bilinear forms, which follows from the fact that symmetries on the whole system, when restricted to an edge, must in fact still be a symmetry for the local generators.
Then, using the classification results for the commutant and invariant bilinear forms, we are able to also classify the quasi-universal cases, thereby re-proving in a natural way the results of \cite{Wiersema_Kokcu_Kemper_Bakalov_2024,Kokcu_Wiersema_Kemper_Bakalov_2024}.
Specifically, we find that the structure of the commutant and bilinear forms depends purely on whether the interaction graph is bipartite or not, or equivalently whether the graph has no loops of odd length.
Hence, combined with the quasi-universal statement, we find precisely the same distinction found in \cite{Kokcu_Wiersema_Kemper_Bakalov_2024}.
However, we can avoid explicit arguments based on the interaction graph by directly proving that any graph in the same class generates the same Lie algebra.

Moreover, in \cite{Diaz_GarciaMartin_Kazi_Larocca_Cerezo_2023} the Lie-algebraic framework for barren plateaus was extended, in a free-fermionic example, to also include the case where the state or observable are chosen arbitrarily.
However, for most of the above mentioned cases, even with a given Lie algebra, a full invariant subspace decomposition has not been discussed, which potentially limits the applicability \cite{Shen_Pielawa_Dunjko_Wang_2026}.
We contribute to this more general case by providing the description of Pauli orbits or checking for orbit intersection, which can also be done efficiently, as stated in Theorem~\ref{thm:full_classification_pauli_lie_algebras}.
Also, we provide a decomposition into invariant subspaces, which may not be irreducible, in Section~\ref{sec:invariant_subspaces}.
There we define a symmetry-adapted basis for the subspaces, which uses both information from the orbits as well as from the commutant.

\subsection{Universality, Computational Power and Classical Simulability}\label{sec:overview:ex:computational:power}

Beyond variational algorithms and optimal control, the question of controllability may be framed as understanding universality or the computational power of a given architecture.
Namely, common questions regarding the computational power of a certain architecture are the following: (1) which architectures realize universal quantum computation? (2) which architectures realize \emph{encoded} universal quantum computation? (3) which architectures realize non-trivial \emph{restricted} models of quantum computation?
These questions are relevant for algorithm design such as constructing circuits out of some given gate set or
for variational quantum algorithms or physical constraints, where the native interactions of the systems limit the realizable circuits for a particular platform.

A known result in the context of universality is the fact that single qubit gates plus any entangling gate realize universal quantum computation on two qubits, hence also on any connected topology over $n$ qubits \cite{DiVincenzo_1995}, which is the foundation for most modern architectures.
Also, similarly Clifford+T is universal, with Clifford gates generated by CNOTs, Hadamard and Phase gates \cite{Bravyi_2005}.
More generally, most known universal gate sets involve \emph{local} gates \cite{Childs_Leung_Mančinska_Ozols,DiVincenzo_1995,Barenco_1995,Kempe_Whaley_2002}, and much effort has gone into understanding the Lie algebras of 2-local generating sets under restricted classes of generators, such as Pauli strings  \cite{Wiersema_Kokcu_Kemper_Bakalov_2024,Kokcu_Wiersema_Kemper_Bakalov_2024}, specific interactions \cite{Kordonowy_Leipold_2026} or in the presence of symmetries \cite{Marvian_2022,Marvian_2024,Kazi_2024}.

Another setting which has received much attention in recent years is that of universality under \emph{global controls} \cite{benjamin_2000,benjamin_2001,benjamin_2003,Benjamin_Bose_2004,LloydQAOA,hu2025universal,morales2020universality,gargiulo2026obstructionsuniversalitygloballycontrolled} in both analog \cite{kornjača2024largescalequantumreservoirlearning,Calliari2026} and digital \cite{menta2024globally,cioni2024conveyorbelt,aiudi2026} settings, possibly aided by few local controls.
This also overlaps with variational algorithms whenever the circuit ansatz involves only global controls, such as standard QAOA \cite{Kazi_Larocca_Farinati_Coles_Cerezo_Zeier_2025,Mao_Yuan_Allcock_Zhang_2025}, as well as optimal control \cite{Zeier_2011,Zimboras_Zeier_SchulteHerbruggen_Burgarth_2015}.
Many other settings exist to obtain universality, which produce less traditional native generating sets \cite{Smith_Klaver_Nautrup_Lechner_Briegel_2025,Van_den_Nest_2006,Briegel_2009}.

One such example is the model of computation known as Parity Quantum Computing \cite{Lechner_2015,Fellner_2022}, which we consider in Section~\ref{sec:parity_computing}.
Their corresponding generating set consists of single qubit rotations (which we can take as Xs and Zs) plus Z-strings on some set of vertices.
In this setting, universality was considered in \cite{Smith_Klaver_Nautrup_Lechner_Briegel_2025}, where they provide sufficient conditions for the universality.
This relies on \emph{breaking} a certain invariant of single qubits, namely, the Pauli weight of a Pauli string or the number of qubits on which it is non-zero modulo two.
Here we show that this invariant is precisely an invariant quadratic form, and we use our framework to provide necessary and sufficient conditions for universality.
Indeed, one finds that for any non-trivial choice of Z-strings, the generating set is quasi-universal, and then the necessary condition for strict universality is the breaking of the invariant quadratic form, which can be done whenever there is a Z-string of even weight.
Hence, when the number of qubits $n$ is even, precisely one Z-string over the entire chain suffices to obtain universality, whereas if $n$ is odd, one obtains only the orthogonal or symplectic group of gates with a single global Z-string.
Hence, two Z-strings, of which at least one is even, are required to obtain universality in such a platform.
Independently, this example was also discussed in \cite{Cuypers_2026} along similar lines, while
we focus on a formulation in the Pauli language.

Related questions of universality and computational power have also been discussed in infinite-dimensional systems, i.e., bosonic systems \cite{Lloyd_1999,Chabaud2026bosonicquantum}.
Here the Weyl algebra stands as an infinite-dimensional analogue of the Pauli strings, as it contains arbitrary monomials of bosonic creation and annihilaiton operators (instead of monomials in Paulis).
In particular, in the context of Lie-theory and factorization of dynamics, much effort has gone into characterizing
finite-dimensional Lie algebras \cite{Edward_Bruschi_2024,Heib_2025},
with criteria heavily based on how the Lie algebra interfaces with the Weyl algebra and (real) symplectic geometry.

Beyond \emph{strict} universality, one can perform universal quantum computation on a restricted set of $n'$ qubits in an $n\geq n'$ qubit system (or even $n$ qudit system, with additional \emph{auxiliary} levels), which leads to the concept of encoded universal quantum computation \cite{Lidar_2000,Kempe_Bacon_DiVincenzo_Whaley_2001,Kempe_Whaley_2002}.
For instance, this has been used in the context of exchange-only qubits \cite{Sala_2017} in semiconducting platforms.
Here the logical subspace over $n'$ qubits is a constrained subspace of the full physical space over $n$ spin-1/2 spins, which is also protected agains certain forms of decoherence \cite{Kempe_Bacon_Lidar_Whaley_2001}, and universality on this subspace can be achieved purely by the exchange interaction \cite{Kempe_Bacon_DiVincenzo_Whaley_2001}.

Another setting where encoded universality appears is measurement-based quantum computation (MBQC)
\cite{raussendorfOneWayQuantumComputer2001}, which realizes quantum computation by means of adaptive single-qubit measurements on an entangled many-qubit system. MBQC encodes $n'$ logical qubits into $n$ physical qubits, with $n' \ll n$, and furthermore the Lie group of gates that MBQC can implement is generally universal on only $n'' < n'$ logical qubits. The Lie group for a particular MBQC scheme is determined by the pre-measurement state and is often generated by Paulis. This Lie group has also been shown to be an invariant of certain subsystem-symmetry-protected and subsystem-symmetry-enriched topological phases, in the sense that every state in the same phase can be used by MBQC to implement the same group of gates \cite{miyakeQuantumComputationEdge2010, darmawanMeasurementbasedQuantumComputation2012a, elseSymmetryProtectedPhasesMeasurementBased2012, raussendorfSymmetryprotectedTopologicalPhases2017, stephenComputationalPowerSymmetryProtected2017, devakulUniversalQuantumComputation2018, raussendorfComputationallyUniversalPhase2019, stephenSubsystemSymmetriesQuantum2019, danielComputationalUniversalitySymmetryprotected2020, raussendorfMeasurementbasedQuantumComputation2023a, Herringer_2025, yangMeasurementbasedQuantumComputation2026}.

Even if an architecture does not realize universal quantum computation, it may still not be classically simulable.
Moreover, even if the architecture is simulable for certain input states and fixed measurement basis, it may become non-simulable for alternative input states or measurements, while remaining efficiently implementable on a given device.
Several examples are known in the literature: Clifford or stabilizer circuits \cite{Veitch_Hamed_Mousavian_Gottesman_Emerson_2014}, diagonal or commuting circuits \cite{Shepherd_2009}, matchgate circuits \cite{Hebenstreit_2020}, the one clean qubit model \cite{Knill_1998}, (noisy) shallow circuits \cite{Bravyi_2020}, and many more \cite{Aaronson_2011,Bravyi_2008}.
For such restricted models it becomes essential to understand which circuits are classically simulable (in a strong or weak sense) and which ones lead to non-trivial quantum computation, which strongly depends on the gate set, the input state and the measurement basis.
In this context, \emph{resource theory} \cite{Chitambar_2019} provides a useful framework in understanding the computational power of certain models of computation.
Here one can identify \emph{free states} and \emph{free operations} in those cases where classical simulation is efficient, and \emph{resourceful} states and operations where it is not, as has largely been explored in the Clifford \cite{Veitch_Hamed_Mousavian_Gottesman_Emerson_2014,Howard_2017} and matchgates/free-fermions setting \cite{Hebenstreit_2020,Projansky_Necaise_Whitfield_2025,Oszmaniec_Dangniam_Morales_Zimboras_2022,Sierant_2026}.

Interestingly, it was also shown in \cite{Cerezo_2025} that classical simulability and trainability of a parametrized architecture are intimately related, providing a form of no-free lunch theorem where an architecture cannot be both trainable and computationally non-trivial (i.e.\ not classical simulable).
These arguments based on randomly initialized circuits need to be contrasted with smart initialization techniques \cite{larocca2024review,blekos2024,zhou2020quantum,FarhiQuantum2022,boulebnane2021,Basso2022TQC}.
Hence, it is essential for such contexts to understand when a given circuit can be classically simulated or produces non-trivial computational power.

Then, it is clear that over arbitrary Pauli generating sets, we produce in general only \emph{restricted quantum computation}, though potentially this can be reduced to encoded quantum computation.
In the context of computational power, we will use the Pauli and Clifford perspective on these groups to highlight where classical simulability can appear and necessary requirements to avoid this.
We discuss in depth the implications for classical simulability in Section~\ref{sec:classical_simulability}.
In particular, we discuss this from the point of view of existing frameworks: $\lieg$-sim, which is based on Lie algebras and invariant subspaces; Pauli propagation, which is based on the Pauli orbits; the stabilizer formalism, which is based on the stabilizer orbits.

Moreover, we apply our framework on various examples of restricted quantum computation, and highlight the differences between an arbitrary symmetry analysis, based on Lie algebras and invariant subspaces, and a restricted Pauli analysis, based on Pauli orbits.

The first family is that of uniformly controlled single qubit gates \cite{Bergholm_Vartiainen_Mottonen_Salomaa_2005}, which typically appear as a subroutine in many quantum algorithms.
This appears naturally as the connected case with maximal number of abelian symmetries, up to isomorphism, and results in the Lie algebra $\su(2)^{\oplus 2^{n-1}}$.
From the Pauli point of view, this admits a minimal generating set of size $n+1$, and all its non-trivial orbits are of size $3\cdot 2^{n-1}$.
Hence, even if there is only a single logical qubit, the large amount of symmetries may still implement non-trivial quantum computation, which highlights the importance of abelian symmetries in computational power.
A similar treatment can also be given to the family of One Clean Qubit (DQC1) models of computation \cite{Knill_1998,Myers_Fahmy_Glaser_Marx_2001,Ambainis_Schulman_Vazirani_2006,Fahmy_Marx_Bermel_Glaser_2008}, which consists of singly-controlled multi-qubit gates where the control qubit is initialized in $\ket{+}$ and the remaining qubits in the maximally mixed state.
Hence, the initial state is described as a mixed state by a single Pauli, which lives in an orbit of exponential size.

Given the importance of symmetries, the second we family we consider is that of diagonal gates, or Instantaneous Quantum Polynomial (IQP) family \cite{Shepherd_2009}, which consists of a polynomial depth of diagonal gates applied to an initial $\ket{+}^{\otimes n}$ state and the X-measurement basis.
Here there there is no logical qubit, and all qubits function as phase qubits.
Nevertheless, such an architecture has been shown to solve problems beyond the power of classical computers \cite{Bremner_2010,Bremner_2017,Nakata_2014,buzet2026iqpcircuits2forrelation} (under reasonable complexity assumptions).
We mostly focus on a 2-local IQP ansatz, consisting of single Zs on all qubits and ZZ-interactions over some interaction graph.
From the point of view of the Lie algebra, these generating sets are trivial, given that each gate contributes a distinct abelian term, $\lieg = \lieu(1)^{\oplus \abs{\pgens}}$.
Hence, it is essential in this case to examine the action of this Lie algebra on the entire matrix space, and specifically on the Pauli strings themselves.
From the point of view of (adjoint) symmetries and Pauli orbits, which are purely determined by the diagonal Clifford group, we find that a 2-local generating set with all-to-all interactions produces the same properties as the full set of diagonal gates.
In particular, all Pauli orbits are either trivial (i.e. the invariant Z-strings) or lie in an orbit of size $2^n$, such that the only conserved quantity in an orbit is the X-part of a Pauli string.
However, the invariant subspaces are always one-dimensional, in both state and operator space.
This is to be compared with the fact that the corresponding Lie algebra generated by 2-local Paulis is only polynomial, $\lieg = \lieu(1)^{\oplus n(n+1)/2}$, whereas the full Lie algebra of diagonal gates has exponential size $\lieu(1)^{\oplus 2^n}$.

\subsection{Symmetries in Many-Body Systems}\label{sec:overview:many:body}

Symmetries play a fundamental role in physics, as a framework for understand phases of matter and phase transitions, conserved quantities in dynamics, chaos and thermalization, and more \cite{Girvin_Yang_2019,Ryu_Schnyder_Furusaki_Ludwig_2010,Altland_Simons_2010,Parameswaran_2018}.
A particularly useful framework for understand symmetries in large classes of models has been that of bond algebras \cite{Nussinov_2009,Cobanera_Ortiz_Nussinov_2010,Cobanera01102011} and their corresponding dual commutant algebras \cite{Moudgalya_2023,Moudgalya_Motrunich_2024,Moudgalya_Motrunich_2022}, in particular in the context of non-standard symmetries, as in the case of many body scars.
Namely, for a given family of Hamiltonians, given as a sum of local terms $H=\sum_{i=1}^N c_ih_i$, one is interested in the matrix algebra and its commutant
\begin{equation}
    \matalg = \algclosure{\{h_i\}_{i=1}^N},\qquad \commalg = \commutant(\{h_i\}_{i=1}^N)
\end{equation}
which are dual to each other via von Neumann's double commutant theorem as $\commutant(\commalg) = \matalg$.
Also, knowledge of these algebras provides information regarding how the Hilbert space partitions into invariant subspaces.
From the point of view of time-independent Hamiltonians, this gives information about degeneracies and general relations in the spectrum of Hamiltonians \cite{Itzykson_1980,hamermesh2012group}, which influence correlation functions and response functions.

Namely, for local generators given by Pauli strings $h_i\in\PP_n$, both the bond and commutant algebras are easily described in terms of Pauli matrix algebras (as defined in Section~\ref{sec:pauli:matrix}).
As mentioned in Section~\ref{sec:quasi_universal_description}, such algebras are completely classified, up to isomorphism, in terms of a tripartition of the system into $m$ logical qubits, $r$ phase qubits and $\ell$ uncontrollable qubits.
Then, under this partition, there is natural Hilbert space structure corresponding to a given bond algebra generated by Pauli strings, i.e.,
\begin{align}
        \HS &\cong \HS_{\rm phys} \otimes \HS_{\rm aux},\nonumber\\
        \HS_{\rm phys} &\cong \HS_L^{\oplus 2^r},\, \HS_L = (\C^2)^{\otimes m} \label{eq:Hilbert_space_partition},\\
        \HS_{\rm aux} &\cong (\C^2)^{\otimes \ell}, \nonumber
\end{align}
where $\HS_L$ is the logical subspace, of which there are $2^r$ copies depending on the state of the phase qubits, and $\HS_{\rm aux}$ is the uncontrollable or \emph{auxiliary} subspace.
Refer to Section~\ref{sec:pauli:matrix} for a precise definition of the invariant subspace decomposition of the state space and the block decomposition of matrices in Pauli matrix algebras.
Also note that the phase qubits correspond to the abelian symmetries, which lie in the center of both algebras $\ZZ(\matalg) = \ZZ(\commalg) = \Span{\{Z_{m+i}\}_{i=1}^r}$.
On the other hand, non-abelian symmetries are supported on the uncontrollable qubits, and the logical qubits correspond to the non-abelian part of the bond algebra.
Hence, we see that there is a complete description of the commutant which consists of the on-site discrete symmetries $P = \bigotimes_i P_i\in\PP_n$.
Also, this provides a group theoretic description of the symmetries via the centraliser of the local terms in the Pauli group $\commalg = \Span[\C]{\cent_{\Pgroup_n}(\{h_i\})}$.
In the binary formalism, this subgroup projects to a binary subspace of rank $2\ell$ and with a radical of dimension $r$, hence may always be computed efficiently for arbitrary sets of Pauli strings, independently of their locality.

In practice one is interested in finding such symmetries for \emph{sequences} of Hamiltonians, i.e. for arbitrarily large system sizes.
We develop in Section~\ref{sec:from_local_to_global} a useful framework for finding such symmetries in local one dimensional models which are `translationally invariant' in the sense that the local terms are translations of each other.
In particular, for a set of k-local Pauli strings $\calA\subseteq\PP_k$, each starting at a vertex $i$, we consider open and periodic boundary conditions
\begin{align*}
        \pgens_n^{\circ}(\calA) &= \{ h_i \mid A\in\calA, i\in[n]\} \\
        \pgens_n(\calA) &= \{ h_i \mid h\in\calA, i\in\{1,\cdots,n-k\}\}
\end{align*}
which correspond to the Hamiltonians
\begin{equation}
    H(\{c_{h,i}\}) = \sum_{i=1}^n \sum_{h\in\calA} c_{h,i} h_i.
\end{equation}
Refer to Section~\ref{sec:from_local_to_global} for more details and more general boundary conditions.

For such models we show, under certain conditions, that one can explicitly compute the commutant for arbitrary values of $n$.
Specifically, we construct a directed graph from the \emph{local commutant} $\commutant(\calA)$ by imposing a certain matching condition, which comes up naturally in the context of tensor product symmetries.

Beyond \emph{unitary} symmetries, anti-unitary symmetries have also played an important role in many body physics, and in particular in the classification of phases of matter \cite{Altland_Zirnbauer_1997,Ryu_Schnyder_Furusaki_Ludwig_2010,Uhlmann_2016} and impose constraints on the spectrum of Hamiltonians \cite{Wigner1932}.
There, a known result is the \emph{ten fold way}, which characterizes certain sets of anti-unitary symmetries in ten distinct classes.
Given the identification between anti-unitary (or more generally anti-linear) and invariant bilinear forms, discussed in Section~\ref{sec:quadratic_bilinear_definitions}, we are also able to classify such symmetries when the interactions $h_i$ are restricted to Pauli strings. Explicitly, if $B$ is an invariant bilinear form, an anti-unitary symmetry is defined as $\calT = \calK B$, with $\calK$ the conjugation operator in the computational basis.
In this case, the set of invariant bilinear forms, denoted as $\bilinear(\pgens)$, acquires a particularly simple description via the commutant.
It is either empty or it is closely linked to the commutant, i.e.,
\begin{equation*}
    \bilinear(\pgens) =
        B\cdot\commutant(\pgens)
\end{equation*}
for one of its elements $B\in\bilinear(\pgens)$.
Moreover, we are also able to describe these symmetries in the local case using the tools from Section~\ref{sec:from_local_to_global}, given their tensor product structure.

We refer to a single antiunitary symmetry $\calT$ as of real or $+$ type if $\calT^2=\id$, and of $-$ or quaternionic type if $\calT^2 = -\id$.
We find that three isomorphism classes for sets of such symmetries exist (see Section~\ref{sec:affine_subspaces_quadratic_forms_isomorphism_classes} and Lemma~\ref{lem:isomorphism_class_affine_subspaces_quadratic_forms}):
\begin{enumerate}
    \item Most anti-unitary symmetries are of real type (type $+$).
    \item Most anti-unitary symmetries are of quaternionic type (type $-$).
    \item Anti-unitary symmetries of real and quaternionic type come in the same amount (type $0$).
    \item There are cases with no anti-unitary symmetries.
\end{enumerate}
This provides \emph{four} distinct symmetry classes.
In the absence of symmetries this reduces to the real, complex and quaternionic ensembles, as described in Dyson's threefold way, and corresponding to the classical Lie algebras $\so(2^n)$, $\su(2^n)$ and $\usp(2^n)$.
However, in the presence of (abelian) symmetries an additional class appears, which may also be described as the intersection of the real and quaternionic case, and instead corresponds to $\su(2^{n-1})$, with a distinct representation than the full special unitary algebra $\su(2^n)$.
These representations are described in Proposition~\ref{prop:pauli_lie_isometry_block_forms}.

As an example, we apply the above symmetry analysis to a many-body model in Section~\ref{sec:ffd_free_fermions_in_disguise}.
In particular, we consider a representation of \emph{free-fermions in disguise} as studied in \cite{Vernier_Piroli_2026}, re-proving some of their results and expanding on others.
This allows us to more clearly understand the periodic nature of the symmetries in system size, by looking at the local symmetries.
Moreover, we also comment on the lack of free-fermionic mapping of Jordan-Wigner type as a necessary condition for quasi-universal Lie algebras, which shows that arbitrary dynamics generated by this model cannot be efficiently simulated or exactly solvable by some free-fermionic mapping, unlike its time-independent counterpart.

\subsection{Random Circuits and Clifford Designs}\label{sec:overview:random_clifford_designs}

Random circuits are a fundamental tool in quantum information, appearing in numerous settings, such as quantum advantage for sampling tasks \cite{Bouland_Fefferman_Nirkhe_Vazirani_2019,Movassagh_2023}, randomized benchmarking \cite{Nakata_Zhao_Okuda_Bannai_Suzuki_2021}, estimation of observables \cite{Huang_2020} and cryptography \cite{Kretschmer_2021,Kretschmer_2023}.
Moreover, they also appear in the context of deep variational circuits, as they provide a basis for the Lie-algebraic theory of barren plateaus \cite{Ragone_2024,Fontana_Rudolph_Duncan_Rungger_Cîrstoiu_2025}, as well as in chaotic dynamics and thermalization \cite{Nahum_2018,Sekino_2008}.
In all such cases \emph{true randomness} over the unitary group is represented by the Haar distribution \cite{bourbaki2004integration}, which is defined as the unique measure which is invariant under left and right multiplication.
Hence, it potentially involves implementing exponentially long circuits for an exact implementation.
Thus, in many contexts it is often more practical to deal with \emph{finite} ensembles of circuits which either exactly or approximately reproduce the moments of the Haar distribution over the unitary group.
Such ensembles are known as (exact or approximate) $t$-designs for the unitary group \cite{Singal_Hsieh_2022,Mitsuhashi_Yoshioka_2023,Helsen_Roth_Onorati_Werner_Eisert_2022,Harrow_Mehraban_2023,Ambainis_Emerson_2007,Belkin_Allen_Ghosh_Kang_Lin_Sud_Chong_Fefferman_Clark_2024}, and typically involve circuits of polynomial or linear depth in system size.
In recent years there has been much effort to find realization of such designs in sub-linear depth \cite{Schuster_Haferkamp_Huang_2025}, which runs counter-intuitively to quantum advantage in random sampling.
Moreover, recent works have also dealt with the problem of $t$-designs \emph{beyond} the unitary group \cite{West_Mele_Larocca_Cerezo_2024,Grevink_Haferkamp_Heinrich_Helsen_Hinsche_Schuster_Zimboras_2025}.

We contribute to this area by showing that, for a given Pauli Lie group, the corresponding Clifford transvection group provides an exact 3-design.
Conversely, we also prove that Clifford transvection groups are \emph{not} 4-designs for Pauli Lie groups.
We combine this result with the classification of Clifford transvection groups for connected generating sets, i.e. the quasi-universal cases subject to some Pauli commutant or invariant bilinear forms, as well as three free-fermionic cases in presence of symmetries.
Then, we can also view this result as an alternative proof for the fact the Clifford \cite{Webb_2016,Zhu_2017}
and real Clifford group \cite{Calderbank_PRL_1997} are 3-designs for the corresponding unitary and special orthogonal group (since they are both Pauli Lie groups).
Similarly, it was known that the Clifford group under (linear) Pauli symmetries is a 3-design for the corresponding Lie group \cite{Mitsuhashi_Yoshioka_2023}, as well as for the (even) matchgate group \cite{Wan_2023}.
In the connected case, this was not known (to our knowledge) for the symplectic group and the two other matchgate groups, as well as all these invariants in the presence of symmetries.
Hence, we can provide a unified framework which shows that Clifford transvection groups approximate Pauli Lie groups (in the sense of $t$-designs), just like the Clifford group approximates the unitary group, even beyond the connected case.

\section{Discussion and Outlook}\label{sec:discussion}

Using tools from the binary formalism, the theory of transvections and graph theory, we develop and discuss various tools aimed at understanding dynamical properties of Pauli strings, in the context of Lie theory and Clifford groups.
Namely, we focus on understanding the fundamental invariant properties of a given connected Pauli generating set with regard to its generated Pauli Lie algebra
or group and Clifford transvection group.
These invariants provide a classification in terms of quasi-universal and free-fermionic sets, which split further into subclasses depending on the specific symmetry properties.
We also describe reachability properties through Pauli orbits and
invariant subspaces.
In addition, we study superoperator, or adjoint, symmetries together with
higher-order replica symmetries, which clarify how Clifford transvection
groups approximate Pauli Lie groups in the sense of $t$-designs.
We provide efficient algorithms and simple criteria for identifying these
properties.
We then apply them to examples from the literature, obtaining a unified Pauli
framework that was previously either unknown or available only through
case-by-case arguments.
In the case of local Pauli strings, we also complement this analysis with simple criteria which build the symmetries for the global generating set out of the local symmetries, by imposing certain matching conditions depending on the size of the system or its connectivity.

This analysis becomes particularly relevant in the context of classical simulability and computational power.
For instance, in variational algorithms, Pauli orbits and invariant
subspaces describe which operator directions are accessible from a chosen
initial state or observable.
This information is directly relevant to trainability, especially for deep
circuits.
Also, for proving universality, our tools give simple criteria, which we use in non-trivial examples, as in the case of Parity Quantum Computing or certain 2-local Pauli generating sets.
We also apply these tools to understand certain restricted models of computation within the Pauli setting, such as IQP, DQC1 and multi-controlled single qubit gates.
Beyond the examples mentioned here, such tools can be applied to arbitrary Pauli generating sets, as well as to obtain Pauli information about arbitrary non-Pauli architectures, in a quantum information or many-body physics context.

On the other hand, in the context classical simulability, we discuss when Pauli orbits and invariant subspaces can be of polynomial size (only partially for the latter), which provide clear bounds for when one does not expect quantum advantage, via algorithms such as $\lieg$-SIM and Pauli backpropagation.

For stabilizer-based algorithms, the action of Clifford transvection groups on stabilizer states is less clear, and is not discussed in this work, though Pauli orbits provide the first step in this direction.
Specifically, a natural generalization is understanding the action of the Clifford groups described here on stabilizer states, i.e.\ the stabilizer orbits under such groups, which would provide crucial information for stabilizer-based algorithms.

Additionally, we do not establish maximality of these Clifford groups in Pauli Lie groups as well as the role played by symmetry-preserving magic gates for Clifford transvection groups, though we conjecture that this holds for all connected generating sets.
Understanding Clifford transvection groups as Pauli Lie groups would point to general symmetry-informed non-Clifford resources \emph{inside} Pauli Lie groups (much like one distinguishes between real and complex magic for the real Clifford group \cite{magni2025anticoncentrationstatedesigndoped}).

In terms of understanding reachability, with the exception of the diagonal case, it remains to determine the Pauli orbits for arbitrary disconnected generating sets, hence their action on the operator space.
This includes the classical case of local unitaries and local Cliffords, with corresponding generating set $\{X_i,Z_i\}$ (as Pauli rotations or transvections), and more generally generating sets which at most share phase qubits (up to isomorphism).

Since the corresponding Clifford transvection groups provide exact 3-designs for Pauli Lie groups, they provide examples of groups of gates which can be used in benchmarking protocols \cite{Helsen_Roth_Onorati_Werner_Eisert_2022}, possibly adapted to the symmetries of the original circuit \cite{Hashagen2018realrandomized,Helsen_2022}.
Additionally, one could also construct Clifford-dominated approximate $t$-designs for Pauli Lie groups by injecting arbitrary symmetry-preserving magic gates, as has been done in the case of the full unitary group \cite{Zhang_2026}.

Beyond the description of orbits and invariant subspaces, it would be interesting to investigate short-time reachability, which has been used in the context of many-body chaos \cite{West_Dowling_Southwell_Sevior_Usman_Modi_Quella_2025} and shallow depth approximate $t$-designs \cite{West_GarciaMartin_Diaz_Cerezo_Larocca_2025} by taking advantage of the structure of the commutator graph.
Hence, a natural question would be whether this can be re-formulated, in part or completely, in terms of colorings of the frustration graph, which describe the Pauli orbits in the matrix algebra of the generators.

Finally, given that transvections are also natural objects for qudits, one could investigate the graph-theoretic formalism and classification of generalized Pauli Lie algebras and Clifford transvection groups.

\ManuscriptPart{Examples and Applications}{part:applications}

We now begin the more technical part of the paper, starting with a more detailed treatment of specific applications and examples, which highlight our framework for the dynamical properties of Pauli strings. Strictly speaking, this part is based on all later parts so some aspects can only be fully understood
after referring to the relevant details covered later. However, this part focuses on the physical applications and the motivation for the mathematical
developments in the later parts. So, the chosen inverted order of presentation enables the reader to first appreciate the implications before delving further
into the methods.

In particular, will be using extensively Theorem~\ref{thm:full_classification_pauli_lie_algebras} (and related corollaries and lemmas) to classify the Lie algebras, possibly iterating upon the connected components (though only for the isomorphism class in this case), as in Lemma~\ref{lem:classification_lie_algebras_disconnected}.
For the quasi-universal cases, we can use the commutant and invariant bilinear forms alone for a classification, via Theorem~\ref{thm:full_classification_pauli_lie_algebras}, combined with the isomorphism class of the commutant (or Pauli matrix algebras, as binary subspaces, Lemma~\ref{lem:subspaces:iso} and surrounding discussion) and invariant bilinear forms (Lemma~\ref{lem:isomorphism_class_affine_subspaces_quadratic_forms}).
For convenience, we use almost exclusively use the language of invariant bilinear forms rather than anti-unitary or anti-linear symmetries.

We also recall here that vertex colorings (see Def.~\ref{def:coloring_to_vector_map}) of a frustration graph (see Def.~\ref{defn:frustration}) correspond to vectors in the matrix algebra generated by $\pgens$, as $P_{\coloring} \simeq \prod_{i\in\frustration{\pgens}} P_i^{\coloring(i)}$ with $\coloring(i)\in\F_2$.
Moreover, if $\frustration{\pgens}=L(\Delta)$ is a line graph of a multigraph $\Delta$ (see Def.~\ref{def:line_graphs}),
we can also view the vertex colorings on the frustration graph as edge colorings of its root multigraph $\Delta$.
In particular, for the purpose of classification, we shall be interested in the center of the matrix algebra, which is also spanned by Paulis. 
In the line graph case, these Paulis correspond to either colorings of cycles of the root multigraph, or,
if the number of vertices in the root multigraph is even, a T-join (see Def.~\ref{def:T-join}).
Here a T-join is an edge coloring of the root multigraph for which each vertex is adjacent to an odd number of colored edges.
Notice that for any coloring which provides a T-join, adding the coloring of a cycle in the root graph provides another T-join, hence T-joins are defined up to cycles symmetries.
Also, we say that a coloring is an \emph{algebraic dependency} (see Def.~\ref{def:dependencies})
whenever the corresponding Pauli $P_{\coloring}$ is proportional to the identity.
We say that a T-join symmetry is \emph{trivial} whenever it is proportional to a cycle symmetry or an algebraic dependency, and non-trivial otherwise.
Using this criterion, we shall distinguish between the possible free-fermionic cases using a mix of symmetry and graph-theoretic tools.
We refer to Lemma~\ref{lem:radical_via_cycles_and_Tjoin} and
Corollary~\ref{cor:distinguishing_line_graph_cases} and the surrounding discussions for more details on cycle symmetries and T-join in the free-fermionic case.

\section{Hardware Efficient Ans\"{a}tze in the Pauli Setting}\label{sec:example:2-local_paulis}

In this section we consider a general family of hardware efficient ans\"{a}tze in the Pauli setting, i.e., a circuit composed of single- and two-qubit gates parametrized by Pauli strings, over arbitrary connectivities.
First, we say that $\calA\subseteq\PP_2$ is a set of Pauli strings over two qubits which is \emph{swap-symmetric} if $\SWAP\cdot P \cdot \SWAP \in \calA$ for all $P\in\calA$.
This does not necessarily imply that only swap-symmetric interactions are permitted, since one could
have local gates of the form $e^{\im\theta_1 X_1Y_2} e^{\im\theta_2 X_2Y_1}$ with $\theta_1\neq \theta_2$.
Instead, this implies that for every gate in our generating set, we can also choose its swapped version (e.g.\ apply
$\text{CNOT}_{1\to 2}$ or $\text{CNOT}_{2\to 1}$).

\begin{figure}
    \centering
    \includegraphics[width=0.8\linewidth]{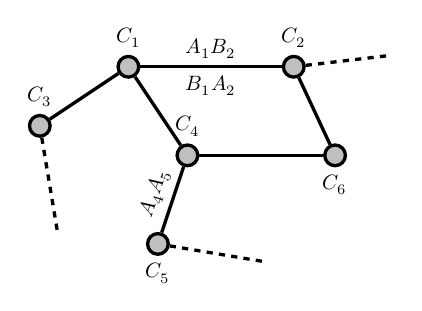}
    \caption{Example of swap-symmetric 2-local ansatz on some connected graph $\graphH$, with vertex operators $C_u$ and edge operators $A_uB_v$ or $A_uA_v$.}
    \label{fig:pauli-2-local}
\end{figure}

Now, let $\graphH$ be the interaction graph and $\calA\subseteq\PP_2$ a swap-symmetric 2-local set of Paulis,
we define the generating set (see Fig.~\ref{fig:pauli-2-local})
\begin{equation}\label{eq:Pauli_graph_ansatz}
    \pgens_{\graphH}(\calA) = \{ P_{uv} \}_{uv\in\edges(\graphH), P\in\calA}
\end{equation}
where, the Pauli string $P_{uv}$ for
$P=P_1\otimes P_2\in \PP_2$ has $P_1$ at site $u$, $P_2$ at site $v$, and $\id$ everywhere else, i.e.,
\begin{equation}
    P_{uv} = \underbrace{\id\otimes\cdots\otimes\id}_{u-1\text{ times}} \otimes P_1 \otimes \underbrace{\id\otimes\cdots\otimes\id}_{u-v-1\text{ times}} \otimes P_2\underbrace{\id\otimes\cdots\otimes\id}_{n-v\text{ times}}.
\end{equation}
This also includes single qubit terms, which are redundant when considered over adjacent edges.
With respect to the frustration graph $\frustration{\pgens_{\graphH}(\calA)}$, note instead that arbitrary subgraphs
of the interaction graph (i.e.\ by removing vertices and/or edges) correspond to
induced subgraphs of the frustration graph (i.e.\ by removing the corresponding Paulis on the vertices/edges).
This will be in particular relevant for the $\calE_6$-criterion (see Def.~\ref{def:e6} and Thm.~\ref{thm:E6_free_characterization_line_graphs}),
since one can first remove both vertices and edges from the interaction graph to obtain an induced subgraph in $\calE_6$.

Within this setting falls for instance the multi-angle QAOA ansatz, which was discussed \cite{Kazi_Larocca_Farinati_Coles_Cerezo_Zeier_2025}, with Xs on single qubits and ZZ interactions.
More generally, this family was discussed in \cite{Wiersema_Kokcu_Kemper_Bakalov_2024,Kokcu_Wiersema_Kemper_Bakalov_2024}, as well as certain non swap-symmetric interactions over one dimensional topologies with open and closed boundary conditions.
We shall consider here two specific examples which showcase how one can straightfowardly apply our tools to compute the Lie algebra, which provides much information about the deep-circuit behaviour of variational algorithms with gate sets as in Eq.~\eqref{eq:Pauli_graph_ansatz}.

Namely, the objective for classification is the following, for arbitrary connected $\graphH$:
\begin{enumerate}
    \item Compute the global commutant and invariant bilinear forms, starting from their local counterparts, which determines the quasi-universal cases.
    \item Determine for which interaction graphs the frustration graph $\frustration{\pgens_{\graphH}}$ is a line graph or not.
    \item If $\frustration{\pgens_{\graphH}}$ is a line graph, determine the number of vertices of the root multigraph and whether it has a \emph{T-join} symmetry to distinguish between. the free-fermionic classes
\end{enumerate}
For the computation of the symmetries we shall use the tools from Section~\ref{sec:classification:groups_lie_algebras}. 
Specifically certain \emph{directed} graphs defined by the local commutant and invariant bilinear forms, as well as Lemma~\ref{lem:local_matching_graphs}.
We summarize our main result here:
\begin{result}\label{result:2-local-lie_algebras}
Let $\pgens = \pgens_{\graphH}(\calA)$ for some connected interaction graph $\graphH$ and 2-local interaction $\calA$ either of Ising or QAOA type or XY-type.
The Lie algebras listed in Table~\ref{tab:ising-quasi-universal} and Table~\ref{tab:xy-quasi-universal} can be obtained by applying our framework:
(i) checking the $\calE_6$-condition on induced subgraphs of $\graphH$ of size six; (ii) computing the commutant and invariant bilinear forms via the tools Section~\ref{sec:from_local_to_global}; and (iii) checking for certain algebraic dependencies in the free-fermionic case.
\end{result}

\subsection{Ising Interactions and QAOA}\label{sec:qaoa}

\begin{figure}
    \centering
    \includegraphics[width=0.8\linewidth]{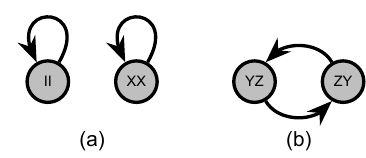}
    \caption{(a) Local Commutant of the Ising model/multi-angle QAOA and its directed graph $\graphD_\commutant(\{\text{XI, IX, ZZ}\})$. (b) Local Invariant Bilinear Forms of the Ising model/multi-angle QAOA and its directed graph $\graphD_\bilinear(\{\text{XI, IX, ZZ}\})$.
    There is a directed edge $A\to B$ whenever $A$ and $B$ \emph{locally match}, i.e., they are of the form $A = A_1A_2$ and $B = B_1B_2$ with $A_2 = B_1$.}
    \label{fig:directed-QAOA}
\end{figure}

\begin{figure*}
    \centering
    \includegraphics{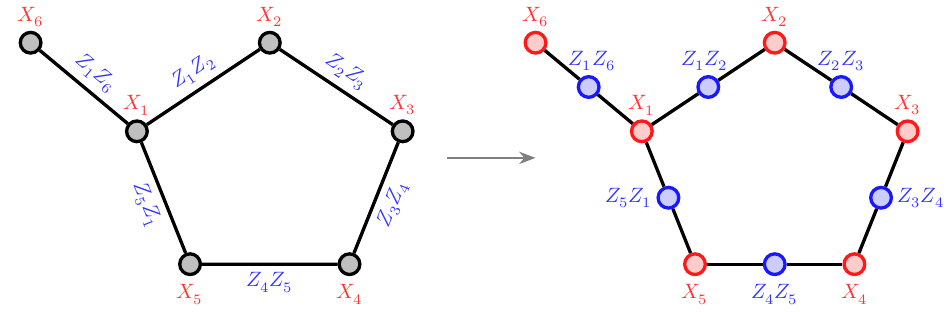}
    \caption{$\graphH$ is the \emph{interaction graph} of a qubit system which interacts via an Ising interaction and subject to local transversal fields, as in the multi-angle QAOA setting. The terms $X_u$ live on the vertices and $Z_uZ_v$ on the edges of $\graphH$. Its frustration graph is the subdivision graph of the interaction graph.}
    \label{fig:qaoa:subdivision}
\end{figure*}

As the first example we consider the aforementioned case of Ising interactions, or multi angle ansatz of QAOA \cite{Herrman_Lotshaw_Ostrowski_Humble_Siopsis_2022}.
This provides a short proof of Theorem~1 in \cite{Kazi_Larocca_Farinati_Coles_Cerezo_Zeier_2025}.
In the notation of Eq.~\eqref{eq:Pauli_graph_ansatz}, we have $\calA = \{ \text{IX, XI, ZZ}\}$, hence the generating set over some interaction graph
is given by (see also Figure~\ref{fig:qaoa:subdivision}):
\begin{equation}
    \pgens_{\graphH} = \{X_u\}_{u\in V(\graphH)}\cup\{Z_uZ_v\}_{\{u,v\}\in E(\graphH)}
\end{equation}
It is immediate to check that the local commutant and invariant bilinear forms consist of:
\begin{equation*}
        \baslong{\commutant(\calA)} = \{ \text{II,\! XX} \}\;\text{ and }\;
        \baslong{\bilinear(\calA)} = \{ \text{YZ,\! ZY} \}.
\end{equation*}
For the commutant, we construct as in Section~\ref{sec:from_local_to_global} directed graphs for the local commutant and invariant bilinear forms. 
The directed graph $\graphD_\commutant(\calA)$ has two disjoint self-loops, whereas for the bilinear forms $\graphD_\bilinear(\calA)$ has a single loop of length $2$ (see Figure~\ref{fig:directed-QAOA}).

Then, we can apply Lemma~\ref{lem:local_matching_graphs} to obtain the Pauli basis for the global commutant and invariant bilinear forms, depending on $\graphH$:
\begin{enumerate}
    \item $\commutant(\pgens_{\graphH})=\Span{\In, X^{\otimes n}}$ for any connected graph of size $n$.
    \item $\bilinear(\pgens_{\graphH})=\Span{Y^{\otimes \bipleft}Z^{\otimes \bipright}, Z^{\otimes \bipleft}Y^{\otimes \bipright}}$ (up to permutations) for any connected bipartite graph with bipartition $(\bipleft,\bipright)$.
    \item $\bilinear(\pgens_{\graphH})=\{0\}$ for any non-bipartite graph.
\end{enumerate}
With respect to the commutant $\commalg$, we immediately find $\rank(\commalg)=0$ (see Eq.~\eqref{eq:rank} and Eq.~\eqref{eq:def:rank_and_nullity_Pauli_matrix_algebra}), $\nullity(\commalg)=1$, hence $\rank(\pgens_\graphH) = 2(n-1)$ for all cases (see Lemma~\ref{lem:Canonical_Form_Arbitrary_Adjacent_Contractions_Frustration_Graph}).
Equivalently, in all cases there is a single phase qubit, no uncontrollable qubits and $n-1$ logical qubits in the sense of Section~\ref{sec:classification:pauli:matrix}.

Moreover, we can classify the isomorphism classes of bilinear forms by invoking Lemma~\ref{lem:isomorphism_class_affine_subspaces_quadratic_forms}, and specifically by viewing the isomorphism classes via their \emph{democratic} invariants, i.e. simply counting how many are symmetric and skew-symmetric:
\begin{enumerate}
    \item If $\bipleft,\bipright$ are both even, $\bilinear(\pgens_{\graphH})$ is of $+$ type
    \item If $\bipleft,\bipright$ are both odd, $\bilinear(\pgens_{\graphH})$ is of $-$ type
    \item If $\bipleft+\bipright$ is odd, $\bilinear(\pgens_{\graphH})$ is of $0$ type
\end{enumerate}

We can now discuss properties of the frustration graph, to distinguish between free-fermionic and quasi-universal cases, as well as bipartite and non-bipartite.

A basic but fundamental observation links the frustration graph $\frustration{\pgens_{\graphH}}$ of $\pgens_{\graphH}$ and the \emph{interaction graph} $\graphH$:
\begin{lem}
Let $S(\graphH)$ be the \emph{subdivision graph} of $\graphH$, created by taking any edge of $\graphH$ and inserting a new vertex of degree 2 directly into the middle of it.
For any graph $\graphH$, the frustration graph $\frustration{\pgens_{\graphH}}$ is $S(\graphH)$ (see Fig.~\ref{fig:qaoa:subdivision}).
\end{lem}

\begin{figure}
    \centering
    \includegraphics{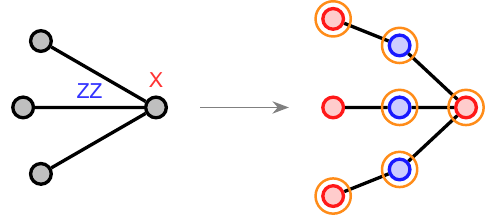}
    \caption{Any graph with a vertex of degree $\geq 3$ possess a claw as subgraph (possibly by removing edges). Its subdivision graph contains $\graphE_6$, hence it is not a line graph of a multigraph.}
    \label{fig:claw:e6}
\end{figure}

Then, we collect some facts about subdivision graphs:
\begin{enumerate}[label=(\arabic*)]
    \item If $\graphH$ is connected, $\frustration{\pgens_{\graphH}}$ is connected as well.
    \item The subdivision graph of $\graphP_n$ is $\graphP_{2n-1}$.
    \item The subdivision graph of $\graphC_n$ is $\graphC_{2n}$.
    \item Any graph with a vertex of degree $3$ contains a claw as a (not necessarily induced) subgraph
    \item The subdivision graph of a claw contains $\graphE_6$ as an induced graph (see Figure~\ref{fig:claw:e6}).
    \item The subdivison graph of a connected graph $\graphH$ is a line graph of a multigraph if and only if $\graphH$ is either the path or the cycle graph.
\end{enumerate}

Hence, we are already able to classify the Lie algebras $\lieg = \lie{\pgens_{\graphH}}$ where $\graphH$ possess a vertex of degree $\geq 3$, using Theorem~\ref{thm:full_classification_pauli_lie_algebras}. Let $\commalg$ be the commutant and, in the bipartite case, $B$ a reference invariant bilinear form, then the classification is shown in Table~\ref{tab:ising-quasi-universal}.
See also Theorem~\ref{prop:pauli_lie_isometry_block_forms} for a description of their representations in block form.

\begin{table}[t]
\caption{\label{tab:ising-quasi-universal}Lie algebras for the Ising interaction.
Quasi-universality is guaranteed for interaction graphs $\graphH$ with a vertex
of degree at least $3$.}
\centering
\footnotesize
\renewcommand{\arraystretch}{1.15}
\begin{tabular*}{\columnwidth}{@{\hspace{1mm}}l@{\extracolsep{\fill}}l@{\hspace{2mm}}l@{\hspace{1mm}}}
\hline\hline
\\[-3.5mm]
Graph $\graphH$ & Parity condition & Lie algebra $\lieg$ \\[0.5mm]
\hline
\\[-3mm]
Non-bipartite
& ---
& $\lieiso(\commalg)\cong \su(2^{n-1})^{\oplus 2}$ \\
Bipartite
& $\bipleft,\bipright$ both even
& $\lieiso(\commalg,B) \cong \so(2^{n-1})^{\oplus 2}$ \\
Bipartite
& $\bipleft,\bipright$ both odd
& $\lieiso(\commalg,B)\cong \usp(2^{n-1})^{\oplus 2}$ \\
Bipartite
& $\bipleft+\bipright$ odd
& $\lieiso(\commalg,B)\cong \su(2^{n-1})$ \\
Path
& ---
& $\so(2n)$ \\
Cycle
& ---
& $\so(2n)^{\oplus 2}$ \\[0.75mm]
\hline\hline
\end{tabular*}
\end{table}

Finally, it remains to classify the line graph cases. First notice that both frustration graphs are line graphs whose root graphs have an even number of vertices:
\begin{enumerate}
    \item If $\graphH = \graphP_n$, we have that $S(\graphP_n) = \graphP_{2n-1} = L(\graphP_{2n})$
    \item If $\graphH = \graphC_n$, we have that $S(\graphC_n) = \graphC_{2n} = L(\graphC_{2n})$
\end{enumerate}
We now need to check the possible cycle and T-join symmetries on root graphs, to see which are trivial or not.

In the path case, there is no cycle symmetry, hence it can only admit a T-join symmetry. 
One can check that that this produces $\prod_{i=1}^n X_i = X^{\otimes n}\neq \In$, which is not algebraic dependency (nor a cycle symmetry).
Hence, it has a non-trivial T-join symmetry over the root-graph (see Corollary~\ref{cor:distinguishing_line_graph_cases} and the surrounding discussion), and no cycle symmetries. 
Then, by Theorem~\ref{thm:full_classification_pauli_lie_algebras}, the Lie algebra is $\so(2n)$.
In terms of free-fermionic mapping, by Lemma~\ref{prop:free_fermionic_mapping_alg_dep} we can simply choose $\{\gamma_i\gamma_{i+1}\}_{i=1}^{2n-1}$, where $n_\Delta=2n$.
In particular we can apply Lemma~\ref{prop:full_Fn_PPn_Majorana_basis_free_fermionic_mapping}.
See also Examples~\ref{ex:free_fermionic_mappings_paths} and \ref{ex:free_fermionic_mappings_cycles} for all cases of free-fermionic mappings when the frustration graph is a path or a cycle and their symmetries.
 
For the cycle case, there is a single cycle symmetry and two possible T-join symmetries, corresponding to an alternating coloring over either the odd or even edges in the root graph $R(\frustration{\pgens_\graphH})$ (or equivalently odd and even vertex colorings in the frustration graph $\frustration{\pgens_\graphH}$).
Over the cycle, we find the non-trivial symmetry $\prod_iX_i\prod_iZ_iZ_{i+1} \simeq X^{\otimes n}\neq \In$.
Since the cycle symmetry together with the identity already spans the center of the matrix algebra, the even or odd coloring cannot produce additional elements, hence they are trivial T-join symmetries.
Indeed, one can check that the even coloring results in $\prod_iZ_iZ_{i+1} = \In$.
Hence, by Theorem~\ref{thm:full_classification_pauli_lie_algebras} the Lie algebra is $\so(2n)^{\oplus 2}$.
Also, we can find the free-fermionic mapping $\{\gamma_i\gamma_{i=1}\}_{i=1}^{2n-3}\cup\{\gamma_{2n-2}\}\cup\{\Gamma\}\cup\{ (\gamma_{2n-1}\gamma_{2n})\Gamma\gamma_1\}$.
Notice that this is \emph{not} the usual free-fermionic mapping for the Ising model in the case of periodic boundary conditions, due to the specific choice we made for free-fermionic mappings (see Section~\ref{sec:Free-Fermionic_Mappings} for more details on free-fermionic mappings).

This new approach to this example was independently obtained in \cite{Cuypers_2026}; we had announced this result in \cite{Natal_2026,Dresden_2026}.

\subsection{XY Interactions}\label{sec:XY_interactions}

\begin{figure}
    \centering
    \includegraphics[width=0.8\linewidth]{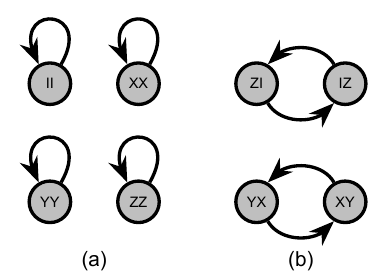}
    \caption{(a) Local Commutant of the XY model and its directed graph $\graphD_\commutant(\{\text{XX, YY}\})$. (b) Local Invariant Bilinear Forms of the XY model and its directed graph $\graphD_\bilinear(\{\text{XX, YY}\})$.
    There is a directed edge $A\to B$ whenever $A$ and $B$ \emph{locally match}, i.e., they are of the form $A = A_1A_2$ and $B = B_1B_2$ with $A_2 = B_1$.}
    \label{fig:directed-XY}
\end{figure}

For the second example, we consider the set for the XY-interaction with $\calA = \{ \text{XX, YY}\}$, hence
\begin{equation}
    \pgens_{\graphH} = \{X_uX_v, Y_uY_v\}_{\{u,v\}\in E(\graphH)}
\end{equation}
which admits as local commutant and invariant bilinear forms as
\begin{align*}
        \baslong{\commutant(\calA)} &= \{ \text{II, XX, YY, ZZ} \} \text{ and}\\
        \baslong{\bilinear(\calA)} &= \{ \text{IZ, ZI, XY, YX} \}.
\end{align*}
Then, $\graphD_\commutant(\calA)$ consists of four self-loops, and $\graphD_\bilinear(\calA)$ consists of two loops of length $2$ (see Fig.~\ref{fig:directed-XY}).
Again, if we apply Lemma~\ref{lem:local_matching_graphs}, we have:
\begin{enumerate}
    \item $\commutant(\pgens_{\graphH})=\Span{\In, X^{\otimes n}, Y^{\otimes n}, Z^{\otimes n}}$ for any connected graph of size $n$.
    \item $\bilinear(\pgens_{\graphH})=\spanempty\{X^{\otimes \bipleft}Y^{\otimes \bipright}, Y^{\otimes \bipleft}X^{\otimes \bipright}, I^{\otimes \bipleft}Z^{\otimes \bipright},\\ Z^{\otimes \bipleft}I^{\otimes \bipright}\}$ (up to permutations) for any connected bipartite graph with bipartition $(\bipleft,\bipright)$.
    \item $\bilinear(\pgens_{\graphH})=\{0\}$ for any non-bipartite graph.
\end{enumerate}
With respect to the commutant $\commalg$, the rank and nullity are
\[
(\rank(\commalg),\nullity(\commalg)) =
\begin{cases}
(0,2) & \text{if } n \text{ is even},\\
(2,0) & \text{if } n \text{ is odd}.
\end{cases}
\]
In the even case all terms commute.
Hence, the symmetries are purely abelian or purely non-abelian depending on the parity of the system size.

Regarding the isomorphism classes of the bilinear forms for bipartite graphs, we obtain:
\begin{enumerate}
    \item If $\bipleft,\bipright$ are both even, $\bilinear(\pgens_{\graphH})$ is of $+$ type.
    \item If $\bipleft,\bipright$ are both odd, $\bilinear(\pgens_{\graphH})$ is of $0$ type.
    \item If $\bipleft+\bipright$ is odd, $\bilinear(\pgens_{\graphH})$ is of $+$ type.
\end{enumerate}
The graph-theoretic properties instead slightly differ, due to the fact that the frustration graph is now no longer the subdivision graph of the interaction graph $\graphH$.
Instead, we have a closer construction to a line graph with edges in $\graphH$ corresponding to (two) vertices in $\frustration{\pgens_{\graphH}}$:
\begin{lem}
Let $\graphH$ be a graph. For each edge in $\graphH$, there are two vertices in $\frustration{\pgens_{\graphH}}$, for the different types of  XX and YY. Two vertices in $\frustration{\pgens_{\graphH}}$ are adjacent iff their edges in $\graphH$ are adjacent \emph{and} they are of different type
(i.e., $X_uX_v$ anti-commutes with $Y_vY_{u'}$ but not with $X_vX_{u'}$).
\end{lem}

\begin{figure}
    \centering
    \includegraphics[width=0.8\linewidth]{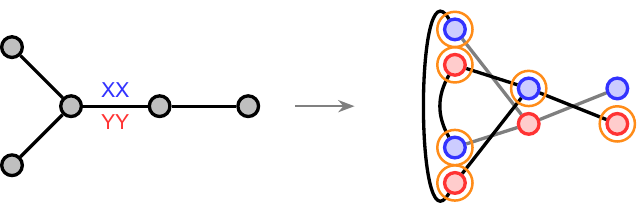}
    \includegraphics[width=0.8\linewidth]{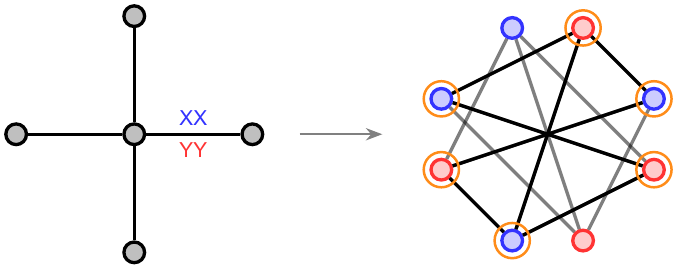}
    \caption{Left: the possible subgraphs of a connected graph with at least four vertices. Right: the corresponding frustration graphs with XY-interactions, which contain induced subgraphs in $\calE_6$, highlighted in orange.}
    \label{fig:long-claw-XY}
\end{figure}

Then, we have the following:
\begin{enumerate}
    \item If $\graphH = \graphP_n$, $\frustration{\pgens_{\graphH}} = \graphP_{n-1} \cup \graphP_{n-1} = L(\graphP_n)\cup L(\graphP_n)$
    \item If $\graphH = \graphC_n$, if $n=2m$ is even $\frustration{\pgens_{\graphH}} = \graphC_m \cup \graphC_m = L(\graphC_m)\cup L(\graphC_m)$, and if $n=2m+1$ is odd $\frustration{\pgens_{\graphH}} = \graphC_{2n}$
    \item If $\graphH$ has a vertex of degree $\geq 3$ and $n\geq 5$, it contains either a `long claw' $\graphL_4$ or a four-leaf star graph $\graphS_4$ (see Fig.~\ref{fig:long-claw-XY}) as a subgraph
    \item If $\graphH$ is the long claw or the four-leaf star graph, $\frustration{\pgens_{\graphH}}$ contains an induced subgraph from $\calE_6$ (see Fig.~\ref{fig:long-claw-XY})
    \item If $\graphH$ is connected with $n\geq 5$, $\frustration{\pgens_{\graphH}}$ is the line graph of a multigraph if and only if $\graphH$ is either a path graph or a cycle graph
\end{enumerate}
Notice that the long claw is denoted as $\Sigma$ in \cite{Kokcu_Wiersema_Kemper_Bakalov_2024}.

Then, we can characterize the non-path and non-cycle cases (over $n\geq 5$ vertices), since they are quasi-universal. This is shown in Table~\ref{tab:xy-quasi-universal}.

\begin{table}[t]
\caption{\label{tab:xy-quasi-universal}Lie algebras for the $XY$ interaction.
Quasi-universality is guaranteed for interaction graphs $\graphH$ with a vertex
of degree at least $3$ and $n\geq 5$.}
\centering
\footnotesize
\renewcommand{\arraystretch}{1.15}
\begin{tabular*}{\columnwidth}{@{\hspace{1mm}}l@{\extracolsep{\fill}}l@{\hspace{2mm}}l@{\hspace{1mm}}}
\hline\hline
\\[-3.5mm]
Graph $\graphH$ & Parity condition & Lie algebra $\lieg$ \\[0.5mm]
\hline
\\[-3mm]
Non-bipartite
& $n$ even
& $\lieiso(\commalg)=\su(2^{n-2})^{\oplus 4}$ \\
Non-bipartite
& $n$ odd
& $\lieiso(\commalg)=\su(2^{n-1})\otimes \id_2$ \\
Bipartite
& $\bipleft,\bipright$ both even
& $\lieiso(\commalg,B)=\so(2^{n-2})^{\oplus 4}$ \\
Bipartite
& $\bipleft,\bipright$ both odd
& $\lieiso(\commalg,B)=\su(2^{n-2})^{\oplus 2}$ \\
Bipartite
& $\bipleft+\bipright$ odd
& $\lieiso(\commalg,B)=\so(2^{n-1})\otimes \id_2$ \\
Path
& ---
& $\so(n)^{\oplus 2}$ \\
Cycle
& $n$ even
& $\so(n)^{\oplus 4}$ \\
Cycle
& $n$ odd
& $\so(2n)\otimes\id_2$\\[0.75mm]
\hline\hline
\end{tabular*}
\end{table}

\begin{figure}
    \centering
    \includegraphics[width=0.8\linewidth]{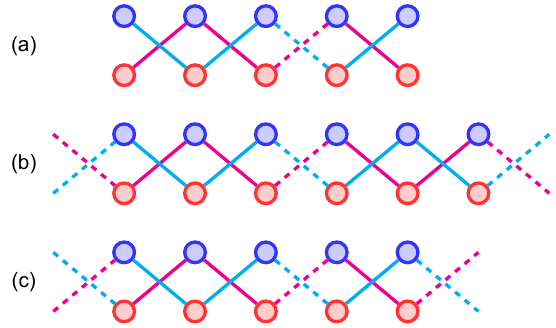}
    \caption{Frustration graphs for the XY-interaction on (a) the path graph, (b) the even cycle graph and (c) the odd cycle graph. 
    Blue vertices are the XX interactions and red vertices are the YY interactions.}
    \label{fig:frustration-XY}
\end{figure}

For the path and cycle cases, we need to work with the explicit generating sets to understand algebraic dependencies, cycle and T-join symmetries. For the path graph we have two connected components $\pgens_1$ and $\pgens_2$, with alternating XX and YY terms (see Fig.~\ref{fig:frustration-XY}(a)), which are themselves path graphs, i.e.,
\begin{align*}
        \pgens_1 &= \{ X_{2i-1}X_{2i}\}_{i=1}^{2\lfloor n/2\rfloor}\cup \{Y_{2i}Y_{2i+1}\}_{i=1}^{2\lceil (n-2)/2\rceil},\\
        \pgens_2 &= \{ Y_{2i-1}Y_{2i}\}_{i=1}^{2\lfloor n/2\rfloor}\cup \{X_{2i}X_{2i+1}\}_{i=1}^{2\lceil (n-2)/2\rceil}.
\end{align*}
One can easily check that in either case there are no algebraic dependencies for any $n$.
Hence, up to isomorphism, the Lie algebra is $\lie{\pgens_1\cup\pgens_2} \cong \so(n)\oplus \so(n)$.
In particular, if $n$ is even, $\pgens_1$ generates the abelian symmetry $X^{\otimes n}$, and $\pgens_2$ generates $Y^{\otimes n}$, which are T-joins for both.
If $n$ is odd instead they have no abelian symmetries (as expected), but they both share the non-abelian symmetries $X^{\otimes n}, Y^{\otimes n}$ and $Z^{\otimes n}$.

In the cycle case, if $n=2m$ is even (see Fig.~\ref{fig:frustration-XY}(b)), we have instead the generating sets (with indices modulo $n$)
\begin{align*}
        \pgens_1 &= \{ X_{2i-1}X_{2i}, Y_{2i}Y_{2i+1}\}_{i=1}^{2m},\\
        \pgens_2 &= \{ Y_{2i-1}Y_{2i}, X_{2i}X_{2i+1}\}_{i=1}^{2m}.
\end{align*}
Then, the alternating colorings over the cycle are $X^{\otimes n}$ and $Y^{\otimes n}$ for both cycles (which provide non-trivial T-join symmetries), and the non-trivial cycle symmetry is $Z^{\otimes n}$.
Given that the root graphs are over $n$ even vertices, the Lie algebra is $\lie{\pgens_1\cup\pgens_2} \cong \so(n)^{\oplus 2}\oplus \so(n)^{\oplus 2} \cong \so(n)^{\oplus 4}$.

If $n$ is odd, there is only one cycle of length $2n$ (see Fig.~\ref{fig:frustration-XY}(c)).
Also, both the alternating colorings contain respectively all XX and all YY terms, $\prod_i X_iX_{i+1} \simeq \prod_iY_iY_{i+1} \simeq I$, hence lead to an algebraic dependency.
Hence, the cycle coloring is also an algebraic dependency, and in particular a Lie-algebraic dependency.
Then, since there are no non-trivial T-join and cycle symmetries (together with a non-abelian commutant of rank $2$), we have $\lie{\pgens} = \so(2n)\otimes\id_2$.

\section{Universality and Computational Power}\label{sec:universality_and_computational_power}

In this section we discuss questions of universality and computational power in the context of Pauli strings.
We consider three examples for which we determine either conditions of universality (Section~\ref{sec:parity_computing}), or, if non-universal, its action on Pauli strings and general operators (Sections~\ref{sec:uniformly_controlled_many_symmetries} and \ref{sec:IQP}). 
These latter examples showcase the role of \emph{phase} qubits in the computational power, by providing exponentially large Pauli orbits even in the absence of a large amount of logical qubits, and in particular either a single or no logical qubits.

\subsection{Parity Computing}\label{sec:parity_computing}

\begin{figure}
    \centering
    \includegraphics[width=0.9\linewidth]{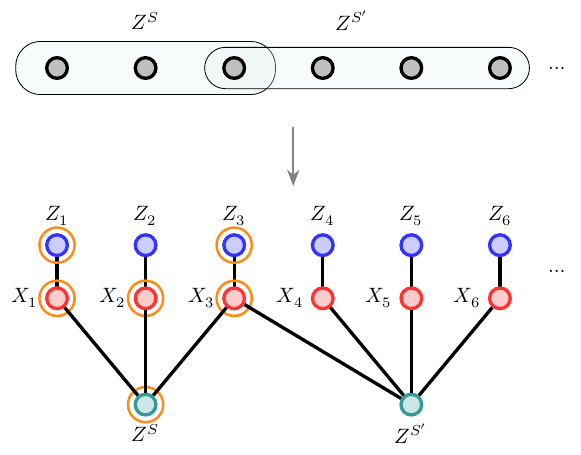}
    \caption{Example of interaction hypergraph in Parity Quantum Computing (top) and its corresponding frustration graph (bottom).
    The induced subgraph highlighted in orange is $E_6$.}
    \label{fig:parity-computing}
\end{figure}

The first example we consider is that of \emph{parity quantum computing} \cite{Fellner_2022}. 
The natural gate set for this model of computation is given by the single qubit gates on all qubits, i.e.,
\begin{equation}
    \pgens_1 = \{ X_j, Z_j \mid j\in[n]\} \iff \vgens_1 = \{\basel_j,\basel_{n+j} \mid j \in[n]\}
\end{equation}
together with a set of Z-strings, which each live over some subset of the of the $n$ qubits (see Fig.~\ref{fig:parity-computing}):
\begin{equation*}
        \parity \subseteq P(\{1,\ldots, n\})\;\text{ and }\;
        \pgens_{\parity} = \qty{ \prod_{i\in S} Z_i \mid S\in\parity}.
\end{equation*}
In \cite{Smith_Klaver_Nautrup_Lechner_Briegel_2025}, the authors studied sufficient conditions under which this set provides universal quantum computation.
We discuss it here in our framework, and provide sufficient and necessary conditions.
Specifically, we prove the following (see also \cite{Cuypers_2026}):
\begin{lem}\label{lem:pqc_universality}
Let $\parity \subseteq P(\{1,\ldots, n\})$ and $\pgens = \pgens_1 \cup\pgens_{\parity}$. Then, $\lie{\pgens} = \su(2^n)$ iff
both conditions hold:
\begin{enumerate}
    \item The underlying hypergraph of $\parity$ is connected and the support of each strings covers $n$, $\cup_{S\in\parity}S = \{1,\cdots,n\}$.
    \item There is a string $S$ of even weight in the set $\parity$.
\end{enumerate}
\end{lem}
This is precisely the setting of Corollary~\ref{cor:Symmetry_Breaking_su_Universality_with_basis}, which deals with universality once given an algebraically independent generating set together with additional elements.
The single-qubit terms already give $\commutant(\pgens_1)=\C\In$ and it follows that
$\pgens_1$ has a unique invariant bilinear form (it is algebraically independent and commutant has dimension $1=2^0$).

Then, $\pgens = \pgens_1 \cup \pgens_{\parity}$, for some set of strings $\parity$ is universal with $\lie{\pgens} = \su(2^n)$ iff the three condition hold:
\begin{enumerate}
    \item Its frustration graph is connected.
    \item The frustration graph is not a line graph of a multigraph (contains an induced subgraph in $\calE_6$).
    \item There is a term in $\pgens_{\parity}$ which breaks the unique invariant bilinear form.
\end{enumerate}

First, we deal with the invariant bilinear form. By Eq.~\eqref{eq:explicit_w_vector_from_quadratic_form}, the invariant bilinear form $B = \iso{w}$ corresponds to the unique binary vector $w\in\Fn$ such that
\begin{equation}
    w = \sum_i \QQ_w(f_i)e_i + \QQ_w(e_i)f_i = \sum_i e_i + f_i
\end{equation}
By definition, $\QQ_w(f_i)=\QQ_w(e_i)=1$ hold as $\QQ_w$ is invariant here.
In the Pauli language, we have $B = Y^{\otimes n} = \iso{w}$.
Hence, by Lemma~\ref{lem:quadratic_forms_over_Fn_as_vectors}, we have that the invariant quadratic form has form
\begin{align*}
    \QQ^*(v) &= \sum_{j=1}^n v_j v_{j+1} + \sum_{j=1}^n v_j + \sum_{j=1}^n v_{n+j} \\
    &= \sum_{j=1}^n (1+(1+v_j)(1+v_{n+j})) = \wt(v)\bmod 2
\end{align*}
or the \emph{Pauli weight} of the Pauli string $P=\iso{v}$ with respect to the single qubit Pauli strings.
By abuse of notation, we also use $\wt(P)$ to denote the weight of the Pauli string $P$.
Then, we have that $\pgens_{\parity}$ breaks this invariant quadratic form if and only if there is at least one Z-string of even weight.
Indeed, the breaking of such an invariant was also used in \cite{Smith_Klaver_Nautrup_Lechner_Briegel_2025} as a necessary ingredient for universality.

Then, we notice that any parity quantum generating set $\pgens$ such that the frustration graph is connected is also \emph{not} the line graph of a multigraph.
To show this, notice that the frustration graph is connected whenever the underlying hypergraph of $\parity$ is connected and the union of the supports
of the strings $S\in\parity$ covers all $n$ qubits.
We can now distinguish between two (possibly overlapping) cases: (1) there exists a single Z-string which connects the entire graph (i.e. $Z^{\otimes n}$), (2) there are two non-trivial Z-strings $P_1,P_2$ ($\wt(P_i)\geq 2$) which overlap on at least one shared site $j$.
In case (2), $P_1$ and $P_2$ have support (at least) on sites $\{k_1,j\}$ and $\{k_2,j\}$. Thus, the subgraph with vertices $\{Z_{k_1}, P_1, Z_j, X_j, P_2, \}$ coincides with $\graphE_6$.
In case (1), if $n\geq 3$, we can instead choose the subgraph with vertices $\{X_1, Z_1, X_2, Z_2, X_3, Z^{\otimes n}\}$, which again is $\graphE_6$ (see Fig.~\ref{fig:parity-computing}).
This proves Lemma~\ref{lem:pqc_universality}.

In particular, case (1) provides the minimally connected generating set $\pgens_{\parity}$. However, it results in a universal generating set if and only if $n$ is even, such that $Z^{\otimes n}$ has odd weight.

If $n$ is odd, one requires at least two strings such that the frustration graph is connected, e.g., any two Z-strings 
$\prod_{i\leq j}Z_i$ and $\prod_{i\geq j}Z_i$
which overlap at precisely one site and cover the entire chain.

Finally, we also highlight that purely odd Z-strings (together with single qubits) can still implement \emph{almost} universal quantum computation.
Namely, if the underlying hypergraph $\parity$ is connected and covers the entire chain, the set is still quasi-universal.
Hence, one can implement the orthogonal gates $\so(2^n)$ if $n$ is even ($Y^{\otimes n}$ is symmetric), and the symplectic gates $\usp(2^n)$ if $n$ is odd ($Y^{\otimes n}$ is skew-symmetric).
One can then also implement \emph{encoded} universal quantum computation by restricting to the subalgebra $\su(2^{n-1})$ (which lives in both $\so(2^n)$ and $\usp(2^n)$), hence on $n-1$ logical qubits \cite{rudolph20022rebitgateuniversal}.

\subsection{Abelian Symmetries and Controlled Gates}\label{sec:uniformly_controlled_many_symmetries}

\begin{figure}
    \centering
    \includegraphics[width=0.5\linewidth]{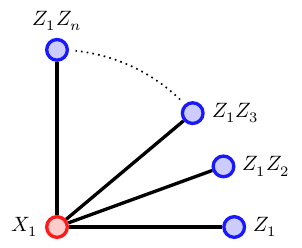}
    \caption{Frustration graph of the generating set in Eq.~\eqref{eq:controlled_single_qubit}.}
    \label{fig:controlled-graph}
\end{figure}

As mentioned in Section~\ref{sec:overview:orbits}, for a \emph{connected generating set}, the sole presence of a large amount of \emph{abelian} symmetries guarantees a Lie algebra and (non-trivial) orbits of exponential size.
In order to highlight this feature, we consider a minimally connected example with the largest amount of abelian symmetries allowed in a system of $n$ qubits.
Namely, this corresponds to the unique Lie algebra where there is a single logical qubit, $\rank(\pgens)=2$ (see Eq.~\eqref{eq:rank}),
and $n-1$ phase qubits, $\nullity(\pgens)=n-1$.
Explicitly, there is a canonical generating set given by:
\begin{equation}\label{eq:controlled_single_qubit}
    \pgens = \{X_1,Z_1\}\cup\{Z_1Z_i\}_{i=2}^{n-1}
\end{equation}
whose frustration graph is a star graph with $n-1$ legs of length one, or in the canonical notation of Theorem~\ref{thm:classes:humphries}, the blown-up path graph $\graphP_{1,n}$ (see Fig.~\ref{fig:controlled-graph}).

\begin{figure}
    \centering
    \includegraphics[width=0.9\linewidth]{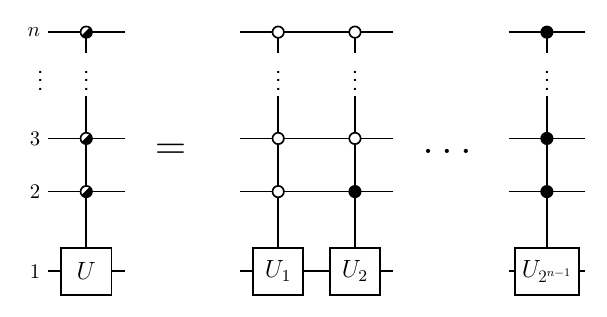}
    \caption{Uniformly-controlled single qubit gates \cite{Bergholm_Vartiainen_Mottonen_Salomaa_2005}, acting on qubit $1$ with controls on qubits $2,\ldots,n$. 
    Each elementary gate is either $0$-controlled (white dot) or $1$-controlled (black-dot) on each of the control qubits. An arbitrary product of these gates is said to be a uniformly-controlled gate.}
    \label{fig:controlled-circuit}
\end{figure}

By definition, this generating set and Lie algebra is of \emph{free-fermionic} type. Indeed it is the line graph of a path graph over $3$ vertices with $n$ parallel edges (hence $q=n-1$ loops) at one end.
Hence, it also admits a generalized free-fermionic mapping over $2$ logical modes and $2(n{-}1)$ phase modes, which we can choose as $\{\gamma_1,\gamma_2\}\cup\{(\gamma_{2i-1}\gamma_{2i})\gamma_2\}_{i=2}^{n-1}$.
Thus the Lie algebra
\begin{subequations}\label{eq:pauli_basis_uniformly_controlled}
    \begin{align}
 \lieg_c^0 &= \so(3)^{\oplus 2^{n-1}} = \su(2)^{\oplus 2^{n-1}}
\intertext{is a direct sum of $\so(3) = \su(2)$ (in its $2$-dimensional representation) and admits the Pauli basis}
        \bas{\lieg_c^0} &= \{X,Y,Z\}\otimes \{I,Z\}^{\otimes (n-1)}.
    \end{align}
\end{subequations}
We can also write the Lie algebra in the block diagonal basis as follows:
\begin{align*}
        \bas{\lieg_c^0}' &=
        \begin{aligned}[t]
        &\{X, Y, Z\}\\
        &\otimes \qty{ \prod_{i=2}^n\frac{1+s_iZ_i}{2} \text{ for } \{s_i\}_{i=2}^n\in\{\pm 1\}^{\times (n-1)} }.
        \end{aligned}
\end{align*}
Hence, the corresponding Lie algebra implements precisely the set of (special)
\emph{uniformly controlled single-qubit gates}
with $n-1$ control qubits and a single target
(see Fig.~\ref{fig:controlled-circuit}), i.e.,
\begin{equation}
    U(\theta)  = \exp( \im\theta\prod_{i=2}^n\frac{1+s_iZ_i}{2} P_1).
\end{equation}
Here the control condition is given by the string $\{s_i\}_{i=2}^n$, and it acts on the first qubit as the Pauli rotation $\exp(\im\theta P_1)$.
This is a set of gates commonly used in several known quantum algorithms, possibly to construct gates beyond this set \cite{Bergholm_Vartiainen_Mottonen_Salomaa_2005}, via controls and targets which can vary in the circuit. Indeed the Toffoli gate (a multi-controlled gate) plus the Hadamard gate on a connected topology are a universal generating set \cite{aharonov2003simpleprooftoffolihadamard}.
We now discuss the algebraic and symmetry properties of this set of circuits, with fixed set of control qubits, from the point of view of the Pauli strings.

We also highlight the distinction between the \emph{special} subgroup of uniformly controlled gates ($\det(U)=+1$) and the \emph{full} group of uniformly controlled gates, which also implements the controlled \emph{relative phase} gates.
Explicitly, we have that the Lie algebra
\begin{align*}
\lieg_c &= \lieu(2)^{\oplus 2^{n-1}} = \lieg_c^0 \oplus \lieu(1)^{\oplus 2^{n-1}}
\intertext{acquires an additional abelian part, also of exponential size (for all possible controls), and it has the basis}
        \bas{\lieg_c} &= \{I,X,Y,Z\}\otimes \{I,Z\}^{\otimes (n-1)}.
\end{align*}
Hence, it also additionally implements the diagonal gates on the last $n-1$ qubits, resulting in full \emph{relative} phase control between all qubits in the computational basis.
As such, the corresponding generating set cannot be connected, since the Lie algebra contains an abelian part.

For $\lieg_c^0$ the commutant is abelian of dimension $2^{n-1}$ and admits the Pauli generating set $\{Z_i\}_{i=2}^n$.
Then, as shown in Section~\ref{sec:Orbits_Full_Space_Fn_Quasi_Universal_Case}, the orbits may be divided into three types, depending on the given Pauli string $P$:
\begin{enumerate}
    \item $P$ is a Z-string with support on the control qubits, i.e., $P = \prod_{i=2}^nZ_i^{\delta_i}$ for some $\delta_i\in\F_2$;
    \item $P$ has no X-part over the control qubits, i.e., it commutes with the commutant and but is not contained in it, hence it is in the Lie algebra, i.e.,
    \begin{equation}
            P\simeq X_1^{\alpha_1}Z_1^{\beta_1} \prod_{i=2}^nZ_i^{\delta_i}\;\text{ for }\;
            P\in\bas{\lieg_c^0}
    \end{equation}
    for some $\alpha_1,\beta_1\in\F_2$ not both zero and $\delta_i\in\F_2$.
    \item $P$ has a non-trivial X-part $\prod_{i=2}^nX_i^{\lambda_i}$ over the control qubits, which is the only invariant in the orbit, i.e.,
    \begin{equation}
            P\simeq X_1^{\alpha_1}Z_1^{\beta_1} \prod_{i=2}^nX_i^{\lambda_i} Z_i^{\delta_i}\;\text{ for }\;
            P\in \prod_{i=2}^nX_i^{\lambda_i}\bas{\lieg_c}
    \end{equation}
    for some $\alpha_1,\beta_1,\delta_i,\lambda_i\in\F_2$ with $\lambda_i$ not all zero.
\end{enumerate}
Thus, excluding the constants of motion, all orbits are of size either $3\cdot 2^{n-1}$ or $4\cdot 2^{n-1}$, and thus they are exponential in system size (i.e.\ $\BigO(2^n)$).

From the point of invariant subspaces (see Sec.~\ref{sec:invariant_subspaces}), we find instead that the (irreducible) invariant subspaces are given by each $2\times 2$ block, spanned by some $P_1\prod_{i=2}^n[(1{+}s_iZ_i)/{2}]$ where $s_i\in\{\pm 1\}$ determine the state of the control and phase qubits and $P_1\in\{X_1,Y_1,Z_1\}$.
Hence, even if the orbits are of exponential size, one can still find operator invariant subspaces of constant size, over which time evolution is greatly restricted.
As also discussed in Section~\ref{sec:invariant_subspaces}, this distinction between Paulis and arbitrary observables lies in the fact that in the presence of a large amount of abelian symmetries, Pauli strings are supported on an exponentially large amount of blocks, see also Fig.~\ref{fig:full-block-decomposition}.

For $\lieg_c$, notice that the orbits do not change under the addition of the diagonal gates.
Namely, the transvection action of the Z-strings on Pauli strings which have a non-trivial X-part on the control qubits is simply to add or remove the Z-part.
Hence, it does not affect the symmetry properties, which is not surprising, given the discussion about the \emph{pathological transvections} in Section~\ref{sec:classification_transvection_groups}.

Another interesting consequence of using Pauli strings as natural generators of the set of uniformly controlled gates is that now the group admits a natural generating set of size which is only \emph{polynomial} in size, $\abs{\pgens} = n+1$.
This is to be compared with a full decomposition in the block diagonal basis, as done in \cite{Bergholm_Vartiainen_Mottonen_Salomaa_2005}, which instead requires a circuit of exponential size for an arbitrary circuit made of uniformly controlled gates.

\begin{figure}
    \centering
    \includegraphics{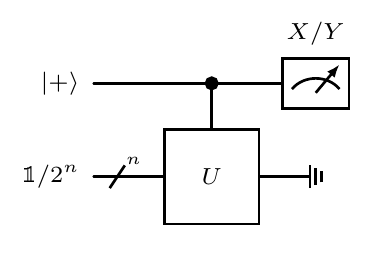}
    \caption{The One Clean Qubit (DQC1) model of computation implements controlled gates on a starting $\ket{+}$ state on the control and the maximally mixed state on the targets, with single qubit measurements on the control qubit.}
    \label{fig:DQC1}
\end{figure}

We end this section by also briefly mentioning another setting of controlled gates which achieve non-trivial restricted quantum computation, known as Deterministic One Clean Qubit (DQC1) \cite{Knill_1998} (see Fig.~\ref{fig:DQC1}). 
In this case the circuit consists of a single qubit initialized in the $\ket{+}$ state, which is the clean qubit, and the rest of the system initialized in the maximally mixed state $\rho = \id/2^n$, which are the dirty qubits. 
Hence, the global (mixed) state may be represented as $(1+X_1)/2$. 
The allowed operations are gates which act on the $n$ dirty qubits, controlled on the clean qubit, and measurements in the $X$ and $Y$ basis over the clean qubit.
The Lie algebra 
\begin{align*}
 \lieg &= \su(2^{n-1})^{\oplus 2}
\intertext{for such controlled gates has a basis}
    \bas{\lieg} &= \qty{ \frac{1+Z_1}{2},  \frac{1-Z_1}{2}}\times \{ I\otimes (\PP_{n-1}\setminus\{\In\})\}\\
\intertext{which can be re-arranged into a Pauli basis} 
    \bas{\lieg}' &= \qty{I,Z_1}\times \{ I\otimes (\PP_{n-1}\setminus\{\In\})\}.
\end{align*}
Thus $\lieg$ may be viewed as a Pauli Lie algebra.
By Corollary~\ref{cor:orbits_full_Fn_strictly_universal_case}, there are are four Pauli orbits:
\begin{enumerate}
    \item the singleton orbits $I$ and $Z_1$;
    \item the Lie algebra itself $\bas{\lieg}'$ if $P$ commutes with $Z_1$ but is not a symmetry;
    \item $\{X,Y\}\otimes \PP_{n-1}$ if $P$ does not commute with $Z_1$.
\end{enumerate}
Hence, we find that under arbitrary dynamics with DQC1, the initial state $(\In +X_1)/2^n$ can explore an exponentially large operator space, since the orbit for $X_1$ has exponential size.
Again, this provides evidence for its computational power, though only as a necessary condition, since specific circuit constructions may still be classically simulable.

\subsection{Instantaneous Quantum Polynomial and Diagonal Gates}\label{sec:IQP}

Given that a large amount of symmetries is strictly related to the presence of large orbits for \emph{connected} Pauli generating sets, a natural question to ask is whether this continues to hold also for \emph{disconnected} generating sets.
In Section~\ref{sec:uniformly_controlled_many_symmetries}, we saw that the strictly diagonal part of the uniformly controlled gates do not change the structure of the orbits, which motivates the study of the actual role played by diagonal Pauli gates in terms of reachability.

In this section we provide a complementary view of these different sets and discuss \emph{completely} disconnected, or commuting, Pauli generating sets. 
Specifically, this constitutes the model of computation known as Instantaneous Quantum Polynomial (IQP)
\cite{Shepherd_2009}, which consists of a polynomial depth of diagonal gates applied to an initial $\ket{+}^{\otimes n}$ state, with measurements in the X-basis.
Despite the restricted gate set, such an architecture has been shown to solve problems beyond the power of classical computers \cite{Bremner_2010,Bremner_2017,Nakata_2014,buzet2026iqpcircuits2forrelation} (under reasonable complexity assumptions).
These circuits live in the Lie group of diagonal gates, whose Lie algebra is spanned by Z-strings over the entire chain, i.e.,
\begin{equation}
    \pgens\subseteq \pgens_Z = \{I,Z\}^{\otimes n}.
\end{equation}

\begin{figure}
    \centering
    \includegraphics[width=0.9\linewidth]{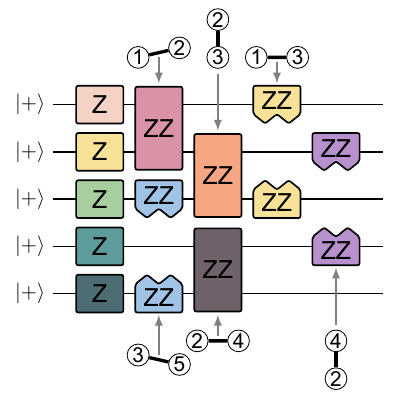}
    \caption{Example of Instantaneous Quantum Polynomial (IQP) circuit with 2-local ZZ rotations chosen over some interaction graph.}
    \label{fig:IQP-circuit}
\end{figure}

Unlike the connected cases, here the orbit with respect to the generators contains only the generators themselves, or, equivalently, the Lie algebra is spanned by the generators themselves $\pgens\subseteq\pgens_Z$ and is a trivial abelian Lie algebra:
\begin{equation}
    \lie{\pgens} = \Span{\im\pgens} \cong \lieu(1)^{\oplus \abs{\pgens}}
\end{equation}
A typical IQP ansatz \cite{Shen_Pielawa_Dunjko_Wang_2026} involves 1- and 2-qubit diagonal gates determined by some connectivity, or equivalently by the edges of a graph $\graphH$ as
(see Fig.~\ref{fig:IQP-circuit} for an example)
\begin{equation}
    \pgens = \{ Z_i\}_{i=1}^n \cup \{Z_iZ_j\}_{\{i,j\}\in\edges(\graphH)}.
\end{equation}
In order to understand the full reachability of diagonal gates in the Pauli setting, we first discuss the largest possible case,
i.e.\ the full set of Z-strings $\pgens_Z = \isolong{V_Z}$, where $V_Z = \Span{\{e_i\}_{i=1}^n}$.
From the Lie algebra perspective, this is exponentially sized, $\lieg \cong \lieu(1)^{\oplus 2^n}$, hence one trivially expects large reachability and non-trivial quantum computation.

For the case with a finite generating set, we now consider the point of view of transvections, starting from the binary picture.
In Sections~\ref{sec:kernel_criteria_pointwise_symplectic_stabilizers}
and~\ref{sec:commutant_symmetric_clifford_groups_lie_algebras} we showed that
the diagonal Clifford group $\calD_n$ has binary image $\DD{V_Z}$, where
$V_Z=\rad(V_Z)$ is the subspace containing the Z-strings as binary vectors.
Also, by Proposition~\ref{prop:full_space_symplectic_symmetric_Clifford_groups},
any Clifford transvection group whose binary image is $\DD{V_Z}$ coincides
with $\calD_n$ (possibly up to phases).
Moreover, by Lemma~\ref{lem:symplectic_diagonal_kernel_structure}, the binary group $\DD{V_Z}$ is isomorphic to the additive group of binary symmetric matrices.
In particular, this means that one can understand minimal generating sets of this group purely in a linear algebraic approach, given that an $\F_2$ matrix additive group has a natural structure as a vector space.
In particular, one can show that for a basis $\{h_i\}_{i=1}^n$ of $V_Z=\rad(V_Z)$, this binary group is minimally generated by the transvections $\{h_i\}_{i=1}^n\cup\{h_i+h_j\}_{i<j=1}^n$, hence it is in fact a (binary) transvection group
\begin{equation}
    \DD{V_Z} = \tvgrouplong{\{h_i\}_{i=1}^n\cup\{h_i+h_j\}_{i<j=1}^n}.
\end{equation}
Then, if we take $Z_i = \iso{h_i}$ as a basis, the transvections with centers $Z_i$ for each qubit and $Z_iZ_j$ for each pair generate the group of diagonal Clifford gates (possibly up to phases), which is a Clifford transvection group
\begin{equation}
    \calD_n = \cltvgrouplong{\{Z_i\}_{i=1}^n\cup\{Z_iZ_j\}_{i<j=1}^n}.
\end{equation}
Thus the one- and two-qubit diagonal generators already generate the same
diagonal Clifford transvection group as the full set of diagonal Pauli strings.
Consequently, the Pauli orbits for the polynomial-dimensional diagonal Lie algebra
$\lieu(1)^{\oplus n(n+1)/2}$ generated by the single- and two-qubit diagonal
Pauli strings coincide with the Pauli orbits for the entire diagonal Lie algebra
$\lieu(1)^{\oplus 2^n}$ generated by all diagonal Pauli strings.
The former has polynomial, in fact quadratic, dimension in $n$, while the
latter has exponential dimension in $n$.

This agreement of Pauli orbits is mirrored by the ordinary commutants, which
coincide for the two diagonal generating sets.
However, as we will show in Section~\ref{sec:square_adjoint_commutants_pauli_lie_algebras}, the operator invariant subspaces of the two Lie groups and
algebras depend purely on the orbits, hence coincide for the single- and two-qubit diagonal gates and the entire set of diagonal gates, despite the vast difference in dimension.
This goes against the usual picture of \emph{semisimple} Lie algebras, where it is known that a subalgebra of a semisimple Lie algebra coincides with the full Lie algebra if and only if they have the same adjoint or square commutant \cite{Zeier_2015}.
Of course, this is not surprising given that it is known that the abelian part requires an additional rank condition, which is not covered by the symmetry properties \cite{Zimboras_Zeier_SchulteHerbruggen_Burgarth_2015}.

Explicitly, the $4^n$ Pauli strings decompose into two types of orbits:
\begin{enumerate}
    \item the Z-strings themselves are singleton orbits, of which there are $2^n$;
    \item any Pauli-string 
    \begin{equation*}
        P\simeq \prod_iX_i^{\lambda_i}\pgens_Z
    \end{equation*}
    with $\lambda_i$ not all $0$ and
    a non-trivial X-part is conjugate to any other Pauli string with the same X-part.
    Hence there are $2^n-1$ such orbits, all of size $2^n$.
\end{enumerate}
The total number of orbits is $2^n+(2^n-1)=2^{n+1}-1$.

In the Pauli picture, this means that the orbits are $V_Z$ and $v+V_Z$ for $v=\sum_i\lambda_if_i$, for $\lambda_i$ not all $0$.
Explicitly, $\DD{V_Z}$ contains, for $u,u'\in V_Z$,
two types of operations (see Lemma~\ref{lem:symplectic_diagonal_kernel_transvection_generators})
\begin{equation*}
    \tau_uv = v + \symp{u}{v}u \text{ and } g_{u,u'}v = v + \symp{u}{v}u' + \symp{u'}{v}u.
\end{equation*}
Let $v\in\Fn$ have non-trivial X-part, so the linear functional
$u\mapsto\symp{u}{v}$ on $V_Z$ is nonzero.
Choose a basis $\{h_i\}_{i=1}^n$ of $V_Z$ such that
$\symp{h_1}{v}=1$ and $\symp{h_i}{v}=0$ for $i>1$.
We show that for every $w\in V_Z$ there is an element $g\in \DD{V_Z}$ such that
$g(v)=v+w$.
First, if $w\in\SpanS[\F_2]{\{h_i\}_{i>1}}$, then
$\symp{h_1+w}{v}=1$, hence the transvection
$\tau_{h_1+w}$ sends $v$ to $v+h_1+w$.
Second, since $\symp{w}{v}=0$, the element $g_{h_1,w}$ acts as
\[
    v \mapsto v+\symp{h_1}{v}w+\symp{w}{v}h_1=v+w.
\]
Thus both $w$ and $h_1+w$ can be added to $v$, for arbitrary
$w\in\SpanS[\F_2]{\{h_i\}_{i>1}}$.
Therefore the orbit of $v$ is the full coset $v+V_Z$, or equivalently,
up to phases, $\calD_n\cdot\iso{v}\simeq\iso{v}\cdot\pgens_Z$.

As mentioned, $\DD{V_Z}$ is \emph{minimally} generated by the transvections $\{Z_i\}\cup\{Z_iZ_j\}$ with Zs on each vertex and ZZ-interactions on all pairs of qubits.
Hence, if we remove even a single element, the corresponding transvection group (and, trivially, Lie algebra) gets smaller.
Hence, unlike the analogous 2-local non-commuting versions, such as the QAOA or Ising set $\{X_i\}_{i\in\vertices(\graphH)}\cup\{Z_iZ_j\}_{\edges(\graphH)}$ (which falls into six cases), the IQP generating set $\{Z_i\}_{i\in\vertices(\graphH)}\cup\{Z_iZ_j\}_{\edges(\graphH)}$ does not produce the same Lie algebra and symmetries for certain large classes of graphs.
Specifically, under the isomorphism of $\DD{V_Z}$ to the vector space of $\F_2$-symmetric $n\times n$ matrices, each set $\{Z_i\}_{i\in\vertices(\graphH)}\cup\{Z_iZ_j\}_{\edges(\graphH)}$ spans a distinct subspace (subgroup) of dimension $\abs{\graphH} = \abs{\vgens(\graphG)} + \abs{\edges(\graphH)} \leq n(n+1)/2$.
Moreover, thanks to the Pauli orbit and commutants analysis, this shows that the group of all 1- and 2-qubit diagonal gates is a 3-design for the full diagonal group, and more specifically the Clifford transvection group, generated by 1- and 2-qubit gates, is a 3-design for the full diagonal group.
Notice that this set has a large amount of algebraic dependencies and is much larger (in size) than the set of \emph{non-commuting} Paulis required to obtain universality, $2n+1$ \cite{Smith_Cautres_Stephen_Nautrup_2025}, which also generate the diagonal gates themselves.

\begin{figure}
    \centering
    \includegraphics[width=0.7\linewidth]{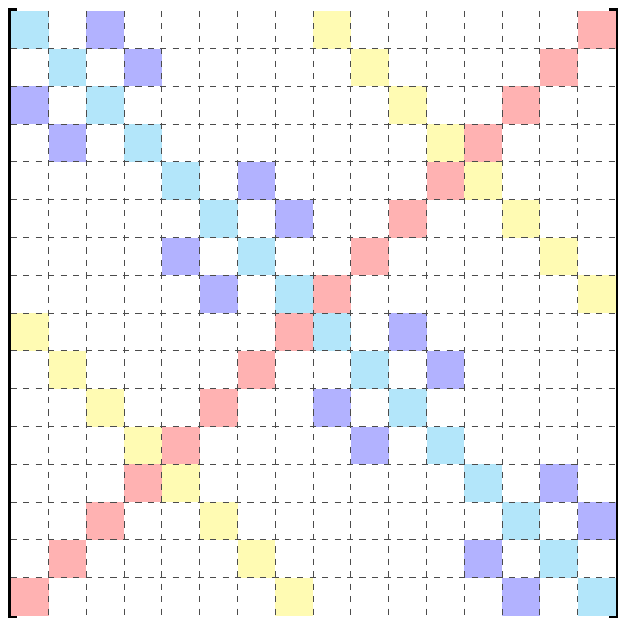}
    \caption{Block decomposition of the operator space under a Pauli Lie algebra with large amount of abelian symmetries.
    The blocks in the same color correspond to the invariant subspaces where certain Pauli strings have non-zero overlap.
    In particular, all commuting Pauli strings live purely on the diagonal blocks.
    In the case of diagonal gates or Instantaneous Quantum Polynomial (IQP), each block has size $1\times 1$.
    In the case of uniformly controlled single qubit gates, each block has size $2\times 2$.}
    \label{fig:full-block-decomposition}
\end{figure}

If we now shift from the purely Pauli picture to the entire operator space or matrix algebra $\matalg(2^n,\C)$, the \emph{invariant subspaces} have instead a completely different form compared to the orbits.
Indeed, we clearly have that the adjoint operators $\ad_H$ for $H\in\pgens$ mutually commute, which means that there is a basis of simultaneous eigenstates for all the generators, or equivalently, $4^n$ one-dimensional (irreducible) invariant subspaces, which we can choose as all of the projectors onto the computational basis $\ket{\{s_i\}_{i=1}^n}\bra{\{s_i'\}_{i=1}^n}$.
The mismatch between the two pictures is due to the fact that all Pauli strings are supported on an exponential number of these projectors, such that one goes from finite-size invariant subspace to exponentially large orbits, see Fig.~\ref{fig:full-block-decomposition}.
This shows how a Lie-algebraic symmetry analysis based only on commutants and
invariant subspaces may provide less information than the Pauli-adapted one,
which makes use of Pauli orbits.
We provide a more general discussion of invariant subspaces and adjoint commutants in Sections~\ref{sec:square_adjoint_commutants_pauli_lie_algebras} and \ref{sec:invariant_subspaces}.

\section{Implications for Classical Simulability}\label{sec:classical_simulability}

We now discuss the implications of our analysis for classical simulability of
Pauli architectures.
The examples in Section~\ref{sec:universality_and_computational_power} show
that phase qubits can produce exponentially large Pauli orbits even when there
are few, or no, logical qubits.
On the opposite side of computational power and universality stands \emph{classical simulability}.
Knowledge of such methods is of great importance for quantum computing, since it provides necessary resources for non-trivial quantum computation, aids in the design of better quantum algorithms, and highlights the limitations of current quantum advantage experiments \cite{Krinitsin_Tausendpfund_Heyl_Rizzi_Schmitt_2025,vovrosh2025simulatingdynamicstwodimensionaltransversefield,pawlowski2025closingquantumclassicalscalinggap,Bermejo_Braccia_Rudolph_Holmes_Cincio_Cerezo_2026,Stoudenmire_2024,larose2024briefhistoryquantumvs,Breach_Placke_Claeys_Parameswaran_2025,tindall2025dynamicsdisorderedquantumsystems,Schuster_Yin_Gao_Yao_2025,Liu_Clark_2026,GonzalezGarcia2025paulipath,Tindall_Fishman_Stoudenmire_Sels_2024,Cerezo_2025}.
Indeed, even though arbitrary quantum computation is expected to be beyond the capabilities of classical computers, under limited resources one may still be able to practically simulate certain quantum algorithms.
Many classical simulation methods exist which have more or less applicability depending on the specific scenario, such as noisy shallow circuits, restricted gate sets, limited entanglement, connectivity and more.
Within the context of Pauli Lie algebras and groups as well as corresponding Clifford transvection groups, we focus on three possible frameworks which come up naturally:
\begin{enumerate}
    \item $\lieg$-sim \cite{Goh_Larocca_Cincio_Cerezo_Sauvage_2023};
    \item Pauli backpropagation \cite{Martinez_Angrisani_Pankovets_Fawzi_Franca_2025,Rudolph_Jones_Teng_Angrisani_Holmes_2025,Rall_Liang_Cook_Kretschmer_2019,Begusic_Hejazi_Chan_2025,Fontana_Rudolph_Duncan_Rungger_Cîrstoiu_2025,Ugale_Master};
    \item stabilizer or Clifford-based algorithms \cite{Aaronson_Gottesmann_2004}.
\end{enumerate}
We briefly explain how these methods work and how they apply to our setting.

The approach of $\lieg$-sim is a classical simulation method developed in \cite{Goh_Larocca_Cincio_Cerezo_Sauvage_2023} which tracks the evolution of operators in the Heisenberg picture under the time evolution by a Lie algebra $\lieg$, produced by the generators $\{H_j\}$ of the problem (say, in a continuous or digital setting).
The algorithm relies on the fundamental fact the operator space gets partitioned into invariant subspaces $\matalg(2^n,\C) = \bigoplus_\lambda \invV_\lambda$ such that one cannot escape the action of a given invariant subspace under time evolution $\ad_\lieg\invV_\lambda\subseteq\invV_\lambda$.
Given a basis over these subspaces $\{B_i^\lambda\}$, any observable $O$ or mixed state $\rho$ may be decomposed as $\sum_{i,\lambda} c_i^\lambda B_i^\lambda$.
Hence, by definition of invariant subspace, the coefficients $c_i^\lambda$ evolve under some linear representation of $\lieg$ such that they do not mix between different subspaces, or values of $\lambda$.
Specifically, one reduces the evolution of matrices of size $2^n$, with time
evolution of the form $U(t)OU^\dagger(t)$, to the evolution of coefficient
vectors of size $\dim(\invV_\lambda)$, where the time evolution takes the form
$\tilde{U}_\lambda(t)\ket{O}_\lambda$.
The algorithm is then efficient whenever the initial coefficients of the observable or state are non-zero only over a subspace of polynomial dimension, and may be solved naively in $\BigO(\dim(\invV_\lambda)^3)$-time (for each $\lambda$), viewed as basic linear algebra steps.
A natural application setting for this algorithm is provided by variational
quantum algorithms, since
the cost observable is often chosen there to belong to the generated Lie algebra~\cite{Kazi_Larocca_Farinati_Coles_Cerezo_Zeier_2025}.

Then, the invariant subspace over which the observable or state lives is provided by the Lie algebra itself.
Hence, whenever the Lie algebra has polynomial size and the initial coefficients are given, $\lieg$-sim provides efficient classical simulation \cite{bärligea2026enablingliealgebraicclassicalsimulation}.
One of the most well known settings where this happens is when the Lie algebra is free-fermionic $\lieg\cong \so(2n)$.
More generally, for a fixed size of abelian symmetries $r=\BigO(1)$, all free-fermionic cases (even, odd and exceptional)
fall into this setting, as discussed in Section~\ref{sec:Majorana_Strings_Free_Fermionic_Lie_Algebras}).
For these cases, $\lieg \cong \so(k)^{\oplus 2^q}$ for some $q\leq r = \BigO(1)$ and $k = \BigO(n)$, hence the dimension of the Lie algebra is still quadratic $\dim(\lieg) = \BigO(n^2)$.
Moreover, the Pauli orbits themselves provide invariant subspaces of polynomial size whenever they commute with the cycle symmetries and they have
a Majorana length which does not scale with system size (see Section~\ref{sec:Orbits_Full_Space_Fn_Free_Fermionic_Case}).
In particular, such spaces (which need not be irreducible) have a size which scales as $\binom{2n}{L} = \BigO(n^L)$ for $L$ the Majorana length (depending on the specific case).

One could also think of contrived quasi-universal cases where $m$
or $r$ grow at most poly-logarithmically with system size $n$, hence $\dim\lieg$ grows quasi-polynomially.
However, it is not clear whether these naturally appear from known generating sets.
Outside the connected setting, such polynomially sized Pauli Lie algebras can
also arise from disconnected generating sets, such as the local unitaries
$\lieg\cong \su(2)^{\oplus n}$ or the $k$-local diagonal gates for constant
$k$, for which $\dim\lieg = \binom{n}{k} = \BigO(n^k)$
(see also Section~\ref{sec:IQP}).
In particular, in the case of diagonal gates the \emph{irreducible} invariant subspaces are all $1$-dimensional and reduce to simultaneous eigenstates in the Heisenberg picture.
Of course, even having only $1$-dimensional subspaces does not guarantee that efficient simulation is always possible, since a given state or observable may still be supported on an exponential number of subspaces.

Viceversa, notice an exponentially sized Lie algebra does not necessarily guarantee that all (irreducible) invariant subspaces have exponential size (excluding the trivial one-dimensional subspaces corresponding to symmetries).
Namely, this can happen whenever the Lie algebra is a direct sum of simple summands $\lieg \cong \bigoplus_i \lieg_i$, where at least one of the simple summands has polynomial dimension.
Reference~\cite{bärligea2026enablingliealgebraicclassicalsimulation} discusses one such case in the context of a $\lieU(1)$ on-site symmetry, which is not covered in the Pauli setting.
In the context of connected Pauli Lie algebras, this happens whenever the amount of abelian symmetries scales with size $r=\BigO(n)$ and the number of logical qubits is fixed $m = \BigO(1)$.
The extremal case of uniformly controlled single qubit gates with $n-1$ controls, $\lieg = \su(2)^{\oplus 2^{n-1}}$ is discussed in Section~\ref{sec:uniformly_controlled_many_symmetries}.
Hence each simple summand $\su(2)$ provides an invariant subspace of dimension three, despite the entire Lie algebra having exponential size.
For observables outside the Lie algebra, similar examples can still produce many
low-dimensional invariant subspaces, as shown by
Theorem~\ref{thm:symmetry_adapated_orbit_subspaces}.
In particular, irreducible subspaces have a dimension of at most $2^{2m}$, hence a small number of logical qubits always implies small irreducible subspaces.

Another framework which has received much attention in recent years, also based on the Heisenberg picture, is that of \emph{Pauli backpropagation} 
\cite{Martinez_Angrisani_Pankovets_Fawzi_Franca_2025,Rudolph_Jones_Teng_Angrisani_Holmes_2025,Rall_Liang_Cook_Kretschmer_2019,Begusic_Hejazi_Chan_2025,Fontana_Rudolph_Duncan_Rungger_Cîrstoiu_2025,Ugale_Master}.
Generally speaking, this approach is based on expanding an observable or mixed state \emph{in the Pauli basis}, $O=\sum_Pc_PP$, independently of the problem, unlike $\lieg$-sim, where the basis is problem-dependent.
At a high level, the underlying idea is to track only the non-zero coefficients in the Pauli string basis, such that the algorithm is efficient whenever the starting state or observable has a polynomial amount of non-zero coefficients, and remains as such throughout the entire time evolution.
Unlike $\lieg$-sim, this constitutes more of a family of techniques than a specific algorithm.
Nevertheless, we can distinguish between two distinct approaches, at least in the unitary time evolution picture (since this has also been applied in the presence of noise).
In the continuous setting, we view the time evolution via $U(t) = \exp(-\im Ht)$, as a power-series expansion in commutators as
\begin{align*}
        &U(t)OU^\dagger(t) = O + t\comm{-\im H}{O} + \frac{t^2}{2}\comm{-\im H}{\comm{-\im H}{O}} + \cdots\\
        &\text{with } H = \sum_P c_P(H) P  \text{ and }
        O = \sum_P c_P(O) P,
\end{align*}
which can also be adapted to the time-dependent Hamiltonian setting.
Then, one uses the symplectic picture of the Pauli strings to efficiently compute commutators by expanding both $H$ and $O$ in the Pauli basis.
In this case both the number of commutators and number of non-zero Pauli coefficients need be small so that the algorithm remains efficient.
In the digital setting, one instead expands non-Clifford gates into Clifford
gates and simple non-Clifford primitives.
These primitives send individual Pauli strings into linear combinations of only
a small number of Pauli strings, as in Pauli rotations $\exp(\im\theta P)$,
multi-controlled gates, measurements, or Pauli noise channels.
This avoids the limitations due to the repeated commutators, though the exponential cost in non-zero coefficients also cannot be avoided in general.

If the circuit architecture or control or interaction Hamiltonians produce universal quantum computation, a single Pauli may propagate into a linear combination of \emph{all} (non-identity) Pauli strings, rendering the classical simulation method inefficient.
However, in the presence of symmetries or other restricted settings, reachability is potentially limited and efficiency can be improved, which is precisely where our framework gives information.
Namely, one can view the effect of arbitrary symmetries, as restricting the space that a single Pauli string can explore.
For an arbitrary circuit architecture or controls, such a restriction may not be trivial to examine, given that Pauli strings may not easily decompose into invariant subspaces.
Now suppose that the generators of the time evolution are Pauli strings, so
that the time-evolution operator lies in a Pauli Lie group.
Then a Pauli orbit $O$ under the corresponding Clifford transvection group gives
the largest set of Pauli strings into which an initial Pauli string $Q$ may
propagate. Thus
\begin{equation}
    U(t)QU^\dagger(t) \subseteq \invV^O = \Span{P | P\in O} \text{ with } Q\in O.
\end{equation}
In particular, we have that the only connected cases where Pauli orbits are of polynomial size are either: (1) the symmetries themselves, which do not evolve under time evolution; (2) when the number of logical and phase qubits is small $m+r = \BigO(1)$; (3) in the free-fermionic cases, when the Pauli $P$ commutes with the cycle symmetries or phase Majorana modes and has fixed Majorana length over the logical modes, $L=\BigO(1)$.
In all other cases, including the free-fermionic ones, the orbits have exponential size in either $m$ or $r$ (or both).
Surprisingly, this also includes the set of diagonal gates, as mentioned before and discussed in Section~\ref{sec:IQP}.
Thus, from the Pauli-orbit perspective, polynomial-size orbits occur only for
trivial orbits, such as those fixed by symmetries or associated with Lie
algebras of size independent of $n$, and for generating sets of
free-fermionic type.

We also mention that the Pauli Lie group setting is not the only case where one has additional information about the restricted reachability.
For instance, this was investigated in the presence of permutational symmetries in \cite{teng2026leveragingsymmetrymergingpauli}, e.g. under full permutation invariance or translation invariance.
In this case, if both the initial state/observable and time evolution satisfy the same permutation invariance, each Pauli string related by the action of the group symmetry evolves with the same coefficient, hence it is sufficient to track the coefficient of a single Pauli for each orbit, throughout the time evolution.
Of course, in general there might be exponentially many orbits, so that one may still have to track a large number of coefficients.

The cases discussed above concern arbitrary long-time or large-depth
reachability.
This is usually outside the standard scope of classical simulation, which
typically focuses on short-depth circuits or short-time evolution.
Then, it is clear that this strictly depends on the given generating set and their algebraic relations, hence its frustration graph.
Moreover, the generating-set dependent \emph{commutator graph} \cite{Diaz_GarciaMartin_Kazi_Larocca_Cerezo_2023,West_Dowling_Southwell_Sevior_Usman_Modi_Quella_2025} as in Def.~\ref{defn:comm:graph} also plays an important role in short time reachability, since two Paulis are adjacent if and only if there is a commutator by a generator which connects them, or equivalently, the action of a transvection.
Indeed, the question of short-time reachability was addressed in \cite{West_Dowling_Southwell_Sevior_Usman_Modi_Quella_2025} in the context of \emph{graph complexity} of a Pauli string under chaotic time evolution, which lower bounds the Krylov complexity.
This provides a measure of operator spreading in the Pauli picture, namely how
quickly an operator can reach all Pauli strings in its orbit, and is therefore
closely related to the efficiency of Pauli-propagation methods.
In the context of this work, we have not explored the structure of such graphs for arbitrary generating sets, since we were only concerned with the specific orbits, hence the entire connected components.
Lemma~\ref{lem:coloring:orbits} gives a graph-theoretic description of Pauli
reachability, analogous to the one for elements of Pauli Lie algebras.
It shows that reachability can be studied purely on the frustration graph by
using a specific coloring update rule, which corresponds to commutators or,
equivalently, to transvections.
Hence, it might also be possible to apply the coloring approach even in the short-time setting.

Finally, we consider possible implications for the well-known method of \emph{stabilizer-based} classical simulation.
Again, a series of stabilizer-based simulation algorithms exist, but we focus here on the main features and discuss what our description of Pauli Lie groups can tell us about their efficiency, and possibly if this information can lead to improved algorithms.
In such an algorithm, a stabilizer state is specified by a maximal set of $n$ commuting Pauli strings $S=\{S_i\}$, as $S_i\ket{\psi_S} = \ket{\psi_S}$, which can be encoded in a \emph{tableau} containing $(2n{+}1)\times(2n{+}1)$ binary variables.
For instance, this includes computational bases states, which are stabilized by $\{s_i Z_i\}_{i=1}^n$ for some $s_i\in\{\pm 1\}$.
Clifford gates act by updating this tableau, thus sending stabilizer states into stabilizer states.
Specifically, every stabilizer state admits a computational-basis decomposition
of the form \cite{Gross_Van_den_Nest_2008}
\begin{equation}\label{eq:stabilizer_states}
    \ket{\psi_S}
    =
    \frac{1}{\sqrt{\abs{V}}}
    \sum_{x\in V} \im^{d^T x}(-1)^{\QQ(x)}\ket{x+t}.
\end{equation}
Here $V\subseteq\F_2^n$ is a binary subspace, $d,t\in\F_2^n$, and
$\QQ$ is a binary quadratic form on $\F_2^n$, restricted to $V$ in
Eq.~\eqref{eq:stabilizer_states}.
The action of a Clifford on such states consists then in updating the subspace $V$, the vectors $d,t$ and the quadratic form $\QQ$.
Non-Clifford gates instead send stabilizer states into linear combinations of stabilizer states, similar as for Pauli strings.
Suppose that the circuit is decomposed into Clifford gates and non-Clifford
gates whose action admits a known expansion into stabilizer states, as in
Clifford+$T$ decompositions or Clifford decompositions of the unitaries
themselves.
Then simulation remains efficient as long as the resulting linear combination
contains only a small number of stabilizer states.
Obtaining such a decomposition may also be a non-trivial task for arbitrary circuits.
The minimal number of stabilizer states required to represent an arbitrary state is called the \emph{stabilizer rank}, and is one of many nonstabilizerness or \emph{magic} measures \cite{Bravyi_Smith_Smolin_2016}.
In general, such measures dictate the efficiency of stabilizer- or Clifford-based simulation algorithms \cite{Seddon_Regula_Pashayan_Ouyang_Campbell_2021}, though they may themselves be a quantity which is computationally hard to access, such that in practice one does not have precise guarantees for convergence and efficiency.
Under a similar approach, instead of decomposing states into stabilizers, one decomposes \emph{gates} into sums of Cliffords \cite{Bravyi2019simulationofquantum}, which works efficently in Clifford-dominated circuits.

In the cases of Pauli Lie groups and transvection groups, we have examined in detail the reachability of Pauli strings, though we have not considered the reachability of stabilizer states themselves.
Nevertheless, we can comment on potential implications.
For the real Clifford group, which is represented inside an orthogonal group,
reachability is more constrained.
Indeed, real Clifford operations preserve the class of real stabilizer states,
corresponding to the case $d=0$ in Eq.~\eqref{eq:stabilizer_states}.
Hence, one does not need to keep track of the corresponding $n$-dimensional binary vector.
More generally, it is clear that restricted Clifford groups lead to restricted reachability, which can be exploited for improved efficiency in a stabilizer based algorithm.
As another example, under the diagonal Clifford gates, any computational basis state remains invariant, whereas an initial state stabilized by $\{s_iX_i\}$ (e.g. $\ket{+}^{\otimes n}$) can be sent into any other state stabilized by $\{s_i'X_i\}$, for arbitrary $s_i,s_i'\in\{\pm 1\}$.

In this context, a certain family of diagonal, or IQP, circuits was considered in \cite{Codsi_2025}, generated by controlled-phase and T-gates.
This lead to improved simulation times to challenge quantum supremacy experiments, where they used a stabilizer-based decomposition, aided by ZX-calculus \cite{Kissinger_van_de_Wetering_2022}.
This was also examined in the stabilizer and sum-over-Clifford context in \cite{camillo2025symmetryacceleratedclassicalsimulationclifforddominated} for the diagonal, real and real-diagonal Clifford groups, highlighting significant advantages in simulation by taking advantage of the restricted reachability.

Purely diagonal Clifford+$T$ circuits can be computationally non-trivial
\cite{Bremner_2016}, but the restriction to discrete phases prevents them from
realizing arbitrary elements of the full Pauli Lie group of diagonal gates.
For instance, the phase gate S and the non-Clifford gate T do not generate the single qubit diagonal gates, but only the power of the T-gate, given that $T=S^2$.
Hence, no trivial `magic injection' can make diagonal Clifford gates universal in the diagonal gates.
On the other hand, the Clifford and real Clifford groups are maximal in their respective Pauli Lie groups (i.e. the $\SU(2^n)$ and $\SO(2^n)$), hence any gate which is in the Pauli Lie group but not in the underlying Clifford group leads to the full Lie group. 
Here one can show maximality using the fact that they are maximal \emph{finite} subgroups \cite{Nebe_Rains_Sloane_2001} and the 2-design property of these groups \cite{Sawicki_2017}.
Due to the non-commutativity of connected generating sets, together with the 3-design property, we give the following conjecture:
\begin{conj}
Let $\pgens\subseteq\PP_n$ be an arbitrary connected Pauli generating set and $e^{\lieg}$ its corresponding Pauli Lie group.
Then, $\cltvgroup{\pgens}$ is a maximal subgroup of $e^{\lieg}$ and the topological closure of the group generated by $\cltvgroup{\pgens}$ and $\{ \exp(\im\pi/8 P) | P\in\pgens\}$ coincides with $e^{\lieg}$.
\end{conj}
Under this conjecture, it is also clear that one can also generalize the Clifford+T decompositions to arbitrary Pauli Lie groups, by taking $\exp(\im\pi/8 P)$ as the generalized non-Clifford resources which naturally lie in the Pauli Lie group.
Correspondingly, one could combine this with a restriced Clifford-based algorithm for improved efficiency of simulation of arbitrary circuits or unitaries in Pauli Lie groups.

\section{Symmetries in Many-Body Systems}

In this section we discuss the Pauli framework in the context of many body systems, for which we are interested in characterizing symmetries, in the sense of commutant and invariant bilinear forms (or anti-unitary symmetries).
Specifically, we are interested here in $k$-\emph{local} generating sets, i.e.\ whose terms (Pauli strings here) live on $k$ sites.
In Section~\ref{sec:from_local_to_global}, we develop tools for computing symmetries on the entire system, using the \emph{local} symmetries, i.e.\ the symmetries of terms which live over $k$ consecutive sites.
In particular, we focus on certain structured families of 2-local generating sets with arbitrary connectivities (which are used in Section~\ref{sec:example:2-local_paulis}) and $k$-local generating sets in one dimension with various boundary conditions.
Then, we apply these to a non-trivial 3-local example, i.e.\ a representation of the free-fermions in disguise algebra \cite{Fendley_2019}.

\subsection{From Local to Global Symmetries}\label{sec:from_local_to_global}

In order to highlight the usefulness of our methods for structured sets, we show how, for general families of \emph{local} Pauli strings, one can directly compute the commutant and invariant bilinear forms.
Hence, combined with the local checking of the $\calE_6$-condition, this provides a full set of local conditions which determine the Lie algebra.

Namely, we shall be interested in the following families of local Pauli strings:
\begin{enumerate}\label{list:local_families}
    \item The \emph{2-local graph ansatz} for swap-symmetric $\calA\subseteq\PP_2$ and connected graphs $\graphH$ has the generating set
    \begin{equation}
        \pgens_{\graphH}(\calA) = \{ A_{uv} \mid A\in\calA, \{u,v\}\in\edges(\graphG)\}.
    \end{equation}
    \item The \emph{$k$-local translationally invariant ansatz} for arbitrary $\calA\subseteq\PP_k$ with $n\geq k\geq 2$ has the generating set
    \begin{equation}
        \pgens_n^{\circ}(\calA) = \{ A_i \mid A\in\calA, i\in[n]\},
    \end{equation}
    where $A_i = I^{\otimes (i-1)} \otimes A\otimes I^{\otimes (n-i)}$ is the Pauli string which starts at site $i$ and coincides with $A$ at sites $i$ to $i+k-1$, with indices defined modulo $n$.
    \item The \emph{k-local translationally invariant ansatz with open boundaries} for arbitrary $\calA_L,\calA_B,\calA_R\subseteq\PP_k$ with $n-2 \geq k\geq 2$ has the generating set
    \begin{align*}
            \pgens_n^{LR}(\calA) &=
            \begin{aligned}[t]
            &\{ A_i \mid A\in\calA_B, i\in\{1,\cdots,n-k\}\}\\
            &\cup \{ \calA_L\otimes I^{\otimes (n-k)}, I^{\otimes(n-k)}\otimes\calA_R\}.
            \end{aligned}
    \end{align*}
    Here $\calA_L$ and $\calA_R$ are the left and right \emph{boundary} terms (which are possibly empty), and $\calA_B$ corresponds to the \emph{bulk} elements. For compactness, we use the notation $\pgens_n^{LR}(\calA)$, where it is clear from context that there are three sets $\calA_L,\calA_B,\calA_R\subseteq\PP_k$ which define the ansatz. If $\calA_L=\calA_R=\emptyset$, we simply set $\calA = \calA_B$ and write the correspoding set as $\pgens_n(\calA)$.
\end{enumerate}
Here translationally invariant is in the sense that the interaction \emph{types} remain the same, but potentially not the \emph{coupling constants}. 
Concretely, we address Hamiltonians of the form
\begin{equation}\label{eq:k-local_hamiltonians}
    H = \sum_i \sum_{h\in\calA} c_{h,i}h_i,
\end{equation}
where $\calA$ is the same set of local interactions, but the coefficients $c_{h,i}$ potentially change from site to site.

We also make no assumption of the form of $\calA$ in the translationally invariant cases. 
This points to a potential lack of \emph{reflection symmetry} in the set, unlike the 2-local case.
Indeed, for $\calA=\{\text{ZX}\}$, $\pgens_n^{\circ}(\calA)$ consists of $\{Z_1X_2, Z_2X_3, \ldots, Z_nX_1\}$ but does not contain $X_1Z_2$ and its translates.
Hence, even though the cycle possess a reflection symmetry, the generating set $\pgens_n^{\circ}(\calA)$ does \emph{not}.
On the other hand, such a symmetry appears for
$\calA = \{\text{ZX, XZ}\}$.
In this case $\pgens_n^{\circ}(\calA) = \pgens_{\graphC_n}(\calA)$, so a
reflection symmetry of the cycle graph also preserves the generating set.
For $k=2$, these families specialize to the settings treated in
\cite{Wiersema_Kokcu_Kemper_Bakalov_2024,Kokcu_Wiersema_Kemper_Bakalov_2024}.
We keep this case in the discussion because it provides a useful benchmark for
the framework developed here and recovers the full symmetry data, namely both
the commutant and the invariant bilinear forms.
The same framework applies without conceptual change to the more general
$k$-local families above, including the cases $k\geq 3$.
We summarize the main result of this section as follows:
\begin{result}[Local-to-global symmetries, informal version; see
Propositions~\ref{prop:local_matching_cycles} and~\ref{prop:local_matching_paths}]
\label{result:from_local_to_global_symmetries}
Let $\pgens\subseteq\PP_n$ be one of the Pauli generating sets in (a)--(c),
specified by $k$-local data $\calA\subseteq\PP_k$ and by either an interaction
graph $\graphH$ or a system size $n$.
Suppose that the corresponding local commutant and local invariant bilinear
forms over $\PP_k$ are known.
Under the technical assumptions specified below, there is a procedure that
computes the global commutant $\commutant(\pgens)$ and the global invariant
bilinear forms $\bilinear(\pgens)$.
These global objects describe the symmetries of the Hamiltonians in
Eq.~\eqref{eq:k-local_hamiltonians} for arbitrary system size $n$ or
interaction graph~$\graphH$.
\end{result}

\begin{figure}
    \centering
    \includegraphics[width=0.8\linewidth]{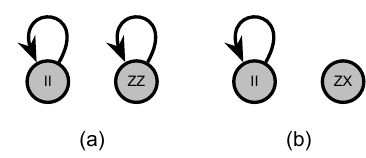}
    \caption{(a) $\graphD_\commutant(\{\text{IZ, ZI, XX}\})$. (b) $\graphD_\commutant(\{\text{IX, ZI, XX}\})$.}
    \label{fig:directed-example}
\end{figure}

In order to compute the commutant and bilinear forms for these families, we have the following basic characterization, which simply amounts to the fact that a global symmetry must restrict to a local symmetry:
\begin{prop}\label{prop:basic_local_matching}
Let $S=\bigotimes_{i=1}^nS_i$ and $B=\bigotimes_{i=1}^nB_i$ be two Pauli strings in $\PP_n$.
Consider a 2-local graph ansatz with swap-symmetric $\calA\subseteq\PP_2$ and connected graph $\graphH$. Then, $S\in\commutant(\pgens_\graphH(\calA))$ and $B\in\bilinear(\pgens_\graphH(\calA))$ iff
\begin{equation*}
         S_u\otimes S_v\in\commutant(\calA) \;\text{ and }\;
         B_u\otimes B_v\in\bilinear(\calA)
\end{equation*}
hold for all edges $\{u,v\}\in\edges(\graphH)$.
Given an arbitrary $k$-local translationally invariant ansatz with $\calA\subseteq\PP_k$, 
$S\in\commutant(\pgens_\graphH(\calA))$ and $B\in\bilinear(\pgens_\graphH(\calA))$ if and only if
\begin{align*}
        &S_i\otimes \cdots\otimes S_{i+k-1}\in\commutant(\calA) \text{ and}\\
        &B_i\otimes \cdots\otimes B_{i+k-1}\in\bilinear(\calA)
\end{align*}
hold for all sites $i\in[n]$.
Finally, given an arbitrary $k$-local translationally invariant ansatz with open boundaries and $\calA_L,\calA_B,\calA_R\subseteq\PP_k$,
$S\in\commutant(\pgens_\graphH(\calA))$ and $B\in\bilinear(\pgens_\graphH(\calA))$ if and only if
\begin{align*}
        &S_i\otimes \cdots\otimes S_{i+k-1}\in\commutant(\calA_B) \text{ and}\\
        &B_i\otimes \cdots\otimes B_{i+k-1}\in\bilinear(\calA_B)
\end{align*}
hold for all bulk terms $i\in[n{-}k]$
and
\begin{align*}
        &S_1\otimes \cdots S_k\in\commutant(\calA_L\cup\calA_B),\\
        &S_{n-k}\otimes \cdots S_n\in\commutant(\calA_R\cup\calA_B),\\
        &B_1\otimes \cdots B_k\in\bilinear(\calA_L\cup\calA_B),\\
        &B_{n-k}\otimes \cdots B_n\in\bilinear(\calA_R\cup\calA_B)
\end{align*}
hold for the left and right boundary terms.
\end{prop}
\begin{proof}
The proof follows trivially by using the fact that the commutation relation and invariant bilinear form relation factorize over tensor products, i.e.,
\begin{align*}
        &\comm{S_1\otimes S_2}{A_1\otimes I} = \comm{S_1}{A_1}\otimes S_2,\\
        &(B_1\otimes B_2)(A_1^T\otimes I) + (A_1\otimes I)(B_1\otimes B_2)\\
        &= (B_1A_1^T+A_1B_1)\otimes B_2. \qedhere
\end{align*}
\end{proof}
Given this observation, our objective is now to construct the \emph{global} symmetries starting from the purely \emph{local} symmetries.
Specifically, for a swap-symmetric set $\calA\subseteq\PP_2$ we say that $\commutant(\calA)$ and $\bilinear(\calA)$ are the \emph{local} commutant and invariant bilinear forms, whereas the $\commutant(\pgens_\graphH(\calA))$ and $\bilinear(\pgens_\graphH(\calA))$ are the \emph{global} commutant and invariant bilinear forms.
Similarly, this also applies for $\calA\subseteq\PP_k$ and the other families, including the left and right commutants and invariant bilinear forms for $\calA_L$ and $\calA_B$.
The following definition turns out to be particularly useful, and intimately related to constructing such global symmetries by going along paths or cycles:
\begin{defn}[Local Matching Graphs]\label{def:local_matchin_graphs}
We say that a $k$-local Pauli string, $k\geq2$, $A=A_1\otimes A_2\otimes\cdots\otimes A_k$ \emph{locally matches} with $B=B_1\otimes B_2\otimes\cdots\otimes B_k$ in $\PP_k$, $A \to B$, if $A_2\otimes\cdots\otimes A_k = B_1\otimes B_2\otimes\cdots\otimes B_{k-1}$.
This defines a relation on $\PP_k$, which is not symmetric in general.
Moreover, the restriction of this relation on any subset $\calB\subseteq\PP_k$ defines a directed graph, whose vertices are the elements of $\calB$ and where a directed edge between $A$ and $B$ exists if $A \to B$.
If we take $\calB = \cent_{\PP_n}(\calA) = \baslong{\commutant(\calA)}$, we denote the corresponding directed graph as $\graphD_\commutant(\calA)$. If $\calB=\bilinear_{\PP_n}(\calA)$, we denote it as $\graphD_\bilinear(\calA)$.
\end{defn}
Notice that the directed graph in this setting may also have \emph{self-loops}, when $A \to A$.
To see how this applies to our setting, consider the example of $\calA=\{\text{IZ, ZI, XX}\}$ on the path graph $\graphH=\graphP_n$.
Then, the local commutant is spanned by II and ZZ.
The corresponding directed graph $\graphD_\commutant(\calA)$ then has two self-loops corresponding to II and ZZ (see Fig.~\ref{fig:directed-example}(a)).
Hence at each pair of sites we must have that an element $S$ in the commutant restricts to either II or ZZ.
If the restriction at sites $1$ and $2$ is ZZ, it is clear that the restriction at sites $2$ and $3$ must also be ZZ, and so on for the entire chain.
This highlights the basic idea of how to `locally match' locally commuting terms to form a single globally commuting term, and the significance of the graphs $\graphD_\commutant(\calA)$ and $\graphD_\bilinear(\calA)$.
Indeed, when going from a site to another, the locally commuting terms must precisely match with respect to the relation defined in Def.~\ref{def:local_matchin_graphs}.

As another example, consider $\calA=\{\text{IX, ZI, XX}\}$ in the case with open boundary conditions (no boundary terms).
The local commutant is now spanned by II and ZX (see Fig.~\ref{fig:directed-example}(b)).
Now, if the restriction at sites $1$ and $2$ is ZX, the restriction at sites $2$ and $3$ cannot be either II or ZX, hence no such term exists in the global commutant.
This comes down to the fact that there is no directed edge from ZX to other elements in $\graphD_\commutant(\calA)$.
In particular, a strict requirement for some element to exist in the commutant, which restricts to some Pauli at a certain edge, is that one can construct arbitrary sequences of directed edges inside $\graphD_\commutant(\calA)$.
Since this has finite size and we are interested in arbitrary sizes $n$, then such arbitrary elements will be possible if and only if the local commutant graph has (directed) \emph{loops}.
As a special case, one always has the self-loop II, which produces the global identity $\In$ in the commutant.

Hence, given Proposition~\ref{prop:basic_local_matching} and the above examples, we see that finding the global commutant, starting from the local one, amounts to precisely finding all ways such that local symmetries \emph{match} on the overlap.

\begin{figure}
    \centering
    \includegraphics[width=\linewidth]{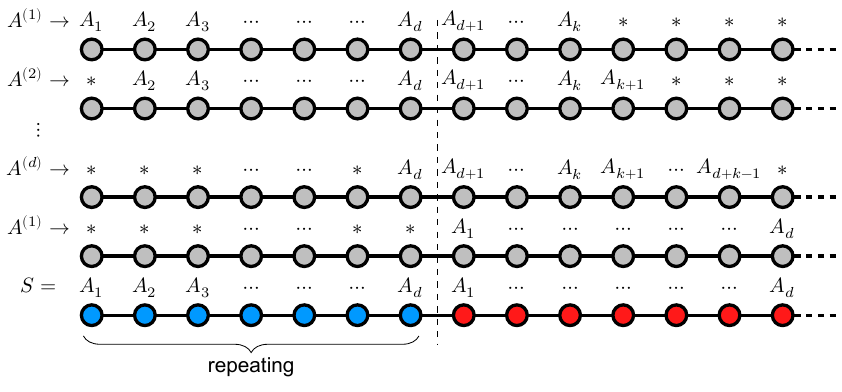}
    \caption{Minimal repeating word defined by a directed loop $(A^{(1)},A^{(2)},\ldots,A^{(d)})$ when $d\leq k$.
    Since $A^{(1)} \to A^{(2)}$, the first $k-1$ letters of $A^{(2)}$ must be of the form $A_2\cdots A_k$, and the $k$-th latter of $A^{(2)}$ is denoted as $A_{k+1}$.
    Letters in the following words are denoted in the same way by adding to $A^{(i+1)}$ a new independent letter at its $k$-th position, compared to the $k-1$ letters of $A^{(i)}$.
    By the loop constraint, $A^{(d)} \to A^{(1)}$, hence $A_{d+1}\cdots A_{d+k-1}=A_1\cdots A_{k-1}$.
    An element in the global commutant or invariant bilinear forms consists of repetitions of this word along the chain, subject to boundary conditions.}
    \label{fig:loop_d_leq_k}
\end{figure}

Now, for an arbitrary directed graph over $N\leq \abs{\PP_k} = 4^k$ vertices, there are two possibilities:
\begin{enumerate}
    \item There exist self-intersecting directed loops.
    \item There exist only disjoint directed loops, which have at most size $N$.
\end{enumerate}
Then, consider the elements which are in a directed loop $(A^{(1)},A^{(2)},\cdots,A^{(d)})$ of length $d$ with $A^j\in\PP_k$.
If $d\leq k$, then we have that the loop is completely specified by a smaller Pauli string $A_1\otimes \cdots\otimes A_d\subseteq\PP_d$, given the constraints imposed by the loop, see Fig.~\ref{fig:loop_d_leq_k}
In particular, the first $A^{(1)}$ already contains the minimal repeating word as follows 
$$(A_1\otimes \cdots \otimes A_d)^{\lfloor d/k\rfloor }\otimes (A_1\otimes \cdots \otimes  A_{k\bmod d}),$$ 
i.e. the smaller string of length $d$ repeats inside the first term as many times as it fits inside $k$.
For instance, if $k=3$ and $d=2$, the loop contains elements of the form $\{A_1A_2A_1, A_2A_1A_2\}$, with the minimal word being $A_1A_2$ (or $A_2A_1$, depending on the starting point of the loop).
Notice that the element $A_1\cdots A_d$ is defined depending on the initial reference element, but if we start at any other point along the loop we can simply take the cyclic translation of $A_1A_2\cdots A_d$, such as $A_2\cdots A_dA_1$ and so on.

\begin{figure}
    \centering
    \includegraphics[width=\linewidth]{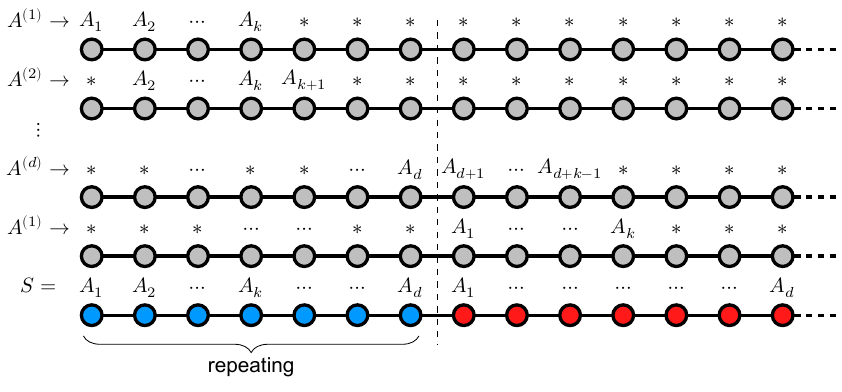}
    \caption{Minimal repeating word defined by a directed loop $(A^{(1)},A^{(2)},\ldots,A^{(d)})$ when $d>k$.
    The notation and construction follows Fig.~\ref{fig:loop_d_leq_k}.}
    \label{fig:loop_d_geq_k}
\end{figure}

If $d>k$, we have instead that the `minimal word' $A_1\otimes \cdots\otimes A_d$ of length $d$
is defined via multiple words in the loop, see Fig.~\ref{fig:loop_d_geq_k}.
For instance, the loop contains in the case of $k=2$ and $d=3$ the elements of the form
$$\{A_1A_2, A_2A_3, A_3A_1\}.$$
Here we can take the minimal word $A_1A_2A_3$ by arbitrarily choosing an initial element in the loop.

Let us first discuss the 2-local case, which lives on arbitrary connected graphs.
In this case, with swap-symmetric $\calA\subseteq\PP_2$, one can also check (for the commutant, see e.g.\ \cite{Wiersema_Kokcu_Kemper_Bakalov_2024}) that we have the following restricted setting for the directed graphs $\graphD_\commutant(\calA)$ and $\graphD_\bilinear(\calA)$, independently of the size of the commutant/invariant bilinear forms:
\begin{enumerate}
    \item There exist self-intersecting directed loops
    \item There exist only disjoint directed loops, which have at most size $2$
\end{enumerate}
Then, we can prove the following lemma when only disjoint loops exist:
\begin{lem}\label{lem:local_matching_graphs}
Let $\calA\subseteq\PP_2$ be swap-symmetric and consider a connected graph $\graphH$ on $n\geq 4$ vertices.
Let $C_1\subseteq\commutant(\calA)$ be the set of vertices which are self-loops in $\graphD_\commutant(\calA)$, and $C_2\subseteq\commutant(\calA)$ the set of vertices which are in loops of size $2$.

If $\graphD_\commutant(\calA)$ has only directed loops of length at most two (in particular, they do not intersect), the Pauli basis of the global commutant, $\commutant(\pgens_{\graphH}(\calA))$, is the following:
\begin{enumerate}
    \item $\{ A^{\otimes n} \}_{A\otimes A\in C_1}$ if $\graphH$ is non-bipartite.
    \item $\{ A^{\otimes n} \}_{A\otimes A\in C_1} \cup \{ A^{\otimes \bipleft}\otimes B^{\otimes \bipright} \}_{A\otimes B\in C_2}$ if $\graphH$ is $(\bipleft,\bipright)$-bipartite.
\end{enumerate}
Similarly, denote with $C_1^b$ and $C_2^b$ the vertices of directed loops of length one and two in $\graphD_\bilinear(\calA)$.
Then, $\bilinear(\pgens_{\graphH}(\calA))$ admits the following Pauli basis:
\begin{enumerate}
    \item $\{ A^{\otimes n} \}_{A\otimes A\in C_1^b}$ if $\graphH$ is non-bipartite.
    \item $\{ A^{\otimes n} \}_{A\otimes A\in C_1^b} \cup \{ A^{\otimes \bipleft}\otimes B^{\otimes \bipright} \}_{A\otimes B\in C_2^b}$ if $\graphH$ is $(\bipleft,\bipright)$-bipartite.
\end{enumerate}
\end{lem}
\begin{proof}
We first consider the Pauli basis of the commutant, i.e. $S = \bigotimes_{v=1}^nS_v\in\commutant(\pgens_{\graphH}(\calA))$.
Denote with $\commalg_2$ the local commutant of $\calA$.

Since $\graphH$ is connected, consider a connected path starting at vertex $1$ and ending at some other vertex $u\neq 1$, via a sequence of edges $(u_1,u_2)(u_2,u_3)\cdots(u_{p-1},u_p)$ with $u_1=1$ and $u_p=u$.
The restriction of $S$ to this path must then take the form $S_{u_1}S_{u_2}\cdots S_{u_p}$ where $S_{u_i}S_{u_{i+1}} \in\commalg_2$ by Proposition~\ref{prop:basic_local_matching}.
Then, if $S_{u_1}\otimes S_{u_2} = A\otimes B\in\commalg_2$, we must have $S_{u_2} \otimes S_{u_3} = B\otimes C\in\commalg_2$.
Equivalently, they need to locally match $S_{u_1}\otimes S_{u_2} \to S_{u_2}\otimes S_{u_3}$.
Then we have two possibilities: (1) $S_{u_1}\otimes S_{u_2}$ forms a self-loop, hence $S_{u_2} = S_{u_1}$; (2) $S_{u_1} \otimes S_{u_2}$ and $S_{u_2}\otimes S_{u_3}$ form a loop of length two inside $\graphD$.

If (1) $S_{u_1}\otimes S_{u_2} = A\otimes A$ forms a self-loop, then $S$ restricts to $A^{\otimes p}$ on the path, and it is clear that this extends to all vertices of the graph.
Hence, $A^{\otimes n}\in\baslong{\commalg}$ for $A\in C_1$.
If (2) $S_{u_1}\otimes S_{u_2} = A\otimes B$ is in a loop of length two, we distinguish between bipartite and non-bipartite graphs.

If $\graphH$ is non-bipartite, then it admits a loop of odd length $(u_1,u_2)(u_2,u_3)\cdots (u_{2p},u_{2p+1})$ with $u_{2p+1}=u_1$.
If $S$ restricts to $A\otimes B$ at $u_1,u_2$, then for $i\in[2p]$, the restriction at $u_i,u_{i+1}$ must be of the form $A\otimes B$ if $i$ is even and $B\otimes A$ if $i$ is odd, and in particular $S_1 = A$ and $S_{2p+1} = B$.
Clearly, this is possible if and only if $A=B$, or $A\otimes B$ forms a self-loop, which we have already discussed.
This proves the non-bipartite case (a).

For the bipartite case, such a obstruction is not possible. In particular, if we choose any connected path $(u_1,u_2)(u_2,u_3)\cdots (u_{d-1},u_d)$ with $u_1=1$ and $u_d=u$, the odd vertices $u_{2i-1}$ and even vertices $u_{2i}$ must lie respectively in two distinct partitions of the the graph.
Hence, if $A\otimes B$ lies in a loop of length $2$, $A^{\otimes \bipleft}\otimes B^{\otimes \bipright}$ is in the commutant.
No other element can exist, which proves the non-bipartite case (b).

For the bilinear forms, we can again apply Proposition~\ref{prop:basic_local_matching} and repeat the same argument, with the restriction lying in the local invariant bilinear forms $\bilinear(\calA)$.
This concludes the proof.
\end{proof}

We now prove the analogous statement for the translationally invariant case (periodic boundary condition):
\begin{prop}\label{prop:local_matching_cycles}
Let $\calA\subseteq\PP_k$ and assume $\graphD_\commutant(\calA)$, $\graphD_\bilinear(\calA)$ consists of disjoint directed loops of maximal length $d_{\max}$ and $d_{\max}'$, respectively.

Let $C_d\subseteq\commutant(\calA)$ be the set of vertices which are in directed loops of size $d$ in $\graphD_\commutant(\calA)$, $d\leq d_{\max}$.
Then, for each element $S^{\loc}\in C_d$ in a loop of length $d$, denote with $S_1S_2\cdots S_d$ the minimal repeating word as defined in Figs.~\ref{fig:loop_d_leq_k} and~\ref{fig:loop_d_geq_k}.
Then, for $n\geq d_{\max}$, the Pauli basis of the global commutant is given by
\begin{equation}\label{eq:local_to_global_on_cycle}
    \commutant(\pgens_n^\circ(\calA)) = \bigcup_{C_d, d|n}\{ (S_1S_2\cdots S_d)^{\otimes n/d} \}_{S^{\loc}\in C_d},
\end{equation}
where $d|n$ means that $d$ is a divisor of $n$.

Similarly, let $C_d^b\subseteq\bilinear(\calA)$ be the set of vertices which are in directed loops of size $d\leq d_{\max}'$ in $\graphD_\bilinear(\calA)$.
Then, for each element $B^{\loc}$ in a loop of length $d$, denote with $B_1B_2\cdots B_d$ the minimal repeating word as defined in Figs.~\ref{fig:loop_d_leq_k} and~\ref{fig:loop_d_geq_k}.
Then, for $n\geq d_{\max}'$, the Pauli basis of the global invariant bilinear forms is given by
\begin{equation}
    \bilinear(\pgens_n^\circ(\calA)) = \bigcup_{C_d^b, d|n}\{ (B_1B_2\cdots B_d)^{\otimes n/d} \}_{B^{\loc}\in C_d^b},
\end{equation}
where $d|n$ denotes that $d$ is a divisor of $n$.
\end{prop}
Hence, we see that the distinction between bipartite and non-bipartite in this setting generalizes to whether the (here unique) cycle has a length such that the repeating minimal words of length $d$ fit precisely inside the cycle.
\begin{proof}
We first consider the commutant, the Pauli basis of the commutant, i.e.\ $S = \bigotimes_{v=1}^nS_v\in\commutant(\pgens_{\graphH}(\calA))$.
Denote with $\commalg_k$ the local commutant of $\calA$.

By definition, the elements in Eq.~\eqref{eq:local_to_global_on_cycle} restrict to elements in the local commutant for each consecutive set of $k$ sites, hence they lie in the commutant.
We now show that these are also unique.

Indeed, assume that $S$ restricts to some $S^{\loc}$ at the first $k$ sites.
Then, since there is only one loop of length $d$ which contains $S^{\loc}$, the only possible successive elements along the cycle must be precisely those coming from the loop.
If $n$ is a multiple of $d$, then $S$ is as in Eq.~\eqref{eq:local_to_global_on_cycle}.
If $n$ is not a multiple of $d$, consider the directed loop $(S^{\loc}, S_2^{\loc}, \ldots S_d^{\loc})$ in $\graphD_\commutant(\calA)$ which starts with $S^{\loc}$.
If $n$ is not a multiple of $d$, then the local restriction starting at site $n$ is not $S_d^{\loc}$.
At the periodic boundary, the restriction for $n \in \{1,\ldots,k-1\}$ must match the
restriction for $n \in\{1,\ldots,k\}$.
This forces the predecessor of $S^{\loc}$ in the directed loop to be
$S_d^{\loc}$, which fails when $d\nmid n$.
Hence no such global element exists.

One repeats the analogous argument for the invariant bilinear forms, which concludes the proof.
\end{proof}

Finally, we prove the case with open boundary conditions:
\begin{prop}\label{prop:local_matching_paths}
Let $\calA_B\subseteq\PP_k$ and assume $\graphD_\commutant(\calA_B)$, $\graphD_\bilinear(\calA_B)$ consists purely of disjoint directed loops.

Let $C_d\subseteq\commutant(\calA_B)$ be the set of vertices which are in directed loops of size $d$ in $\graphD_\commutant(\calA_B)$.
Then, for each element $S^{\loc}$ in a loop of length $d$, denote with $S_1S_2\cdots S_d$ the minimal repeating word as defined in Figs.~\ref{fig:loop_d_leq_k} and~\ref{fig:loop_d_geq_k}, with respect to its unique loop.

Then, if $\calA_L=\calA_R = \emptyset$, for $n\geq k$, the Pauli basis of the global commutant $\commutant(\pgens_n(\calA_B))$ is
\begin{equation}\label{eq:local_to_global_on_path}
    \bigcup_d\{ (S_1S_2\cdots S_d)^{\otimes \lfloor n/d \rfloor} (S_1S_2\cdots S_{{n\bmod d}} \}_{S^{\loc}\in C_d}.
\end{equation}
In particular, if the local commutant has dimension $2^{2m+r}$ for some $m$, $r$, so does the global commutant.

Similarly, let $C_d^b\subseteq\bilinear(\calA_B)$ be the set of vertices which are in directed loops of size $d$ in $\graphD_\bilinear(\calA)$.
Then, for each element $B^{\loc}$ in a loop of length $d$, denote with $B_1B_2\cdots B_d$ the minimal repeating word as defined in Figs.~\ref{fig:loop_d_leq_k} and~\ref{fig:loop_d_geq_k}, and denote with $B_d^{\loc}$.
Then, the Pauli basis of the global invariant bilinear forms $\bilinear(\pgens_n(\calA))$ is
\begin{equation}
    \bigcup_d\{ (B_1B_2\cdots B_d)^{\otimes \lfloor n/d \rfloor} (B_1B_2\cdots B_{{n\bmod d}} \}_{B^{\loc}\in C_d^b}.
\end{equation}

If $\calA_L$, $\calA_B$ are not empty, then the Pauli basis of the commutant for $\pgens_n^{LR}(\calA)$ consists of all elements $S$ in the case without boundary elements, subject to the condition that there are some elements $S_L\in\commutant(\calA_L\cup\calA_B)$ and $S_R\in\commutant(\calA_R\cup\calA_B)$ such that the restriction of $S$ to the left and right boundary is $S_L$ and $S_R$, respectively.
From the point of view of the local matching relation, it means that $S_L$ and $S_R$ must be in the same loop in $\graphD_\commutant(\calA_B)$, and the corresponding element starts in $S_L$ and ends in $S_R$.

The corresponding statement holds for the invariant bilinear forms $\bilinear(\pgens_n^{LR}(\calA))$.
\end{prop}
\begin{proof}
It suffices to notice that in the case of open boundary conditions, the last elements in the chain need not locally match the first ones (unlike the periodic boundary case).
Consider first the case with no boundary terms. 
For each possible starting element $S^{\loc}\in C_d$, for some $d$, there is a unique corresponding directed loop, given that directed loops are disjoint.
We can then require that $S\in\commutant(\pgens_n(\calA))$ restricts to $S^{\loc}$ on the first $k$ sites.
Then, we can move along the path, which results in the word $S_1S_2\cdots S_d$ repeating a certain number of times, plus a remainder on the right side of the chain.
The resulting unique element associated to $S^{\loc}$ is as in Eq.~\eqref{eq:local_to_global_on_path}.
Additionally, if there are boundary terms, the first $k$ terms must lie in $\commutant(\calA_L\cup\calA_B)$, and the last $k$ terms must lie in $\commutant(\calA_R\cup\calA_B)$.
The proof follows the same idea for the invariant bilinear forms.
\end{proof}

\subsection{Free-Fermions in Disguise}\label{sec:ffd_free_fermions_in_disguise}

For free-fermionic mappings beyond line graphs \cite{Chapman_Flammia_2020}, a new family of free-fermionic mappings was discovered \cite{Fendley_2019,Elman_Chapman_Flammia_2021,Fendley_Pozsgay_2024}, with the original example being referred to as \emph{free fermions in disguise} (FFD).
The techniques used to derive the free-fermionic mapping for free fermions in disguise go beyond simple Pauli-to-Majorana mappings and yield \emph{Hamiltonian and parameter-dependent} free-fermionic mappings.
In particular, \cite{Elman_Chapman_Flammia_2021} provided a generalization of the specific model in \cite{Fendley_2019} to the set of graphs which are free of \emph{claws} and \emph{even-holes}.
Unlike line graphs of multigraphs (or line graphs of trees), this property is not invariant under $t$-equivalence, or sequences of contractions.
As an example, a cycle graph $\graphC_n$ of odd length, which is claw and even-hole free, is $t$-equivalent to the blown-up path graph $\graphP_{n-2,2}$, which instead contains a claw.
In particular, this shows that free-fermionic mappings of the FFD type cannot be generalized to the entire Lie algebra and Lie group.
Indeed, versions of the FFD mapping for circuits have only worked for specific circuit structures \cite{Fukai_Pozsgay_2025}.

In order to better understand the gap between these mappings and our graph- and symmetry-based approach, we now use our toolset to analyse the FFD example in its original representation.
Namely, we show the \emph{lack} of free-fermionic mapping in the sense of Definition~\ref{defn:free_fermionic_mappings}, and study the symmetry structure, a problem which was also studied in \cite{Vernier_Piroli_2026} in the context of eigenvalue degeneracies and correlation functions.
Moreover, we also consider the \emph{anti-unitary} symmetries, viewed as invariant bilinear forms.
Such symmetries have played an important role in the classification of phases of matter \cite{Altland_Zirnbauer_1997,Ryu_Schnyder_Furusaki_Ludwig_2010,Uhlmann_2016} and impose constraints on the spectrum of Hamiltonians \cite{Wigner1932}.
Hence, we also expect them to play a role in constraining the possible spectral and dynamical properties of the free-fermionic solution, though we only deal here with the problem of classification.
Indeed, in the context of the Pauli representation of the FFD algebra, we shall show that they provide quasi-universal Lie algebras, as a consequence of the lack of `ordinary' free-fermionic mappings.

Let us first write explicitly the algebra of FFD operators $\{h_i\}_{i=1}^n$, which is specified by the relations:
\begin{subequations}\label{eq:FFD_algebra}
    \begin{align}
        h_i^2 &= 1\quad \text{for } i\in [n],\\
        \acomm{h_i}{h_j} &= 0\quad \text{for } 0< \abs{j-i} \leq 2,\\
        \comm{h_i}{h_j} &= 0\quad \text{for } \abs{j-i} > 2.
    \end{align}
\end{subequations}
Specifically, we focus on the representation of this algebra over $n$ qubits as discussed in \cite{Fendley_2019,Vernier_Piroli_2026}, i.e.,
\begin{equation}\label{eq:standard_FFD_representation}
        \pgens_{\mathrm{FFD}} = \{ Z_{i-2}Z_{i-1}X_i\}_{i=1}^n.
\end{equation}
We consider three possible boundary conditions: (i) open boundary \emph{with} cutting, by setting $Z_{-1} = Z_0 = I$, hence cutting off any term over the left boundary; (ii) periodic boundary, by setting $Z_i = Z_{i+n}$ and indices taken modulo $n$; (iii) open boundary \emph{without} cutting, i.e.\ by ignoring $Z_{i-2}Z_{i-1}X_i$ for $i\in\{1,2\}$.
Reference~\cite{Vernier_Piroli_2026} only deals with (i) the open boundary case with cut left boundary, though our framework allows us to conveniently deal with all three settings.
We shall however not delve into the details of Hilbert space structure and free-fermionic mappings, since we are only interested in the symmetry properties, i.e.\ finding a set of generators for the commutant (and invariant bilinear forms).

\begin{figure}
    \centering
    \includegraphics[width=0.7\linewidth]{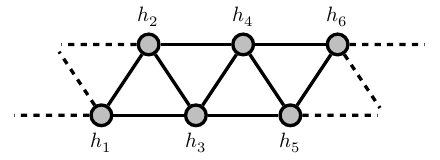}
    \caption{Frustration graph of the FFD ansatz contains a graph in the $\calE_6$ class for any boundary condition, assuming $n\geq 6$.}
    \label{fig:ffd-frustration-graph}
\end{figure}

By definition of FFD algebra in Eq.~\eqref{eq:FFD_algebra}, one can check that any six consecutive terms 
in any Pauli representation
realize a frustration graph in the $\calE_6$ class, i.e.\ one of the $32$ forbidden subgraphs for
line graphs of multigraphs (refer to Thm.~\ref{thm:E6_free_characterization_line_graphs} and Figs.~\ref{fig:ffd-frustration-graph} and~\ref{fig:e6-class-atlas}).
The only exception is in the case of periodic boundary conditions with $n\leq 6$, in which case there are additional connections.
Hence, the FFD set does not admit a free-fermionic mapping in the sense of Definition~\ref{defn:free_fermionic_mappings} for any $n\geq 6$, which was expected.
Indeed, since the commutation relations do not change while going along the graph, any sequence of six consecutive terms in $\pgens_{\mathrm{FFD}}$ produces the same graph in $\calE_6$.
Then, it is clear that such mappings in the sense of \cite{Fendley_Pozsgay_2024,Elman_Chapman_Flammia_2021} cannot be invariant under $t$-equivalence, or invariant properties of the Lie algebra or group, since otherwise one would be able to efficiently compute the spectral properties of exponentially large matrices.

We now look at the symmetries, hence the matrix algebra (which is an example of bond algebra \cite{Nussinov_2009,Cobanera01102011,Moudgalya_2023}) and its commutant.
Notice that the FFD generators are all algebraically independent in any representation, given that they overlap over different sites.
This also includes the boundary terms, since it is clear that products of the 3-local bulk generators cannot cancel out to obtain something purely on the boundary.
Hence, for all $n$, the matrix algebra must have dimension $2^n$ (periodic and open boundary with cutting) or $2^{n-2}$ (open boundary without cutting), though the rank and dimension of the center do not necessarily have the same dimension for all $n$.

With regards to the symmetries, it was shown in \cite{Fendley_Pozsgay_2024} that the Pauli strings in Eq.~\eqref{eq:standard_FFD_representation} admit an \emph{ancillary} set of Pauli strings which also realize the FFD algebra.
Explicitly, these are just the reflection, i.e.,
\begin{equation}
    \pgens_{\mathrm{FFD}}^r = \{ X_iZ_{i+1}Z_{i+2}\}_{i=1}^n,
\end{equation}
where the boundary terms $i\in\{n-1,n\}$ depend on the boundary conditions.
Since this generating set is also connected, it cannot lie in the center of $\pgens_{\mathrm{FFD}}$, though it lies in its commutant.
Notice that even though each individual FFD set is algebraically independent, the combined set is not in general, hence they can share abelian symmetries.

We describe here the remaining elements in the commutant of $\pgens_{\mathrm{FFD}}$ for all boundary cases, which must in fact lie in the joint commutant of the two sets.
This can be seen more clearly using the picture of Fig.~\ref{fig:partition-commutant}, up to isomorphism. 
We have that $\pgens_{\mathrm{FFD}}$ is generated by some logical and phase part of Pauli strings in the chain, whereas $\pgens_{\mathrm{FFD}}^r$
lies in the commutant and is generated by some uncontrollable pairs and phase terms shared with $\pgens_{\mathrm{FFD}}$. 
The shared commutant then contains all phase terms plus the shared uncontrollable pairs. 
Hence, the only redundancy in describing the joint symmetries of $\pgens_{\mathrm{FFD}}$ and $\pgens_{\mathrm{FFD}}^r$ must come from the center.

The combined generating set $\pgens_{\mathrm{FFD}}\cup \pgens_{\mathrm{FFD}}^r$ then fits within the families described in Section~\ref{sec:from_local_to_global}, depending on the specific boundary conditions.
Namely, we have the bulk, left, and right boundary terms
\begin{equation*}
        \calA_B = \{\text{ZZX,\! XZZ}\},
        \calA_L = \{\text{XII,\! XZI}\},
        \calA_R = \{\text{IZX,\! IIX}\}.
\end{equation*}
Thus we can now compute the symmetries of the combined generating sets, depending on the boundary conditions, using the notation of Section~\ref{sec:from_local_to_global}: periodic boundary $\pgens_n^\circ(\calA_B)$; open boundary without cutting $\pgens_n(\calA_B)$; open boundary with cutting $\pgens_n^{LR}(\calA)$.

\begin{figure}
    \centering
    \includegraphics[width=\linewidth]{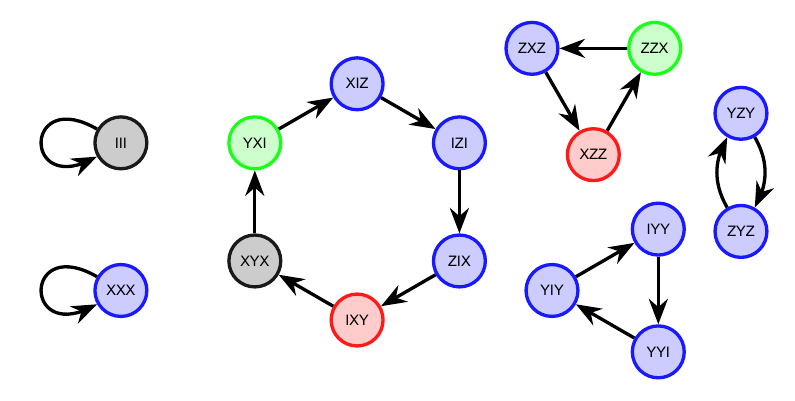}
    \caption{The local commutant of the doubled FFD ansatz and its directed graph $\graphD_\commutant(\{\text{XZZ, ZZX}\})$.
    All vertices are contained in the commutant of the bulk term, $\commutant(\text{XZZ, ZXX})$.
    Vertices in red are those which also belong to the commutant of the cut left boundary.
    Vertices in green for the right boundary.
    Vertices in black belong to the bulk and both boundaries. Vertices in blue belong purely to the bulk term.
    There is a directed edge $A\to B$ whenever $A$ and $B$ \emph{locally match}, i.e., they are of the form $A = A_1A_2A_3$ and $B = B_1AB_2B_3$ with $A_2A_2 = B_1B_2$.}
    \label{fig:ffd-directed-graph}
\end{figure}

Hence, the construction of the joint commutant reduces to \emph{locally} matching the local commutant along the chain, depending on the boundary conditions.
Specifically, one finds that the local commutant of the bulk part $\calA_B$ has dimension $16$ and the corresponding directed `local-matching' graph $\graphD_\commutant(\calA_B)$ (defined in Def.~\ref{def:local_matchin_graphs}) is precisely made of six disjoint loops.
Refer to Fig.~\ref{fig:ffd-directed-graph} for the notation.

One can check that this satisfies the hypotheses of Propositions~\ref{prop:local_matching_cycles} and~\ref{prop:local_matching_paths}, which allow us to construct the global commutant purely from the global commutant.
The corresponding four minimal repeating words are then the following (up to cyclic translations):
\begin{align*}
    S^{(1)} &= \text{I}, & S^{(2)} &= \text{X},\\
    S^{(3)} &= \text{ZXZ}, & S^{(4)} &= \text{YIY},\\
    S^{(5)} &= \text{ZY}, & S^{(6)} &= \text{XIZIXY}.
\end{align*}
These words automatically provide the joint commutant of the two FFD copies in either the periodic boundary or open boundary.
The freest case is the open boundary case without cut terms $\pgens_n(\calA)$, since, for each word $S = S_1\cdots S_d$ of length $d$, we can find a corresponding Pauli string in the commutant as
\begin{equation}
    S^{\otimes \lfloor n/d\rfloor} S_1\cdots S_{n\bmod d},
\end{equation}
and, for all $n\geq 3$, we obtain
\begin{equation}
    \dim(\commutant(\pgens_n(\calA))) = 16.
\end{equation}
In the periodic boundary case instead we have a dependence on $n$ based on how each word or loop length $d\in\{1,2,3,6\}$ divides $n$.
The case $n=0\bmod 6$ is the freest case for the periodic boundary, since it coincides with the open boundary case, and each element in the joint commutant has
the form $S^{\otimes n/6}$, where $S$ is one of the possible $16$ words coming from the local commutant of the bulk.
More generally, for $n\geq 6$, the joint commutant dimensions in the periodic
case are summarized in Table~\ref{tab:ffd-commutant-dimensions}.

\begin{table}[t]
\caption{Joint commutant dimensions for the doubled free fermions in disguise
(FFD) ansatz, grouped by $n\bmod 6$. The second column gives the periodic
case $\pgens_n^\circ(\calA)$ on $n$ qubits, and the third column gives the
open boundary case with cut boundary terms $\pgens_n^{LR}(\calA)$ on $n$
qubits, namely with the boundary terms $\text{XZI}$ and $\text{IZX}$. 
For the open boundary case without boundary terms, $\pgens_n(\calA)$, the commutant has dimension $16$ for any $n\geq 3$.}
\label{tab:ffd-commutant-dimensions}
\centering
\footnotesize
\renewcommand{\arraystretch}{1.15}
\begin{tabular*}{\columnwidth}{@{\hspace{1mm}}c@{\extracolsep{\fill}}c@{\hspace{3mm}}c@{\hspace{1mm}}}
\hline\hline
\\[-3mm]
$n\bmod 6$
& $\dim(\commutant(\pgens_n^\circ(\calA)))$
& $\dim(\commutant(\pgens_n^{LR}(\calA)))$ \\[0.75mm]
\hline
\\[-3mm]
0 & 16 & 1\\
1 & 2  & 2\\
2 & 4  & 1\\
3 & 8  & 2\\
4 & 4  & 4\\
5 & 2  & 2\\[0.75mm]
\hline\hline
\end{tabular*}
\end{table}

Finally, we discuss the open-boundary case with cut boundary terms, which was the one covered in \cite{Vernier_Piroli_2026}.
We go through each possible case, by applying the possible initial and final positions, as denoted in Fig.~\ref{fig:ffd-directed-graph}.
Clearly, we can never get the the elements from the self-loop XXX (indeed, the boundary terms XZI and IZX break the parity symmetry), as well as from the loop of length two and one of the loops of length three.
From the remaining loop of length three we get only possible element, which must start with XZZ and finish with ZZX.
Hence, it can only appear when $n=1\bmod 3$ and $\text{XZZ}^{\otimes (n-1)/3}\text{X}$, since the other cases end with either XZZ (if $n{=}0\bmod 3$) or XZX (if $n{=}2\bmod 3$).
From the loop of length six we can potentially get up to four terms, depending on $n$, since we can choose two initial terms IXY and XYX
and two final terms XYX and YXI. 
We get the element which starts with XYX and ends with XYX only when $n=3\bmod 6$, given that the corresponding repeating word is XYXIZI, which results in $\text{XYXIZI}^{\otimes (n-3)/6}\text{XYX}$.
The element which starts with XYX and ends with YXI instead appears when $n=4\bmod 6$ and $\text{XYXIZI}^{\otimes (n-4)/6}\text{XYXI}$.
The element which starts with IXY and ends with XYX also appears when $n=4\bmod 6$ and $\text{IXYXIZ}^{\otimes (n-4)/6}\text{IXYX}$.
The remaining element which starts with IXY and ends in YXI instead appears when $n=5\bmod 6$, $\text{IXYXIZI}^{\otimes (n-5)/6}\text{IXYXI}$.
Thus, including the identity, the joint commutant dimensions for all $n\geq 3$
are summarized in Table~\ref{tab:ffd-commutant-dimensions}.

We highlight that the only irreducible cases happen with $n \bmod 6 \in \{0,2\}$ for the open boundary case with cutting.
Given that the two FFD copies are isomorphic and commuting, this means that the commutant of a single copy is spanned by the other, as well as the fact that they do not generate a center.
Hence, $\algclosure{\pgens_{\mathrm{FFD}}}$ is conjugate, after an invertible change of basis, to the matrix algebra $\matalg(2^{n/2},\C)\otimes I^{\otimes n/2}$.
Thus, in the decomposition of Eq.~\eqref{eq:Hilbert_space_partition},
$\pgens_{\mathrm{FFD}}$ naturally acts on a Hilbert space of dimension
$2^{n/2}$.
This does not directly match with the case $n \bmod 6 \in \{0,2\}$ in \cite{Vernier_Piroli_2026} given that they remove the operators `centered' at the right boundary, hence this case is $M \bmod 6 \in \{1,3\}$ in their notation.

\begin{figure}
    \centering
    \includegraphics[width=\linewidth]{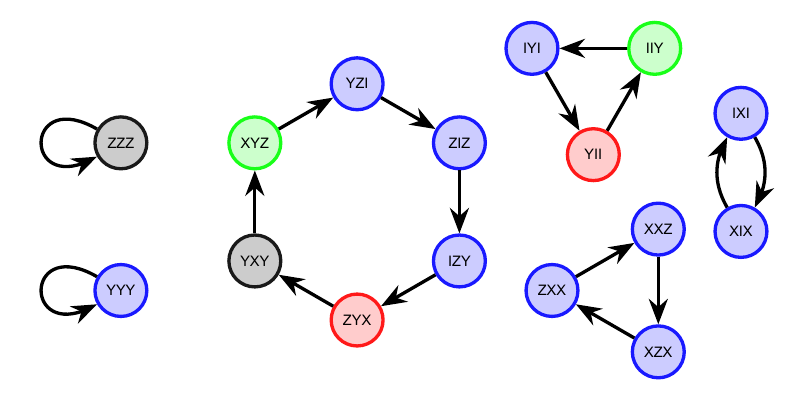}
    \caption{The local invariant bilinear forms of the doubled FFD ansatz $\bilinear(\text{XZZ, ZXX})$ and its directed graph $\graphD_\bilinear(\{\text{XZZ, ZZX}\})$. Notation is as in Fig.~\ref{fig:ffd-directed-graph}.}
    \label{fig:ffd-directed-bilinear}
\end{figure}

We also discuss the invariant bilinear forms for FFD, or its antiunitary symmetries.
As for the commutant, we can apply the local matching lemmas from Section~\ref{sec:from_local_to_global}.
The resulting graph for the local invariant bilinear forms is given in Fig.~\ref{fig:ffd-directed-bilinear}, where one finds the same structure.
This is not surprising, given that the bulk generators are Pauli strings of odd size, hence conserve the same invariant bilinear form as the single qubits, i.e.\ the one associated to the Pauli weight modulo $2$.
The corresponding (local) Pauli bilinear form is YYY, and all other bilinear forms are obtained by multiplying this by the (local) commutant.
Hence, YYY does not affect the matching relations.
Notice that this does not happen for arbitrary sets, as seen in the QAOA example (see Fig.~\ref{fig:directed-QAOA}).

At the level of the global invariant bilinear forms, we also have the same description in terms of the words coming from the loops.
Notice that $Y^{\otimes n}$ (hence the Pauli weight modulo $2$) is not conserved in the presence the cut boundary terms.
However, for all $n$ and all boundary case, $Z^{\otimes n}$ is still an invariant bilinear form, and it suffices to multiply the commutant by $Z^{\otimes n}$ to obtain all other bilinear forms.
Since this element again lies in a self-loop, it does not affect the local matching structure of the invariant bilinear forms compared to the commutant.
In particular, from the dynamical point of view, the standard FFD algebra representations can never produce strictly universal quantum computation over some number of qubits, given that they always lie in some orthogonal or symplectic Lie algebra, even in the absence of (linear) symmetries.

Then, for the invariant bilinear forms, it remains to establish the type of isomorphism class, as described in Corollary~\ref{cor:Canonical_Forms_Invariant_Bilinear_Forms} and Lemma~\ref{lem:isomorphism_class_affine_subspaces_quadratic_forms}.
Given that there are only finitely many bilinear forms, we can apply the
criterion based on the democratic Arf invariant from
Eq.~\eqref{eq:def:democratic_arf_affine_subspace}, as in the 2-local setting.
We start with the open boundary case without cutting.
For any $n$, the global bilinear forms arising from the self-loop at ZZZ and
from the loops containing ZXX and IXI are symmetric.
Hence, six bilinear forms are always symmetric.
By examining each case for $n\bmod 6$, one verifies that the bilinear forms are
of $+$ type except when $n\bmod 6=3$, i.e.,
\begin{equation*}
    \type(\bilinear(\pgens_n(\calA))) =
    \begin{dcases}
        0 & \text{if } n \bmod 6 = 3,\\
        + & \text{otherwise}.
    \end{dcases}
\end{equation*}
In the periodic boundary case we find instead:
\begin{equation*}
    \type(\bilinear(\pgens_n^\circ(\calA))) =
    \begin{dcases}
        + & \text{if } n \bmod 2 = 0,\\
        0 & \text{if } n \bmod 2 = 1.
    \end{dcases}
\end{equation*}
Finally, in the open boundary case with cutting, we find:
\begin{equation*}
    \type(\bilinear(\pgens_n^{LR}(\calA))) =
    \begin{dcases}
        0 & \text{if } n \bmod 6 = 1,\\
        + & \text{otherwise}.
    \end{dcases}
\end{equation*}
Then, in the open boundary case with cutting, when $n \bmod 6 \in \{0,2\}$, the Lie algebra generated by $\pgens_{\mathrm{FFD}}$ must in fact be $\so(2^{n/2})\otimes I^{\otimes n/2}$, given that its commutant is $I^{\otimes n/2}\otimes\matalg(2^{n/2},\C)$ and its invariant bilinear forms are of $+$ type, by Theorem~\ref{thm:full_classification_pauli_lie_algebras}.

For brevity, we do not compute here the full Lie algebra in each case and the corresponding symmetry decomposition.
Indeed, unlike \cite{Vernier_Piroli_2026}, we have not computed the precise contribution to the commutant and matrix algebra of the abelian and non-abelian part for either copy (as well as their relationship to parameter-dependent constants of motion and transfer matrices for the exact solution).
Moreover, given that the two generated sets are disconnected, albeit isomorphic, we cannot fully describe the Lie algebra purely from the joint commutant and invariant bilinear forms.
Nevertheless, we have shown how our tools can provide much information about a non-trivial but structured generating set, using a mix of our standard toolset combined with the specific structure of a local generating set to obtain precise information about the symmetries.
For instance, we have not distinguished between the contribution to the symmetries coming from the center of the FFD ansatz or its ancillary copy.
A graph-theoretic description was discussed in the original paper \cite{Fendley_Pozsgay_2024}.

\section{Clifford \texorpdfstring{$t$}{t}-Designs for Pauli Lie Groups}\label{sec:clifford_3_designs_groups}

We consider in this section the relation between the symmetries of a Pauli Lie group and Clifford transvection groups, in the language of t-designs.
In particular, we show for \emph{any} Pauli generating set that the corresponding Clifford transvection group is a 3-design for the generated Pauli Lie group.
Also, we show that this Clifford group is \emph{not} a 4-design for the corresponding Pauli Lie group.

Before delving into the results, we set some notation.
Also, we recall the corresponding definitions of the \emph{replica} or $t$-tensor power representation 
\begin{align*}
        &g^{\otimes t} = \underbrace{g \otimes g \otimes \cdots g}_{t\text{ times}} \;\text{ and }\;
        H^{\hotimes t} = \sum_{i=1}^t H_i\\
        &\text{with }\;
        H_i = \underbrace{\id \otimes \cdots \otimes\id}_{i-1\text{ times}} \otimes H\otimes \underbrace{\id \otimes \cdots\otimes \id }_{t-i\text{ times}}
\end{align*}
for a group and a Lie algebra.
Then, we have the following definitions of commutants for sets of such objects and immediate corollaries (see also Section~\ref{sec:commutant}):
\begin{defn}[Commutants]\leavevmode
\begin{enumerate}
    \item Tensor power commutant of a set of Lie algebra generators $\commutant_{\mathrm{Lie}}^{(t)}(\pgens) = \commutant(\{ H^{\hotimes t} | H\in\pgens\})$. We have that $\commutant_{\mathrm{Lie}}^{(t)}(\pgens) = \commutant_{\mathrm{Lie}}^{(t)}(\lie{\pgens})$.
    \item Tensor power commutant of a set of group generators $\commutant_{\mathrm{group}}^{(t)}(S) = \commutant(\{ g^{\otimes t} | g\in S\})$. We have that $\commutant_{\mathrm{group}}^{(t)}(S) = \commutant_{\mathrm{group}}^{(t)}(\groupclosure{S})$.
\end{enumerate}
\end{defn}
For (connected) Lie groups and their Lie algebras, we additionally have $\commutant_{\mathrm{Lie}}^{(t)}(\lieg) = \commutant_{\mathrm{group}}^{(t)}(e^{\lieg})$.
This follows from the fact that $\lieg$ and $e^{\lieg}$ share the same
irreducible representations, see
\cite[Proposition~4.5]{hall2015} or
\cite[Ch.~III, \S 6, no.~5, Prop.~13, Cor.~1]{BourbakiLie1989}.
Equivalently, this follows from the fact that $e^{\lieg}$ in any representation $\Phi$ is a power series in the induced representation $\phi$ of $\lieg$, hence $\Phi(e^{\lieg})$ commutes with the same operators as $\phi(\lieg)$. 
Viceversa, differentiating $\comm{\Phi(e^{t X})}{C}=0$ at $t=0$ gives that $\phi(X)$ commutes with $C$, hence the Lie algebra and Lie group have precisely the same commutant in any representation.

Given that we are only interested in exact $t$-designs, we can take the following characterization of a $t$-design $\calE\subseteq \lieG$ as a definition:
\begin{defn}[\cite{Ambainis_Emerson_2007,Gross_Audenaert_Eisert_2007,Roy_Scott_2009}]
Let $\lieG\subseteq\lieU(d)$ be a subgroup of the unitary group.
An ensemble $\calE\subseteq\lieG$ is said to be a unitary (exact) $t$-design for $\lieG$ if
\begin{equation}
    \commutant_{\mathrm{group}}^{(t)}(\calE) = \commutant_{\mathrm{group}}^{(t)}(\lieG)
\end{equation}
\end{defn}
An immediate consequence is that if $\calE$ is a $t$-design, it is also a $t'$-design for all $t'\leq t$.
Finally, we also recall the definition of a transvection with center $P\in\PP_n$ (see Section~\ref{sec:transvections_in_clifford_group}) as
\begin{equation}
    \tvu{P} = (I + \im P)/{\sqrt{2}}.
\end{equation}
We can now state the main result of this section:
\begin{thm}\label{thm:transvection_3_design_property}
Consider a set of Pauli strings $\pgens\subseteq\PP_n$, which generates a Pauli Lie algebra $\lieg = \lie{\pgens}$ and Pauli Lie group $e^{\lieg}$.
The Clifford transvection group $\cltvgroup{\pgens} = \groupclosure{\{\tvu{G}\}_{G\in \pgens}}$ is a 3-design for the Pauli Lie group $e^{\lieg}$.
\end{thm}
For the full Clifford group, Theorem~\ref{thm:transvection_3_design_property}
recovers the known unitary 3-design property \cite{Webb_2016,Zhu_2017}.
It also places related 3-design results for the real Clifford group
\cite{Hashagen2018realrandomized}, Clifford groups with Pauli symmetries
\cite{Mitsuhashi_Yoshioka_2023}, and the even matchgate Clifford group
\cite{Wan_2023} into a common generator-level framework.
The idea of the proof is straightforward, and is based on the duality between
matrix algebras and their commutant algebras, as well as the fact that
commutants of connected Lie groups and their Lie algebras coincide.
Hence, it suffices to show each generator of the Lie algebra
$\im G\in\im\pgens$ and each generator of the Clifford transvection group
$\cltvgroup{\pgens}$ produce the same matrix algebra in the tensor-power
representations.
For completeness, we also show the proofs for the 1- and 2-design property,
though this is automatically implied by the 3-design property.
We give the proof here instead of just stating the result since it does not
directly rely on arguments in \cite{Webb_2016} or \cite{Zhu_2017},
but only on the generators being Clifford transvections or Pauli rotations for the Pauli Lie group.
However, explicit knowledge of the commutants for the Pauli Lie group and
Clifford transvection groups does require knowledge of the Pauli orbits, as we
show in Section~\ref{sec:comms_and_invariant_subspaces}.
Indeed, it is a distinct proof precisely because no knowledge of the precise
groups is required, but only of their generators.
As such, it only relies on the knowledge of simple algebraic properties of
Pauli strings and basic properties of commutants.

\begin{proof}
Fix a generator $G\in\pgens$.
In the $t$-fold tensor-power representation, we write $G_i$ for the operator
acting as $G$ on the $i$th tensor factor and as $\id$ on all other tensor
factors.
Thus $G^{\hotimes t}=\sum_{i=1}^t G_i$.
The operators $G_i$ commute with one another and satisfy $G_i^2=\id$.
Let $S_j(x_1,\ldots,x_t)$ denote the elementary $j$th symmetric polynomial in
$t$ variables. When the variables are clear from context, we write simply
$S_j$. Then, we have that
\begin{align}
G^{\hotimes t} &= \sum_{i=1}^t G_i =  S_1(G_1,\ldots,G_t) \text{ and}
\label{eq:tensor:power:g}
\\
\tvu{G}^{\otimes t} &= \prod_{i=1}^t \frac{\In +\im G_i}{\sqrt{2}}
= \frac{1}{2^{t/2}}\sum_{\{n_i\}\in\{0,1\}^{\times t}} \im^{\sum_{i=1}^t n_i} \prod_{i=1}^t G_i^{n_i} 
\nonumber
\\
&= \frac{1}{2^{t/2}}\sum_{j=0}^t \im^j S_j(G_1,\ldots,G_t).
\label{eq:tensor_power_transvection_symmetric_polynomials}
\end{align}
If $t\leq 3$, we now show that, for each fixed $G\in\pgens$, the
unital algebras generated by $G^{\hotimes t}$ and by
$\tvu{G}^{\otimes t}$ both contain the elementary symmetric polynomials in
the commuting binary variables $\{G_i\}_{i=1}^t$.
Consequently, the sets $\{G^{\hotimes t},\id\}_{G\in\pgens}$ and
$\{\tvu{G}^{\otimes t},\id\}_{G\in\pgens}$ also generate the same matrix algebra.
By the double commutant theorem \cite{bresar2014,lorenz2008}, recalled in
Lemma~\ref{lem:general_commutant}, this is equivalent to showing that the
commutants of the two sets of matrices $S$ and $S'$ coincide, i.e.,
\begin{align*}
&\commutant(S) = \commutant(S') \\
\Longleftrightarrow \;&
\commutant(\commutant(S)) = \commutant(\commutant(S'))\\
\Longleftrightarrow \;&
\algclosure{S,\id} = \algclosure{S',\id}.
\end{align*}

For $t=1$, the equality of generated matrix algebras follows from
\begin{equation*}
    \tvu{G}=\frac{\id+\im G}{\sqrt 2}
    \; \text{ and } \;
    G=-\im(\sqrt 2\,\tvu{G}-\id).
\end{equation*}
Thus $\tvu{G}$ lies in the unital algebra generated by $G$, and conversely
$G$ lies in the unital algebra generated by $\tvu{G}$.
For $t=2$, write
\[
    S_1(G_1,G_2)=G_1+G_2
    \;\text{ and }\;
    S_2(G_1,G_2)=G_1G_2.
\]
The unital algebra generated by $G^{\hotimes 2}=S_1$ also contains $S_2$,
because
\[
    [S_1(G_1,G_2)]^2=(G_1+G_2)^2=2\id+2G_1G_2=2\id+2S_2.
\]
By Eq.~\eqref{eq:tensor_power_transvection_symmetric_polynomials} for $t=2$,
and since this algebra is unital and closed under complex linear combinations, it
contains the linear combination 
\[
    \tvu{G}^{\otimes 2}
    =
    \frac12[\id+\im S_1(G_1,G_2)-S_2(G_1,G_2)]
\]
of $\id$, $S_1$, and $S_2$,
which proves
$\algclosureS{\tvu{G}^{\otimes 2},\id}\subseteq
\algclosureS{G^{\hotimes 2},\id}$.
Conversely,
\[
    (\tvu{G}^{\otimes 2})^2
    =
    (\tvu{G}^2)\otimes(\tvu{G}^2)
    =
    (\im G_1)(\im G_2)
    =
    -S_2.
\]
Here $\tvu{G}^2=  [(\id+\im G)/{\sqrt2}]^2  =\im G$ follows by applying $G^2=\id$.
Thus $S_2\in\algclosureS{\tvu{G}^{\otimes 2},\id}$.
Moreover, by Eq.~\eqref{eq:tensor_power_transvection_symmetric_polynomials}
for $t=2$,
\[
    2\tvu{G}^{\otimes 2}=\id+\im S_1-S_2.
\]
Hence $S_1\in\algclosureS{\tvu{G}^{\otimes 2},\id}$ as well.
Therefore $\tvu{G}^{\otimes 2}$ generates the same unital algebra as
$S_1=G^{\hotimes 2}$, i.e.,
$\algclosureS{\tvu{G}^{\otimes 2},\id}
    = \algclosureS{G^{\hotimes 2},\id}$.

For $t=3$, the symmetric polynomials are
\begin{align*}
        S_1(G_1,G_2,G_3) &= G_1 + G_2 + G_3,\\
        S_2(G_1,G_2,G_3) &= G_1G_2 + G_1G_3 + G_2G_3,\\
        S_3(G_1,G_2,G_3) &= G_1G_2G_3.
\end{align*}
Then, using that the $G_i$ commute and satisfy $G_i^2=\id$, one checks the equations
\begin{align*}
        &S_1(G_1,G_2,G_3)^2
        = 3\id + 2S_2(G_1,G_2,G_3),\\
        &S_1(G_1,G_2,G_3) S_2(G_1,G_2,G_3)
        \\
        &= 2S_1(G_1,G_2,G_3) + 3S_3(G_1,G_2,G_3).
\end{align*}
Therefore the unital algebra generated by the element $G^{\hotimes 3}=S_1(G_1,G_2,G_3)$ also contains
$S_2(G_1,G_2,G_3)$ and $S_3(G_1,G_2,G_3)$. By
Eq.~\eqref{eq:tensor_power_transvection_symmetric_polynomials} for $t=3$,
it contains $\tvu{G}^{\otimes 3}$, which proves
$\algclosureS{\tvu{G}^{\otimes 3},\id}\subseteq
\algclosureS{G^{\hotimes 3},\id}$.
Finally, consider the tensor power of the Clifford transvection
\begin{align*}
        \tvu{G}^{\otimes 3} &= \frac{1}{2\sqrt{2}}(\id+\im G_1)(\id+\im G_2)(\id+\im G_3) \\
        &=
        \frac{1}{2\sqrt{2}}
        \begin{aligned}[t]
        &[\id + \im S_1(G_1,G_2,G_3) - S_2(G_1,G_2,G_3)\\
        & - \im S_3(G_1,G_2,G_3)],
        \end{aligned}
        \\
        (\tvu{G}^{\otimes 3})^2
        &=
        (\tvu{G}^2)\otimes(\tvu{G}^2)\otimes(\tvu{G}^2)\\
        &=
        (\im G_1)(\im G_2)(\im G_3)
        =
	        -\im S_3(G_1,G_2,G_3).
\end{align*}
Hence, up to non-zero scalar multiples,
$(\tvu{G}^{\otimes 3})^2$ provides the generator $S_3$.
Combining this with the expansion of $\tvu{G}^{\otimes 3}$ and the available
identity gives the generator $\im S_1-S_2$.
It remains to recover $S_1$ and $S_2$.

Since the $G_i$ commute and satisfy $G_i^2=\id$, multiplication by
$S_3=G_1G_2G_3$ exchanges $S_1$ and $S_2$, i.e.,
$S_3S_1=S_2$ and $S_3S_2=S_1$.
Therefore
\begin{equation}
    \im S_3(\im S_1-S_2)=-S_2-\im S_1.
\end{equation}
The two generated linear combinations $\im S_1-S_2$ and
$-S_2-\im S_1$ are linearly independent over $\C$.
Hence the generated algebra contains both $S_1$ and $S_2$, and it already
contains $S_3$. This proves $\algclosureS{\tvu{G}^{\otimes 3},\id}
    =
    \algclosureS{G^{\hotimes 3},\id}$.
\end{proof}

The restriction to $t\leq3$ in the proof is essential.
For these values of $t$, the elementary symmetric polynomials in the commuting
binary variables $G_i$ can be recovered from the tensor-power Lie generator
$G^{\hotimes t}=S_1$ and from the tensor powers of the Clifford transvection
$\tvu{G}^{\otimes t}$.
Equivalently, for each Pauli generator $G$, the two constructions generate the
same unital algebra in the $t$-fold tensor power.
For $t=4$, this recovery mechanism no longer accounts for the additional
fourth-order invariant of the Clifford action.
We make this obstruction explicit below through the Pauli fourth-moment element
$\Omega_4$.

We can also complement this discussion with the equivalent connection between the \emph{adjoint} commutants, which we can instead leverage to understand the reachability of time evolution in the Heisenberg picture.
To be specific, first recall the definition of \emph{adjoint} representation for an invertible operator $g\in \GL(d,\C)$, as a group element, $\Ad_g(M) = gMg^{-1}$ and for an arbitrary matrix $H\in\matalg(d,\C)$, as a Lie algebra element, $\ad_H(M) = \comm{H}{M}$.
We also have the corresponding commutants
\cite{Zeier_2015,Zimboras_Zeier_SchulteHerbruggen_Burgarth_2015}, which consist of superoperators commuting with the adjoint representations $\Ad_g$ or $\ad_H$:
\begin{enumerate}
    \item Adjoint commutant of a set of Lie algebra generators $\commutant_{\ad}(\pgens) = \commutant(\{ \ad_H | H\in\pgens\})$. We have that $\commutant_{\ad}(\pgens) = \commutant_{\ad}(\lie{\pgens})$.
    \item Adjoint commutant of a set of group generators ($S\subseteq\GL(2^n,\C)$) $\commutant_{\Ad}(S) = \commutant(\{ \Ad_g | g\in S\})$. We have that $\commutant_{\Ad}(S) = \commutant_{\Ad}(\groupclosure{S})$
\end{enumerate}
as well as $\commutant_{\ad}(\lieg) = \commutant_{\Ad}(e^{\lieg})$.
Then, we can prove the following:
\begin{lem}\label{lem:adjoint_commutant_lie_and_clifford}
Consider a set of Paulis $\pgens\subseteq\PP_n$.
Then the adjoint commutant of $\lieg = \lie{\pgens}$ coincides with the adjoint commutant of the Clifford transvection group $\cltvgroup{\pgens}$.
\end{lem}
\begin{proof}
For convenience, we vectorize matrices.
For every $X\in\matalg(2^n,\C)$, we use
\[
    \mathrm{vec}(AXB)=(B^T\otimes A)\mathrm{vec}(X).
\]
Thus, for $g\in\GL(2^n,\C)$ and $M\in\matalg(2^n,\C)$,
\begin{align*}
    \mathrm{vec}(\Ad_g(X))
    &=
    (g^{-T}\otimes g)\mathrm{vec}(X),\\
    \mathrm{vec}(\ad_M(X))
    &=
    (-M^T\otimes\id+\id\otimes M)\mathrm{vec}(X).
\end{align*}
Since vectorization is reversible, the polynomial identities between the vectorized
representatives are equivalent to the corresponding identities between the
original linear maps.
Fix $G\in\pgens$, and write
$\widehat{\ad}_G$, $\widehat{\Ad}_G$, and
$\widehat{\Ad}_{\tvu{G}}$ for the vectorized representatives of
$\ad_G$, $\Ad_G$, and $\Ad_{\tvu{G}}$, respectively:
\begin{align*}
    \mathrm{vec}(\ad_G(X)) &= \widehat{\ad}_G\mathrm{vec}(X),\\
    \mathrm{vec}(\Ad_G(X)) &= \widehat{\Ad}_G\mathrm{vec}(X),\\
    \mathrm{vec}(\Ad_{\tvu{G}}(X)) &= \widehat{\Ad}_{\tvu{G}}\mathrm{vec}(X).
\end{align*}
Equivalently,
\[
    \widehat{\ad}_G=-G^T\otimes\id+\id\otimes G
    \;\text{ and }\;
    \widehat{\Ad}_G=G^T\otimes G.
\]
We first prove $\algclosureS{\ad_G,\id} \subseteq \algclosureS{\Ad_{\tvu{G}},\id}$.
Using $\tvu{G}^{-1}=(\id-\im G)/\sqrt2$ and $G^2=\id$, we get
\begin{align*}
    \widehat{\Ad}_{\tvu{G}}
    &=
    \frac12(\id-\im G^T)\otimes(\id+\im G)\\
    &=
    \frac12(\id\otimes\id-\im G^T\otimes\id
    +\id\otimes\im G+G^T\otimes G)\\
    &=
    \frac12(\id\otimes\id-\im \widehat{\ad}_G+\widehat{\Ad}_G).
\end{align*}
Moreover, $\tvu{G}^2=\im G$, so
\[
    \widehat{\Ad}_{\tvu{G}}^2=\widehat{\Ad}_G
    \;\text{ and }\;
    \widehat{\Ad}_{\tvu{G}}^4=\id\otimes\id.
\]
Therefore
\[
    -\im \widehat{\ad}_G
    =
    2\widehat{\Ad}_{\tvu{G}}
    -\widehat{\Ad}_{\tvu{G}}^2
	    -\widehat{\Ad}_{\tvu{G}}^4.
\]
By reversibility of vectorization, this gives the identity
\[
    -\im \ad_G
    =
    2\Ad_{\tvu{G}}
    -\Ad_{\tvu{G}}^2
    -\Ad_{\tvu{G}}^4
\]
for linear maps on $\matalg(2^n,\C)$.
Thus $\ad_G\in\algclosureS{\Ad_{\tvu{G}},\id}$, proving the inclusion.

Conversely, we prove that $\algclosureS{\Ad_{\tvu{G}},\id} \subseteq \algclosureS{\ad_G,\id}$.
Starting from $\widehat{\ad}_G$, we find
\begin{align*}
    \widehat{\ad}_G^2
    =
    (-G^T\otimes\id+\id\otimes G)^2
    =
    2(\id\otimes\id-\widehat{\Ad}_G).
\end{align*}
Hence $\widehat{\Ad}_G=\id\otimes\id-\frac12\widehat{\ad}_G^2$, and therefore
\[
    \widehat{\Ad}_{\tvu{G}}
    =
    \frac12(\id\otimes\id-\im \widehat{\ad}_G+\widehat{\Ad}_G)
    =
	    \id\otimes\id-\frac{\im}{2}\widehat{\ad}_G-\frac14\widehat{\ad}_G^2.
\]
By invertibility of vectorization, this gives the identity
\[
    \Ad_{\tvu{G}}
    =
    \id-\frac{\im}{2}\ad_G-\frac14\ad_G^2
\]
for linear maps on $\matalg(2^n,\C)$.
Since the generated matrix algebra is unital, the right-hand side lies in
$\algclosureS{\ad_G,\id}$.
Thus $\Ad_{\tvu{G}}\in\algclosureS{\ad_G,\id}$, proving the reverse inclusion.
The two generated matrix algebras coincide for every $G\in\pgens$, and hence
their commutants coincide.
\end{proof}

Theorem~\ref{thm:transvection_3_design_property} identifies the
tensor-power commutants of Clifford transvection groups with those of the
corresponding Pauli Lie groups through third moments.
Lemma~\ref{lem:adjoint_commutant_lie_and_clifford} gives the analogous
generator-level statement in the adjoint representation.
The preceding discussion also indicates why this agreement should not
persist at fourth order.
Webb proves that any ensemble whose support is contained in the full Clifford
group fails to be a unitary 4-design for the full unitary group
\cite[Sec.~4, Lemma~5]{Webb_2016}.
The result for the Clifford group itself is stated in
\cite[Theorem~3]{Webb_2016} and also follows from
\cite[Theorem~1]{Zhu_2017}.
Here the target group for the $t$-design is instead the generally smaller Pauli Lie group
$\exp(\lie{\pgens})$, whose fourth tensor-power commutant can be larger.
Therefore the result on not being a 4-design
for the Clifford group does not immediately imply 
the one for Clifford transvection groups as stated below.
It does, however, identify the relevant fourth-order obstruction in the fourth tensor power commutant.
This obstruction appears in several equivalent forms
\cite{Zhu_Kueng_Grassl_Gross_2016,Bittel_Eisert_Leone_Mele_Oliviero_2025};
in our notation, it is represented by the Pauli fourth-moment sum $\Omega_4$
defined in Eq.~\eqref{eq:clifford:omega4} below.
The more direct proof below does not require a characterization of either fourth
commutant.
It only uses this single Clifford-invariant element to separate the fourth
tensor-power commutants for an arbitrary Pauli
generating set $\pgens$.
\begin{thm}\label{thm:transvection_not_4_design}
Consider a set of Pauli strings $\pgens\subseteq\PP_n$ containing a
non-identity Pauli string, which generates a Pauli Lie algebra $\lieg =
\lie{\pgens}$ and Pauli Lie group $e^{\lieg}$.
The Clifford transvection group $\cltvgroup{\pgens} = \groupclosure{\{\tvu{G}\}_{G\in \pgens}}$ is not a 4-design for the Pauli Lie group $e^{\lieg}$.
\end{thm}
\begin{proof}
As in the ordinary Clifford case, it suffices to show that an element in the commutant of the Clifford transvection group does not commute with some element in the Pauli Lie group.
Also, since Clifford transvection groups are subgroups of the Clifford group, it suffices to consider the following element
\begin{equation}
    \Omega_4 = \sum_{P\in\PP_n} P^{\otimes 4},\label{eq:clifford:omega4}
\end{equation}
The element $\Omega_4$ is invariant under every $n$-qubit Clifford
unitary $C\in\cl_n$.
Indeed, conjugation by such a $C$ sends each summand $P^{\otimes4}$ to
\[
    C^{\otimes4} P^{\otimes4}(C^\dagger)^{\otimes4}
    =
    (CPC^\dagger)^{\otimes4}.
\]
Since $C\in\cl_n$ normalizes the $n$-qubit Pauli group,
Eq.~\eqref{eq:clifford:plus:minus} gives a $P'\in\PP_n$ with
$CPC^\dagger=\pm P'$, see Eq.~\eqref{eq:clifford:plus:minus}.
The sign disappears in the fourth tensor power, so the summand becomes
$(P')^{\otimes4}$.
Thus conjugation by $C^{\otimes4}$ permutes the summands of $\Omega_4$.

Then, since $e^{\im\theta G}\in e^{\lieg}$ for arbitrary $\theta$ and $G\in\pgens$, it suffices to consider a non-identity generator $G\in\pgens$ and $\theta=\pi/8$, i.e.\ a gate isomorphic to the T-gate.
The action of a Pauli rotation $U=\exp(\im\theta G)$ on another Pauli yields
\begin{equation}
    \Ad_U(P) = \begin{dcases}
        P & \text{if }\comm{G}{P} = 0,\\
        \cos(2\theta) P + \im \sin(2\theta) G P & \text{if }\acomm{G}{P} = 0.
    \end{dcases}
\end{equation}
Then the four-fold tensor power $U^{\otimes 4}$ of
$U=\exp(\im\pi/8 G)$ commutes with $\Omega_4$ iff
$U^{\otimes 4}\Omega_4 (U^\dagger)^{\otimes 4}= \Omega_4$, so we look
at the conjugation action of $U^{\otimes 4}$ on this additional element
\begin{equation}
        \Ad_{U^{\otimes 4}}(\Omega_4) = \sum_{\comm{G}{P}=0}P^{\otimes 4} + \frac{1}{4}\sum_{\acomm{G}{P}=0}(P + \im GP)^{\otimes 4}.
\end{equation}
The commuting part is clearly invariant, so we consider only the anti-commuting part. For $G=\iso{w}$ and $P=\iso{v}$,
\begin{equation}
    GP = \iso{w}\iso{v} = -\im (-1)^{\sign{w}{v}}\iso{v+w}.
\end{equation}
where $\sign{w}{v}$ is a suitable sign function, see Eq.~\eqref{eq:Pauli:sign}.
In the binary picture, the Pauli strings anti-commuting with $G=\iso{w}$ are
exactly those with labels $x\in\Fn$ satisfying $\symp{x}{w}=1$.
Choose one such label $u$ with $\symp{u}{w}=1$.
Then the set of anti-commuting labels is the affine subspace
\[
    u+w^\perp=\{u+y\mid y\in w^\perp\}.
\]
Since $w\in w^\perp$, translation by $w$ preserves the affine subspace
$u+w^\perp$.
Choose a set $R\subseteq u+w^\perp$ containing exactly one element $v$ from each
two-element pair $\{x,x+w\}$.
Then
\begin{align*}
        &\sum_{\acomm{G}{P}=0}(P + \im GP)^{\otimes 4}\\
        &=\sum_{v\in R}(\iso{v} + \im \iso{w}\iso{v})^{\otimes 4} + (\iso{v+w} + \im \iso{w}\iso{v+w})^{\otimes 4}.
\end{align*}
Consider now each term defined by $v\in R$, i.e.,
\begin{align*}
        &[\iso{v} -(-1)^{\sign{w}{v}}\iso{v+w}]^{\otimes 4} + [\iso{v+w} - (-1)^{\sign{w}{v+w}}\iso{v}]^{\otimes 4}\\
        &=[\iso{v} -(-1)^{\sign{w}{v}}\iso{v+w}]^{\otimes 4} + [\iso{v+w} + (-1)^{\sign{w}{v}}\iso{v}]^{\otimes 4}\\
        &=[\iso{v} -(-1)^{\sign{w}{v}}\iso{v+w}]^{\otimes 4} + [\iso{v} + (-1)^{\sign{w}{v}}\iso{v+w}]^{\otimes 4},
\end{align*}
where we used Lemma~\ref{prop:Pauli:sign}(e).
Then, after expanding the tensor products, the terms with an odd number
of factors $\iso{v+w}$ cancel, whereas the terms with an even number of such
factors survive.
This yields
\begin{align*}
&2(
\begin{aligned}[t]
&\iso{v}^{\otimes 4}+\iso{v+w}^{\otimes 4} \\
&+\iso{v}{\otimes}\iso{v}{\otimes}\iso{v+w}{\otimes}\iso{v+w}
+\iso{v}{\otimes}\iso{v+w}{\otimes}\iso{v}{\otimes}\iso{v+w}\\
&+\iso{v}{\otimes}\iso{v+w}{\otimes}\iso{v+w}{\otimes}\iso{v}
+\iso{v+w}{\otimes}\iso{v}{\otimes}\iso{v}{\otimes}\iso{v+w}\\
&
+\iso{v+w}{\otimes}\iso{v}{\otimes}\iso{v+w}{\otimes}\iso{v}
+\iso{v+w}{\otimes}\iso{v+w}{\otimes}\iso{v}{\otimes}\iso{v}).
\end{aligned}
\end{align*}
The mixed tensor words in the last lines are not of the form
$Q^{\otimes4}$.
Moreover, the unordered pair $\{v,v+w\}$ is represented only once by the
choice $v\in R$.
Hence no contribution from a different representative can cancel these mixed
terms.
Therefore $\Ad_{U^{\otimes4}}(\Omega_4)$ contains terms that are absent from
$\Omega_4$, so $U^{\otimes4}$ does not commute with $\Omega_4$.
\end{proof}

Notice that these results make no assumption on the (anti-)commutation relations between the generators $\pgens$, but simply use basic results about their eigenvalue structure.
Hence, it is also not clear a priori to \emph{which} groups and Lie algebras these results apply.
However, they become particularly useful in combination with the classification of Clifford transvection groups and Lie algebras for connected generating sets.
We refer to Section~\ref{sec:classification:groups_lie_algebras} for more details.
We distinguish among seven possible families, depending on their symmetries and quasi-universality or free-fermionicness:
\begin{enumerate}
    \item the unitary (Clifford) group under Pauli symmetries;
    \item the orthogonal (Clifford) group under Pauli symmetries;
    \item the unitary symplectic (Clifford) group under Pauli symmetries;
    \item the unitary (Clifford) group as intersection of the orthogonal and symplectic group, under Pauli symmetries;
    \item the (even) matchgate group, generated by quadratic Majoranas (up to additional non-parity symmetries), whose corresponding Clifford transvection group is an extension of the symmetric group $S_{2m}$ for some $m$;
    \item the \emph{odd} matchgate group, generated by quadratic and linear Majoranas (up to additional non-parity symmetries), whose corresponding Clifford transvection group is an extension of the symmetric group $S_{2m+1}$ for some $m$;
    \item the \emph{exceptional} matchgate group, which is generated by quadratic and linear Majoranas as well as the parity symmetry (up to additional non-parity symmetries), whose corresponding Clifford transvection group is an extension of the symmetric group $S_{2m+2}$ for some $m$.
\end{enumerate}
Here we use matchgates to refer to free-fermionic generating sets, though, strictly speaking, the circuits consisting of (continuous) matchgates \cite{Jozsa_2008} only possess a parity symmetry as for ordinary quadratic Majoranas. 
We consider here also the cases of additional non-parity symmetries, which in the line-graph formalism correspond to \emph{cycle} symmetries of the generators.

In the strictly universal case, with trivial commutant and invariant
bilinear forms, this reproduces the known result that the Clifford group
is a 3-design for the unitary group \cite{Webb_2016,Zhu_2017}.
Analogous 3-design statements were also known for the real Clifford group
\cite{Hashagen2018realrandomized}, for Clifford groups under Pauli symmetries
\cite{Mitsuhashi_Yoshioka_2023}, and for the even matchgate group
\cite{Wan_2023}.
Our framework gives a uniform proof for all these cases.
It also covers cases which, to our knowledge, had not been studied before in
the literature, namely the unitary symplectic group, the unitary group with mixed
representation, and the odd and exceptional matchgate groups.
The corresponding Lie algebras are discussed in
Section~\ref{sec:Majorana_Strings_Free_Fermionic_Lie_Algebras} and
Section~\ref{sec:pauli_lie_algebras_isometries}, while the
free-fermionic Clifford transvection groups are discussed in
Section~\ref{sec:Transvection_Group_Line_Graph_Symmetric_Group}.

Notice that the above list does not exhaust all possible Pauli Lie groups and Clifford transvection groups, due to the possibility of disconnected generating sets.
A trivial well-known example is that of the local unitary $\lieU(2)^{\otimes n}$ and local Clifford $\cl_1^{\otimes n}$ groups, which simply consist of tensor products of the unitary and Clifford group on a single qubit.
We also discuss in Section~\ref{sec:IQP} the set of diagonal gates, for which the diagonal Clifford gates provide a 3-design and the generating set is completely disconnected.
Also, one can consider adding certain Pauli strings in the center of matrix algebra, which commute with all generators.
In this setting, one finds that in certain cases the addition of these elements does not change the Clifford transvection group nor the corresponding 3-design property, whereas in other cases even a single commuting element can strongly affect the structure of the symmetries. 
We also mention this in the case of the uniformly controlled single qubit gates in Section~\ref{sec:uniformly_controlled_many_symmetries}.

Beyond the exact $t\leq 3$ design statements discussed here, one can obtain
higher-order designs by supplementing Clifford circuits with non-Clifford
resources.
Depending on the number of injected magic gates and on the circuit depth, such
constructions can form approximate, and in some regimes exact, $t$-designs for
larger $t$ \cite{Zhang_2026}.
Related constructions based on Clifford transvection gates together with
non-Clifford gates have also been developed for the full unitary group
\cite{Tan_Rengaswamy_Calderbank_2022}.

\section{Commutants and Invariant Subspaces for Paulis}\label{sec:comms_and_invariant_subspaces}

\subsection{Commutants for Pauli Lie Algebras and Clifford Groups}\label{sec:square_adjoint_commutants_pauli_lie_algebras}

In the context of reachability, the study of commutants and invariant subspaces provides fundamental information.
We complement Section~\ref{sec:clifford_3_designs_groups} by characterizing the adjoint commutant of a Pauli Lie group or
algebra in terms of the \emph{Pauli orbits} of its corresponding Clifford transvection group.
This provides an alternative proof to that of \cite{Diaz_GarciaMartin_Kazi_Larocca_Cerezo_2023} based on the Clifford and symplectic properties of Pauli strings, under the identification between orbits and commutator graph, as stated in Lemma~\ref{lem:orbits_and_commutato_graph}.
Specifically, we state the following theorem, which is proven in Appendix~\ref{app:sec:proof_adjoint_commutant}:
\begin{thm}[Orbits and Adjoint Commutant]\label{thm:Adjoint_Commutant_Paulis_Orbits}
Consider a set of Pauli strings $\pgens=\isolong{\vgens}\subseteq\PP_n$ and its orbits $\orb(\PP_n)$ with respect to the action of the Clifford transvection group $\cltvgroup{\pgens}$.
Then, the adjoint commutant has a basis given by the linear maps
$$ M\mapsto \sum_{P\in O} P M CP$$
for each of the orbits $O\in\orb(\PP_n)$ and $C\in\cent_{\PP_n}(\pgens)$.
In particular, the adjoint commutant has dimension $\dim(\commutant(\pgens)) \cdot \abs{\orb(\PP_n)}$.
\end{thm}

We cannot prove this type of result for arbitrary Clifford subgroups, since we do not have the guarantee that the non-zero coefficients are parametrized by a single element in an orbit together with the some Pauli in the fixed points of the subgroup.

Moreover, Theorem~\ref{thm:Adjoint_Commutant_Paulis_Orbits} applies to arbitrary generating sets, independently of their connectedness.
We restrict ourselves now to connected generating sets and use the results of Section~\ref{sec:classification:orbits}.
Let the commutant have size $\dim(\commutant(\pgens))=2^{2\ell+r}$ with $2^r$ the dimension of the center.
Also, denote with $q$ the number of independent \emph{cycle symmetries} of the free-fermionic case (see also Corollary~\ref{cor:free_fermionic_arbitrary_generating_sets_classification}), which satisfies either $q=r$ or $q=r-1$.
For fixed commutant, we have the following possible cases
\begin{enumerate}
    \item For the quasi-universal case with no invariant bilinear forms, there are $2^{2\ell+r+1}$ orbits.
    \item For the quasi-universal case with invariant bilinear forms, there are $3\cdot 2^{2\ell+r}$ orbits
    (independently of the type of a set of bilinear forms).
    \item For the even free-fermionic case, there are $2^{2\ell}(2m+1 + 6(2^q-1))$ orbits.
    \item For the odd free-fermionic case, there are $2^{2\ell}(m+1 + 3(2^q-1))$ orbits.
    \item For the exceptional free-fermionic case, there are $2^{2\ell}(\lfloor (m{+}1)/2\rfloor+1 + 2(2^q-1))$ orbits.
\end{enumerate}
Additionally, we also have the case of diagonal gates, hence disconnected generated set, which are discussed in Section~\ref{sec:IQP}:
\begin{enumerate}[resume]
    \item For the `universal' diagonal case, there are $2^{n+1}-1$ orbits.
\end{enumerate}
Hence, we have determined the dimension for the adjoint (hence square) commutants for all connected Pauli Lie algebras, which completes the discussion started in \cite{West_Dowling_Southwell_Sevior_Usman_Modi_Quella_2025}, at least in the context of connected generating sets.

\subsection{Operator Invariant Subspaces for Paulis}\label{sec:invariant_subspaces}

Finally, we can also comment on the structure of invariant subspaces.
We do not derive a full decomposition of the matrix space into irreducible subspaces, but provide some coarser decomposition using basic facts about orbits and symmetries.
Moreover, as also shown in \cite{Diaz_GarciaMartin_Kazi_Larocca_Cerezo_2023} for the case of even free-fermions, the choice of irreducible subspaces need not be unique, given the presence of degeneracies.
Equivalently, this is due to the presence of non-abelian \emph{superoperator symmetries} in the adjoint commutant, which can also appear independently of non-abelian symmetries in the ordinary commutant.

A basic but fundamental fact is that Pauli orbits span invariant subspaces.
Indeed, conjugation by a Pauli rotation $U(\theta)=e^{\im\theta G}$ either
fixes a Pauli string $P$, or mixes it only with the Pauli string obtained from
$P$ by the corresponding Clifford transvection, i.e.,
\begin{equation}
    \Ad_{U(\theta)}(P) = \begin{dcases}
        P &\text{if }\comm{G}{P} = 0,\\
        \cos(2\theta)P 
        &\text{if } \acomm{G}{P} = 0.\\
        + \sin(2\theta)\Ad_{\tvu{G}}P
    \end{dcases}
\end{equation}
Thus the Pauli-spanned invariant subspaces for the adjoint Pauli Lie group
action are controlled by the orbits of the corresponding Clifford transvection
group.
Hence, we always have that under time evolution $\ad_\lieg\invV_O \subseteq\invV_O$ with $\invV_O = \Span[\C]{P | P\in O}$ and $O$ an orbit for the Pauli Lie algebra $\lieg$.
As shown in \cite{Diaz_GarciaMartin_Kazi_Larocca_Cerezo_2023}, such a decomposition is not necessarily \emph{irreducible}, in the sense that there may be smaller subspaces $\invV \subsetneq \invV_O$ which are still invariant under time evolution.

In order to discuss a finer decomposition, let us first consider purely the effect of the commutant and invariant bilinear forms, which in fact completely determine the quasi-universal cases, possibly up to some central elements.
Indeed, purely by the commutant constraints we have that the adjoint maps $\Ad_C(H) = CHC^{-1}$ by elements of the commutant are also in the adjoint commutant of the Lie algebra. Thus, for $H\in\lieg$, we obtain
\begin{equation}
    \Ad_C(H) = H \iff \Ad_C\ad_H = \ad_H\Ad_C.
\end{equation}
which is true for any set of generators, not necessarily Pauli strings.
Similarly, we also have for the $B$-conjugated negative transpose $\theta_B(H) = -B^{-1} H^T B$ that
\begin{equation}
    \theta_B(H) = H \iff \theta_B\ad_H = \ad_H\theta_B.
\end{equation}
Then, \emph{up to central elements}, the quasi-universal Lie algebras consist of the intersection of $\lieu(2^n)$ and the simultaneous $+1$ eigenspace of $\Ad_{C_i}$ for a generating set $C_i$ of $\commalg$ and possibly a $\theta_B$ for $B\in\bilinear(\pgens)$.
Hence, the basis of the Lie algebra (which is a real vector space) is also a basis for a complex invariant subspace, as expected.
However, given that $\Ad_C$ and $\theta_B$ mutually commute and commute with $\ad_\pgens$, by standard linear algebra \emph{all} joint eigenspaces of these operators provide invariant subspaces also for $\ad_\pgens$.
Since $\Ad_C$ and $\theta_B$ are involutive maps, the eigenvalue condition corresponds to (anti-)commutation relations and $B$-(skew-)symmetry, i.e.,
\begin{align*}
        \Ad_C(M) &= \pm M \iff CM \pm MC = 0,\\
        \theta_B(M) &= \pm M \iff M^TB \pm BM = 0,
\end{align*}
where now $M$ is some element in the operator space $\matalg(2^n,\C)$.
In particular, it suffices to take a generating set $\{C_i\}_{i=1}^{2\ell+r}$ for the commutant and a single invariant bilinear form $B\in\bilinear(\pgens)$ (if this exists) to completely determine this decomposition.
As we show in Section~\ref{sec:Orbits_Full_Space_Fn_Quasi_Universal_Case}, these conditions, up to some central elements, are already satisfied by all Pauli strings, which provide bases for these eigenspaces.
Then, these conditions do not provide a finer decomposition than that already given by the Pauli orbits (in fact, it is coarser, given that it does not differentiate between elements in the commutant or not).

Also, in the notation of Theorem~\ref{thm:Adjoint_Commutant_Paulis_Orbits}, notice the adjoint maps $\Ad_C$ correspond to the singleton orbits coming from the commutant, namely to the two-sided multiplication maps $\multmap{C}{C}$ with $\multmap{A}{B}(M)=AMB$.
With respect to the doubled Pauli basis of the superoperator space, we can instead write the map $\theta_B$ as
\begin{align*}
        \theta_B
        &= \sum_{v\in\Fn} (-1)^{\QQ_0(v)} \multmap{\iso{v}}{\iso{v}}\\
        &= \id + \sum_{\substack{\theta_B(P)=-P\\ P\neq I}} \multmap{P}{P}
        - \sum_{\theta_B(P)=P} \multmap{P}{P}.
\end{align*}
This expression also provides a decomposition into Pauli orbits whenever $\lieg$ has no symmetries (i.e. the orthogonal or symplectic case, up to isomorphism).
However, in the presence of symmetries $\theta_B$ will decompose into contribution from multiple orbits with respect to $\lieg$ or $\cltvgroup{\pgens}$.

Then, to find a finer decomposition it is clear that we must use additional elements from the adjoint commutant.
The simplest such elements are the left and right multiplication maps $M\mapsto CM$ and $M\mapsto MC$, with $C\in\commutant(\lieg)$.
Hence, we proceed in the same way as in \cite{Lastres_Pollmann_Moudgalya_2024}, in the presence of multiple `parity' symmetries.
It is clear that these are again involutions with eigenvalues $\pm 1$.
However, their eigenspaces are in general not spanned by Pauli strings,
given that multiplication by a non-trivial Pauli $\iso{u}\neq \In$ acts transitively on the Paulis with
\begin{equation}
    \iso{u}\iso{v} \simeq \iso{u+v} \neq \iso{v},\;\forall v\in\Fn.
\end{equation}
Consider the simple case of $C=Z$ and $n=1$, for which the (non-semisimple) Lie algebra which commutes with $Z$ in $\matalg(2,\C)$ is precisely spanned by $\{I,Z\}$, hence has orbits $\{I\}$, $\{Z\}$ and $\{X,Y\}$.
Instead the corresponding joint eigenspaces in $\matalg(2,\C)$ for the left and right multiplication map are one dimensional and given by the projectors onto the eigenstates of $Z$ and the ladder operators
\begin{align*}
    \invV_{++} &= \Span{I+Z}, & \invV_{--} &= \Span{I-Z},\\
    \invV_{+-} &= \Span{X+\im Y}, & \invV_{-+} &= \Span{X-\im Y}
\end{align*}
with $\invV_{s,s'}$ being the eigenspace with eigenvalue $s$ under left multiplication and $s'$ under right multiplication.
Then, these eigenspaces (which are clearly irreducible) provide a decomposition of $\matalg(2,\C)$ precisely into the diagonal and off-diagonal blocks with respect to the computational basis.
Indeed, they also satisfy, as matrix subspaces, $(1+sZ)\cdot\invV_{s,s'}\cdot(1+s'Z) = \invV_{s,s'}$.
Also, if we take $X$ as reference, notice that $X\pm \im Y = (1\pm Z)X$.
For a single commutant Pauli $C=\iso{u}\in\PP_n$, consider the left and
right multiplication maps by $C$.
They decompose $\matalg(2^n,\C)$ into four joint eigenspaces, labelled by
$s,s'\in\{\pm1\}$, each of dimension $2^{2n-2}$.
Choosing any $w\in\Fn$ with $\symp{w}{u}=1$, these eigenspaces can be written
as
\begin{align*}
        \invV_{s,s} &= \Span{(\id+s\iso{u})\iso{v}}_{v\in u^\perp/u},\\
        \invV_{s,-s} &= \Span{(\id+s\iso{u})\iso{v+w}}_{v\in u^\perp/u}.
\end{align*}
Here $u^\perp/u$ denotes any choice of representatives for
$u^\perp/\SpanS[\F_2]{\{u\}}$.
In particular, every basis element above has the form
$(\id+s\iso{u})\iso{v}$ or $(\id+s\iso{u})\iso{v+w}$,
with $v\in u^\perp/u$.
Thus each basis element is a linear combination of precisely two Pauli strings.
The factor $(\id+s\iso{u})$ is proportional to the Hilbert-space projector
onto the $s$-eigenspace of $C=\iso{u}$.
Given the orthogonal projectors in $\matalg(2^n,\C)$ onto the eigenspaces of $C$ of eigenvalues $s\in\{\pm 1\}$, we have
\begin{equation*}
    R_s = \frac{\id+sC}{2},\; R_sR_t = \delta_{s,t} R_s,\; \sum_{s\in\{\pm 1\}} R_s = \id.
\end{equation*}
These Hilbert-space projectors induce orthogonal projectors on the operator
space by left and right multiplication with $s,s',t,t'\in\{\pm 1\}$:
\begin{align*}
    \calR_{s,s'}(M) &= R_sMR_{s'}, &  \calR_{s,s'}\calR_{t,t'} &= \delta_{s,t}\delta_{s',t'}\calR_{s,s'},\\
    \sum_{s,s'} \calR_{s,s'} &= \id, & \Im(\calR_{s,s'}) &= \invV_{s,s'}.
\end{align*}

\begin{figure}
    \centering
    \includegraphics[width=0.8\linewidth]{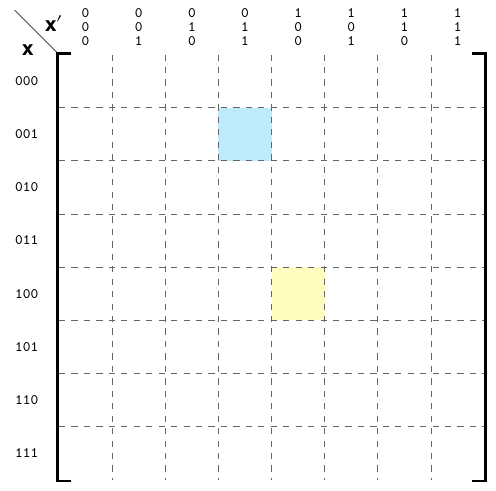}
    \caption{Operator invariant subspaces of a Pauli Lie algebra as specified by commuting Pauli symmetries $\{C_i\}_{i=1}^{\ell+r}$ (here $\ell+r=3$ for convenience), viewed as a full block decomposition of the matrix algebra $\matalg(2^n,\C)$ into diagonal and off-diagonal blocks of size $2^m\times 2^m$.
    The subspaces correspond to matrices which live inside a single block.
    Here we label each block as a pair of bitstrings $\textbf{x}, \textbf{x}'$, such that $s_i = (-1)^{x_i}$ and $s_i' = (-1)^{x_i'}$. Diagonal blocks (yellow) correspond to simultaneous eigenspaces of $C_i$ and off-diagonal blocks (blue) send simultaneous eigenspaces into one another.}
    \label{fig:invariant-subspaces}
\end{figure}

In the presence of multiple (linear) symmetries, we can no longer choose an arbitrary generating set $C_i$ for $\commutant(\pgens)$.
This is due to the fact that left and right multiplication maps by some $C,C'\in\PP_n$ do \emph{not} always commute,
but they commute if and only if $C$ and $C'$ commute as ordinary Pauli strings.
However, we can still \emph{choose} a set of Pauli strings which generate a maximally commuting subset of the Pauli strings.
In binary notation, such a maximally commuting subset is represented by a
maximal isotropic subspace of $\vgens^\perp$.
If $\rank(\vgens^\perp)=2\ell$ and $\nullity(\vgens^\perp)=r$, then this
subspace has dimension $\ell+r$.
Denote such a subspace by
\[
    W=\SpanS[\F_2]{\{u_i\}_{i=1}^{\ell+r}} = \rad(W),
    \qquad C_i=\iso{u_i}.
\]
In the picture of the partition of the system into $m$ logical, $r$ phase and $\ell$ uncontrollable qubits, this corresponds to choosing the operators $\{Z_{m+i}\}_{i=1}^{\ell+r}$ inside the commutant.
We precisely show in Appendix~\ref{sec:app:operator_invariant_subspaces} the natural generalization of the discussion above for invariant subspaces in the presence of multiple symmetries.
The matrix algebra $\matalg(2^n,\C)$ is partitioned into $4^{\ell+r}$ eigenspaces, labelled by the eigenvalues $\textbf{s}, \textbf{s}'\in\{\pm1\}^{\times (\ell+r)}$, with a basis whose elements are a linear combination of $2^{\ell+r}$ Pauli strings:
\begin{align*}
        \invV_{\textbf{s}, \textbf{s}'} &=
        \spanempty\Big(\Big\{
        \begin{aligned}[t]
        &\prod_{i=1}^{\ell+r}(1+s_i\iso{u_i})\isolong{w_{\textbf{ss}'}+v} 
        \\
        &
        \text{for } v\in W^\perp/W \Big\}\Big),
        \end{aligned}\\
        \matalg(2^n,\C) &= \bigoplus_{\{s_i,s_i'\}\in\{\pm 1\}^{\times 2(\ell+r)}} \invV_{\textbf{s}, \textbf{s}'}.
\end{align*}
For each pair of signs $\textbf{s},\textbf{s}'$, set
\[
    \epsilon_i(\textbf{s},\textbf{s}')
    :=
    [(1{-}s_is_i')/2]\bmod 2,
\]
which records whether the left and right
eigenvalue labels for $C_i$ differ.
The vector $w_{\textbf{s},\textbf{s}'}$ is chosen so that
\[
    \symp{w_{\textbf{s},\textbf{s}'}}{u_i}
    =
    \epsilon_i(\textbf{s},\textbf{s}')
\]
for all $i$, equivalently it commutes with $u_i$ when $s_i=s_i'$ and
anticommutes with $u_i$ when $s_i\neq s_i'$.
By taking $v\in W^\perp/W$, we avoid double counting.
In the matrix picture, we obtain a full decomposition of $\matalg(2^n,\C)$ into $2^{2(\ell+r)}$ blocks of size $2^m\times 2^m$; see also Fig.~\ref{fig:invariant-subspaces}.

In particular, the invariant subspaces satisfy
\[
    \prod_i\frac{\id+s_iC_i}{2}
    \invV_{\textbf{s}, \textbf{s}'}
    \prod_i\frac{\id+s_i'C_i}{2}
    =
    \invV_{\textbf{s}, \textbf{s}'}.
\]
In the canonical picture $C_i=Z_{m+i}$, we may choose
\begin{equation}
    \isolong{w_{\textbf{s},\textbf{s}'}}
    =
    \prod_{i=1}^{\ell+r} X_{m+i}^{\epsilon_i(\textbf{s},\textbf{s}')}.
\end{equation}
The elements of $W^\perp/W$ instead correspond to arbitrary Pauli strings over
the $m$ logical qubits,
$\prod_{i=1}^m X_i^{\alpha_i}Z_i^{\beta_i}$, for arbitrary
$\alpha_i,\beta_i\in\F_2$.
Thus the block takes the canonical form
\begin{equation}
    \invV_{\textbf{s}, \textbf{s}'}
    =
    \SpanL{\PP_m
    \prod_{i=1}^{\ell+r}
    (1+s_iZ_{m+i})X_{m+i}^{\epsilon_i(\textbf{s},\textbf{s}')}},
\end{equation}
which has dimension $4^m$.
Hence, we automatically have that any irreducible invariant subspace for \emph{any} Pauli generating set with a given commutant $\commutant(\pgens)$ (which can be chosen inside $\invV_{\textbf{s}, \textbf{s}'}$), also has dimension at most $4^m$, i.e. it is independent of the dimension of the commutant.
This is in strong contrast to the orbits for the connected cases (and disconnected as well, per Section~\ref{sec:IQP}), which instead all are of exponential size $\BigO(2^n)$ (excluding the trivial singleton orbits), as long as the dimension of the center of the commutant also has size exponential in $n$, i.e.\ $r=\BigO(n)$.

This block decomposition is still not necessarily irreducible, even in the
quasi-universal case, and it need not be unique.
The reason is that the adjoint commutant can have a non-abelian structure.
For example, each diagonal block $\invV_{\textbf{s},\textbf{s}}$ contains the
one-dimensional subspace spanned by the joint projector for the commuting
Pauli strings $\{C_i\}_{i=1}^{\ell+r}$.
The remaining part of the same block is the corresponding traceless subspace,
so we have the additional splitting
\begin{align*}
        \invV_{\textbf{s}, \textbf{s}}^d &= \SpanL{\prod_{i=1}^{\ell+r}(1+s_i\iso{u_i})},\\
        \invV_{\textbf{s}, \textbf{s}}^0 &= \spanempty\Big(\Big\{
        \begin{aligned}[t]
        &\prod_{i=1}^{\ell+r}(1+s_i\iso{u_i})\isolong{v} \\
        & \text{for } v\in W^\perp/W,\; v\neq 0 \Big\}\Big).
        \end{aligned}
\end{align*}
It is now also straightforward to combine this with the orbit structure, by requiring that the basis elements be symmetrized under the Clifford transvection group $\cltvgroup{\pgens}$, given that it also commutes with $C_i\in\commutant(\pgens)$.
Equivalently, we can take the intersection of the subspaces defined from the
commutant alone and the subspaces defined by the orbits.
Hence, for each orbit $O\in\orb(\PP_n)$, we obtain an invariant subspace
\begin{align*}
        \invV_{\textbf{s}, \textbf{s}'}^O &= \invV_{\textbf{s}, \textbf{s}'} \cap \invV_O.
\end{align*}
Some of these intersections can be zero.
For instance, take the singleton orbit $O=\{C_i\}$.
If the block labels satisfy $s_i=-s_i'$, then
$\invV_{\textbf{s},\textbf{s}'}$ is off-diagonal with respect to the
$C_i$-eigenspaces.
By contrast, $C_i$ has support only in diagonal blocks with respect to these
eigenspaces.
Thus $C_i$ has no component in the block $\invV_{\textbf{s},\textbf{s}'}$, and
$\invV_{\textbf{s},\textbf{s}'}\cap\invV_O=0$.
Also, the subspaces $\invV_{\textbf{s}, \textbf{s}'}^O$ may be non-distinct in general for different choices of signs and orbits. 
For instance, this happens for the singleton orbits $O = \{C_i\}$ for some $i$, which all produce the same one-dimensional subspace for a given choice of $\{s_i,s_i'\}_{i=1}^{\ell+r}$ and $s_i'=s_i$.
Consequently, these intersections do not in general give
$4^{\ell+r}\abs{\orb(\PP_n)}$ nonzero invariant subspaces.
We summarize the above discussion in the following:
\begin{thm}\label{thm:symmetry_adapated_orbit_subspaces}
Consider a set of Pauli strings $\pgens\subseteq\PP_n$ and its orbits
$\orb(\PP_n)$ with respect to the Clifford transvection group
$\cltvgroup{\pgens}$.
Let $\rank(\vgens^\perp)=2\ell$ and $\nullity(\vgens^\perp)=r$.
Let $\{C_i\}_{i=1}^{\ell+r}$, $C_i=\iso{u_i}$, be a maximal algebraically
independent subset of mutually commuting Pauli strings in
$\commutant(\pgens)$, and set
\[
    W=\SpanS[\F_2]{\{u_i\}_{i=1}^{\ell+r}} .
\]
Then the restriction of the symplectic form to $W$ is zero, so
$W=\rad(W)$.
Moreover, the operator space $\matalg(2^n,\C)$ has a
direct-sum decomposition into the joint eigenspaces of left and right
multiplication by the $C_i$.
We denote these eigenspaces by $\invV_{\textbf{s},\textbf{s}'}$, where
$\textbf{s},\textbf{s}'\in\{\pm1\}^{\times(\ell+r)}$.
Explicitly,
\begin{align}
        \invV_{\textbf{s},\textbf{s}'} =
        \spanempty\Big(\Big\{
        \begin{aligned}[t]
        &\prod_{i=1}^{\ell+r}(1{+}s_i\iso{u_i})
        \isolong{w_{\textbf{s},\textbf{s}'}{+}v} \\
        & \text{for representatives } v\in W^\perp/W\Big\}\Big),
        \end{aligned}
\end{align}
where $w_{\textbf{s},\textbf{s}'}$ is any vector satisfying
\[
    \symp{w_{\textbf{s},\textbf{s}'}}{u_i}
    =
    [(1{-}s_is_i')/{2}]\bmod 2
\]
for all $i$, so that $w_{\textbf{s},\textbf{s}'}$ commutes with $u_i$ when
$s_i=s_i'$ and anticommutes with $u_i$ otherwise.
Independently, the orbit decomposition gives invariant subspaces
\[
    \invV_O=\SpanL{\{\iso{v}\mid v\in O\}}
    \qquad \text{with } O\in\orb(\PP_n).
\]
For each orbit $O$ and each pair of block labels
$\textbf{s},\textbf{s}'$, the intersection
\[
    \invV_{\textbf{s},\textbf{s}'}^O
    =
    \invV_O\cap\invV_{\textbf{s},\textbf{s}'}
\]
is again invariant.
After discarding zero and repeated intersections, these subspaces give a
common refinement of the orbit decomposition and the block decomposition.
Moreover, if $\dim(\algclosure{\pgens})=2^{2m+r}$, then
each block has dimension $2^{2m}$ and
\[
    \dim \invV_{\textbf{s},\textbf{s}'}^O
    \leq
    \min\{2^{2m},\abs{O}\}.
    \label{eq:invariant_subspace_dimension_bound}
\]
\end{thm}
We give a detailed proof in Section~\ref{sec:app:operator_invariant_subspaces}.
We also conjecture that, by choosing a decomposition into distinct subspaces, the subspaces $\invV_{\textbf{s}, \textbf{s}'}^O$ are in fact also a choice of \emph{irreducible subspaces} for any Pauli Lie group (algebra) or Clifford transvection group (which may not be unique due to degeneracies).
This result already shows that, even though orbits may be of exponential size due to a large amount of abelian symmetries, the sizes of invariant subspaces do not grow with the number of abelian symmetries.

The free-fermionic case provides a useful illustration of the dimension
bound in Eq.~\eqref{eq:invariant_subspace_dimension_bound}.
Although the corresponding Pauli orbits may be enlarged by many abelian symmetries, the invariant subspaces obtained from their intersections with the
blocks $\invV_{\textbf{s},\textbf{s}'}$ can remain small.
As described in Proposition~\ref{prop:orbits_in_V_line_graphs_Majorana} and
Section~\ref{sec:Orbits_Full_Space_Fn_Free_Fermionic_Case}, these orbits are
built from a logical Majorana and phase part (in the language of Fig.~\ref{fig:partition-Majorana}); in the
full Pauli space one also allows affine translates represented by
$\majiso{\tilde{w}}$, which may contain uncontrollable terms.
After intersecting with a block $\invV_{\textbf{s},\textbf{s}'}$, the
phase part of the orbit is fixed by the block labels and contribute only
scalar phases.
Consequently, the dimension of the resulting invariant subspace is controlled
by the logical part of the Majorana strings in the orbit, not by the number of
phase modes.
For logical Majorana modes growin with system size $m=\BigO(n)$, this gives invariant subspaces of
polynomial size in the system size $\abs{\invV_{\textbf{s},\textbf{s}'}^O} = \BigO(\poly(n))$.

\ManuscriptPart{Pauli Framework}{part:theory}

Now we transition from application to theory and develop tools for the classification of Pauli Lie algebras, binary and Clifford transvection groups and their orbits, taking advantage of the binary and graph-theoretic formalisms.
Specifically, the following sections provide the necessary background on the binary formalism, the Clifford group, the Pauli group as a symplectic space, transvections and transvection groups, as well as their relation to to Pauli Lie (and matrix) algebras.

\section{Pauli Strings and the Binary Symplectic Space\label{sec:pauli}}

We start in this section by recalling Pauli strings and the binary symplectic space.
We collect a set of useful properties and we continue this discussion
in Section~\ref{sec:clifford:transvections}.

We recall the $2\times 2$ identity operator $I\in \C^{2\times 2}$
and
the Pauli operators
\begin{equation*}
X =
\left[
\begin{smallmatrix}
0 & 1\\
1 & 0
\end{smallmatrix}
\right],\;
Y =
\left[
\begin{smallmatrix}
0 & -\im\\
\im & 0
\end{smallmatrix}
\right],\,
Z =
\left[
\begin{smallmatrix}
1 & 0\\
0 & -1
\end{smallmatrix}
\right] \in \C^{2\times 2}.
\end{equation*}
The set $\PP_n:=\{I,X,Y,Z\}^{\otimes n}$ of (hermitian) Pauli strings of length $n$
consists of all $4^n$ tensor-product operators of the form
$\bigotimes_{o=1}^n A_o$ with  $A_o
\in \{X, Y, Z, I\}$. Two Pauli strings $P,Q \in \PP_n$
either commute ($\comm{P}{Q} = PQ - QP =0$)
or anticommute ($\acomm{P}{Q} = PQ + QP =0$), see, e.g., Lemma~\ref{lem:symp:pauli}(a).
We sometimes suppress the tensor product
by substituting, e.g., $X \otimes Z \otimes I$ with
$\mathrm{XZI}$.
Let us also introduce
$X_j, Y_j, Z_j \in  \C^{2^n \times 2^n}$ as a notation for Pauli strings
that only act on the $j$th qubit,
e.g., $X_j:= I^{\otimes (j-1)} \otimes X  \otimes I^{\otimes (n-j)}$.
The Pauli strings of length $n$ generate the $n$-qubit Pauli group
\begin{equation*}
\Pgroup_n = \groupclosure{\PP_n} = \{ \im^\varpi P \,\text{ for }\, \varpi \in\{0,1,2,3\}, P \in \PP_n\},
\end{equation*}
which is closed under matrix multiplication.
Also, the Pauli strings are a basis for the matrix algebra of operators over $n$, $\mat{2^n}{\C}\cong\mat{2}{\C}^{\otimes n} = \Span{\PP_n}$.
We say that two elements $P$ and $Q$ of the Pauli group
are equivalent up to a phase ($P\simeq Q$) if
$P = \im^\varpi Q$  for  some $\varpi \in \{0,1,2,3\}$, or in other words, if $P= z Q$ for an element $z\in \ZZ(\Pgroup_n)$ from its center
[see Eq.~\eqref{eq:center}]
\begin{equation*}
\ZZ(\Pgroup_n) = \{ \In, -\In, \im \In, -\im \In\}.
\end{equation*}
The relation $P\simeq Q$ is clearly an equivalence relation
(that is reflexive, symmetric, and transitive). We identify the
canonical representative in each equivalence class
$P \ZZ(\Pgroup_n)$ as the Pauli string $P$
via the projection
\begin{equation}\label{eq:def:herm_projection}
\herm(Pz) = \herm(zP) := P \,\text{ for }\, z \in \ZZ(\Pgroup_n) \,\text{ and }\, P \in \PP_n.
\end{equation}
Thus we have introduced a homomorphism from the Pauli group $\Pgroup$
to the quotient group $\Pabel_n = \Pgroup_n/\ZZ(\Pgroup_n)$ [see Eq.~\eqref{eq:right:coset}].
The group $\Pabel_n$ is
obviously abelian (as $X_j Y_j \simeq Y_j X_j \simeq Z_j$)
and all its elements besides the identity have order two.
In particular, $\Pabel_n$ is an elementary abelian 2-group
isomorphic to the abelian group $\Z_2^{2n}= \Z_2 \times \cdots \times \Z_2$ of order $4^n$
\cite{aschbacher_finite_group_theory}.
When restricted to the equivalence classes of the abelian group $\Pabel_n$,
the projection $\herm$ induces a bijective map from $\Pabel_n$ to
the set $\PP_n$ of Pauli strings, i.e., $\herm(\Pabel_n) = \PP_n$
and $\Pabel_n = \herm^{-1}(\PP_n)$.
Occasionally we shall also be interesting in performing products of elements $P,Q\in\Pgroup_n$ in the Pauli group which return a Pauli string, we denote this with:
\begin{equation}
    P \setprod Q = \herm(PQ) \in \PP_n.
\end{equation}
Additionally, we also define this as a set operation $A\diamond B$ for $A,B\subseteq\Pgroup_n$, which produces the set with elementwise $\setprod$ product:
\begin{equation}\label{eq:setprod_in_paulis}
    \begin{split}
        A\setprod B &:= \{ \herm(PQ) | P\in A, Q\in B\}\subseteq\PP_n.
    \end{split}
\end{equation}

The abelian group $\Pabel_n$ is interpreted as a vector space over the binary field $\F_2$
by defining the scalar multiplication $m(x,P):= P^x$
for $x \in \F_2$ and $P \in  \PP_n$
(where $P^0 = \In$ and $P^1 = P$).
We introduce the map $\isoempty$ from the
binary vector space $\Fn$  of dimension $2n$
to the set of Pauli strings $\PP_n$ as
\begin{equation}\label{eq:symp:iso}
\isolong{v} = \iso{v}:= \prod_{j=1}^n \im^{\component{v}{j}\component{v}{n+j}}  X_j^{\component{v}{j}} Z_j^{\component{v}{n+j}} \in \PP_n,
\end{equation}
where $v\in \Fn$ and $\component{v}{j}$ is its $j$th component. 
\begin{lem} The map $\isoempty$ from Eq.~\eqref{eq:symp:iso}
induces a vector-space isomorphism from $\Fn$ to
$\Pabel_n$, i.e., $\iso{xv} = (\iso{v})^x$ and $\iso{v+w} \simeq \iso{v} \iso{w} \simeq \iso{w} \iso{v}$
for $v,w\in \Fn$ and $x\in\F_2$.
\end{lem}

Let $\basel_{j}$ and $\basel_{n+j}$ with $j\in\{1,\ldots,n\}$
denote the canonical basis vectors of $\Fn$.
The unique inverse of $\isoempty$ maps Pauli strings to
vectors $v \in \Fn$
and is thus given by
\begin{align}
\inviso(\iso{v}):=\isoempty^{-1}(\iso{v})
&= \sum_{j=1}^n \component{v}{j} \basel_j +  \component{v}{n+j} \basel_{n+j} \label{eq:pauli:binary}
\\
&=
\sum_{j=1}^n \component{v}{j} \inviso(X_j) + \component{v}{n+j}\inviso(Z_j). \nonumber
\end{align}
The vector space $\Fn$ has a natural symplectic form $\sympempty$
from $\Fn\times \Fn$ to $\F_2$ where
\begin{gather}\label{eq:symp:scalar}
\symp{v}{w} := \sum_{j=1}^n \component{v}{j} \component{w}{n+j} + \component{v}{n+j} \component{w}{j},\\
\symp{\basel_i}{\basel_j} = \symp{\basel_{n+i}}{\basel_{n+j}} = 0,\,
\symp{\basel_i}{\basel_{n+j}} = \symp{\basel_{n+j}}{\basel_i} = \delta_{ij}. \nonumber
\end{gather}
We denote the value $\symp{v}{w}$ of the symplectic form on the vectors
$v$ and $w$
as their symplectic product.
The symplectic form is alternating [$\symp{v}{v}=0$], bilinear
[$\symp{v+\tilde{v}}{w} = \symp{v}{w} + \symp{\tilde{v}}{w}$ and
$\symp{v}{w+\tilde{w}} = \symp{v}{w} + \symp{v}{\tilde{w}}$], and nondegenerate
[$v=0 \in \Fn$ if $\symp{v}{w}=0$ holds for all $w$].
This implies that the symplectic form is also symmetric
[$\symp{v}{w}=\symp{w}{v}$].
We obtain:
\begin{lem}[Symplectic properties of Pauli strings]\label{lem:symp:pauli}
Given the Pauli strings $P, Q, R, \iso{v}, \iso{w} \in \PP_n$, we have:\\
(a) $\iso{v} \iso{w} = (-1)^{\symp{v}{w}} \iso{w} \iso{v}$.\\
(b) $R$ commutes with $PQ$ iff it commutes or\\
\phantom{(b)} anticommutes
with both $P$ and $Q$.\\
(c) $R$ anticommutes with $PQ$ iff it commutes
with either \\
\phantom{(c)} $P$ or $Q$ and
anticommutes
with the other one.\\
(d) $\symp{v}{w} = 0$ iff $\comm{\iso{v}}{\iso{w}}=0$.\\
(e) $\symp{v}{w} = 1$ iff $\acomm{\iso{v}}{\iso{w}}=0$.
\end{lem}
\begin{proof}
(a) easily follows from the one-qubit case.
All other results are a direct consequence of (a).
\end{proof}

In the following, we will need precise information on the phase difference between
the product
$\iso{v} \iso{w}$
of Pauli strings
and the corresponding Pauli string $\iso{v+w}$ for vectors $v,w \in \Fn$.
We now detail this phase difference:

\begin{prop}[Phase in the Pauli-string product]\label{prop:Pauli:sign}
Given vectors $v,w \in \Fn$, consider the formula
\begin{equation}\label{eq:Pauli:sign}
\iso{v} \iso{w} = \im^{\symp{v}{w}} (-1)^{\sign{v}{w}}  \iso{v+w}.
\end{equation}
Here, $\signempty$ denotes a boolean function from $\Fn \times \Fn$ to $\F_2$,
which has the following properties:
\begin{align*}
&\text{(a) } \sign{v}{w}+\sign{w}{v} = \symp{v}{w}. \\
&\text{(b) } \signempty \text{ is alternating: } \sign{v}{v}=0.\\
&\text{(c) } \sign{0}{v}=\sign{v}{0}=0.\\
&\text{(d) } \sign{v}{\tilde{v}} + \sign{v+\tilde{v}}{w} + \sign{\tilde{v}}{w} + \sign{v}{\tilde{v}+w}\\
&\phantom{\text{(d) }}=\symp{v}{\tilde{v}}\symp{v+\tilde{v}}{w} + \symp{\tilde{v}}{w}\symp{v}{\tilde{v}+w}.\\
&\text{(e) } \sign{v}{w} + \sign{v+w}{w} = \symp{v}{w}.\\
&\text{(f) } \sign{v}{w} = \signtwo{v}{w} + \signthree{v}{w} + \signfour{v}{w} \text{ where}\\
& \begin{aligned}[t]
\quad &\signtwo{v}{w} &&  \hspace{-3.5mm} =  \sum_{j=1}^{n} \component{v}{j} \component{w}{n+j}, \\
&\signthree{v}{w} && \hspace{-3.5mm} = \sum_{j=1}^{n}
\component{v}{j} \component{v}{n+j} \component{w}{j}
+ \component{v}{j} \component{v}{n+j} \component{w}{n+j}\\
&&& \hspace{-3.5mm} \phantom{= }\hspace{2.5mm} + \component{v}{j} \component{w}{j} \component{w}{n+j}
+ \component{v}{n+j} \component{w}{j} \component{w}{n+j},
\\
&\signfour{v}{w} && \hspace{-3.5mm} =
\sum_{j=1}^{n}  \component{v}{j} \component{v}{n+j} \component{w}{j} \component{w}{n+j}\\
&&& \hspace{-3.5mm} \phantom{= }\hspace{1mm} +  \hspace{-3.5mm} \sum_{1 \le i < j \le n}  \hspace{-2mm}
(\component{v}{i} \component{v}{j} \component{w}{n+i} \component{w}{n+j}
+ \component{v}{i} \component{v}{n+j} \component{w}{j} \component{w}{n+i}\\
&&& \hspace{-3.5mm} \phantom{= + } \hspace{5.75mm} + \component{v}{j} \component{v}{n+i} \component{w}{i} \component{w}{n+j}
+ \component{v}{n+i} \component{v}{n+j} \component{w}{i} \component{w}{j}).
\end{aligned}
\end{align*}
\end{prop}

Before we start with the proof of Prop.~\ref{prop:Pauli:sign}, we present two explicit
examples for the polynomials $\sign{v}{w}$:
\begin{align}
&\sign{(\component{v}{1},\component{v}{2})^T}{(\component{w}{1},\component{w}{2})^{T}} =  \component{v}{1} \component{w}{2}  {+}
\component{v}{1} \component{v}{2} \component{w}{1} {+} \component{v}{1} \component{v}{2} \component{w}{2} \nonumber \\
&{+} \component{v}{1} \component{v}{2} \component{w}{1} \component{w}{2} {+} \component{v}{1} \component{w}{1} \component{w}{2} {+} \component{v}{2}
\component{w}{1} \component{w}{2},
\\
&\sign{(\component{v}{1},\component{v}{2},\component{v}{3},\component{v}{4})^T}{(\component{w}{1},\component{w}{2},\component{w}{3},
\component{w}{4})^{T}} = \component{v}{1} \component{w}{3} {+} \component{v}{2} \component{w}{4}
 \nonumber \\
& {+} \component{v}{1} \component{v}{3} \component{w}{1} {+} \component{v}{1} \component{v}{3} \component{w}{3} {+}
\component{v}{1} \component{w}{1} \component{w}{3}   {+} \component{v}{2} \component{v}{4} \component{w}{2} {+}
\component{v}{2} \component{v}{4} \component{w}{4}
 \nonumber \\
&  {+} \component{v}{2} \component{w}{2} \component{w}{4} {+} \component{v}{3} \component{w}{1} \component{w}{3} {+}
\component{v}{4} \component{w}{2} \component{w}{4} {+} \component{v}{1} \component{v}{2} \component{w}{3} \component{w}{4}
{+} \component{v}{1} \component{v}{3} \component{w}{1} \component{w}{3}
\nonumber \\
&    {+} \component{v}{1} \component{v}{4} \component{w}{2} \component{w}{3}  {+}
\component{v}{2} \component{v}{3} \component{w}{1} \component{w}{4} {+} \component{v}{2} \component{v}{4} \component{w}{2} \component{w}{4}
{+}  \component{v}{3} \component{v}{4} \component{w}{1} \component{w}{2}.
\end{align}

\begin{proof}
The validity of Eq.~\eqref{eq:Pauli:sign} for a suitable boolean function $\signempty$
is easily verified. (a) follows by applying Eq.~\eqref{eq:Pauli:sign}
also to $(-1)^{\symp{v}{w}} \iso{w} \iso{v}$.
Clearly, (b) holds as $\iso{v}^2 = \In$
and (c) follows as $
\In  \iso{v} = \iso{v} \In =\iso{v}$.
For $a_j \in \F_2$, we recall the relation
\begin{equation}\label{eq:magic:i}
\prod_{j=1}^{\gamma} \im^{a_j} = \im^{\sum_{j=1}^{\gamma}a_j} \prod_{1\leq i < j \le \gamma} (-1)^{a_i a_j}.
\end{equation}
We compute
\begin{align*}
&(\iso{v} \iso{\tilde{v}}) \iso{w}
= \im^{\symp{v}{\tilde{v}}}  (-1)^{\sign{v}{\tilde{v}}}  \iso{v+\tilde{v}} \iso{w}\\
&=\im^{\symp{v}{\tilde{v}}}\im^{\symp{v+\tilde{v}}{w}}  (-1)^{\sign{v}{\tilde{v}}+\sign{v+\tilde{v}}{w}}  \iso{v+\tilde{v}+w}\\
&=\im^{\symp{v}{\tilde{v}}+\symp{v+\tilde{v}}{w}}  (-1)^{\symp{v}{\tilde{v}}\symp{v+\tilde{v}}{w} + \sign{v}{\tilde{v}}+\sign{v+\tilde{v}}{w}}  \iso{v+\tilde{v}+w}
\intertext{and}
&\iso{v}  (\iso{\tilde{v}} \iso{w})
= \im^{\symp{\tilde{v}}{w}}  (-1)^{\sign{\tilde{v}}{w}}  \iso{v} \iso{\tilde{v}+w}\\
&= \im^{\symp{\tilde{v}}{w}}\im^{\symp{v}{\tilde{v}+w}} (-1)^{\sign{\tilde{v}}{w}+\sign{v}{\tilde{v}+w}}  \iso{v+\tilde{v}+w}\\
&= \im^{\symp{\tilde{v}}{w} + \symp{v}{\tilde{v}+w}} (-1)^{\symp{\tilde{v}}{w}\symp{v}{\tilde{v}+w} + \sign{\tilde{v}}{w}+\sign{v}{\tilde{v}+w}}  \iso{v+\tilde{v}+w}\\
\end{align*}
where the sums are over $\F_2$. (d) follows by comparison, and (e) as the special case with $\tilde{v} = w$.
We set $x_j := \component{v}{j}$, $z_j := \component{v}{n+j}$, $\tilde{x}_j:=\component{w}{j}$,  $\tilde{z}_j:=\component{w}{n+j}$
for $j \in \{1,\ldots,n\}$.
We compute $\iso{v} \iso{w}$ and $\iso{v+w}$:
\begin{align*}
& \iso{v} \iso{w} = \prod_{j=1}^n \im^{x_jz_j} \im^{\tilde{x}_j \tilde{z}_j} (-1)^{z_j \tilde{x}_j} X_j^{x_j + \tilde{x}_j} Z_j^{z_j+ \tilde{z}_j},
\\
& \iso{v+w} = \prod_{j=1}^n \im^{(x_j+\tilde{x}_j)(z_j+\tilde{z}_j)} X_j^{x_j + \tilde{x}_j} Z_j^{z_j+ \tilde{z}_j}.
\end{align*}
Thus $\im^{\symp{v}{w}} (-1)^{\sign{v}{w}}$ agrees with $\lambda= $
\begin{align*}
&\prod_{j=1}^n \frac{\im^{x_jz_j}\im^{\tilde{x}_j \tilde{z}_j}  (-1)^{z_j \tilde{x}_j}}{\im^{x_jz_j+x_j\tilde{z}_j + \tilde{x}_j z_j + \tilde{x}_j \tilde{z}_j}}
\!=\! \prod_{j=1}^n \frac{\im^{x_jz_j}\im^{\tilde{x}_j \tilde{z}_j}  (-1)^{z_j \tilde{x}_j}}{\im^{x_jz_j}\im^{x_j\tilde{z}_j} \im^{\tilde{x}_j z_j} \im^{\tilde{x}_j \tilde{z}_j}
(-1)^{r_j}
} \\
& \quad \text{where }
\begin{aligned}[t]
 r_j & = (x_j z_j x_j\tilde{z}_j + x_j z_j \tilde{x}_j z_j + x_j z_j \tilde{x}_j z_j \\
& \hspace{1.75mm} +
x_j \tilde{z}_j \tilde{x}_j z_j + x_j \tilde{z}_j \tilde{x}_j z_j + \tilde{x}_j z_j \tilde{x}_j \tilde{z}_j)\\
& = x_jz_j\tilde{z}_j + x_j z_j \tilde{x}_j + x_j \tilde{x}_j \tilde{z}_j + z_j \tilde{x}_j \tilde{z}_j.
\end{aligned}
\end{align*}
Here, we have used, e.g., $x_j^2 = x_j$. This simplifies to
\begin{align*}
\lambda & = \prod_{j=1}^n \im^{x_jz_j}\im^{\tilde{x}_j \tilde{z}_j} (-1)^{x_j \tilde{z}_j} (-1)^{\tilde{x}_jz_j} (-1)^{z_j\tilde{x}_j} (-1)^{r_j}\\
&= \prod_{j=1}^n \im^{x_jz_j+\tilde{x}_j \tilde{z}_j}
(-1)^{x_jz_j \tilde{x}_j \tilde{z}_j}
(-1)^{x_j \tilde{z}_j} (-1)^{r_j}
\end{align*}
by applying Eq.~\eqref{eq:magic:i}. We again use Eq.~\eqref{eq:magic:i} and obtain
\begin{align*}
\lambda  & =
\begin{aligned}[t]
 &\im^{\sum_{j=1}^n x_j \tilde{z}_j + \tilde{x}_j z_j} \prod_{1\leq i < j \leq n} (-1)^{s_{ij}}
\\
&\times
\prod_{j=1}^n (-1)^{x_jz_j \tilde{x}_j \tilde{z}_j}
(-1)^{x_j \tilde{z}_j} (-1)^{r_j}
\\
& \begin{aligned}[t]
&\text{where } s_{ij}  = (x_i \tilde{z}_i + \tilde{x}_i z_i)(x_j \tilde{z}_j + \tilde{x}_j z_j)\\
& = x_i x_j \tilde{z}_i  \tilde{z}_j+
x_i \tilde{z}_i  \tilde{x}_j z_j +
x_j z_i  \tilde{x}_i \tilde{z}_j +
z_i z_j \tilde{x}_i \tilde{x}_j,
\end{aligned}
\end{aligned}
\end{align*}
which completes the proof of (f).
\end{proof}

We apply the previous discussion to commutators of the form
$\comm{-\im \iso{v}/2}{-\im \iso{w}/2}$ and obtain formulas which
are essentially equivalent to the ones in
\cite{Gupta_Master}:

\begin{prop}[Prefactor of Pauli-string commutators]\label{prop:Pauli:comm}
Given vectors $v,w \in \Fn$, we obtain
\begin{subequations}
\label{eq:Pauli:comm}
\begin{align}
&\comm{-\im \iso{v}/2}{-\im \iso{w}/2} =  \tilde{\kappa}(v,w)\, (-\im \iso{v+w}/2)
\intertext{where $\tilde{\kappa}(v,w) = - \tilde{\kappa}(w,v)$ and}
&\tilde{\kappa}(v,w) =
\begin{dcases}
0 &  \text{if } \symp{v}{w} = 0,\\
(-1)^{\kappa(v,w)} & \text{if }  \symp{v}{w}=1.
\end{dcases}
\end{align}
\end{subequations}
\end{prop}

\begin{proof}
We apply Proposition~\ref{prop:Pauli:sign}
and compute
\begin{align*}
&\comm{-\im \iso{v}}{-\im \iso{w}} = \im^2 ( \iso{v} \iso{w} - \iso{w} \iso{v}) \\
&=  [1 - (-1)^{\symp{v}{w}}]\, (-\iso{v} \iso{w})\\
&=
\begin{dcases}
0 &  \text{if } \symp{v}{w} = 0,\\
(-1)^{\kappa(v,w)} \, (-\im 2 \iso{v+w}) & \text{if }  \symp{v}{w}=1.
\end{dcases}
\end{align*}
Proposition~\ref{prop:Pauli:sign}(a) finally implies $\tilde{\kappa}(v,w) = - \tilde{\kappa}(w,v)$.
\end{proof}

\section{From the Clifford Group to Symplectic Transvections\label{sec:clifford:transvections}}

Building on Section~\ref{sec:pauli}, we discuss the Clifford group
and its relation to the binary symplectic group (see Section~\ref{sec:clifford}).
These structures are quite standard in
quantum information theory \cite{Calderbank_PRL_1997,Calderbank_IEEE_1998,Gottesman_PRA_1998,Gottesman_PhD,Nielsen_Chuang_2008,Dehaene_PRA_2003,Aaronson_Gottesmann_2004}.
In Section~\ref{sec:transvections}, we recall the notion of a symplectic transvection
as particular elements of the symplectic group \cite{omeara_1978,grove_2002}.

\subsection{Properties of the Clifford Group\label{sec:clifford}}

We recall the Clifford group $\Cl_n$ as the normalizer of the Pauli group $\Pgroup_n$ in the unitary group $\lieU(2^n)$, i.e.,
\begin{alignat}{2}\label{eq:clifford}
\Cl_n &:= \{ U \in \lieU(2^n) \,|\, U \Pgroup_n U^{\dagger} = \Pgroup_n\}\\
&\phantom{:}= \{
\begin{aligned}[t]
& U \in \lieU(2^n) \text{ such that for each } P \in \PP_n  \\[-1mm]
& \text{exists } Q\in \PP_n \text{ with } U P U^{\dagger} = \pm Q \label{eq:clifford:plus:minus}
\}.
\end{aligned}
\end{alignat}
To verify Eq.~\eqref{eq:clifford:plus:minus},
we apply the definition of the Clifford group in Eq.~\eqref{eq:clifford} and get
$Q = \alpha\, U\! P U^{\dagger}$ with $\alpha = \im^{\varpi} \in \C$ and $\varpi\in\{0,1,2,3\}$.
Thus $\In = Q^2 = \alpha^2 U\! P^2 U^{\dagger} = \alpha^2 \In$ implies
$\alpha = \pm 1$. Clearly, $\Cl_n$ is an infinite group as all the phases
$\exp(2\pi\im\theta) \In \in \Cl_n$ for $\theta \in \R$ are possible, but they are physically irrelevant.
Similarly to the Paulis, we say that two unitaries $U,U'\in\lieU(2^n)$ coincide up to phases ($U\simeq U'$) if $U = \exp(2\pi\im\theta) U'$ for some $\theta\in\R$. 
Additionally, we say that two unitary subgroups $G,H\subseteq \lieU(2^n)$ coincide up to phases ($G\simeq H$) if $\groupclosure{\ZZ(\lieU(2^n))\cup G} = \groupclosure{\ZZ(\lieU(2^n))\cup H}$, where $\ZZ(\lieU(2^n)) = \{ \exp(2\pi\im\theta) \In | \theta\in\R\}\cong\lieU(1)$ is the set of phases.

Any Clifford unitary $U \in \Cl_n$ induces
a map $\SY_U$ from $\Fn$ to $\Fn$
such that (for $v \in \Fn$)
\begin{align}\label{eq:proj:conj}
[\SY(U)](v) = \SY_U(v) = \SY_U v
&:= [\inviso \circ \herm \circ \Ad_U \circ \isoempty](v) \\
&\phantom{:}= \inviso[\herm(U\iso{v}U^{\dagger})]. \nonumber
\end{align}
The binary symplectic group $\Sp(2n,\F_2)$ is
the subgroup of the general linear group $\GL(2n,\F_2) \subset \mat{2n}{\F_2}$
that leaves the symplectic form invariant \cite{omeara_1978,grove_2002}, i.e.,
\begin{align}
\Sp(2n,\F_2) := \{
&  g\in\GL(2n,\F_2) \text{ such that } \symp{gv}{gw} \nonumber \\[-1mm]
&= \symp{v}{w} \text{ for all } v,w\in \Fn\}. \label{eq:symp:group}
\end{align}
We collect some properties of the map $\SY_U$:
\begin{lem}[Symplectic structure of the Clifford group]\label{lem:symp:clifford}
Given a Clifford unitary $U\in \Cl_n$ and its in\-duced
map $\SY_U $ on $\Fn$ as defined in Eq.~\eqref{eq:proj:conj}, we observe:
\begin{enumerate}
\item\label{lem:symp:clifford:a}
$\SY_U$ is a linear map, i.e.,\\
$\SY_U(xv) = x\SY_U(v)$ and
$\SY_U(v+w) = \SY_U(v)+\SY_U(w)$.
\item\label{lem:symp:clifford:b}
$\SY_U$ is invertible, i.e., $(\SY_U)^{-1} = \SY_{U^{\dagger}}$.
\item\label{lem:symp:clifford:c}
$\SY_U  \in \GL(2n,\F_2)$.
\item\label{lem:symp:clifford:d}
$\SY_U  \in \Sp(2n,\F_2)$.
\item\label{lem:symp:clifford:e}
$\SY(U\!P) = \SY(PU) = \SY(U)$ for any $P\in\PP_n$.
\item\label{lem:symp:clifford:f}
$\SY$ is an isomorphism from $\Cl_n/\Pgroup_n$ to $\Sp(2n,\F_2)$.
Equivalently, $\SY$ fits into a short exact sequence
\begin{equation*}
    1\to \Pgroup_n \to \Cl_n \xrightarrow{\SY} \Sp(2n,\F_2)\to 1.
\end{equation*}
\item\label{lem:symp:clifford:g}
For any subgroup $H$ of $\Cl_n$, restricting $\SY$ gives
\begin{equation*}
    1\to H\cap\Pgroup_n \to H \xrightarrow{\SY} \SY(H)\to 1.
\end{equation*}
\item\label{lem:symp:clifford:h}
If the subgroups $H,H'$ of $\Cl_n$ satisfy
\begin{equation*}
    H\subseteq H',
    \;
    \SY(H)=\SY(H'),
    \;
    H\cap\Pgroup_n=H'\cap\Pgroup_n,
\end{equation*}
then $H=H'$.
\item\label{lem:symp:clifford:i}
If the subgroups $H,H'$ of $\Cl_n$ satisfy
\begin{gather*}
    H\subseteq H',
    \;
    \SY(H)=\SY(H'),\\
    H\cap\Pgroup_n \simeq
    H'\cap\Pgroup_n,
\end{gather*}
then they agree up to scalar phases,
\begin{equation*}
    H \simeq H'.
\end{equation*}
\end{enumerate}
\end{lem}

\begin{proof}
We prove (a) by computing
\begin{align*}
& \SY_U(xv) =  \inviso[\herm(U\iso{xv}U^{\dagger})]
= \inviso[\herm(U \iso{v}^x U^{\dagger})]\\
& =
    \begin{dcases}
        \inviso[\herm(U \iso{v} U^{\dagger})] = 1 \SY_U(v) &  \text{if }\, x = 1\\
        \inviso[\herm(U \In U^{\dagger})] = \inviso(\In)
        = 0 \SY_U(v)
         & \text{if }\,  x=0
    \end{dcases}
\intertext{and by also applying Eq.~\eqref{eq:Pauli:sign} to obtain}
& \SY_U(v+w) =  \inviso[\herm(U\iso{v+w}U^{\dagger})] \\
&=
\inviso[\herm(U \im^{\symp{v}{w}} (-1)^{\sign{v}{w}} \iso{v} \iso{w}U^{\dagger})]\\
&= \inviso[\herm(
U \iso{v} U^{\dagger} U \iso{w}U^{\dagger})] = \inviso[\herm(  \pm \isolong{\SY_U(v)} \isolong{\SY_U(w)} )]\\
&= \inviso[\herm(\im^{\symp{\SY_U(v)}{\SY_U(w)}} (-1)^{\sign{\SY_U(v)}{\SY_U(w)}} \iso{\SY_U(v)+\SY_U(w)} )]\\
& = \inviso[\isolong{\SY_U(v)+\SY_U(w)}] = \SY_U(v)+\SY_U(w).
\end{align*}
In order to prove (b), we set
$U^{\dagger} \iso{v} U = \pm \iso{w}$ [see Eq.~\eqref{eq:clifford:plus:minus}]
and compute
\begin{align*}
\SY_U(\SY_{U^{\dagger}} v) &=
\inviso[ \herm( U \isolong{\inviso[\herm(U^{\dagger} \iso{v} U  )    ]}         U^{\dagger}   )  ]\\
&= \inviso[ \herm( U \herm(U^{\dagger} \iso{v} U  )         U^{\dagger}   )  ]
= \inviso[ \herm( U \herm(\pm \iso{w}  )         U^{\dagger}   )  ]\\
&= \inviso[ \herm( U  \iso{w}         U^{\dagger}   )  ] =
\inviso[ \herm( \pm U  U^{\dagger} \iso{v} U       U^{\dagger}   )  ]\\
&= \inviso[ \herm(  \pm \iso{v}    )  ] = \inviso(\iso{v})  = v.
\end{align*}
The case of $\SY_{U^{\dagger}}(\SY_U w)$ is similar. This proves (b).
(a) and (b) imply (c).
Note that $U\iso{v}U^\dagger$ commutes with $U\iso{w}U^\dagger$ iff $\iso{v}$ and $\iso{w}$ commute.
We infer that $\symp{v}{w}=0$ iff $\symp{\SY_Uv}{\SY_Uw}=0$
for all $v,w \in \Fn$
which follows via Lemma~\ref{lem:symp:pauli}. This then implies (d).
Given a Pauli string $P = \iso{\tilde{v}} \in \PP_n$,
we analyze $\SY$ using Eq.~\eqref{eq:proj:conj}
and compute
\begin{align*}
&\herm((U\!P)\iso{v}(U\!P)^{\dagger})
=\herm(U\!P \iso{v} P U^{\dagger})\\
&= \herm((-1)^{\symp{v}{\tilde{v}}} U\iso{v}U^{\dagger})
= \herm(U\iso{v}U^{\dagger}).
\intertext{Setting set $Q = \iso{\tilde{w}} = \herm(U\iso{v}U^{\dagger})  \in \PP_n$,
we also obtain}
&\herm((PU)\iso{v}(PU)^{\dagger})
=\herm(PU \iso{v} U^{\dagger} P)\\
&= \herm( \pm  PQ  P) = \herm(\pm (-1)^{\symp{\tilde{w}}{\tilde{v}}} Q)
= \herm(U\iso{v}U^{\dagger}).
\end{align*}
These two computations then imply
(e) and (f). Part (g) follows by restricting the short exact sequence in (f) to $H$.
Part (h) is Lemma~\ref{lem:maximality_of_quotient_group} applied to $G=\Cl_n$, $N=\Pgroup_n$, and $\pi=\SY$.
For (i), set $\hat{H}=\groupclosure{\ZZ(\Pgroup_n)\cup H}$ and $\hat{H}'=\groupclosure{\ZZ(\Pgroup_n)\cup H'}$.
First, $H\subseteq H'$ implies $\hat{H}\subseteq\hat{H}'$.
Second, since $\ZZ(\Pgroup_n)\subseteq\Pgroup_n=\ker(\SY)$, adjoining scalar phases does not change the binary image:
\begin{equation*}
    \SY(\hat{H})=\SY(H)=\SY(H')=\SY(\hat{H}').
\end{equation*}
Third, the Pauli kernels of the enlarged groups are
\begin{align*}
    \hat{H}\cap\Pgroup_n
    &=
    \groupclosure{\ZZ(\Pgroup_n)\cup (H\cap\Pgroup_n)},\\
    \hat{H}'\cap\Pgroup_n
    &=
    \groupclosure{\ZZ(\Pgroup_n)\cup (H'\cap\Pgroup_n)}.
\end{align*}
The assumed equality of the phase-adjoined Pauli kernels in (i) therefore gives $\hat{H}\cap\Pgroup_n=\hat{H}'\cap\Pgroup_n$.
Thus $\hat{H}$ and $\hat{H}'$ satisfy the assumptions of Lemma~\ref{lem:maximality_of_quotient_group}, again with $G=\Cl_n$, $N=\Pgroup_n$, and $\pi=\SY$, and therefore $\hat{H}=\hat{H}'$.
\end{proof}

We define the sign function $\eta_U$ from $\Fn$ to
$\F_2$
using
\begin{subequations}
\label{eq:eta}
\begin{align}
U \iso{w} U^{\dagger} &= (-1)^{\eta_U(w)}\, \herm(U\iso{w}U^{\dagger})\\
&= (-1)^{\eta_U(w)}\, \isolong{\SY_U w}.
\end{align}
\end{subequations}
Thus conjugating Pauli strings with a given Clifford unitary $U\in\Cl_n$ is uniquely specified by
$\SY_U$ and $\eta_U$. We first analyze $\eta_U$ for
$U = \iso{v} \in \PP_n$, i.e., $\iso{v} \iso{w} \iso{v} = (-1)^{\symp{v}{w}} \iso{w}$ and
$\eta_{\iso{v}}(w) = \symp{v}{w}$.
Recall that $\symp{v}{\basel_j}=v_{n+j}$ and $\symp{v}{\basel_{n+j}}=v_{j}$. Hence any sign combination for $X_j=\iso{\basel_j}$ and $Z_j=\iso{\basel_{n+j}}$ with
$j\in\{1,\ldots,n\}$ is possible by choosing
$v$.
We describe the general result:
\begin{lem}[Signs of a Clifford conjugation]\label{lem:sign:clifford}
Recall $\signempty$ from Eq.~\eqref{eq:Pauli:sign} and $\SY_U$ from Eq.~\eqref{eq:proj:conj}.
Properties of the sign function
$\eta_U$ [see Eq.~\eqref{eq:eta}] for conjugating with a Clifford unitary $U\in \Cl_n$
are summarized by:
\begin{enumerate}
    \item\label{lem:sign:clifford:a}
    $\eta_{\In}(w)=0$ for all $w\in \Fn$
and $\eta_U(w) =0$ for $w=0$.
    \item\label{lem:sign:clifford:b}
    $\eta_{U^{\dagger}}(w) = \eta_U(\SY_{U^{\dagger}} w)$.
    \item\label{lem:sign:clifford:c}
    $\eta_{[U\iso{v}]}(w) = \eta_{U}(w) + \symp{v}{w}$.
    \item\label{lem:sign:clifford:d}
    $\eta_{[\iso{v}U]}(w) = \eta_{U}(w) + \symp{v}{\SY_{U} w}$.
    \item\label{lem:sign:clifford:e}
    For every $M \in \Sp(2n,\F_2)$, a Clifford
unitary $U'$ with $M = \SY_{U'}$ exists such that the signs can be chosen as $U'  \isolong{M^{-1} \basel_j} (U')^{\dagger}  =\iso{\basel_j} $, i.e., $\eta_{(U')^{\dagger}}(\basel_j) = 0$.
    \item\label{lem:sign:clifford:f}
    $\eta_U(v+w) + \eta_U(v)+\eta_U(w) =  \sign{v}{w}  + \sign{\SY_Uv}{\SY_Uw}$.
    \item\label{lem:sign:clifford:g}
    The function $\eta_U$ is uniquely determined by its values $\eta_U(\basel_j)$ on the $2n$ basis vectors $\basel_j \in \Fn$
and $\SY_U$.
    \item\label{lem:sign:clifford:h}
    $\eta_{UU'}(v) = \eta_{U'}(v) + \eta_U(\SY_U'v)$
    \item\label{lem:sign:clifford:i}
    Conjugation with $U$ is uniquely specified by $\SY_U$ and the values $\eta_U(\basel_j)$ on the $2n$ basis vectors $\basel_j \in \Fn$.
    \item\label{lem:sign:clifford:j}
    $\eta_U$ is in general \emph{neither} linear \emph{nor}
a quadratic form.
\end{enumerate}
\end{lem}

\begin{proof}
(a) follows from
$\In \iso{w} \In= \iso{w}$ and
$U \In U^{\dagger} = \In$.
Setting $v = \SY_{U^{\dagger}} w$ and applying Eq.~\eqref{eq:eta}, we have
\begin{align*}
(-1)^{\eta_{U^{\dagger}}(w)} \iso{w}
&= U [  (-1)^{\eta_{U^{\dagger}}(w)}  U^{\dagger} \iso{w} U ] U^{\dagger}\\
&= U \iso{v} U^{\dagger} = (-1)^{\eta_U(v)} \iso{w}
\end{align*}
which proves (b).
The discussion preceding this lemma directly implies (c) and (d).
A Clifford unitary $u$ with $M = \SY_u$ exists due to
Lemma~\ref{lem:symp:clifford}\ref{lem:symp:clifford:f}.
Thus we can choose $v$ such that $U' = \iso{v} U$, where
\begin{equation*}
\iso{v} U \isolong{M^{-1} \basel_j}
U^{\dagger} \iso{v} = (-1)^{\eta_U(M^{-1} \basel_j) + \symp{v}{\basel_j}} \iso{\basel_j} =\iso{\basel_j}
\end{equation*}
proves (e).
Lemma~\ref{lem:symp:pauli}(a) and Proposition~\ref{prop:Pauli:sign} imply
\begin{align*}
&U \iso{v} \iso{w} U^{\dagger}
= \im^{\symp{v}{w}} (-1)^{\sign{v}{w}} (-1)^{\eta_U(v+w)}\, \isolong{\SY_u (v+w)}
\intertext{and}
&U \iso{v} U^{\dagger} U \iso{w} U^{\dagger}
= (-1)^{\eta_U(v)} \isolong{\SY_U v} (-1)^{\eta_U(w)}\, \isolong{\SY_U w} \\
&= \im^{\symp{\SY_Uv}{\SY_Uw}} (-1)^{\sign{\SY_Uv}{\SY_Uw}}  \\
&\phantom{=}\, \times
(-1)^{\eta_U(v)+\eta_U(w)}\,
\isolong{\SY_U (v+w)}.
\end{align*}
The proof of (f) is now obtained by comparison.
For $U,U'\in\Cl_n$ we have:
\begin{align*}
        UU'(\iso{v})U'^\dagger U^\dagger &= (-1)^{\eta_{U'}(v)} U\iso{\SY_{U'}v}U^\dagger\\
        &= (-1)^{\eta_{U'}(v) + \eta_U(\SY_{U'}v)}\iso{\SY_{UU'}v}
\end{align*}
which proves (g).
(h) and (i) are a direct consequence of (f). To verify (j), we choose
$U=\mathrm{CNOT}(1,2,2)$ (see Table~\ref{tab:ex:symplectic})
as an example. The polynomial in (f) is then given by
\begin{align*}
& \component{v}{1} \component{w}{4} + \component{v}{4} \component{w}{1} + \component{v}{1} \component{v}{2} \component{w}{4}
+ \component{v}{1} \component{v}{3} \component{w}{4} + \component{v}{1} \component{v}{4} \component{w}{2} +
\component{v}{1} \component{v}{4} \component{w}{3}  \\
&+ \component{v}{2} \component{v}{4} \component{w}{1} + \component{v}{3} \component{v}{4} \component{w}{1} +
\component{v}{1} \component{w}{2} \component{w}{4} + \component{v}{1} \component{w}{3} \component{w}{4}
+ \component{v}{2} \component{w}{1} \component{w}{4}  \\
&+ \component{v}{3} \component{w}{1} \component{w}{4} + \component{v}{4} \component{w}{1} \component{w}{2} +
\component{v}{4} \component{w}{1} \component{w}{3},
\end{align*}
which is clearly neither zero (which shows that
$\eta_U$ is not linear)
nor bilinear (which shows that $\eta_U$
is not a quadratic form).
\end{proof}

Similarly as in \cite{Calderbank_IEEE_1998}, the finite Clifford group
\begin{align}\label{eq:clifford:finite}
&\cl_n := \Cl_n \cap \mat{2^n}{\Q(z_8)} \\
&\text{with} z_8 :=\exp(2\pi\im/8) = (1{+}\im)/{\sqrt{2}} \nonumber
\end{align}
is defined by restricting matrix entries to the cyclotomic field $\Q(z_8)$
using the primitive 8th root $z_8$ of unity, i.e., $z_8^8=1$ and no smaller
positive power equals $1$.
This also adds $\im= z_8^2$ and $\sqrt{2}=z_8 - z_8^3$ to the rational numbers $\Q$.
Hence the global phases $z_8^{\varpi}$ in the center
\begin{align*}
\ZZ(\cl_n) = \{  z_8^{\varpi} \In \text{ for }   \varpi \in \{0,\ldots,7\} \}
\end{align*}
of the finite group $\cl_n$ are restricted to eight values.

A complete set of generators for $\cl_n$ consists of $X_j$, $Z_j$,
\begin{align*}
\mathrm{H}_j &= I^{\otimes (j-1)}
\otimes
\tfrac{1}{\sqrt{2}}
\left[
\begin{smallmatrix}
1 & \phantom{+} 1\\
1 & -1\\
\end{smallmatrix}
\right]
\otimes I^{\otimes (n-j)},\\
\mathrm{S}_j &=
I^{\otimes (j-1)}
\otimes
\left[
\begin{smallmatrix}
1 & 0\\
0 & i\\
\end{smallmatrix}
\right]\otimes I^{\otimes (n-j)}, \text{ and}
\\
\mathrm{CNOT}(c,t,n)
&=
\tfrac{1}{2}[ \In + Z_c + X_t - Z_c X_t]
\end{align*}
for $j,c,t \in\{1,\ldots,n\}$ and $c \ne t$.
Table~\ref{tab:ex:symplectic} shows the corresponding unitary matrices $U$
for the case of two qubits
together
with their symplectic
matrices $\SY_U$ and sign values $\eta_U(\basel_j)$ for $j\in \{1,\ldots,2n\}$.

\begin{table}[t]
\caption{\label{tab:ex:symplectic}Clifford unitaries $U$ for two qubits with their symplectic
matrices $\SY_U$ and sign values $\eta_U(\basel_j)$ for $1\leq j \leq 2n$.}
\centering
\begin{tabular}{@{\hspace{1mm}} l
@{\hspace{3mm}} c
@{\hspace{4mm}} c
@{\hspace{4mm}} c
@{\hspace{1mm}}}
\\[-3mm]
\hline\hline
\\[-3mm]
Name & Unitary & Symplectic & Sign values \\
& matrix $U$ & matrix $\SY_U$ & $\eta_U(\basel_j)$
\\[0.5mm] \hline
\\[-2.5mm]
$\mathrm{CNOT}(1,2,2)$
&
$
\left[
\begin{smallmatrix}
1 & 0 & 0 & 0\\
0 & 1 & 0 & 0\\
0 & 0 & 0 & 1\\
0 & 0 & 1 &0
\end{smallmatrix}
\right]
$
&
$
\left[
\begin{smallmatrix}
1 & 0 & 0 & 0\\
1 & 1 & 0 & 0\\
0 & 0 & 1 & 1\\
0 & 0 & 0 & 1
\end{smallmatrix}
\right]
$
&
$
\left[
\begin{smallmatrix}
0 \\ 0 \\ 0 \\ 0
\end{smallmatrix}
\right]
$
\\[3mm]
$\mathrm{CNOT}(2,1,2)$
&
$
\left[
\begin{smallmatrix}
1 & 0 & 0 & 0\\
0 & 0 & 0 & 1\\
0 & 0 & 1 & 0\\
0 & 1 & 0 &0
\end{smallmatrix}
\right]
$
&
$
\left[
\begin{smallmatrix}
1 & 1 & 0 & 0\\
0 & 1 & 0 & 0\\
0 & 0 & 1 & 0\\
0 & 0 & 1 & 1
\end{smallmatrix}
\right]
$
&
$
\left[
\begin{smallmatrix}
0 \\ 0 \\ 0 \\ 0
\end{smallmatrix}
\right]
$
\\[3mm]
$\mathrm{H}_1$
&
$
\tfrac{1}{\sqrt{2}}
\left[
\begin{smallmatrix}
1 & 0 & \phantom{-}1 & \phantom{-}0\\
0 & 1 & \phantom{-}0 & \phantom{-}1\\
1 & 0 & -1 & \phantom{-}0\\
0 & 1& \phantom{-}0 & -1
\end{smallmatrix}
\right]
$
&
$
\left[
\begin{smallmatrix}
0 & 0 & 1 & 0\\
0 & 1 & 0 & 0\\
1 & 0 & 0 & 0\\
0 & 0 & 0 & 1
\end{smallmatrix}
\right]
$
&
$
\left[
\begin{smallmatrix}
0 \\ 0 \\ 0 \\ 0
\end{smallmatrix}
\right]
$
\\[3mm]
$\mathrm{H}_2$
&
$
\tfrac{1}{\sqrt{2}}
\left[
\begin{smallmatrix}
1 & \phantom{-}1 & 0 & \phantom{-}0\\
1 & -1 & 0 & \phantom{-}0\\
0 & \phantom{-}0 & 1 & \phantom{-}1\\
0 & \phantom{-}0& 1 & -1
\end{smallmatrix}
\right]
$
&
$
\left[
\begin{smallmatrix}
1 & 0 & 0 & 0\\
0 & 0 & 0 & 1\\
0 & 0 & 1 & 0\\
0 & 1 & 0 & 0
\end{smallmatrix}
\right]
$
&
$
\left[
\begin{smallmatrix}
0 \\ 0 \\ 0 \\ 0
\end{smallmatrix}
\right]
$
\\[3mm]
$\mathrm{S}_1$
&
$
\left[
\begin{smallmatrix}
1 &0 & 0 & 0\\
0 & 1 & 0 & 0\\
0 & 0 & \im & 0\\
0 & 0 & 0 & \im
\end{smallmatrix}
\right]
$
&
$
\left[
\begin{smallmatrix}
1 & 0 & 0 & 0\\
0 & 1 & 0 & 0\\
1 & 0 & 1 & 0\\
0 & 0 & 0 & 1
\end{smallmatrix}
\right]
$
&
$
\left[
\begin{smallmatrix}
0 \\ 0 \\ 0 \\ 0
\end{smallmatrix}
\right]
$
\\[3mm]
$\mathrm{S}_2$
&
$
\left[
\begin{smallmatrix}
1 &0 & 0 & 0\\
0 & \im & 0 & 0\\
0 & 0 & 1 & 0\\
0 & 0 & 0 & \im
\end{smallmatrix}
\right]
$
&
$
\left[
\begin{smallmatrix}
1 & 0 & 0 & 0\\
0 & 1 & 0 & 0\\
0 & 0 & 1 & 0\\
0 & 1 & 0 & 1
\end{smallmatrix}
\right]
$
&
$
\left[
\begin{smallmatrix}
0 \\ 0 \\ 0 \\ 0
\end{smallmatrix}
\right]
$
\\[3mm]
$X_1$
&
$
\left[
\begin{smallmatrix}
0 & 0 & 1 & 0\\
0 & 0 & 0 & 1\\
1 & 0 & 0 & 0\\
0 & 1 & 0 & 0
\end{smallmatrix}
\right]
$
&
$
\left[
\begin{smallmatrix}
1 & 0 & 0 & 0\\
0 & 1 & 0 & 0\\
0 & 0 & 1 & 0\\
0 & 0 & 0 & 1
\end{smallmatrix}
\right]
$
&
$
\left[
\begin{smallmatrix}
0 \\ 0 \\ 1 \\ 0
\end{smallmatrix}
\right]
$
\\[3mm]
$X_2$
&
$
\left[
\begin{smallmatrix}
0 & 1 & 0 & 0\\
1 & 0 & 0 & 0\\
0 & 0 & 0 & 1\\
0 & 0 & 1 & 0
\end{smallmatrix}
\right]
$
&
$
\left[
\begin{smallmatrix}
1 & 0 & 0 & 0\\
0 & 1 & 0 & 0\\
0 & 0 & 1 & 0\\
0 & 0 & 0 & 1
\end{smallmatrix}
\right]
$
&
$
\left[
\begin{smallmatrix}
0 \\ 0 \\ 0 \\ 1
\end{smallmatrix}
\right]
$
\\[3mm]
$Z_1$
&
$
\left[
\begin{smallmatrix}
1 & 0 & \phantom{-}0 & \phantom{-}0\\
0 & 1 & \phantom{-}0 & \phantom{-}0\\
0 & 0 & -1 & \phantom{-}0\\
0 & 0 & \phantom{-}0 & -1
\end{smallmatrix}
\right]
$
&
$
\left[
\begin{smallmatrix}
1 & 0 & 0 & 0\\
0 & 1 & 0 & 0\\
0 & 0 & 1 & 0\\
0 & 0 & 0 & 1
\end{smallmatrix}
\right]
$
&
$
\left[
\begin{smallmatrix}
1 \\ 0 \\ 0 \\ 0
\end{smallmatrix}
\right]
$
\\[3mm]
$Z_2$
&
$
\left[
\begin{smallmatrix}
1 & \phantom{-}0 & 0 & \phantom{-}0\\
0 & -1 & 0 & \phantom{-}0\\
0 & \phantom{-}0 & 1 & \phantom{-}0\\
0 & \phantom{-}0 & 0 & -1
\end{smallmatrix}
\right]
$
&
$
\left[
\begin{smallmatrix}
1 & 0 & 0 & 0\\
0 & 1 & 0 & 0\\
0 & 0 & 1 & 0\\
0 & 0 & 0 & 1
\end{smallmatrix}
\right]
$
&
$
\left[
\begin{smallmatrix}
0 \\ 1 \\ 0 \\ 0
\end{smallmatrix}
\right]
$
\\[3mm]
\hline\hline
\end{tabular}
\end{table}

Based on Lemma~\ref{lem:symp:clifford}\ref{lem:symp:clifford:f}, the projection $\SY$ induces the same binary symplectic quotient for the finite Clifford group $\cl_n$.
We record this together with the corresponding subgroup criterion.
\begin{lem}[Finite Clifford subgroups over binary images]\label{lem:finite_clifford_subgroups_over_binary_images}
\leavevmode
\begin{enumerate}
\item\label{lem:finite_clifford_subgroups_over_binary_images:a}
The map $\SY$ induces an isomorphism from $\cl_n/\Pgroup_n$ to $\Sp(2n,\F_2)$.
Thus $\SY$ fits into a short exact sequence
\begin{equation}
    1\to \Pgroup_n \to \cl_n \xrightarrow{\SY} \Sp(2n,\F_2)\to 1.
\end{equation}
\item\label{lem:finite_clifford_subgroups_over_binary_images:b}
For any subgroup $H$ of $\cl_n$, restricting $\SY$ gives
\begin{equation}
    1\to H\cap\Pgroup_n \to H \xrightarrow{\SY} \SY(H)\to 1.
\end{equation}
\item\label{lem:finite_clifford_subgroups_over_binary_images:c}
If the subgroups $H,H'$ of $\cl_n$ satisfy
\begin{equation}
    H\subseteq H',
    \;
    \SY(H)=\SY(H'),
    \;
    H\cap\Pgroup_n=H'\cap\Pgroup_n,
\end{equation}
then $H=H'$.
\item\label{lem:finite_clifford_subgroups_over_binary_images:d}
If the subgroups $H,H'$ of $\cl_n$ satisfy
\begin{gather*}
    H\subseteq H',
    \;
    \SY(H)=\SY(H'),\\
    H\cap\Pgroup_n\simeq H'\cap\Pgroup_n,
\end{gather*}
then they agree after adjoining scalar phases,
\begin{equation*}
    H\simeq H'.
\end{equation*}
\end{enumerate}
\end{lem}
\begin{proof}
Lemma~\ref{lem:symp:clifford}\ref{lem:symp:clifford:f} gives the corresponding quotient statement for $\Cl_n$.
The finite Clifford generators listed above project under $\SY$ onto generators of $\Sp(2n,\F_2)$, so $\SY$ remains surjective after restriction to $\cl_n$ and has kernel $\Pgroup_n$ there.
This proves (a).
Part (b) follows by restricting the short exact sequence in (a) to $H$.
Part (c) is Lemma~\ref{lem:maximality_of_quotient_group} applied to $G=\cl_n$, $N=\Pgroup_n$, and $\pi=\SY$.
Part (d) follows by applying the same argument after adjoining scalar phases, exactly as in the proof of Lemma~\ref{lem:symp:clifford}\ref{lem:symp:clifford:i}.
\end{proof}

\subsection{Symplectic Transvections\label{sec:transvections}}

We recall the notion of a symplectic transvection.
The literature differs on whether the
identity matrix should be a symplectic transvection,
but we explicitly allow it.

\begin{defn}[Symplectic transvection, its center  \cite{omeara_1978,grove_2002}]
An element $\tau\in \Sp(2n,\F_2)$ from the binary symplectic group is a
\emph{symplectic transvection} (or \emph{transvection} for brevity) if either $\tau=\idn$
or the image of
$\tau-\idn$ is one dimensional, i.e.,
$\dim[\Im(\tau-\idn)]=1$.
Its \emph{center} $v$
is the zero vector $v=0\in\Fn$
if $\tau=\idn$ and the nonzero vector $v\in \Im[(\tau-\idn)]$
otherwise.
\end{defn}

Let $v\in\Fn$ and $\tau_v$ be the linear map defined as $\tau_vw = w + \symp{v}{w}v$.
Thus $\tau_v$  is a transvection with center $v$.
We show that $\tau_v$ is the unique transvection with center $v$:
\begin{lem}[Explicit form of a transvection \cite{omeara_1978,grove_2002}]
\label{lem:explict:transvections}
Given a transvection $\tau$ with center $v$, then $\tau=\tau_v$ with
\begin{align*}
    \tau_vw &= w + \symp{v}{w}v \\
    &=
    \begin{dcases}
        w &  \text{if } \symp{v}{w} = 0,\\
        w+v & \text{if }  \symp{v}{w}=1.
    \end{dcases}
\end{align*}
\end{lem}
\begin{proof}
The statement is trivially true for $v=0$.
If $\tau\neq\idn$ (i.e.\ $v\neq 0$), then by \cite[Proposition 1.4.12]{omeara_1978}, a transvection $\tau$ in $\Sp(2n,\F_2)$
may always be written as $\tau = \sigma_{a,v}$ with $\sigma_{\lambda,v}w \equiv w + \lambda \symp{v}{w}$ where $\lambda\in\F_2\setminus\{0\}$ and $v$ a nonzero vector $v\in \Im[(\tau-\idn)]$.
Clearly, for $\F_2$, we may only have $\lambda=1$ and $v$ is the unique non-zero vector in $\Im[(\tau-\idn)]$.
Hence, we have $\tau=\tau_v=\sigma_{1,v}$.
\end{proof}
This provides a bijection between vectors
$v\in\Fn$ and transvections $\tau_v$.
For a nonempty set $\vgens \subseteq \Fn$, let
\begin{equation*}
\tvlong{\vgens} = \tv{\vgens} :=\{\tau_v \text{ for } v \in \vgens\}
\end{equation*}
denote the corresponding set of transvections, and
\begin{equation}\label{eq:tv:group}
\tvgroupempty = \tvgroup{\vgens}:=\groupclosure{\tv{\vgens}} \subseteq \Sp(2n,\F_2)
\end{equation}
is the generated \emph{transvection group}.
In general, not all elements of a transvection group are transvections
as additional elements are generated.
To characterize all transvections
in a group $\tvgroup{\vgens}$, one introduces the set
\begin{equation}
\tvcenter{\tvgroup{\vgens}} := \{ w \text{ for transvections } \tau_w \in \tvgroup{\vgens}\}
\supseteq \vgens
\label{eq:tv:center}
\end{equation}
of their centers. We observe the following properties:
\begin{lem}\label{lem:transvection:conjugation}
Given a nonempty set $\vgens \subseteq \Fn$, we have:\\
(a) $\tau_v^2 = 1$ for $v\in \vgens$ and $0 \in \tvcenter{\tvgroup{\vgens}}$.\\
(b) $\tau_v \tau_w = \tau_w \tau_v$ iff $\symp{v}{w}=0$ iff $\comm{\iso{v}}{\iso{w}}=0$.\\
(c) $\tau_v  \tau_w  \tau_v = \tau_{\tau_v w}$.\\
(d) If $v,w \in \tvcenter{\tvgroup{\vgens}}$ and $\symp{v}{w}=1$, then $v+w \in \tvcenter{\tvgroup{\vgens}}$.
\end{lem}

\begin{proof}
(a) is clear and (c) implies (d). For (b), compute
$$\tau_{v} \tau_{w} \tilde{w} - \tau_{w} \tau_{v} \tilde{w}
=\symp{v}{w} ( \symp{w}{\tilde{w}} v + \symp{v}{\tilde{w}} w),$$
where $( \symp{w}{\tilde{w}} v + \symp{v}{\tilde{w}} w)=0$ implies
$v=w$ or
$\symp{w}{\tilde{w}}= \symp{v}{\tilde{w}} =0$ for all $\tilde{w}$,
where $v=w=0$ holds
in the latter case
as the symplectic form is nondegenerate on $\Fn$.
But then $\symp{v}{w}=0$ and
(b) follows by also applying
Lemma~\ref{lem:symp:pauli}(d).
For (c), note that
\begin{equation*}
\tau_v  \tau_w  \tau_v \tilde{w}
=  \tilde{w} + \symp{w + \symp{v}{w} v}{\tilde{w}} (w + \symp{v}{w} v) = \tau_{\tau_v w}. \mbox{\qedhere}
\end{equation*}
\end{proof}

Starting from the generating set $\tv{\vgens}$, one can obtain new transvections in the transvection group $\tvgroup{\vgens}$
(which are not already contained in $\tv{\vgens}$)
by conjugation
as Lemma~\ref{lem:transvection:conjugation}(d) shows. We obtain a (possibly) new transvection
$\tau_{\tau_v w} = \tau_v  \tau_w  \tau_v$ from the transvections $\tau_v$ and $\tau_w$
if $\symp{v}{w}=1$.
However, there can be transvections in the transvection group
that are not conjugated to elements from the generating set.
\begin{defn}(Pathological transvection) \label{def:pathological}
A transvection $\tau_v$ in the transvection group $\tvgroup{\vgens}$ generated by the transvections $\tv{\vgens}$
is called \emph{pathological} if $v\neq 0$, $\tau_v \notin \tv{\vgens}$, and $\tau_v$ is not conjugated
to an element of $\tv{\vgens}$, i.e., there are no $\tau_w \in \tv{\vgens}$ and $g \in \tvgroup{\vgens}$ with
$\tau_v = g^{-1} \tau_w g$. Let $\tvpathol{\tvgroup{\vgens}}\subseteq \tvcenter{\tvgroup{\vgens}}$ denote the set of pathological transvections
in the transvection group
$\tvgroup{\vgens}$.
\end{defn}

We prove in Appendix~\ref{app:pathological} the following characterization of pathological transvections:
\begin{lem}\label{lem:pathological}
Each pathological transvection $\tau_v$ in a transvection group
$\tvgroup{\vgens}$ is contained in the center $\ZZ(\tvgroup{\vgens})$
of $\tvgroup{\vgens}$, i.e., $\tau_v \in \ZZ(\tvgroup{\vgens})$.
Moreover, a transvection $\tau_v$ with $v\neq 0$ and $\tau_v \notin \tv{\vgens}$
is pathological iff $\tau_v \in \ZZ(\tvgroup{\vgens})$.
\end{lem}

\begin{table}[t]
\caption{Transvection groups $\tvgroup{\vgens}$
with and without pathological transvection $\tau_w$
and a factorization of $\tau_w$
into generating transvections $\tau_{v_j}$.
Here, $v_j$ denotes the $j$th vector in $\vgens$.}
\label{tab:ex:pathological}
\centering
\begin{tabular}{@{\hspace{1mm}} l
@{\hspace{2mm}} l
@{\hspace{2mm}} l
@{\hspace{6mm}} l
@{\hspace{4mm}} l
@{\hspace{1mm}}}
\\[-3mm]
\hline\hline
\\[-3mm]
Ex.\ & Paulis & Vectors & Basis of
&Centers $w$
\\
& $\pgens \subseteq \PP_n$
& $\vgens \subseteq \Fn$
& $\rad(\vgens)$
& of pathol.\ $\tau_w$
\\[0.5mm] \hline
\\[-2.5mm]
(a)
&
$\mathrm{XI}$
&
$(1000)^T$
&
$(1100)^T$
\\
&
$\mathrm{IX}$
&
$(0100)^T$
\\
&
$\mathrm{ZZ}$
&
$(0011)^T$
\\[0.5mm]
(b)
&
$\mathrm{XII}$
&
$(100000)^T$
&
$(100000)^T$
&
$(111000)^T$
\\
&
$\mathrm{IXI}$
&
$(010000)^T$
&
$(010000)^T$
\\
&
$\mathrm{IIX}$
&
$(001000)^T$
&
$(001000)^T$
\\
&
$\mathrm{XXI}$
&
$(110000)^T$
\\
&
$\mathrm{XIX}$
&
$(101000)^T$
&
\multicolumn{2}{l}{
$\tau_w =
\tau_{v_1} \tau_{v_2} \tau_{v_3} \tau_{v_4} \tau_{v_5} \tau_{v_6}$}
\\
&
$\mathrm{IXX}$
&
$(011000)^T$
\\[0.5mm]
(c)
&
$\mathrm{XZX}$
&
$(101010)^T$
&
$(100000)^T$
&
$(100000)^T$
\\
&
$\mathrm{IXI}$
&
$(010000)^T$
\\
&
$\mathrm{XXZ}$
&
$(110001)^T$
&
\multicolumn{2}{l}{
$\tau_w=
\tau_{v_1}  \tau_{v_3} \tau_{v_5} \tau_{v_2} \tau_{v_3}  $ }
\\
&
$\mathrm{IYY}$
&
$(011011)^T$
&
\multicolumn{2}{l}{
$
\times\, \tau_{v_4} \tau_{v_6} \tau_{v_1} \tau_{v_5} \tau_{v_2} \tau_{v_3} $ }
\\
&
$\mathrm{IYI}$
&
$(010010)^T$
&
\multicolumn{2}{l}{
$\times\, \tau_{v_4} \tau_{v_6}
\tau_{v_1} \tau_{v_5} \tau_{v_2} \tau_{v_4} \tau_{v_6}$ }
\\
&
$\mathrm{IIZ}$
&
$(000001)^T$
\\[0.5mm]
(d)
&
$\mathrm{XII}$
&
$(100000)^T$
&
$(100000)^T$
\\
&
$\mathrm{IXI}$
&
$(010000)^T$
&
$(010000)^T$
\\
&
$\mathrm{IIX}$
&
$(001000)^T$
\\
&
$\mathrm{IIZ}$
&
$(000001)^T$
\\[0.5mm]
(e)
&
$\mathrm{XII}$
&
$(100000)^T$
&
$(100000)^T$
&
$(110000)^T$
\\
&
$\mathrm{IXI}$
&
$(010000)^T$
&
$(010000)^T$
\\
&
$\mathrm{IIX}$
&
$(001000)^T$
\\
&
$\mathrm{IIZ}$
&
$(000001)^T$
&
\multicolumn{2}{l}{
$\tau_w =\tau_{v_1} \tau_{v_2} \tau_{v_3} \tau_{v_4} \tau_{v_6}
$}
\\
&
$\mathrm{XIX}$
&
$(101000)^T$
&
\multicolumn{2}{l}{
$
\times\, \tau_{v_3} \tau_{v_5} \tau_{v_4} \tau_{v_6} \tau_{v_5}$}
\\
&
$\mathrm{IXZ}$
&
$(010001)^T$
\\[0.5mm]
(f)
&
$\mathrm{IIZY}$
&
$(00010011)^T$
&
$(00001000)^T$
&
$(00001000)^T$
\\
&
$\mathrm{IIYI}$
&
$(00100010)^T$
\\
&
$\mathrm{IZYI}$
&
$(00100110)^T$
&
\multicolumn{2}{l}{
$\tau_w =
\tau_{v_1} \tau_{v_2} \tau_{v_1} \tau_{v_3} \tau_{v_4}$}
\\
&
$\mathrm{IIYZ}$
&
$(00100011)^T$
&
\multicolumn{2}{l}{
$\times\,
\tau_{v_{10}} \tau_{v_{11}} \tau_{v_1} \tau_{v_2} \tau_{v_{13}}$}
\\
&
$\mathrm{IIIY}$
&
$(00010001)^T$
&
\multicolumn{2}{l}{
$\times\,
\tau_{v_4} \tau_{v_5}
\tau_{v_{12}} \tau_{v_4} \tau_{v_{13}}
$}
\\
&
$\mathrm{IYII}$
&
$(01000100)^T$
\\
&
$\mathrm{ZIXI}$
&
$(00101000)^T$
\\
&
$\mathrm{ZIIX}$
&
$(00011000)^T$
\\
&
$\mathrm{ZXII}$
&
$(01001000)^T$
\\
&
$\mathrm{ZZII}$
&
$(00001100)^T$
\\
&
$\mathrm{ZIIZ}$
&
$(00001001)^T$
\\
&
$\mathrm{ZIZI}$
&
$(00001010)^T$
\\
&
$\mathrm{ZZZI}$
&
$(00001110)^T$
\\
&
$\mathrm{ZIZZ}$
&
$(00001011)^T$
\\[1mm]
\hline\hline
\end{tabular}
\end{table}

We will discuss concrete examples of pathological transvections.
Now, we recall the definition of the radical of $\vgens\subseteq\Fn$ as  \cite{omeara_1978,grove_2002,Bourbaki1959}
\begin{equation}\label{eq:radical}
    \rad(\vgens) := \{ u\in\Span[\F_2]{\vgens} \text{ with } \symp{u}{v} = 0 \text{ for all } v\in\vgens\}
\end{equation}
which corresponds to the Pauli strings in the center of $\groupclosure{\isolong{\vgens}}$, hence the Pauli string basis of $\ZZ(\algclosure{\isolong{\vgens}})$. 
We also write $\nullity(\vgens) = \dim(\rad(\vgens))$.
More importantly, we have $w \in \rad(\vgens)$
for transvections $\tau_w$ in
center $\ZZ(\tvgroup{\vgens})$ of the transvection group $\tvgroup{\vgens}$. But not all elements in the radical $\rad(\vgens)$
correspond to transvections in the center $\ZZ(\tvgroup{\vgens})$, only the ones
related to pathological transvections (as well as the identity and possibly generating transvections from $\tv{\vgens}$).
This is highlighted in
Table~\ref{tab:ex:pathological} which shows various examples of transvection groups
with and without pathological transvections, despite a nontrivial radical.
In summary, the transvection groups $\tvgroup{\vgens}$ provide a more refined characterization
than the corresponding subgroups $\groupclosure{\isolong{\vgens}}$ of the Pauli group,
which necessarily contain
all Pauli strings $\isolong{\rad(\vgens)}$ arising from
the radical
$\rad(\vgens)$. This distinction will be crucial in Section~\ref{sec:pauli:Lie}.

\subsection{Symplectic Transvections Viewed as Elements in the Clifford Group}\label{sec:transvections_in_clifford_group}

In Sec.~\ref{sec:transvections}, we have recalled transvections and some of their
properties. We now detail how transvections relate to elements of the Clifford group:

\begin{lem}\label{lem:clifford_transvection_binary_action}
For a Pauli string $\iso{v}\in \PP_n$,
we introduce
\begin{align}\label{eq:tvu}
\tvulong{\iso{v}} & = \tvu{\iso{v}} = \tvu{v} := \frac{\In{+}\im \iso{v}}{\sqrt{2}}\\
& = \exp( \pi \im \iso{v} /4) = \cos(\pi/4) \In + \im \sin(\pi/4) \iso{v}
\nonumber
\end{align}
and observe the following properties:\\
(a) $\tvu{v}$ is unitary with $\tvu{v}^{\dagger} = (\In{-}\im \iso{v})/{\sqrt{2}}$.\\
(b) $\tvu{v} \in \Cl_n$ is a Clifford unitary and the induced\\
\phantom{(b)} symplectic matrix observes $\SY(\tvu{v}) = \tau_v$.
\end{lem}

\begin{proof}
(a) is obvious. For (b), we compute [see Eq.~\eqref{eq:Pauli:sign}]
\begin{align*}
&\tfrac{1}{2} ( \In{+}\im \iso{v} ) \iso{w} (\In{-}\im \iso{v}) = \tfrac{1}{2} (\iso{w} {+} \im \iso{v} \iso{w}  )  (\In{-}\im \iso{v})\\
&= \tfrac{1}{2} ( 1 {+} (-1)^{\symp{v}{w}} ) \iso{w}  + \tfrac{1}{2} ( 1 {-} (-1)^{\symp{v}{w}} ) \im \iso{v} \iso{w}\\
&=
\begin{dcases}
\iso{w} &  \text{if } \symp{v}{w} = 0,\\
\im\iso{v}\iso{w} = (-1)^{\sign{v}{w}+1} \iso{w+v} & \text{if }  \symp{v}{w}=1.\mbox{\qedhere}
\end{dcases}
\end{align*}
\end{proof}

Thus the symplectic matrix $\SY(\tvu{v})$ corresponding to the Clifford unitary $\tvu{v}$ from Eq.~\eqref{eq:tvu}
is equal to the transvection $\tau_v$.
Recall from Lemma~\ref{lem:symp:clifford}\ref{lem:symp:clifford:e} that all the Clifford unitaries $\tvu{v}\iso{w}$ with $v,w \in \Fn$
lead to the same transvection $\tau_v =\SY(\tvu{v}\iso{w}) =\SY(\tvu{v}) $.

We denote the Clifford subgroup generated by the Clifford transvections with centers $\iso{v} = P\in\pgens$ by
\begin{equation}\label{eq:def:clifford:transvections}
    \cltvgroupempty = \cltvgroup{\pgens} = \cltvgroup{\vgens} := \groupclosure{\{\tvu{v}\}_{v\in\vgens}} \subseteq \cl_n.
\end{equation}
The projection from the Clifford group to the binary symplectic group sends $\cltvgroup{\vgens}$ onto the transvection group $\tvgroup{\vgens}$.
However, the lift from binary symplectic subgroups to the Clifford group may in general not be unique.
Thus, $\cltvgroup{\vgens}$ and $\Pgroup_n\cdot\cltvgroup{\vgens}$ may be different Clifford subgroups, although both project to the same subgroup $\tvgroup{\vgens}$.
Consequently, an identification of a Clifford transvection group inside $\cl_n$ requires not only the binary transvection group $\tvgroup{\vgens}$, but also the Pauli kernel $\Pgroup_n\cap\cltvgroup{\vgens}$.
\begin{lem}[Clifford transvection groups over binary transvection groups]\label{lem:clifford_transvection_group_binary_image_kernel}
Let $\vgens\subseteq\Fn$ and let $\cltvgroup{\vgens}\subseteq\cl_n$ be the Clifford transvection group from Eq.~\eqref{eq:def:clifford:transvections}.
\begin{enumerate}
\item The map $\SY$ from Eq.~\eqref{eq:proj:conj} induces an isomorphism
\begin{equation}
    \cltvgroup{\vgens}/(\Pgroup_n\cap\cltvgroup{\vgens})
    \cong \tvgroup{\vgens}.
\end{equation}
\item Equivalently, $\SY$ fits into a short exact sequence
\begin{equation}\label{eq:quotient_clifford_transvection_groups_to_binary}
    1 \to \Pgroup_n\cap \cltvgroup{\vgens}
    \to \cltvgroup{\vgens}
    \xrightarrow{\SY} \tvgroup{\vgens} \to 1.
\end{equation}
\item If the subgroup $H'$ of $\cl_n$ satisfies
\begin{equation}
    \cltvgroup{\vgens}\subseteq H',
    \;
    \SY(H')=\tvgroup{\vgens},
    \;
    H'\cap\Pgroup_n=\cltvgroup{\vgens}\cap\Pgroup_n,
\end{equation}
then $H'=\cltvgroup{\vgens}$.
\item If the subgroup $H'$ of $\cl_n$ satisfies
\begin{gather*}
    \cltvgroup{\vgens}\subseteq H',
    \;
    \SY(H')=\tvgroup{\vgens},\\
    \cltvgroup{\vgens}\cap\Pgroup_n \simeq H'\cap\Pgroup_n,
\end{gather*}
then the two subgroups agree after adjoining scalar phases,
\begin{equation}
    \cltvgroup{\vgens} \simeq H'.
\end{equation}
\end{enumerate}

\end{lem}
\begin{proof}
Since $\SY(\tvu{v})=\tau_v$ for all $v\in\vgens$, we have $\SY(\cltvgroup{\vgens})=\tvgroup{\vgens}$.
Items (a) and (b) are therefore the quotient-isomorphism and short-exact-sequence formulations of Lemma~\ref{lem:finite_clifford_subgroups_over_binary_images} applied to $H=\cltvgroup{\vgens}$.
Items (c) and (d) are the equality and phase-adjoined equality criteria from Lemma~\ref{lem:finite_clifford_subgroups_over_binary_images}, again applied to $H=\cltvgroup{\vgens}$.
\end{proof}
Regarding the intersection of the Clifford transvection group with the Pauli group, the identity $\tvu{v}^2 = \im \iso{v}$ gives the lower bound
\begin{equation}\label{eq:lower_bound_paulis_in_transvection_groups}
    \groupclosure{\im\pgens} \subseteq \Pgroup_n\cap\cltvgroup{\vgens}.
\end{equation}
Later, in Proposition~\ref{prop:full_space_symplectic_symmetric_Clifford_groups} and the preceding discussion, this intersection is identified, up to phases, with the Pauli subgroup corresponding to the span of the generators.

\section{Pauli Lie Algebras and Transvection Groups}\label{sec:connection_Pauli_Lie_algebras_transvection_groups}

\subsection{Lie Subalgebras of the Unitary Lie Algebra\label{sec:general:Lie}}

We start by recalling a general class of Lie algebras and
some related properties.
A nonempty set $\pgens \subset \C^{d \times d}$
of hermitian matrices $H_j \in \pgens$ serves as a \emph{generating set} for
the (real) \emph{Lie algebra}
\begin{align}\label{eq:lie:def}
\lieg &= \lieg(\pgens) := \lie{\pgens} \\
&= \Span[\R]{\comm{\im H_{j_1}}{\comm{\im H_{j_2}}{\comm{\im H_{j_3}}{\cdots}}} \text{ for } H_{j_1},H_{j_2},\ldots \in \pgens}
\nonumber
\end{align}
that is given by real-linear combinations of nested commutators of $\im H_{j_1},\im H_{j_2},\ldots\in \im\pgens$
\cite{hall2015,jacobson1979,deGraaf2000,dalessandro2022}.
The commutator of two matrices $A,B \in \C^{d\times d}$ is defined as $[A,B]:=AB-BA$.
Finitely many commutators are sufficient in Eq.~\eqref{eq:lie:def} and
$\lieg$ has a finite-dimensional basis (and specifically $\dim\lieg\leq d^2$).
The Lie algebra $\lieg\subseteq\lieu(d)$ is a subalgebra of the unitary Lie algebra
$\lieu(d)\subset \C^{d\times d}$ consisting of skew-hermitian matrices.
The (connected) \emph{Lie group} $\lieG:=\exp(\lieg)$
is infinitesimally generated by $\lieg$ and it is contained in the
unitary group $\lieU(d)\subset \C^{d\times d}$.
Equivalently, $\lieG$ is the unique connected matrix Lie subgroup
of $\lieU(d)$ with Lie algebra $\lieg$.

Any Lie algebra $\lieg\subseteq \lieu(d)$ has
a \emph{reductive decomposition}
\begin{equation}\label{eq:reductive}
\lieg \isom \ZZ(\lieg) \lieoplus [\lieoplus_{j \in\calJ} \lies_j]
= \ZZ(\lieg) \lieoplus \lies_{j_1}  \lieoplus \cdots  \lieoplus \lies_{j_{\abs{\calJ}}},
\end{equation}
where $\ZZ(\lieg) \subseteq \lieg$
is the center of $\lieg$ [see Eq.~\eqref{eq:center}] and $\lies:= \lieoplus_{j \in\calJ} \lies_j$
is its semisimple part \cite{BourbakiLie1989,Bourbaki2008b}.
The center $\ZZ(\lieg)$ contains the elements from $\lieg$ that commute with all
elements of $\lieg$, i.e., $[\ZZ(\lieg),\lieg]=0$.
Moreover,
$\lieg_1\! \lieoplus \lieg_2$ denotes the direct sum of two commuting Lie subalgebras $\lieg_j \subseteq \lieg$
with
$\lieg_1 \cap \lieg_2 = \{0\}$
and $\comm{\lieg_1}{\lieg_2}=0$.
We use the symbol $\lieoplus$ to distinguish it from other types of direct sums
(such as block diagonal decompositions of matrices).
The semisimple part $\lies=\lieoplus_{j \in\calJ} \lies_j$
is a direct sum of simple Lie algebras $\lies_j$ (such as $\su(q)$, $\so(q)$, and $\usp(q)$
for suitable $q \geq 2$), where
$\lies_j$ is not abelian and there exists \emph{no} finer direct sum decomposition
with $\lies_j \isom \lieoplus_{i\in \calK(j)} \lies_i$ (i.e.,
$\lies_j$ has only itself and the
zero ideal as ideals \cite{hall2015}).
Finally, note that $\comm{\lies}{\lies}=\lies$ \cite[Chapter 1, §6.1, Thm.~1]{BourbakiLie1989}.

\subsection{Pauli Lie Algebras\label{sec:pauli:Lie}}

We now limit ourselves to Pauli-string generators:

\begin{defn}[Pauli Lie algebra and group]\label{def:pauli:lie}
A \emph{Pauli Lie algebra} is a Lie algebra
$\lieg:=\lie{\pgens}$
that is generated
by a set $\pgens \subseteq \PP_n$ of (hermitian) Pauli strings of length $n$,
i.e., $H_j=\bigotimes_{o=1}^n A_o \in \pgens$ with  $A_o
\in \{X, Y, Z, I\}$
(see Section~\ref{sec:pauli}).
A \emph{Pauli Lie group} is the connected matrix Lie subgroup
$\lieG=\exp(\lieg)\subseteq\lieU(2^n)$ associated with a Pauli Lie algebra
$\lieg$. Equivalently, it is generated by the one-parameter subgroups
$\{\exp(\im\theta H_j) \text{ for } \theta\in\R\}$ with $H_j\in\pgens$.
\end{defn}

We collect some basic properties of Pauli Lie algebras:
\begin{lem}[Basic properties of Pauli Lie algebras]\label{lem:pauli:lie}
Consider a Pauli Lie algebra $\lieg:=\lie{\pgens}$
generated by a set $\pgens \subseteq \PP_n$ of Pauli strings.
We obtain:\\
(a) $\lieg \cap \im\PP_n$ is a Pauli-string basis of $\lieg = \SpanS[\R]{\lieg \cap \im \PP_n}$.\\
(b) $\ZZ(\lieg) \cap \im\PP_n$ is a Pauli-string basis of the center $\ZZ(\lieg)$.\\
(c) Any $\im\iso{\tilde{v}} \in  \lieg \cap \im\PP_n$
is either contained in the center \\
\phantom{(c)} $\ZZ(\lieg)$ or in the semisimple part $\lies$ of $\lieg$.\\
(d) The generators $\pgens$ directly determine a basis
$$\{\im g \in \im\pgens \text{ with }  [g,\tilde{g}] = 0 \text{ for all } \tilde{g} \in  \pgens\}$$
\phantom{(d)} of the center $\ZZ(\lieg)$ of $\lieg$.
\end{lem}

\begin{proof}
Proposition~\ref{prop:Pauli:comm} implies that commutators
\begin{equation*}
\comm{\im \iso{v}/2}{\im \iso{w}/2} =
\begin{dcases}
0 &  \text{if } \symp{v}{w} = 0,\\
\pm (\im \iso{v+w}/2) & \text{if }  \symp{v}{w}=1.
\end{dcases}
\end{equation*}
of Pauli strings (multiplied by $\im$) are either zero or yield---up to a real factor---other Pauli strings (multiplied by $\im$).
But one never gets a sum of Pauli strings. Thus (a) follows.
We set $\pgens = \{ \iso{v} \text{ for } v \in \vgens\}$.
For (b), we
assume
\begin{equation*}
z = \sum_{w \in \calW}  c_w \im \iso{w} \in \ZZ(\lieg).
\end{equation*}
It follows that $\comm{z}{\iso{v}}=0$ for all $v \in \vgens$, or equivalently,
\begin{align*}
z = \iso{v} z \iso{v} = \sum_{w  \in \calW, \symp{v}{w}=0}
\hspace{-5mm} c_w \im \iso{w}
- \sum_{w  \in \calW, \symp{v}{w}=1} \hspace{-5mm}  c_w \im \iso{w}
\end{align*}
holds for all $v \in \vgens$.
This implies that $c_w=0$ for all $v \in \vgens$ with $\symp{v}{w}=1$
and $z$ is spanned by some $\iso{w}$ that commute with all $\iso{v}$ for $v \in \vgens$.
We then obtain that $\ZZ(\lieg)$ has a basis consisting of $\iso{w}$ commuting with all
$\iso{v}$ for $v \in \vgens$. But $\ZZ(\lieg) \subseteq \lieg$ and $\ZZ(\lieg)$ consequently
has a Pauli-string basis which is given by $\ZZ(\lieg) \cap \im\PP_n$. This proves (b) and (c).
For (d), the proposed basis for the center is clearly contained in the center.
Let us assume that there exists a $\im\iso{\tilde{v}} \in \ZZ(\lieg)$ that
is not contained in the proposed basis of the center, i.e., $\tilde{v} \notin \vgens$.
One concludes that $\im\iso{\tilde{v}}$ agrees
(up to a real factor)
with a multi-commutator of
elements from $\im\pgens \cap \lies$ by using (a) and Prop.~\ref{prop:Pauli:comm}.
But then
$\im\iso{\tilde{v}} \in \lies$ follows from
$\comm{\lies}{\lies}=\lies$ \cite[Theorem~1 in §6.1 of Chapter 1]{BourbakiLie1989}. This is a contradiction which proves (d).
\end{proof}

Lemma~\ref{lem:pauli:lie}(d) shows that the center of a Pauli Lie algebra can be efficiently
computed in the binary symplectic space. Moreover, we always have a
Pauli-string basis for a Pauli Lie algebra due to Lemma~\ref{lem:pauli:lie}(a).
Based on Lemma~\ref{lem:pauli:lie}, we define the Pauli-string bases
for $\lieg \isom \ZZ(\lieg) \lieoplus \lies$, its center $\ZZ(\lieg)$, and its semisimple
part $\lies$ as
\begin{subequations}
\label{eq:pauli:bases}
\begin{alignat}{3}
\bas{\lieg} &= \baslong{\lieg} &&:= \lieg \cap \im\PP_n,\\
\bas{\ZZ(\lieg)} &= \baslong{\ZZ(\lieg)} &&:= \ZZ(\lieg) \cap \im\PP_n,\\
\bas{\lies} &= \baslong{\lies} &&:= \lies \cap \im\PP_n.
\end{alignat}
\end{subequations}
We can compute the Lie closure for a given
set $\pgens$ of Pauli strings by computing the symplectic product $\symp{v}{w}$
for the Pauli strings $\iso{v}$ and $\iso{w}$
to efficiently decide whether to add the Pauli string $\iso{v+w}$ (see Proposition~\ref{prop:Pauli:comm}).
However, an exponential number of steps might be required starting from a small set
of generators as the
Lie algebra could have an exponential dimension $\dim\lieg\leq 4^n$.

We shortly recall from Eq.~\eqref{eq:pauli:binary} the mapping
\begin{align*}
\iso{v} \mapsto \inviso(\iso{v}) = v
\end{align*}
which maps Pauli strings $\iso{v}$ to binary vectors $v\in\Fn$.
We develop a mapping from bases of Pauli Lie algebras
to centers of transvections in  transvection groups:

\begin{prop}[Linking Pauli Lie algebras to transvection groups]\label{prop:pauli:transvection}
Consider a Pauli Lie algebra $\lieg=\lies\oplus\ZZ(\lieg)\subseteq\lieu(2^n)$ and the
Pauli-string bases
$\bas{\lieg}$ and $\bas{\ZZ(\lieg)}$
from Eqs.~\eqref{eq:pauli:bases}.
The set
$\calW = \inviso(\bas{\lieg}) \subseteq \Fn$ of binary vectors induces
transvections $\tau_w$ with $w\in \calW$,
which generate
the transvection group $\tvgroup{\calW}$
[see Eq.~\eqref{eq:tv:group}]. We observe:\\
(a) $\inviso$ induces an injective map from $\bas{\lieg}$ to the set $\tvcenter{\tvgroup{\calW}}$ \\
\phantom{(a)} of centers of transvections in $\tvgroup{\calW}$; $\inviso(\bas{\lieg}) \subseteq \tvcenter{\tvgroup{\calW}}$.\\
(b) $\inviso(\bas{\ZZ(\lieg)}) \subseteq \tvcenter{\ZZ(\tvgroup{\calW})}$\\
(c) $\tvcenter{\tvgroup{\calW}} = \inviso(\bas{\lieg}) \cup \tvpathol{\tvgroup{\calW}}$
and $\inviso(\bas{\lieg}) \cap \tvpathol{\tvgroup{\calW}} = \emptyset$ \\
\phantom{(c)}  where $\tvpathol{\tvgroup{\calW}}$
is the set
of centers of pathological\\
\phantom{(c)}  transvections in $\tvgroup{\calW}$ (see Def.~\ref{def:pathological}).\\
(d) $\tvcenter{\ZZ(\tvgroup{\calW})} = \inviso(\bas{\ZZ(\lieg)}) \cup \tvpathol{\tvgroup{\calW}}$.
\end{prop}

\begin{proof}
(a) follows from Lemma~\ref{lem:transvection:conjugation}(d), while
Lemma~\ref{lem:transvection:conjugation}(b) implies (b).
Moreover, (c) is a consequence of Definition~\ref{def:pathological} and (a).
Lemma~\ref{lem:pathological} proves (d).
\end{proof}

We neglect the pathological transvections  by ignoring
the centers $\ZZ(\lieg)$ in $\lieg$ and $\ZZ(\tvgroup{\calW})$ in $\tvgroup{\calW}$
and the statements in
Proposition~\ref{prop:pauli:transvection} verify the following bijection:

\begin{thm}[Linking Pauli Lie algebras to transvection groups]\label{thm:pauli:transvection}
Consider a Pauli Lie algebra $\lieg=\lies\oplus\ZZ(\lieg)\subseteq\lieu(2^n)$, the
Pauli-string bases
$\bas{\lieg}$, $\bas{\ZZ(\lieg)}$, and $\bas{\lies}$
from Eqs.~\eqref{eq:pauli:bases},
and the set $\calW = \inviso(\bas{\lieg}) \subseteq \Fn$ of binary vectors.
Hence $\inviso$ induces a bijective map from the Pauli-string basis  of $\lieg$
except for its center
to the set of centers of transvections in the
transvection group except for its center, i.e.,
from $\bas{\lies} = \bas{\lieg} \setminus \bas{\ZZ(\lieg)}$ to $\tvcenter{\tvgroup{\calW}} \setminus \tvcenter{\ZZ(\tvgroup{\calW})}$.
\end{thm}

We extend the statement of Theorem~\ref{thm:pauli:transvection} to describe the orbits
$\tvgroup{\vgens}\cdot \vgens$ of the transvection group $\tvgroup{\vgens}$ acting on
$\vgens$ for a generating set $\pgens= \isolong{\vgens} \subseteq \PP_n$ of Pauli strings:

\begin{cor}[Pauli Lie algebras and transvection group orbits]\label{cor:pauli:transvection}
A set $\pgens= \isolong{\vgens} \subseteq \PP_n$ of Pauli strings for binary vectors $\vgens \subseteq \Fn$
generates the Pauli Lie algebras $\lieg$. The transvection group $\tvgroup{\vgens}$
acts on the vectors $\vgens$ such that the orbits $\tvgroup{\vgens}\cdot \vgens$ corresponds
to the Pauli-string basis $\bas{\lieg} = \isolong{\tvgroup{\vgens} \cdot \vgens}$ of $\lieg$.
\end{cor}

\begin{proof}
Recall from Lemma~\ref{lem:explict:transvections} that
\begin{equation*}
    \tau_vw =    \begin{dcases}
        w &  \text{if } \symp{v}{w} = 0,\\
        w+v & \text{if }  \symp{v}{w}=1.
    \end{dcases}
\end{equation*}
Thus the orbit $\tvgroup{\vgens} \cdot \vgens$ contains all vectors $\tilde{v}$ in $\Fn$ that correspond to elements $i\iso{\tilde{v}}$
reachable in $\lieg$ via commutator. This
is based on Prop.~\ref{prop:pauli:transvection}(a) and
also relies on Proposition~\ref{prop:Pauli:comm}
and on the fact that
$\symp{v}{w}=1$ iff $\comm{\iso{v}}{\iso{w}}\neq 0$
following Lemma~\ref{lem:symp:pauli}(d) and (e).
Clearly, $\ZZ(\tvgroup{\vgens})$ acts trivially on $\vgens$.
No other elements (not in $\vgens$) appear in the orbit $\tvgroup{\vgens} \cdot \vgens$
based on Thm.~\ref{thm:pauli:transvection}
and  Prop.~\ref{prop:pauli:transvection}.
\end{proof}

Given the above statements, it is clear how to use transvection groups in order to classify Pauli Lie algebras.
We can determine the orbit of the transvection group $\tvgroup{\vgens}$ acting on
$\vgens$ for a generating set $\pgens= \isolong{\vgens}$ consisting of Pauli strings. Afterwards, one finds
an isomorphism between the orbit and a Pauli Lie algebra.
As we are dealing with finite groups, the task is much more restricted than for general Lie algebras.
This will allows us to take advantage of several known results regarding the classification of transvection groups and their orbits.
Indeed, beyond identifying the Pauli Lie algebra, we are also able to infer its action on the operator space by determining all of the possible orbits
under the action of the transvection group. Notice that the above steps do not necessarily require knowing the transvection group explicitly,
though in most cases, understanding the structure of the orbits also provides automatically the type of the considered transvection group.

We also highlight however that, if one is only interested in understanding the Pauli Lie algebra, one can also restrict the analysis to the subspace spanned by the generators.
Namely, we have the following:
\begin{lem}\label{lem:restriction_transvection_group_subspace}
Let $\vgens\subseteq\Fn$ be a set of binary vectors and $V=\Span{\vgens}$. Then, $\tvgroup{\vgens}\cdot\vgens\subseteq V$.
\end{lem}
\begin{proof}
Clearly, $\vgens\subseteq V$.
Furthermore, for any $u\in V$ and $v\in\vgens$, we have $\tau_vu = u+\symp{v}{u} v\in V$, hence $\tvgroup{\vgens}\cdot V \subseteq V$, which proves the statement.
\end{proof}
As an immediate consequence, consider
\begin{equation*}
\tvgroup{\vgens}|_V = \{ g|_V \text{ for }  g\in\tvgroup{\vgens}\},
\end{equation*}
which restricts the action of the elements of the transvection group to $V$.
Then, it suffices to describe $\tvgroup{\vgens}|_V$ or the transvections $\tau_v|_V$ to obtain the Pauli basis of the Lie algebra generated by $\pgens = \isolong{\vgens}$.

We also comment on the relationship between the Clifford transvection groups and the Pauli Lie groups.
Given a set of Pauli strings $\pgens\subseteq\PP_n$, we have
\begin{equation}\label{eq:clifford:transvection:inclusion}
    \cltvgroup{\pgens} \subseteq \cl_n\cap \exp(\lie{\pgens}).
\end{equation}
This is a consequence of the fact that transvections $\tvu{G} = \exp(\im\pi/4 G)$ are discrete rotations with respect to Pauli strings. 
For the quasi-universal cases,
the inclusion in Eq.~\eqref{eq:clifford:transvection:inclusion}
is later strengthened, up to phases, to an equality; see Eq.~\eqref{eq:transvections_are_clifford_cap_lie_group} and the surrounding discussion.

\section{Pauli Matrix Algebras}\label{sec:pauli:matrix}

\subsection{Complex Matrix Algebras Generated by Hermitian Matrices \label{sec:general:matrix}}

We recall that a nonempty set of complex matrices $\pgens\subset \C^{d \times d}$ generates a \emph{complex matrix algebra}
\begin{align}\label{eq:mat:alg}
\matalg &= \matalg(\pgens) := \algclosure{\pgens} \\
&= \Span[\C]{ \id_d, \textstyle\prod_{i=1}^m A_{j_i}  \text{ for } A_{j_i} \in \pgens \text{ and } 1 \leq m < \infty}
\nonumber
\end{align}
which contains all complex-linear combinations of arbitrary products of its generators $A_{j_i}\in \pgens$.
Finitely many products are sufficient in Eq.~\eqref{eq:mat:alg} and $\matalg$ has a finite-dimensional basis of size $\dim\matalg\leq d^2$.
As in Sec.~\ref{sec:general:Lie}, we now assume that all generators are hermitian.
\begin{lem}
For a nonempty set of hermitian generators, the generated complex matrix algebra $\matalg$ is semisimple.
\end{lem}

\begin{proof}
Clearly, $M\in\matalg$ implies $M^{\dagger}\in\matalg$, i.e., $\matalg$ is $\dagger$-closed.
The following proof only relies on the fact that $\matalg$ is $\dagger$-closed.
We use some elementary facts about finite-dimensional matrix algebras \cite{lorenz2008}.
The Jacobson radical $\calJ$ of $\matalg$ is the intersection of all maximal left ideals of  $\matalg$
and it is a two-sided ideal of $\matalg$.
In finite dimensions,
$\calJ$ is nilpotent, i.e., all its elements are nilpotent.
For $B \in \calJ$, we have $B^\dagger \in \matalg$
and $B^\dagger B \in \calJ$ (as $\calJ$ is a left ideal).
Thus $B^\dagger B$ is nilpotent and has only zero eigenvalues.
Note that $B^\dagger B$ is hermitian positive semidefinite, hence diagonalizable
with nonnegative real eigenvalues. We conclude that $B^\dagger B=0$ and then $B=0$
(since $\trace(B^\dagger B) = \sum_{jk} \abs{B_{jk}}^2 = 0$).
In summary, $\calJ=0$ and $\matalg$ is semisimple.
\end{proof}

As we have verified that $\matalg$ is semisimple for hermitian generators,
a basis change transforms  $\matalg$
into a block-diagonal decomposition
\begin{equation}\label{eq:block_diagonal_decomposition_matrix_algebras}
\matalg  \conjugated \bigoplus_{\lambda=1}^{\dim[\ZZ(\matalg)]}
\id_{m_{\lambda}} \otimes \C^{d_{\lambda}\times d_{\lambda}}
\end{equation}
Here and below, $\calA\conjugated\calB$ for two matrix or Lie subalgebras means that there is an invertible change of basis $S$ on the underlying vector space such that $S\calA S^{-1}=\calB$.
Thus this records a change of basis of the action, not only an algebra isomorphism class.
for dimensions $d_{\lambda}$ and multiplicities $m_{\lambda}$
with $\sum_\lambda d_{\lambda} m_{\lambda} = d$.
Here, $\ZZ(\matalg)$ denotes the center of $\matalg$, which
contains every element from $\matalg$ that commutes with all elements of $\matalg$ [see Eq.~\eqref{eq:center}].
Note that $\matalg$
isomorphic to
\begin{equation}\label{eq:isomorphic_matrix_algebras}
\matalg  \cong  \bigoplus_{\lambda=1}^{\dim[\ZZ(\matalg)]} \C^{d_{\lambda}\times d_{\lambda}}.
\end{equation}

\subsection{Pauli Matrix Algebras}

Given the symplectic properties of Pauli strings and the isomorphism between the abelian Pauli group $\Pabel_n$ (viewed as a symplectic space) and the binary symplectic space $\Fn$, this also results in isomorphisms of subspaces.
Specifically, a subspace $W$ in $\Fn$ spanned by some vectors in $\vgens = \{v_j \text{ for } j \in\{1,\ldots s\}\}$ corresponds to a subgroup of $\Pabel_n$ generated by $\pgens=\isolong{\vgens}$. All elements of this subgroup are of the form $\iso{v_1}^{x_1}\cdots \iso{v_s}^{x_s}$ with  $x_j\in\F_2$
and a Pauli string can always be chosen as a representative for these elements.
Since the product of two Pauli strings is proportional to another Pauli string and Pauli strings are linearly independent, they are closed under products and complex linear combinations (in $\Pabel_n$). We now clarify this correspondence and
start by introducing the Pauli matrix algebra.
\begin{defn}[Pauli matrix algebra]\label{def:pauli:matrix_algebra}
A \emph{Pauli matrix algebra} is a complex matrix algebra
$\matalg$
that is generated
by a set $\pgens \subseteq \PP_n$ of (hermitian) Pauli strings of length $n$,
i.e., $H_j=\bigotimes_{o=1}^n A_o \in \pgens$ with  $A_o
\in \{X, Y, Z, I\}$
(see Section~\ref{sec:pauli}).
\end{defn}

The matrix-algebra closure can be determined by including $\iso{v+w}$
for each pair of $\iso{v}$ and $\iso{w}$ independently of the symplectic product
$\symp{v}{w}$
(see Proposition~\ref{prop:Pauli:sign}).
Hence
the Pauli matrix algebra might contain Pauli strings not available in the
corresponding Lie algebra. In the symplectic space $\Fn$,
the matrix closure for a given
set $\pgens$ of Pauli strings corresponds to linear combinations of vectors $u,v$ for $\iso{v},\iso{u}\in\pgens$.
An exponential number of steps might be required starting from a small set
of generators as the
matrix algebra could have an exponential dimension $\dim\matalg\leq 4^n$.
We collect some basic properties:
\begin{lem}[Basic properties of Pauli matrix algebras]\label{lem:pauli:matrix}
Consider a Pauli matrix algebra $\matalg:=\algclosure{\pgens}$
generated by a set $\pgens \subseteq \PP_n$ of Pauli strings.
We obtain:
\begin{enumerate}
\item $\matalg \cap \PP_n$ is a Pauli-string basis of the matrix algebra $\matalg = \Span[\C]{\matalg \cap \PP_n}$.
\item $\ZZ(\matalg) \cap \PP_n$ is a Pauli-string basis of the center $\ZZ(\matalg)$.
\end{enumerate}
\end{lem}
\begin{proof}
Prop.~\ref{prop:Pauli:sign} implies that products of Pauli strings yield other Pauli string, multiplied by some complex phase factor.
Also, Pauli strings are linearly independent, which proves (a).

Now denote with $\calW$ the Pauli-string basis of $\matalg$ and let $\vgens=\inviso(\pgens)$. Consider a matrix in the center:
\begin{equation*}
    M = \sum_{w\in\calW} c_w\iso{w}\in\ZZ(\matalg).
\end{equation*}
It follows that $\comm{M}{\iso{v}}=0$ for all $v \in \vgens$, or equivalently,
\begin{align*}
M = \iso{v} M \iso{v} = \sum_{w  \in \calW, \symp{v}{w}=0}
\hspace{-5mm} c_w \iso{w}
- \sum_{w  \in \calW, \symp{v}{w}=1} \hspace{-5mm} c_w \iso{w}
\end{align*}
holds for all $v \in \vgens$.
This implies that $c_w=0$ for all $v \in \vgens$ with $\symp{v}{w}=1$
and $M$ is spanned by some $\iso{w}$ that commute with all $\iso{v}$ for $v \in \vgens$.
Thus $\ZZ(\matalg)$ has a Pauli-string basis consisting of $\iso{w}$ commuting with all
$\iso{v}$ for $v \in \vgens$, given by $\ZZ(\matalg) \cap \PP_n$. This proves (b).
\end{proof}

Based on Lemma~\ref{lem:pauli:matrix}, we define the Pauli-string bases
for $\matalg$ and its center $\ZZ(\matalg)$ as
\begin{subequations}
\label{eq:pauli:bases:matrix}
\begin{alignat}{3}
\bas{\matalg} &= \baslong{\matalg} &&:= \matalg \cap \PP_n,\\
\bas{\ZZ(\matalg)} &= \baslong{\ZZ(\matalg)} &&:= \ZZ(\matalg) \cap \PP_n.
\end{alignat}
\end{subequations}
We shortly recall from Eq.~\eqref{eq:pauli:binary} the mapping
\begin{align*}
\iso{v} \mapsto \inviso(\iso{v}) = v
\end{align*}
which maps Pauli strings $\iso{v}$ to binary vectors $v\in\Fn$.
This mapping naturally provides a bijection from a basis of a Pauli matrix algebra
to elements of a subspace of $\Fn$, as well as between generating sets and spanning sets:
\begin{prop}[Linking Pauli matrix algebras to subspaces]\label{prop:pauli:matrix_algebra_subspaces}
Consider a Pauli matrix algebra $\matalg\subseteq\mat{2^n}{\C}$
generated by a set $\pgens$ of Pauli strings,
the
Pauli-string bases
$\bas{\matalg}$ and $\bas{\ZZ(\matalg)}$
from Eqs.~\eqref{eq:pauli:bases:matrix}, and the set $V = \inviso(\bas{\matalg})$.
We observe:
\begin{enumerate}
    \item $\calB_{\matalg} = \groupclosure{\calB_{\matalg}}\cap\PP_n$ and $V$ is a subspace of $\Fn$.
    \item
    $\inviso$ induces a bijective map from $\pgens$ to a spanning set $\vgens= \{v_j \text{ for } j \in\{1,\ldots s\}\}$ of $V$.
    \item $\inviso(\ZZ(\matalg)) = \{ u\in V \text{ with } \symp{u}{v} = 0 \text{ for all } v\in V\}$.
    \item $\ZZ(\matalg) = \ZZ(\groupclosure{\calB_{\matalg}})\cap\PP_n$.
\end{enumerate}
\end{prop}
\begin{proof}
(a) follows from the fact that $\matalg$ is closed under products, and that the product of two Pauli strings is proportional to another Pauli string, as well as from the isomorphism $\inviso$, which maps products to linear combinations.
Regarding (b), for any $P\in\matalg\cap\PP_n$, we have
\begin{equation}\label{eq:pauli:matrix_algebra_subspaces}
P\simeq \iso{v_1}^{x_1} \iso{v_2}^{x_2}\cdots \iso{v_s}^{x_s} \text{ with } \iso{v_j} \in\pgens \text{ and } x_j\in\F_2,
\end{equation}
hence any vector in $V$ can be written as $\sum_{j=1}^s x_j v_j$ with $v_j=\inviso(\isolong{v_j})$.
Moreover, (c) follows as in the proof of Lemma~\ref{lem:pauli:matrix}(b).
Clearly,  $\ZZ(\matalg) \subset \ZZ(\groupclosure{\calB_{\matalg}})$
and we obtain (d) using (c) and by comparing with Eq.~\eqref{eq:pauli:matrix_algebra_subspaces}.
\end{proof}

\begin{defn}[Nullity and rank of binary symplectic subspaces]\label{def:rank_nullity_binary_symplectic_subspace}
Let $V\subseteq\Fn$ be a subspace of the binary symplectic space, and let $\vgens\subseteq\Fn$ be a spanning set with $V=\SpanS[\F_2]{\vgens}$.
We recall from Eq.~\eqref{eq:radical} that
\begin{equation*}
\rad(\vgens)=\rad(V)=
\{ v\in V \mid
\symp{v}{w} = 0 \text{ for all } w \in V \}.
\end{equation*}
We define the \emph{nullity} of $V$ by
\begin{equation}\label{eq:nullity}
    \nullity(V):=\dim\rad(V).
\end{equation}
Let $W$ be any supplementary subspace of $\rad(V)$ in $V$.
Then
\begin{equation}\label{eq:supplementary}
V = W \oplus \rad(V) \text{ with } \rad(W)=\{0\},
\end{equation}
since $W \cap \rad(V) = \{0\}$ implies $\rad(W) = W\cap \rad(V) = \{0\}$.
Following \cite[Ch.~IX, \S 1, no.~7, Def.~7, p.~21]{Bourbaki1959},
we define the \emph{rank} of $V$ by 
\begin{equation}\label{eq:rank}
\rank(V) := \dim V - \nullity(V) = \dim W.
\end{equation}
Whenever $V=\SpanS[\F_2]{\vgens}$, we also write $\rank(\vgens)=\rank(V)$ and $\nullity(\vgens)=\nullity(V)$.
\end{defn}
Thus $\rank(V)$ is not the dimension of the subspace $V$, but the rank of the symplectic form restricted to $V$.
Equivalently, $\rank(V)=\dim V$ holds exactly when the restricted form is non-degenerate, i.e., when $\rad(V)=\{0\}$.
By Lemma \ref{eq:pauli:matrix_algebra_subspaces}, for a Pauli generating set $\pgens=\isolong{\vgens}$, we also have $\ZZ(\algclosure{\pgens}) = \Span[\C]{\isolong{\rad(\vgens)}}$.

\subsection{Classification of Pauli Matrix Algebras\label{sec:classification:pauli:matrix}}

Given the correspondence between subspaces of $\Fn$ and Pauli matrix algebras with Proposition~\ref{prop:pauli:matrix_algebra_subspaces}, we are now able to completely characterize, up to isomorphisms, the matrix algebras generated by sets of Pauli strings, purely using the symplectic properties of the Pauli strings.
Namely, we can provide canonical generating sets based on the Witt decomposition:
\begin{lem}[\cite{Bourbaki1959}]\label{lem:symplectic_basis}
Consider a subspace $V$ of $\Fn$, with $2m=\rank(V)$ and $r=\dim\rad(V)$.
One can always choose a basis  of $V= W\oplus\rad(V)$ of the form
\begin{align*}
&\vecbas{V} = \vecbas{W} \cup \vecbas{\rad(V)} \text{ with }\\
&\vecbas{W} =  \{e_j,f_j \}_{j=1}^m  \text{ and } \vecbas{\rad(V)} =  \{h_j\}_{j=1}^r
\end{align*}
such that $\symp{e_j}{f_j} = 1$ and all other symplectic products are zero.
In particular, $\vecbas{\rad(V)}$ is a basis for the radical $\rad(V)$ and $\vecbas{W}$ is a basis for a supplementary subspace $W$ of $\rad(V)$ in $V$
with $\rad(W)=\{0\}$.
\end{lem}
Given Proposition~\ref{prop:pauli:matrix_algebra_subspaces} and the existence of a symplectic basis for subspaces, we obtain
for a Pauli matrix algebra $\matalg$ and a suitable vector space $V \subseteq \Fn$
that
\begin{align*}
\dim(\matalg) &= \abs{V} = 2^{2m+r}\\
\dim(\ZZ(\matalg)) &= \abs{\rad(V)} = 2^r
\end{align*}
with
$\vecbas{V} =\inviso(\bas{\matalg})$, $2m=\rank(V)$, and $r=\nullity(V)$.
For ease of notation in the Pauli setting, we now define the rank of a Pauli matrix algebra $\matalg=\Span[\C]{\isolong{V}}=\algclosure{\pgens}$ as the rank of its underlying binary vector space, and the nullity as the dimension of the radical:
\begin{subequations}\label{eq:def:rank_and_nullity_Pauli_matrix_algebra}
    \begin{align}
        \rank(\pgens) = \rank(\vgens) &:= \rank(\matalg) =\rank(V)\\
        \nullity(\pgens) = \nullity(\vgens)  &:= \nullity(\ZZ(\matalg)) = \nullity(V).
    \end{align}
\end{subequations}

Furthermore, we have a generating set with canonical (anti)commutation relations:
\begin{cor}\label{cor:pauli_matrix_algebra_classification}
For any Pauli matrix algebra $\matalg$ with $2m=\rank(\matalg)$ and $r=\nullity(\matalg)$, there exists a generating set of the form
\begin{align*}
\pgens = \{A_j,B_j\}_{j=1}^m \cup \{C_j\}_{j=1}^r
\end{align*}
such that $\acomm{A_j}{B_j}=0$ and all other pairs mutually commute.
Also, $\pgens_2$ is a generating set for $\ZZ(\matalg)$
and the matrix algebra generated by $\pgens_1$ is simple.
\end{cor}
Notice that this set is also \emph{minimal}, in the sense that removing any element from it will not generate the same matrix algebra. In Section~\ref{sec:graphs_to_lie_algebras}, we shall study in more detail non-minimal generating sets and how to deal with redundancies.
Then, we can also write explicitly a basis of (not necessarily hermitian) Pauli strings for $\matalg$:
\begin{alignat}{3}\label{eq:symplectic_basis_pauli_matrix_algebra}
        \matalg & = &&\;
        \SpanL[\C]{ \iso{v} \text{ for }  \\
        &&&\quad v = \sum_{i=1}^m \component{v}{i} e_i + \component{v}{m+i}f_i + \sum_{j=1}^r \component{v}{2m+j}h_j \in\F_2^{2m+r} } \nonumber \\
        &=&&\; \SpanL[\C]{ \prod_{i=1}^m A_i^{\component{v}{i}}B_i^{\component{v}{m+i}} \prod_{j=1}^r C_j^{\component{v}{2m+j}} \text{ for } v\in\F_2^{2m+r}  }
        \nonumber
\end{alignat}
where the phases in the products are irrelevant, since we are considering a basis over a complex vector space.
In order to prepare for the classification of Pauli matrix algebras, we identify the invariants of subspaces of $\Fn$:
\begin{lem}\label{lem:subspaces:iso}
Consider two subspaces $V,V'\subseteq\Fn$. Then $V$ and $V'$ are isomorphic if and only if $\dim(V)=\dim(V')$ and $\rank(V)=\rank(V')$. If $V,V'$ are isomorphic, there is a $g\in\Sp(2n,\F_2)$ such that $V'=g\cdot V$.
\end{lem}
\begin{proof}
Assume first they are isomorphic via the map $\phi(V)=V'$. Clearly, $\dim(V) = \dim(V')$. Then, choose a basis $\vecbas{\rad(V)}$ of $\rad(V)$ and a basis $\vecbas{\rad(V')}$ $\rad(V')$. Since $\phi$ is an isomorphism, we must have $\symp{\phi(u)}{\phi(v)} = 0$ for each $u\in\rad(V)$ and $v\in V$, hence for each $v'\in V$ (simply choose $v' = \phi(v)$). Thus $\phi(\rad(V))\subseteq \rad(V')$. Repeat this for $\phi^{-1}$ to find that $\phi(\rad(V)) = \rad(V')$, which implies $\rank(V)=\rank(V')$.
Viceversa, assume $\dim(V)=\dim(V')=2m+r$ and $\rank(V)=\rank(V')=2m$. 
Consider symplectic bases $\vecbasempty, \vecbasempty'$ for $V,V'$. 
Then we can define a linear map over $\Fn$ such that $g\cdot e_i = e_i'$, $g\cdot f_i = f_i'$ and $g\cdot h_j = h_j'$ for all $i\in[m]$ and $j\in[r]$. 
This is also symplectic, since it preserves the symplectic products, hence an isomorphism.
Then, we can uniquely specify $g$ as an element of $\Sp(2n,\F_2)$ by choosing any basis for $\Fn$ which includes $\vecbasempty$.
\end{proof}
This also determines the isomorphism classes of Pauli matrix algebras up to Cliffords via their rank and dimension, or equivalently their rank and nullity.
In particular, for any subspace $V$ of rank $2m$ and dimension $2m+r$, we can always choose as a representative in $\Fn$, up to isomorphism, the subspace spanned by the symplectic basis $\vecbas{V}$ with $\isolong{e_i}=X_i$, $\isolong{f_i} = Z_i$ and $\isolong{h_j} = Z_{m+j}$.
Then, any Pauli matrix algebra, up to isomorphism, is spanned by Paulis of the form:
\begin{align}\label{eq:Canonical_Basis_Pauli_Matrix_Algebra}
        \matalg &\conjugated \SpanL[\C]{\prod_{i=1}^m X_i^{\component{v}{i}}Z_i^{\component{v}{m+i}} \prod_{j=1}^r Z_{m+j}^{\component{v}{2m+j}}
        \text{ for } v\in\F_2^{2m+r} }
        \nonumber \\
        &= \SpanL[\C]{\PP_m\otimes\{I,Z\}^{\otimes r}\otimes I^{\otimes (n-m-r)}}.
\end{align}
Notice that the center $\ZZ(\matalg)$ is spanned purely by the diagonal $Z$-strings, $\prod_{j=1}^r Z_{m+j}^{\component{w}{j}}$ for $w\in\F_2^r$.
Given that we can always define a $g\in\Sp(2n,\F_2)$ which realizes this basis change, then we can also find a Clifford $U\in\cl_n$ which sends the given matrix algebra $\matalg$ into the canonical one from Eq.~\eqref{eq:Canonical_Basis_Pauli_Matrix_Algebra}.

This characterization gives us already a lot of information about the possible dynamics generated by Paulis, up to isomorphism. Namely, with respect to a Pauli matrix algebra, we can partition the qubits into three sets:
\begin{enumerate}
    \item Over $n-m-r$ qubits the Paulis act trivially, hence these are \emph{uncontrollable}
    \item Over $m$ qubits they act arbitrarily as complete Pauli strings, hence these are the \emph{logical} qubits
    \item Over the remaining $r$ qubits they act only by changing phases in the computational basis, hence these are named \emph{phase} qubits
\end{enumerate}
Then, following \cite{Aguilar_Cichy_Eisert_Bittel_2024}, we refer to the first $m$ as the \emph{logical} qubits, whereas the next $r$ serve as \emph{phase} qubits.
Since any Lie algebra generated by Pauli strings (see Section~\ref{sec:pauli:Lie}) sits inside some Pauli matrix algebra, it must also respects the partition with respect to the qubits.
However, the action on the logical qubits will differ depending on the specific Lie algebra, e.g.,
$\su(d)$, $\usp(d)$, and $\so(d)$ (see Section~\ref{sec:classification:groups_lie_algebras}) or tensor products of these.

Finally, given Eq.~\eqref{eq:Canonical_Basis_Pauli_Matrix_Algebra}, it is immediate state a block-diagonal decomposition as in Eq.~\eqref{eq:block_diagonal_decomposition_matrix_algebras}, with respect to the computational basis, which completes the classification of Pauli matrix algebras:
\begin{lem}\label{lem:Classification_Pauli_Matrix_Algebras}
Consider a Pauli matrix algebra $\matalg\subseteq \C^{2^n\times 2^n}$
with $\rank(\matalg)=2m$ and $\nullity(\matalg)=r$. 
Up to isomorphism, the isotypical projectors are
of the form
\begin{align}
        \prod_{j=1}^r \frac{1+s_jZ_{m+j}}{2}
        = \frac{1}{2^r} \sum_{\{n_j\}\in\{0,1\}^{r}} \prod_{j=1}^r (s_jZ_{m+j})^{n_j}
\end{align}
and $\matalg$ has the form of $2^r$ copies of the $2^m\times 2^m$ simple matrix algebra, each with degeneracy $2^{n-m-r}$ as in
\begin{equation}
    \matalg \conjugated \bigoplus_{\lambda=1}^{2^r} \id_{2^{n-m-r}}\otimes \C^{2^m \times 2^m}.
\end{equation}
\end{lem}
Notice that for each Pauli matrix algebra $\matalg$ with $\rank(\matalg)=2m$ and $\nullity(\matalg)=r$, up to isomorphism, there is a minimal
embedding into a smaller matrix algebra of size $2^{m+r}\times 2^{m+r}$.

Then, under this decomposition, we have the corresponding subspace decomposition of $\C^{2^n}$:
\begin{equation}
    \C^{2^n} = \bigoplus_{\lambda=1}^{2^r} \C^{2^{n-m-r}}\otimes \C^{2^m}
\end{equation}
where each irreducible subspace has dimension $2^{n-m-r}$ and occurs with degeneracy $2^m$, inside $2^r$ isotopycal subspaces of size $2^{n-r}$.

\ManuscriptPart{Graph-Theoretic Framework}{part:graph_theoretic_framework}

In this section we introduce several graph-theoretic methods for the analysis of Pauli Lie algebras and transvection groups. 
We will mainly be concerned with the frustration graph which characterizes the Lie algebra, transvection group, and (Lie) algebraic dependencies, and the commutator graph which helps characterize the orbits. 
The graph-theoretic approach is powerful because there are certain graph transformations which preserve Lie-algebraic properties, allowing for classifications built around canonical graphs. 
However, we will also highlight certain cases where graphs do not give the complete picture.

\section{Graphs and Contractions}\label{sec:graph_theoretic_formalism}

\subsection{Graphs and Sequences\label{sec:graph:def}}

A graph
$\graphG$  is in this work assumed to be undirected, unweighted,
and without loops \cite{diestel2017}, and it
is specified by its vertex set $\vertices=\vertices(\graphG)=\{1,\ldots,\abs{\vertices}\}$ and its edge set $\edges=\edges(\graphG)$
where $\edges$ contains
unordered pairs $\{\vva,\vvb\}$ of vertices $\vva,\vvb \in \vertices$ with $\vva\neq\vvb$.
Two graphs are considered as equal if there is a permutation of the vertex set of the first graph mapping it
to the vertex set of the second graph such that the edges agree.
For clarity, $\vertices \cap \edges = \emptyset$.
Let $\neighbors(\vva) \subset \vertices$ denote the set of neighbors of the vertex
$\vva\in \vertices$,
i.e., $\neighbors(\vva)$ contains all vertices $\vvb\in \vertices$ adjacent to $\vva$ or, equivalently, $\{\vva, \vvb\} \in \edges$.
The set $\incidents(\vva)\subseteq \edges$ of incident edges of $\vva \in \vertices$ is given by all edges
$\{\vva,\vvb\} \in \edges$ with $\vvb \in \vertices$.

If $U\subseteq\vertices(\graphG)$ is a subset of the vertices of $\graphG$, then $\graphG[U]$ denotes the induced subgraph
of $\graphG$ on $U$, i.e.,
\begin{subequations}
\label{eq:induced_subgraph}
\begin{align}
\vertices(\graphG[U])&=U
\text{ and}\\
\edges(\graphG[U])
&=\bigl\{\{u,v\}\in\edges(\graphG) \text{ for } u,v\in U\bigr\}.
\end{align}
\end{subequations}
We write $\deg_{\graphG}(\vva)=\abs{\neighbors(\vva)}$ for the degree of a vertex $\vva$ of $\graphG$ and define the \emph{core}
\begin{equation}\label{eq:core}
\graphcore{\graphG}:=\graphG\bigl[\{\vva\in\vertices(\graphG) \text{ for } \deg_{\graphG}(\vva)\geq 2\}\bigr]
\end{equation}
as the induced subgraph on the vertices of degree at least two.

Until now, we have considered sets $\vgens = \{v_1,\ldots,v_s\}$ of binary vectors with $s= \abs{\vgens}$  and
the corresponding generating sets $\pgens = \{ \iso{v_1}, \ldots, \iso{v_s}\}$ of Pauli operators.
In the following, we will need to also consider ordered sequences $\vgens =\seq{v_1,\ldots,v_s}$ where
elements can repeat (and similarly for Pauli operators). As for sets, $\abs{\vgens}$ denotes the length of the sequence $\vgens$.
Occasionally, we describe the $a$-th element in a sequence $\vgens$ by $\vgens[\vva]$
and we use
\begin{equation*}
\set{\vgens} = \setlong{\vgens} := \{ \vgens[\vva] \text{ for } \vva \in \{1,\ldots,\abs{\vgens}\} \}
\end{equation*}
to denote the set of all its elements. For usual sets $\vgens$, we set $\set{\vgens}= \setlong{\vgens} :=\vgens$ for convenience.
Every set can be considered as a sequence with respect to some order.
Maps applied to a sequence will act on each component, i.e., $\pgens = \isolong{\vgens}$ means
that $\pgens=\seq{ \isolong{\vgens[1]}, \ldots, \isolong{\vgens[\abs{\vgens}]}  }$.

This enables us to define a labeled graph $\graphG$ as a graph complemented with a sequence
$\labels$ of vertex labels where $\labels[\vva]$ denotes
the $\vva$-th element in $\labels$.

\subsection{Frustration and Commutation Graphs}

From the transvection group literature we find a graph-theoretic approach which is equivalent to that used for Lie-algebras and other properties of certains sets of Pauli strings, which starts by defining the frustration or anti-commutation graph (see also Fig.~\ref{fig:frustration-graph}):
\begin{defn}[\cite{Gintz,Wajnryb_1980,Janssen_1983,humphries_1985,Chapman_Flammia_2020,Planat_Saniga_2007}]
\label{defn:frustration}
Consider a sequence $\pgens$ of Pauli strings where $\pgens[\vva]$ is its $\vva$-th element.
The (labeled) frustration graph $\frustration{\pgens}$ of $\pgens$
has the vertex set $\vertices = \{1,\ldots,\abs{\pgens}\}$
and $\pgens$ as its sequence of labels. Its edge set
\begin{equation*}
\edges = \{ \{\vva,\vvb\} \text{ for } \vva,\vvb \in \vertices \text{ and } \acomm{ \pgens[\vva]}{ \pgens[\vvb]} = 0 \},
\end{equation*}
contains all edges for which the Pauli strings $\pgens[\vva]$ and $\pgens[\vvb]$
anticommute (or, equivalently, $ \comm{ \pgens[\vva]}{ \pgens[\vvb]} \neq 0$).
The unlabeled frustration graph is obtained by removing the vertex labels.
Similarly, the frustration graph $\frustration{\vgens}$ for a sequence $\vgens$ of binary vectors is defined by
the respective condition $\symp{\vgens[\vva]}{\vgens[\vvb]}=1$
for binary vectors $\vgens[\vva]$ and $\vgens[\vvb]$.
\end{defn}

\begin{figure}
    \centering
    \includegraphics[width=\linewidth]{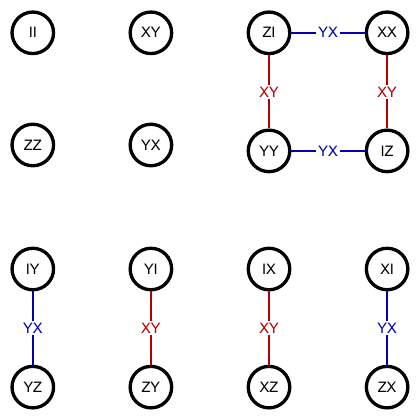}
    \caption{Example of commutator graph over $\PP_2$ for the Pauli generating set $\{\text{XY, YX}\}$.}
    \label{fig:commutator-graph}
\end{figure}

Moreover, the \emph{adjacency matrix} $A(\frustration{\pgens})$ of $\frustration{\pgens}$ has the entries
$A(\frustration{\pgens})_{ij} = \symp{v_i}{v_j}$ with $\vgens = \seq{ v_1,\ldots, v_{\abs{\vgens}}}$.
We have stated the more general version for sequences $\pgens$ in Definition~\ref{defn:frustration} so that we can later address some complications
with graph transformations.
Another graph-theoretic object which has emerged in study of dynamical systems evolving under Pauli dynamics is that of the commutator graph, which has been used to study symmetries and invariant subspaces in the Heisenberg picture (see also Fig.~\ref{fig:commutator-graph} and \ref{fig:commutator-and-frustration-graphs}(b)):

\begin{figure}
    \centering
    \includegraphics[width=0.5\linewidth]{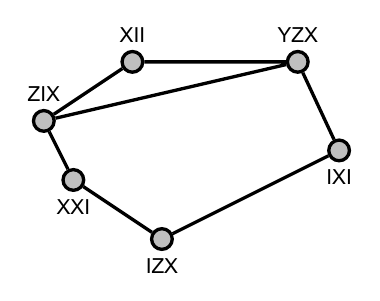}
    \caption{Example of the frustration graph for a Pauli generating set.}
    \label{fig:frustration-graph}
\end{figure}

\begin{defn}[\cite{Diaz_GarciaMartin_Kazi_Larocca_Cerezo_2023,West_Dowling_Southwell_Sevior_Usman_Modi_Quella_2025}]
\label{defn:comm:graph}
Given a set $\pgens\subseteq\PP_n$ of Pauli strings with $\PP_n$ considered as an ordered sequence.
The labeled commutator graph $\commutatorgraph{\pgens}$ has the vertex set $\vertices = \{1,\ldots,\abs{\PP_n}\}$,
$\PP_n$ as its label sequence, and the
edge set
\begin{equation*}
\edges = \{ \{\vva,\vvb\} \text{ for } \vva,\vvb \in \vertices, \comm{P}{\PP_n[\vva]} \simeq \PP_n[\vvb] \text{ for } P \in \pgens \}.
\end{equation*}
The commutator graph $\commutatorgraph{\vgens}$ for binary vectors $\vgens$ is defined on
all binary vectors $\Fn$ and uses
the condition $\tau_v\, \Fn[\vva] = \Fn[\vvb]$ for $v \in \vgens$; $\Fn[\vva]$ is $\vva$-th binary vector.
\end{defn}

Let us shortly highlight differences between the Definitions~\ref{defn:frustration} and \ref{defn:comm:graph} assuming that $\pgens$ is a set.
It is usually much more practical to work with the frustration graph $\frustration{\pgens}$ from Def.~\ref{defn:frustration}
as it is much smaller than the commutator graph $\commutatorgraph{\pgens}$ from  Def.~\ref{defn:comm:graph}
with a vertex set corresponding to all Paulis strings. Clearly, the frustration and commutator graph agree if the generating set contains
all Pauli strings, i.e., $\pgens = \PP_n$.
In general, the commutator graph reveals more relevant information as it is larger and
the condition in Def.~\ref{defn:comm:graph} is stronger
than the one in Def.~\ref{defn:frustration}. Indeed, it is surprising
how far one can get using only the frustration graph.

It is easy to determine if two Paulis are adjacent in a commutator graph, as one simply needs to check if they anti-commute and their product is a Pauli in the generating set.
However, understanding their connectivity, i.e., which Paulis are in the same connected component in the commutator graph, has not been extensively studied beyond a few examples \cite{Diaz_GarciaMartin_Kazi_Larocca_Cerezo_2023,West_Dowling_Southwell_Sevior_Usman_Modi_Quella_2025};
but see \cite{Fischer1971,Aschbacher1997}.
Therefore, in order to understand the full power of these tools, we shall also develop a more general theory, which allows either a computational
or a conceptual
approach (which, in certain cases, is exact)
to study the properties of the defined graphs for different sets of Pauli strings.

We can now connect properties of frustration and commutator graphs with the orbits of the transvection group, which clearly motivates the study of these objects. We relate the connectedness of frustration graphs to the action of the generators under the transvection group:

\begin{lem}\label{lem:pauli:single_orbit_equivalent_to_connectedness}
Consider a set $\pgens$ of Pauli strings with
$\pgens = \isolong{\vgens}$ for a set $\vgens$ of binary vectors.
The frustration graph $\frustration{\vgens}$ is connected iff
$\vgens$ is contained in a single orbit of the transvection group
$\tvgroup{\vgens}$ generated by $\vgens$.
If $\frustration{\pgens}$ is connected, then the Pauli basis $\bas{\lieg}$ of the Lie algebra $\lieg$
generated by $\pgens$ is contained in a single orbit of $\tvgroup{\vgens}$.
\end{lem}

Note that $\tvgroup{\vgens} \cdot \vgens$ is in general a union of orbits.
Having a \emph{single} orbit is equivalent to saying that
$\tvgroup{\vgens}$ acts transitively on this set,
i.e., for all $u,v\in \tvgroup{\vgens} \cdot \vgens$ there exists
$g\in\tvgroup{\vgens}$ with $gu=v$.

\begin{proof}
Any pair $(\tilde{v},\tilde{w})$ of distinct vectors in $\vgens$ corresponds to an edge
of the frustration graph $\frustration{\pgens}$.
If $\frustration{\pgens}$ is connected, then for any $v_0, v_h \in\vgens$ with $v_0 \neq v_h$, there exists a path
$(v_0,v_1)(v_1,v_2)\cdots(v_{h-1},v_h)$
in $\frustration{\pgens}$ such that $v_{j-1} \neq  v_j$ and $\symp{v_{j-1}}{v_j}=1$ for all $j$.
It suffices to show that adjacent vertices $u=v_{j-1}$ and $v=v_j$ are in the same orbit.
Indeed, let $w = u + v$, so that $\symp{w}{u}=\symp{w}{v}=1$.
Then
$v = \tau_w u = \tau_{\tau_{u} v} u = \tau_{u} \tau_{v} \tau_{u} u = \tau_{u} \tau_{v} u$ and hence $u$ and $v$
lie in the same orbit. Thus all connected pairs do as well.

Assume now that $u,v\in\vgens$ with $u\neq v$ are in the same orbit, i.e., there exists $g\in\tvgroup{\set{\vgens}}$ such that $v=g u$.
Since the group is generated by transvections $\tau_{\tilde{v}}$ with  $\tilde{v} \in \set{\vgens}$,
we may also write this as  $v = \tau_{w_h} \cdots \tau_{w_1}(u)$ for vectors
$w_j \in \set{\vgens}$.
Without loss of generality, we remove all elements $\tau_{w_j}$ that act trivially on their argument, i.e.,
$\tau_{w_j} \tau_{w_{j-1}} \cdots \tau_{w_1}(u) = \tau_{w_{j-1}} \cdots \tau_{w_1}(u)$.
Equivalently, if
$\symp{u+ \sum_{\nu=1}^{j-1}w_\nu}{w_j}=0$, then $\tau_{w_j}$ can be omitted.
Hence
$v = \tau_{w_h} \cdots \tau_{w_1}(u) = u + w_1 + \cdots + w_h$.
Clearly, $\symp{u}{w_1}=1$, yielding the edge $(u,w_1)$.
More generally,
$\symp{u+ \sum_{\nu=1}^{j-1}w_\nu}{w_j}=1$ implies that
$w_j$ is adjacent to one element of $\{u, w_1, \ldots, w_{j-1} \}$.
If $v$ is not adjacent to $u$, let $j_0$ be the smallest index such that $v$ is adjacent to $w_{j_0}$.
Then let $j_1$ be the smallest index such that $w_{j_0}$ is adjacent to $w_{j_1}$ (so $j_1<j_0$).
Iterating this construction, we eventually reach $w_{j_{s+1}}=w_1$, which is adjacent to $u$.
This yields a path
$(u,w_{j_{s+1}})(w_{j_{s+1}},w_{j_s})\cdots(w_{j_0},v)$ and proves the first claim.
Together with Corollary~\ref{cor:pauli:transvection} this then implies the second statement.
\end{proof}

It is then trivial to show a version for multiple connected components (including isolated vertices):
\begin{cor}
Consider a set $\vgens$ of binary vectors.
Each connected component $\frustration{\vgens_i}$
of the frustration graph $\frustration{\vgens}$
for sets $\vgens_i$ with $\vgens = \cup_i \vgens_i$
corresponds to a distinct orbit $\tvgroup{\vgens}\cdot \vgens_i$
under the transvection group with $(\tvgroup{\vgens}\cdot \vgens_i)
\cap (\tvgroup{\vgens}\cdot \vgens_j) = \emptyset$ for $i\neq j$.
\end{cor}

And a similar proof describes the relation between the commutator graph and the orbits:
\begin{lem}\label{lem:orbits_and_commutato_graph}
Consider a set $\pgens$ of Pauli strings with $\pgens = \isolong{\vgens}$ for a set $\vgens$ of binary vectors.
The vertices of the connected components of the commutator graph $\commutatorgraph{\pgens}$ coincide with the elements of the orbits
of the transvection group $\tvgroup{\vgens}$ generated by $\vgens$.
\end{lem}
\begin{proof}
Assuming that two elements $u=v_0$ and $v=v_h$ are in the same connected component of $\commutatorgraph{\vgens}$, hence there is a connected path $(v_0,v_1)(v_1,v_2)\cdots (v_{h-1},v_h)$ such that $v_j = \tau_{w_j}v_{j-1}$ with $w_j\in\vgens$.
Clearly, this means that $v,u$ are in the same orbit, with $v = \tau_{w_h}\cdots\tau_{w_2}\tau_{w_1}u = gv$.

If $u=v_0$ and $v=v_h$ are in the same orbit, then $v=gu$.
As before, we can expand $g=\tau_{w_h}\cdots\tau_{w_2}\tau_{w_1}$ into a product of transvections and choose a subsequence such that each transvection does not act trivially on its argument, hence $\symp{w_j}{v_{j-1}}=1$ for $v_j = \tau_{w_j}v_{j-1}$.
Then, we get a connected path $(v_0,v_1)(v_1,v_2)\cdots (v_{h-1},v_h)$.
\end{proof}

We can use the above result to clarify the significance of a restricted transvection group $\tvgroup{\vgens}|_V$ to the subspace spanned by $\vgens$.
Recall the correspondence between the subspace $V$ spanned by $\vgens$ and the matrix algebra $\matalg$ generated by the Pauli strings $\pgens$.
Then, this result combined with Lemma~\ref{lem:restriction_transvection_group_subspace} implies: (1) the intersection of the Pauli strings with $\bas{\matalg}$ with the commutator graph produces connected components; (2) the action of $\tvgroup{\vgens}|_V$ on $V$ describes these components/orbits.
As we shall see, knowledge of these orbits will be useful also for describing \emph{all} of the orbits/connected components of the Pauli Lie algebra.

\subsection{Contractions and \texorpdfstring{$t$}{t}-equivalence}

In both \cite{Aguilar_Cichy_Eisert_Bittel_2024} and the transvection group literature, a fundamental concept in the graph-theoretic approach is how to transform generating sets, while preserving all relevant properties. We have the following operation on sequences of generators:
\begin{defn}[Contractions on Generators \cite{Gintz,Wajnryb_1980,Janssen_1983,humphries_1985,Aguilar_Cichy_Eisert_Bittel_2024}]
Given a sequence $\pgens$ of Pauli strings, the \emph{contraction} of the $\vva$-th Pauli string $\pgens[\vva]$ onto
the $b$-th Pauli string $\pgens[\vvb]$ for $\vva\neq\vvb$ maps $\pgens[\vvb]$ to $\herm(\pgens[\vva]\pgens[\vvb])$ and keeps all other ones unchanged.
We also say that a contraction is \emph{valid} if the Paulis $\pgens[\vva]$ and $\pgens[\vvb]$ anticommute.
For a sequence $\vgens$ of binary vectors, $\vgens[\vvb]$ is similarly mapped to $\vgens[\vvb] + \vgens[\vva]$.
\end{defn}
We also refer to $\pgens[\vva]$ as the \emph{source} of the contraction and $\pgens[\vvb]$ as the \emph{target}.

In practice, we often avoid the general language of sequences of generators.
Instead, for $P$ and $Q$ from a generating set $\pgens$, the contraction of $P$ onto $Q$ replaces $Q$ by $\herm(PQ)$ and yields the new generating set
\begin{equation}\label{eq:set:contraction}
\pgens' = ( \pgens \setminus \{ Q \} ) \cup \{ \herm(PQ) \}.
\end{equation}
This operation is clearly reversible as contracting $P$ onto the new generator $\herm(PQ)$ recovers
$Q \simeq P\herm(PQ)$.
In the binary formalism, this corresponds to $u+(u+v)=v$.
Since $\herm(PQ)$ already lies in the matrix algebra generated by $\pgens$, we have
$\pgens' \subseteq \algclosure{\pgens}$.
By reversibility, the original generator $Q$ lies in the matrix algebra generated by $\pgens'$, so
$\pgens \subseteq \algclosure{\pgens'}$.\
Hence the generated algebras coincide:
\begin{lem}\label{lem:contractions_conserve_matrix algebras}
For \emph{any} contraction mapping a generating set $\pgens$
to generating set $\pgens'$, the generated matrix algebra is preserved, i.e.,
$\algclosure{\pgens} = \algclosure{\pgens'}$ (in the binary formalism, $\SpanS[\F_2]{\vgens} = \SpanS[\F_2]{\vgens'}$).
\end{lem}

As we will be studying invariant properties under \emph{valid} contractions
throughout the manuscript, we formalize this notion:
\begin{defn}[$t$-Equivalence \cite{humphries_1985}]
\label{defn:t:equivalence}
Two sequences of Pauli strings (or binary vectors) are $t$-equivalent
if a sequence of \emph{valid} contractions sends one into another.
\end{defn}

In the binary formalism, valid contractions of $u$ onto $v$ are
equivalent to applying the transvection $\tau_u$ to $v$.
Hence we obtain transformation of sequences which map $\seq{u,v,\ldots}$ to $\seq{u,\tau_u v,\ldots}$.
Notice that by using transvections of a generators onto another as the \emph{definition} of valid contractions, then arbitrary pairs are allowed (even a vertex onto itself), since anyway only connected vertices result in a non-trivial transformation.
Valid contractions correspond to commutators $\comm{P}{Q} \simeq PQ$
as the product of two anti-commuting Paulis is their commutator; we can write $\tau_{\tau_uv} = \tau_u\tau_v\tau_u$).
We can collect invariant properties on $t$-equivalence:
\begin{lem}[\cite{Aguilar_Cichy_Eisert_Bittel_2024,humphries_1985,Janssen_1983}]\label{lem:properties:preserved:t:equivalence}
Consider a set $\pgens$ of Pauli-strings with $\pgens=\isolong{\vgens}$ for a set $\vgens$ of binary vectors
that is $t$-equivalent to $\pgens'=\isolong{\vgens'}$. Then we have:
\begin{enumerate}
\item $\algclosure{\pgens} = \algclosure{\pgens'}$.
\item $\rad(\vgens) = \rad(\vgens')$.
\item $\tvgroup{\vgens} = \tvgroup{\vgens'}$.
\item $\tvgroup{\vgens}\cdot\vgens = \tvgroup{\vgens'}\cdot\vgens'$.
\item $\lie{\pgens} = \lie{\pgens'}$.
\end{enumerate}
\end{lem}

\begin{proof}
(a) follows from Lemma~\ref{lem:contractions_conserve_matrix algebras} and (b)
is a direct consequence of $\rad(\vgens) = \rad( \Span[\F_2]{\vgens})$.

We consider the valid contraction of $u$ onto $v$ for elements
from $\vgens$ with $\symp{u}{v}=1$.
For (c), $\tau_{v+u}=\tau_u\tau_v\tau_u \in \tvgroup{\vgens}$
and $\tvgroup{\vgens'}\subseteq \tvgroup{\vgens}$. By symmetry,
$v=(v+u)+u$ and $\symp{u}{v+u}=1$, we obtain
$\tau_v=\tau_u\tau_{v+u}\tau_v\in \tvgroup{\vgens'}$ and (c).

For (d), $v+u=\tau_u(v)\in \tvgroup{\vgens}\cdot\vgens$, and similarly
$v=\tau_u(v+u)\in \tvgroup{\vgens'}\cdot\vgens'$.
Since all other vectors are unchanged, the orbit sets coincide and (d) holds.
Statement (e) follows by applying
Corollary~\ref{cor:pauli:transvection} to (d).
\end{proof}

\begin{figure}
    \centering
    \includegraphics{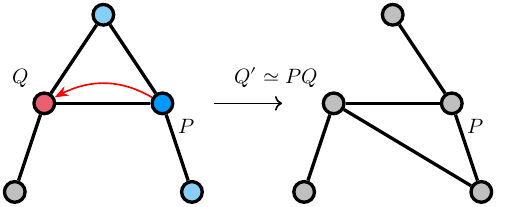}
    \caption{Effect of a valid contraction of $P$ onto $Q$ on the frustration graph (Lemma~\ref{lem:valid:contraction:graph}):
    the target vertex flips its adjacency to each neighbor of the source vertex
    other than the target vertex itself (shown in light blue), while all other adjacencies are unchanged.}
    \label{fig:contraction}
\end{figure}

If valid contractions suitably conserve the algebraic properties of the generating set, a generating
set can be classified by transforming it into a canonical representative by a series of valid contractions.
The canonical representative is usually chosen such that proofs are especially easy.
This approach acquires particular meaning by realizing the impact of these transformation on frustration graphs
$\frustration{\pgens}$
for generators $\pgens$.
Clearly, a contraction of one vertex onto another one is valid whenever the vertices
are adjacent.
We show its effects in the following lemma (see Fig.~\ref{fig:contraction}):

\begin{lem}[\cite{humphries_1985,Aguilar_Cichy_Eisert_Bittel_2024}]\label{lem:valid:contraction:graph}
Consider a sequence $\pgens'$ of generators obtained by
a valid contraction of $\pgens[\vva]$ onto $\pgens[\vvb]$ for the sequence $\pgens$ of generators and $\vva\neq\vvb$.
Then the frustration graph $\frustration{\pgens}$
transforms into $\frustration{\pgens'}$
where $\pgens'[\vvb]=\herm(\pgens[\vva]\pgens[\vvb])$ while all other generators are unchanged.
For each vertex $\vvc \neq \vvb$ adjacent to $\vva$, the edge
$\{\vvb,\vvc\}$ is removed if it exists and added otherwise.
\end{lem}

We can restate the conclusion of Lemma~\ref{lem:valid:contraction:graph}
more formally. The edges $\edges$ of $\frustration{\pgens}$ are mapped to the edges
\begin{equation*}
    \edges' = (\edges \setminus \incidents(\vvb) )\cup \{ \{\vvb,\vvc\} \text{ for } \vvc \in (\neighbors(\vva) \triangle \neighbors(\vvb))\setminus\{\vvb\}\}.
\end{equation*}
of $\frustration{\pgens'}$.
Here, $\incidents(\vvb) = \{\{\vvb,\vvd\} \in \edges
\text{ for } \vvd \in \vertices\}$ are the incident edges of $\vvb$,
$\neighbors(\vva)$ are the neighbors of $\vva$,
and
\begin{align*}
\neighbors(\vva) \triangle \neighbors(\vvb)
&= (\neighbors(\vva) \setminus \neighbors(\vvb)) \cup (\neighbors(\vvb) \setminus \neighbors(\vva)\\
&= (\neighbors(\vva) \cup \neighbors(\vvb)) \setminus (\neighbors(\vva) \cap \neighbors(\vvb)
\end{align*}
is the symmetric difference of the neighbors $\vva$ and $\vvb$.

Also, with respect to the the adjacency matrix $A(\graphG)$, a contraction is equivalent to performing row addition \cite{Aguilar_Cichy_Eisert_Bittel_2024,Viswanathan_Adjoua_Feniou_Badreddine_Piquemal_2025}, possibly conditioned on the adjacency if it needs to be valid.
Thus we can write a contraction in matrix terms.
Let $\graphG'$ the graph obtained by contracting a vertex $\vva$ onto $\vvb\neq \vva$ in $\graphG$, then (over $\F_2$) \cite{Viswanathan_Adjoua_Feniou_Badreddine_Piquemal_2025}
\begin{gather}
    A(\graphG') = M_{\vva\vvb}^T A(\graphG) M_{\vva\vvb}
    \label{eq:adjacency_matrix_contraction}
\intertext{where $M_{\vva\vvb}$ performs the row addition and is defined as}
    (M_{\vva\vvb})_{ij} = \delta_{ij} + \delta_{i\vvb}\delta_{j\vva}
    \text{ with } M_{\vva\vvb}^T e_i = e_i + \delta_{i\vvb} e_{\vva}.
\end{gather}

Given that contractions can be stated in a purely graph-theoretic language, the search for canonical generating sets
can now be greatly aided by separating the purely graph-theoretic properties from the algebraic ones.

\begin{figure}
    \centering
    \includegraphics[width=0.5\linewidth]{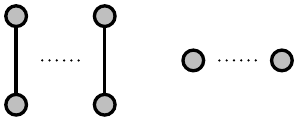}
    \caption{Any graph is equivalent through a series of arbitrary contractions to $m$ pairs of connected vertices plus $r$ isolated vertices.}
    \label{fig:canonical-arbitrary}
\end{figure}

Indeed, the typical approach in the transvection group literature \cite{Janssen_1983,humphries_1985,Seven_2005} makes use of the graph-theoretic properties by finding special representatives in the $t$-equivalence classes.
This approach was also used for Pauli Lie algebras by \cite{Aguilar_Cichy_Eisert_Bittel_2024}, as well as in the literature of the lit-only sigma game \cite{Vorstermans}.
Before considering such representatives, we first state equivalence when we consider arbitrary contractions, which is particularly simple:
\begin{lem}[\cite{Khovanova_2008}]\label{lem:Canonical_Form_Arbitrary_Adjacent_Contractions_Frustration_Graph}
A graph $\graphG$ with $n$ vertices is equivalent via contractions to a graph with $m$ isolated edges $\graphK_2$ and $r$ isolated vertices $\graphK_1$
with $2m+r=n$ (see Fig.~\ref{fig:canonical-arbitrary}). Furthermore, $2m=\rank(A(\graphG))$ and $r=\nullity(A(\graphG))= n-2m$
where $A(\graphG)$ is the adjacency matrix of $\graphG$.
\end{lem}
Accordingly, we also refer to graphs as \emph{non-degenerate} if $\nullity(A(\graphG))=0$ and degenerate otherwise.
In the binary picture, we can also see this as a natural consequence of being able to construct a symplectic basis from any given basis, since taking linear combinations is the same as taking arbitrary contractions.
Specifically, this construction may be seen as a form of Gaussian elimination, where we iteratively remove orthogonal components of previous isolated edges.
We will see precisely how the graph-theoretic characterization matches the binary one in Section~\ref{sec:graphs_to_lie_algebras}, see also Lemma~\ref{lem:subspaces:iso}.

Under valid contractions reachability is greatly restricted. 
Indeed one finds that connectivity is preserved:
\begin{lem}
Valid contractions preserve connectivity, i.e., they map connected graphs
to connected graphs.
\end{lem}

\begin{proof}
On the Pauli side this may be seen as the fact the transvection group and orbit containing the generators is unchanged, hence the generating set must still have connected frustration graph by Lemma~\ref{lem:pauli:single_orbit_equivalent_to_connectedness}.
On the graph-theoretic side, we first notice that valid contractions of $\vva$ onto $\vvb$ only affect the neighborhood of the single vertex $\vvb$.
Given a neighbor $\vvc$ of the vertex $\vva$, there are two possibilities: (i) If $\vvb$ and $\vvc$ are also neighbors, then $\vvb$ disconnects from $\vvc$, while still being a neighbor of $\vva$ and hence conserving connectivity of $\vvb$ to $\vvc$ through $\vva$. (ii) If $\vvb$ and $\vvc$ are not neighbors, then $\vvb$ connects to $\vvc$.
In either case, connectivity is preserved.
\end{proof}

\subsection{Trees and the  \texorpdfstring{$\calE_6$}{E6} Condition\label{sec:trees}}

As valid contractions preserve connectivity, how far can connected graphs be simplified using only valid contractions?
The following result shows that every connected graph admits a tree in its $t$-equivalence class.

\begin{prop}[$t$-equivalent trees {\cite[Theorem~3.3]{Brown_Humphries_1986a}}]\label{prop:t_equivalent_to_tree_algorithm}
A connected graph $\graphG$ is $t$-equivalent to a tree, which
can be obtained by a finite sequence of valid contractions.
\end{prop}

\begin{figure*}[t]
    \centering
    \includegraphics{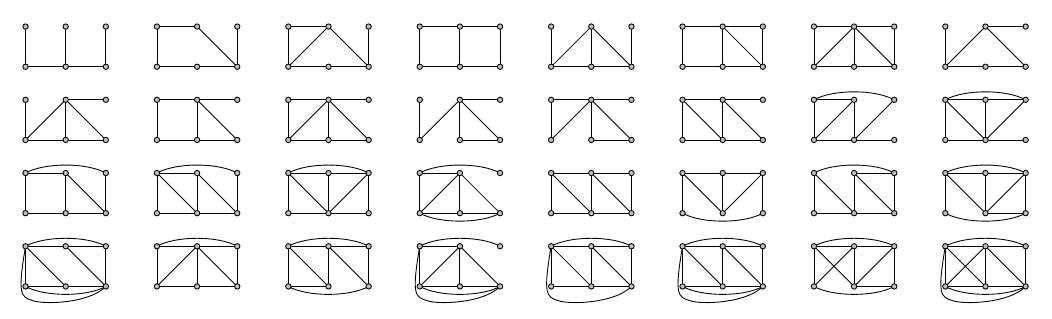}
    \caption{The $32$ unlabeled six-vertex graphs in the set $\calE_6$. These are the isomorphism classes obtained from the $t$-equivalence orbit of the labeled $\graphE_6$.
    The orbit contains $11097$ labeled graphs.
    \label{fig:e6-class-atlas}}
\end{figure*}

\begin{proof}
Our constructive proof follows the algorithm from \cite[Theorem~3.3]{Brown_Humphries_1986a}.
We choose a root vertex $\vva$ and a connected component $C$ of $\graphG\setminus\{\vva\}$.
A vertex of $C$ is root-adjacent if it is adjacent to $\vva$.
Since $\graphG$ is connected, $C$ has at least one root-adjacent vertex.

We first prove the component-pruning step: we reduce to the case where $\vva$ has exactly one
neighbor in $C$ by a sequence of
valid contractions of a vertex $\vvb$ onto a vertex $\vvc$ for $\vvb, \vvc \in C$.
If the root-adjacent vertices $\vvb$ and $\vvc$ are adjacent in $C$, then $\vvb+\vvc$ replaces $\vvc$
and $\vvb+\vvc$ is not adjacent to $\vva$. Thus the number of root-adjacent vertices in $C$ decreases by one.

If no two root-adjacent vertices are adjacent, choose root-adjacent vertices $\vvb$ and $\vvc$
with a minimal distance in $C$. Let $\vvd_1$ be the next vertex after $\vvb$ on the shortest path
$(\vvb,\vvd_1,\vvd_2,\ldots,\vvc)$
from $\vvb$ to $\vvc$.
Then $\vvd_1$ is not root-adjacent by the minimality of the chosen pair.
Contracting $\vvd_1$ onto $\vvb$ is a valid contraction which
keeps $\vvb$ adjacent to $\vva$ as $\vvd_1$ is not root-adjacent. Thus the set of root-adjacent vertices is unchanged.
This also shortens the distance between $\vvb$ and $\vvc$ in $C$, i.e., $\vvb$ is adjacent to $\vvd_2$
after the contraction as
\begin{equation*}
\symp{\vvb+\vvd_1}{\vvd_2}=\symp{\vvb}{\vvd_2}+\symp{\vvd_1}{\vvd_2}=0+1=1.
\end{equation*}
Repeating this shortening step eventually produces an adjacent pair of root-adjacent vertices,
and the preceding paragraph then reduces the number of root-adjacent vertices.
Hence, after finitely many steps, $\vva$ has a unique neighbor in $C$.

Apply this component-pruning step to every component of $\graphG\setminus\{\vva\}$.
The components remain connected and no edges between distinct components are created, so each component is now attached to $\vva$ by exactly one edge.
Declare that unique neighbor to be the child of $\vva$ in the corresponding component and repeat the same construction recursively inside that component, rooted at this child.
The recursion stops because the vertex sets strictly decrease.
At the end, every non-root vertex has exactly one parent edge, and the graph is
connected; hence the output is a tree $t$-equivalent to $\graphG$.
\end{proof}

The proof gives an algorithm which performs
a polynomial number of valid contractions
in terms of the number of vertices of the graph.
We emphasize that this is a reduction to a tree, not a reduction inside the class of trees.
Indeed, a valid contraction on a tree preserves the tree property only in the trivial case where
the source vertex of the contraction is a leaf, and then the graph is unchanged.
We recall a six-vertex graph which will play a prominent role
in this work:

\begin{figure}[t]
    \centering
    \includegraphics{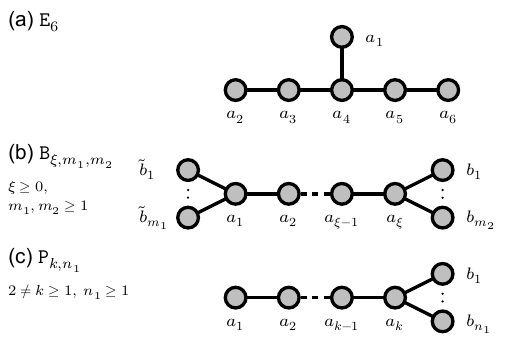}
    \caption{(a) The unlabeled six-vertex graph $\graphE_6$, (b) the double broom $\graphB_{\xi,m_1,m_2}$, and (c) the blown-up path graph $\graphP_{k,n_1}$.
    \label{fig:e6-double-broom-blown-up-path}}
\end{figure}

\begin{defn}[$\graphE_6$  and $\calE_6$]\label{def:e6}
We denote the
unlabeled six-vertex graph in Fig.~\ref{fig:e6-double-broom-blown-up-path}(a)
by $\graphE_6$. We say a graph is $\graphE_6$-free if none of its induced subgraphs is
isomorphic to $\graphE_6$.
The set $\calE_6$ consists of the
unlabeled six-vertex graphs in Fig.~\ref{fig:e6-class-atlas} (including $\graphE_6$).
We say a graph is $\calE_6$-free if none of its induced subgraphs is isomorphic to a graph in $\calE_6$.
\end{defn}

Let $\graphB_{\xi,m_1,m_2}$ with $\xi\geq 0$ and $m_1,m_2\geq 1$ denote a double-broom
graph whose core $\graphcore{\graphB_{\xi,m_1,m_2}}$ is a path graph on $\xi$ vertices
and whose two core ends carry $m_1$ and $m_2$ leaves
(see Fig.~\ref{fig:e6-double-broom-blown-up-path}(b)).
For $\xi=0$ we only allow $m_1=m_2=1$, giving a single edge; for $\xi=1$ the two
core ends coincide.
Similarly, a blown-up path graph (or broom)
$\graphP_{k,n_1}$ with $k\geq 1$ and $n_1 \geq 1$ is constructed from a path graph
by adding at least one leaf to one of its ends. It is shown in
Fig.~\ref{fig:e6-double-broom-blown-up-path}(c) and we assume that it
has at least two vertices.
We collect some elementary properties:

\begin{prop}[cf.~\cite{Brown_Humphries_1986a,Brown_Humphries_1986b,Seven_2005}]\label{prop:trees:e6}\leavevmode
\begin{enumerate}
\item A six-vertex graph is $t$-equivalent to $\graphE_6$ if and only if it is isomorphic to a graph in $\calE_6$.
Moreover, $\graphE_6$ is the only tree in $\calE_6$ and
$\calE_6$ is preserved under valid contractions.
\item A tree is $\calE_6$-free if and only if it is $\graphE_6$-free.
\item Let $\graphG$ be a graph. The property that $\graphG$ contains an
induced subgraph isomorphic to one of the graphs in $\calE_6$ is preserved
under valid contractions.
\item A tree is $\graphE_6$-free if and only if every vertex of degree at least three has at most one non-leaf neighbor.
\item Every $\graphE_6$-free tree with at least two vertices is a double broom $\graphB_{\xi,m_1,m_2}$.
\item A connected graph with at least two vertices is $\calE_6$-free iff each of its $t$-equivalent trees is a double broom $\graphB_{\xi,m_1,m_2}$.
\item Let $\graphG$ be a connected graph and let
$x$, $y$, and $z$ be vectors labeling three distinct vertices. Assume that the vertices
labeled by $x$ and $y$ are not adjacent and have the same neighbor set.
Then there is a sequence of valid contractions, applied only to the vertex initially
labeled by $x$, after which this vertex has the same neighbors
as the vertex labeled by $z$.
\item Every double broom $\graphB_{\xi,m_1,m_2}$ is
$t$-equivalent to a blown-up path graph $\graphP_{k,n_1}$.
\item Every tree which is $t$-equivalent to a blown-up path graph is $\graphE_6$-free.
\item A connected graph with at least two vertices
 is $\calE_6$-free iff it is $t$-equivalent to a blown-up path graph $\graphP_{k,n_1}$.
\end{enumerate}
\end{prop}

\begin{proof}
A finite computation shows (a): Starting from one labeled copy of
$\graphE_6$, one performs a breadth-first search in the space of labeled
six-vertex graphs by applying all admissible valid contractions. Canonicalizing
the resulting labeled graphs up to isomorphism gives exactly the $32$
unlabeled graphs shown in Fig.~\ref{fig:e6-class-atlas}. The same computation
checks that exactly one of these $32$ graphs is a tree, namely $\graphE_6$
itself. Since the labeled graphs are constructed by closure under valid
contractions, $\calE_6$ is preserved under valid contractions.
Statement (b) is a consequence of (a), as $\graphE_6$ is the only tree in
$\calE_6$.

We prove (c). This is \cite[Lemma~3.7]{Seven_2005};
we describe a brute-force proof by enumerating all possibilities.
Let $\graphG$ contain an induced six-vertex subgraph
isomorphic to one of the graphs in $\calE_6$ with vertex set $U$ and consider
one valid contraction of $\graphG$. If the valid contraction does
not change a vertex in $U$, then the induced subgraph on $U$ is unchanged. If
both vertices involved in the valid contraction lie in $U$, then the induced subgraph
on $U$ is changed by a valid contraction and remains in $\calE_6$ by (a).
It remains to consider the boundary case where the changed vertex lies in $U$
and the other vertex does not. Then the induced subgraph on these seven
vertices is obtained from a graph in $\calE_6$ by adding one extra vertex
adjacent to at least one vertex of $U$.
A finite seven-vertex check verifies that,
after any valid boundary contraction, one of the induced six-vertex
subgraphs is again in $\calE_6$. This proves (c).

We now prove (d):
Let $\graphT$ be a tree. If a vertex $v$ of $\graphT$ has degree at
least three and has two distinct neighbors $u,w$ which are not leaves, then
choosing a third neighbor $z$ of $v$, a neighbor $u'\neq v$ of $u$, and a
neighbor $w'\neq v$ of $w$ gives the induced subgraph on the vertices
$\{u',u,v,w,w',z\}$.
For a tree $\graphT$, the only edges among these six vertices are
$$
\{ \{u',u\}, \{u,v\}, \{v,w\}, \{w,w'\}, \{v,z\}\},
$$
so this induced subgraph is $\graphE_6$. Conversely, an induced subgraph isomorphic to
$\graphE_6$ contains a vertex of degree three with two non-leaf neighbors in
$\graphT$. This proves (d).

We prove (e). This uses elementary graph properties which were then used in
\cite[Proposition~4.1]{Brown_Humphries_1986b} and
\cite[Proposition~2.4]{Seven_2005}.
Let $\graphT$ be a tree with at least two vertices such that no induced subgraph is isomorphic to $\graphE_6$.
By Eq.~\eqref{eq:core}, $\graphcore{\graphT}$ is the induced subgraph on the non-leaf vertices. If $\vertices(\graphcore{\graphT})$ is empty, then $\graphT$ is a single edge
and hence equal to $\graphB_{0,1,1}$. Otherwise, the graph $\graphcore{\graphT}$
is connected, since
removing leaves from a tree does not disconnect the remaining non-leaf
vertices. By (d), $\graphcore{\graphT}$
has maximum degree at
most $2$: a vertex with three neighbors in $\vertices(\graphcore{\graphT})$, or with two neighbors in
$\vertices(\graphcore{\graphT})$ and an additional leaf neighbor, would have degree at least $3$ in $\graphT$
and two non-leaf neighbors. Thus $\graphcore{\graphT}$ is a connected induced subgraph of a
tree with maximum degree at most $2$, hence a path.

Finally, no leaf can be attached to an interior vertex of this path. Indeed,
such an interior vertex has two non-leaf neighbors in $\graphcore{\graphT}$, and the extra
leaf is excluded by (d). Therefore all leaves of
$\graphT$ are attached to the two ends of the path $\graphcore{\graphT}$. This is exactly a
double broom $\graphB_{\xi,m_1,m_2}$ with $\xi=\abs{\vertices(\graphcore{\graphT})}$.
This completes the proof of (e).

By (c), a connected graph is $\calE_6$-free if and only if all of its
$t$-equivalent trees are $\calE_6$-free. By (b) and (e), this is equivalent to
all of its $t$-equivalent trees being double brooms. This proves (f).

We prove (g), following the leaf transfer from
\cite[Proposition~3.4]{Brown_Humphries_1986a}.
Choose a path $(z_0,z_1,\ldots,z_s)$ of vectors labeling vertices
with $z_0=y$ and $z_s=z$ which does not contain the vertex initially labeled by $x$.
Such a path exists because the vertices labeled by $x$ and $y$ have the same
neighbor set, so any path segment passing through the former can be rerouted
through the latter.
Assume inductively that the vertex carrying the current label $x_j$ has the
same neighbors as the vertex labeled by $z_j$. Since the vertex labeled by
$z_j$ is adjacent to the vertex labeled by $z_{j+1}$, the vertex carrying
$x_j$ is also adjacent to the vertex labeled by $z_{j+1}$, so the first contraction
\begin{equation*}
x_j\mapsto x_j+z_{j+1}
\end{equation*}
is valid. After this first contraction, the new label $x_j+z_{j+1}$ is adjacent
to the vertex labeled by $z_j$, because $x_j$ is not adjacent to $z_j$ by the
induction hypothesis while $z_{j+1}$ is adjacent to $z_j$. Hence the second
contraction
\begin{equation*}
x_j+z_{j+1}\mapsto x_j+z_j+z_{j+1}
\end{equation*}
is valid and replaces the label $x_j$ by $x_{j+1}=x_j+z_j+z_{j+1}$.
Only the moving label changes. Thus, to compare the final neighbor set with
that of the vertex labeled by $z_{j+1}$, it suffices to test adjacency to each
unchanged vertex label $u$. For such $u$,
\begin{align*}
\symp{x_{j+1}}{u}
&=\symp{x_j+z_j+z_{j+1}}{u}\\
&=\symp{x_j+z_j}{u}+\symp{z_{j+1}}{u}
=\symp{z_{j+1}}{u},
\end{align*}
since the induction hypothesis gives $\symp{x_j}{u}=\symp{z_j}{u}$.
Therefore the vertex carrying $x_{j+1}$ has the same neighbors as the vertex
labeled by $z_{j+1}$. Iterating along the path proves (g).

We prove (h) by applying the leaf transfer from (g) as in
\cite[Proposition~3.4]{Brown_Humphries_1986a}.
If $\xi=0$, then $\graphB_{\xi,m_1,m_2}=\graphP_{1,1}$. If $\xi>0$ and at least one of $m_1$ and $m_2$ is equal to $1$,
then $\graphB_{\xi,m_1,m_2}$ is already a
blown-up path graph: the unique leaf on that side is included into the path
spine, while the leaves on the other side form the blown-up endpoint. If
$m_1,m_2\geq 2$, choose two leaves $x,y$ at one end and a leaf $z$ at the
other end. The leaves $x$ and $y$ have the same neighbor set and are not
adjacent, so (g) transfers $x$ to the other end. Hence
$
\graphB_{\xi,m_1,m_2}
$
is $t$-equivalent to
$\graphB_{\xi,m_1-1,m_2+1}$.
Iterating this step gives a double broom with one leaf on one
side, hence a blown-up path graph which proves (h).

We prove (i). A blown-up path graph is a double broom, so every vertex of degree
at least three has at most one non-leaf neighbor; hence it is $\graphE_6$-free by (d).
By (c), every tree which is
$t$-equivalent to a blown-up path graph is $\calE_6$-free, and hence
$\graphE_6$-free by (b). This proves (i).

We prove (j). First assume that $\graphG$ is connected, has at least two
vertices, and is $\calE_6$-free. By
Proposition~\ref{prop:t_equivalent_to_tree_algorithm}, $\graphG$ is
$t$-equivalent to a tree $\graphT$. By (f), this tree is a double broom
$\graphB_{\xi,m_1,m_2}$, and by (h) it is $t$-equivalent to a blown-up path graph.
Hence $\graphG$ is $t$-equivalent to a blown-up path graph. Conversely, assume
that $\graphG$ is $t$-equivalent to a blown-up path graph. Then every
$t$-equivalent tree is $\graphE_6$-free by (i), hence a double broom by (e). By
(f), $\graphG$ is $\calE_6$-free. This proves (j).
\end{proof}

Proposition~\ref{prop:trees:e6} gives a graph-theoretic criterion for when a
connected graph is $t$-equivalent to a blown-up path graph. After reduction to
a tree, this is exactly the $\graphE_6$-free case. Thus the $\calE_6$ condition
separates the blown-up path family from the other canonical graph families that
appear in the classification below.

\subsection{Canonical Graphs}

The discussion in the previous section gives a partial characterization of canonical graphs via the blown-up path graph and the $\calE_6$-condition.
It remains to understand what are the canonical representatives which are \emph{not} $\calE_6$-free, as well as how to determine for each given connected graph its $t$-equivalence graph.

\begin{figure}
    \centering
    \includegraphics{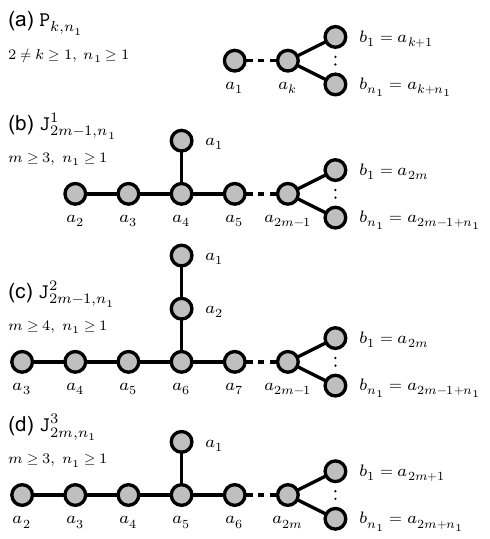}
    \caption{The four classes in Theorem~\ref{thm:classes:humphries} for $t$-equivalent representatives
    of any given connected graph \cite{Brown_Humphries_1986a,Brown_Humphries_1986b,humphries_1985,Janssen_1983,Seven_2005}.}
    \label{fig:humphries_classes}
\end{figure}

We now state two such choices of canonical representatives in $t$-equivalence classes for connected graphs:
\begin{thm}[\cite{Brown_Humphries_1986a,Brown_Humphries_1986b,humphries_1985,Janssen_1983,Seven_2005}]\label{thm:classes:humphries}
A connected graph on $n\geq 3$ vertices is $t$-equivalent to one and only one graph in the following four canonical graphs classes (see Fig.~\ref{fig:humphries_classes}):
\begin{enumerate}
    \item the blown-up path graph $\graphP_{k,n_1}$\\
    with a path of length $k$ attached to $n_1$ length-one leaves where $k+n_1=n$, $2 \neq k\geq 1$, and $n_1\geq 1$;
    \item one of three possible extensions of blown-up path graphs, which also contain $n_1\geq 1$ length-one leaves attached to the path, as well as:
    \begin{nestedcaseenum}
        \item $\graphX_{2m-1,n_1}^1$: one length-one leaf attached at the third vertex along the path of length $2m-1$\\ with $n_1+2m-1=n$, $m\geq 3$,
        \item $\graphX_{2m-1,n_1}^2$: a leg of length $2$ at the fourth vertex along the path of length $2m-2$\\ with $n_1+2m-1=n$, $m\geq 4$,
        \item $\graphX_{2m,n_1}^3$: one length-one leaf attached at the fourth vertex along the path of length $2m$\\ with $n_1+2m=n$,
        $m\geq 3$.
    \end{nestedcaseenum}
\end{enumerate}
\end{thm}

We have that $\graphP_{n-1,1} = \graphP_n$ is the usual path graph on $n$ vertices.
The conditions on the values $m$, $n_1$ serve to make the representatives well-defined and unique (up to isomorphism and $t$-equivalence), especially at small sizes.
Clearly, $\graphX_{2m-1,n_1}^1$ is only well defined for $m\geq 3$. Similarly $m\geq 4$ is required for $\graphX_{2m-1,n_1}^2$ and $m\geq 3$ for
$\graphX_{2m,n_1}^3$.

The existence and uniqueness of canonical graphs under $t$-equivalence, as established in Theorem~\ref{thm:classes:humphries} and illustrated in Figure~\ref{fig:humphries_classes}, are conceptually convenient, as they simplify many arguments. However, establishing this theorem is somewhat delicate without relying on results that appear later in this work. We therefore do not provide an independent proof, but use it freely in what follows.
The use of $t$-equivalent graphs was pioneered in \cite{Brown_Humphries_1986a,Brown_Humphries_1986b}, and the corresponding result for the special case of non-degenerate spaces was treated in \cite{humphries_1985}. Within a different framework, and explicitly identifying the graphs shown in Figure~\ref{fig:humphries_classes}, the general result was established in \cite{Janssen_1983}. We refer to \cite{Seven_2005} for a useful summary.

\begin{figure}
    \centering
    \includegraphics{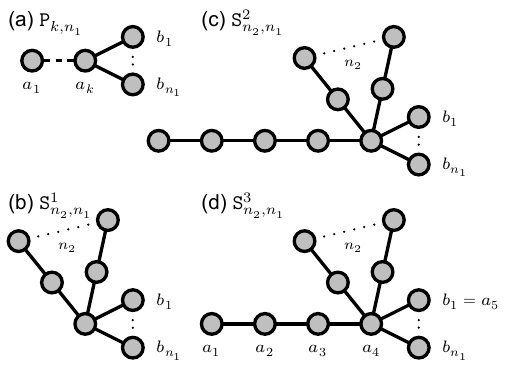}
    \caption{The four classes in \cite{Aguilar_Cichy_Eisert_Bittel_2024} for $t$-equivalent representatives of any given connected graph.}
    \label{fig:eisert_classes}
\end{figure}

A complementary approach to canonical graphs under t-equivalence was developed in \cite{Gintz}. Based on purely combinatorial arguments, this approach seeks to identify a different class of canonical graphs, namely star (and path) graphs, and to provide an algorithm that determines, for each connected graph, a sequence of valid contractions transforming it into a star. This program was completed in \cite{Aguilar_Cichy_Eisert_Bittel_2024}, which provides an algorithm that transforms any connected graph into one of the canonical graphs shown in Figure~\ref{fig:eisert_classes}. This algorithm can also be applied to the canonical graphs in Figure~\ref{fig:humphries_classes},
thereby providing a bridge between these two classes of canonical graphs.

\begin{thm}[{\cite[Theorem~1]{Aguilar_Cichy_Eisert_Bittel_2024}}]\label{thm:classes:eisert}
A connected graph on $n\geq 3$ vertices is $t$-equivalent to one and only one graph in the following four canonical graphs classes (see Fig.~\ref{fig:eisert_classes}):
\begin{enumerate}
    \item the blown-up path graph $\graphP_{k,n_1}$\\
    with $k+n_1=n$, $2 \neq k \geq 1$, and $ n_1\geq 1$;
    \item one of three possible star graphs, containing $n_1\geq 1$ length-one leaves, $n_2$ legs of length two, as well as:
    \begin{nestedcaseenum}
        \item $\graphS_{n_2,n_1}^1$: nothing else \\ with $n_1+2n_2+1=n$ and $n_2\geq 2$,
        \item $\graphS_{n_2,n_1}^2$: one leg of length four \\ with $n_1+2n_2+5=n$ and $n_2\geq 1$,
        \item $\graphS_{n_2,n_1}^3$: one leg of length three \\ with $n_1+2n_2+4=n$ and $n_2\geq 1$.
    \end{nestedcaseenum}
\end{enumerate}
\end{thm}

Again, we impose the conditions on $m,n_1$ such that they are well-defined and there is no redundancy in the representatives.
Namely, we have $\graphS_{0,n_1}^1 = \graphP_{1,n_1}$, $\graphS_{0,n_1}^2 = \graphP_{5,n_1}$ and $\graphS_{0,n_1}^3 = \graphP_{4,n_1}$, which imposes $n_2\geq 1$ for all star graphs.
Also, $\graphP_{0,n_1}$ is not well-defined and $\graphP_{1,n_1} = \graphP_{2,n_1-1}$ (for $n_1\geq 1$), so that we can choose $m\geq 1$, $m\neq 2$, for uniqueness.
Also, $\graphS_{n_2,0}^i$ for $i\in\{1,2,3\}$ is $t$-equivalent to one of $\graphS_{n_2-1,2}^j$, which implies $n_1\geq 1$.
Finally, $\graphS_{1,n_1}^1 = \graphP_{3,n_1}$, which imposes $n_2\geq 2$ (for any $n_1\geq 1$) for star graphs of type 1.
Notice that there is redundancy in \cite[Theorem~1]{Aguilar_Cichy_Eisert_Bittel_2024}, since they give only $n_2\geq 1$ for each graph.

Also, later we shall connect the two classes $\graphX_{2m-1,n_1}^1$, $\graphX_{2m-1,n_1}^2$, $\graphX_{2m,n_1}^3$, and $\graphS_{n_2,n_1}^j$ via the use of certain invariants of graphs under $t$-equivalence (see Prop.~\ref{prop:isomorphism_classes_quadratic_forms_arf_invariant_canonical}).

In view of the classification of Lie algebras, transvection groups and orbits (see Section~\ref{sec:classification:groups_lie_algebras}) we highlight some differences between the two choices of representatives (excluding the common choice of blown-up path graph):
The classes in Thm.~\ref{thm:classes:humphries} are commonly used in the transvection group literature to provide graphs of \emph{alternating} isomorphism classes, e.g., $\graphX_{2m-1,n_1}^1$ alternates between the $\so(d)$- and $\usp(d)$-types depending on $m$. On the other hand, they include path graphs which commonly appear in generating sets. In particular, we shall see that they admit simple labels in the language of Majorana strings.
The classes from Thm.~\ref{thm:classes:eisert} have been used in \cite{Aguilar_Cichy_Eisert_Bittel_2024} to provide sequences (with increasing $n_1$ or $n_2$) which naturally respect the same isomorphism classes of \emph{quadratic forms}, e.g. in the Lie-algebra language, $\graphS_{n_2,n_1}^1$ is always of $\usp(d)$-type, $\graphS_{n_2,n_1}^2$ of $\so(d)$-type and $\graphS_{n_2,n_1}^3$ of $\su(d)$-type. However, they do not match to common generating sets in many-body or circuit models, which often involve regular lattices, hence include \emph{paths}.

\section{From Graphs to (Lie) Algebras}\label{sec:graphs_to_lie_algebras}

\subsection{Algebraic and Lie-Algebraic Dependencies}

The classification of the $t$-equivalence classes of graphs will be critical, including
how they are related to the generating set.
Then, before delving into the results of such a graph-theoretic classification, we first highlight some limitations of such an approach for a (Lie-)algebraic analysis.
These limitations are summarized in Fig.~\ref{fig:graph-only-limitations} and spelled out in the following examples:
\begin{figure}
    \centering
    \includegraphics{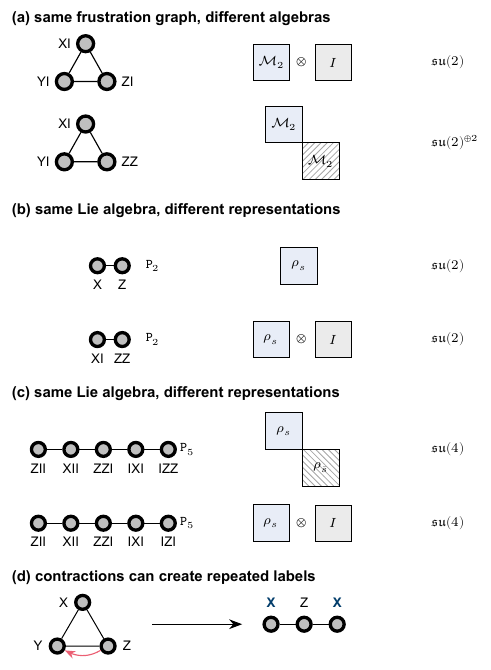}
    \caption{Three limitations of graph-only data.
    (a) The same frustration graph can arise from generating sets with different associative and Lie algebras, as in Example~\ref{ex:contraction:one}, motivating the later distinction between graph data and algebraic dependencies.
    (b) and (c) Even when the Lie algebra is fixed up to isomorphism, the representation can differ, as in Example~\ref{ex:contraction:two}.
    (d) Graph contractions on labelled generating sequences can create repeated Pauli labels, as in Example~\ref{ex:contraction:three}, motivating the use of sequences instead of sets.}
    \label{fig:graph-only-limitations}
\end{figure}

\begin{ex}\label{ex:contraction:one}
There exist generating sets with the same frustration graph but distinct Lie algebras or associative algebras,
i.e., $\frustration{\pgens_1} \cong \frustration{\pgens_2} \cong \graphC_3$ for the generating sets
$\pgens_1 = \{ \mathrm{XI}, \mathrm{YI}, \mathrm{ZI}\}$ and $\pgens_2 = \{ \mathrm{XI}, \mathrm{YI}, \mathrm{ZZ}\}$.
This is illustrated in Fig.~\ref{fig:graph-only-limitations}(a).
Moreover,
\begin{align*}
&\su(2) \cong \lie{\pgens_1}  \neq \lie{\pgens_2} \cong \su(2)^{\oplus 2} \text{ and}\\
& \matalg_2\otimes\id_2 = \algclosure{\pgens_1}  \neq \algclosure{\pgens_2} = \matalg_2^{\oplus 2}.
\end{align*}
\end{ex}

\begin{ex}\label{ex:contraction:two}
There exist minimal generating sets with the same frustration graph which generate the same
Lie algebra up to isomorphism, but with (unitarily) inequivalent representations.
This is illustrated in Fig.~\ref{fig:graph-only-limitations}(b) and (c).
(i)~For instance, they can live in spaces of different size, e.g.,
$ \frustration{\pgens_1} \cong \frustration{\pgens_2} \cong \graphP_2$
for $\pgens_1 = \{ \mathrm{X, Z}\}$ and $\pgens_2 = \{ \mathrm{XI, ZZ}\}$.
In particular,
\begin{gather*}
           \lie{\pgens_1}  \conjugated \su(2)     \neq  \su(2)\otimes I \conjugated \lie{\pgens_2}.
\end{gather*}
The representation is given by $\rho_s$ for $\lie{\pgens_1}$
and by $\rho_s\otimes I$ for $\lie{\pgens_2}$, where
$\rho_s$ is the standard representation of $\su(2)$.
(ii)~But also when they live in the same space, e.g.,
$\frustration{\pgens_1} \cong \frustration{\pgens_2} \cong \graphP_5$
for $\pgens_1 = \{\mathrm{ZII, XII, ZZI, IXI, IZZ}\} $ and
$\pgens_2 = \{\mathrm{ZII, XII, ZZI, IXI, IZI}\}$. Moreover,
\begin{gather*}
 \lie{\pgens_1}  \conjugated \su(4)   \neq \su(4)\otimes I \conjugated \lie{\pgens_2}.
\end{gather*}
The representation decomposes as $\rho_s\oplus\rho_{\bar{s}}$ for $\lie{\pgens_1}$
and as $\rho_s\otimes I$ for $\lie{\pgens_2}$, where
$\rho_s$ is the standard representation of $\su(4)$ and $\rho_{\bar{s}}$ is its dual
\cite{BrauerWeyl1935,Kaufman1949,Boerner1969,SW86,Kazi_Larocca_Farinati_Coles_Cerezo_Zeier_2025}.
\end{ex}

\begin{ex}\label{ex:contraction:three}
Contractions of sequences of generators
where no Pauli can appear twice can produce seqeunces
where Paulis appear multiple times.
For example, if Z is contracted onto Y in $\pgens_1 = \seq{\text{X, Y, Z}} $, then $\pgens_1$
is mapped to $\pgens_2 =\seq{\text{X, X, Z}}$.
This is illustrated in Fig.~\ref{fig:graph-only-limitations}(d).
\end{ex}
We note that in Examples~\ref{ex:contraction:one} and \ref{ex:contraction:three} the ambiguity of the graph-theoretic approach
is related to the presence of redundant generators (i.e. one of the generators from $\pgens_1$ can be removed while preserving Lie-algebraic properties).
In Example~\ref{ex:contraction:two} instead the ambiguity in the representation is resolved when specifying the number of qubits
(which is straightforward) and symmetry properties of the generating sets (i.e.\ $\pgens_1$ has an abelian commutant, whereas
$\pgens_2$ has a non-abelian commutant with trivial center). These symmetry properties are further discussed in Sec.~\ref{sec:commutant}.
In particular, from Example~\ref{ex:contraction:three}, to establish a close connection between graph-theoretic
and algebraic properties one needs to remove redundant generators.
Let us make this precise by giving the following definition:
\begin{defn}[Algebraic and Lie-algebraic Dependencies \cite{Aguilar_Cichy_Eisert_Bittel_2024}]
\label{def:dependencies}
A set of Pauli strings $\pgens$ is said to be \emph{algebraically dependent} if there exists an element $G\in\pgens$ such that $\algclosure{\pgens} = \algclosure{\pgens\setminus\{G\}}$ and it is algebraically independent if there is no such element.
A set of Pauli strings is said to be \emph{Lie-algebraically dependent} if there exists an element $G\in\pgens$ such that $\lie{\pgens} = \lie{(\pgens\setminus\{G\})}$ and it is Lie-algebraically independent or minimal if there is no such element.
\end{defn}

We collect some immediate consequences:
\begin{lem}\label{lem:dependency}
Consider a sequence $\pgens = \isolong{\vgens}$ of Pauli strings with $\vgens \subseteq \Fn$.
We obtain:
\begin{enumerate}
\item $\pgens$ is algebraically dependent iff there is a subsequence of Pauli strings $S\subseteq\pgens$ such that $\prod_{G\in S} G \simeq I$.
\item An algebraic dependency in the binary picture corresponds to a linear dependency such that $\sum_{j=1}^m v_{i_j} = 0$.
\item $\pgens$ is Lie-algebraically dependent iff there is a finite sequence of
Pauli strings $\pgens[i_j]\in\pgens$ with $j\in\{1,\ldots,m+1\}$ and
$i_{m+1}\not\in\{i_1,\ldots,i_m\}$ such that
\[
G_{i_{m{+}1}} \simeq
\comm{\pgens[i_m]}{\comm{\pgens[i_{m-1}]}{\ldots,
\comm{\pgens[i_{2}]}{\pgens[i_1]}}}.
\]
\item A Lie-algebraic dependency as in (c) is witnessed in the binary picture by
$$
v_{i_{m{+}1}}
=\tau_{v_{i_m}}\tau_{v_{i_{m-1}}}\cdots\tau_{v_{i_2}}v_{i_1}.
$$
after removing all transvection steps that fix their argument.
For a word obtained from a non-zero nested commutator, the remaining steps
are precisely those corresponding to non-zero commutators.
The associated algebraic dependency is obtained by keeping exactly those
vectors among $v_{i_1},\ldots,v_{i_m},v_{i_{m+1}}$ that occur an odd number of times:
the iterated commutator is represented by the binary vector $\sum_{j=1}^m v_{i_j}$.
It is proportional to $\pgens[i_{m+1}]$ precisely when
$v_{i_{m+1}}=\sum_{j=1}^m v_{i_j}$.
\item If $\pgens$ is Lie-algebraically dependent, then it is also algebraically dependent.
\end{enumerate}
\end{lem}

\begin{proof}
For (a), suppose first that $\pgens$ is algebraically dependent in the sense of Definition~\ref{def:dependencies}.
Then some $G\in\pgens$ lies in the matrix algebra generated by $\pgens\setminus\{G\}$.
Since the Pauli strings form a basis of the Pauli matrix algebra, this means that $G$ is proportional to a product of elements from $\pgens\setminus\{G\}$.
Thus there is a subsequence $S\subseteq\pgens$ with $\prod_{H\in S}H\simeq I$.
For the converse, if such a product exists, then any element of $S$ is proportional to the product of the remaining elements and can be removed from the
generating set.
Writing the Pauli strings in $S$ as $\pgens[i]\simeq\iso{v_i}$, multiplication corresponds, up to phase, to addition of binary vectors:
$$
\prod_{i\in S}\pgens[i]\simeq \isoempty\qty(\sum_{i\in S}v_i).
$$
Hence this product is proportional to the identity iff
$\sum_{i\in S}v_i=0$.
This is exactly a linear dependency of the corresponding binary vectors, which proves (b).

For (c), suppose first that $\pgens$ is Lie-algebraically dependent following Definition~\ref{def:dependencies}.
Then some $G\in\pgens$ lies in the Lie algebra generated by $\pgens\setminus\{G\}$.
By Lemma~\ref{lem:pauli:lie}(a), this Lie algebra has a Pauli-string basis given by its Pauli strings,
while by definition it is spanned by iterated commutators of generators in $\pgens\setminus\{G\}$.
Since every non-zero iterated commutator of Pauli strings is again proportional to a Pauli string,
linear independence of Pauli strings implies that $G$ is proportional to one such iterated commutator.
The converse is immediate from the definition of the generated Lie algebra.
To translate this condition to the binary picture, use Proposition~\ref{prop:Pauli:comm}.
Therefore an iterated non-zero commutator
$$
\comm{\pgens[i_m]}{\comm{\pgens[i_{m-1}]}{\ldots,
\comm{\pgens[i_2]}{\pgens[i_1]}}}
$$
corresponds to a binary vector
$v_{i_1}+v_{i_2}+\cdots+v_{i_m}$.
At the binary level, this is obtained by the transvection word
$$
\tau_{v_{i_m}}\tau_{v_{i_{m-1}}}\cdots\tau_{v_{i_2}}v_{i_1},
$$
after omitting trivial transvection steps.
For a word coming from a non-zero nested commutator, the corresponding
commutator steps are precisely the non-trivial transvection steps.
Thus the iterated commutator is proportional to $\pgens[i_{m+1}]$
if and only if
$$
v_{i_{m+1}}
=\tau_{v_{i_m}}\tau_{v_{i_{m-1}}}\cdots\tau_{v_{i_2}}v_{i_1}.
$$
This proves the stated transvection formulation of Lie-algebraic dependencies.

Finally, if such a Lie-algebraic dependency exists, then
$v_{i_{m+1}}=\sum_{j=1}^m v_{i_j}$.
Hence the sum of all vectors among
$v_{i_1},\ldots,v_{i_m},v_{i_{m+1}}$ that occur an odd number of times is zero.
These odd-occurrence vectors therefore form an algebraic dependency.
This proves (d), and (e) follows immediately from (d).
\end{proof}
Notice that (Lie-)algebraic dependencies include the trivial case of repeated elements $G_1 = G_2$. Hence all
sequences with repeating generators have trivial dependencies.

We call the product $\prod_{G\in S} G$ in Lemma~\ref{lem:dependency}(a) an \emph{algebraic dependency} of $\pgens$. Similarly,
we say that the commutator
$\pgens[m+1] \simeq \comm{\pgens[m]}{\comm{\pgens[m-1]}{\ldots \comm{\pgens[2]}{\pgens[1]}}}$ in Lemma~\ref{lem:dependency}(c)
is a \emph{Lie-algebraic dependency} of $\pgens$.
Clearly, $\pgens[m+1] \pgens[m] \pgens[m-1]\cdots \pgens[1]$ is an algebraic dependency.
In the binary formalism, an algebraic dependency is equivalent to a linear dependency of the generators $\vgens$, $\sum_{v\in S} v = 0$ with $S\subseteq\vgens$. We highlight that the algebraic dependency of $\pgens\subseteq\PP_n$ is over a $\C$-matrix algebra, whereas the linear dependency of $\vgens$ is over an $\F_2$-vector space.
A Lie-algebraic dependency instead is equivalent to a sequence of vectors $v_{i_j}\in\pgens$, $j\in\{1,\ldots,m+1\}$, such that $v_{i_{m+1}} = \tau_{v_{i_m}}\tau_{v_{i_{m-1}}}\cdots\tau_{v_{i_2}}v_{i_1}$, up to deleting trivial transvection steps.
Notice that we do not require $i_j\neq i_{j'}$, hence the same vector may also appear multiple times in the product.

Given an algebraic dependency in the binary picture
with sequence $S$ and $w,w'\in S$, it is also clear that the generating sequences $\seq{w,\ldots,w',\ldots}$ and $\seq{w',\ldots,w',\ldots}$
are related via the series of (possibly arbitrary) contractions of various $v$ onto $w$ with
$w + \sum_{v\in S\setminus\{w,w'\}}  v = w'$.
Similarly, in the case of a Lie-algebraic dependency, the generating sequences $\seq{v_1,\ldots,v_{m+1},\ldots}$ and $\seq{v_{m+1},\ldots,v_{m+1},\ldots}$ are $t$-equivalent via the sequence of contractions $\tau_{v_m}\tau_{v_{m-1}}\cdots\tau_{v_2}$ applied to $v_1$.
Then, we also obtain the characterization that a (Lie-)algebraically dependent generating \emph{set} written as a sequence is one which is equivalent to another sequence with a single element appearing multiple times via a series of (valid) contractions.
This justifies a posteriori why we considered the possibility of elements being repeated, and the use of sequences.

Notice that a minimal generating set for a Pauli Lie algebra is not necessarily a minimal generating set for its transvection group, due to the possible presence of pathological transvections.
Consider for instance the example in Table~\ref{tab:ex:pathological}(b).
Its generating set is minimal for both its Lie algebra $\lieu(1)^{\oplus 6}$ and transvection group.
However, the generating set plus its pathological element is also minimal for $\lieu(1)^{\oplus 7}$, while not being minimal for its transvection group.
Hence, minimal generating sets for Lie algebras are not in bijection with minimal generating sets for transvection groups.
In \cite{Aguilar_Cichy_Eisert_Bittel_2024}, non-minimal generating sets are first considered at a graph-theoretic level, by finding suitable representatives, where Lie-algebraic dependencies can then be easily checked and removed.

Then, in order to understand the algebraic properties of such generating sets, we have two possible roads forward: (1) exploit $t$-equivalence and graph contractions, and consider explicit representative graphs and generating sets
(see also Sections~\ref{sec:limits_dependencies} and \ref{sec:two:colors}), (2) use only properties of the graphs and generating sets which are invariant under $t$-equivalence and isomorphism, and avoid graph transformations (see e.g. Section~\ref{sec:invariant_quadratic_forms_graphs}).
We will also see that this extends \emph{for the most part} even in the presence of (Lie-)algebraic dependencies.
The first approach is particularly convenient, since, as we shall see, there exist simple labellings in terms of which one is able to prove the various algebraic properties of a $t$-equivalent generating set (up to isomorphism).
Indeed, this approach is discussed in \cite{Aguilar_Cichy_Eisert_Bittel_2024}.
In the transvection group literature instead, a mixed approach has been typically used, which uses $t$-equivalence for proving general results, but which takes advantage of the invariants of the generating sets and graphs to classify generating sets (e.g.\ describing the orbits \cite{Seven_2005}).
Depending on the application and specific generating set, it may prove difficult to transform the graph, and easier to understand directly the invariants. Indeed, purely by algebraic properties (e.g. solving systems of linear equations) we will be able to describe certain invariants of the graph.
We shall discuss the results of such a classification in Section~\ref{sec:classification:groups_lie_algebras}.

\subsection{Absence of Algebraic Dependencies\label{sec:absence}}

Before further discussing algebraic and Lie-algebraic dependencies and the resulting
challenges, we first discuss the absence of algebraic dependencies,
which provides a bijection between algebraically independent generating sets and frustration graphs, up to isomorphisms:
\begin{lem}[\cite{humphries_1985}]\label{lem:Bijection_Graphs_Linearly_Independent_Sets}
For any graph $\graphG$ there is an $n\in\N$ and a linearly independent set $\vgens\subseteq\Fn$ such that $\graphG = \frustration{\vgens}$.
\end{lem}
\begin{proof}
First, we define an abstract binary vector space $V$ formally spanned by the basis indexed by the vertices
$\tilde{\vgens} = \{ \tilde{v}_j \text{ for } j\in \vertices(\graphG)\}$, hence $\dim(V) = \abs{\vertices(\graphG)}$.
This space also is canonically a symplectic space with the symplectic form $\Omega$
that maps $V\times V$ to $\F_2$ such that $\Omega(\tilde{v}_i,\tilde{v}_j) = A(\graphG)_{ij}$
with $A(\graphG)$ being the adjacency matrix of $\graphG$.

We can perform Gaussian elimination, or through arbitrary contractions, to obtain a symplectic basis
\begin{equation*}
\tilde{\vgens}_s = \{ \tilde{v}_i^s\}_{i=1}^{2m+r} =
 \{\tilde{e}_i,\tilde{f}_i\}_{i=1}^m \cup\{\tilde{h}_j\}_{j=1}^r
\end{equation*}
Since this is a change of basis, there is an invertible linear map $T\in\GL(V)$ such that, element-wise, $T(\tilde{\vgens}_s) = \tilde{\vgens}$.
Since constructing a symplectic basis is the same as obtaining the canonical graph in Lemma~\ref{lem:Canonical_Form_Arbitrary_Adjacent_Contractions_Frustration_Graph} via arbitrary contractions, we have that $2m=\rank(A(\graphG)) = \rank(V)$ (see Eq.~\eqref{eq:rank}) and $r = \nullity(A(\graphG)) = \nullity(V) = \abs{\vertices(\graphG)} - 2m$.

Now there is an isomorphism $\phi$ from $V$ to a subspace $W$ of $\Fn$ with $n\geq m+r$, which identifies a symplectic basis of $V$ with a symplectic basis of $W$. Indeed, this has a natural linear extension and satisfies $\Omega(\tilde{v},\tilde{u}) = \symp{\phi(\tilde{v})}{\phi(\tilde{u})}$ by bilinearity of a symplectic form.
Namely, one can choose the canonical symplectic basis
\begin{equation*}
\{\basel_i,\basel_{n+i}\}_{i=1}^m  \cup \{\basel_{n+m+j}\}_{j=1}^r
\end{equation*}
which corresponds to
\begin{equation*}
\{X_i,Z_i\}_{i=1}^m \cup\{Z_j\}_{j=1}^r.
\end{equation*}
Since $\phi$ is an isomorphism of symplectic spaces and $T$ is linear, we have that $\phi\circ T$
maps from $V$ to $W$ so that $\tilde{\vgens}_s$ is mapped to a basis
$\vgens = \{ v_i\}_{i=1}^{2m+r}$ of $W$ which has $\graphG$ as a frustration graph:
\begin{align*}
        \symp{v_i}{v_j} &= \symp{\phi(T(\tilde{v}_i^s))}{\phi(T(\tilde{v}_j^s))}
        = \Omega(T(\tilde{v}_i^s),T(\tilde{v}_j^s)) \\
        &= \Omega(\tilde{v}_i,\tilde{v}_j) = A(\graphG)_{i,j}.
\end{align*}
Hence, for any $\graphG$ with $\abs{\vertices(\graphG)}\leq 2n$, there is a linearly independent subset $\vgens$ of $\Fn$ with
$\graphG \cong \frustration{\vgens}$.
\end{proof}

Now it is straightforward to show that linearly independent sets with the same frustration graph are in fact isomorphic:
\begin{cor}\label{cor:two:linear:independent}
Consider two linearly independent sets $\vgens,\vgens'\subseteq\Fn$ such that
$\frustration{\vgens }= \frustration{\vgens'}$. Then, there exists some $g\in\Sp(2n,\F_2)$ such that $\vgens' = g\cdot\vgens$.
\end{cor}
\begin{proof}
We can apply the same procedure as in Lemma~\ref{lem:Bijection_Graphs_Linearly_Independent_Sets}, by changing to a symplectic basis
$\vgens_s,\vgens_s'$, for any two bases with the same frustration graph, which allows one to define an isomorphism which maps
$\vgens$ to $\vgens'$.
Then, this isomorphism can be extended to a symplectic transformation $g\in\Sp(2n,\F_2)$ on $\Fn$ by extending $\vgens_s,\vgens_s'$ to symplectic bases for $\Fn$.
\end{proof}
Using the above lemma as well as Lemma~\ref{lem:Canonical_Form_Arbitrary_Adjacent_Contractions_Frustration_Graph}, it is then immediate to show that the rank and nullity of $\graphG$ respectively coincide with the rank and radical dimension of $\vgens$, which further identifies algebraic, graph-theoretic, or combinatorial properties.

Notice that the statement of Corollary~\ref{cor:two:linear:independent} is no longer true when we relax $\vgens,\vgens'$ to just be minimal generating sets. This can seen with the generating sets in Example~\ref{ex:contraction:two}(ii), which have the same frustration graph,
but are not isomorphic with regard to a symplectic map (hence Clifford inequivalent).
Indeed, one has $\nullity(\vgens_1)=1$ and $\nullity(\vgens_2)=2$, with $\pgens_i = \isolong{\vgens_i}$.

However, it still true for linearly dependent sets if guarantee that the linear dependencies have the \emph{same form} with respect to both.
Namely, given two sets $\vgens = \{v_i\}_{i=1}^s$ and $\vgens' = \{v_i'\}_{i=1}^s$ such that: (1) there are linearly independent spanning subsets $\vgens = \{v_{i_j}\}_{j=1}^t$ and $\vgens' = \{v_{i_j}'\}_{j=1}^t$ that have the same frustration graph; (2) they have the same algebraic dependencies $J\subseteq[s]$, or $\sum_{i\in J} v_i = 0 = \sum_{i\in J}v_i'$.
Then, the basis change $g\cdot\vgens = \vgens'$ naturally extends also to all linearly dependent vectors, $g\vgens = \vgens'$.

Hence, we see that as long as we consider only linearly or respectively algebraically independent sets, the labelling of a frustration graph is unique up to isomorphism, and possibly the size of symplectic space or the number of qubits.
Specifically, we say that a labelling or generating set for a given frustration graph is \emph{minimal} if it lies inside $\Fn$ or $\PP_n$ with $n=\rank(A(\graphG))/2 + \nullity(A(\graphG))$.
This is minimal since a smaller $n$ does not allow for a symplectic basis with $m=\rank(A(\graphG))/2$ pairs and $r=\nullity(A(\graphG))$ isolated vertices.
In particular, if one is able to classify the transvection group, Lie algebra, and orbits for these representative generating sets, then
the classification is also known for any algebraically independent set, up to isomorphism and number of qubits.
Next, we shall see that in fact, up to isomorphism, it suffices to speak purely of graph-theoretic properties, without any explicit labelling, by making use of \emph{colorings} and induced subgraphs.

\subsection{Graph Colorings\label{sec:two:colors}}

We now study a useful tool that highlights graph-theoretic properties to perform an algebraic classification and to understand (Lie-)algebraic dependencies:
\begin{defn}[Colorings and their vector realizations \cite{humphries_1985,Janssen_1983,Aguilar_Cichy_Eisert_Bittel_2024}]\label{def:coloring_to_vector_map}
Let $\graphG$ be a graph. The coloring space of $\graphG$ is
\[
\cspace(\graphG):=\F_2^{\vertices(\graphG)}
\]
and its elements are called $2$-colorings, or simply colorings.
Thus a coloring $\coloring\in\cspace(\graphG)$ is a binary coordinate vector
indexed by $\vertices(\graphG)$. Equivalently, the associated coloring map
from $\vertices(\graphG)$ to $\F_2$ is given by
\[
\vva\mapsto\coloring(\vva)=\coloring_{\vva},
\]
which provides the value of $\coloring$ at the vertex $\vva$.

Now let $\pgens=\isolong{\vgens}$ be a sequence of Paulis with vectors
$\vgens\subseteq\Fn$, and let $\frustration{\pgens}$ be the corresponding
frustration graph with vertex labels $\pgens[\vva]$ and vector labels
$v_{\vva}:=\vgens[\vva]$ for $\vva\in\vertices(\frustration{\pgens})$.
The realization map $\coltovec$  is the linear map from $\cspace(\frustration{\vgens})$
to $\SpanS[\F_2]{\vgens}$ given by
\begin{equation}\label{eq:coloring}
\coltovec(\coloring)
=\coltovec_{\vgens}(\coloring)
:=\sum_{\vva\in\vertices(\frustration{\vgens})}\coloring(\vva)\,v_{\vva}.
\end{equation}
\end{defn}
Hence, a coloring $\coloring$ is simply a choice of coordinates for $v$ over $V$. Let $v,v'\in V$ have colorings $\coloring,\coloring'$ respectively, then, by viewing colorings as column vectors:
\begin{equation}\label{eq:symplectic_product_w_colorings}
    \symp{v}{v'} = \symp{\coltovec(\coloring)}{\coltovec(\coloring')} = \coloring^TA(\graphG)\coloring_{v'}
\end{equation}

If $\vgens$ is linearly independent, then the map
$\coltovec$ is an isomorphism.
In general, $\coltovec$ is surjective, and different
colorings may have the same realization in $\SpanS[\F_2]{\vgens}$.

\begin{figure}
    \centering
    \includegraphics[width=0.9\linewidth]{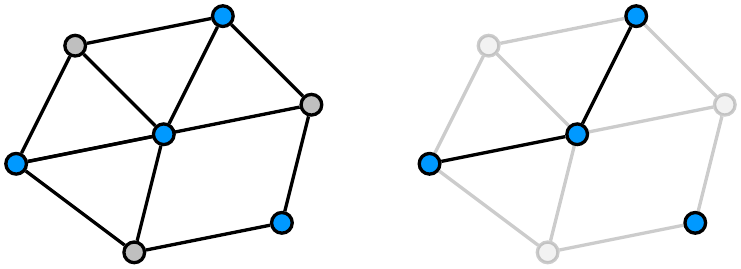}
    \caption{Example of a graph $2$-coloring, its support, and the corresponding induced subgraph.}
    \label{fig:graph-coloring}
\end{figure}

We use the induced-subgraph notation from Eq.~\eqref{eq:induced_subgraph}.
For a vertex subset $U\subseteq\vertices(\graphG)$ of a graph $\graphG$, the induced subgraph $\graphG[U]$ on $U$ keeps precisely the
vertices in $U$ and all edges of $\graphG$ whose endpoints both lie in $U$ (see Fig.~\ref{fig:graph-coloring}).
For a coloring $\coloring\in\cspace(\graphG)$, we write
$$\supp(\coloring)
:=\{u\in\vertices(\graphG) \text{ with } \coloring(u)=1\}$$
and call
$\graphG[\coloring]=\graphG[\supp(\coloring)]$ the induced subgraph
associated with $\coloring$.
For the frustration graph $\frustration{\pgens}$, a coloring is also associated with
the Pauli product
\[
G=\prod_{\vva\in\vertices(\frustration{\pgens})}
\pgens[\vva]^{\coloring(\vva)}\in\algclosure{\pgens}.
\]
Equivalently, an induced subgraph $\graphH$ of $\frustration{\pgens}$ has a
coloring $\coloring$ as its characteristic function. In the binary formalism,
this same coloring has realization $\coltovec(\coloring)$ via
Eq.~\eqref{eq:coloring}. Hence, any element in the multiplicative closure
$\algclosure{\pgens}$ is the Pauli product associated with a coloring
$\coloring$ or, equivalently, with an induced subgraph $\graphH$ of
$\frustration{\pgens}$.
A coloring $\coloring$ of $\frustration{\pgens}$ is \emph{valid} \cite{Aguilar_Cichy_Eisert_Bittel_2024}
if $\im\isolong{\coltovec(\coloring)} \in \lieg = \lie{\pgens}$ (or, equivalently,
$\coltovec(\coloring)\in\tvgroup{\vgens}\cdot\vgens$).

Thus finding valid colorings will be important for a graph-theoretic classification of the Lie algebra or general orbits.
For this purpose, we have the following general rule, which bridges the action of commutators and the transvections
with the graph-theoretic picture, in the language of $2$-colorings:
\begin{lem}[\cite{Aguilar_Cichy_Eisert_Bittel_2024,Kokcu_Wiersema_Kemper_Bakalov_2024}]\label{lem:valid_colorings_as_flips}
Consider a Pauli generating set $\pgens=\isolong{\vgens}$ for vectors $\vgens \subseteq \Fn$ that generates the Lie algebra $\lieg=\lie{\pgens}$.
Then a Pauli $P\in\algclosure{\matalg}$ (vector $v\in\Span{\vgens}$) is in the Lie algebra (orbit $\tvgroup{\vgens}\cdot\vgens$) if and only if there is a sequence of color flips on vertices connected to an odd number of colored vertices which brings a coloring with a single colored vertex into a coloring $\coloring$ whose associated Pauli product is $P=\isolong{\coltovec(\coloring)}$ (see Fig.~\ref{fig:valid-colorings}).
\end{lem}

Also notice that this can be easily extended to consider the orbits in $\algclosure{\pgens}$ (or in $\SpanS[\F_2]{\vgens}$), for which we provide a proof:
\begin{lem}\label{lem:coloring:orbits}
Consider a Pauli generating set $\pgens=\isolong{\vgens}$ for vectors $\vgens = \{v_i\} \subseteq  \Fn$.
Then, two Pauli strings $\iso{v},\iso{w}\in\algclosure{\pgens}$ (or $v,w\in V=\SpanS[\F_2]{\vgens}$)
with colorings $\coloring$ and $\coloring'$ realizing $v=\coltovec(\coloring)$ and $w=\coltovec(\coloring')$, belong to the same connected component of the commutator
graph (or $v$ and $w$ to the same orbit of $\tvgroup{\vgens}$) if and only if there is a sequence of
color flips on vertices connected to an odd number of colored vertices which brings $\coloring$ into
$\coloring'$ (see Fig.~\ref{fig:orbit-colorings}).
\end{lem}
\begin{proof}
It suffices to show the case where two vectors $v,w\in V$ with possibly non-unique colorings $\coloring,\coloring'$ satisfying $v=\coltovec(\coloring)$ and $w=\coltovec(\coloring')$ are adjacent (i.e. there exists $v_i\in\vgens$ such that $w=\tau_{v_i}v$) iff there is a color flip on $\coloring$ at the vertex $i$ connected to an odd number of colored vertices which sends it to $\coloring'$.
Then, sequences of color flips are along connected paths, hence in the same orbit/connected component, and viceversa connected paths determine such sequences of color flips, which proves the statement.

\begin{figure}
    \centering
    \includegraphics[width=0.9\linewidth]{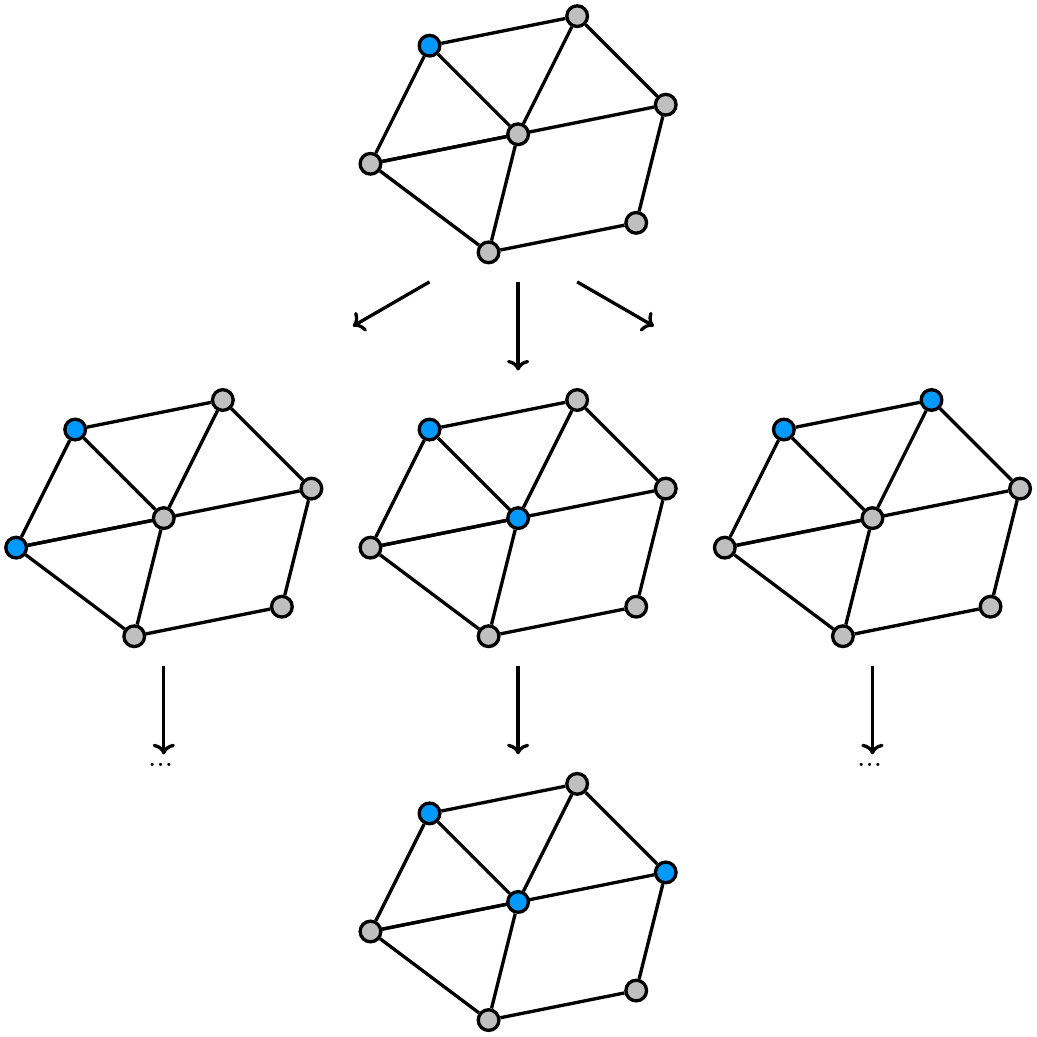}
    \caption{Example of \emph{valid} colorings for a generating set of a Pauli Lie algebra, obtained as a sequence of color flips on vertices with an odd number of colored neighbours.}
    \label{fig:valid-colorings}
\end{figure}

\begin{figure}
    \centering
    \includegraphics[width=0.9\linewidth]{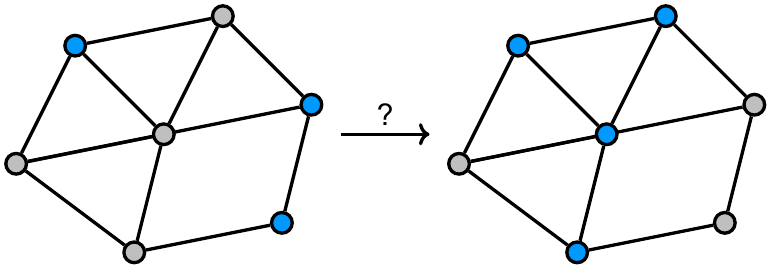}
    \caption{The classification of orbits in $\algclosure{\pgens}$ or $V=\Span{\vgens}$ in the graph theoretic picture amounts to the knowledge of when a coloring can be brought into another through valid color flips.}
    \label{fig:orbit-colorings}
\end{figure}

Now, we highlight that an arbitrary color flip on vertex $i$ is equivalent to the operation which sends a coloring $\coloring$ into another with $\coloring'(i) = \coloring(i) + \delta_{ij}\bmod 2$.
Now, assume $v=\coltovec(\coloring)$. The effect of the transvection on $v$ has the following effect:
\begin{align*}
        w &= \tau_{v_i}v
        = \tau_{v_i}\sum_{j\in\vertices}\coloring(j)v_j
        = \sum_{j\in\vertices} \coloring(j)\tau_{v_i}v_j\\
        &= \sum_{j\in\vertices} \coloring(j)[v_j + A(\graphG)_{ij}v_j] \\
        &= \sum_{j\neq i\in\vertices} \coloring(j) v_j + [\coloring(i) + \sum_j A(\graphG)_{ij}\coloring(j)]v_i \\
        &= \sum_{j\neq i\in\vertices} \coloring(j) v_j + [\coloring(i) + \sum_{j | \coloring(j)=1} A(\graphG)_{ij}]v_i
\end{align*}
or, more explicitly
\begin{equation}
    \coloring'(i) = \coloring(i) + \sum_{j | \coloring(j)=1} A(\graphG)_{ij}.
\end{equation}
Hence $w$ is realized by the coloring obtained from $\coloring$ by flipping the coordinate at $i$ precisely when the number of colored neighbors of $i$ is odd, since $\sum_{j\in J} A(\graphG)_{ij}$ counts the number of neighbors of $i$ in $J$.
Indeed, if it is even, then the transvection does nothing, i.e. $v=w$.
This concludes the proof.
\end{proof}
Under the color-flipping rules specified by Lemma~\ref{lem:valid_colorings_as_flips} and Lemma~\ref{lem:coloring:orbits}, one can define a graph-theoretic game which has been studied in the literature from a combinatorial perspective \cite{Reeder_2005,evenor2017solutionsreederspuzzle}, as well as with applications to certain finite groups. 
One can define a definition of this game in \cite{Reeder_2005}, from which it is known as Reeder's game.

\begin{lem}\label{lem:euler_characteristic_color_flip_invariant}
Consider a generating set $\pgens=\isolong{\vgens}$ for vectors
$\vgens \subseteq \Fn$ with frustration graph $\frustration{\vgens}$.
Identify a coloring of $\frustration{\vgens}$ with its support
$C\subseteq\vertices(\frustration{\vgens})$.
Under the color flips of Lemma~\ref{lem:coloring:orbits}, the Euler characteristic
\[
\chi(C):=\abs{C}-\abs{\edges(\frustration{\vgens}[C])}
\]
of the induced subgraph on the colored vertices is invariant modulo $2$.
\end{lem}
\begin{proof}
A color flip at a vertex $v$ is allowed precisely when $v$ has an odd number $d$ of neighbors in $C$, measured before the flip.
The flip replaces $C$ by the symmetric difference $C\triangle\{v\}$, i.e.\ it either adds $v$ to $C$ or removes $v$ from $C$, so $\abs{C}$ changes by $1$ modulo $2$.
Only edges incident to $v$ can change in the induced subgraph.
If $v\notin C$, then the flip adds $v$, and the new induced edges are exactly the $d$ edges from $v$ to its neighbors in $C$.
If $v\in C$, then the flip removes $v$, and exactly these $d$ induced edges are deleted.
Hence $\abs{\edges(\frustration{\vgens}[C])}$ changes by $d$ modulo $2$ in either case.
Thus, modulo $2$, the change of $\chi(C)$ is $1-d$, which is zero because $d$ is odd.
\end{proof}

\begin{figure}
    \centering
    \includegraphics{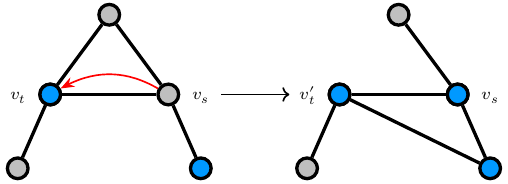}
    \caption{Effect of a contraction on colorings. Contracting the
    source $\sourceidx$ onto the target $\targetidx$ replaces $v_{\targetidx}$ by $v_{\targetidx}+v_{\sourceidx}$, while the
    coloring representing the same vector changes by the reciprocal rule
    $\coloring'(\sourceidx)=\coloring(\sourceidx)+\coloring(\targetidx)$. Thus a colored target toggles
    the source color, while all other colors are unchanged.}
    \label{fig:contraction-coloring}
\end{figure}

We can also talk about the effect of (arbitrary) contractions over colorings,
viewed as a simple change of basis; see Fig.~\ref{fig:contraction-coloring}
for the reciprocal update rule.

\begin{lem}\label{lem:transformation_colorings_contractions}
Consider a generating set $\vgens' = \{v_j'\}$ with frustration graph
$\graphG'$ obtained from $\vgens = \{v_j\}$ with frustration graph $\graphG$
by contracting a source vector $v_{\sourceidx}$ onto a target vector $v_{\targetidx}$, so that
$v_{\targetidx}'=v_{\targetidx}+v_{\sourceidx}$ and $v_j'=v_j$ for $j\neq \targetidx$.
Both sets span the same subspace $V$.
For any coloring $\coloring$ of $\graphG$, define a coloring $\coloring'$
of $\graphG'$ in $\F_2$ by
\begin{equation*}
\coloring'(j)=
\begin{cases}
\coloring(\sourceidx)+\coloring(\targetidx) & \text{if } j=\sourceidx,\\
\coloring(j) & \text{if } j\neq \sourceidx.
\end{cases}
\end{equation*}
Then $\coltovec_{\vgens}(\coloring)=\coltovec_{\vgens'}(\coloring')$.
Conversely, this rule is invertible and gives the corresponding coordinate
transformation between the two generating sets.
\end{lem}
\begin{proof}
Using the definition of the contraction and the transformation rule for
colorings, we compute
\begin{align*}
\coltovec_{\vgens'}(\coloring')
&= \sum_{j\neq \targetidx}\coloring'(j)v_j' + \coloring'(\targetidx)v_{\targetidx}'\\
&= \sum_{j\neq \sourceidx,\targetidx}\coloring'(j)v_j'
    +\coloring'(\sourceidx)v_{\sourceidx}' + \coloring'(\targetidx)v_{\targetidx}'\\
&= \sum_{j\neq \sourceidx,\targetidx}\coloring(j)v_j
    +[\coloring(\sourceidx){+}\coloring(\targetidx)]v_{\sourceidx}+\coloring(\targetidx)(v_{\targetidx}{+}v_{\sourceidx})\\
&= \sum_j\coloring(j)v_j
=\coltovec_{\vgens}(\coloring),
\end{align*}
where the two occurrences of $\coloring(\targetidx)v_{\sourceidx}$ cancel over $\F_2$.
The same formula is its own inverse, since replacing
$\coloring(\sourceidx)$ by $\coloring(\sourceidx)+\coloring(\targetidx)$ twice recovers
$\coloring(\sourceidx)$.
\end{proof}
Hence, arbitrary contractions of a source vertex $\sourceidx$ onto a target vertex $\targetidx$ coincide precisely with color flips on $\sourceidx$, conditioned on the coloring of $\targetidx$.

Note that, in general, there will be multiple colorings
for a given vector $v \in \SpanS[\F_2]{\vgens}$ for vectors $\vgens \subseteq \Fn$.
In particular, there may be valid colorings which are \emph{not} obtained through color flips, but whose realizations still lie in the Lie algebra or a suitable orbit.
Of course, there must also exist a coloring as in Lemma~\ref{lem:valid_colorings_as_flips}.

Hence, this redundancy prevents us to use just colorings to determine the Lie algebra and transvections for general generating sets.

We can however completely characterize \emph{all} colorings whose realizations lie in the radical (and whose Pauli products lie in the center of the matrix algebra).

\begin{figure}
    \centering
    \includegraphics[width=0.6\linewidth]{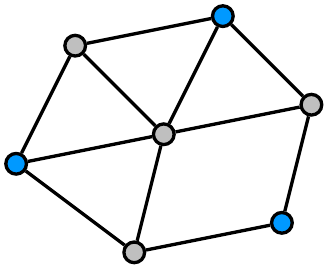}
    \caption{Example of a coloring whose associated Pauli product lies in the center of the matrix algebra, or whose realization lies in the radical.
    Each vertex is adjacent to an even number of colored vertices.}
    \label{fig:even-graph-coloring}
\end{figure}

\begin{lem}[\cite{Khovanova_2008,Aguilar_Cichy_Eisert_Bittel_2024}]\label{lem:Commutant_Algebraic_Dependencies_Graph_Coloring_Subgraph}
A vector $v$ is in $\rad(\vgens)$ (or $\iso{v}$ is in $\ZZ(\algclosure{\pgens})$)
iff every induced subgraph $\graphH$ associated with a coloring $\coloring$
satisfying $v=\coltovec(\coloring)$ has the property that every vertex of
$\frustration{\pgens}$ is adjacent to an even number of vertices of $\graphH$
(see Fig.~\ref{fig:even-graph-coloring}).
Moreover, if $\vgens$ is linearly independent, $\rad(\vgens) = \coltovec(\ker(A(\graphG)))$.
\end{lem}
\begin{proof}
Let $\vgens = \{v_j \text{ for } j \in \vertices  \}$, let
$\coloring\in\cspace(\frustration{\vgens})$, and set
$v=\coltovec(\coloring)$. The coloring $\coloring$ determines the induced
subgraph $\graphH=\frustration{\pgens}[\coloring]$, with
$\vertices(\graphH)=\supp(\coloring)$. Equivalently,
$v = \sum_{i\in \vertices(\graphH)}v_i$. Then, we have that:
\begin{equation}\label{eq:edges:vertices}
        \symp{v}{v_j} = \sum_{i\in \vertices(\graphH)} \symp{v_i}{v_j} = \sum_{i\in\neighbors(j)\cap \vertices(\graphH)} 1,
\end{equation}
which is equal to the number of edges of $\frustration{\pgens}$ modulo two from the vertex $j$
for some vector $v_j\in \vgens$ to vertices of $\graphH$.
Hence $v\in V$ is in the radical $\rad(V)$ with $\symp{v}{v_j}=0$ for all $j \in \vertices(\frustration{\pgens})$ if and only if $\graphH$ shares an even number of
edges with all of $\frustration{\pgens}$.
Also, for $v,v'\in V$ we have by Eq.~\eqref{eq:symplectic_product_w_colorings}:
\begin{equation}
    \symp{v}{v'} = \coloring^TA(\graphG)\coloring'
\end{equation}
Hence $v\in\rad(\vgens)$ iff all of its corresponding colorings are in the kernel $\coloring\in \ker(A(\graphG))$.
In particular, if $\vgens$ is linearly independent, $\coltovec$ provides an isomorphism and we can write $\rad(\vgens) = \coltovec(\ker(A(\graphG)))$.

If $\vgens$ is not linearly independent, then every subgraph $\graphH$
associated with a coloring realizing $v$ shares an even number of edges with
$\frustration{\pgens}$ whenever one does.
\end{proof}
As a special case, the colorings encoding linear dependencies are precisely those colorings
$\coloring$ (or associated subgraphs $\graphH$) whose realizations vanish, i.e.
$$\coltovec(\coloring)=\sum_{j\in\vertices}\coloring(j)v_j=0.$$
Notice however that a coloring alone does not \emph{a priori} characterize Lie-algebraic dependencies, since those in general may require a specific \emph{ordering} of vectors appearing in the sum, as $v_{i_{m+1}} = \tau_{v_{i_m}}\tau_{v_{i_{m-1}}}\cdots\tau_{v_{i_2}}v_{i_1}$, where trivial transvection steps may be omitted.
As we shall see however, we may in fact reduce Lie-algebraic dependencies to algebraic dependencies subject to certain additional conditions, hence also to colorings (see Sections~\ref{sec:limits_dependencies} and \ref{sec:Lie_Algebraic_Dependencies_in_general}).
In particular, we may have $i_s=i_{s'}$ for $s\neq s'$, hence, in the sum $v_{i_1} + \cdots + v_{i_{m+1}}=\coltovec(\coloring)=\sum_j\coloring(j)v_j = 0$, the indices for which $\coloring(j)=1$ may be a proper subset of $\{i_1,\cdots, i_{m+1}\}$.
Hence, it is not sufficient to specify a coloring to determine a Lie-algebraic dependency, since: (1) $\coloring$ ignores those vectors appearing an even number of times in the sum; (2) $\coloring$ does not specify the order in which the sum can be rearranged to produce a Lie-algebraic dependency.
In other words, an algebraic dependency alone does not uniquely determine a Lie-algebraic dependency.

In \cite{Aguilar_Cichy_Eisert_Bittel_2024}, they produce a minimal generating set for the Lie algebra in a constructive way. Namely, they first find representatives in $t$-equivalence class of the frustration graph, then they prove that Lie-algebraic dependencies can only occur in limited cases and hence are able to efficiently remove elements to obtain a minimal generating set.
We also discuss this limitations below (see Lemma~\ref{lem:Canonical_t_equivalent_Graph_Radical})
In the transvection group literature, this issue is mostly avoided as they either assume that the given generating set is already minimal, or they even have the stronger assumption that there are no algebraic dependencies.
In any case, it might however also be possible to purely rely on graph-theoretic information if one is only interested in
a partial analysis of the generating set, hence avoiding any further algebraic analysis.
Namely, as we shall see in Sec.~\ref{sec:classification:groups_lie_algebras}, it will be possible to determine the Lie algebra \emph{type} (up to isomorphism)
restricted to an irreducible subspace, purely by graph-theoretic means.

We emphasize again that the above Lemmas~\ref{lem:coloring:orbits} and \ref{lem:Commutant_Algebraic_Dependencies_Graph_Coloring_Subgraph} do not require a bijection between colorings and Paulis or vectors.
Indeed, in the case when we have algebraically independent generating sets (hence also no Lie-algebraic dependencies), the relation between graph-theoretic and algebraic properties becomes much tighter since colorings and subgraphs are in bijection with Paulis:
\begin{lem}\label{lem:bijection_colorings_algebraically_independent_sets}
If $\pgens= \isolong{\vgens} \subseteq\PP_n$ for vectors $\vgens$
is an algebraically independent set, then the map
$\coltovec$ from $\cspace(\frustration{\vgens})$ to $V=\SpanS[\F_2]{\vgens}$
in Eq.~\eqref{eq:coloring}
is an isomorphism. Consequently, colorings and induced subgraphs of
$\frustration{\pgens}$ are bijectively mapped to the Pauli products in
$\algclosure{\pgens}$.
\end{lem}

This motivates the following definition:
\begin{defn}[Induced subgraph relative to a vector]\label{def:induced:frustration}
Let $\vgens \subseteq \Fn$ be a linearly independent set of binary vectors, let
$V=\Span[\F_2]{\vgens}$ denote its span, and let $\frustration{\vgens}$ be its frustration graph.
For a vector $v\in V$, let $\coloring\in\cspace(\frustration{\vgens})$ be the
unique coloring satisfying $v=\coltovec(\coloring)$. Then
we define the induced subgraph $\inducedfrustration{\vgens}{v}$ of $\frustration{\vgens}$ by
\begin{align*}
\vertices(\inducedfrustration{\vgens}{v})
&:= \{\vva \in \vertices(\frustration{\vgens}) \text{ with } \coloring(\vva)=1\}, \\
\edges(\inducedfrustration{\vgens}{v})
&:= \bigl\{\{\vva,\vvb\}\in \edges(\frustration{\vgens}) \text{ for }
\vva,\vvb \in \vertices(\inducedfrustration{\vgens}{v})\bigr\}.
\end{align*}
\end{defn}

In the linearly independent case, the isomorphism $\coltovec$ is compatible
with the usual vector-space structure on colorings.
Namely, we can define the addition $(\coloring+\coloring')(i) := \coloring(i)+\coloring'(i)\bmod 2$ (which generalizes color flips on single vertices) and the symplectic product $\symp{\coloring}{\coloring'} := \sum_{i,j\in\vertices(\graphG)}\coloring(i)\coloring'(i)A(\graphG)_{i,j}$.
A basis of $\cspace(\graphG)$ is given by the single-vertex colorings
$\mathbf{1}_j$, defined by $\mathbf{1}_j(i)=\delta_{ij}$, and the isomorphism $\coltovec$
identifies this coloring space with $V$ for a linearly independent
$\vgens=\{v_j\}_{j\in\vertices(\graphG)}$.
Hence, $\cspace(\graphG)$ provides a canonical realization of a linearly
independent generating set with unlabelled frustration graph $\graphG$.
In view of this structure and the transformation rule \ref{lem:transformation_colorings_contractions}, a contraction of a source vector $v_{\sourceidx}$ onto a target vector $v_{\targetidx}$ in $\vgens$ corresponds, under $\coltovec$, to the \emph{reciprocal} contraction of the single-vertex coloring $\mathbf{1}_{\targetidx}$ onto $\mathbf{1}_{\sourceidx}$.
This will be especially important in understanding what happens to (Lie-)algebraic dependencies under contractions and their description in terms of colorings.

Therefore, in the linearly independent case, these tools allow one to
characterize algebraic properties through graph-theoretic means.
Of course, given the isomorphism
$\coltovec$ from $\cspace(\graphG)$ to $V=\SpanS[\F_2]{\vgens}$, a purely
graph-theoretic characterization will be able to determine all properties in
the subspace, or equivalently, the Pauli matrix algebra
$\matalg = \Span[\C]{\isolong{V}}$.

A complementary viewpoint identifies the same coloring space with the
dual space and connects it to dual transvection groups and the lit-only sigma game
\cite{Sjöstrand_2025,Vorstermans}. In the linearly
independent case, a coloring $\coloring\in\cspace(\graphG)$ can be identified
not only with the vector $\coltovec_{\vgens}(\coloring)\in V$, but also with
the functional $\widehat{\coloring}\in V^*$ defined by
$\widehat{\coloring}(v_i)=\coloring(i)$ and extended linearly. Hence
$\cspace(\graphG)\cong V^*$ as soon as $\vgens=\{v_i\}$ is a basis of $V$,
and a linear map $g\colon V\to V$ acts on colorings by the dual action
$g^*(\widehat{\coloring})=\widehat{\coloring}\circ g$. This is the setting of
dual symplectic transvections over $\F_2$ studied in \cite{Sjöstrand_2025}.
Their graph-theoretic form is the lit-only sigma game, where toggling a vertex $j$
flips the colors of its neighbors precisely when $\coloring(j)=1$, and the
toggle acts trivially otherwise.

The corresponding valid contraction
interpretation is obtained by fixing a target vertex $\targetidx$ and considering the
neighborhood functional
$\lambda_{\targetidx}\in\cspace(\graphG[\vertices(\graphG)\setminus\{\targetidx\}])$ defined by
$\lambda_{\targetidx}(j)=A(\graphG)_{\targetidx j}=\symp{v_{\targetidx}}{v_j}$ for $j\neq \targetidx$. By
Lemma~\ref{lem:valid:contraction:graph}, or equivalently by the adjacency
matrix update in Eq.~\eqref{eq:adjacency_matrix_contraction}, a contraction
of the source $\sourceidx$ onto the target $\targetidx$ updates the target row by adding the
source row over $\F_2$. Thus, if the
contraction sends $v_{\targetidx}$ to $v_{\targetidx}+v_{\sourceidx}$, then on the induced subgraph
$\graphG[\vertices(\graphG)\setminus\{\targetidx\}]$ one has
$\lambda_{\targetidx}'(j)=\lambda_{\targetidx}(j)+A(\graphG)_{\sourceidx j}$.
A contraction is valid precisely if $\lambda_{\targetidx}(\sourceidx)=1$. In this case, toggling $\sourceidx$
is a nontrivial lit-only sigma-game move and flips the colors of the neighbors
of $\sourceidx$ in the induced subgraph. This is
the combinatorial viewpoint on valid contractions
used in \cite{Gintz,Aguilar_Cichy_Eisert_Bittel_2024}.

After fixing a target vertex, this provides the dual counterpart of
Lemma~\ref{lem:transformation_colorings_contractions}. That lemma gives the
reciprocal coordinate-change rule for colorings representing fixed realized
vectors under a contraction of the generating set. Here, instead, the coloring
is the functional $\lambda_t$ recording adjacency to the fixed target vertex.
Applying the dual viewpoint to this special functional turns the same
contraction into the corresponding lit-only sigma-game move on
$\graphG[\vertices(\graphG)\setminus\{t\}]$. In particular, this dual-orbit
interpretation applies only to colorings in the above mathematical sense;
other colors used in figures are visual annotations and may carry a different
interpretation.

We also note that the lit-only sigma game and Reeder's game \cite{Reeder_2005} were also described as \emph{dual} to each other in the sense of \cite{huang2010litonlysigmagamedualgame}, which shows a strong relation between understanding canonical graphs for $t$-equivalence and orbits under the transvection group.

\subsection{Limits to (Lie-)Algebraic Dependencies\label{sec:limits_dependencies}}

We are now ready to tackle the issues arising from (Lie-)algebraic dependencies and various way to deal with them, which will be necessary for a graph-theoretic analysis.
We start by giving a structural characterization of algebraic dependencies, using colorings and \emph{extensions} in a larger space. Let us make this precise:

\begin{defn}[Extension of Generating Sequences and Sets \cite{Brown_Humphries_1986a,Janssen_1983}]\label{def:extension_generating_set_subspace_projection}
Consider two generating sequences (or sets) $\vgens$ and $\tilde{\vgens}$ contained inside the symplectic spaces $V$ and $\tilde{V}$, respectively.
We say that $\tilde{\vgens}$ is an \emph{extension} of $\vgens$ if there exist a projection map $\projection$  from $\tilde{V}$ to $V$
and an inclusion map $\inclusion$ from $V$ to $\tilde{V}$
such that $\projection$
is a surjective linear map that preserves the symplectic form (i.e.,
$\symp{\tilde{u}}{\tilde{v}}=\symp{\projection(\tilde{u})}{\projection(\tilde{v})}$
for all $\tilde{u},\tilde{v}\in\tilde{V}$), $\projection$ induces a bijection between the generating sequences (or sets)
with $\projection(\tilde{\vgens}[j]) = \vgens[j]$ (or $\projection(\tilde{\vgens}) = \vgens$), and $\projection\circ \inclusion$ is the identity map
on $V$ with $\mathrm{Im}(\inclusion) \cong \tilde{V}/\ker(\projection)$.
\end{defn}

We can collect some properties of extensions:

\begin{lem}\label{lem:extensions}
Consider generating sequences (or sets) $\vgens$ and $\tilde{\vgens}$
with $V=\SpanS[\F_2]{\vgens}$ and  $\tilde{V}=\SpanS[\F_2]{\tilde{\vgens}}$
and let $\graphG=\frustration{\vgens}$ with adjacency matrix $A(\graphG)$.
Then we obtain:
\begin{enumerate}
\item $\frustration{\vgens} = \frustration{\tilde{\vgens}}$ for any extension $\tilde{\vgens}$  of $\vgens$.
\item For $\vgens$, there exists a minimal extension $\tilde{\vgens}\subseteq\Fn$
with $n = \rank(A(\graphG))/2+\nullity(A(\graphG))$, and $\tilde{\vgens}$ being linearly independent.
\item The minimal linearly independent extension in (b) is unique up to isomorphism.
\end{enumerate}
\end{lem}

If $\vgens$ in (b) of Lemma~\ref{lem:extensions} is linearly independent, then
$\rank(A(\graphG))=\rank(V)$ and $\nullity(A(\graphG))=\nullity(V)$.
In this case the minimal value of $n$ can equivalently be written as
$n=\rank(V)/2+\nullity(V)$.

\begin{proof}
For (a), Definition~\ref{def:extension_generating_set_subspace_projection}
implies that $\projection(\tilde{\vgens}[j])=\vgens[j]$ and that $\projection$
preserves the symplectic form. Hence the two generating sequences have the
same adjacency matrix.
For (b), we apply the construction in
Lemma~\ref{lem:Bijection_Graphs_Linearly_Independent_Sets} to
$\graphG=\frustration{\vgens}$. This gives a linearly independent
$\tilde{\vgens}\subseteq\Fn$ with
\begin{equation}
    n=\rank(A(\graphG))/2+\nullity(A(\graphG)).
\end{equation}
Since $\tilde{\vgens}$ is linearly independent,
$\projection(\tilde{\vgens}[j])=\vgens[j]$ extends to a surjective linear map
$\projection$. Thus $\frustration{\tilde{\vgens}}=\frustration{\vgens}$ implies that
$\projection$ preserves the symplectic
form. Surjectivity of $\projection$ gives an inclusion map
$\inclusion$ with $\projection\circ\inclusion=\id_V$.
We conclude that $\tilde{\vgens}$ is an extension of $\vgens$.
Finally, (c) follows from Corollary~\ref{cor:two:linear:independent}.
\end{proof}

Notice that the inclusion is only the \emph{right} inverse of the projection, in general. Hence, even though $\projection(\tilde{\vgens}) = \vgens$ (element-wise), it may not be true that $\inclusion(\vgens) = \tilde{\vgens}$.
We show some basic relations between the projection, the transvection, the radical and algebraic dependencies, which also connects the Pauli language to the transvection group literature (based on \cite{Brown_Humphries_1986a,Janssen_1983}):

\begin{lem}\label{lem:projection_extension_basic_properties}
Let $\tilde{\vgens}\subseteq\tilde{V}$ an extension of $\vgens\subseteq V$, with projective map $\projection$, inclusion $\inclusion$ and shared (unlabelled) frustration graph $\graphG$.
Write $\vgens=(v_j)_{j\in\vertices(\graphG)}$ and $\tilde{\vgens}=(\tilde{v}_j)_{j\in\vertices(\graphG)}$, with $\projection(\tilde{v}_j)=v_j$.
\begin{enumerate}
    \item\label{lem:projection_extension_basic_properties:a}
    We have that $\ker(\projection) \subseteq \rad(\tilde{V})$ and $\projection(\rad(\tilde{V})) = \rad(V)$.
    Equivalently, the restriction of $\projection$ to $\rad(\tilde{V})$ induces a short exact sequence
    \begin{equation}
        0 \to \ker(\projection) \to \rad(\tilde{V}) \xrightarrow{\projection} \rad(V) \to 0.
    \end{equation}
    Thus $\rad(V)\cong\rad(\tilde{V})/\ker(\projection)$.
    \item\label{lem:projection_extension_basic_properties:b}
    If $\tilde{\vgens}$ is linearly independent, each coloring $\coloring$ determines
    the unique realization
    \begin{equation}\label{eq:unique:coloring}
        \tilde{v}_{\coloring}:=\coltovec_{\tilde{\vgens}}(\coloring)= \sum_{j\in\vertices(\graphG)}\coloring(j)\tilde{v}_j\in\tilde{V}.
    \end{equation}
    The vector $\tilde{v}_{\coloring}$ lies in $\ker(\projection)\subseteq \rad(\tilde{V})$ iff the realization of the same coloring with respect to $\vgens$ vanishes, i.e.,
    $\coltovec_{\vgens}(\coloring)=\sum_{j \in\vertices(\graphG)}\coloring(j)v_j=0$.
    \item\label{lem:projection_extension_basic_properties:c}
    If $\vgens'$ and $\tilde{\vgens}'$ are obtained from $\vgens$ and $\tilde{\vgens}$ by applying the same arbitrary contraction, then $\tilde{\vgens}'$ is an extension of $\vgens'$ with the same linear projection map $\projection$.
    \item\label{lem:projection_extension_basic_properties:d}
    The projection induces a surjective homomorphism
    $\tvproj$ from $\tvgroup{\tilde{\vgens}}$ to $\tvgroup{\vgens}|_V$, defined by
    $ \tvproj(\tilde{g})(v):=\projection(\tilde{g}\tilde{v})$ for any
     $\tilde{v}\in\tilde{V}$ with $\projection(\tilde{v})=v$.
    Its kernel is
    \begin{equation}
        \tvprojker:=\ker(\tvproj)
        =
        \{\tilde{g}\in\tvgroup{\tilde{\vgens}} \text{ with } \projection\circ\tilde{g}=\projection\},
    \end{equation}
    and one obtains the short exact sequence
    \begin{equation}
        1\to \tvprojker\to\tvgroup{\tilde{\vgens}}\xrightarrow{\tvproj}\tvgroup{\vgens}|_V\to 1.
    \end{equation}
    \item\label{lem:projection_extension_basic_properties:e}
    Using the homomorphism $\tvproj$ from \ref{lem:projection_extension_basic_properties:d}, $\projection$ induces the well-defined surjective map on orbit spaces
    from $\tilde{V}/\tvgroup{\tilde{\vgens}}$ to $V/\tvproj(\tvgroup{\tilde{\vgens}})=V/(\tvgroup{\vgens}|_V)$ as
    \begin{equation*}
        \tvgroup{\tilde{\vgens}}\cdot\tilde{v}
        \mapsto
        \tvproj(\tvgroup{\tilde{\vgens}})\cdot\projection(\tilde{v}).
    \end{equation*}
    In particular,
    \begin{equation}
        \projection(\tvgroup{\tilde{\vgens}}\cdot\tilde{\vgens})
        =
        \tvproj(\tvgroup{\tilde{\vgens}})\cdot\vgens
        =
        \tvgroup{\vgens}|_V\cdot\vgens
        =
        \tvgroup{\vgens}\cdot\vgens.
    \end{equation}
\end{enumerate}
\end{lem}

\begin{proof}
We start with the proof of (a):
If $\tilde{u}\in\ker(\projection)$, then
$\symp{\tilde{u}}{\tilde{v}}=\symp{\projection(\tilde{u})}{\projection(\tilde{v})}=0$
for all $\tilde{v}\in\tilde{V}$, hence $\ker(\projection)\subseteq\rad(\tilde{V})$.
If $\tilde{u}\in\rad(\tilde{V})$, then
$\symp{\projection(\tilde{u})}{v}=0$ for all $v\in V$ by surjectivity of $\projection$, so
$\projection(\rad(\tilde{V}))\subseteq\rad(V)$.
Conversely, if $u\in\rad(V)$ and $\projection(\tilde{u})=u$, then
$\symp{\tilde{u}}{\tilde{v}}=\symp{u}{\projection(\tilde{v})}=0$ for all $\tilde{v}\in\tilde{V}$, hence
$\tilde{u}\in\rad(\tilde{V})$ and $u\in\projection(\rad(\tilde{V}))$.
Thus $\projection(\rad(\tilde{V}))=\rad(V)$.
The short exact sequence follows by restricting $\projection$ to $\rad(\tilde{V})$, whose kernel is $\ker(\projection)$, and therefore
$\rad(V)\cong\rad(\tilde{V})/\ker(\projection)$.

For (b), we assume that $\tilde{\vgens}$ is linearly independent.
Then every coloring $\coloring$ determines the unique realization
$\tilde{v}_{\coloring}$ as in Eq.~\eqref{eq:unique:coloring}.
Using $\projection(\tilde{v}_j)=v_j$, we have
$
    \projection(\tilde{v}_{\coloring})
    =
    \coltovec_{\vgens}(\coloring)
$.
Hence $\tilde{v}_{\coloring}\in\ker(\projection)$ if and only if
$\coltovec_{\vgens}(\coloring)=0$, which is precisely the condition that $\coloring$ encodes a linear dependency among the generators $\vgens$.
The inclusion $\ker(\projection)\subseteq\rad(\tilde{V})$ from (a) gives the stated radical containment.

For (c), it suffices to check one arbitrary contraction.
Suppose that $v_{j_1}$ is replaced by $v_{j_1}'=v_{j_1}+v_{j_2}$ and, simultaneously, $\tilde{v}_{j_1}$ is replaced by $\tilde{v}_{j_1}'=\tilde{v}_{j_1}+\tilde{v}_{j_2}$, with all other generators unchanged.
Then the linearity of $\projection$ implies
$\projection(\tilde{v}_{j_1}') =v_{j_1}'$.
For all other indices the projection relation is unchanged.
Thus the same linear map $\projection$ sends $\tilde{\vgens}'$ element-wise to $\vgens'$, and the remaining extension properties are unchanged.

For (d), first note that each generating transvection
commutes with $\projection$, i.e.,
$
    \projection(\tau_{\tilde{v}_j}\tilde{x})
    =
    \tau_{v_j}\projection(\tilde{x})
$
for all $\tilde{x}\in\tilde{V}$, since $\projection(\tilde{v}_j)=v_j$ and $\projection$ preserves the symplectic form.
Thus any $\tilde{g}\in\tvgroup{\tilde{\vgens}}$ induces a well-defined map $\tvproj(\tilde{g})\in\tvgroup{\vgens}|_V$ by
$\tvproj(\tilde{g})(v)=\projection(\tilde{g}\tilde{v})$ for $\projection(\tilde{v})=v$,
which shows that the definition of $\tvproj$ is independent of $\tilde{v}$.
The images of the generators $\tau_{\tilde{v}_j}$ are the generators $\tau_{v_j}|_V$, so $\tvproj$ is surjective.
Moreover, $\tilde{g}\in\ker(\tvproj)$ iff $\projection(\tilde{g}\tilde{v})=\projection(\tilde{v})$ for all $\tilde{v}\in\tilde{V}$, i.e., iff $\projection\circ\tilde{g}=\projection$.
This identifies $\tvprojker=\ker(\tvproj)$ and gives the stated short exact sequence.

For (e), we apply $\projection(\tilde{g}\tilde{v})=\tvproj(\tilde{g})\projection(\tilde{v})$
 from (d).
Therefore, if two vectors of $\tilde{V}$ lie in the same $\tvgroup{\tilde{\vgens}}$-orbit, their projections lie in the same $\tvproj(\tvgroup{\tilde{\vgens}})$-orbit in $V$.
This gives the well-defined map on orbit spaces.
It is surjective because $\projection$ is surjective, and $\tvproj(\tvgroup{\tilde{\vgens}})=\tvgroup{\vgens}|_V$ by (d).
Applying this to the orbit of the generating sequence gives
\begin{equation*}
    \projection(\tvgroup{\tilde{\vgens}}\cdot\tilde{\vgens})
    =
    \tvproj(\tvgroup{\tilde{\vgens}})\cdot\vgens
    =
    \tvgroup{\vgens}|_V\cdot\vgens
    =
    \tvgroup{\vgens}\cdot\vgens. \qedhere
\end{equation*}
\end{proof}

We start by noticing that this property is invariant under $t$-equivalence:
\begin{lem}\label{lem:Contractions_conserve_Lie_algebraic_dependencies}
Arbitrary contractions preserve algebraic dependencies. Valid contractions
preserve Lie-algebraic dependencies. Since contractions are reversible, this
preservation is bijective between a generating sequence and its contracted sequence.
\end{lem}
\begin{proof}
We apply Lemma~\ref{lem:projection_extension_basic_properties}.
Choose a linearly independent extension $\projection(\tilde{\vgens})=\vgens$.
By Lemma~\ref{lem:projection_extension_basic_properties}\ref{lem:projection_extension_basic_properties:b}, colorings
encode linear dependencies of $\vgens$ exactly when their realizations
$\tilde v_\coloring=\coltovec_{\tilde{\vgens}}(\coloring)$ lie in
$\ker(\projection)$.
If $\vgens'$ is obtained from $\vgens$ by an arbitrary contraction and $\tilde{\vgens}'$ by the same contraction, then Lemma~\ref{lem:projection_extension_basic_properties}\ref{lem:projection_extension_basic_properties:c} shows that the same projection map is still leads to an extension $\projection(\tilde{\vgens}')=\vgens'$.
Thus the relevant kernel is unchanged, and the contraction only rewrites the same kernel vectors in the contracted coloring coordinates; algebraic dependencies are therefore preserved bijectively.
For Lie-algebraic dependencies, valid contractions are transvections and hence invertible changes of generators inside the same transvection action.
Using Lemma~\ref{lem:projection_extension_basic_properties}\ref{lem:projection_extension_basic_properties:d}--\ref{lem:projection_extension_basic_properties:e}, the induced transvection action and its orbits are transported through the projection, so a dependency expressed by a transvection word before the contraction is equivalently expressed after the contraction, and conversely by applying the inverse valid contraction.
\end{proof}

As a first classification of (Lie-)algebraic dependencies, we may look just at those generating sets whose frustration graph is one of the canonical representatives.
Specifically, given that algebraic dependencies correspond to colorings whose
realizations lie in the radical of the linearly independent extension, we can get a first graph-theoretic understanding based on the radical of all canonical graphs (or their unique linearly independent generating sets, up to isomorphism).
This allows to \emph{partially} shift the analysis from algebraic dependencies on generating sets to graphs.
For the canonical graphs in Theorem~\ref{thm:classes:humphries} or \ref{thm:classes:eisert}, we have the following characterization of the radical:

\begin{lem}[\cite{Janssen_1983,Seven_2005,Aguilar_Cichy_Eisert_Bittel_2024}]\label{lem:Canonical_t_equivalent_Graph_Radical}
Let $\tilde{\vgens}$ be the unique linearly independent set for one
of the canonical graph families listed in
Theorems~\ref{thm:classes:humphries} or \ref{thm:classes:eisert}.
For each such canonical graph, the following vectors form a basis of
$\rad(\tilde{\vgens})$:
\begin{enumerate}
    \item\label{lem:Canonical_t_equivalent_Graph_Radical:length-one-leaves}
    If $n_1\geq 2$, the $n_1-1$ vectors
    $b_i+b_{i+1}$ for $1\leq i<n_1$, where $b_i$ is the vector label of
    the $i$th length-one leaf.
    \item If $n_1\geq 1$, there are additional basis elements depending on the specific canonical representative:
    \begin{nestedcaseenum}
        \item\label{lem:Canonical_t_equivalent_Graph_Radical:path-extra}
        For $\graphP_{2m,n_1}$, the vector $\sum_{j=1}^{m+1} a_{2j-1}$.
        \item\label{lem:Canonical_t_equivalent_Graph_Radical:x3-extra}
        For $\graphX_{2m,n_1}^3$, the vector $a_1+a_2+a_4$.
        \item\label{lem:Canonical_t_equivalent_Graph_Radical:s3-extra}
        For $\graphS_{n_2,n_1}^3$, the vector $a_1+a_3+b_1$.
    \end{nestedcaseenum}
\end{enumerate}
\end{lem}

In particular, the canonical representatives split into two classes.
For representatives whose radical is generated only by the vectors supported on length-one leaves, namely
$$
\graphP_{2m-1,n_1},\quad
\graphS_{n_2,n_1}^1,\quad
\graphS_{n_2,n_1}^2,\quad
\graphX_{2m-1,n_1}^1,\quad
\graphX_{2m-1,n_1}^2,
$$
we have $\nullity(\tilde{\vgens})=\nullity(A(\graphG))=n_1-1$.
For representatives with one additional radical element, namely
$$
\graphP_{2m,n_1},\quad
\graphS_{n_2,n_1}^3,\quad
\graphX_{2m,n_1}^3,
$$
we have $\nullity(\tilde{\vgens})=\nullity(A(\graphG))=n_1$.

\begin{proof}
We use colorings only as coordinates for vectors in
$\SpanS[\F_2]{\tilde{\vgens}}$. Since $\tilde{\vgens}$ is linearly
independent, each vector $u\in\SpanS[\F_2]{\tilde{\vgens}}$ has a unique
coloring $\coloring$ with
$u=\coltovec_{\tilde{\vgens}}(\coloring)=\sum_i\coloring(i)\tilde{v}_i$.
By Lemma~\ref{lem:Commutant_Algebraic_Dependencies_Graph_Coloring_Subgraph}, this vector lies in $\rad(\tilde{\vgens})$ precisely when, at every vertex, the sum over the colors of all adjacent vertices is zero in $\F_2$. Equivalently, each vertex has an even number of colored neighbors.
This condition is $A(\graphG) \coloring=0$ over $\F_2$, i.e.\
$\coloring\in\ker A(\graphG)$; under $\coltovec_{\tilde{\vgens}}$, these
colorings realize precisely the vectors in $\rad(\tilde{\vgens})$.
For the $n_1$ length-one leaves, the equation at each $b_j$ forces the common neighbor $c$ to have color $0$, as indicated by the grey vertex in Figure~\ref{fig:radical-coloring-proof}(a).
The colors $\coloring(b_j)$ themselves are not fixed by these local leaf equations.
They enter only through the equation at $c$, where their contribution is the total parity
$\lambda = \sum_{j=1}^{n_1}\coloring(b_j)$.
Thus two colorings supported on length-one leaves with the same total parity differ by an even-parity coloring supported on those leaves, whose realizations are generated by the vectors $b_j+b_{j+1}$ for $1\leq j<n_1$.
Hence the part of the radical resulting from the length-one leaves is spanned by
the vectors $b_i+b_{i+1}$ for $1\leq i<n_1$, where only
the total parity $\lambda$ is still relevant for the remaining equations.
For the rest of the proof,
all length-one leaves may be chosen to be not colored if $\lambda=0$;
we may choose a case with exactly one colored length-one leaf if $\lambda=1$.
It remains to solve the same parity equations on the part not corresponding to length-one leaves of the canonical representative.

For later use, we record the local recursion on pendant paths, as shown in Figure~\ref{fig:radical-coloring-proof}(a).
For $o\geq1$, on a pendant path $c,x_1,x_2,\ldots,x_{2o},x_{2o+1}$, the parity equation at the leaf $x_{2o+1}$ forces $\coloring(x_{2o})=0$.
The equation at $x_{2o}$ gives $\coloring(x_{2o-1})=\coloring(x_{2o+1})$, and continuing inward forces the colors to alternate between $0$ on even positions and the terminal value $\coloring(x_{2o+1})$ on odd positions.
Thus the coloring on such a path is determined by the single terminal value, while the equation at the center $c$ decides whether this value survives.

For $\graphP_{k,n_1}$, the remaining core is just the spine.
Solving the parity equations from the left endpoint gives the alternating pattern shown in Figure~\ref{fig:radical-coloring-proof}(b).
At the other end, the leaf equations at the length-one leaves force the center to have color $0$.
For $\graphP_{2m-1,n_1}$, this condition is incompatible with a nonzero coloring.
For $\graphP_{2m,n_1}$, the two endpoint conditions are compatible, giving exactly the alternating odd-vertex vector $\sum_{j=1}^{m+1}a_{2j-1}$ in \ref{lem:Canonical_t_equivalent_Graph_Radical:path-extra}.

For $\graphX_{2m-1,n_1}^1$ in Figure~\ref{fig:radical-coloring-proof}(c), the spine has the same odd-path obstruction as for $\graphP_{2m-1,n_1}$.
In addition, the extra length-one leaf forces its spine vertex to have color $0$.
This is incompatible with the nonzero alternating spine pattern, so no additional vector survives.

For $\graphX_{2m-1,n_1}^2$ in Figure~\ref{fig:radical-coloring-proof}(d), the extra leg has length two.
The leaf equation forces the middle vertex of this leg to have color $0$, and the next equation identifies the terminal color with the color of the spine vertex.
The core equations force this spine value to vanish in any surviving coloring, so again
only the vectors supported on length-one leaves remain.

For $\graphX_{2m,n_1}^3$ in Figure~\ref{fig:radical-coloring-proof}(e), the parity equations are instead compatible with one nonzero core coloring.
The local recursion forces this coloring to be exactly $a_1+a_2+a_4$ as in \ref{lem:Canonical_t_equivalent_Graph_Radical:x3-extra}, and all other possibilities vanish.

\begin{figure}
    \centering
    \includegraphics{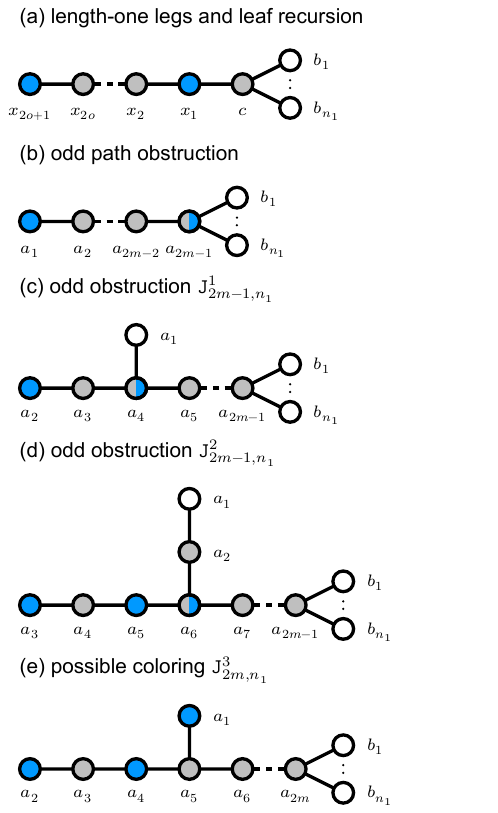}
    \caption{Local parity constraints used to identify colorings of the canonical representatives corresponding to radical elements. Blue vertices have color $1$, grey vertices have color $0$, white vertices are not fixed locally but enter only through the parity equation of Lemma~\ref{lem:Commutant_Algebraic_Dependencies_Graph_Coloring_Subgraph}, and half-filled vertices mark incompatible local color requirements.}
    \label{fig:radical-coloring-proof}
\end{figure}

The star representatives require a separate check at the common center.
The length-one legs force the center to have color $0$, while their endpoint colors enter the center equation only through the parity $\lambda$.
On every leg of length two, the leaf equation forces the middle vertex to have color $0$, and the middle equation then forces the terminal vertex to have color $0$ as well.
Thus, for $\graphS_{n_2,n_1}^1$ and $\graphS_{n_2,n_1}^2$, no additional coloring survives beyond the length-one legs.
For $\graphS_{n_2,n_1}^3$, the leg of length three leaves the coloring $a_1+a_3$ on that leg, and the equation at the center requires $\lambda=1$ on the length-one leaves.
For the length-one leaves, the parity condition $\lambda=1$ may be realized by choosing the leaf $b_1$ to have color $1$; hence the additional radical vector is $a_1+a_3+b_1$ as in \ref{lem:Canonical_t_equivalent_Graph_Radical:s3-extra}.
The displayed vectors are independent, and the preceding parity analysis
exhausts all kernel vectors.
\end{proof}

Thus the radical for each canonical graphs is either trivial or one-dimensional if
there is a single length-one leaf (i.e.\ $n_1=1$).
As (Lie-)algebraic dependencies are invariant under (valid) contractions,
one can aim at identifying graphs that are
$t$-equivalent to canonical graphs with $n_1=1$
(or have almost maximal rank). We provide some initial information
in this direction in the following Section~\ref{sec:Twins}, while
Section~\ref{sec:Lie_Algebraic_Dependencies_in_general} contains
a full description including the (Lie-)algebraic dependencies.
Along similar lines, we discuss the rank of the canonical graphs:
\begin{align*}
&\rank(A(\graphS_{n_2,n_1}^1)) = 2n_2+2,\;
\rank(A(\graphS_{n_2,n_1}^2)) = 2n_2+6\\
&\rank(A(\graphS_{n_2,n_1}^3)) = 2n_2+4,\;
 \rank(A(\graphG)) = 2m \text{ for} \\
&\hspace{0.2em} \graphG \in \{\graphP_{2m-1,n_1}, \graphP_{2m,n_1}, \graphX_{2m-1,n_1}^1, \graphX_{2m-1,n_1}^2, \graphX_{2m,n_1}^3\}.
\end{align*}

For the minimal allowed values of $m$ and $n_2$ in the canonical graphs listed in Theorem~\ref{thm:classes:humphries} and \ref{thm:classes:eisert}, the following  cases have rank at least six:
\begin{align}
&\rank(A(\graphX_{2m-1,n_1}^1)),\;
\rank(A(\graphX_{2m,n_1}^3)) \geq 6, \nonumber\\
&\rank(A(\graphS_{n_2,n_1}^1)),\;
\rank(A(\graphS_{n_2,n_1}^3)) \geq 6, \label{eq:small_rank_cases} \\
&\rank(A(\graphX_{2m-1,n_1}^2)),\;
\rank(A(\graphS_{n_2,n_1}^2)) \geq 8. \nonumber
\end{align}
If $\rank(A(\graphG))\leq 5$, then $\graphG$ is $t$-equivalent to $\graphP_{k,n_1}$  for suitable $k$ and $n_1$.
Conversely, $$\rank(A(\graphP_{2m-1,n_1}))=\rank(A(\graphP_{2m,n_1}))=2m,$$ so the small-rank cases are precisely those with $2m\leq 5$.
In particular, every connected graph with less than six vertices is $t$-equivalent to some $\graphP_{k,n_1}$.
Furthermore, the graphs in $\graphP_{2m,1}$, $\graphS_{n_2,1}^3$, and $\graphX_{2m,1}^3$ cannot be $t$-equivalent to a graph with maximal rank, since there is at least one non-trivial element in their radical.

Finally, all algebraic dependencies whose realized vectors lie in the
subspace $B$ from Eq.~\eqref{eq:canonical-dependency-B} are also
Lie-algebraic dependencies.
This is a necessary and sufficient condition for such graphs. 
If we eliminate all such Lie-algebraic dependencies, we find a minimal generating set even if there is a remaining algebraic dependency:
\begin{prop}[\cite{Janssen_1983,Aguilar_Cichy_Eisert_Bittel_2024}]\label{prop:Limits_Lie_Algebraic_Dependencies_canonical}
Let $\vgens$ be a generating set whose frustration graph $\graphG$ belongs to
one of the canonical graph families listed in
Theorems~\ref{thm:classes:humphries} or \ref{thm:classes:eisert}, and let
$\tilde{\vgens}$ be its unique linearly independent extension with projection map
$\projection$ and $\tilde{V}=\SpanS[\F_2]{\tilde{\vgens}}$.
Moreover,
\begin{equation}\label{eq:canonical-dependency-B}
B:=\Span[\F_2]{b_j+b_{j+1}\text{ for } 1\leq j<n_1}\subseteq\tilde V,
\end{equation}
where $b_j$ denotes the vector label of the $j$th length-one leaf.
For $\tilde{v}\in\ker(\projection)$, viewed as an algebraic dependency of
$\isolong{\vgens}$ through the projection $\projection$, we have:
\begin{enumerate}
    \item If $\tilde{v}\in B$, then the
    corresponding algebraic dependency is Lie-algebraic.
    \item If $\tilde{v}\notin B$, then the
    corresponding algebraic dependency is not Lie-algebraic.
\end{enumerate}
In particular, $\vgens$ is minimal if and only if
$\ker(\projection)\cap B=\{0\}$.
Furthermore, if $\vgens$ is minimal, then it has at most one algebraic dependency.
\end{prop}

Before we present the proof of Proposition~\ref{prop:Limits_Lie_Algebraic_Dependencies_canonical},
we first provide some condition which will be used in the proof.

\begin{lem}\label{lem:local_path_orbit_interval}
Consider a generating set $\vgens$ and vertices
$\vva_1,\ldots,\vva_h$ of $\frustration{\vgens}$ such that the induced
subgraph $\frustration{\vgens}[\{\vva_1,\ldots,\vva_h\}]$ is the path with
edges $\{\vva_o,\vva_{o+1}\}$ for $1\leq o<h$.
Write $v_o$ for the binary vector labeling $\vva_o$.
For the restricted transvection action generated by $v_1,\ldots,v_h$,
the orbit of $v_{\tilde{o}}$ consists exactly of the non-zero vectors
$$
v_i+v_{i+1}+\cdots+v_j \;\text{ for }\; 1\leq i\leq \tilde{o}\leq j\leq h .
$$
Equivalently, these are exactly the non-zero vectors whose supports
inside this induced path are intervals containing $\vva_{\tilde{o}}$.
For this restricted path action, a vector whose support inside the path is
not an interval cannot lie in the orbit of a single path vertex.
\end{lem}
\begin{proof}
This is the color-flip description of transvection orbits from Lemma~\ref{lem:coloring:orbits}, specialized to the induced path
$\frustration{\vgens}[\{\vva_1,\ldots,\vva_h\}]$ and to the transvections generated by its vertices.
For a subset $C\subseteq\{\vva_1,\ldots,\vva_h\}$, write
$x_C=\sum_{\vva_{\hat{o}}\in C}v_{\hat{o}}$ for the corresponding vector.
The transvection generated by $v_o$ gives
$$
\tau_{v_o}(x_C)=x_C+\symp{v_o}{x_C}v_o
=x_C+v_o  \sum_{\vva_{\hat{o}}\in C}\symp{v_o}{v_{\hat{o}}}.
$$
Since $\symp{v_o}{v_{\hat{o}}}=1$ precisely when $\{\vva_o,\vva_{\hat{o}}\}$ is an edge of the induced path, this toggles $\vva_o$ precisely when $\vva_o$ has an odd number of neighbors in $C$ inside the path.
If $C$ is an interval $\{\vva_i,\ldots,\vva_j\}$, then among the existing path vertices only $\vva_{i-1}$, $\vva_i$, $\vva_j$, and $\vva_{j+1}$ can be toggled.
Toggling $\vva_{i-1}$ or $\vva_{j+1}$ extends the interval by one vertex, while toggling $\vva_i$ or $\vva_j$ removes an endpoint.
All other path vertices either have no neighbor in $C$ or two neighbors in $C$ and hence do not change the support.
Thus interval supports are preserved.
Starting from the singleton interval $\{\vva_{\tilde{o}}\}$, this proves by induction that every support reached from $v_{\tilde{o}}$ is an interval.
Conversely, one obtains any interval containing $\vva_{\tilde{o}}$ by successively extending to the left and to the right.
\end{proof}

Proposition~\ref{prop:Limits_Lie_Algebraic_Dependencies_canonical}
also follows from general results discussed later in this work
(see Proposition~\ref{prop:valid_colorings_even_odd_cases_alg_ind} and
Proposition~\ref{prop:valid_colorings_orbits_line_graph_blown_up_path_graph_alg_dep_exceptional_case}
for the blown-up path cases, and Theorem~\ref{thm:classification_quasi_universal_algebraically_independent_E6}
together with Corollary~\ref{cor:quadratic_form_euler_characteristic} for the other cases),
but we now provide a direct proof:

\begin{proof}[Proof of Proposition~\ref{prop:Limits_Lie_Algebraic_Dependencies_canonical}]
We first prove (a).
For $\hat{o}\geq1$, assume that $\tilde{v}=b_{i_1}+\cdots+b_{i_{2\hat{o}}}$ is supported on an even number of length-one leaves.
Writing $v_0$ for their common neighbor, the relation
$\projection(\tilde{v})=0$ gives
\begin{align*}
        \projection(b_{i_1})
        &= \projection(b_{i_2}+\cdots+b_{i_{2\hat{o}}})\\
        &= \projection(v_0+b_{i_2}+\cdots+b_{i_{2\hat{o}-1}}+v_0+b_{i_{2\hat{o}}})\\
        &= \projection(\tau_{v_0}[\prod_{j=2}^{2\hat{o}-1}\tau_{b_{i_j}}]\tau_{v_0}b_{i_{2\hat{o}}}).
\end{align*}
Among the vectors appearing in the transvection word on the last line and its target,
exactly $b_{i_1},\ldots,b_{i_{2\hat{o}}}$ occur an odd number of times, since $v_0$ appears twice.
Thus Lemma~\ref{lem:dependency}(d) shows that the corresponding algebraic dependency is Lie-algebraic.

For (b), it remains to exclude the algebraic dependencies not generated by $B$.
Ignoring the pair vectors in $B$,
Lemma~\ref{lem:Canonical_t_equivalent_Graph_Radical} implies that
the only remaining possibilities are the alternating vector
$\sum_{j=1}^{m+1}a_{2j-1}$ for $\graphP_{2m,n_1}$, the vector
$a_1+a_2+a_4$ for $\graphX_{2m,n_1}^3$, and the vector
$a_1+a_3+b_1$ for $\graphS_{n_2,n_1}^3$.
For the first case, the only vertices outside the spine that can occur are the length-one leaves.
Up to $B$, these length-one leaves contribute only through their total
parity at the common neighbor and do not provide additional spine vertices.
Thus a Lie-algebraic dependency would require the alternating spine support with one colored spine vertex removed to lie in the orbit of a single spine vertex under the restricted spine action.
This is impossible by Lemma~\ref{lem:local_path_orbit_interval}, since that support is not an interval on the spine path.

For the two remaining exceptional vectors, we use a full-graph invariant, so the color-flip sequence may use vertices outside the displayed support.
By Lemma~\ref{lem:euler_characteristic_color_flip_invariant}, the Euler characteristic of the colored induced subgraph is unchanged modulo $2$ under the color-flip dynamics.
A single vertex has odd Euler characteristic.
However, after deleting any one colored vertex from $a_1+a_2+a_4$ or from $a_1+a_3+b_1$, the remaining two colored vertices are non-adjacent.
Their induced subgraph therefore has Euler characteristic $2=0\bmod 2$.
Thus no vector with one of these supports can be reached from a single
vertex by any sequence of color flips, even when the sequence is allowed to use
vertices outside the displayed support.
Thus no algebraic dependency outside $B$ is Lie-algebraic.

The minimality criterion follows because a non-zero element of
$\ker(\projection)\cap B$
is precisely a Lie-algebraic dependency that can be eliminated.
After quotienting by $B$,
Lemma~\ref{lem:Canonical_t_equivalent_Graph_Radical} leaves at most one additional radical generator.
Hence a minimal generating set has at most one algebraic dependency.
\end{proof}

\begin{figure}
    \centering
    \includegraphics{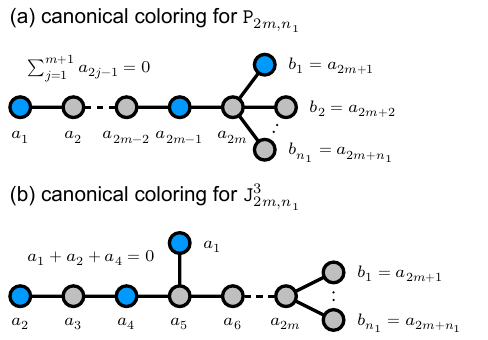}
    \caption{
    Canonical coloring for the algebraic dependency of minimal generating sets whose frustration graph is $\graphP_{2m,n_1}$ and $\graphX_{2m,n_1}^3$.}
    \label{fig:canonical-dependency-coloring}
\end{figure}

Thus Proposition~\ref{prop:Limits_Lie_Algebraic_Dependencies_canonical}
shows that removing Lie-algebraic dependencies from a set may only send the canonical $t$-representative graph
into a representative with a smaller number of length-one leaves.
Moreover, checking for Lie-algebraic dependencies in the graphs of canonical form becomes particularly easy, since
it reduces to checking whether the dependency lies in the subspace $B$ from
Eq.~\eqref{eq:canonical-dependency-B}.
For minimal generating sets, which have a single algebraic dependency, there exists a series of contractions such that only a single length-one leaf
is colored as detailed in the proof of Lemma~\ref{lem:Canonical_t_equivalent_Graph_Radical} (see also \cite{Aguilar_Cichy_Eisert_Bittel_2024}).
For the cases $\graphP_{2m,n_1}$ and $\graphX_{2m,n_1}^3$ we can then
choose a canonical coloring for the algebraic dependency, as illustrated in
Figure~\ref{fig:canonical-dependency-coloring}.
The case $\graphS_{n_2,n_1}^3$ is handled in the proof above by the same
Euler-characteristic obstruction as $\graphX_{2m,n_1}^3$.

Finally, we complement this result with the fact that a linearly dependent set with fixed algebraic dependencies is also uniquely determined, as briefly discussed in Section~\ref{sec:absence}.
In particular, a set with \emph{any} prescribed algebraic dependencies may be realized, when those dependencies are specified by colorings on the frustration graph (which need not be connected).
We formalize this in the following lemma:
\begin{lem}\label{lem:bijection_graphs_linearly_dependent_sets_with_fixed_algebraic_dependencies}
Let $\graphG$ be a graph, $X$ a set of colorings over $\graphG$ and $\tilde{\vgens}$ the unique linearly independent set over $\graphG$ in $\Fn$ with $n\geq\rank(A(\graphG))/2+\nullity(A(\graphG))$.
If the realizations $\coltovec_{\tilde{\vgens}}(\coloring)$, for
$\coloring\in X$, lie in $\rad(\tilde{\vgens})$, there is a linearly dependent set $\vgens = \{v_i\}_{i=1}^{\abs{\vertices(\graphG)}} \subseteq\Fn$ such that $\vgens$ has frustration graph $\frustration{\vgens} = \graphG$ and algebraic dependencies specified by $X$, i.e., $\coltovec_{\vgens}(\coloring)=\sum_i\coloring(i)v_i = 0$ for each coloring $\coloring\in X$. This is also unique up to isomorphism in $\Fn$.
\end{lem}
\begin{proof}
Using unique extensions, set $\tilde{V} = \SpanS{\tilde{\vgens}}$.
Define
\[
R:=\SpanS{\{\coltovec_{\tilde{\vgens}}(\coloring)\text{ for } \coloring\in X\}}
\subseteq\rad(\tilde{V})
\]
and let $\projection\colon \tilde{V} \to \tilde{V}/R$ be the quotient map.
Set $V:=\projection(\tilde{V})$ and
$\vgens=(v_i)_{i\in\vertices(\graphG)}$, where
$v_i:=\projection(\tilde{v}_i)$.
Choosing a supplementary subspace to $R$ in $\tilde{V}$ identifies the
quotient $V=\tilde{V}/R$ with a subspace of $\tilde{V}$.
Hence, after this choice, $V$ may be viewed as a subspace of $\Fn$.
By definition of $\projection$, $\vgens$ satisfies the algebraic dependencies defined by $\coloring\in X$:
for each $\coloring\in X$,
\begin{equation}
    \coltovec_{\vgens}(\coloring)
    =
    \sum_i\coloring(i)v_i
    =
    \projection\!\left(\coltovec_{\tilde{\vgens}}(\coloring)\right)
    =0,
\end{equation}
because $\coltovec_{\tilde{\vgens}}(\coloring)\in R$.
Since $R\subseteq\rad(\tilde V)$, the symplectic form on $\tilde V$
induces a well-defined symplectic form on $V=\tilde V/R$, i.e.,
\begin{equation}
    \symp{v+R}{v'+R}:=\symp{v}{v'}.
\end{equation}
With this induced form, the projected sequence $\vgens$ has the same
frustration graph as $\tilde{\vgens}$. Via the chosen identification of
$V$ with $\tilde V/R\subseteq\tilde V\subseteq\Fn$, $\vgens$ may be regarded as
a sequence in $\Fn$.

For uniqueness, let $\vgens'=(v_i')_{i\in\vertices(\graphG)}$ be another
generating sequence with the same frustration graph such that the colorings in
$X$ impose its algebraic dependencies, i.e.\
\[
\coltovec_{\vgens'}(\coloring)
=
\sum_i\coloring(i)v_i'=0\quad\text{for all }\coloring\in X.
\]
Choose linearly independent subsequences $\vecbasempty\subseteq\vgens$ and
$\vecbasempty'\subseteq\vgens'$ spanning $V$ and $V'$.
Then $g\vecbasempty=\vecbasempty'$ for some $g\in\Sp(2n,\F_2)$ because the
two bases have the same frustration graph.
The relations indexed by $X$ express every generator outside
$\vecbasempty$ in the same coordinates relative to $\vecbasempty$, so $g$
extends to the full sequences and gives $\vgens'=g\vgens$.
\end{proof}
Hence, specifying a graph together with algebraic dependencies is sufficient to completely determine its algebraic properties (e.g.\ its Pauli Lie algebra and transvection group), up to isomorphism.
Furthermore, this also confirms both existence and uniqueness of the minimal generating sets specified by Proposition~\ref{prop:Limits_Lie_Algebraic_Dependencies_canonical}.
Assuming we have a minimal generating set (without Lie-algebraic dependencies), we find that there are $6$ different classes of Pauli Lie algebras or transvection groups generated by connected frustration graphs.
By Lemma~\ref{lem:Bijection_Graphs_Linearly_Independent_Sets} and \ref{lem:bijection_graphs_linearly_dependent_sets_with_fixed_algebraic_dependencies}, these are also well defined and always admit a representation in $\PP_n$.
Let us denote with $\vgens(\graphG)\subseteq\Fn$ the unique algebraically
independent set with frustration graph $\graphG$.
For $\graphG\in\{\graphP_{2m,n_1},\graphX_{2m,n_1}^3\}$, we denote by
$\vgens^D(\graphG)\subseteq\Fn$ the generating set with the
unique algebraic dependency which is not a Lie-algebraic dependency.

\begin{cor}\label{cor:minimal_canonical_sets}
Let $\vgens\subseteq\Fn$ be a set of binary vectors with connected frustration graph so that $\vgens$ is also a minimal generating set.
Then $\vgens$ is $t$-equivalent to exactly one of the following cases (using the classes from Theorem~\ref{thm:classes:humphries}):
$\vgens(\graphP_{k,n_1})$, $\vgens(\graphX_{2m-1,n_1}^1)$, $\vgens(\graphX_{2m-1,n_1}^2)$,
$\vgens(\graphX_{2m,n_1}^3)$, $\vgens^D(\graphP_{2m,n_1})$, and $\vgens^D(\graphX_{2m,n_1}^3)$.
\end{cor}

Then, two tasks remain before a full classification for a generic generating set.
First, whenever a minimal generating set is given, completely determine the corresponding transvection group, Lie algebra and orbits.
Specifically, this becomes particularly easy when the graphs are changed to one of the representatives.
This was approach in \cite{Janssen_1983} for transvection groups with connected generating sets, and in \cite{Aguilar_Cichy_Eisert_Bittel_2024} for Pauli Lie algebras.
In the next section we will also look at graph-theoretic approaches to deal with specific algebraic dependencies.

Second, given a non-minimal generating set, determine which is the corresponding minimal generating set, hence the class of the given set. One option, which was used by \cite{Aguilar_Cichy_Eisert_Bittel_2024}, is to systematically remove all Lie-algebraic dependencies, by first changing the graph to its $t$-equivalent form.
Alternatively, we shall consider criteria to determine this class \emph{without} the use of contractions, as well as to identify possible Lie-algebraic dependencies (see Section~\ref{sec:Lie_Algebraic_Dependencies_in_general}), thus generalizing Proposition~\ref{prop:Limits_Lie_Algebraic_Dependencies_canonical}.
Namely, we will take advantage of its algebraic $t$-invariant properties, using either \emph{symmetry} property of the generators (see Corollary~\ref{cor:distinguishing_quasi_universal_cases}) or, certain $t$-invariant graph-theoretic properties (see Corollary~\ref{cor:distinguishing_line_graph_cases}).

\subsection{Twins and (Lie-)Algebraic Dependencies\label{sec:Twins}}

We now introduce an additional graph-theoretic tool to address the study of radicals and Lie-algebraic dependencies.
Namely, let us introduce the notion of \emph{twins}, which was also studied in the context of free-fermion solvability \cite{Chapman_Flammia_2020,Ruh_Elman_2025}:
\begin{defn}[Twin Vertices (\cite{Cuypers_2021_E6})]
Let $\graphG$ be a simple graph. We say that two vertices are \emph{false twins} if they share the same neighbors (hence they are not adjacent). We denote with $\sim$ the equivalence relation given by vertices which are false twins.
A graph if twin-free if there are no false twins.

We also define the graph $\modout{\graphG}$ as the graph whose vertices are equivalence classes under $\sim$, such that classes are adjacent in $\modout{\graphG}$ if any of their representative vertices are adjacent in $\graphG$.
\end{defn}
Similarly, we say that two elements in a generating set $\pgens$ (or $\vgens$) are false twins if they are false twins in the frustration graph $\frustration{\pgens}$.
Clearly, we see that all twin-free graphs in the canonical classes are precisely those with $n_1=1$, since only the legs of length one provide twins in these graphs.
Hence, natural questions to ask are whether (1) \emph{all} twin-free graphs have a radical of dimension at most $1$ and (2) being twin-free is invariant under $t$-equivalence.

We provide simple examples to show that both statements are false in general. These limit the usefulness of twins in understanding (Lie-)algebraic dependencies. However, they still provide useful tools which we shall use later in the classification.
Regarding question (1), consider the cycle graph $\graphC_6$ on six vertices, which is twin-free. The radical is two dimensional and spanned by the vectors realized by the two colorings supported on alternating vertices of the cycle.
Then, regarding question (2), consider the path graph $\graphP_3$, whose extremal vertices are twins. A valid contraction of the center vertex onto one of the extremal ones
yields the cycle graph $\graphC_3$, which is twin-free.

First, we provide some basic properties of the connection between twins and the radical:
\begin{lem}[\cite{Cuypers_2021_E6}]\label{lem:twins:basic_facts}
Let $\vgens\subseteq\Fn$ be some generating sequence (or set) with frustration graph $\graphG$, and which spans a subspace $V$. Consider the quotient $\overline{V} = V/\rad(V)$ (which consists of the affine spaces $v+\rad(V)$) with the projection map $\projection\colon v\in V\mapsto \tilde{v}\in \overline{V}$. Also, let $\tilde{\vgens}$ be the subsequence (or subset) of $\vgens$ obtained by choosing one representative from each false-twin equivalence class.
\begin{enumerate}
    \item Two $v_1,v_2\in\vgens$ are twins iff $v_1+v_2\in\rad(V)$.
    \item The sets $\overline{\vgens} = \projection(\vgens)$ and $\tilde{\vgens}$ have frustration graph $\modout{\graphG}$.
    \item The set $\tilde{\vgens}$ is an extension of $\overline{\vgens}$ (see Def.~\ref{def:extension_generating_set_subspace_projection}).
    \item The map $p$ induces a bijection between $\vgens$ and $\overline{\vgens}$ (hence their frustration graphs are isomorphic) if and only if $\vgens$ is twin-free.
\end{enumerate}
\end{lem}
\begin{proof}
By definition of projection map, we have that $\projection(v_1)=\projection(v_2)$ (for $v_i\in V$) iff there are some $u_1,u_2$ such $v_1 + u_1 = v_2 + u_2$ or equivalently $v_1+v_2\in\rad(V)$.
Hence, two elements $v_{i_1},v_{i_2}$ in $\vgens$ are mapped to the same element if and only if $v_{i_1}+v_{i_2}\in\rad(V)$, or equivalently
\begin{equation*}
        \symp{v_{i_1}+v_{i_2}}{v_j} = 0 \;\text{ or }\;
        \symp{v_{i_1}}{v_j} = \symp{v_{i_2}}{v_j}
\end{equation*}
for all $v_j\in\vgens$. Hence, $v_{i_1}$ and $v_{i_2}$ have the same neighbors and are (false) twins, which proves (a).
Then, $p$ sends two vectors $v_{i_1},v_{i_2}$ into the same coset $v+\rad(V)$ if and only if they are twin vertices in the frustration graph of $\vgens$.
Hence, the set $\overline{\vgens}$ has frustration graph $\modout{\graphG}$, which proves (b).

Since $p$ acts injectively on vectors in distinct equivalence classes (either as cosets by $\rad(V)$ in $V$, or as twins in $\vgens$), it is invertible on a twinless subsequence $\tilde{\vgens}$, hence bijective to $\overline{\vgens}$. Since $p$ is also linear and symplectic, it provides an isomorphism, which proves (c), as well as (d) as an immediate corollary.
\end{proof}

We highlight how statement (b) in Lemma~\ref{lem:twins:basic_facts} provides a simple condition for checking \emph{trivial} algebraic dependencies from the graph,
i.e., pairs of twins that are actually identical, since
$v_{i_1}+v_{i_2}=0$ iff $v_{i_1}=v_{i_2}$.
These are also Lie-algebraic dependencies.
As a consequence, the differences of false twins span a subspace of
$\rad(V)$.
This generalizes the basis in
Lemma~\ref{lem:Canonical_t_equivalent_Graph_Radical}\ref{lem:Canonical_t_equivalent_Graph_Radical:length-one-leaves}.
Specifically, after ordering each false-twin equivalence class as
$u_1,\ldots,u_{\twincount}$, this subspace
has a basis given by the consecutive sums
$u_a+u_{a+1}$ for $1\leq a<\twincount$, over all classes with $\twincount\geq 2$.
Also, for any algebraic dependency coming from even sets of twins in the same
equivalence class, we can generalize the argument for
Proposition~\ref{prop:Limits_Lie_Algebraic_Dependencies_canonical}(a) by
constructing a Lie-algebraic dependency from an algebraic dependency which
involves an even number of twins in the same equivalence class.
Indeed, one simply replaces $v_0$ with any vector that is adjacent to all
vectors in the equivalence class.

The previous Lemma~\ref{lem:twins:basic_facts} identifies false-twin classes with generators that
coincide after projection to the radical quotient.
The next statement provides the corresponding transvection-group kernel when a
projection collapses one false-twin class.
We use the notation from
Lemma~\ref{lem:projection_extension_basic_properties}\ref{lem:projection_extension_basic_properties:d}: the projection $\projection$ from
$\tilde{V}$ to $V$ induces a surjective homomorphism
$\tvproj$ from
$\tvgroup{\tilde{\vgens}}$ to $\tvgroup{\vgens}|_V$, defined by
$\tvproj(\tilde g)(v)=\projection(\tilde g\tilde v)$ whenever
$\projection(\tilde v)=v$.
We write $\tvprojker:=\ker(\tvproj)$ for its kernel.

\begin{lem}[Kernel for collapsing one false-twin class {\cite[Proposition~6.6]{Brown_Humphries_1986a}}]\label{lem:one_generator_blowup_transvection_kernel_candidate}
Let $\tilde{\vgens}\subseteq\tilde{V}$ be an extension of
$\vgens\subseteq V$ in the sense of
Definition~\ref{def:extension_generating_set_subspace_projection}, with
projection $\projection$.
Assume that the projection collapses exactly one false-twin class
$\{\tilde v_1,\ldots,\tilde v_{\twincount}\}\subseteq\tilde{\vgens}$ to a
generator $v_0\in\vgens$, i.e.,
$\projection(\tilde v_a)=v_0$ for all $a$, and is injective on the remaining
generators.
Assume moreover that the kernel of $\projection$ is generated by the
differences inside this false-twin class,
\[
    W:=\ker(\projection)
    =
    \SpanS[\F_2]{\{\tilde v_a+\tilde v_1\}_{a=2}^{\twincount}}.
\]
Assume that the cosets in the orbit of $v_0+\rad(V)$ under
$\tvgroup{\vgens}|_V$ span $V/\rad(V)$.
Equivalently, this holds if $\frustration{\vgens}$ is connected, since then
$\tvgroup{\vgens}\cdot\vgens=\tvgroup{\vgens}\cdot v_0$ and the orbit of
$v_0$ spans $V$.
Then $\tvprojker$ is elementary abelian and is naturally
identified with the additive group of $\F_2$-linear maps from
$V/\rad(V)$ to $W$.
In particular,
\[
    \dim \tvprojker
    =
    \dim(V/\rad(V))\dim(W).
\]
\end{lem}
\begin{proof}
This proof follows the argument of
\cite[Proposition~6.6]{Brown_Humphries_1986a}, specialized to the case where
one false-twin class is collapsed by the projection and written in the notation
of Definition~\ref{def:extension_generating_set_subspace_projection}.
Lemma~\ref{lem:projection_extension_basic_properties}\ref{lem:projection_extension_basic_properties:d} gives the induced
surjective homomorphism and identifies its kernel as the subgroup acting
trivially after applying $\projection$.
Thus every element $\tilde g\in\tvprojker$ determines a correction map
$\lambda_{\tilde g}$ by
\[
    \tilde g x=x+\lambda_{\tilde g}(x)
    \qquad \text{with }
    \lambda_{\tilde g}(x):=\tilde g x+x.
\]
Since $\projection\circ\tilde g=\projection$, we have
\[
    \projection(\lambda_{\tilde g}(x))
    =
    \projection(\tilde g x+x)
    =
    \projection(x)+\projection(x)
    =
    0.
\]
Thus $\lambda_{\tilde g}(x)\in W=\ker(\projection)$ for all
$x\in\tilde{V}$.
Moreover, $W\subseteq\rad(\tilde{V})$ by
Lemma~\ref{lem:projection_extension_basic_properties}\ref{lem:projection_extension_basic_properties:a}, and every
transvection in $\tvgroup{\tilde{\vgens}}$ fixes $\rad(\tilde{V})$ pointwise.
Thus $\lambda_{\tilde g}$ vanishes on $\rad(\tilde{V})$.
Since $W=\ker(\projection)\subseteq\rad(\tilde{V})$, the value
$\lambda_{\tilde g}(\tilde{x})$ depends only on
$x=\projection(\tilde{x})\in V$.
This defines a linear map from $V$ to $W$.
Lemma~\ref{lem:projection_extension_basic_properties}\ref{lem:projection_extension_basic_properties:a} also gives
$\projection(\rad(\tilde{V}))=\rad(V)$, so this map vanishes on $\rad(V)$.
It therefore induces a linear map from $V/\rad(V)$ to $W$.
The assignment $\tilde g\mapsto\lambda_{\tilde g}$ is injective, since
$\lambda_{\tilde g}=0$ implies $\tilde g=\id_{\tilde{V}}$.
It is a group homomorphism to the additive group of such linear maps, because
every element of $\tvprojker$ fixes $W$ pointwise.

For $w_a:=\tilde v_a+\tilde v_0\in W$, we have
$\tilde v_a=\tilde v_0+w_a$ with $w_a\in\rad(\tilde{V})$.
Hence
\[
    \symp{\tilde v_a}{x}
    =
    \symp{\tilde v_0+w_a}{x}
    =
    \symp{\tilde v_0}{x}.
\]
Also $\symp{\tilde v_a}{\tilde v_0}=0$, since
$w_a=\tilde v_a+\tilde v_0$ is radical and the symplectic form is alternating.
Moreover, $\tau_{\tilde v_a}$ and $\tau_{\tilde v_0}$ have the same image under
$\tvproj$, so $\tau_{\tilde v_a}\tau_{\tilde v_0}\in\tvprojker$.
A direct expansion gives
\[
\begin{aligned}
    \tau_{\tilde v_a}\tau_{\tilde v_0}(x)
    &=
    x+\symp{\tilde v_0}{x}\tilde v_0
    +
    \symp{\tilde v_a}{x+\symp{\tilde v_0}{x}\tilde v_0}\tilde v_a\\
    &=
    x+\symp{\tilde v_0}{x}(\tilde v_0+\tilde v_a)\\
    &=
    x+\symp{\tilde v_0}{x}w_a .
\end{aligned}
\]
Hence all these products realize rank-one maps from $V/\rad(V)$ to $W$.
By the surjectivity of $\tvproj$, every element of $\tvgroup{\vgens}|_V$ has a
lift to $\tvgroup{\tilde{\vgens}}$.
Conjugating by such lifts gives the analogous rank-one maps with $\tilde v_0$
replaced by representatives of the orbit of $v_0+\rad(V)$.
Let $\bar{V}:=V/\rad(V)$.
For an orbit coset $\bar u = u + \rad(V) \in\bar V$ and $w_a\in W$, the corresponding map is
\begin{equation}\label{eq:x:bar:map}
    \bar x\longmapsto \symp{\bar u}{\bar x}\,w_a = \symp{u}{x}\,w_a.
\end{equation}
The form induced by $\symp{\cdot}{\cdot}$ on $\bar V$ is non-degenerate.
Hence the assignment
$\bar u\mapsto(\bar x\mapsto\symp{\bar u}{\bar x})$ identifies $\bar V$ with
its dual space $\bar V^*$.
By the spanning hypothesis, the orbit cosets $\bar u$ span $\bar V$; therefore
the corresponding linear functionals span $\bar V^*$.
Since the vectors $w_a$ span $W$, the rank-one maps from Eq.~\eqref{eq:x:bar:map} span
$\bar V^*\otimes W\cong\operatorname{Hom}_{\F_2}(\bar V,W)$.
Thus every $\F_2$-linear map from $V/\rad(V)$ to $W$ is realized, giving the
claimed identification and dimension formula.
\end{proof}

Altogether, twins provide a graph-theoretic way to detect part of the
radical and some of the resulting algebraic and Lie-algebraic dependencies.
For the canonical blown-up path families, this recovers the radical vectors
coming from parallel legs; in general, it should be viewed as a useful local
test rather than a complete description of the entire radical.

\ManuscriptPart{Commutants and Invariant Forms}{part:commutants_invariant_forms}

Before delving into the classification of Lie algebra, transvection and orbits, we now introduce some fundamental \emph{symmetry} tools, which go alongside the graph-theoretic formalism described until now, which uses $t$-equivalence and coloring transformations to describe orbits/Lie algebras.
Namely, as we shall see, such a description will be sufficient to describe the generating sets whose graphs are $t$-equivalent to one of $\graphX_{2m-1,n_1}^1$, $\graphX_{2m-1,n_1}^2$ or $\graphX_{2m,n_1}^3$

\section{Commutants and Orthogonal Complements}\label{sec:commutant}

We start by recalling the definition of \emph{commutant} of a set of matrices $\pgens\subseteq\C^{d\times d}$ as the set of matrices which commute with it:
\begin{align}
        \commutant(\pgens) &= \cent_{\C^{d\times d}}(\pgens) \\
        &= \{ C\in\C^{d\times d} \text{ with } CG = GC \text{ for all } G\in\pgens \} \nonumber
\end{align}
as well as some basic facts about the commutant:
\begin{lem}[\cite{bresar2014,lorenz2008,jacobson1985,jacobson1989,lux2010,curtisreiner1962,curtisreiner1981}]\label{lem:general_commutant}
Consider a set of hermitian matrices $\pgens\subseteq\C^{d\times d}$, $H^\dagger\in\pgens$, and its generated matrix algebra $\matalg$. Then:
\begin{enumerate}
    \item\label{lem:general_commutant:a}
    $\commutant(\pgens) = \commutant(\matalg)$
    \item\label{lem:general_commutant:b}
    $\commutant(\matalg)$ is a matrix algebra
    \item\label{lem:general_commutant:c}
    $\commutant(\commutant(\matalg)) = \matalg$
    \item\label{lem:general_commutant:d}
    $\ZZ(\matalg) = \ZZ(\commutant(\matalg))$
    \item\label{lem:general_commutant:e}
    $\commutant(\matalg\cup\commutant(\matalg)) = \ZZ(\matalg)$
    \item\label{lem:general_commutant:f}
    If $\commutant(\matalg)$ has the decomposition \ref{eq:block_diagonal_decomposition_matrix_algebras}
    \begin{equation}
        \commutant(\matalg) \conjugated \bigoplus_{\lambda=1}^{\dim[\ZZ(\matalg)]} \C^{m_\lambda\times m_\lambda}\otimes\id_{d_\lambda}
    \end{equation}
    then $\matalg$ has the decomposition with exchanged dimensions and multiplicities
    \begin{equation}
        \matalg \conjugated \bigoplus_{\lambda=1}^{\dim[\ZZ(\matalg)]} \id_{m_\lambda}\otimes\C^{d_\lambda\times d_\lambda}
    \end{equation}
    and viceversa.
\end{enumerate}
\end{lem}
Condition (e) in particular allows us to use commutants of sets $\pgens$ to obtain information about the matrix algebras generated by $\pgens$, including its isomorphism class, as well as its invariant and irreducible subspaces.
This is especially useful when the commutant is small compared to the matrix algebra, $\dim(\matalg)\gg \dim(\commutant(\matalg))$, since it allows for more efficient computations or more compact descriptions of certain properties of the system.
Also, under unitary dynamics generated by some parametrized or controlled Hamiltonians $\{H_i\}\subseteq\C^{d\times d}$, the commutant provides all (linear) constants of motion $O\in\C^{d\times d}$, i.e.,
\begin{equation*}
    \expval{O(t)} = \expval{O(0)} \;\text{ iff }\; O\in\commutant(\{H_i\}).
\end{equation*}

In the context of Pauli matrix algebras, such a duality is further exemplified in the binary picture, where we have:
\begin{lem}\label{lem:pauli:commutant}
Consider a set of Pauli strings $\pgens\subseteq\PP_n$, which generates a Pauli matrix algebra $\matalg = \isolong{U}$ with $U\subseteq\Fn$. Then:
\begin{enumerate}
    \item $\commalg = \commutant(\matalg)$ is a Pauli matrix algebra
    \item $\bas{\commalg} = \commalg\cap\PP_n = \cent_{\PP_n}(\pgens)$
    \item $\inviso(\bas{\commalg}) = \{ v\in \Fn \mid \symp{v}{u} = 0,\,\forall u\in U\} $
\end{enumerate}
\end{lem}
\begin{proof}
The proof of (a) follows by the same argument used in the proof of Lemma~\ref{lem:pauli:matrix}(b), which trivially proves (b). Then (c) follow from the fact that that the commutator of two Pauli strings $\iso{v},\iso{w}$ is zero if and only if $\symp{v}{w} = 0$.
\end{proof}
Then, thank to Lemma~\ref{lem:pauli:commutant}(a), commutants of Pauli matrix algebras are also Pauli matrix algebras and possess all their properties.
The orthogonal complement of a set of binary vectors $\vgens\subseteq\Fn$ is the subspace that is orthogonal to it, i.e.,
\begin{equation}\label{eq:def:orthogonal_complement}
   \vgens^\perp = \{ v\in\Fn \text{ with } \symp{v}{w} = 0 \text{ for all } w \in\pgens\}
\end{equation}
which corresponds, by Lemma~\ref{lem:pauli:commutant}, to the Paulis which span the commutant of $\pgens=\isolong{\vgens}$.
Furthermore, the orthogonal complement respects a similar duality with subspace of its generators as the commutants of a set of matrices is dual to the matrix algebra (Lemma~\ref{lem:general_commutant}\ref{lem:general_commutant:c}):
\begin{equation}
    (\vgens^\perp)^\perp = V = \Span[\F_2]{\vgens}
\end{equation}
Also, we have the common radical $\rad(V\cup V^\perp) = \rad(V)$ (as in Lemma~\ref{lem:general_commutant}\ref{lem:general_commutant:d}) which also coincides with the common complement $(V\cup V^\perp)^\perp = \rad(V)$ (as in Lemma~\ref{lem:general_commutant}\ref{lem:general_commutant:e}).
Notice that a change of basis for $\vgens$ acts naturally on its orthogonal complement, i.e. $(g\vgens)^\perp = g\cdot\vgens^\perp$.

Furthermore, up to isomorphism, we can make use of the canonical representation of Pauli matrix algebras to show that there is the following duality:
\begin{subequations}\label{eq:canonical_form_pauli_matrix_algebra_and_commutant}
    \begin{align}
        \bas{\matalg} &= \PP_m\otimes\{I,Z\}^{\otimes r}\otimes I^{\otimes(n-m-r)} \\
        \bas{\commalg} &= I^{\otimes m}\otimes\{I,Z\}^{\otimes r}\otimes \PP_{n-m-r}
    \end{align}
\end{subequations}
Then, uncontrollable qubits in the matrix algebra correspond to non-abelian symmetries in the commutant, whereas the phase qubits correspond to the center of the commutant as well.
This provides a particularly simple case of the general commutant-matrix algebra duality in Lemma~\ref{lem:general_commutant}\ref{lem:general_commutant:f}.

Notice that a supplementary subspace $W$ of $K = \Span{V\cup V^\perp}$ in $\Fn$ has dimension $\nullity(\vgens)$. Also, there is a choice of $W$ and a symplectic basis such the combined basis of $W$ and $K$ is a symplectic basis for $\Fn$, hence in particular $W=\rad(W)$.
Indeed, one simply needs to consider a basis which includes a symplectic basis for $K$ and then use contractions of the basis of $K$ onto the other elements to obtain another symplectic basis.
Furthermore, given that $V\cap V^\perp = \rad(V)$, we can also define a non-degenerate subspace $\tilde{V}^\perp \cong V^\perp/\rad(V)$ which is supplementary of $\rad(V)$ in $V^\perp$.
Then, for a given subspace $V$ ($2m=\rank(V)$ (see Eq.~\eqref{eq:rank}), $r=\dim\rad(V)$), we have the following (non-canonical) direct sum decomposition of the entire space and corresponding symplectic basis:
\begin{subequations}\label{eq:full_Fn_decomposition_symplectic_basis}
    \begin{align}
        \Fn &= V \oplus \tilde{V}^\perp \oplus W\\
        \text{and } \vecbasempty &= \vecbas{V}\cup\vecbas{\tilde{V}^\perp}\cup\vecbas{W} \text{ where} \\
        \vecbas{V} &= \{ e_i, f_i\}_{i=1}^m\cup\{h_j\}_{j=1}^r,\\
        \vecbas{\tilde{V}^\perp} &= \{\tilde{e}_j, \tilde{f}_j\}_{j=1}^{n-m-r},\\
        \vecbas{W} &= \{ \tilde{h}_j\}_{j=1}^r.
    \end{align}
\end{subequations}
In the Pauli string formalism, the bases correspond to generating sets for Pauli matrix algebras, and their union is a generating set for the full matrix algebra $\C^{2^n\times 2^n}$.
Explicitly, if we choose the matrix algebra as in Eq.~\eqref{eq:Canonical_Basis_Pauli_Matrix_Algebra}, we have:
\begin{subequations}\label{eq:full_Pn_Pauli_decomposition_single_qubit_basis}
    \begin{align}
        \bas{\matalg} &= \PP_m \otimes\{I,Z\}^{\otimes r}\otimes I^{\otimes (n-m-r)}\\
        \bas{\commalg} &= I^{\otimes m} \otimes\{I,Z\}^{\otimes r}\otimes \PP_{n-m-r}\\
        \bas{\isolong{W}} &= I^{\otimes m}\otimes\{I,X\}^{\otimes r}\otimes I^{\otimes (n-m-r)}
    \end{align}
\end{subequations}
which, combined, clearly generate the entire matrix algebra $\matalg(2^n,\C)$.

This will later be useful in the description of the action of transvection groups on the entire space, not just on the generated subspaces and its orthogonal complement.

\section{Quadratic and Bilinear Forms}\label{sec:quadratic_bilinear_forms_intro}

\subsection{Definitions and Basic Properties}\label{sec:quadratic_bilinear_definitions}

Alongside the commutant and orthogonal complement, we now discuss our second symmetry tool for the classification of Pauli Lie algebras.
We now provide a general overview of quadratic and bilinear forms, as well as their isometries, over general fields $\F$. Specifically, we shall highlight differences over $\F_2$ (for which $\characteristic(\F)=2$) and either $\F=\R$ or $\F=\C$ (for which $\characteristic(\F)=0$).
In our setting, we will mostly consider either $V=\C^{2^n}$, the Hilbert space over which our qubit system lives, or $V=\Fn$, the binary symplectic space over which the Pauli strings live.
Let us now give the two main definitions:
\begin{defn}[Bilinear and Quadratic Forms \cite{grove_2002}]
A bilinear form over a vector space $V$ over a field $\F$ is a function $\genbilempty\colon V\times V\mapsto \F$ which is linear in both its arguments. A bilinear form is said to be symmetric if $\genbil{u}{v}=\genbil{v}{u}$. A bilinear form is said to be skew-symmetric if $\genbil{u}{v}=-\genbil{v}{u}$. It is said to be alternating or symplectic if $\genbil{v}{v}=0$.

A quadratic form over a vector space $V$ over a field $\F$, is any scalar function $\QQ\colon V \mapsto \F$ such that $\genbilempty_\QQ(u, v)\equiv \QQ(u+v)-\QQ(u)-\QQ(v)$ is a bilinear form and $\QQ(\alpha v) = \alpha^2 \QQ(v)$ for any $\alpha\in\F,v\in V$. In particular, $\genbilempty_\QQ$ is symmetric.
\end{defn}
We also say that a vector is isotropic for $\QQ$ if $\QQ(v)=0$ and anisotropic otherwise.
A subspace $W\subseteq V$ is said to be isotropic if there is some non-zero $w\in W$ such that $\QQ(w)=0$ and $W$ is totally isotropic if $\QQ(w)=0$ for all $w\in W$.
Notice that there cannot be a totally anisotropic subspace, since $\QQ(0)=0$.

A bilinear (quadratic) form $\B$ ($\QQ$) is said to be degenerate if there is some non-zero vector $v$ such that $\genbil{v}{u}=0$ for all $u\in V$ ($\QQ(v)=0$), and it is called non-degenerate otherwise.

We start by discussing some general properties \cite{grove_2002,Bourbaki1959}.
We notice that an alternating bilinear form is always skew-symmetric. Over $\characteristic(\F)\neq 2$, we also have that a bilinear form is skew-symmetric if and only if it is alternating.
On the other hand, when $\characteristic(\F)=2$ (hence $+1=-1$), a symmetric bilinear form is skew-symmetric and viceversa, but a skew-symmetric bilinear form may not be alternating. Indeed, this is the case for the standard Euclidean product over $\F_2^d$, $\genbil{a}{b} = \sum_{i=1}^d a_ib_i$, which is skew-symmetric but not alternating $\genbil{a}{a} = \sum_{i=1}^d a_i^2 = \sum_{i=1} a_i$ and coincides with the parity of the non-zero coefficients (hence not always zero).

For a fixed basis (hence under the isomorphism $V\cong \F^d$), for any bilinear form $\genbilempty$ there exists a matrix $B$ such that $\genbil{u}{v}=u^TBv$, with matrix elements $B_{ij} = \genbil{\basel_i}{\basel_j}$. For any given matrix $B$ (and fixed basis for $V$), there is a unique bilinear form, which we denote by $\genbilempty_B$. We also refer to a matrix $B$ as bilinear form.
For instance, given the canonical basis $\{\basel_j\}_{j=1}^{2n}$ of $\Fn$ (with $X_j = \isolong{\basel_j}$ and $Z_j = \isolong{\basel_{n+j}}$), the matrix corresponding to the symplectic form $\symp{v}{w} = v^T\Omega w$ is the following, in block form
\begin{equation}
    \Omega = \mqty( 0_n & \id_n\\ \id_n & 0_n )
\end{equation}
where $\id_n$ is the identity matrix and $0_n$ is the zero matrix.
Over different bases, there are different matrices, which are related by \emph{congruence} $B' = \activemap^{-T} B\activemap^{-1}$ for $\activemap\in\GL(V)$.
We also notice that the form $\genbilempty_B$ is (skew-)symmetric if and only $B$ is (skew-)symmetric.

We can provide an alternative description of bilinear forms when $V\cong\C^d$ is a Hilbert space, equipped with a \emph{hermitian} form, i.e. a non-degenerate \emph{sesquilinear} form $\braket{\cdot}{\cdot}$ \cite{Uhlmann_2016}. We recall that anti-linear operators are maps $\calT\colon V\mapsto V$ such that $\calT(\alpha v + \beta w) = \alpha^*\calT(v) + \beta^*\calT(w)$ where $\alpha^*$ is the complex conjugate of $\alpha\in\C$. Also, anti-unitary operators are those such that $\braket{\calT u}{\calT v} = \braket{u}{v}^* = \braket{v}{u}$ for all $u,v\in V$.
For a fixed basis $\{\basel_i\}$ over $V\cong\C^d$ (hence the canonical hermitian form), let $\calK$ be the conjugation operator (acting as $\calK(\basel_i)=\basel_i$ and extended by anti-linearity, such that $\calK^2 = \id$), then we can define the anti-linear operator $\calT = \calK \cdot B$ such that:
\begin{equation}
    \genbil{u}{v} = u^TBv = \mel{u}{\calK B}{v}^* = \mel{u}{\calT}{v}^*
\end{equation}
Viceversa, we can also write the hermitian product as bilinear-like $\braket{u}{v} = u^T\calK v$ (in the canonical basis).
Whenever $B$ is unitary $B^\dagger = B^{-1}$, $\calT = \calK B$ is anti-unitary, which connects to the typical physics language.
Furthermore, we have in this case that $B$ is symmetric iff $B^*B = \id$ iff the anti-unitary operator squares to the identity $\calT^2 = \id$, and $B$ is skew-symmetric iff $B^*B = -\id$ iff the anti-unitary operator squares to $\calT^2=-\id$.

Over a field of characteristic $\neq 2$, a symmetric bilinear form $\B$ determines a unique quadratic form as $\QQ_\B(v) = \genbil{v}{v}$. Hence, there is a unique correspondence between quadratic forms and symmetric bilinear forms when $\characteristic(\F)\neq 2$.
For a fixed basis and arbitrary field, we can also represent a quadratic form using a (possibly non-unique) matrix $B$, as $\QQ(v) = v^TBv$, which results in the symmetric bilinear form $\B_{\QQ} = \genbilempty_{B+B^T}$ with the symmetrized matrix $B+B^T$. Also, over $\characteristic(\F)\neq 2$, $\QQ$ is completely determined by the symmetric matrix $B+B^T$.

In the binary symplectic case $V=\Fn$, we shall specifically consider only quadratic forms over the canonical non-degenerate symplectic form, i.e. such that $\QQ(u+v)+\QQ(u)+\QQ(v) = \symp{u}{v}$.
In this case, one can again give a matrix representation of the quadratic form as $\QQ(v) = v^TBv$ where $B$ \cite{Sjöstrand_2025} is lower-triangular.
However, since (skew)symmetric bilinear forms over $\characteristic(\F)=2$ do not uniquely determine a quadratic form, $B+B^T$ (or the symmetric form determine by $\QQ$) does uniquely define a quadratic form $\QQ$, as one also needs the diagonal elements of $B$, $B_{ii} = \QQ(\basel_i)$.
On the other hand, the (skew)symmetric bilinear form defined by $\QQ$ over $\characteristic(\F)=2$ is also alternating, since $\QQ(v+v)-\QQ(v)-\QQ(v)=0$, hence always symplectic.
When treating quadratic forms over the binary symplectic space we shall not consider explicit matrices representing quadratic forms, since it will suffice to only give certain diagonal elements, given that we always fix a reference symplectic form $\QQ(u+v)+\QQ(u)+\QQ(v) = \symp{u}{v}$.

\subsection{Isomorphism Classes for (Skew-)Symmetric Bilinear Forms and Quadratic Forms}\label{sec:isomorphism_classes_bilinear_quadratic_forms}

We say that two bilinear forms $\genbilempty,\genbilempty'$ are isomorphic if and only if their matrices $B,B'$ (over bases) are congruent.
Thus a congruence is written as $\activemap^{-T} B\activemap^{-1} = B'$ for some invertible linear map $\activemap\in\GL(V)$ acting on vectors,
such that $\genbilempty'(\activemap u,\activemap v) = \genbil{u}{v}$ for all $u,v\in V$.

Similarly, two quadratic forms $\QQ,\QQ'$ over $V$ are said to be isomorphic if their coordinate expressions are related by such a change of basis, or equivalently if there is an invertible linear map $\activemap\in\GL(V)$ acting on vectors such that $\QQ'(\activemap v) = \QQ(v)$ for all $v\in V$, or $\QQ'=\QQ\circ\activemap^{-1}$.
In the binary symplectic case of $\Fn$, we shall consider only isomorphisms which respect the canonical symplectic form, hence $\activemap\in\Sp(V)=\Sp(2n,\F_2)$.

Then, a classification of bilinear and quadratic forms amounts to a classification of \emph{isomorphism classes} (congruence classes) of matrices, and specifically finding canonical representatives, which depends on various features of the field, beyond just the characteristic.
Specifically, we recall the results for (skew-)symmetric bilinear forms/matrices over $\F=\R$, $\C$ and $\F_2$, as well as for quadratic forms (over $\F_2$, as otherwise the classification is equivalent to that of symmetric bilinear forms).
For skew-symmetric matrices $A^T=-A$ with zero diagonal over \emph{any} field \cite[§5]{Bourbaki1959}, the isomorphism class is determined purely by its (even) rank, $\rank(A) = 2m$, and size: there exists $\activemap\in\GL(d,\F)$ such that
\begin{equation}\label{eq:canonical_form_skew_symmetric_matrix_congruence}
    \activemap^{-T} A\activemap^{-1} = \underbrace{\mqty(0 & 1\\ -1 & 0) \boxplus \cdots \boxplus \mqty(0 & 1\\ -1 & 0)}_{m\text{ times}} \boxplus\, 0_{d-2m}
\end{equation}
with $0_n$ the $n\times n$ zero matrix.
When $\characteristic(\F)=2$ and $+1=-1$, we recover precisely the existence of a sympletic basis (Lemma~\ref{lem:symplectic_basis}), with respect to which the symplectic form (as a matrix) is equivalent to that of Eq.~\eqref{eq:canonical_form_skew_symmetric_matrix_congruence}.
This is for instance the case of adjacency matrices of graphs, which serve as the symplectic forms for the subspaces spanned by linearly independent sets in the case $\Fn$.
This fact was also used in \cite{Viswanathan_Adjoua_Feniou_Badreddine_Piquemal_2025} to understand invariants of frustration graphs of generating sets of Pauli strings under contractions.
Furthermore, when $\characteristic(\F)=2$, $A^T=-A=A$ and skew symmetric matrices and symmetric matrices coincide.
Notice that if we also consider (skew-)symmetric with non-zero diagonal, corresponding to non-alternating symmetric bilinear forms, then under equivalence we find a different representative, as detailed in \cite{Kim_Seo_2008}.

For symmetric matrices instead there are significant differences over different fields, with $\characteristic(\F)\neq 2$ \cite[Chapter~4]{grove_2002}.
In the case $\F=\C$, we have that the isomorphism class of symmetric matrices $S^T=S$ is again purely determined by the rank $m$ (which need not be even), and size: there exists $\activemap\in\GL(d,\C)$ such that
\begin{equation}\label{eq:canonical_form_complex_symmetric_matrix_congruence}
    \activemap^{-T} S\activemap^{-1} = \mqty( \id_m \\ & 0_{d-m} )
\end{equation}
Correspondingly, complex quadratic forms in this basis will be of the form $\QQ(v) = \sum_{i=1}^m v_i^2$.
Over $\F=\R$, we have additional possibilities, related to the fact that $\im\not\in\R$. Namely, the isomorphism class is determined not just by the rank, but by the number of positive and negative eigenvalues (as well as size), $(p_+,p_-,p_0)$: there exists $\activemap\in\GL(d,\R)$ such that
\begin{equation}\label{eq:canonical_form_real_symmetric_matrix_congruence}
    \activemap^{-T} S\activemap^{-1} = \mqty( \id_{p_+} \\ & -\id_{p_-} \\ && 0_{d-m} )
\end{equation}
Correspondingly, complex quadratic forms in this basis will be of the form $\QQ(v) = \sum_{i=1}^{p_+}v_i^2 - \sum_{i=p_++1}^{p_-}v_i^2$.
In the non-degenerate case $p_0=0$, for symmetric bilinear forms or quadratic forms we also talk about the \emph{signature} $(p,q)=(p_+,p_-)$.

Finally, we give the classification for quadratic forms over $\F_2$, which does not reduce to that of symmetric bilinear forms.
The needed classification separates the two non-degenerate Arf
classes and the degenerate radical case, and is the form used in the
transvection literature; see
\cite[\S 2.13]{Janssen_1983} and \cite[\S 8.7]{Brown_Humphries_1986b}.
When $\QQ|_{\rad(V)}=0$, and
$\{e_i,f_i\}_{i=1}^m\cup\{h_j\}_{j=1}^r$ is a symplectic basis for
$V$, we use the Arf invariant
\begin{equation}\label{eq:arf_invariant_binary_quadratic_form}
    \Arf(\QQ)=\sum_{i=1}^m\QQ(e_i)\QQ(f_i).
\end{equation}
Since the zero vector $0$ lies in $\rad(V)$ and $\QQ(0)=0$, the
value $0$ always occurs on the radical; saying that $\QQ|_{\rad(V)}$ is not identically zero is
therefore equivalent to saying that its image is all of $\F_2$. We also
write $\abs{\QQ^{-1}(a)}$ for the number of vectors $v\in V$ with
$\QQ(v)=a$.
We state the classification now and give further references and a proof in
Appendix~\ref{app:binary-quadratic-forms}.
\begin{prop}[\cite{Bourbaki1959,Browder_1972}]\label{prop:Quadratic_Form_Isomorphism_Classes}
Let $\QQ$ be a quadratic form over a symplectic vector space $V$ with associated (possibly degenerate) symplectic form $\genbilempty_\QQ$. Let $\{e_i,f_i\}_{i=1}^m\cup\{h_j\}_{j=1}^r$ be a symplectic basis for $V$, with respect to $\genbilempty_\QQ$.
When $\QQ|_{\rad(V)}=0$, the Arf invariant in
Eq.~\eqref{eq:arf_invariant_binary_quadratic_form} is well-defined, and
independent of the choice of the symplectic basis.
Also, $\QQ$ belongs to one of the following isomorphism classes:
\begin{enumerate}
    \item $\QQ|_{\rad(V)} = 0$, $\Arf(\QQ) = 0$ (type $+$).
    \item $\QQ|_{\rad(V)} = 0$, $\Arf(\QQ) = 1$ (type $-$).
    \item $\QQ(\rad(V))=\F_2$ (type $0$), equivalently
    $\QQ|_{\rad(V)}$ is not identically zero.
\end{enumerate}
Equivalently, the type is determined by the \emph{typical} (an)isotropy of a vector:
\begin{enumerate}
    \item $\abs{\QQ^{-1}(0)} > \abs{\QQ^{-1}(1)}$ (type $+$).
    \item $\abs{\QQ^{-1}(0)} < \abs{\QQ^{-1}(1)}$ (type $-$).
    \item $\abs{\QQ^{-1}(0)} = \abs{\QQ^{-1}(1)}$ (type $0$).
\end{enumerate}
Each class has the following canonical representatives, respectively:
\begin{enumerate}[itemsep=0.1cm]
    \item $\QQ(v) = \sum_{i=1}^m \component{v}{i}\component{v}{m+i}$ (type $+$).
    \item $\QQ(v) = \sum_{i=1}^m \component{v}{i}\component{v}{m+i} + \component{v}{m}^2 + \component{v}{2m}^2$ (type $-$).
    \item $\QQ(v) = \sum_{i=1}^m \component{v}{i}\component{v}{m+i} + \component{v}{2m+1}^2$ (type $0$).
\end{enumerate}
\end{prop}

We now use the type labels of these equivalent
characterizations and package them in the following notation:

\begin{defn}[Type of a quadratic form]\label{def:type}
Proposition~\ref{prop:Quadratic_Form_Isomorphism_Classes} defines the \emph{type} of a quadratic form $\QQ$ as
\[
    \type(\QQ)\in\{+,-,0\},
\]
according to the common label of these equivalent conditions.
\end{defn}
In the case $\rad(V)=0$, then $\Arf(\QQ)$ is the unique invariant for isomorphism classes of quadratic forms.
Notice that a quadratic form with $\type(\QQ)=0$ is only well defined when the radical is non-trivial.
On the other hand, when the space is fully degenerate $V=\rad(V)$, the only quadratic form which is well-defined is one with $\type(\QQ)=0$.

Since the radical
[see Eq.~\eqref{eq:radical}]
$$\rad(\vgens) = \{ u\in\Span[\F_2]{\vgens} \text{ with } \symp{u}{v} = 0 \text{ for all } v\in\vgens\}$$
of a set of vectors $\vgens \subset \Fn$
plays a central role in the classification of quadratic forms,
we now study the action of a quadratic form $\QQ$ on it:
\begin{defn}[Isotropic and anisotropic radical]\label{def:isotropic:radical}
Consider a symplectic space $V$ with a quadratic form $\QQ$.
We define the
isotropic and the anisotropic radical of $\vgens$ as
\begin{align}
\radzero(V,\QQ) &:= \rad(V)\cap\QQ^{-1}(0) \label{eq:isotropic:radical}
\\
\radone(V,\QQ) &:= \rad(V)\cap\QQ^{-1}(1).  \label{eq:anisotropic:radical}
\end{align}
\end{defn}

We collect some of their elementary properties:
\begin{lem}\label{lem:isotropic:radical}
Let $\radzero(V,\QQ)$ and $\radone(V,\QQ)$ denote the
isotropic and the anisotropic radical for a
symplectic space $V$ with a quadratic form $\QQ$.
We observe the following properties:
\begin{enumerate}
\item $\QQ$ acts as a linear functional on the radical:\\
$\QQ(u{+}u') = \QQ(u) + \QQ(u')$ for $u, u' \in  \rad(\vgens)$.
\item $\radzero(V,\QQ)$ is a subspace of $\rad(V)$.
\item $\dim[\rad(V)] - \dim[\radzero(V,\QQ)] \leq 1$.
\item $\rad(V) = \radzero(V,\QQ)$ for $\type(\QQ)\in\{+,-\}$.
\item $\dim[\rad(V)] = \dim[\radzero(V,\QQ)]+1$ for $\type(\QQ)=0$.
\item $\radone(V,\QQ)$ is an \emph{affine} subspace of $\rad(V)$ over $\radzero(V,\QQ)$.
\item $\radone(V,\QQ)=\emptyset$ for $\type(\QQ)\in\{+,-\}$.
\item $\radone(V,\QQ)$ has an affine dimension of $\dim[\rad(V)]-1$ for $\type(\QQ)=0$.
\end{enumerate}
\end{lem}

\begin{proof}
Recall that $\QQ(u{+}v)=\QQ(u)+\QQ(v)+\genbil{u}{v}$
for all $u,v\in V$ and the associated symplectic form $\genbilempty$.
For (a), we have $\QQ(u{+}v)=\QQ(u)+\QQ(v)+\genbil{u}{v}
=\QQ(u)+\QQ(v)$ for $u,v\in \rad(V)$. For (b), $\radzero(V,\QQ)$ is the kernel
of $\QQ|_{\rad(V)}$ and thus a subspace.
By (a), the image of $\QQ|_{\rad(V)}$ is either $\{0\}$ or all of $\F_2$.
Thus $\dim[\rad(V)]-\dim[\radzero(V,\QQ)]
=\dim \QQ(\rad(V)) \le 1$ and (c) follows.
If $\type(\QQ)\in\{+,-\}$, then Proposition~\ref{prop:Quadratic_Form_Isomorphism_Classes}
implies $\QQ|_{\rad(V)}=0$ and this proves (d).
If $\type(\QQ)=0$, then Proposition~\ref{prop:Quadratic_Form_Isomorphism_Classes}
implies that the image of $\QQ|_{\rad(V)}$ is equal to $\F_2$.
Thus its kernel $\radzero(V,\QQ)$ has codimension one in $\rad(V)$, proving (e).
For (f), the result is immediate if $\radone(V,\QQ)=\emptyset$. Otherwise
choose $v \in\radone(V,\QQ)$ with $\QQ(v)=1$. We obtain
$\QQ(v{+}u)=\QQ(v)+\QQ(u)=1+\QQ(u)$. Hence
$v+u\in \radone(V)$ iff $\QQ(u)=0$ iff $u\in\radzero(V,\QQ)$
and $\radone(V,\QQ)=v+\radzero(V,\QQ)$ which proves (f).
If $\type(\QQ)\in\{+,-\}$, (d) implies (g)
as  there is no vector $v \in \rad(V)$ with $\QQ(v)=1$.
If $\type(\QQ)=0$, then $\radone(V,\QQ)\neq\emptyset$ because
$\QQ(\rad(V))=\F_2$. By (f), the affine dimension of $\radone(V,\QQ)$
is given by $\dim[\radzero(V,\QQ)]$. Now (e) implies (h).
\end{proof}

Notice that in the literature an alternative definition of the Arf invariant exists \cite{Cimasoni_Simian_2026} as a \emph{democratic invariant}, which is manifestly basis-independent and measures the average value of the quadratic form over the vector space:
\begin{equation}
    \Arf_D(\QQ) = \frac{1}{\abs{\rad(V)}\sqrt{\abs{V/\rad(V)}}}\sum_{v\in V} (-1)^{\QQ(v)}
\end{equation}
We provide a short proof for the relationship between the two invariants:
\begin{lem}
Let $\QQ$ be a quadratic form over a symplectic space $V$. Then, $\Arf_D(\QQ)$ has value $+1$ if $\type(\QQ)=+$, $-1$ if $\type(\QQ)=-$ and $0$ if $\type(\QQ)=0$. When $\QQ|_{\rad(V)}=0$ we also have the relation between the two definitions as $\Arf_D(\QQ) = (-1)^{\Arf(\QQ)}$.
\end{lem}
\begin{proof}
Consider a symplectic basis $\{e_i,f_i\}\cup\{h_j\}$ for $V$ and write vectors as $v=\sum_i\alpha_i e_i + \beta_if_i + \sum_j \gamma_jh_j\in V$.
We can now write, using $\QQ(\alpha_ie_i+\beta_if_i) = \alpha_i\QQ(e_i)+\beta_i\QQ(f_i)+\alpha_i\beta_i$:
\begin{align*}
        \Arf_D(\QQ) &=&& \tfrac{1}{\abs{\rad(V)}\sqrt{\abs{V/\rad(V)}}}
        \left[\sum_{\{\gamma_j\}} \prod_j (-1)^{\gamma_j \QQ(h_j)}\right] \\
         &&&\times \left[\sum_{\{\alpha_i,\beta_i\}}\prod_i (-1)^{\alpha_i\QQ(e_i)+\beta_i\QQ(f_i)+\alpha_i\beta_i}\right].
\end{align*}
If $\QQ|_{\rad(V)} = \F_2$ then, without loss of generality we can choose $h_1$ such that $\QQ_0(h_1)=1$ and transform the other vectors in the basis of the radical as $h_j \mapsto h_j + \QQ_0(h_j)h_1$.
Now the only anisotropic vector in the basis of the radical is $h_1$, which immediately shows that $\Arf_D(\QQ)$ is zero:
\begin{align*}
        & \Arf_D(\QQ) \\
        & = \tfrac{1}{\abs{\rad(V)}\sqrt{\abs{V/\rad(V)}}} \left[\sum_{\{\gamma_j\}} (-1)^{\gamma_1}\right]
         \left[\sum_{v\in V/\rad} (-1)^{\QQ(v)}\right]  \\
        &=  \tfrac{1}{\abs{\rad(V)}\sqrt{\abs{V/\rad(V)}}}  \\
        & \times \left[\sum_{\{\gamma_{j>1}\}}1 - \sum_{\{\gamma_{j>1}\}}1\right]
         \left[\sum_{v\in V/\rad} (-1)^{\QQ(v)}\right] = 0.
\end{align*}
We now assume that $\QQ|_{\rad(V)}=0$, which reduces the problem to the non-degenerate subspace $V/\rad(V)$:
\begin{align*}
        &\Arf_D(\QQ) \\
        &=  \frac{1}{\sqrt{\abs{V/\rad(V)}}}
         \prod_i \sum_{\alpha_i,\beta_i}(-1)^{\alpha_i\QQ(e_i)+\beta_i\QQ(f_i)+\alpha_i\beta_i}
\end{align*}
Now, for each $i$, we have three possible cases: $\QQ(e_i)=\QQ(f_i)=0$, $\QQ(e_i)=1+\QQ(f_i)$ and $\QQ(e_i)=\QQ(f_i)=1$.
One can check that in the first two cases, for which $\QQ(e_i)\QQ(f_i)=0$, there are three isotropic vector and one anisotropic one, hence the sum equals $2$.
Viceversa, when $\QQ(e_i)=\QQ(f_i)=1$, or $\QQ(e_i)\QQ(f_i)=1$, there is one isotropic vector and three anisotropic ones, hence the sum is $-2$.
We can then simplify the sum as $(-2)^{\QQ_0(e_i)\QQ(f_i)}$.
Finally notice that the subspace spanned by $\{e_i,f_i\}_{i=1}^m$ is naturally identified with $V/\rad$ and hence has the same size $2^{2m}$:
\begin{align*}
        &\Arf_D(\QQ) = \frac{1}{\sqrt{\abs{V/\rad(V)}}} \prod_i  (-2)^{\QQ_0(e_i)\QQ(f_i)} \\
        &= \frac{1}{\cancel{\sqrt{\abs{V/\rad(V)}}}} \cancel{2^m} \prod_i  (-1)^{\QQ_0(e_i)\QQ(f_i)}
        = (-1)^{\Arf(\QQ)}
\end{align*}
which concludes the proof.
\end{proof}
Then one only needs $\Arf_D$ to distinguish the three isomorphism classes of quadratic forms.
However, using this definition is clearly not computationally scalable, since it requires evaluation on all vectors (hence scales as $2^{\dim(V)}$), whereas the criterion in Proposition~\ref{prop:Quadratic_Form_Isomorphism_Classes} requires evaluation only on a symplectic basis of $V$.

Also, $\Arf_D$ provides us with the number of isotropic and anisotropic vectors in $V$, given that $\Arf_D = (N_0 - N_1)/(2^m2^r)$ and $N_0+N_1 = 2^{2m+r}$, where $N_0 = \abs{\QQ^{-1}(0)}$ and $N_1 = \abs{\QQ^{-1}(1)}$.
If $\Arf_D(\QQ)=0$, then we clearly have $N_0=N_1 = 2^{2m+r-1}$.
If $\Arf_D(\QQ) = (-1)^s$, then we have:
\begin{equation}
    N_0 = 2^r 2^{m-1}(2^m+s),\; N_1 = 2^r2^{m-1}(2^m-s)
\end{equation}
Moreover, we will later be interested in the non-trivial isotropic and anisotropic vectors, i.e. those in $V\setminus\rad(V)$.
If $\Arf_D=0$, the radical is half isotropic and half anisotropic, hence:
\begin{equation}
    \tilde{N}_0 = N_0 - 2^{r-1},\; \tilde{N}_1 = N_1 - 2^{r-1}
\end{equation}
If $\Arf_D=(-1)^s$, the radical is always isotropic, hence:
\begin{equation}
    \tilde{N}_0 = N_0 - 2^r,\; \tilde{N}_1 = N_1
\end{equation}

\subsection{Isometry Groups and Lie Algebras}

Having classified the relevant invariant bilinear and quadratic forms, we now turn to the linear maps which preserve them.
General background on classical groups, symplectic and orthogonal groups, and their relation to bilinear and quadratic forms can be found in \cite{Bourbaki1959,Carter_1972,Dieudonne_1963,omeara_1978,Dickson_1901,grove_2002,Taylor_1992,Dieudonne_1973,Eichler_1952,VanDerWaerden_1935}.
We look at \emph{isometries} for such forms, i.e., invertible linear maps which preserve a bilinear or quadratic form. Equivalently, from the opposite perspective, the form is invariant under those maps. This lets us characterize sets of isometries for a given form, and conversely determine invariant forms for given sets of operators.
This is the point at which the classification of invariant forms becomes a classification of classical symmetry groups: different isomorphism classes of invariant forms give the corresponding orthogonal, symplectic, and finite-field isometry groups. Later, we will use the same viewpoint in the reverse direction: given a Pauli Lie algebra or a transvection group, we will identify the bilinear and quadratic forms that it preserves.
This will be important for the classification of transvection groups and Lie algebras.

A linear isometry is an invertible linear map preserving the relevant bilinear or quadratic form; see \cite[Ch.~IX, \S 1, no.~8, Prop.~8, Cor.]{Bourbaki1959} and \cite[Ch.~7]{Taylor_1992}.
An invertible operator $\activemap\in\GL(d,\F)$ is said to conserve a bilinear form if $\genbil{\activemap u}{\activemap v} = \genbil{u}{v}$ for all $u,v$ and it conserves a quadratic form if $\QQ(\activemap v) = \QQ(v)$ for all $v$. Equivalently, we say that $\activemap$ is an isometry for $\genbilempty$ or $\QQ$.
Equivalently, we say that $\genbilempty$ ($\QQ$) is an invariant bilinear (quadratic) form for $\activemap$. This extends to sets and groups.
A bilinear form $\genbilempty$ with corresponding matrix $B$ is invariant for $\activemap$ if and only if $\activemap^T B\activemap = B$.

Given a bilinear form $\genbilempty$, or matrix $B$, we define its set of isometries as:
\begin{align*}
        \lieISOGL{\B}
        &:=
        \{ \activemap\in\GL(d,\F) \mid
        \genbil{\activemap u}{\activemap v} = \genbil{u}{v}\\
        &\hspace{4em}
        \text{for all }u,v\} \\
        &\phantom{:}=
        \{ \activemap\in\GL(d,\F) \mid
        \activemap^T B\activemap = B\}
\end{align*}
When referring to the representing matrix $B$, we also write this group as $\lieISOGL{B}$.
Given a quadratic form we define its set of isometries, or its orthogonal group, as (see, e.g., Appendix~\ref{app:witt-extension-f2})
\begin{equation}
\label{eq:orthogonal_group_quadratic_form}
    \lieO(\QQ) := \{ \activemap\in\GL(d,\F) | \QQ(\activemap v) = \QQ(v),\forall v \}.
\end{equation}
If only the type of $\QQ$ is fixed, and not a specific representative, then the group is determined only up to conjugacy, hence up to isomorphism. 
Consequently, the notation attached to a type should be understood after choosing one representative of that type.
Indeed, if $\QQ'=\QQ\circ\activemap^{-1}$, then
\begin{equation*}
    \lieO(\QQ')=\activemap\lieO(\QQ)\activemap^{-1}.
\end{equation*}
The isometry sets $\lieISOGL{\B}$ and $\lieO(\QQ)$ are groups. Moreover, an orthogonal group conserves the associated symmetric bilinear form $\B_\QQ(u,v) = \QQ(u+v)-\QQ(u)-\QQ(v)$, i.e. $\lieO(\QQ)\subseteq\lieISOGL{\B_\QQ}$. Over $\characteristic(\F)\neq 2$, this inclusion is an equality. Moreover, if $\B$ is symmetric, we also have the reciprocal statement $\lieISOGL{\B} = \lieO(\QQ_\B)$. If $\B$ is alternating (when $\characteristic(\F)\neq 2$, this is the same as skew-symmetric), then we call $\lieISOGL{\B}$ the symplectic group with respect to $\B$, $\Sp(\B)$ \cite[Ch.~IX, \S 3, no.~4, \S 5, no.~3, and \S 6, no.~2]{Bourbaki1959}.
Under change of basis, the corresponding matrix groups are conjugate.
Of course, up to the usual change-of-basis equivalence, the isomorphism classes of these groups are determined by those of their bilinear and quadratic forms.

\begin{rem}
\label{rem:forms_to_isometry_groups_dictionary}
The following standard names should be read as the group-theoretic counterpart of the classification of invariant forms recalled above, in particular the binary quadratic classification in Proposition~\ref{prop:Quadratic_Form_Isomorphism_Classes}. Up to change of basis, the isomorphism type of an isometry group is determined by the isomorphism type of the form it preserves. Thus symmetric, alternating, and binary quadratic invariant forms give rise to orthogonal, symplectic, and finite-field orthogonal groups \cite[Chs.~7, 8, and 11]{Taylor_1992}.
\end{rem}

For reference, the relevant dictionary is as follows.
For non-degenerate forms over $V\cong \R^d$, we have the orthogonal groups $\lieO(p,q,\R)$ for symmetric bilinear and quadratic forms with signature $(p,q)$ with $p+q=d$ (where $\lieO(d,0,\R) = \lieO(d)$ is the usual orthogonal group) and the symplectic group $\Sp(2n,\R)$ when $d=2n$ for the skew-symmetric or alternating form.
For $V\cong \C^d$ instead we have the single orthogonal group $\lieO(d,\C)$ and symplectic group $\Sp(2n,\C)$.

Over $V\cong \F_2^d$, the alternating-form classification in Eq.~\eqref{eq:canonical_form_skew_symmetric_matrix_congruence}, equivalently the existence of a symplectic basis from Lemma~\ref{lem:symplectic_basis}, implies that in the non-degenerate case we always have $d=2n$, and we have the symplectic group $\Sp(2n,\F_2)$ over the unique alternating form. Also, we have the orthogonal groups $O^+(2n,\F_2)$ and $O^-(2n,\F_2)$ for the quadratic forms $\QQ$ with types $\type(\QQ)=+$ and $\type(\QQ)=-$, respectively, in the notation of Proposition~\ref{prop:Quadratic_Form_Isomorphism_Classes}
and Appendix~\ref{app:binary-quadratic-forms}.
In the degenerate case over $V\cong \F_2^{2n+r}$, the same alternating-form classification gives the symplectic group $\Sp(2n,r,\F_2)$. The three degenerate quadratic types $\type(\QQ)=+,-,0$ from Proposition~\ref{prop:Quadratic_Form_Isomorphism_Classes} then correspond to the orthogonal isometry groups $O^+(2n,r,\F_2)$, $O^-(2n,r,\F_2)$ and $O^0(2n,r,\F_2)$. 
These characteristic-two quadratic cases are classified in \cite[Ch.~IX, \S 6, Ex.~27]{Bourbaki1959}, where the corresponding notation $\lieO(\QQ)$ is used in Ex.~27(d).
See Table~\ref{tab:isometry-groups} for a summary of the main matrix groups discussed here.

\begin{table*}[t]
\caption{Classical matrix groups appearing in the discussion of isometries.
The compact columns apply to the complex Lie groups.}
\label{tab:isometry-groups}
\footnotesize
\begin{tabular*}{\textwidth}{@{\hspace{1mm}}l@{\extracolsep{\fill}}c@{\hspace{2mm}}l@{\hspace{2mm}}c@{\hspace{2mm}}c@{\hspace{1mm}}c@{\hspace{1mm}}}
\hline\hline
\\[-2.5mm]
$V$ & group & invariant form & case & compact group & compact Lie algebra
\\[0.5mm] \hline
\\[-2.0mm]
$\C^d$ & $\GL(d,\C)$ & -- & -- & $\lieU(d)$ & $\lieu(d)$\\[1.5mm]
$\C^d$ & $\SL(d,\C)$ & -- & -- & $\SU(d)$ & $\su(d)$\\[1.5mm]
$\C^d$ & $\lieO(d,\C)$ & symmetric bilinear & non-degenerate & $\lieO(d)$ & $\so(d)$\\[1.5mm]
$\C^d$ & $\SO(d,\C)$ & symmetric bilinear & non-degenerate & $\SO(d)$ & $\so(d)$\\[1.5mm]
$\C^{2n}$ & $\Sp(2n,\C)$ & alternating bilinear & non-degenerate & $\USp(2n)$ & $\usp(2n)$\\[1.5mm]
\hline
\\[-2.0mm]
$\F_2^{2n}$ & $\Sp(2n,\F_2)$ & alternating bilinear & non-degenerate & -- & --\\[1.5mm]
$\F_2^{2n+r}$ & $\Sp(2n,r,\F_2)$ & alternating bilinear & degenerate & -- & --\\[1.5mm]
$\F_2^{2n}$ & $O^+(2n,\F_2)$ & $\type(\QQ)=+$ & non-degenerate & -- & --\\[1.5mm]
$\F_2^{2n}$ & $O^-(2n,\F_2)$ & $\type(\QQ)=-$ & non-degenerate & -- & --\\[1.5mm]
$\F_2^{2n+r}$ & $O^+(2n,r,\F_2)$ & $\type(\QQ)=+$ & degenerate & -- & --\\[1.5mm]
$\F_2^{2n+r}$ & $O^-(2n,r,\F_2)$ & $\type(\QQ)=-$ & degenerate & -- & --\\[1.5mm]
$\F_2^{2n+r}$ & $O^0(2n,r,\F_2)$ & $\type(\QQ)=0$ & degenerate & -- & --\\[1.5mm]
\hline\hline
\end{tabular*}
\end{table*}

For complex groups, there is a second distinction which matters for Pauli Lie algebras: the full complex isometry group need not be compact, while the physical unitary realization is obtained by intersecting with the unitary group \cite[Ch.~1, \S\S 1.2.2--1.2.4]{hall2015}.
When $\F=\C$, we can also talk about the \emph{unitary} subgroups, which we denote by
\begin{equation}\label{eq:def:compact_bilinear_isometry_intersection}
    \lieISO(\B)=\lieISOin{\lieU(d)}{\B}:=\lieISOGL{\B}\cap \lieU(d).
\end{equation}
These are the \emph{compact forms} of $\lieISOGL{\B}$.
For instance, we find that the compact form of the complex orthogonal group is just the real orthogonal group with positive definite signature $\lieO(d,\C)\cap \lieU(d) = \lieO(d)$. However, the compact form of the complex symplectic group is \emph{not} the real symplectic group (which is in fact non-compact), and we denote it as $\USp(2n) = \Sp(2n,\C)\cap \lieU(2n)$.

Over $\R$ or $\C$, these isometry groups are matrix Lie groups, and their Lie algebras are obtained by differentiating the group-level invariance equations along one-parameter subgroups.
Then, in the context of Lie theory, we also say that a matrix (as a Lie algebra element) conserves a bilinear form with matrix $B$ iff $M^TB + BM = 0$, obtained by differentiating $\activemap(t)^T B\activemap(t)=B$ at $t=0$ for $\activemap(t) = \exp(tM)$ \cite[Props.~3.24--3.25]{hall2015}.
We can talk about the Lie algebras $\lieisogl{B}$ of these isometry groups, which lie in $\gl(d,\F)=\mat{d}{\F}$:
\begin{equation}
    \lieisogl{B} := \{ M\in\gl(d,\F) | M^TB + BM = 0\}
\end{equation}
Again, up to the usual change-of-basis equivalence, the isomorphism classes of these Lie algebras are determined by those of their bilinear/quadratic forms.
Also, since the full orthogonal group and its determinant-one subgroup $\SO(B,\F)=\lieO(B,\F)\cap\det^{-1}(1)$ have the same Lie algebra over $\R$ or $\C$, we talk about the \emph{special} orthogonal Lie algebras $\so(B,\F) = \lieisogl{B}$ (with $B$ non-degenerate symmetric).
Over $V\cong \R^d$, up to isomorphism, we have the special orthogonal Lie algebras $\so(p,q,\R)$ for symmetric bilinear/quadratic forms of signature $(p,q)$ with $p+q=d$ (where $\so(d,0,\R) = \so(d)$ is the usual orthogonal Lie algebra) and the symplectic Lie algebra $\fsp(2n,\R)$ when $d=2n$ for the skew-symmetric/alternating form.
For $V\cong \C^d$ instead we have the single orthogonal Lie algebra $\so(d,\C)$ and symplectic Lie algebra $\fsp(2n,\C)$.
For $\F=\C$, we can again talk about the compact forms of these, as the skew-hermitian parts, given by $\so(d,\R)$ and $\usp(2n)$ ($d=2n$).
Also, it is known that the (real) special orthogonal and unitary symplectic Lie groups are compact and connected, hence we have $\SO(d) = \exp[\so(d)]$ and $\USp(2n) = \exp[\fsp(2n)]$.
The compact groups and Lie algebras are also summarized in Table~\ref{tab:isometry-groups}.

\begin{rem}
\label{rem:isometries_as_fixed_points_orientation}
The fixed-point formulation below packages the preceding examples in the form that will be used later. For a non-degenerate form, preserving the form is equivalent to being fixed by an involution of the ambient general linear group. The infinitesimal preservation condition is the corresponding fixed-point condition in the ambient Lie algebra.
\end{rem}

Notice that when the form is non-degenerate, $B$ is invertible and we can define the group automorphism over $\GL(V)$ as $\Theta_B(g) = B^{-1}g^{-T}B$. Similarly, we can define the Lie algebra automorphism over $\gl(d,\F)$ as $\theta_B(M) = -B^{-1}M^TB$.
Equivalently, the two involutions are
\begin{equation}\label{eq:def:bilinear_isometry_involutions}
    \Theta_B(g):=B^{-1}g^{-T}B,
    \quad
    \theta_B(M):=-B^{-1}M^TB.
\end{equation}
For a chosen ambient Lie group $\lieG$ and Lie algebra $\lieg$, we use the notation
\begin{equation}\label{eq:def:restricted_bilinear_isometry_fixed_points}
    \lieISOin{\lieG}{B}:=\lieG^{\Theta_B},
    \qquad
    \lieisoin{\lieg}{B}:=\lieg^{\theta_B}.
\end{equation}
In particular, the full general-linear objects are
\begin{align}
        \lieISOGL{B} &= \GL(d,\F)^{\Theta_B},
        \label{eq:def:full_bilinear_isometry_group_fixed_points}\\
        \lieisogl{B} &= \gl(d,\F)^{\theta_B}.
        \label{eq:def:full_bilinear_isometry_lie_algebra_fixed_points}
\end{align}
Then, when isometries are viewed as subgroups (subalgebras), we see that there is a connection to the space of \emph{adjoint symmetries}, in the sense of the maps over its ambient group $\GL(V)$ (Lie algebra $\gl(V)$).
This is a simple instance of general statements regarding Lie (sub)algebras and \emph{adjoint} commutants \cite{Zeier_2015,Zimboras_Zeier_SchulteHerbruggen_Burgarth_2015}.
Also, if we restrict to the case of compact Lie groups (algebras), the relevant fixed points are those of the unitary group (resp. Lie algebra).
For non-degenerate $B$, the compact group from Eq.~\eqref{eq:def:compact_bilinear_isometry_intersection} can equivalently be written in fixed-point form.
Together with its Lie algebra version, this gives
\begin{align}
    \lieISO(B)
    =
    \lieISOin{\lieU(d)}{B}
    &:=
    \lieU(d)^{\Theta_B},
    \label{eq:def:compact_bilinear_isometry_group_fixed_points}\\
    \lieiso(B)
    =
    \lieisoin{\lieu(d)}{B}
    &:=
    \lieu(d)^{\theta_B}.
    \label{eq:def:compact_bilinear_isometry_lie_algebra_fixed_points}
\end{align}
For we take the fully degenerate case $B=0$, we also write $\lieiso(0) = \lieu(d)$ and $\lieISO(0) = \lieU(d)$, since the trivial bilinear form $B=0$ is conserved by all operators.

Later we shall consider certain groups/Lie algebras obtained by requiring that they are sets of isometries for multiple non-degenerate bilinear or quadratic forms.
In the non-degenerate bilinear form case, this is equivalent to requiring simultaneous fixed points under the corresponding automorphisms $\Theta_B$ or $\theta_B$.
We denote such objects as:
\begin{subequations}\label{eq:def:multiple_bilinear_isometry_fixed_points}
    \begin{align}
        \lieISOGL{X_B}
        &:=
        \bigcap_{B\in X_B}\lieISOGL{B}\\
        &=
        \GL(d,\F)^{\{\Theta_B\}_{B\in X_B}}\\
        \lieisogl{X_B}
        &:=
        \bigcap_{B\in X_B}\lieisogl{B}\\
        &=
        \gl(d,\F)^{\{\theta_B\}_{B\in X_B}}\\
        \lieO(X_Q) &= \bigcap_{\QQ\in X_Q}\lieO(\QQ)
    \end{align}
\end{subequations}
where it is clear from context whether the definition applies to individual or sets of bilinear/quadratic forms.
\begin{rem}[Pauli compact isometry notation]\label{rem:pauli_compact_isometry_notation}
From now on, in the Pauli setting over $\C$, the notation $\lieISO(\cdot)$ and $\lieiso(\cdot)$ refers to the compact or unitary fixed-point objects from Eqs.~\eqref{eq:def:compact_bilinear_isometry_group_fixed_points}--\eqref{eq:def:compact_bilinear_isometry_lie_algebra_fixed_points}, with $d=2^n$.
The argument specifies the imposed constraints: a Pauli bilinear form $B$, a Pauli commutant algebra $\commalg$ (as specified later in Def.~\ref{def:commutant_pauli_lie_algebra_isometries}), or both.
\end{rem}
Furthermore, in the context of representation theory, we highlight that these are in fact \emph{specific} representations of these Lie groups (algebras) as \emph{matrix} Lie groups (algebras).
Their natural definitions also provide the so-called \emph{standard} representations of the abstract Lie groups or Lie algebras.
Indeed, we shall see later examples of other (non-isomorphic) representations of these Lie algebras in larger spaces than those in which they are naturally defined.

\subsection{Symmetries for Sets of Operators}

We can also ask the reciprocal question, i.e. for a given set of operators, determine what are the bilinear/quadratic forms for which these are isometries.
Namely, for a set of invertible operators $S\subseteq\GL(d,\F)$ (for any $\F$), we define the set of invariant bilinear forms as those matrices such that:
\begin{equation*}
    \bilinear(S) = \{ B\in\mat{d}{\F} \mid \activemap^T B\activemap = B \text{ for all } \activemap\in S\}
\end{equation*}
By abuse of notation, we also consider the set of invariant bilinear forms for general matrices $\pgens\subseteq\gl(d,\F)$ (viewed as Lie algebra generators):
\begin{equation*}
    \bilinear(\pgens) = \{ B\in\mat{d}{\F} \mid G^TB + BG = 0 \text{ for all } G\in\pgens\}
\end{equation*}

Furthermore, they are vector spaces, hence are closed under $\F$-linear combinations, or $B,B'\in\bilinear$ implies $\alpha B+\beta B'\in\bilinear$.
Knowing this object gives a lot of information about the group generated by $S$ (Lie algebra generated by $\pgens$), since it simultaneously conserves all invariant bilinear forms:
\begin{align*}
        \groupclosure{S}
        &\subseteq
        \lieISOGL{\bilinear(S)},\\
        \lie{\pgens}
        &\subseteq
        \lieisogl{\bilinear(\pgens)}.
\end{align*}
Since they are vector spaces, it suffices to look only at a basis to obtain all relevant information.
Under a change of basis of the generators, the invariant bilinear forms change under the inverse change of basis, or:
\begin{equation*}
    \bilinear(\activemap S\activemap^{-1}) = \activemap^{-T}\bilinear(S)\activemap^{-1},\,
    \bilinear(\activemap\pgens \activemap^{-1}) = \activemap^{-T}\bilinear(\pgens)\activemap^{-1}.
\end{equation*}

When $\F=\C$, we may equivalently talk about the set of anti-linear symmetries, defined as:
\begin{align*}
        \antilinear(S) &= \{ \calT \text{ with } g^\dagger \calT g = \calT \text{ for all } g\in S\}\\
        \antilinear(\pgens) &= \{ \calT \text{ with } G^\dagger \calT + \calT G = 0 \text{ for all } G\in\pgens\}
\end{align*}
such that $\antilinear(S) = \calK\cdot\bilinear(S)$ (and likewise for $\pgens$).
The study of spaces of such symmetries (or subgroups within them) has been used to classify various objects in physics, such as the well-known ten-fold way for phases of matter \cite{Altland_Zirnbauer_1997,Ryu_Schnyder_Furusaki_Ludwig_2010}.
We prefer here to deal with bilinear and quadratic forms using linear operators/matrices rather than anti-linear operators, which allows to only deal with the usual matrix multiplication.
However, there is an entirely equivalent treatment one can do using anti-unitary symmetries and complex conjugation, instead of invariant bilinear forms and transposition.

As before, we can also talk about isomorphism classes of these spaces, simply extending the congruence from individual elements to the entire space.
We shall specifically discuss this in the case of Pauli strings as generators (of Lie algebras).

Finally, we also define the invariant quadratic forms, for a set of invertible operators, as those such that:
\begin{align}\label{eq:invariant_quadratic_forms_general_group_subsets}
    \quadratic(S) = \{ &\QQ\text{ quadratic form over }V
    \\
    &  \text{such that } \QQ\circ\activemap = \QQ \text{ for all } \activemap\in S\}. \nonumber
\end{align}
Over $\characteristic(\F)\neq 2$, this is in one-to-one correspondence with the symmetric part of the bilinear forms in $\bilinear(S)$. However, given that in this case the bilinear forms also describe the quadratic ones, we shall only consider this space when $\characteristic(\F)=2$, and specifically over the binary symplectic space $\Fn$.
Again, this information provides an upper bound on the group generated by $S$ (over any field), since:
\begin{equation}
    \groupclosure{S} \subseteq \lieO(\quadratic(S)).
\end{equation}
Like bilinear forms, for any field, these also change inversely as the generators, or:
\begin{equation}
    \quadratic(\activemap S\activemap^{-1}) = \quadratic(S)\circ \activemap^{-1}.
\end{equation}

\section{Invariant Quadratic and Bilinear Forms for Pauli Strings}\label{sec:quadratic_bilinear_forms_for_pauli}

\subsection{Connecting \texorpdfstring{$\C$}{C}-Bilinear Forms and \texorpdfstring{$\F_2$}{F2}-Quadratic Forms}\label{sec:from_F2_quadratico_to_C_bilinear}

We now focus our attention purely on the Pauli case. We start with the following definition, which looks at the bilinear forms inside the Pauli strings:
\begin{equation}
    \bilinear_{\PP_n}(\pgens) = \{ B\in\PP_n | G^TB + BG = 0,\forall G\in\pgens\}.
\end{equation}
Notice that here the choice of basis is the computational basis, with respect to which we have defined transposition, the Pauli matrices and Pauli strings as matrices in $\mat{d}{2^n} = \C^{2^n\times 2^n}$.
It is immediate to show, using the same techniques in Lemma~\ref{lem:pauli:matrix}(b), that $\bilinear(\pgens)$ is spanned by Pauli strings and the basis are the Pauli bilinear forms $\bas{\bilinear(\pgens)} = \bilinear(\pgens)\cap\PP_n = \bilinear_{\PP_n}(\pgens)$. This follows from the fact that $G^TB+BG$ is either $0$ or proportional to the Pauli $BG$, for $G\in\PP_n$.
Similarly, one can also show that, for a given Pauli bilinear form $B$, the Lie algebra of isometries of $B$ is spanned by Paulis $\lieiso(B) = \Span{ G\in\PP_n | G^TB + BG = 0 }$. This extends trivially to multiple invariant bilinear forms.

Also, given that we are interested in generating sets of transvections, which are uniquely defined in terms of vectors, we also abuse notation and write for $\F_2$-quadratic forms:
\begin{align*}
    \quadratic(\vgens) = \{ & \QQ\text{ quadratic form over }\Fn \\
    &\text{such that } \QQ(v)=1 \text{ for all } v\in\vgens\},
\end{align*}
which coincides with $\quadratic(\{\tau_v\}_{v\in\vgens})$, since
$$\QQ(u) = \QQ(\tau_vu) = \QQ(u) + \symp{u}{v}\QQ(v) + \symp{u}{v}$$
for all $u\in\Fn$, if and only if the center of the transvection is anisotropic $\QQ(v)=1$ (or $v=0$).
As for general groups, this object is independent of the choice of generators, and only depends on the group, hence $\quadratic(\tvgroup{\vgens}\cdot\vgens) = \quadratic(\vgens)$.
Knowledge of this object then gives a significant constraint, since we must have that the group is in the joint orthogonal group of all quadratic forms:
\begin{equation}
    \tvgroup{\vgens} \subseteq \lieO(\quadratic(\vgens)),
\end{equation}
and the transvection centers are jointly anisotropic vectors for all $\QQ$:
\begin{equation}
    \tvcenter{\tvgroup{\vgens}} \subseteq \bigcap_{\QQ\in\quadratic(\vgens)} \QQ^{-1}(1)\cap\Span{\vgens}.
\end{equation}
Similarly, we must also have that the Lie algebra is in the joint Lie algebras of isometries:
\begin{equation}
    \lie{\pgens} \subseteq \lieiso(\bilinear(\pgens)).
\end{equation}
Later, this will allows us to characterize the isomorphism classes of  anisotropic vectors with certain Lie algebras (see Section~\ref{sec:pauli_lie_algebras_isometries}).
Also, we shall prove that this setting is exactly the one described by algebraically independent Pauli strings whose graph is $t$-equivalent to one of $\graphX_{2m-1,n_1}^1$, $\graphX_{2m-1,n_1}^2$ or $\graphX_{2m,n_1}^3$.

Namely, our objective in this and the following sections shall be to find the possible cases for each of these objects, up to isomorphism (see Theorem~\ref{thm:derived_pauli_lie_algebras_isometries} and Proposition~\ref{prop:pauli_lie_isometry_block_forms}).
Also, for given generating sets $\pgens=\isolong{\vgens}$, we shall consider convenient criterions to identify the corresponding isomorphism class (see Theorem~\ref{thm:classification_of_affine_subspaces_isomorphism_classes} and Lemma~\ref{lem:isomorphism_class_affine_subspaces_quadratic_forms}).

We now collect some basic facts about quadratic forms in the binary symplectic space.
\begin{lem}\label{lem:quadratic_forms_over_Fn_as_vectors}
Consider the canonical basis $\basel_j$, $\basel_{n+j}$ for $j \in \{1,\ldots,n\}$ of $\Fn$ corresponding to $X_j$, $Z_j$.
Define $\QQ_0(v)=\sum_{j=1}^n \component{v}{j} \component{v}{n+j}$. Then:
\begin{enumerate}
    \item $\type(\QQ_0)=+$;
    \item A Pauli string $P$ is (skew)symmetric in the computational basis if and only if $\QQ_0(v)=0$ ($\QQ_0(v)=1$);
    \item For any quadratic forms $\QQ,\QQ'$ over $\Fn$, there exists a unique vector $w\in\Fn$ such that $\QQ(v)+\QQ'(v) = \symp{w}{v}$.
    \item There is a bijection between quadratic forms $\QQ$ and vectors $w\in\Fn$, such that $\QQ(v) = \QQ_0(v) + \symp{w}{v}$.
\end{enumerate}
\end{lem}
\begin{proof}
Clearly, $(a)$ holds by inspection of $\QQ_0(v)$ and the canonical quadratic form with $\type(\QQ)=+$. By definition $\QQ_0(v) = \#\text{of }Y\text{ in }P=\iso{v}$, which proves (b).

In order to prove (c), we notice by definition that any quadratic form satisfies:
\begin{equation}
    \QQ(u+v) = \QQ(u) + \QQ(v) + \symp{u}{v}
\end{equation}
and we can subtract/add the corresponding identity for $\QQ'$ to get:
\begin{equation}
    \QQ(u+v) - \QQ'(u+v) = (\QQ(u) - \QQ'(u)) + (\QQ(v) - \QQ'(v))
\end{equation}
which shows that $\QQ-\QQ'$ is a linear functional, hence, since $\sympempty$ is non-degenerate, there is a unique $w$ such that $(\QQ-\QQ')(v) = \symp{w}{v}$, hence $\QQ$ may be written as $\QQ(v) = \QQ'(v) + \symp{w}{v}$.
Viceversa, it is immediate to see that $\QQ'(v) + \symp{w}{v}$ is also a quadratic form for any $w\in\Fn$, which proves (c).
For (d), just take $\QQ' = \QQ_0$.
\end{proof}
Given the bijection between quadratic forms and vectors from Lemma~\ref{lem:quadratic_forms_over_Fn_as_vectors}, we introduce the following notation.
\begin{defn}[Quadratic form associated with a vector]\label{def:quadratic-form-associated-vector}
Let $\QQ_0$ be the canonical quadratic form on $\Fn$ from Lemma~\ref{lem:quadratic_forms_over_Fn_as_vectors}. For $w\in\Fn$, we denote by $\QQ_w$ the quadratic form over $\Fn$ defined by
\begin{equation}\label{eq:quadratic-form-associated-vector}
    \QQ_w(v):=\QQ_0(v)+\symp{w}{v}.
\end{equation}
Equivalently, $\QQ_w$ is the unique quadratic form associated with $w$ under the bijection from Lemma~\ref{lem:quadratic_forms_over_Fn_as_vectors}.
\end{defn}
Also notice that we have this bijection only by choosing a global reference quadratic form such that all others are obtained by adding a linear functional. Equivalently, we have chosen a specific symplectic basis with respect to which $\QQ_0(\basel_i)=0$.
This is very much related to the choice of basis over $\C^{2^n}$, which fixes $X$ and $Z$ to be symmetric and $Y$ to be skew-symmetric.
Under this choice of basis, we can also write $\QQ(v)$ as
\begin{subequations}
\begin{align}
        \QQ(v) &= \QQ\qty(\sum_i\alpha_ie_i + \beta_if_i) \\
        &= \sum_i\alpha_i\QQ(e_i) + \beta_i\QQ(f_i) + \sum_i\alpha_i\beta_i \\
        &= \QQ_0(v) + \sum_i\alpha_i\QQ(e_i) + \beta_i\QQ(f_i),
\end{align}
\end{subequations}
which describes the vector
\begin{equation}\label{eq:explicit_w_vector_from_quadratic_form}
w = \sum_i \QQ(f_i)e_i + \QQ(e_i)f_i  \;\text{ for which }\; \QQ = \QQ_w.
\end{equation}
Using Definition~\ref{def:quadratic-form-associated-vector}, we represent the space of invariant quadratic forms by the corresponding space of associated vectors and write
\begin{equation*}
    \quadratic(\vgens) = \{ w\in\Fn \mid \QQ_w(v)=1 \text{ for all } v\in\vgens\}.
\end{equation*}

Furthermore, given this choice, we are now also able to provide a bijection between quadratic forms over $\Fn$ and the Pauli bilinear forms over $\C^{2^n}$:
\begin{lem}\label{lem:bilinear_pauli_and_binary_quadratic_forms}
Pauli bilinear forms of $\pgens = \iso{\vgens}$ as matrices $B$ are in bijection with invariant quadratic forms $\QQ_w$ of $\vgens$ as vectors via $B = \iso{w}$, hence $\bilinear(\pgens) = \isolong{\quadratic(\vgens)}$.
\end{lem}
\begin{proof}
For $Q\in\bilinear(\pgens)$ and $v = \inviso(P)\in\pgens$, we have
\begin{gather*}
        P^T = -QPQ,
        (-1)^{\QQ_0(v)} P = (-1)^{1+\symp{w}{v}}P, \\\
        \QQ_0(v) = \symp{w}{v} + 1, \text{ and }
        \QQ_w(v) = 1.\qedhere
\end{gather*}
\end{proof}
Hence the Pauli bilinear forms $Q\in\bilinear_{\PP_n}(\pgens)$ coincide precisely with those $w\in\quadratic(\vgens)$ such that $\QQ_w$ is an invariant quadratic form for $\vgens$ over the symplectic space.
As an immediate corollary, we also have a correspondence between the anisotropic vectors for $\QQ_w$ and the Lie algebras of isometries for $B=\iso{w}$, i.e.,
\begin{equation}
    \lieiso(\iso{w}) = \Span{\isolong{\QQ_w^{-1}(1)}}
\end{equation}
and similarly for multiple invariant quadratic forms.

\begin{rem}\label{rem:interchange-vector-quadratic-form}
From now on, we will refer interchangeably to $w$ and $\QQ_w$ as invariant quadratic forms for $\vgens$, depending on the context.
\end{rem}

We can further highlight the structure of the invariant quadratic forms by connecting to the commutant/orthogonal complement of the generators by using other \emph{affine} properties of the quadratic forms.
We first recall that an \emph{affine space} $A\subseteq\Fn$ is a set of vectors with an \emph{origin} $w$ and underlying vector space $V$, such that it contains all vectors of the form $\{w + v\}_{v\in V}$.
We denote it with $w+V$. It is closed with respect to \emph{affine} linear combinations, i.e. if $u,v\in w+V$ then $\alpha u + \beta v \in w+V$ iff $\alpha+\beta=1$. If $w\in V$, then clearly $w+V = V$ is a regular vector space.
\begin{lem}\label{lem:affine_structure_invariant_quadratic_form_subspace}
Consider a Pauli generating set with $\quadratic(\vgens)\neq\emptyset$, and let $w^*\in\quadratic(\vgens)$ be a reference vector for the affine space $\quadratic(\vgens)$, i.e., the associated vector of one invariant quadratic form in the notation of Definition~\ref{def:quadratic-form-associated-vector}. Then, $\QQ_w$ is invariant for $\vgens$ if and only if $w\in w^*+\vgens^\perp$, or $\quadratic(\vgens) = w^* + \vgens^\perp$.
\end{lem}
\begin{proof}
The condition for $\QQ_w$ to be a quadratic form is given by the set of equations $\QQ_w(v)=1,v\in\vgens$.
We can expand this into a set of linear non-homogeneous equations in $w$: $\symp{w}{v} = 1 + \QQ_0(v), v\in\vgens$.
Given that the solutions to the homogeneous part $\symp{w}{v}=0$ coincide with $\vgens^\perp$, then, for any solution $w$ such that $\QQ_w(v)=1,\forall v\in\vgens$, the set of solutions to the full system is the affine space $w+\vgens^\perp$.
\end{proof}
Hence, in order to describe the space of invariant quadratic forms for some generating set, it suffices to find the orthogonal complement and a reference vector $w^*\in\quadratic(\vgens)$.
Also, if $\{w_i\}\in\quadratic(\vgens)$, the affine linear combinations $\sum_i\alpha_iw_i$ with $\sum_i\alpha_i=1$ are in $\quadratic(\vgens)$ as well.
The even linear combinations $\sum_i\alpha_iw_i$ with $\sum_i\alpha_i=0$ lie instead in the orthogonal complement $\vgens^\perp$.

In the Pauli formalism, where sums becomes products, we may also write $\bilinear(\pgens) = B\cdot\commutant(\pgens)$, where $B$ is any reference Pauli bilinear form in $\bilinear(\pgens)\neq\{0\}$.

We highlight that, unlike the correspondence between subspaces and their orthogonal complements (which is also a duality), there is not always a correspondence between sets of anisotropic vectors for an affine space and sets of invariant quadratic forms.

To be precise, there exist affine spaces $w^*+W$, with reference vector $w^*$ and subspace $W$, such that their anisotropic vectors $\vgens = \cap_w\QQ_w^{-1}(1) = \QQ_{w^*}^{-1}(1)\cap W^\perp$ have \emph{larger} invariant quadratic forms than the given affine space.
Indeed, we have in general that $w^*+W\subseteq\quadratic(\vgens)$ as well as $W\subseteq\vgens^\perp$.
In fact, since $w^*\in\quadratic(\vgens)$, the only way for $\quadratic(\vgens)$ to not be $w^*+W$ is that the orthogonal complement of $\vgens$ is larger than $W$.

As a consequence, we have that there doesn't exist a set $\vgens$ such that $\quadratic(\vgens)= w^*+W$ for all affine subspaces $w^*+W$. Indeed, by contradiction, assume that such a $\vgens$ existed, then $\vgens^\perp \supsetneq W$, which implies that $\quadratic(\vgens) = w^* + \vgens^\perp \supseteq w^* + W$, which is an absurd.

Let us give an explicit example where the reference vector $w^*$ and subspace $W$ are chosen as follows:
\begin{equation}\label{eq:example_affine_space_is_not_space_of_invariant_quadratic_forms}
    w^* = \basel_{n+1},\quad W = \Span{\{\basel_j\}_{j=2}^n}.
\end{equation}
We have $\QQ_0(w^*)=0$, and $w$ is in the affine subspace if it is of the form $\basel_{n+1}+\sum_{j=2}^n\alpha_j\basel_j$.
In the Pauli picture, we are asking whether there is a set $\pgens\subseteq\PP_n$ such that $B^* = X_1\in\quadratic(\pgens)$ and $\commutant(\pgens) = \Span{\{Z_j\}_{j=2}^n}$.
Now notice that $(\cap_w\QQ_w^{-1}(1))^\perp \supseteq W$ hence $\cap_w\QQ_w^{-1}(1) \subseteq W^\perp$ and $\cap_w\QQ_w^{-1}(1)) = \cap_w(\QQ_w^{-1}(1)\cap W^\perp)$.
Hence, to find the anisotropic vectors it suffices to look within $W^\perp = \Span{\basel_1,\basel_{n+1}}+W$. Notice that that $\QQ_0|_W=0$.
For any $\basel_{n+1}+w\in w^*+W$ and $v_1+w'\in \Span{\basel_1,\basel_{n+1}}+W$, we obtain
\begin{align*}
        &\QQ_{\basel_{n+1}+w}(v_1+w') = \QQ_0(v_1+W) + \symp{\basel_{n+1}+w}{v_1+w'} \\
        &= \QQ_0(v_1) + \symp{\basel_{n+1}}{v_1}
        = \begin{dcases}
            0 & v_1\in \{ 0, \basel_{n+1}, \basel_1+\basel_{n+1}\}\\
            1 & v_1 = \basel_1
        \end{dcases}
\end{align*}
which implies that $\cap_w\QQ_w^{-1}(1)\cap W^\perp = \basel_1+W^\perp$, whose span is $V = \Span{\{\basel_j\}_{j=1}^n}$, hence strictly larger than $W^\perp$ (as well as the larger invariant quadratic forms with $\QQ_{\basel_1}(\cap_w\QQ_w^{-1}(1)) = 1$).
Later, given a classification of equivalence classes of affine subspaces over $\F_2$ we shall see precisely which of these correspond to a space of invariant quadratic forms and which not.

Then, one should be careful when \emph{assigning} symmetries, as not all choice of affine spaces $w^*+W$ may have generating sets $\pgens$ such that $\quadratic(\pgens) = w^* + W$.
Also, this shows that a classification of affine spaces is in fact \emph{not} a classification of spaces of invariant quadratic forms.

\subsection{Invariant Quadratic Forms for Graphs}\label{sec:invariant_quadratic_forms_graphs}

An alternative point of view to study quadratic forms of transvections is one which looks purely to the subspace spanned by the generators.
We highlight this connection with the following Lemma:
\begin{lem}\label{lem:unique_restricted_invariant_quadratic_form}
Consider a Pauli generating set $\pgens=\isolong{\vgens}$ which spans a subspace $V$ of $\Fn$. Then, the restriction of any quadratic form of $\vgens$ to $V$ is unique $\QQ_w|_V = \QQ_{w'}|_V \equiv \QQ^*$ for all $w,w'\in\quadratic(\vgens)$.
\end{lem}
\begin{proof}
By Lemma~\ref{lem:affine_structure_invariant_quadratic_form_subspace}, $\quadratic(\vgens)$ is an affine space with subspace $\vgens^\perp=V^\perp$. Hence any two $w,w'\in\quadratic(\vgens)$ satisfy $w+w'\in V^\perp$. Using Eq.~\eqref{eq:quadratic-form-associated-vector}, for all $v\in V$ the statement follows from
\begin{align*}
        \QQ_w(v) + \QQ_{w'}(v) &= \QQ_0(v) + \symp{w}{v} + \QQ_0(v) + \symp{w'}{v}\\
        &= \symp{w+w'}{v} = 0. \qedhere
\end{align*}
\end{proof}
This still holds already if we require closure under affine linear combinations, which implies $\QQ_w+\QQ_{w'}|_{W^\perp} = 0$, even without a given generating set $\vgens$.

We can also consider the reverse problem, where we are given a subspace $V$ of $\Fn$ and some restricted non-trivial quadratic form $\QQ^*\colon V\mapsto \F$.
Then, to find a $w\in\Fn$ such that $\QQ_w|_V = \QQ^*$ it is necessary and sufficient to solve the system of linear equations $(w,v) = \QQ^*(v) + \QQ_0(v)$, which can be evaluated on any basis of $V$.
This reduces to the usual set of inhomogeneous linear equations if the vectors $v\in V$ for which this is evaluated is invariant under $\QQ^*$, i.e. $\QQ^*(\vgens)=1$ hence $(w,v) = 1 + \QQ_0(v)$ for all $v\in\vgens$ for any $\QQ_w|_V = \QQ^*$.

Hence, the invariant quadratic forms of a set of Paulis collapses to a single one in the restricted subspace, though it is always possible to recover the full quadratic forms over $\Fn$, given an explicit basis of $V$ as a subspace of $\Fn$. This highlights that in fact there should be a strict correspondence between subspace-restricted quadratic forms and sets of quadratic forms over the full space $\Fn$.

Given that all quadratic forms restrict to a single one in the generated subspace, it is clear that the isomorphism class of \emph{spaces} of quadratic forms should be closely related with isomorphism classes of \emph{single} quadratic forms. We shall show this in Theorem~\ref{thm:classification_of_affine_subspaces_isomorphism_classes}.

This perspective is particularly useful if one is purely interested in understanding the orbits inside the subspace of the generators, which is sufficient to describe the Lie algebra as well as the components of the commutator graph in the corresponding matrix algebra.
Indeed, this is the common approach in the transvection group literature \cite{Janssen_1983,humphries_1985,Seven_2005}, where one works inside the subspace of the generators.
Then, Lemma~\ref{lem:unique_restricted_invariant_quadratic_form} provides a clear connection between these sets over $\Fn$ and the single invariant quadratic form over the subspace of the generators.

Moreover, in terms of transvection groups, we have:
\begin{equation}
    \tvgroup{\vgens}|_V \subseteq \lieO(\QQ^*)
\end{equation}
We will come back to the relationship between this restricted transvection group on $V$ and the full transvection group in Section~\ref{sec:orthogonal_diagonal_kernels_maximality}.
This also highlights that $\QQ^*$ is independent of the specific generating set of $\tvgroup{\vgens}$.

In the context of algebraically independent sets, one is also able to forget about quadratic forms of the type $\QQ_w$ and build this function directly from the adjacency matrix of the graph.
To be precise, we have the following:
\begin{lem}[cf.~\cite{Bourbaki1959}, Ch.~IX, \S 3, no.~4]\label{lem:uniqueness_quadratic_form_for_linearly_dep_set}
Let $\vgens=\{v_1,\ldots,v_s\}\subseteq \Fn$ be a set of binary vectors and let
$V=\SpanS[\F_2]{\vgens}$ denote its span.
Given $\vgens$, define the function $F$ from $\F_2^s$ to $\F_2$ by
\begin{equation}\label{eq:existence_quadratic_form_for_set_relations}
F(c)
:=
\sum_{j=1}^s c_j
+
\sum_{1\le i<j\le s} c_i c_j\,\symp{v_i}{v_j}
\end{equation}
where $c=(c_1,\ldots,c_s)$. There exists a quadratic form $\QQ$ on $V$ with associated symplectic product
$\sympempty$ such that $\QQ(v)=1$ for all $v\in \vgens$ if and only if
$F(c)=0$ for every linear relation
$\sum_{j=1}^s c_jv_j=0$ over $\F_2$ among the elements of $\vgens$.
If such a quadratic form exists, then it is unique and, for any representation
$u=\sum_{j=1}^s c_j v_j$, is given by
\begin{equation}\label{eq:uniqueness_quadratic_form_for_set}
\QQ(u)
=
F(c).
\end{equation}
Moreover, $\QQ$ is preserved by every element $g$ of the transvection group
$\tvgroup{\vgens}$,
i.e., $\QQ(g u) = \QQ(u)$ for all $u \in V$.
\end{lem}

\begin{proof}
Assume first that such a quadratic form $\QQ$ exists.
Then $\QQ(0)=0$ and, by repeatedly applying the polarization identity
$\QQ(u+u') = \QQ(u)+\QQ(u')+\symp{u}{u'}$
to any representation $u=\sum_{j=1}^s c_jv_j$, we obtain
Eq.~\eqref{eq:uniqueness_quadratic_form_for_set}.
If $\sum_{j=1}^s c_jv_j=0$ is a linear relation, then the left-hand side of
Eq.~\eqref{eq:uniqueness_quadratic_form_for_set} is $\QQ(0)=0$.
Hence its right-hand side gives $F(c)=0$.

Conversely, assume that $F(c)=0$ for every linear relation
$\sum_{j=1}^s c_jv_j=0$ over $\F_2$ among the elements of $\vgens$.
If two coefficient vectors $c,d\in\F_2^s$ represent the same vector in $V$,
then $c+d$ is the coefficient vector of the linear relation
$\sum_{j=1}^s(c_j+d_j)v_j=0$. The polarization identity for $F$ gives
\[
F(c)+F(d)
=
F(c+d)+\symp{\sum_{j=1}^s c_jv_j}{\sum_{j=1}^s d_jv_j}.
\]
Here $F(c+d)=0$ by assumption,
and the symplectic term is zero because the two represented vectors are equal.
Thus $F(c)=F(d)$, so $\QQ(u):=F(c)$ is well-defined for
$u=\sum_{j=1}^s c_jv_j$. Moreover, the same displayed polarization identity gives
\[
\QQ(u+u')=\QQ(u)+\QQ(u')+\symp{u}{u'}.
\]
Hence $\QQ$ is a quadratic form with associated symplectic product
$\sympempty$, and $\QQ(v_j)=1$ for all $j$.

For uniqueness, let $\QQ$ and $\QQ'$ be two quadratic forms on $V$ with associated
symplectic product $\sympempty$ such that $\QQ(v)=\QQ'(v)=1$ for all $v\in\vgens$.
Then $L:=\QQ+\QQ'$ is linear, because the bilinear terms cancel in
$L(u+u')$.
Since $\QQ(v)=\QQ'(v)=1$ for every $v\in\vgens$, we have
$L(v)=1+1=0$ for all $v\in\vgens$.
As $\vgens$ spans $V$ and $L$ is linear, it follows that $L=0$, hence
$\QQ=\QQ'$.

Moreover, $\QQ$ is preserved by every
generating transvection $\tau_v$ with $v\in \vgens$, since for any $u\in V$,
\begin{align*}
& \QQ(\tau_v(u))
 =
\QQ(u+\symp{u}{v}v)\\
& =
\QQ(u)+\symp{u}{v}\QQ(v)+\symp{u}{\symp{u}{v}v}
=
\QQ(u),
\end{align*}
where $\QQ(v)=1$ and $\symp{u}{\symp{u}{v}v}=\symp{u}{v}^2=\symp{u}{v}$.
Hence $\QQ$ is preserved by all elements of $\tvgroup{\vgens}$.
\end{proof}

As an immediate corollary, we have for the linearly independent case:
\begin{cor}\label{cor:uniqueness_existence_quadratic_form_for_basis}
Let $\vgens \subseteq \Fn$ be a linearly independent set of binary vectors and let
$V=\SpanS[\F_2]{\vgens}$ denote its span.
Then there exists a unique quadratic form $\QQ$ on $V$ with associated
symplectic product $\sympempty$ such that
$\QQ(v)=1$ for all $v\in \vgens$.
Moreover, $\QQ$ is preserved by every
element $g$ of the transvection group $\tvgroup{\vgens}$,
i.e., $\QQ(g u) = \QQ(u)$ for all $u \in V$.
\end{cor}
\begin{proof}
Since $\vgens$ is linearly independent, the only linear relation among its
elements is the zero relation. Hence $F(0)=0$, and the claim follows from
Lemma~\ref{lem:uniqueness_quadratic_form_for_linearly_dep_set}.
\end{proof}

We use the result of Corollary~\ref{cor:uniqueness_existence_quadratic_form_for_basis}
to recall the standard quadratic form as follows:

\begin{defn}[Standard quadratic form for a basis]\label{def:standard:quadratic}
Let $\vgens \subseteq \Fn$ be a linearly independent set of binary vectors and let
$V=\Span[\F_2]{\vgens}$ denote its span. We denote the standard
quadratic on $V$ from Corollary~\ref{cor:uniqueness_existence_quadratic_form_for_basis}
by $\QQstandard{\vgens}$.
\end{defn}

Following Lemma~\ref{lem:bijection_colorings_algebraically_independent_sets} and Definition~\ref{def:induced:frustration},
we identify
vectors $v\in V=\Span[\F_2]{\vgens}$ with induced subgraphs $\inducedfrustration{\vgens}{v}$ of the frustration graph
$\frustration{\vgens}$.

\begin{cor}[\cite{humphries_1985}]\label{cor:quadratic_form_euler_characteristic}
Let $\vgens \subseteq \Fn$ be a linearly independent set of binary vectors, let
$V=\Span[\F_2]{\vgens}$ denote its span, and let $\frustration{\vgens}$ be its frustration graph.
The value $\QQstandard{\vgens}(v)$ of
the standard quadratic form from Definition~\ref{def:standard:quadratic}
for a vector $v \in V$
has a graph-theoretic interpretation in terms of the Euler characteristic modulo $2$
of the induced subgraph
$\inducedfrustration{\vgens}{v}$ of $ \frustration{\vgens}$ (see Definition~\ref{def:induced:frustration}).
More explicitly,
\begin{equation}\label{eq:quadratic_form_euler_characteristic}
\QQstandard{\vgens}(v)=
\big[\abs{\vertices(\inducedfrustration{\vgens}{v})}
-
\abs{\edges(\inducedfrustration{\vgens}{v})}\big]\bmod 2.
\end{equation}
\end{cor}
\begin{proof}
Let $\vgens=\{v_1,\dots,v_s\}$ and write $v=\sum_{i=1}^s \coloring(i)v_i \in V$ with $\coloring(i)\in\F_2$.
By Corollary~\ref{cor:uniqueness_existence_quadratic_form_for_basis},
$$
\QQstandard{\vgens}(v)
=
\sum_{i=1}^s \coloring(i)
+
\sum_{1\le i<j\le s} \coloring(i)\coloring(j)\symp{v_i}{v_j}.
$$
The first term counts the number of vertices of the induced subgraph
$\inducedfrustration{\vgens}{v}$ modulo $2$.
The second term counts the number of edges of $\inducedfrustration{\vgens}{v}$ modulo $2$, since
$\symp{v_i}{v_j}=1$ precisely when $\{i,j\}\in \edges(\frustration{\vgens})$.
Since addition and subtraction agree modulo $2$, this proves
Eq.~\eqref{eq:quadratic_form_euler_characteristic}.
\end{proof}

Hence, we find the binary counterpart to the graph-theoretic statement in Lemma~\ref{lem:euler_characteristic_color_flip_invariant}, which already showed the invariance of the Euler characteristic of a coloring under transvections.

In the presence of algebraic dependencies, the existence of an invariant quadratic form is not always guaranteed.
As an example, consider $\pgens = \{\text{ZI, XI, IZ, IX, ZZ}\}$ or equivalently $\vgens = \{e_1,f_1,e_2,f_2,e_1+e_2\}$ with $\symp{e_i}{f_i} = 1$ and the remaining products are zero.
Then, if $\QQ(e_i)=\QQ(f_i)=1$, $\QQ(e_1+e_2)=\QQ(e_1)+\QQ(e_2)=0$, which means that there is no invariant quadratic form for this set.

Lemma~\ref{lem:uniqueness_quadratic_form_for_linearly_dep_set} gives a coordinate expression for $\QQ$ in terms of a coefficient vector $c\in\F_2^{\abs{\vgens}}$ whose realization is $\coltovec_{\vgens}(c)\in V$. Namely,
\begin{equation}\label{eq:matrix_expression_restricted_quadratic_form}
    \QQ(\coltovec_{\vgens}(c)) = c^T(A(\graphG)^u+I)c,
\end{equation}
where $A(\graphG)^u$ denotes the upper triangular part of the adjacency matrix. 

This also allows us to speak directly of the unique invariant quadratic form for a \emph{graph}, as well as its isomorphism class.
Indeed, it is clear from the vector point of view that this is in fact an invariant of the graph under $t$-equivalence.
As we now show, this invariant is in fact sufficient to distinguish between the non-path graph classes.
Namely, we have the following for the classes in Theorem~\ref{thm:classes:humphries} (as shown by \cite{Janssen_1983}) and for the classes in Theorem~\ref{thm:classes:eisert}:
\begin{prop}\label{prop:isomorphism_classes_quadratic_forms_arf_invariant_canonical}
Let $\vgens$ be the unique linearly independent set for one of the canonical
graph families listed in
Theorems~\ref{thm:classes:humphries} or \ref{thm:classes:eisert},
with $V=\SpanS[\F_2]{\vgens}$ and unique invariant quadratic form
$\QQ=\QQstandard{\vgens}$ on $V$.
For each such canonical graph, $\QQ$ has the following type
(see Def.~\ref{def:type}):
\begin{enumerate}
    \item For $\graphX_{2m-1,n_1}^1$,
    \[
    \type(\QQ)=
    \begin{cases}
    + & \text{if }\,m\bmod 4\in\{0,1\},\\
    - & \text{if }\,m\bmod 4\in\{2,3\}.
    \end{cases}
    \]
    \item For $\graphX_{2m-1,n_1}^2$,
    \[
    \type(\QQ)=
    \begin{cases}
    - & \text{if }\,m\bmod 4\in\{0,1\},\\
    + & \text{if }\,m\bmod 4\in\{2,3\}.
    \end{cases}
    \]
    \item For $\graphX_{2m,n_1}^3$, $\type(\QQ)=0$.
    \item For $\graphS_{n_2,n_1}^1$, $\type(\QQ)=-$.
    \item For $\graphS_{n_2,n_1}^2$, $\type(\QQ)=+$.
    \item For $\graphS_{n_2,n_1}^3$, $\type(\QQ)=0$.
    \item For the odd blown-up path graph $\graphP_{2m-1,n_1}$, $m\geq 1$,
    \[
    \type(\QQ)=
    \begin{cases}
    + & \text{if }\,m\bmod 4\in\{0,3\},\\
    - & \text{if }\,m\bmod 4\in\{1,2\}.
    \end{cases}
    \]
    \item For the even blown-up path graph $\graphP_{2m,n_1}$, $m\geq 2$,
    \[
    \type(\QQ)=
    \begin{cases}
    0 & \text{if }\,m\bmod 4\in\{0,2\},\\
    - & \text{if }\,m\bmod 4=1,\\
    + & \text{if }\,m\bmod 4=3.
    \end{cases}
    \]
\end{enumerate}
The radical data are detailed in the following (refer to Figures~\ref{fig:humphries_classes} and \ref{fig:eisert_classes}
for the vertex labels):
\begin{enumerate}[resume]
    \item The isotropic radical is
    \[
    \radzero(V,\QQ)=
    \begin{cases}
    \rad(V) &
    \text{for } \graphP_{2m,n_1}\\[-1mm]
    & \text{with odd } m,\\[1mm]
    \Span{\{b_i{+}b_{i+1}\}_{i=1}^{n_1-1}} & \text{otherwise}.
    \end{cases}
    \]
\end{enumerate}
The anisotropic radical is given as follows:
\begin{enumerate}[resume]
    \item $\radone(V,\QQ)= (a_1{+}a_2{+}a_4)+\radzero(V,\QQ)$ for $\graphX_{2m,n_1}^3$.
    \item $\radone(V,\QQ)= (a_1{+}a_3{+}b_1)+\radzero(V,\QQ)$ for $\graphS_{n_2,n_1}^3$.
    \item $\radone(V,\QQ) = \sum_{i=1}^{m+1} a_{2i-1} + \radzero(V,\QQ)$ for $\graphP_{2m,n_1}$ with even $m$.
    \vspace{0.5ex}
    \item Apart from $\graphX_{2m,n_1}^3$, $\graphS_{n_2,n_1}^3$, and $\graphP_{2m,n_1}$ with even $m$, no anisotropic radical coset occurs; equivalently, $\radone(V,\QQ)=\emptyset$ and $\radzero(V,\QQ)=\rad(V)$.
\end{enumerate}
\end{prop}

\begin{lem}\label{lem:canonical_quadratic_radical_values}
In the setting of Proposition~\ref{prop:isomorphism_classes_quadratic_forms_arf_invariant_canonical}, write, as in Eq.~\eqref{eq:canonical-dependency-B},
\begin{equation}
B=\Span[\F_2]{b_i{+}b_{i+1}\mid 1\leq i<n_1},
\label{eq:canonical-quadratic-B}
\end{equation}
and set
\begin{equation*}
\nu_1:=\sum_{j=1}^{m+1}a_{2j-1},\,
\nu_2:=a_1+a_2+a_4,\,
\nu_3:=a_1+a_3+b_1.
\end{equation*}
For $\graphP_{2m,n_1}$, $\graphX^3_{2m,n_1}$, and
$\graphS^3_{n_2,n_1}$, these are the additional generators $\nu_j$ shown in
panels (c), (d), and (b) of Fig.~\ref{fig:quadratic-form-radical-proof}, respectively.
Then the $\nu_j$ and their values under $\QQ$ are those in
Table~\ref{tab:canonical-quadratic-radical-generators}, and the corresponding
isotropic and anisotropic radicals are those in
Table~\ref{tab:canonical-quadratic-radical-data}.
\end{lem}

\begin{table}[t]
\caption{For canonical graphs, additional radical elements $\nu_j$
that are not obtained from length-one leaf pairs and their values $\QQ(\nu_j)$ under the standard
quadratic form $\QQ=\QQstandard{\vgens}$.}
\label{tab:canonical-quadratic-radical-generators}
\footnotesize
\begin{tabular*}{\columnwidth}{@{\hspace{1mm}}l@{\extracolsep{\fill}}c@{\hspace{2mm}}c@{\hspace{1mm}}}
\hline\hline
\\[-2.5mm]
canonical graph & extra $\nu_j$ & $\QQ(\nu_j)$
\\[0.5mm] \hline
\\[-2.0mm]
$\graphP_{2m-1,n_1}$, $\graphX^1_{2m-1,n_1}$, $\graphX^2_{2m-1,n_1}$ & none & --\\[1.5mm]
$\graphS^1_{n_2,n_1}$, $\graphS^2_{n_2,n_1}$ & none & --\\[1.5mm]
$\graphP_{2m,n_1}$ & $\nu_1$ & $(m{+}1)\bmod 2$\\[1.5mm]
$\graphX^3_{2m,n_1}$ & $\nu_2$ & $1$\\[1.5mm]
$\graphS^3_{n_2,n_1}$ & $\nu_3$ & $1$\\[1.5mm]
\hline\hline
\end{tabular*}
\end{table}

\begin{table}[t]
\caption{Isotropic and anisotropic radical data based on
Table~\ref{tab:canonical-quadratic-radical-generators}. Here $B$ is defined in
Eq.~\eqref{eq:canonical-quadratic-B}. In the row with no extra $\nu_j$,
$\rad(V)=\radzero(V,\QQ)=B$. For $\graphP_{2m,n_1}$ with $m$ odd, $\nu_1$ is
isotropic and $\rad(V)=\radzero(V,\QQ)=B+\SpanS[\F_2]{\{\nu_1\}}$. In the
remaining rows with an extra generator,
$\rad(V)=\radzero(V,\QQ)+\SpanS[\F_2]{\{\nu_j\}} =
B+\SpanS[\F_2]{\{\nu_j\}}$. Here $\nu_j+B$ denotes the
affine translate of $B$ by $\nu_j$, whereas $B+\SpanS[\F_2]{\{\nu_1\}}$ is a
linear sum of subspaces.}
\label{tab:canonical-quadratic-radical-data}
\footnotesize
\begin{tabular*}{\columnwidth}{@{\hspace{1mm}}l@{\extracolsep{\fill}}c@{\hspace{2mm}}c@{\hspace{1mm}}}
\hline\hline
\\[-2.5mm]
case & $\radzero(V,\QQ)$ & $\radone(V,\QQ)$
\\[0.5mm] \hline
\\[-2.0mm]
no extra $\nu_j$ & $B$ & $\emptyset$\\[1.5mm]
$\graphP_{2m,n_1}$, $m$ odd & $B+\SpanS[\F_2]{\{\nu_1\}}$ & $\emptyset$\\[1.5mm]
$\graphP_{2m,n_1}$, $m$ even & $B$ & $\nu_1+B$\\[1.5mm]
$\graphX^3_{2m,n_1}$ & $B$ & $\nu_2+B$\\[1.5mm]
$\graphS^3_{n_2,n_1}$ & $B$ & $\nu_3+B$\\[1.5mm]
\hline\hline
\end{tabular*}
\end{table}

\begin{proof}
For canonical graphs and a linearly independent $\vgens$, set $\QQ=\QQstandard{\vgens}$.
Hence Corollary~\ref{cor:quadratic_form_euler_characteristic} gives, for
$u\in V$,
\begin{equation}\label{eq:proof:quadratic_form_euler_characteristic}
\QQ(u)=
\big[\abs{\vertices(\inducedfrustration{\vgens}{u})}
-
\abs{\edges(\inducedfrustration{\vgens}{u})}\big]\bmod 2.
\end{equation}
By Lemma~\ref{lem:Canonical_t_equivalent_Graph_Radical}, the radical is
generated by the length-one leaf pairs $b_i+b_{i+1}$, together with the listed
extra generator when such a generator occurs.
Panel (a) of Figure~\ref{fig:quadratic-form-radical-proof} shows that each
$b_i+b_{i+1}$ has support on two length-one leaves. These two leaves are not
adjacent, so the induced subgraph has two vertices and no edges. Hence
Eq.~\eqref{eq:proof:quadratic_form_euler_characteristic} gives
$\QQ(b_i{+}b_{i+1})=2-0=0$.

\begin{figure}[t]
    \centering
    \includegraphics{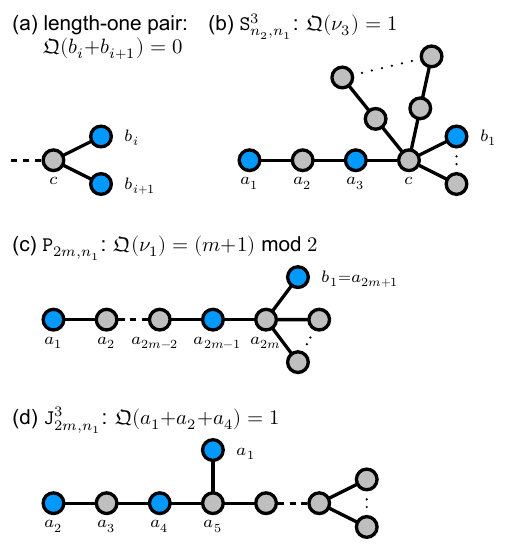}
    \caption{Radical computations used in Lemma~\ref{lem:canonical_quadratic_radical_values}. Blue vertices mark the support of the displayed radical element. (a) length-one leaf pairs; (b) the radical element for $\graphS^3_{n_2,n_1}$; (c) the radical element for $\graphP_{2m,n_1}$, where $\QQ$ does not vanish on the radical for even $m$; (d) the radical element for $\graphX^3_{2m,n_1}$.}
    \label{fig:quadratic-form-radical-proof}
\end{figure}

It remains only to evaluate the possible extra generators.
For $\graphP_{2m,n_1}$, panel (c), with $b_1=a_{2m+1}$, gives the
alternating coloring
$\nu_1=\sum_{j=1}^{m+1}a_{2j-1}$. The colored vertices are pairwise
non-adjacent along the displayed path, including the terminal leaf
$a_{2m+1}$, so the induced subgraph has $m+1$ vertices and no edges. Hence
Eq.~\eqref{eq:proof:quadratic_form_euler_characteristic} gives
$\QQ(\nu_1)=(m{+}1)\bmod 2$.
For $\graphX^3_{2m,n_1}$, panel (d) shows that the support of
$\nu_2=a_1+a_2+a_4$ consists of three colored vertices with no edges among
them in the induced subgraph. Hence
Eq.~\eqref{eq:proof:quadratic_form_euler_characteristic} gives
$\QQ(\nu_2)=(3{-}0)\bmod 2=1$.
For $\graphS^3_{n_2,n_1}$, panel (b) shows that
$\nu_3=a_1+a_3+b_1$ is supported on three vertices with no edges among them in
the induced subgraph. Hence Eq.~\eqref{eq:proof:quadratic_form_euler_characteristic}
gives $\QQ(\nu_3)=(3{-}0)\bmod 2=1$.
This proves Table~\ref{tab:canonical-quadratic-radical-generators}.

Lemma~\ref{lem:isotropic:radical} then gives the radical data in
Table~\ref{tab:canonical-quadratic-radical-data}. Indeed,
$B\subseteq\radzero(V,\QQ)$, and each listed extra generator $\nu_j$ either
enlarges $\radzero(V,\QQ)$ if $\QQ(\nu_j)=0$, or gives the anisotropic coset
$\nu_j+B$ if $\QQ(\nu_j)=1$. Thus, for $\graphP_{2m,n_1}$, the generator
$\nu_1$ is isotropic, i.e.\ $\QQ(\nu_1)=0$, exactly when $(m{+}1)\bmod 2=0$, i.e.\ when $m$ is odd;
this is precisely the case $\graphP_{2m,n_1}$ with $m$ odd. In the remaining extra
generator cases, $\QQ(\nu_j)=1$, so $\radzero(V,\QQ)=B$ and
$\radone(V,\QQ)=\nu_j+B$.
\end{proof}

We use the Arf invariant from
Proposition~\ref{prop:Quadratic_Form_Isomorphism_Classes}: when $\QQ$ vanishes on
$\rad(V)$, it is computed on $V/\rad(V)$ as
$\Arf(\QQ)=\sum_j\QQ(e_j)\QQ(f_j)$ for any symplectic basis
$\{e_j,f_j\}$ of $V/\rad(V)$.

\begin{figure*}[t]
    \centering
    \includegraphics{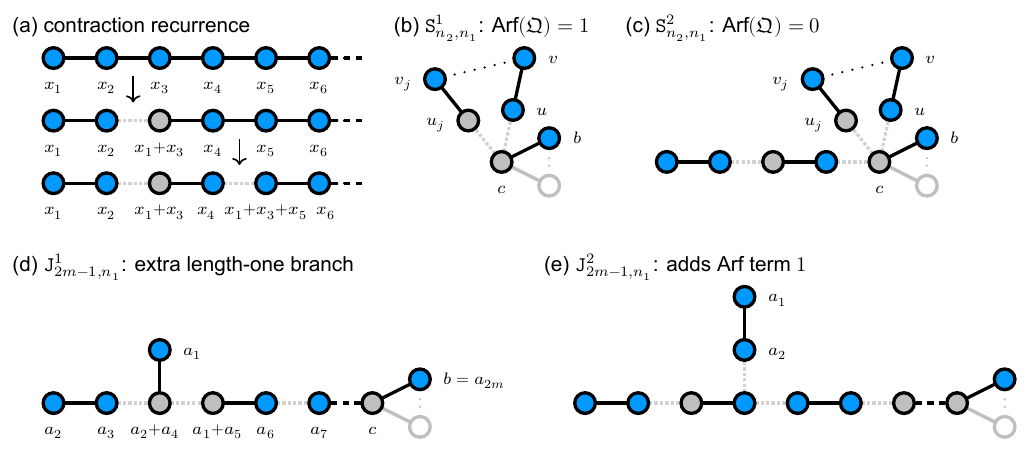}
    \caption{Local Arf-invariant computations used in Lemma~\ref{lem:canonical_quadratic_arf_counts}. Blue and grey vertices mark $\QQ$-values $1$ and $0$ after the indicated basis changes. Isolated pairs are used in the Arf sum, and pale dotted edges mark the removed edges. Panel (a) gives the contraction recurrence; panels (b) and (c) give the star computations for $\graphS^1_{n_2,n_1}$ and $\graphS^2_{n_2,n_1}$; panels (d) and (e) show the disconnected pairs for $\graphX^1_{2m-1,n_1}$ and $\graphX^2_{2m-1,n_1}$. In panels (b)--(e), a blue length-one leaf represents the common image of the length-one leaves after quotienting by $B$, while a white leaf with grey outline and grey edge symbolizes the remaining length-one leaves. Vertices marked by $c$ are the centers.}
    \label{fig:quadratic-form-arf-peeling-proof}
\end{figure*}

\begin{lem}\label{lem:canonical_quadratic_arf_counts}
For the cases in Proposition~\ref{prop:isomorphism_classes_quadratic_forms_arf_invariant_canonical}
where $\QQ$ vanishes on the radical, the Arf invariants $\Arf(\QQ)$ and the
corresponding types $\type(\QQ)$ are those in
Table~\ref{tab:canonical-quadratic-arf-types}.
\end{lem}

\begin{table}[t]
\caption{Arf invariants $\Arf(\QQ)$ and types $\type(\QQ)$ for the cases where $\QQ$ vanishes on the radical.}
\label{tab:canonical-quadratic-arf-types}
\footnotesize
\begin{tabular*}{\columnwidth}{@{\hspace{1mm}}l@{\extracolsep{\fill}}c@{\hspace{2mm}}l@{\hspace{1mm}}}
\hline\hline
\\[-2.5mm]
canonical graph & $\Arf(\QQ)$ & $\type(\QQ)$
\\[0.5mm] \hline
\\[-2.0mm]
$\graphP_{2m-1,n_1}$ & $\lfloor(m{+}1)/2\rfloor$ & $+$ if $m\bmod 4\in\{0,3\}$\\
 & & $-$ if $m\bmod 4\in\{1,2\}$\\[2mm]
$\graphP_{2m,n_1}$, $m$ odd & $\lfloor(m{+}1)/2\rfloor$ & $-$ if $m\bmod 4=1$\\
 & & $+$ if $m\bmod 4=3$\\[2mm]
$\graphX^1_{2m-1,n_1}$ & $\lfloor m/2\rfloor$ & $+$ if $m\bmod 4\in\{0,1\}$\\
 & & $-$ if $m\bmod 4\in\{2,3\}$\\[2mm]
$\graphX^2_{2m-1,n_1}$ & $\lfloor m/2\rfloor+1$ & $-$ if $m\bmod 4\in\{0,1\}$\\
 & & $+$ if $m\bmod 4\in\{2,3\}$\\[2mm]
$\graphS^1_{n_2,n_1}$ & $1$ & $-$\\[2mm]
$\graphS^2_{n_2,n_1}$ & $0$ & $+$\\[1.5mm]
\hline\hline
\end{tabular*}
\end{table}

\begin{proof}
By Proposition~\ref{prop:Quadratic_Form_Isomorphism_Classes}, the Arf invariant is
computed on the non-degenerate quotient $V/\rad(V)$.
Thus the following computation only concerns the induced non-degenerate
quadratic space on $V/\rad(V)$. In particular, the subspace
$B\subseteq\rad(V)$ is zero in the quotient and hence adds no terms to
the Arf sum; the radical data have already been determined in
Lemma~\ref{lem:canonical_quadratic_radical_values}.
Here we use (not necessarily valid)
contractions as bookkeeping for basis changes.
This is allowed by Proposition~\ref{prop:Quadratic_Form_Isomorphism_Classes}, since
the Arf invariant is independent of the chosen symplectic basis.

Panel (a) of Figure~\ref{fig:quadratic-form-arf-peeling-proof} displays the first two steps on an
initially all-$1$ path corresponding to the standard quadratic form. In the first step, replacing $x_3$ by
$x_1+x_3$ isolates the symplectic pair $(x_1,x_2)$: indeed
$\symp{x_2}{x_1+x_3}=1+1=0$, while $x_1$ remains paired only with $x_2$.
Since $x_1$ and $x_3$ are non-adjacent and both have $\QQ$-value $1$, the new
left endpoint has value $\QQ(x_1+x_3)=1+1=0$.
In the second step, replacing $x_5$ by $x_1+x_3+x_5$ isolates the pair
$(x_1+x_3,x_4)$, whose $\QQ$-values are $(0,1)$.

Repeating this basis change leads to a decomposition  into symplectic pairs.
Let $(u_j,v_j)$ be the $j$th isolated pair and set
\[
q_j:=\QQ(u_j)\QQ(v_j).
\]
Then $q_j$ is the corresponding Arf term. The first two terms shown in
panel (a) of Figure~\ref{fig:quadratic-form-arf-peeling-proof} are $q_1=1$ and $q_2=0$. At each later
step, the second vector in the isolated pair is again an unchanged original
generator with $\QQ$-value $1$, while the first vector is the current left
endpoint. Passing to the next step adds one more non-adjacent original generator
of $\QQ$-value $1$, so the Arf terms satisfy
\[
q_{j+1}=(1{+}q_j)\bmod 2.
\]
Thus $q_j$ starts with $q_1=1$ and then alternates between $0$ and $1$.

For $\graphP_{2m-1,n_1}$, the quotient has $m$ such isolated pairs along the
spine, so this gives $\sum_{j=1}^m q_j=\lfloor(m{+}1)/2\rfloor$ for the Arf sum.
Panel (c) of
Figure~\ref{fig:quadratic-form-radical-proof} provides the additional radical
generator $\nu_1$.
By Lemma~\ref{lem:canonical_quadratic_radical_values},
$\QQ$ vanishes on the radical exactly when $m$ is odd. In that case
$\nu_1$ is zero in $V/\rad(V)$. After quotienting by $B$, all length-one
leaves are identified, and the relation $\nu_1=0$ eliminates the remaining
leaf variable. Thus the quotient is represented by the
$2m$ spine generators $a_1,\ldots,a_{2m}$ with the same all-$1$ path
recurrence as in panel (a) of
Figure~\ref{fig:quadratic-form-arf-peeling-proof}. Hence the Arf sum is again
$\sum_{j=1}^m q_j=\lfloor(m{+}1)/2\rfloor$.

For the star cases, passing to the quotient by $B$ again makes the
subspace spanned by the length-one leaf pairs $b_i+b_{i+1}$ vanish.
Thus all length-one leaves have one common remaining image in
$V/\rad(V)$, denoted by $b$. In panels (b) and (c), the blue leaf labeled $b$
represents this common image. The white leaf with grey outline and grey edge
is not an additional Arf-sum vector but indicates that
the original star may have further length-one legs, all differing from
$b$ by elements of $B$.

For $\graphS^1_{n_2,n_1}$ there are $n_2$ length-two legs. Panel (b)
displays one chosen length-two leg and one representative of the remaining
$n_2-1$ length-two legs. Let the chosen leg have middle vertex $u$ and
terminal vertex $v$. We first isolate the pair $(u,v)$ by replacing the center
$c$ by $c+v$; this removes the edge between the center and $u$ because
$\symp{u}{c+v}=\symp{u}{c}+\symp{u}{v}=1+1=0$.
Both $u$ and $v$ have $\QQ$-value $1$, so this isolated pair contributes Arf
term $1$. This is the only length-two leg kept in its original blue-blue form.
The new center has $\QQ(c+v)=1+1+0=0$, since $c$ and $v$ are non-adjacent.
Pairing the new center
with the chosen length-one representative $b$ gives the grey-blue pair displayed in
panel (b). To isolate this pair from the remaining length-two legs, consider
any such leg with middle vertex
$u_j$ adjacent to the center and terminal vertex $v_j$, as labeled in panels
(b) and (c). Contracting $b$ onto $u_j$ is implemented by the
basis replacement $u_j\mapsto b+u_j$:
it removes the center pairing because
$\symp{c}{b+u_j}=\symp{c}{b}+\symp{c}{u_j}=1+1=0$, while $b+u_j$ still pairs
with $v_j$. Since $b$ and $u_j$ are non-adjacent and both have $\QQ$-value $1$,
the new middle vector has $\QQ$-value $0$. Thus each remaining length-two leg
is isolated as the displayed grey-blue pair and has Arf term $0$, and the
new center--$b$ pair is isolated with Arf term $0$. Hence the
total Arf sum is $1$.

For $\graphS^2_{n_2,n_1}$, first disconnect the length-four leg shown in panel (c).
This produces one blue-blue pair and one grey-blue pair, hence contributes one
Arf term $1$. The basis change which removes the last edge to the center leaves
the resulting center with $\QQ$-value $1$ at this intermediate stage. Thus,
after this length-four leg has been split off, the remaining star is handled
exactly as in the $\graphS^1_{n_2,n_1}$ case. The ordinary length-two leg
labeled $u,v$ is isolated by the same replacement of the center as above and
contributes the single Arf term $1$, while the chosen length-one
representative $b$ and every remaining ordinary length-two leg give only grey-blue
isolated pairs of Arf term $0$. After these replacements, the center is changed to $\QQ$-value $0$,
as displayed in panel (c). Thus the two non-zero Arf terms give $1+1=0$ in the
Arf sum.

For $\graphX^1_{2m-1,n_1}$ and $\graphX^2_{2m-1,n_1}$, passing to the
quotient by $B$ makes the subspace spanned by the length-one leaf pairs
$b_i+b_{i+1}$ vanish.
Panels (d) and (e) of Figure~\ref{fig:quadratic-form-arf-peeling-proof}
display respectively the corresponding disconnected pairs near the extra branch. After quotienting
by $B$, the blue length-one leaf represents the common image of all length-one
leaves, while the white length-one leaves differ from it by elements of $B$ and
hence do not give additional Arf-sum vectors.
In panel (d), the displayed pairs give Arf terms
$1$, $0$, $0$,
namely from $(a_2,a_3)$, $(a_1,a_2+a_4)$, and $(a_1+a_5,a_6)$. The
displayed center--leaf pair $(c,b)$ records the remaining length-one leaf image
and has Arf term $0$. After this local block,
the usual alternating recurrence from panel (a) resumes. Hence the Arf sum is
$\lfloor m/2\rfloor$ for $\graphX^1_{2m-1,n_1}$.
In panel (e), the extra length-two branch contributes one additional
blue-blue pair of Arf term $1$, and the remaining spine follows the common
alternating recurrence. Hence the Arf sum is $\lfloor m/2\rfloor+1$ for $\graphX^2_{2m-1,n_1}$.
\end{proof}

\begin{proof}[Proof of Proposition~\ref{prop:isomorphism_classes_quadratic_forms_arf_invariant_canonical}]
We apply Lemma~\ref{lem:canonical_quadratic_radical_values} to prove the asserted
statements about $\radzero(V,\QQ)$ and $\radone(V,\QQ)$.
This also identifies the cases for which $\QQ$ is non-zero on the radical.
Now by Proposition~\ref{prop:Quadratic_Form_Isomorphism_Classes}, these are precisely
type-$0$ cases: $\graphX^3_{2m,n_1}$, $\graphS^3_{n_2,n_1}$, and
$\graphP_{2m,n_1}$ for $m$ even.

In all remaining cases, $\QQ$ vanishes on the radical.
Lemma~\ref{lem:canonical_quadratic_arf_counts} computes the Arf invariant on
$V/\rad(V)$.
Again by Proposition~\ref{prop:Quadratic_Form_Isomorphism_Classes}, Arf invariant
$0$ gives $\type(\QQ)=+$ and Arf invariant $1$ gives $\type(\QQ)=-$.
The mod-$4$ alternatives in
Table~\ref{tab:canonical-quadratic-arf-types} are exactly the parity
reductions of the listed Arf sums, so they give the remaining type
statements in the proposition.
\end{proof}

Hence, under the assumption that $\graphG$ is not $t$-equivalent to some blown-up path graph $\graphP_{k,n_1}$, we can completely determine the $t$-equivalent representative by the isomorphism class of its invariant quadratic form together with rank and nullity.
Here we apply the classification into canonical graphs based on
Theorems~\ref{thm:classes:humphries}
and~\ref{thm:classes:eisert}.
Namely, we have the following:
\begin{cor}\label{cor:distinguishing_non_path_canonical_representatives}
Let $\graphG$ be a graph not $t$-equivalent to $\graphP_{k,n_1}$, and let $\QQ$ be its unique invariant quadratic form.
Also, let $\rank(A(\graphG)) = 2m$ and $\nullity(A(\graphG)) = r$. Here $n_1$ denotes the number of length-one leaves;
equivalently, $n_1=r+1$ for $\graphS_{n_2,n_1}^1,\graphS_{n_2,n_1}^2,\graphX_{2m-1,n_1}^1,\graphX_{2m-1,n_1}^2$,
and $n_1=r$ for $\graphS_{n_2,n_1}^3,\graphX_{2m,n_1}^3$.
For the star representatives, the rank condition $\rank(A(\graphG))=2m$
fixes $n_2=m-1$ for $\graphS_{n_2,n_1}^1$, $n_2=m-3$ for
$\graphS_{n_2,n_1}^2$, and $n_2=m-2$ for $\graphS_{n_2,n_1}^3$.
Then $\graphG$ is $t$-equivalent to the following representative(s):
\begin{enumerate}
    \item If $\type(\QQ)=+$, then $\graphG$ is $t$-equivalent to $\graphS_{m-3,n_1}^2$; moreover, it is $t$-equivalent to $\graphX_{2m-1,n_1}^1$ when $m\bmod 4\in\{0,1\}$, and to $\graphX_{2m-1,n_1}^2$ when $m\bmod 4\in\{2,3\}$.
    \item If $\type(\QQ)=-$, then $\graphG$ is $t$-equivalent to $\graphS_{m-1,n_1}^1$; moreover, it is $t$-equivalent to $\graphX_{2m-1,n_1}^1$ when $m\bmod 4\in\{2,3\}$, and to $\graphX_{2m-1,n_1}^2$ when $m\bmod 4\in\{0,1\}$.
    \item If $\type(\QQ)=0$, then $\graphG$ is $t$-equivalent to both of the representatives $\graphS_{m-2,n_1}^3$ and $\graphX_{2m,n_1}^3$.
\end{enumerate}
\end{cor}
In Section~\ref{sec:Line_Graphs_E6_Condition}, we shall consider conditions which are equivalent to stating $t$-(in)equivalence to a blown-up path graph, which provide witness without directly showing $t$-(in)equivalence through a series of valid contractions.

\subsection{Quadratic Forms and Algebraic Dependencies}\label{sec:quadratic_forms_with_alg_dep}

Unlike the linearly or algebraically independent case, when the generating set has algebraic dependencies, there is no guarantee of existence of quadratic forms.
This is also reflected in the loss of invariant quadratic forms under projection, despite the fact that there is always an extension with non-trivial invariant quadratic forms.

However, we can predict what happens depending on the setting,
\begin{lem}\label{lem:quadratic_forms_under_projections_extensions}
Let $\vgens\subseteq\Fn$ be a set of binary vectors, with $V=\Span{\vgens}$, and $\tilde{\vgens}$ its unique linearly independent extension with the same frustration graph. Denote with $\projection$ the projection map and $\inclusion$ the inclusion map.
Also, let $\tilde{\QQ} = \QQ_{\tilde{\vgens}}$ be the unique invariant quadratic form for $\tilde{\vgens}$ over $\tilde{V} = \Span{\vgens}$.
Then:
\begin{enumerate}
    \item $\vgens$ has an invariant quadratic form $\QQ$ over $V$ if and only if $\ker(\projection)\subseteq\radzero(\tilde{V},\tilde{\QQ})$
    \item If $\vgens$ has an invariant quadratic form $\QQ$ over $V$, then $\type(\QQ)=\type(\tilde{\QQ})$
\end{enumerate}
Furthermore:
\begin{itemize}
    \item If $\vgens$ allows for an invariant quadratic form $\QQ$,  then $\projection(\radzero(\tilde{V},\tilde{\QQ})) = \radzero(\tilde{V},\QQ)$ holds.
    \item $\vgens$ has no invariant quadratic form if and only if $\type(\tilde{\QQ})=0$ and $\QQ|_{\ker(\projection)} = \F_2$.
    \item If $\type(\tilde{\QQ})\in\{+,-\}$, any restriction will also have an invariant quadratic form, with the same type.
\end{itemize}
\end{lem}
\begin{proof}
Applying Lemma~\ref{lem:projection_extension_basic_properties}\ref{lem:projection_extension_basic_properties:a}, we have $\ker(\projection)\subseteq\rad(\tilde{V})$ and $\projection(\rad(\tilde{V})) = \rad(V)$.
By the definition of the inclusion map, we have $i(\vgens) = \tilde{\vgens}$.

Assume first that $\QQ$ exists. By invariance of $\QQ$ and $\tilde{\QQ}$, for each $v\in \vgens$, we have $\QQ(v) = \tilde{\QQ}(i(v)) = \tilde{\QQ}(\tilde{v})=1$. 
By the coefficient formula in Lemma~\ref{lem:uniqueness_quadratic_form_for_linearly_dep_set}, the value of an invariant quadratic form on a linear combination of generators is determined by the coefficients and the pairwise symplectic products of the generators. 
Since $\tilde{\vgens}$ and $\vgens$ have the same frustration graph, we have
\begin{equation*}
    \QQ\Big(\sum_j c_jv_j\Big)
    =
    \tilde{\QQ}\Big(\sum_j c_j\tilde{v}_j\Big)
\end{equation*}
for every coefficient vector $c$.
Then, consider $\tilde{u}=\sum_jc_j\tilde{v}_j\in\ker(\projection)$, we have
\begin{equation}
        0 = \QQ(0) = \QQ\Big(\sum_jc_jv_j\Big) = \tilde{\QQ}\Big(\sum_jc_j\tilde{v}_j\Big)
\end{equation}
which implies $\tilde{u}\in\radzero(\tilde{V},\tilde{\QQ})$.
Also, since $\projection$ is surjective, for any $u\in V$ there is a $\tilde{u} = \sum_jc_j\tilde{v}_j$ such that $\projection(\tilde{u}) = u$, hence $u=\sum_jc_jv_j$.
Thus, $\QQ(\projection(\tilde{u})) = \tilde{\QQ}(u)$, which implies
\begin{equation}\label{eq:projected_QQ}
    \QQ\circ \projection = \tilde{\QQ}|_{\tilde{V}/\ker(\projection)} \then \QQ = \tilde{\QQ}\circ\inclusion,
\end{equation}
where we used the fact that $\projection$ acts injectively on $\inclusion(V)\subseteq \tilde{V}$, which is naturally identified as (a representative of) the subspace $\tilde{V}/\ker(\projection)$, such that $\inclusion \circ \projection|_{\tilde{V}/\ker(\projection)} = \id_{\tilde{V}/\ker(\projection)}$.
Since $\ker(\projection)\subseteq \radzero(\tilde{V},\tilde{\QQ})$, quotienting by $\ker(\projection)$ does not remove anisotropic vectors (if any) from the radical, i.e.
\[
    \tilde{\QQ}|_{\rad(\tilde{V})/\ker(\projection)}
    =
    \begin{cases}
        0, & \text{if } \tilde{\QQ}|_{\rad(\tilde{V})}=0,\\
        \F_2, & \text{if } \tilde{\QQ}|_{\rad(\tilde{V})}=\F_2,
    \end{cases}
\]
Also, we have that $\inclusion(\rad(V))$ is naturally identified with (a representative of) $\rad(\tilde{V})/\ker(\projection)$.
Combining this with Eq.~\eqref{eq:projected_QQ}, we have that the action of $\QQ$ on $\rad(V)$ is the same as that of $\tilde{\QQ}$ of $\rad(\tilde{V})$
$$\QQ|_{\rad(V)} = \tilde{\QQ}|_{\inclusion(\rad(V))} = \tilde{\QQ}|_{\rad(\tilde{V})/\ker(\projection)}.$$
Similarly, a non-degenerate subspace of $V$ is naturally identified via $\inclusion$ to a non-degenerate subspace of $\tilde{V}/\ker(\projection)$, on which $\QQ$ and $\tilde{\QQ}$ respectively act the same, as a consequence of Eq.~\eqref{eq:projected_QQ}.
As such, $\QQ$ acts the same on the radical and has the same Arf invariant (which only depends on the non-degenerate part) as $\tilde{\QQ}$.
Thus, $\QQ$ and $\tilde{\QQ}$ have the same type, which proves (b).

Viceversa, assume that there is a $u\in\ker(\projection)\subseteq\rad(\tilde{V})$ such that $\tilde{\QQ}(u)=1$.
This can only happen if $\type(\tilde{\QQ})=0$.
We prove by contradiction and assume that an invariant quadratic form $\QQ$ exists.
Then, for $u=\sum_jc_j\tilde{v}_j$, we have:
\begin{equation*}
        1 = \QQ(\sum_jc_j\tilde{v}_j)
        = \QQ(\sum_jc_jv_j) = \QQ(0) = 0
\end{equation*}
which is absurd.
This proves (a) and the other statements immediately follow, which concludes the proof.
\end{proof}

Now, using Lemma~\ref{lem:quadratic_forms_under_projections_extensions}, it is straightforward to generalize Prop~\ref{prop:isomorphism_classes_quadratic_forms_arf_invariant_canonical} to the case where the generating sets are minimal but have an algebraic dependency:
\begin{prop}\label{prop:quadratic_forms_arf_invariant_canonical_graph_t_representatives_alg_dep_cases}
Let $\graphG$ be one of the canonical graph families
$\graphP_{2m,n_1}$, $\graphX_{2m,n_1}^3$, or $\graphS_{n_2,n_1}^3$.
Let $\vgens^D(\graphG)$ be the minimal unique generating set
(by Lemma~\ref{lem:bijection_graphs_linearly_dependent_sets_with_fixed_algebraic_dependencies})
with frustration graph $\graphG$ and one algebraic dependency.
Then, $\vgens^D(\graphX_{2m,n_1}^3)$, $\vgens^D(\graphS_{n_2,n_1}^3)$ do not admit invariant quadratic forms.
Also, for the even blown-up path graph $\vgens=\vgens^D(\graphP_{2m,n_1})$ with $m\geq 2$, we obtain
\begin{enumerate}
    \item no invariant quadratic forms for $m\bmod 4 \in\{0,2\}$,
    \item an invariant quadratic form of the same type as the algebraically independent case $\vgens(\graphP_{2m,n_1})$ for $m\bmod 4\in\{1,3\}$; equivalently, $\type(\QQ)=-$ if $m\bmod 4=1$ and $\type(\QQ)=+$ if $m\bmod 4=3$.
\end{enumerate}
\end{prop}
\begin{proof}
Consider first $\graphX_{2m,n_1}^3$, $\graphS_{n_2,n_1}^3$ and $\graphP_{2m,n_1}$ ($m\geq 2$) for $m$ even.
By Proposition~\ref{prop:isomorphism_classes_quadratic_forms_arf_invariant_canonical}, the coloring corresponding to the algebraic dependency is in the anistropic radical of the extension.
Hence, we have that $\QQ|_{\ker(p)} = \F_2$ and by Lemma~\ref{lem:quadratic_forms_under_projections_extensions}(a), the linearly dependent generating sets have no invariant quadratic forms.

For $\graphP_{2m,n_1}$ and $m$ odd, the entire radical is isotropic, hence also $\QQ|_{\ker(p)} = 0$ and $\vgens^D(\graphP_{2m,n_1})$ has an invariant quadratic form of the same type as $\vgens^D(\graphP_{2m,n_1})$ by Lemma~\ref{lem:quadratic_forms_under_projections_extensions}(b).
\end{proof}
Notice that Propositions~\ref{prop:isomorphism_classes_quadratic_forms_arf_invariant_canonical}
and~\ref{prop:quadratic_forms_arf_invariant_canonical_graph_t_representatives_alg_dep_cases},
together with Theorems~\ref{thm:classes:humphries}
and~\ref{thm:classes:eisert}, imply a full characterization of
\emph{any} generating set whose frustration graph is \emph{not}
$t$-equivalent to some $\graphP_{k,n_1}$.

\begin{cor}\label{cor:distinguishing_quasi_universal_cases}
Let $\vgens\subseteq\Fn$ be a set of binary vectors with connected frustration graph $\graphG$, which is not $t$-equivalent to some $\graphP_{k,n_1}$.
Also, let $\rank(\vgens)=2m$ (see Eq.~\eqref{eq:rank}) and $\nullity(\vgens)=r$. Then:
\begin{enumerate}
    \item If $\vgens$ has an invariant quadratic form $\QQ$, it has a minimal generating subset which is $t$-equivalent to $\vgens(\graphG)$ with $\graphG$ one of $\graphX_{2m-1,n_1}^1$, $\graphX_{2m-1,n_1}^2$ or $\graphX_{2m,n_1}^3$, depending on the isomorphism class of $\QQ$, and where $n_1=r+1$ in the first two cases and $n_1=r$ in the third case
    \item If $\vgens$ has no invariant quadratic forms, it has a minimal generating subset which is $t$-equivalent to $\vgens^D(\graphX_{2m,n_1}^3)$ where $n_1=r+1$
\end{enumerate}
\end{cor}
\begin{proof}
We first use the unique linearly independent extension, which always has an invariant quadratic form.
By Theorems~\ref{thm:classes:humphries}
and~\ref{thm:classes:eisert}, since $\graphG$ is not $t$-equivalent to some
$\graphP_{k,n_1}$, it must be $t$-equivalent to one of the remaining
$t$-equivalent representatives.
These are completely determined by the isomorphism class of the invariant quadratic form of the linearly independent extension, denoted as $\tilde{\QQ}$, by Proposition~\ref{prop:isomorphism_classes_quadratic_forms_arf_invariant_canonical}.

Then, we can project to our actual set and remove any Lie-algebraic dependencies, which can only remove legs of length $1$.
If the minimal generating set has no algebraic dependencies, it also has an invariant quadratic form, which is of the same type as $\tilde{\QQ}$.
Hence, it is $t$-equivalent to one of the classes in (a), depending on the rank and radical dimension.

If the minimal generating set has algebraic dependencies, it has no invariant quadratic forms by Proposition~\ref{prop:quadratic_forms_arf_invariant_canonical_graph_t_representatives_alg_dep_cases}.
Hence, the graph must be $t$-equivalent to $\graphX_{2m,n_1}^3$ with $n_1=r+1$.
Furthermore, if $\vgens$ has no invariant quadratic forms, it must have an algebraically dependent minimal generating set.
This proves that algebraic dependency is an equivalent condition to lack of quadratic forms, for minimal generating sets.

Conversely, if $\vgens$ has invariant quadratic forms, its minimal generating set must be algebraically independent.
This concludes the proof.
\end{proof}

Indeed, if such a $\vgens$ has no invariant quadratic forms, then it must be $t$-equivalent to some $\graphX_{2m,n_1}^3$, with $m$ and $n_1$ determined by the rank and radical (plus any Lie-algebraic dependencies).
Furthermore, the minimal generating set has a single algebraic dependency.
If such a $\vgens$ has an invariant quadratic form $\QQ$ over $V$, then it is $t$-equivalent to one of $\graphX_{2m-1,n_1}^1$, $\graphX_{2m-1,n_1}^2$ or $\graphX_{2m,n_1}^3$ depending on the isomorphism class of $\QQ$ (or, as we shall see $\quadratic(\vgens)$).
Furthermore, the minimal generating set has no algebraic dependencies.

\subsection{Change of Basis}

By Definition~\ref{def:quadratic-form-associated-vector}, we view vectors $w\in\Fn$ as quadratic forms $\QQ_w$ relative to the reference form $\QQ_0$. It remains to rephrase change of basis and isomorphism classes of quadratic forms in terms of these vectors and their affine subspaces. Since we are interested in invariant quadratic forms, we directly frame the basis change of quadratic forms as $\QQ' = \QQ\circ \activemap^{-1}$, which guarantees $\QQ(v)=1$ iff $\QQ'(\activemap v)=1$.
This is the change-of-basis convention from Section~\ref{sec:isomorphism_classes_bilinear_quadratic_forms}, where isomorphic quadratic forms satisfy $\QQ'(\activemap v)=\QQ(v)$, equivalently $\QQ'=\QQ\circ\activemap^{-1}$.
Then, we can also re-phrase the concept of isomorphism of quadratic forms in terms of their vectors $w$:

\begin{defn}[Affine lift relative to $\QQ_0$]\label{def:affine-lift-relative-Q0}
Let $\activemap\in\Sp(2n,\F_2)$. Relative to the reference quadratic form $\QQ_0$, define the translation vector $\trans{\activemap}\in\Fn$ based on the coordinate formula suggested by Eq.~\eqref{eq:explicit_w_vector_from_quadratic_form}:
\begin{equation}\label{eq:translation-vector-relative-Q0}
    \trans{\activemap}
    := \sum_i \qty(\QQ_0(\activemap^{-1}f_i)e_i + \QQ_0(\activemap^{-1}e_i)f_i).
\end{equation}
The affine transformation $\aff{\activemap}\in\Aff(\Fn)$ is defined by
\begin{equation}\label{eq:affine-lift-relative-Q0}
    \aff{\activemap}w := \trans{\activemap}+\activemap w
    \quad\text{ for } w\in\Fn.
\end{equation}
\end{defn}
With this notation, the change of basis of quadratic forms becomes an affine transformation of their associated vectors.
\begin{lem}\label{lem:change_of_basis_vector_quadratic_form_affine_transformation}
Let $w\in\Fn$, $\activemap\in\Sp(2n,\F_2)$, and $\QQ_{w'} = \QQ_w\circ \activemap^{-1}$. Then
$w' = \aff{\activemap}w$, where $\aff{\activemap}$ is the affine transformation from Definition~\ref{def:affine-lift-relative-Q0}.
\end{lem}
\begin{proof}
We apply the definition of $w'$ using Eq.~\eqref{eq:explicit_w_vector_from_quadratic_form}, together with the fact that the components of $v$ in the $e_i,f_i$ basis are given by $\symp{v}{f_i}$ and $\symp{v}{e_i}$ respectively:
\begin{align*}
        w' &= \sum_i \QQ_{w'}(f_i)e_i + \QQ_{w'}(e_i)f_i \\
        &= \sum_i \QQ_w(\activemap^{-1}f_i)e_i + \QQ_w(\activemap^{-1}e_i)f_i \\
        &=
        \begin{aligned}[t]
        &\sum_i (\QQ_0(\activemap^{-1}f_i)e_i + \QQ_0(\activemap^{-1}e_i)f_i) \\
        &+ \sum_i (\symp{w}{\activemap^{-1}f_i}e_i + \symp{w}{\activemap^{-1}e_i}f_i)
        \end{aligned}
        \\
        &= \trans{\activemap} + \sum_i (\symp{\activemap w}{f_i}e_i + \symp{\activemap w}{e_i}f_i)
        = \trans{\activemap} + \activemap w
\end{align*}
which proves the statement.
\end{proof}
As an immediate corollary, we have that $\QQ_0(\activemap^{-1}v) = \QQ_0(v) + \symp{\trans{\activemap}}{v}$ for any $v\in\Fn$.
In the case of transvections $\activemap=\tau_u=\activemap^{-1}$, we have specifically $\trans{\activemap} = 0$ for $\QQ_0(u)=1$ and $\trans{\activemap} = u$ for $\QQ_0(u)=0$, or compactly $\trans{\activemap} = (1+\QQ_0(u))u$.

Notice that if $\activemap\in \lieO(\QQ_0)$, we have $\trans{\activemap} = 0$ and the change of basis gets pulled back to the vector $\QQ_w\circ \activemap^{-1} = \QQ_{\activemap w}$.
This shows that, at least for orthogonal transformations $\activemap\in \lieO(\QQ_0)$, the vectors $w$ are the natural object to work with, since they change the same way as the generators, namely $\vgens' = \activemap\vgens$ implies $w' = \activemap w$.
More generally, the set of affine symplectic transformations is a group, since it provides a homomorphism of $\Sp(2n,\F_2)$ in $\Aff(\Fn)$:
\begin{align*}
        &\aff{\activemap}\aff{\activemap'}(v)
        = \aff{\activemap}(\trans{\activemap'} + \activemap'v)\\
        &= \trans{\activemap}
        + \activemap \trans{\activemap'}
        + \activemap\activemap'v,\\
        &\aff{\activemap\activemap'}(v)
        = \trans{\activemap\activemap'} + \activemap\activemap'v,\\
        &\trans{\activemap\activemap'}
        = \sum_i \QQ_0((\activemap\activemap')^{-1}f_i)e_i
         + \sum_i\QQ_0((\activemap\activemap')^{-1}e_i)f_i \\
        &=
        \begin{aligned}[t]
        & \sum_i\QQ_0(\activemap^{-1}f_i)e_i
        + \sum_i\QQ_0(\activemap^{-1}e_i)f_i \\
        &+ \sum_i\symp{\trans{\activemap'}}{\activemap^{-1}f_i}e_i
        + \sum_i\symp{\trans{\activemap'}}{\activemap^{-1}e_i}f_i
        \end{aligned}\\
        &= \trans{\activemap} + \activemap \trans{\activemap'}.
\end{align*}
In particular, we have
\begin{align*}
    &\trans{\id}
    = \trans{\activemap\activemap^{-1}}
    = 0 = \trans{\activemap} + \activemap \trans{\activemap^{-1}}\\
    &\text{and }\; \trans{\activemap^{-1}} = \activemap^{-1}\trans{\activemap}.
\end{align*}
Also, for any $\activemap$, $\QQ_0(\trans{\activemap}) = \Arf(\QQ_0) = 0$.
We can also show that the lifts $\aff{\activemap}$ are precisely the affine transformations with linear part $\activemap$ which conserve $\QQ_0$. To see this, let $(b,\activemap)\in\Aff(\Fn)$ with $(b,\activemap)v=b+\activemap v$ satisfy $\QQ_0\circ(b,\activemap)=\QQ_0$. Then:
\begin{align*}
        \QQ_0(v) &= \QQ_0((b,\activemap)v)
        = \QQ_0(b + \activemap v) \\
        &= \QQ_0(b) + \QQ_0(\activemap v) + \symp{b}{\activemap v} \\
        &= \QQ_0(b) + \QQ_0(v) + \symp{\trans{\activemap^{-1}}}{v} + \symp{\activemap^{-1}b}{v}, \\
        \QQ_0(b) &=
        \symp{\trans{\activemap^{-1}}+\activemap^{-1}b}{v}.
\end{align*}
Since this needs to be true for all $v$, the linear functional on the right must be constant.
Hence $\QQ_0(b)=0$ and $\trans{\activemap^{-1}}+\activemap^{-1}b=0$, since no non-zero vector commutes with all $v\in\Fn$. This shows that $b$ and $\activemap$ must satisfy $b = \activemap \trans{\activemap^{-1}} = \trans{\activemap}$, so $(b,\activemap)=\aff{\activemap}$. Thus the orthogonal affine transformations have the form
\begin{equation}
\AO := \{ \aff{\activemap} \text{ for } \activemap\in\Sp(2n,\F_2)\}.
\end{equation}
Thus, isomorphism of quadratic forms can be read as the orbit relation of this affine orthogonal action on the associated vectors.
\begin{lem}
Two quadratic forms $\QQ_w,\QQ_{w'}$ are isomorphic, i.e., $\QQ_w \circ \activemap^{-1} = \QQ_{w'}$ for $\activemap\in\Sp(2n,\F_2)$, if and only if their corresponding vectors $w,w'$ are related by an affine orthogonal transformation $w' = \aff{\activemap}w$, where $\aff{\activemap}$ is defined in Eq.~\eqref{eq:affine-lift-relative-Q0}.
\end{lem}

In order to study the isomorphism classes, we note that the Arf invariant of $\QQ_w$ (when viewed over all of $\Fn$, hence the Arf invariant completely determines the isomorphism classes) written in terms of $w$ and $\QQ_0$:
\begin{align*}
        \Arf(\QQ_w) &= \sum_i \QQ_w(e_i)\QQ_w(f_i) \\
        &=
        \begin{aligned}[t]
        &\sum_i (\QQ_0(e_i)\QQ_0(f_i) + \symp{w}{e_i}\QQ_0(f_i) \\
        &+ \symp{w}{f_i}\QQ_0(e_i) + \symp{w}{e_i}\symp{w}{f_i})
        \end{aligned}
        \\
        &= \Arf(\QQ_0) + \QQ_0(w) = \QQ_0(w)
\end{align*}
Hence two quadratic forms $\QQ_w$ and $\QQ_{w'}$ over $\Fn$ are isomorphic if and only if $\QQ_0(w) = \QQ_0(w')$, for any $w,w'\in\Fn$.

Indeed, for any $w\in\Fn$, we have that $\QQ_0$ is an invariant under change of basis $w'=\aff{\activemap}w$, or $\QQ_0(w) = \QQ_0(w')$, which follows trivially from the fact $\QQ_0\circ \aff{\activemap} = \QQ_0$.
We can then use the properties of this group of transformations to replace the ordinary equivalence classes of non-degenerate quadratic forms, specified by $\Arf(\QQ_w)$, with those in terms of vectors, specified by $\QQ_0(w)$.
We can also highlight how the action of the affine transformations for vectors $w$ precisely matches that of the change of basis for quadratic forms $\QQ_w$, namely that its orbits are precisely the levels sets of $\QQ_0(w)$ (instead of the Arf invariant $\Arf(\QQ_w)$):
\begin{lem}
Let $w,w'\in\Fn$. Then, there is a $\activemap\in\Sp(2n,\F_2)$ such that $w' = \aff{\activemap}w$ if and only if $\QQ_0(w)=\QQ_0(w')$.
\end{lem}
First, we recall the one-dimensional consequence of Witt's extension theorem
from Corollary~\ref{cor:witt-transitivity-fixed-value}. If $v,v'\neq0$ satisfy
$\QQ_0(v)=\QQ_0(v')$, then there exists an element of $\lieO(\QQ_0)$ sending
$v$ to $v'$.
Thus, $\lieO(\QQ_0)$ acts transitively on the \emph{non-zero} isotropic and anisotropic vectors of $\QQ_0$, respectively, i.e. there is some $\activemap\in \lieO(\QQ_0)$ such that $v' = \activemap v$ for some non-zero $v,v'\in\Fn$ if and only if $\QQ_0(v) = \QQ_0(v')$.

We can also provide an alternative argument here using transvections.
For any $n$, if $v,v'$ also anti-commute, we can see this explicitly in terms of transvections, since $v' = \tau_{v+v'}v$ and $\QQ_0(v+v') = \QQ_0(v) + \QQ_0(v') + \symp{v}{v'} = 1$, or $\tau_{v+v'}\in\lieO(\QQ_0)$.
If $v,v'$ do not commute, we can still show this explicitly in terms of transvections whenever $n\geq 3$, or $\dim(\Fn)\geq 6$.
In this case, for any two distinct vectors $v_1,v_2$, we can construct a symplectic basis $\{v_1,f_1,v_2,f_2\}$ where $\symp{v_i}{f_i}=1$ and the other products are zero. Furthermore, the orthogonal complement of this basis is non zero (it is not of maximal rank) and we can choose $v_3$ such that $\symp{v_i}{v_j} = 0$ for all $i,j$ and then we have $v_2 = \tau_{f_1+v_2+f_2+v_3}\tau_{v_1+f_1+f_2+v_3}v_1$.
Notice that this argument does not work when $n=2$, despite the fact that the orthogonal group $\lieO(\QQ_0)$ acts transitively on the non-zero isotropic and isotropic vectors, respectively.
Indeed, one can check that the $\QQ_0$-orthogonal transvections produce two distinct orbits in $\QQ_0^{-1}(1)$ and do not generate the entire orthogonal group $\lieO(\QQ_0)$.
However, in this case one can also define the swap matrix $Sv_1 = v_2$, $Sf_1=f_2$ and $S^2=\id$ over some symplectic basis defined by $v_1,v_2\in\Fn$. This is clearly symplectic (given $\symp{v_1}{v_2}=0$) and orthogonal for $\QQ_0$ (given $\QQ_0(v_1)=\QQ_0(v_2)$, hence connects the quadratic forms $\QQ_{v_1}$ and $\QQ_{v_2}$.

For any $n$, we see that orthogonal transvections $\activemap\in \lieO(\QQ_0)$ are not sufficient to connect all of isomorphic quadratic forms, since they do not connect $\QQ_0$ and $\QQ_w$ with $w\neq 0,\QQ_0(w)=0$.
It is however clear that, for any $\QQ_0(w)=1$, the affine transvection $\aff{\tau_w}$ brings any $w$ into $0$ as $\aff{\tau_w}(w) = w + \tau_ww = 0$. We will use this fact to find representatives of isomorphism classes.

Given the change of basis in Lemma~\ref{lem:change_of_basis_vector_quadratic_form_affine_transformation}, the sets of invariant quadratic forms for some set $\vgens\subseteq\Fn$ change as follows under any $\activemap\in\Sp(2n,\F_2)$:
\begin{equation}
    \quadratic(\activemap\cdot\vgens) = \aff{\activemap}\cdot\quadratic(\vgens).
\end{equation}
Then, the affine transformations $\aff{\activemap}\in\AO$ provide the right equivalence relation for quadratic forms in vector form.
We are now ready to study \emph{equivalence classes} of sets of invariant quadratic forms.
Given Lemma~\ref{lem:unique_restricted_invariant_quadratic_form}, one can also expect a close connection with the classification of single quadratic forms (see Proposition~\ref{prop:Quadratic_Form_Isomorphism_Classes}).
Indeed, we shall prove precisely this.
Furthermore, given that each space of invariant quadratic forms is an affine subspace, we shall first understand the isomorphism classes for affine subspaces and then prove which of these can in fact be obtained from invariant quadratic forms for transvections (see Corollary~\ref{cor:Canonical_Forms_Invariant_Bilinear_Forms}).

First, we provide a preliminary lemma, which always allows us to choose a special basis for $\quadratic(\pgens)$, or more generally affine subspaces:
\begin{lem}\label{lem:convenient_basis_invariant_quadratic_forms}
Consider the affine subspace $w+W$ and a symplectic basis $\calB = \{e_i,f_i\}_{i=1}^{\ell}\cup\{h_j\}_{j=1}^r$ for $W$. There is a $w'\in w+W$ such that one and only one of the following holds:
\begin{enumerate}
    \item $\symp{w'}{W} = 0$
    \item $\symp{w'}{\rad(W)} = \F_2$, and specifically $w'$ only anti-commutes with a subset of $\{h_j\}$
\end{enumerate}
Also, in case (b) one can choose a new symplectic basis for the radical $\{h_j'\}$ such that $w'$ only anti-commutes with $h_1'$: $\symp{h_1'}{w'} = 1$ and $\symp{h_{j>1}'}{w'} = 0$.
\end{lem}
\begin{proof}
Consider some symplectic basis $\calB$ and initial reference $w$.
It's straightforward to show by a series of contractions/symplectic projection of $\{e_i,f_i\}$ onto $w$ that we can choose a $w$ that commutes with all of them, namely $w \mapsto w' = w + \sum_i\symp{w}{e_i}f_i + \symp{w}{f_i}e_i$.
Now we have two cases: the new $w'$ commutes with all $\{h_j\}$, hence also all of $\calB$, which proves case (a); $w'$ anti-commutes with some of $\{h_j\}$, without loss of generality we may choose $h_1$ to be in this set, and proves (b).
Clearly, they do not overlap, since, for any $w' = w+v\in w+W$ and for any $u\in\rad(W)$, we have $\symp{w+v}{u} = \symp{w}{u}$, hence the commutation relations with of $w+W$ with $\rad(W)$ are independent of the choice of representative $w$.
In case (b), we choose a new symplectic basis $\calB'$ where we perform contractions of $h_1$ onto all other elements which anti-commute with $w'$, namely $h_j' = h_j + \symp{w'}{h_j}h_1$.
This concludes the proof.
\end{proof}

\subsection{Isomorphism Classes}\label{sec:affine_subspaces_quadratic_forms_isomorphism_classes}

We now arrive the classification of isomorphism classes of \emph{affine subspaces} with respect to this equivalence relation. Then, we shall consider which of these cases are in fact allowed in general and within the context of transvections:
In the notation of Definition~\ref{def:affine-lift-relative-Q0}, the affine lift $\aff{\activemap}=(\trans{\activemap},\activemap)\in\AO[\QQ_0]$ is the affine orthogonal transformation of the reference form $\QQ_0$ associated with $\activemap\in\Sp(2n,\F_2)$.
\begin{thm}\label{thm:classification_of_affine_subspaces_isomorphism_classes}
Recall the canonical reference quadratic form $\QQ_0$  from Definition~\ref{def:quadratic-form-associated-vector}, and write $\AO=\AO[\QQ_0]$.
Consider an affine subspace $w+W\subseteq\Fn$.
There is a change of basis $\aff{\activemap}=(\trans{\activemap},\activemap)\in\AO$ such that $w'+W' = \aff{\activemap}(w+W)$ where $W'$ is spanned by the non-empty canonical symplectic basis $\{e_i,f_i\}_{i=1}^\ell\cup\{e_{j+\ell}\}_{j=1}^r$.
Write $m=n-\ell-r$.
Then $w'$ can be one and only one of three cases: (1) $w'=0$ (only for $m\geq 1$), (2) $w'=e_{\ell+r+1}+f_{\ell+r+1}$ (only for $m\geq 1$), (3) $w'=f_{\ell+1}$ (only for $r\geq 1$).

Furthermore, we have that the three classes are in bijection with exactly of the restricted quadratic forms $\QQ^* = \QQ_w|_V$ over $V=W^\perp$ (see Proposition~\ref{prop:Quadratic_Form_Isomorphism_Classes}):
\begin{enumerate}
    \item $\type(\QQ_w|_V)=+$ iff $w'=0$
    \item $\type(\QQ_w|_V)=-$ iff $w'=e_{\ell+r+1}+f_{\ell+r+1}$
    \item $\type(\QQ_w|_V)=0$ iff $w'=f_{\ell+1}$
\end{enumerate}
\end{thm}
\begin{proof}
First, we choose a symplectic basis and perform a change of basis such that $W$ has the canonical symplectic basis $\{e_i,f_i\}_{i=1}^\ell\cup\{e_{j+\ell}\}_{j=1}^r$, or equivalently that $W^\perp$ has symplectic basis $\{e_{j+\ell}\}_{j=1}^r\cup \{e_{i+\ell+r},f_{i+\ell+r}\}_{i=1}^m$, which in the Pauli matrix algebra language corresponds to the mutual canonical form in Eq.~\eqref{eq:canonical_form_pauli_matrix_algebra_and_commutant}.
Applying Lemma~\ref{lem:convenient_basis_invariant_quadratic_forms}, we can choose a reference quadratic form $w$ such that either (i) $w$ commutes with $\calB$ or (ii) $w$ anti-commutes only with some elements in the radical basis.

Consider first case (i), i.e. $w\in V=W^\perp$.
Without loss of generality, we can choose $w$ to live in a supplementary subspace of $\rad(W)=\rad(V)$ in $V$, which is isomorphic to $V/\rad(V)$.
Specifically, we consider the subspace spanned by $\{e_{i+\ell+r},f_{i+\ell+r}\}_{i=1}^m$, which is non-degenerate.
Then, let $\aff{\activemap}=(\trans{\activemap},\activemap)\in\AO[\QQ_0|_{V/\rad(V)}]$ be an affine orthogonal transformation in this subspace.
By Corollary~\ref{cor:affine-witt-extension-nondegenerate-summand}, this affine map extends to all of $\Fn$ as an affine orthogonal transformation whose linear part fixes $W$ pointwise.
Then, we have that $w'\in V/\rad(V)$ is conjugate to $w$ if and only if $\QQ_0(w)=\QQ_0(w')$.

Hence, if $\QQ_0(w)=0$, we can choose $w'=0$, while leaving the symplectic basis and the space unchanged.
Similarly, if $\QQ_0(w)=1$, we can choose $w'=e_{m+r+1}+f_{m+r+1}$.
This proves cases (a) and (b).

Consider now case (ii).
If $w$ anti-commutes with some elements in the radical basis $j\in J$. First choose a new basis for the radical such that $w$ only anti-commutes with $e_1$, and the rest are changed to $e_j+\symp{w}{e_j}e_1$.

Then, we perform a transformation $\activemap$ which sends this basis into the original one $\activemap(e_j+\symp{w}{e_j}e_1)=e_j$ for $j\geq 2$, sends $f_1$ into $f_1+\sum_{j>1}\symp{w}{e_j}f_j$, and leaves all other vectors unchanged.
This preserves commutation relations (consider here $1\leq i,j\leq r$), since $\symp{\activemap e_1}{\activemap e_j}=0$, $\symp{\activemap f_i}{\activemap f_j}=0$, $\symp{\activemap e_1}{\activemap f_j} = \delta_{1j}$, $\symp{\activemap e_{i>1}}{\activemap f_1} = \symp{w}{e_i}\symp{e_1}{f_1}+\sum_j\symp{w}{e_j}\symp{e_i}{f_j} = 0$ and $\symp{\activemap e_{i>1}}{\activemap f_{j>1}} = \delta_{ij}$.
Clearly, this is symplectic (preserves commutation relations) and orthogonal (it sends isotropic vectors into isotropic vectors).
Therefore we have $\aff{\activemap}w = \activemap w$ such that the commutation relations remain unchanged $\symp{w}{\calB} = \symp{\activemap w}{\activemap\calB}$.

Without loss of generality, we may now choose either $w = v + f_1$ or $w = v + e_1+f_1$ where $v\in V/\rad(V)$, since $\symp{w}{e_1} = 1$ and $\symp{w}{\calB\setminus\{e_1\}} = 0$
If $\QQ_0(v)=0$, we can perform the change of basis $\aff{\tau_v}w = v + w + \symp{v}{w}v = v + w = f_1$, which leaves $W$ unchanged.
If $\QQ_0(v)=1$, we first choose $w=v + e_1+f_1$ and then perform the change of basis $\aff{\tau_{v+e_1}}w = w + \symp{v+e_1}{w}(v+e_1) = f_1$.

It is now straightforward to show that for all the above cases the restricted quadratic form $\QQ^* = \QQ_w|_{W^\perp}$ is in fact of the type as in the statement of the theorem.
Indeed, write an arbitrary vector $u\in W^\perp$ as
\[
    u=\sum_{j=1}^r \gamma_j e_{\ell+j}
    +\sum_{i=1}^m(\alpha_i e_{\ell+r+i}+\beta_i f_{\ell+r+i})
    \in W^\perp .
\]
For $w'=0$, we obtain
$\QQ_0(u)=\sum_{i=1}^m \alpha_i\beta_i$,
so the restriction vanishes on the radical and has Arf invariant $0$, hence type $+$.
For $w'=e_{\ell+r+1}+f_{\ell+r+1}$, we obtain
\[
    \QQ_{w'}(u)=\sum_{i=1}^m \alpha_i\beta_i+\alpha_1+\beta_1,
\]
so the restriction still vanishes on the radical and has Arf invariant $1$, hence type $-$.
For $w'=f_{\ell+1}$, we have $\QQ_{w'}(e_{\ell+1})=1$ on the radical of $W^\perp$, hence the type is $0$, and proves case (c).

Finally, the isomorphism type of this quadratic form is an invariant under change of basis in the sense that, for any $w'\in w+W$, $\QQ_{w'}|_{W^\perp}$ and $(\QQ_{w'}\circ \activemap^{-1})|_{\activemap W^\perp}$ are in the same isomorphism class, which shows that in fact the three classes are distinct.
This concludes the proof.
\end{proof}

Given the correspondence between $w+W$ and  $\QQ^*=\QQ_w|_{W^\perp}$, we define
\[
    \type(w+W):=\type(\QQ^*).
\]
Then, for a given commutant, we have three distinct possibilities for isomorphism classes of invariant quadratic forms, corresponding to those above.

We can now ask which of these spaces can in fact be a space of invariant
quadratic forms for a transvection group. Since this question is invariant
under change of basis, it suffices to look at the representative classes from
Theorem~\ref{thm:classification_of_affine_subspaces_isomorphism_classes}.
The example in
Eq.~\eqref{eq:example_affine_space_is_not_space_of_invariant_quadratic_forms}
has $r=n-1$ and shows that, in the case $\type(w+W)=+$ with $m=1$, the
common anisotropic vectors need not recover the restricted space: they may span
a fully degenerate subspace of type $0$ rather than all of the restricted space
$V=W^\perp$.
Lemma~\ref{lem:exceptional_plus_affine_space_collapse} proves the same
result for arbitrary radical dimension $r$.

\begin{lem}\label{lem:exceptional_plus_affine_space_collapse}
Let $w+W\subseteq\Fn$ be an affine subspace, where $w\in\Fn$ and
$W\subseteq\Fn$ is a linear subspace. Set $V=W^\perp$ and
$\QQ^*=\QQ_w|_V$. Then the common anisotropic vectors of the quadratic forms in the affine space
$w+W$ are precisely
\[
    \bigcap_{w'\in w+W}\QQ_{w'}^{-1}(1)=\{v\in V \text{ for }\QQ^*(v)=1\}.
\]
Moreover, suppose that $w+W$ is a representative from
Theorem~\ref{thm:classification_of_affine_subspaces_isomorphism_classes} with
$\type(\QQ^*)=+$, and
that the restricted space $V$ has radical dimension $r$ and non-degenerate
dimension $2m$ with $m=1$. Then
\[
    \Span[\F_2]{\bigcap_{w'\in w+W}\QQ_{w'}^{-1}(1)}
    =
    \SpanS[\F_2]{\{e+f\}}\oplus\rad(V)
\]
for some symplectic pair $e,f$ in the non-degenerate part of $V$. In particular,
the span of the common anisotropic vectors is fully degenerate of dimension
$r+1$. Equivalently, if these
vectors are used as transvection generators, their span has type $0$ with
radical dimension $r+1$ and no non-degenerate part.
\end{lem}
\begin{proof}
Write $w'=w+u$ with $u\in W$. Then, for any $v\in\Fn$, recall from
Eq.~\eqref{eq:quadratic-form-associated-vector} that 
$\QQ_{w'}(v)=\QQ_w(v)+\symp{u}{v}$.
Thus $v$ is anisotropic for all forms in $w+W$ if and only if
$\symp{u}{v}=0$ for all $u\in W$ and $\QQ_w(v)=1$. This is equivalent to
$v\in W^\perp=V$ and $\QQ^*(v)=1$.

In the exceptional case, $\QQ^*$ has type $+$ with one symplectic pair and
$\QQ^*$ vanishes on $\rad(V)$. By the normal form for binary quadratic forms, choose
coordinates
$V=\SpanS[\F_2]{\{e,f\}}\oplus\rad(V)$ such that
\[
    \QQ^*(xe+yf+h)=xy \qquad \text{for }
    h\in\rad(V).
\]
Hence $\QQ^*(xe+yf+h)=1$ if and only if $x=y=1$. Equivalently, the image of
$e+f$ is the unique anisotropic vector in $V/\rad(V)$. The common anisotropic
vectors are therefore exactly
\[
    \bigcap_{w'\in w+W}\QQ_{w'}^{-1}(1)=e+f+\rad(V),
\]
and their span is
\[
    \SpanS[\F_2]{\{e+f\}}\oplus\rad(V),
\]
which is fully degenerate and has dimension $r+1$.
\end{proof}

We now show that the case $\type(w+W)=+$ with $m=1$ is the only
representative class for which the common anisotropic vectors fail to span $V=W^\perp$.

\begin{table*}[t]
\caption{Explicit anisotropic spanning sets for $V=W^\perp$ in the
non-exceptional cases of
Lemma~\ref{lem:exceptional_plus_affine_space_unique}. Here
$V=N\oplus\rad(V)$, the non-degenerate part $N$ has symplectic pairs
$e_i,f_i$, and $h_j$ is a basis of $\rad(V)$ with
$\isolong{e_i}=X_i$, $\isolong{f_i}=Z_i$,
$\isolong{h_j}=Z_{\ell+j}$.}
\label{tab:anisotropic-spanning-certificates}
\footnotesize
\begin{tabular*}{\textwidth}{@{\hspace{1mm}}l@{\extracolsep{\fill}}l@{\hspace{4mm}}l@{\hspace{1mm}}}
\hline\hline
\\[-2.3mm]
type & vector spanning set for $V=W^\perp$ & Pauli generating set
\\[0.4mm] \hline
\\[-2.0mm]
$+$, $\ell\geq2$
& $\{e_1{+}f_1,\ e_1{+}e_2{+}f_2\}
\cup\{e_1{+}f_1{+}e_i,\ e_1{+}f_1{+}f_i\}_{i=2}^\ell
\cup\{e_1{+}f_1{+}h_j\}_{j=1}^r$
& $\{Y_1,\ X_1Y_2\}
\cup\,Y_1\{X_i,Z_i\}_{i=2}^\ell
\cup\{Y_1Z_{\ell+j}\}_{j=1}^r$
\\[2mm]
$-$
& $\{e_\ell,\ f_\ell\}
\cup\{e_\ell{+}e_i,\ e_\ell{+}f_i\}_{i=1}^{\ell-1}
\cup\{e_\ell{+}h_j\}_{j=1}^r$
& $\{X_\ell,\ Z_\ell\}
\cup\,X_\ell\{X_i,Z_i\}_{i=1}^{\ell-1}
\cup\{X_\ell Z_{\ell+j}\}_{j=1}^r$
\\[2mm]
$0$
& $\{h_1\}
\cup\{h_1{+}e_i,\ h_1{+}f_i\}_{i=1}^\ell
\cup\{h_1{+}h_j\}_{j=2}^r$
& $\{Z_{\ell+1}\}
\cup\,Z_{\ell+1}\{X_i,Z_i\}_{i=1}^\ell
\cup\{Z_{\ell+1}Z_{\ell+j}\}_{j=2}^r$
\\[1.2mm]
\hline\hline
\end{tabular*}
\end{table*}

\begin{lem}\label{lem:exceptional_plus_affine_space_unique}
In the setting of Lemma~\ref{lem:exceptional_plus_affine_space_collapse}, and
for the representative classes from
Theorem~\ref{thm:classification_of_affine_subspaces_isomorphism_classes}, assume
that we are not in the case of type $+$ with $m=1$. Then the common
anisotropic vectors span the whole restricted space:
\[
    \Span[\F_2]{\bigcap_{w'\in w+W}\QQ_{w'}^{-1}(1)}=V=W^\perp.
\]
Thus the case of type $+$ with $\ell=1$ is the only representative class in which
the common anisotropic vectors span a proper subspace of $V$.
\end{lem}
\begin{proof}
By Lemma~\ref{lem:exceptional_plus_affine_space_collapse}, the common
anisotropic vectors of $w+W$ are exactly the anisotropic vectors of
$\QQ^*=\QQ_w|_V$. Hence the claim is equivalent to saying that, in every
non-exceptional representative class, the anisotropic vectors of $\QQ^*$ span
$V$.

We use the normal forms from
Proposition~\ref{prop:Quadratic_Form_Isomorphism_Classes}. If
$\type(\QQ^*)=0$, then $\QQ^*(\rad(V))=\F_2$. Choose $h_1\in\rad(V)$ with
$\QQ^*(h_1)=1$. Since $h_1\in\rad(V)$, every isotropic vector $v\in V$
satisfies $\QQ^*(v+h_1)=\QQ^*(v)+\QQ^*(h_1)=1$. Thus
$v=h_1+(v+h_1)$ is the sum of two anisotropic vectors and
the anisotropic vectors span $V$.

It remains to consider the cases $\type(\QQ^*)=\pm$, where $\QQ^*$ vanishes on
$\rad(V)$. Once one anisotropic vector $a$ with $\QQ^*(a)=1$ has been found, every
$h\in\rad(V)$ lies in the span of anisotropic vectors: indeed,
$\QQ^*(a+h)=\QQ^*(a)=1$, since $\QQ^*(h)=0$ and $\symp{a}{h}=0$ for
$h\in\rad(V)$. Thus the radical is already in the span. Hence it
remains only to span representatives of the non-degenerate quotient
$V/\rad(V)$. 

For type $-$, choose the normal-form symplectic basis so that
$\QQ^*(e_m)=\QQ^*(f_m)=1$, while $\QQ^*(e_i)=\QQ^*(f_i)=0$ for $i<m$.
Then $e_m$ and $f_m$ are anisotropic. For every other symplectic pair
$e_i,f_i$, the vectors $e_m+e_i$ and $e_m+f_i$ are anisotropic, and hence
$e_i=e_m+(e_m+e_i)$ and $f_i=e_m+(e_m+f_i)$ lie in the span.

For type $+$, choose the normal-form symplectic basis so that
$\QQ^*(e_i)=\QQ^*(f_i)=0$ for all $i$. The case
$m=1$ is exactly the exceptional case of
Lemma~\ref{lem:exceptional_plus_affine_space_collapse}. If $m\geq2$, then
$e_1+f_1$ is anisotropic, and so are $e_1+f_1+e_i$ and $e_1+f_1+f_i$ for
$i\geq2$. This puts $e_i,f_i$ with $i\geq2$ in the span, because
$e_i=(e_1+f_1)+(e_1+f_1+e_i)$ and similarly for $f_i$. Finally,
$e_1+e_2+f_2$ is anisotropic, so
$e_1=(e_1+e_2+f_2)+e_2+f_2$ lies in the span. Then
$f_1=(e_1+f_1)+e_1$ lies in the span as well.
This proves the claim in all non-exceptional cases.
\end{proof}

The same argument can be written in coordinates as explicit anisotropic
spanning sets. Table~\ref{tab:anisotropic-spanning-certificates}
lists these sets for the non-exceptional normal forms, together with the
corresponding Pauli generators. To use the table, one checks anisotropy by
substituting the listed vectors into the normal form of $\QQ^*$. The entries
involving only the symplectic pairs span the non-degenerate quotient
$V/\rad(V)$, while the entries with $h_j$ add the radical vectors. Thus the
listed vectors span $V=N\oplus\rad(V)$, where $N$ denotes the non-degenerate
part. Since each displayed set also has $\dim V$ elements, each row may be
read as an explicit basis of $V$. As the ambient symplectic form is
non-degenerate, $V=W^\perp$ implies $V^\perp=W$.

Then, we can rephrase Theorem~\ref{thm:classification_of_affine_subspaces_isomorphism_classes} purely in the Pauli language, together with the constraint of 'realizable' spaces of bilinear forms for Paulis:
\begin{cor}\label{cor:Canonical_Forms_Invariant_Bilinear_Forms}
Consider some Pauli generating set $\pgens$ with non trivial invariant quadratic forms.
Set $m=n-\ell-r$.
There is a Clifford $U\in\cl_n$, such that $\commalg'=\commutant(\pgens')$, where $\pgens' = \Ad_U\pgens$, has Pauli basis $\bas{\commalg'} = \{I,Z\}^{\otimes r}\otimes\PP_\ell\otimes I^{\otimes m}$ and there is a reference bilinear form $B\in \bilinear(\pgens)$ which is one and only one of the following (see Fig~\ref{fig:partition-commutant}): (1) $I$ (only for $m\geq 2$), (2) $Y_1$ (only for $m\geq 1$), (3) $X_{m+r+1}$ (only for $r\geq 1$).
\end{cor}
We shall see in the next section what these properties imply for transvection groups and Pauli Lie algebras. Also, we shall see which cases are in fact possible (beyond the constraints coming from $\QQ^*$ being well defined).

As in the case of individual quadratic forms, there exist equivalent criterions for evaluating the isomorphism type.
One way is to use the correspondence between the isomorphism classes spaces of invariant quadratic spaces (affine subspaces) $w+W$ and those of the restricted quadratic forms $\QQ^* = \QQ_w|_{W^\perp}$.
Then, it is sufficient to compute the action on the radical $\rad(W)=\rad(W^\perp)$ and possibly the Arf invariant of $\QQ^*$ over $W^\perp$, which only requires evalutation over the symplectic basis.

Alternatively, we focus here on those criterions which work by using purely properties of $w$ as a vector in $\Fn$ as well as the orthogonal complement/spaces of invariant quadratic forms:
\begin{lem}\label{lem:isomorphism_class_affine_subspaces_quadratic_forms}
Let $w+W$ be an affine subspace.
The isomorphism class of $w+W$ up to affine orthogonal transformations coincides with the democratic invariant over $w+W$:
\begin{enumerate}
    \item $\type(w+W)=+$ if and only if\\ $\abs{\QQ_0^{-1}(0)\cap (w + W)} > \abs{\QQ_0^{-1}(1)\cap (w + W)}$
    \item $\type(w+W)=-$ if and only if\\ $\abs{\QQ_0^{-1}(0)\cap (w + W)} < \abs{\QQ_0^{-1}(1)\cap (w + W)}$
    \item $\type(w+W)=0$ if and only if\\ $\abs{\QQ_0^{-1}(0)\cap (w + W)} = \abs{\QQ_0^{-1}(1)\cap (w + W)}$
\end{enumerate}
or equivalently, for a given reference $w$ we have:
\begin{enumerate}
    \item $\type(w+W)=+$ iff $\QQ_w|_{\rad(W)} = 0$, $\Arf(\QQ_w|_W) = \QQ_0(w)$
    \item $\type(w+W)=-$ iff $\QQ_w|_{\rad(W)} = 0$, $\Arf(\QQ_w|_W) = \QQ_0(w)+1$
    \item $\type(w+W)=0$ iff $\QQ_w|_{\rad(W)} = \F_2$
\end{enumerate}
\end{lem}
\begin{proof}
We can define a democratic Arf invariant of $\QQ_0$ over $w+W$ as:
\begin{align*}
        \Arf_D(\QQ_0|_{w+W}) &= \frac{1}{2^m2^r}\sum_{v\in W} (-1)^{\QQ_0(w+v)} \\
        &= \frac{1}{2^m2^r}\sum_{v\in W} (-1)^{\QQ_0(w)+\QQ_0(v)+\symp{w}{v}} \\
        &= (-1)^{\QQ_0(w)}\Arf_D(\QQ_w|_W)
\end{align*}
Then, if $\QQ_w|_{\rad(W)}=\F_2$, we must have $\Arf_D(\QQ_w|_W)=0$ and  $$\abs{\QQ_0^{-1}(0)\cap (w + W)} = \abs{\QQ_0^{-1}(1)\cap (w + W)}$$,
and $\type(w+W)=0$.

If $\QQ_w|_{\rad(W)}=0$, we can link the (ordinary) Arf invariant of $\QQ_w|_{W^\perp}$ to that of $\QQ_w|_W$ by using the direct sum decomposition of $\Fn$ in $W^\perp/\rad(W)\oplus R\oplus W/\rad(W)$ as in Eq.~\eqref{eq:full_Fn_decomposition_symplectic_basis} (here quotients are viewed as subspaces by choosing representatives).
Then, with respect to a joint symplectic basis of these non-degenerate subspaces, we have:
\begin{align*}
        \QQ_0(w) &= \Arf(\QQ_w) \\
        &= \Arf(\QQ_w|_W) + \Arf(\QQ_w|_R) + \Arf(\QQ_w|_{W^\perp}) \\
        &= \Arf(\QQ_w|_W) + \Arf(\QQ_w|_{W^\perp})
\end{align*}
which implies $\Arf(\QQ_w|_W) = \QQ_0(w) + \Arf(\QQ_w|_{W^\perp})$.
This is well defined when $\QQ_w|_{\rad(W)}=0$ and indeed an invariant of
the affine subspace with respect to different representatives. 

Notice that this is non-trivial, since, even though all
three terms are invariant under change of basis $\aff{\activemap}(w+W)$,
both $\QQ_0(w)$ and $\Arf(\QQ_w|_{W^\perp})$ are not \emph{individually}
invariant under different choices of $w,w'\in w+W$.
It remains to check that the criterion does not depend on the chosen
representative $w\in w+W$. Let $w'=w+s$ with $s\in W$. Then
$\QQ_{w'}(v)=\QQ_w(v)+\symp{s}{v}$. For $v\in\rad(W)$ we have
$\symp{s}{v}=0$, so $\QQ_{w'}|_{\rad(W)}=\QQ_w|_{\rad(W)}$. Thus the type
$0$ condition is independent of the representative. If
$\QQ_w|_{\rad(W)}=0$, then evaluating the Arf invariant from
Eq.~\eqref{eq:arf_invariant_binary_quadratic_form} in a symplectic basis of
$W$ gives
$\Arf(\QQ_{w'}|_W)=\Arf(\QQ_w|_W)+\QQ_w(s)$, while
$\QQ_0(w')=\QQ_0(w+s)=\QQ_0(w)+\QQ_w(s)$. Hence
\[
    \QQ_0(w')+\Arf(\QQ_{w'}|_W)
    =
    \QQ_0(w)+\Arf(\QQ_w|_W),
\]
and the distinction between the $+$ and $-$ cases is also independent of the
chosen representative.
\end{proof}

Computationally, Lemma~\ref{lem:isomorphism_class_affine_subspaces_quadratic_forms} provides an alternative criterion to determine the isomorphism class of the invariant quadratic forms for a set of generators. Algorithm~\ref{alg:quadratic_isomorphism_class} summarizes this criterion in a form useful for computation.
\begin{algorithm}[t]
    \caption{Pseudocode for computing the isomorphism class of invariant quadratic forms via the orthogonal complement.
        $\pgens=\isolong{\vgens}$ is the Pauli generating set, with output a string in $\{+, -, 0\}$.
        $\vecbasempty_\perp$ is a symplectic basis for $\vgens^\perp$, $\vecbasempty_\rad$ is a basis for the radical $\rad(\vgens)$.
        $w^*$ is a vector in $\Fn$. $\texttt{ArfInvariant}(w^*,\vecbasempty_\perp)$ computes the Arf invariant for $\QQ_{w^*}$ over the non-degenerate part of the symplectic basis $\vecbasempty_\perp$.}\label{alg:quadratic_isomorphism_class}
    \SetKwProg{Fn}{Function}{}{end}
    \SetKwFunction{FClass}{IsomorphismClass}
    \SetKwFunction{FArf}{ArfInvariant}

    \Fn{\FClass{$w^*$, $\vecbasempty_\perp$, $\vecbasempty_\rad$}}{
        \For{$u$ in $\vecbasempty_\rad$}{
            \If{$\QQ_0(u) + \symp{w^*}{u} = 1$}{
                \Return 0
            }
        }
        $\mathrm{arf}$ $\gets$ \FArf{$w^*$, $\vecbasempty_\perp$}\;

        \Return $(-1)^{\mathrm{arf}+\QQ_0(w^*) \bmod 2}$
    }
\end{algorithm}

\begin{enumerate}
    \item Given $\vgens$, compute a symplectic basis for $\vgens^\perp$, $\vecbasempty = \{e_i,f_i\}\cup\{h_j\}$
    \item Solve for some special $w\in\quadratic(\vgens)$
    \item Evaluate $\QQ_w$ on $\vecbasempty$:
    \begin{nestedcaseenum}
        \item If there is some $h_j$ such that $\QQ_w(h_j) = 1$, then $\type(w+W)=0$
        \item If $\QQ_w(h_j)=0$ for all $j$, compute the Arf invariant of $\QQ_w$ on $\{e_i,f_i\}$
    \end{nestedcaseenum}
\end{enumerate}
Notice that, whenever $\dim(W) = \dim(V^\perp)$ is of constant size $\BigO(1)$ with respect to the size of the system, the first criterion is also efficient and, for a given explicit list of elements (Paulis) in $w+W$, a simple way to check for the isomorphism class.
However, for general orthogonal complements, it will be necessary to use compute either the Arf invariant on either the space of the generators $V=W^\perp$ or its orthogonal complement $V^\perp = W$.

We can also precisely compute the number of invariant quadratic forms $w$ which are isotropic or anisotropic with respect to $\QQ_0$ by using the democractic Arf invariant over $w+W$.
We denote this with $N_0$ and $N_1$ respectively, and write $\rank(W)=2\ell$ and $\nullity(W)=r$.
If $\Arf_D(\QQ_0|_{w+W})=0 \iff \QQ_w|_{\rad(W)} = \F_2$, then we have $N_0 = N_1 = 2^{2\ell+r-1}$.
If $\Arf_D(\QQ_0|_{w+W})=(-1)^s$, we have instead $N_0 = 2^r2^{\ell-1}(2^{\ell}+s)$ and $N_1 = 2^r2^{\ell-1}(2^{\ell}-s)$.

Notice that this applies for any quadratic form $\QQ$ on $\Fn$, when restricted to some affine subspace $w+W$.
More precisely, for $\QQ$ a quadratic form and $\tilde{w}+W$ an affine subspace, we can generalize the definition of democratic Arf invariant as
\begin{equation}\label{eq:def:democratic_arf_affine_subspace}
    \Arf_D(\QQ|_{\tilde{w}+W}) := \frac{1}{2^m2^r} \sum_{v\in W} (-1)^{\QQ(\tilde{w}+v)}
\end{equation}
This quantity is manifestly invariant under the simultaneous change of basis
$\QQ \mapsto \QQ\circ\activemap^{-1}$ and
$\tilde{w}+W \mapsto \activemap(\tilde{w}+W)$ for
$\activemap\in\Sp(2n,\F_2)$.
This brings us to the definition:
\begin{defn}\label{def:types_of_quadratic_form_on_affine_subspaces}
Let $\QQ$ be an invariant quadratic form over $\Fn$ and $\tilde{w}+W$ an affine subspace in $\Fn$. Consider $\Arf_D(\QQ|_{\tilde{w}+W})$ as defined in Eq.~\eqref{eq:def:democratic_arf_affine_subspace}.
Then, we say that:
\begin{enumerate}
    \item $\QQ|_{\tilde{w}+W}$ has type $+$ if $\Arf_D(\QQ|_{\tilde{w}+W})=+1$.
    \item $\QQ|_{\tilde{w}+W}$ has type $-$ if $\Arf_D(\QQ|_{\tilde{w}+W})=-1$.
    \item $\QQ|_{\tilde{w}+W}$ has type $0$ if $\Arf_D(\QQ|_{\tilde{w}+W})=0$.
\end{enumerate}
\end{defn}
If we now let $\QQ=\QQ_w$, we can also write:
\begin{equation}
    \Arf_D(\QQ_w|_{\tilde{w}+W}) = (-1)^{\QQ_w(\tilde{w})} \Arf_D(\QQ_{w+\tilde{w}}|_W).
\end{equation}
Hence, the action of $\QQ_w$ on the affine subspace $\tilde{w}+W$ depends on the value of $\QQ_w(\tilde{w})$ and the action of $\QQ_{w+\tilde{w}}$ on the subspace $W$.
The number of isotropic and anisotropic vectors for $\QQ$ in $\tilde{w}+W$ are given by
\begin{align*}
        N_0 &
        = \abs{\QQ^{-1}(0)\cap (\tilde{w}+W)}\\
        &= \begin{dcases}
            2^{2\ell+r-1} & \Arf_D(\QQ|_{\tilde{w}+W}) = 0\\
            2^r2^{\ell-1}(2^{\ell}+s) & \Arf_D(\QQ|_{\tilde{w}+W}) = (-1)^s
        \end{dcases}\\
        N_1 &
        =\abs{\QQ^{-1}(1)\cap (\tilde{w}+W)}\\
        &= \begin{dcases}
            2^{2\ell+r-1} & \Arf_D(\QQ|_{\tilde{w}+W}) = 0\\
            2^r2^{\ell-1}(2^{\ell}-s) & \Arf_D(\QQ|_{\tilde{w}+W}) = (-1)^s
        \end{dcases}.
\end{align*}
Later we shall also be interested in the case where $\tilde{w}\in W^\perp$ and the action of $\QQ_w$ on the `affine radical' $\tilde{w}+\rad(W)\subseteq\tilde{w} + \rad(W)$, which we shall prove coincides with $W^\perp\cap(\tilde{w}+W)$ in Lemma~\ref{lem:Canonical_Basis_Affine_Subspaces_Parametrization}(c)
\begin{equation}
    \tilde{w}\in W^\perp, \QQ(\tilde{w}+u) = \QQ(\tilde{w}) + \QQ_w(u).
\end{equation}
Then, since $\QQ(\tilde{w})$ is a constant, we have that $\QQ|_{\rad(W)} = 0$ iff $\QQ|_{\tilde{w}+\rad(W)} = \QQ(\tilde{w})$, and $\QQ|_{\rad(W)}=\F_2$ iff $\QQ|_{\tilde{w}+\rad(W)}=\F_2$.
Denote now with $\tilde{N}_0$ and $\tilde{N}_1$ the number of isotropic and anisotropic vectors in $\tilde{w}+(W\setminus\rad(W))$ (with $\tilde{w}\in W^\perp$), respectively.
Then, if $\QQ$ of type $0$ over $\tilde{w}+W$, we have $\QQ|_{\tilde{w}+\rad(W)}=\F_2$, hence
\begin{equation}
    \tilde{N}_0 = \tilde{N}_1 = 2^{2\ell+r-1} - 2^{r-1}.
\end{equation}
Instead, when $\QQ$ is of type $s\in\{\pm\}$ over $\tilde{w}+W$, hence $\QQ|_{\tilde{w}+\rad(W)}=\QQ(\tilde{w})$, the number of isotropic and anistropic vectors for $\QQ$ in $\tilde{w}+W$ depends not only on its type but also on the value of $\QQ(\tilde{w}) = \alpha$:
\begin{align*}
        \tilde{N}_0 &= N_0 - (1-\alpha)2^r,\\
        \tilde{N}_1 &= N_1 - \alpha 2^r.
\end{align*}

In summary, we have:
\begin{lem}\label{lem:isotropic_sizes_over_affine_subspaces}
Let $\QQ$ be an invariant quadratic form over $\Fn$ and $\tilde{w}+W$ an affine subspace in $\Fn$, with $\rank(W)=2\ell$, $\nullity(W)=r$.
Let $N_0 = \abs{\QQ^{-1}(0)\cap (\tilde{w}+W)}$ and $N_1=\abs{\QQ^{-1}(1)\cap (\tilde{w}+W)}$.
Here, $\type(\QQ|_{\tilde{w}+W})$ denotes the type of $\QQ$ over $\tilde{w}+W$.
We have:
\begin{align*}
        N_0 &= \begin{dcases}
            2^{2\ell+r-1} & \text{if } \Arf_D(\QQ|_{\tilde{w}+W}) = 0,\\
            2^r2^{\ell-1}(2^{\ell}+s) & \text{if } \Arf_D(\QQ|_{\tilde{w}+W}) = (-1)^s,
        \end{dcases}\\
        N_1 &= \begin{dcases}
            2^{2\ell+r-1} & \text{if } \Arf_D(\QQ|_{\tilde{w}+W}) = 0,\\
            2^r2^{\ell-1}(2^{\ell}-s) & \text{if } \Arf_D(\QQ|_{\tilde{w}+W}) = (-1)^s.
        \end{dcases}
\end{align*}
Also, if $\tilde{w}\in W^\perp$, let
\begin{align*}
\tilde{N}_0 &= \abs{\QQ^{-1}(0)\cap (\tilde{w}+(W\setminus\rad(W)))},\\
\tilde{N}_1 &=\abs{\QQ^{-1}(1)\cap (\tilde{w}+(W\setminus\rad(W)))},
\end{align*}
and $\alpha = \QQ(\tilde{w})$. Then
\begin{align*}
        \tilde{N}_0 &= \begin{dcases}
            2^{2\ell+r-1}-2^{r-1} & \text{if } \type(\QQ|_{\tilde{w}+W})=0,\\
            N_0 - (1-\alpha)2^r & \text{otherwise},
        \end{dcases}\\
        \tilde{N}_1 &= \begin{dcases}
            2^{2\ell+r-1}-2^{r-1} & \text{if } \type(\QQ|_{\tilde{w}+W})=0,\\
            N_1 - \alpha 2^r & \text{otherwise}.
        \end{dcases}
\end{align*}
\end{lem}

\ManuscriptPart{Cliffords, Line Graphs and Dependencies}{part:symmetry_constrained_groups_algebras}

\section{Clifford Group and Pauli Lie Algebras under Symmetries}\label{sec:symmetry_classification_transvection_groups_pauli_lie_algebras}

We now have the necessary tools to describe which transvection/Clifford groups and Pauli Lie Algebras are possible under given linear symmetries/Pauli commutants or given invariant bilinear forms, their isomorphism classes and representatives.

\subsection{Kernel Criteria and Symplectic Stabilizers}\label{sec:kernel_criteria_pointwise_symplectic_stabilizers}

We are interested in determining groups and Lie algebras from their imposed symmetries, their quotient images, and their kernels.
The common group-theoretic pattern is the following.
Given a group homomorphism $\pi$ with domain $G$ and a subgroup $H\subseteq G$, the restriction to $H$ gives the short exact sequence
\begin{equation*}
    1\to H\cap\ker(\pi)\to H\xrightarrow{\pi}\pi(H)\to 1.
\end{equation*}
Since $\ker(\pi)$ is normal in $G$, Lemma~\ref{lem:maximality_of_quotient_group} can be used with $N=\ker(\pi)$. If $H\subseteq H'$, equality of the images $\pi(H)=\pi(H')$ and equality of the kernels $H\cap\ker(\pi)=H'\cap\ker(\pi)$ imply $H=H'$.
We will apply this to Clifford groups, for the quotient by the Pauli group, as well as to symplectic groups, for quotients induced by restriction to a subspace or by projection.

We can now start to provide a preliminary classification of transvection groups and Pauli Lie algebras subject to the symmetry constraint of a given orthogonal complement or commutant.
We start by defining the $\F_2$-subgroups of $\Sp(2n,\F_2)$, followed by the corresponding Clifford subgroups and finally the (compact) Pauli Lie algebras, while also providing some basic properties.
Then, in Section~\ref{sec:classification:groups_lie_algebras} we shall prove which sets of generators result in which groups or Pauli Lie algebras.

Specifically, in the binary picture we have the constraint $\symp{W}{\vgens}=0$ for a subspace $W\subseteq\Fn$, meaning that $\symp{w}{v}=0$ for all $w\in W$ and $v\in\vgens$.
Thus $\Span{\vgens}\subseteq W^\perp$.
The largest subspace of allowable transvection centers compatible with this constraint is obtained by taking $\Span{\vgens}=W^\perp$, since $\symp{w}{v}=0$ is equivalent to $\tau_v w=w$ for all $w\in W$ and $v\in W^\perp$.
For a general subspace $W$, this pointwise stabilizer is 
\begin{equation}
    \Sp(2n,\F_2)_W = \{ \activemap\in\Sp(2n,\F_2) \mid \activemap u = u \text{ for all } u\in W\}.
    \label{eq:pointwise_symplectic_stabilizer}
\end{equation}

We now specialize to the case in which the maximal compatible transvection generating subspace is denoted by $V=\Span{\vgens}$, so that $W=V^\perp$.
Imposing the same pointwise fixation of $V^\perp$ on arbitrary elements $\activemap\in\Sp(2n,\F_2)$ gives the pointwise stabilizer $\Sp(2n,\F_2)_{V^\perp}$.
We introduce
\begin{equation}\label{eq:def:degenerate_symplectic_group}
    \Sp^\#(V)
    =
    \{\activemap\in\Sp(V)\mid \activemap u=u \text{ for all } u\in\rad(V)\}
\end{equation}
as the subgroup of $\Sp(V)$ that fixes the radical $\rad(V)=V\cap V^\perp$ pointwise.
The restriction homomorphism from $\Sp(2n,\F_2)_{V^\perp}$ to the subspace $V$ gives the following short exact sequence \cite{Janssen_1983}.

\begin{lem}[Pointwise symplectic stabilizer sequence]\label{lem:pointwise_symplectic_stabilizer_sequence}
The restriction map $\activemap|_V$ with $\activemap \in \Sp(2n,\F_2)_{V^\perp}$ and $V\subseteq \Fn$
gives the short exact sequence
\begin{equation}\label{eq:quotient_symplectic_group_over_Fn}
    1 \to \DD{\rad(V)} \to \Sp(2n,\F_2)_{V^\perp} \to \Sp^{\#}(V) \to 1,
\end{equation}
and the kernel is
\begin{align*}
    \DD{\rad(V)}
    &=
    \{ \activemap\in \Sp(2n,\F_2)_{V^\perp} \mid \activemap|_V = \id_V\}\\
    &=
    \{ \activemap\in\Sp(2n,\F_2) \mid \activemap|_{V+V^\perp} = \id_{V+V^\perp}\}.
\end{align*}
\end{lem}

\begin{proof}
The restriction map is well-defined because every $\activemap\in\Sp(2n,\F_2)_{V^\perp}$ preserves $V=(V^\perp)^\perp$.
Its image fixes $\rad(V)=V\cap V^\perp$ pointwise, hence lies in $\Sp^\#(V)$.
Conversely, for $\activemap_0\in\Sp^\#(V)$,
define a map on $V+V^\perp$ by
\[
    v+u\mapsto \activemap_0 v+u\qquad \text{for } v\in V,\ u\in V^\perp.
\]
This is well-defined because two decompositions differ by an element of $V\cap V^\perp=\rad(V)$, and $\activemap_0$ fixes $\rad(V)$ pointwise.
For $v,v'\in V$ and $u,u'\in V^\perp$, we have
\begin{align*}
    \symp{\activemap_0 v+u}{\activemap_0 v'+u'}
    &=
    \symp{\activemap_0 v}{\activemap_0 v'}+\symp{u}{u'}\\
    &=
    \symp{v}{v'}+\symp{u}{u'}
    =
    \symp{v+u}{v'+u'}.
\end{align*}
Hence the map preserves the symplectic form restricted to $V+V^\perp$.
By the symplectic extension statement in Corollary~\ref{cor:symplectic-extension-f2}, this isometry of $V+V^\perp$ extends to an element of $\Sp(2n,\F_2)$.
The extension fixes $V^\perp$ pointwise because the map above already does so on $V^\perp$.
Thus the image is $\Sp^\#(V)$.

The kernel consists precisely of those elements which act trivially on $V$ and already fix $V^\perp$ pointwise.
This gives the first displayed description of $\DD{\rad(V)}$.
The second follows because an element of $\Sp(2n,\F_2)_{V^\perp}$ already fixes $V^\perp$ pointwise; imposing $\activemap|_V=\id_V$ is therefore equivalent to imposing $\activemap|_{V+V^\perp}=\id_{V+V^\perp}$.
\end{proof}

We shall now identify the kernel of this restriction map more explicitly.
Its elements act trivially everywhere except on a supplementary subspace of $V+V^\perp$.
When $V=V^\perp=\rad(V)$, we also have the special case $\Sp(2n,\F_2)_{V^\perp} = \DD{\rad(V)}$ and $\Sp^{\#}(\rad(V))$ is the trivial group.

\begin{lem}[Symplectic diagonal kernel]\label{lem:symplectic_diagonal_kernel_structure}
Let $V\subseteq\Fn$ and set $r=\dim\rad(V)$.
The group $\DD{\rad(V)}$ from Lemma~\ref{lem:pointwise_symplectic_stabilizer_sequence} has the following descriptions.
\begin{enumerate}
\item Every element of $\DD{\rad(V)}$ is uniquely of the form
\[
    x\mapsto x+\phi(x)
\]
where the shift map $\phi$ from $\Fn$ to $\Fn$ is linear and satisfies $\ker(\phi)\supseteq V+V^\perp$, $\Im(\phi)\subseteq\rad(V)$, and
\begin{equation}\label{eq:symmetry_phi_kernel}
    \symp{\phi(v)}{u} = \symp{v}{\phi(u)}
\end{equation}
for all $v,u\in\Fn$.
\item Choosing a basis of $\rad(V)$ identifies $\DD{\rad(V)}$ with the additive group of symmetric $r\times r$ matrices over $\F_2$.
\end{enumerate}
\end{lem}

\begin{proof}
We first prove (a), i.e., the characterization of the possible maps $\phi$.
There are two things to show: every element of $\DD{\rad(V)}$ produces a unique shift map with the stated properties, and conversely every such shift map produces an element of $\DD{\rad(V)}$.

Let $\activemap\in \DD{\rad(V)}$ and define
\[
    \phi(v)=\activemap v+v
\]
for $v\in\Fn$.
This is the unique possible shift map, since $\activemap v=v+\phi(v)$.
We verify the three conditions in the statement.
First, $\activemap$ fixes $V+V^\perp$ pointwise, so $\phi(v)=0$ for all
$v\in V+V^\perp$.
Hence $\ker(\phi)\supseteq V+V^\perp$.

Second, $\Im(\phi)\subseteq\rad(V)$.
Indeed, for $x\in V+V^\perp$ and $u\in\Fn$, the symplectic property of
$\activemap$ gives
\[
    \symp{x}{u}
    =
    \symp{\activemap x}{\activemap u}
    =
    \symp{x}{u+\phi(u)}
    =
    \symp{x}{u}+\symp{x}{\phi(u)}.
\]
Thus $\symp{x}{\phi(u)}=0$ for all $x\in V+V^\perp$, and therefore
$\phi(u)\in (V+V^\perp)^\perp=\rad(V)$ for all $u\in\Fn$.

Third, expanding
$\symp{\activemap v}{\activemap u}=\symp{v}{u}$ with
$\activemap=\id+\phi$ for arbitrary $v,u\in\Fn$ gives
\[
    0
    =
    \symp{\phi(v)}{u}
    +
    \symp{v}{\phi(u)}
    +
    \symp{\phi(v)}{\phi(u)}.
\]
The last term vanishes because $\Im(\phi)\subseteq\rad(V)$.
Hence $\symp{\phi(v)}{u}=\symp{v}{\phi(u)}$, which is
Eq.~\eqref{eq:symmetry_phi_kernel}.
This proves the forward direction of (a).

Conversely, suppose that a linear map $\phi$ satisfies
$\ker(\phi)\supseteq V+V^\perp$, $\Im(\phi)\subseteq\rad(V)$, and
Eq.~\eqref{eq:symmetry_phi_kernel}, and define $\activemap=\id+\phi$.
Then $\activemap$ fixes $V+V^\perp$ pointwise.
It is invertible because
$\Im(\phi)\subseteq\rad(V)\subseteq V+V^\perp\subseteq\ker(\phi)$, so
$\phi^2=0$ and $\activemap$ is its own inverse.
Finally, $\symp{\phi(v)}{\phi(u)}=0$ because $\Im(\phi)\subseteq\rad(V)$.
Together with Eq.~\eqref{eq:symmetry_phi_kernel}, the same expansion shows
that $\activemap$ preserves the symplectic form.
Thus $\activemap\in\DD{\rad(V)}$, which proves the intrinsic characterization
in (a).

We now prove (b).
Again, there are two things to show: the shift maps from (a) are parametrized by symmetric matrices after choosing a basis of $\rad(V)$, and this parametrization identifies composition with addition of matrices.

Choose a supplementary subspace $T$ of $V+V^\perp$ with basis $\{\tilde{h}_j\}_{j=1}^r$ symplectically dual to a basis $\{h_j\}_{j=1}^r$ of $\rad(V)$, as in Eq.~\eqref{eq:full_Fn_decomposition_symplectic_basis}.
$\Fn = (V+V^\perp)\oplus T$ and we can write any vector in $\Fn$ as $v+\tilde{w}$, with $v\in V+V^\perp$ and $\tilde{w}\in T$.
Then
\begin{equation*}
    \phi(v+\tilde{w}) = \phi(v) + \phi(\tilde{w}) = \phi(\tilde{w})
\end{equation*}
Using the chosen dual bases, we expand $\phi(\tilde{w})$ in
$\{h_j\}_{j=1}^r$ and $\tilde{w}$ in $\{\tilde{h}_j\}_{j=1}^r$
with $\symp{\tilde h_i}{h_j}=\delta_{ij}$.
Define the matrix coefficients
$B_{ij}=\symp{\phi(\tilde{h}_i)}{\tilde{h}_j}$.
Since $\tilde{w} = \sum_{i=1}^r \symp{\tilde{w}}{h_i} \tilde{h}_i$ and
$\phi(\tilde{w})\in\rad(V)$, linearity gives
\begin{equation}\label{eq:symplectic_diagonal_kernel_phi_formula}
        \phi(\tilde{w})
        = \sum_{i,j=1}^r B_{ij}\symp{\tilde{w}}{h_i} h_j.
\end{equation}

Evaluating Eq.~\eqref{eq:symmetry_phi_kernel} on the dual basis vectors $\tilde h_i,\tilde h_j$ shows that $B_{ij}=B_{ji}$.
Thus $B_{ij}$ must be a $r\times r$ symmetric matrix.
Conversely, any symmetric matrix defines a suitable $\phi$ by setting $\phi|_{V+V^\perp}=0$ and using Eq.~\eqref{eq:symplectic_diagonal_kernel_phi_formula} on $T$.
The formula depends only on the $T$-component because for $v+\tilde{w}\in \Fn = (V+V^\perp)\oplus T$, we have $\symp{v}{\rad(V)} = 0$.
Hence the elements of $\DD{\rad(V)}$ are parametrized by symmetric $r\times r$ matrices.
This completes the parametrization part of (b).

It remains to check that this correspondence is compatible with the group law.
For two elements represented by $\phi$ and $\phi'$, we have
\begin{align*}
        \activemap\activemap' v &= \activemap(v + \phi'(v)) \\
        &= v + \phi'(v) + \phi(v) + \phi(\phi'(v))
        = v + \phi'(v) + \phi(v)
\end{align*}
where $\phi|_{V+V^\perp}=0$ and $\Im(\phi')\subseteq\rad(V)$.
Thus composition in $\DD{\rad(V)}$ corresponds to addition of the associated symmetric matrices.
This proves (b).
		\end{proof}
	
We next translate the symmetric-matrix basis from Lemma~\ref{lem:symplectic_diagonal_kernel_structure} back into transvections.

\begin{lem}[Transvection generators of the symplectic diagonal kernel]\label{lem:symplectic_diagonal_kernel_transvection_generators}
Let $V\subseteq\Fn$ and let $\{h_i\}_{i=1}^r$ be a basis of $\rad(V)$.
Then $\DD{\rad(V)}$ is generated by the transvections with centers $h_i$ and $h_i+h_j$ for $1\leq i<j\leq r$.
\end{lem}

\begin{proof}
By Lemma~\ref{lem:symplectic_diagonal_kernel_structure}, it suffices to realize a basis of the corresponding symmetric matrices.
We do this in two steps: first the diagonal matrix units, and then the off-diagonal symmetric matrix units.
In the basis of $\rad(V)$, these transformations are
\begin{align*}
        \phi_{ii}(v) &= \symp{v}{h_i}h_i,\\
        \phi_{ij}(v) &= \symp{v}{h_j}h_i + \symp{v}{h_i}h_j.
\end{align*}

The diagonal terms are the shifts produced by the transvections $\tau_{h_i}$ with centers in the chosen radical basis, since $\tau_{h_i}(v)=v+\phi_{ii}(v)$.
The off-diagonal terms are produced by the following products for $u,u'\in\rad(V)$:
\begin{align*}
        \tau_{u+u'}\tau_u\tau_{u'}(v) &= 
        v + \symp{u}{v}u + \symp{u'}{v}u' 
        + \symp{u{+}u'}{v}(u{+}u')\\
        &= v + \symp{v}{u}u' + \symp{v}{u'}u.
\end{align*}
Taking $u=h_i$ and $u'=h_j$ realizes the off-diagonal transformation $\phi_{ij}$.
Thus the listed transvections realize all diagonal and off-diagonal symmetric matrix units, which form a basis of the symmetric matrices.
Hence they generate $\DD{\rad(V)}$ by Lemma~\ref{lem:symplectic_diagonal_kernel_structure}.
\end{proof}

We shall use Lemmas~\ref{lem:symplectic_diagonal_kernel_structure} and~\ref{lem:symplectic_diagonal_kernel_transvection_generators}, together with Lemma~\ref{lem:maximality_of_quotient_group}, to study $\DD{\rad(V)}$ as a transvection group, and to prove that in fact a certain transvection group coincides with $\Sp(2n,\F_2)_{V^\perp}$.
Notice that such transvections in $\rad(V)$ are \emph{pathological} in the sense of Definition~\ref{def:pathological} for a transvection group with connected generating set.

There is also a second quotient kernel associated with the degenerate symplectic group $\Sp^\#(V)$ itself, distinct from the diagonal kernel $\DD{\rad(V)}$ above.
Up to change of basis, the only invariants of a subspace $V$ are its rank $2m$ and the dimension of its radical $r$. Thus
the canonical representative of $\Sp(V)$ on $\F_2^{2m+r}$ is denoted by
\begin{equation*}
    \Sp^{\#}(V)\cong\Sp^{\#}(2m,r,\F_2).
\end{equation*}
With regard to this canonical representative, the group $\Sp^{\#}(2m,r,\F_2)$ can also be interpreted via the natural quotient that maps it to the non-degenerate symplectic group
$\Sp(V/\rad(V))\cong \Sp(2m,\F_2)$.
Thus Eq.~\eqref{eq:quotient_symplectic_group_over_Fn} is complemented by the short sequence
\begin{equation*}
    1 \to K \to \Sp^{\#}(2m,r,\F_2) \to \Sp(2m,\F_2) \to 1.
\end{equation*}
Here $K$ is an internal quotient kernel of the degenerate group $\Sp^\#(V)$, whereas $\DD{\rad(V)}$ is the kernel obtained when the ambient pointwise stabilizer $\Sp(2n,\F_2)_{V^\perp}$ is restricted to $V$.
The following lemma identifies this kernel in an abstract alternating space.

\begin{lem}\label{lem:degenerate_symplectic_quotient_kernel}
Let $V$ be an alternating space over $\F_2$ with radical $R=\rad(V)$, and let $\Sp^{\#}(V)$ denote the isometry group of $V$ fixing $R$ pointwise.
Then the quotient action gives a split exact sequence
\begin{equation*}
    1 \to K \to \Sp^{\#}(V) \to \Sp(V/R) \to 1,
\end{equation*}
where $K$ is the kernel of the induced action on $V/R$.
More explicitly, writing $\overline{\activemap}(v+R)=\activemap v+R$, we have
\begin{equation}
    K = \{ \activemap\in \Sp^{\#}(V) \mid \overline{\activemap} = \id_{V/R}\}.
\label{eq:degenerate_symplectic_kernel}
\end{equation}
The elements of $K$ are naturally identified with the linear maps from $V/R$ to $R$.
In particular, if $\rank(V)=2m$ and $\dim(R)=r$, then $K\cong \F_2^{2mr}$ as an additive group.
\end{lem}

\begin{proof}
We first define the quotient map.
For $\activemap\in\Sp^\#(V)$, define
$\pi$ as the map from $\Sp^{\#}(V)$ to $\Sp(V/R)$ by
\begin{equation*}
    \pi(\activemap)(v+R):=\activemap v+R.
\end{equation*}
This is well-defined, since every $\activemap\in\Sp^{\#}(V)$ fixes $R$ pointwise.
It is a group homomorphism.

We next identify the kernel $K$.
An element $\activemap\in\Sp^\#(V)$ lies in $K$ iff
$\pi(\activemap)=\id_{V/R}$, equivalently iff
\[
    \activemap v+R=v+R
    \qquad\text{for all }v\in V.
\]
Thus $\activemap v+v\in R$ for all $v\in V$, and we may define a map $\phi_{\activemap}$
from $V$ to $R$ by
\[
    \phi_{\activemap}(v):=\activemap v+v.
\]
Since $\activemap$ fixes $R$ pointwise, $\phi_{\activemap}$ vanishes on $R$.
Therefore $\phi_{\activemap}$ factors through a linear map
$\bar{\phi}_{\activemap}$ from $V/R$ to $R$.

Conversely, we construct an element of the kernel from any linear map
$\bar{\phi}$ from $V/R$ to $R$.
Define
\begin{equation*}
    \activemap_{\bar{\phi}}(v)=v+\bar{\phi}(v+R).
\end{equation*}
It fixes $R$ pointwise: if $v\in R$, then $v+R=R$ and hence $\bar{\phi}(v+R)=0$.
Moreover, $\activemap_{\bar{\phi}}$ is its own inverse.
Indeed, since $\bar{\phi}(v+R)\in R$, the vectors $v$ and $\activemap_{\bar{\phi}}(v)$ define the same class in $V/R$.
Thus applying $\activemap_{\bar{\phi}}$ twice adds the same correction term twice, which is zero over $\F_2$.
It also preserves the alternating form, since all correction terms lie in the radical:
\begin{align*}
    \genbil{\activemap_{\bar{\phi}}v}{\activemap_{\bar{\phi}}u}
    &=
    \genbil{v+\bar{\phi}(v{+}R)}{u+\bar{\phi}(u{+}R)}
    =
    \genbil{v}{u}.
\end{align*}
Hence $\activemap_{\bar{\phi}}\in\Sp^\#(V)$, and it lies in $K$ because
$\activemap_{\bar{\phi}}(v)+R=v+R$ for all $v\in V$.
The two constructions are inverse to each other.
Moreover, composition in $K$ corresponds to adding the associated maps
$V/R\to R$.
This proves the claimed identification of $K$.

It remains to prove that the sequence is split.
Choose a symplectic basis as in Lemma~\ref{lem:symplectic_basis}, so that
$V$ is identified with $V_{\rm nd}\oplus R$, where $V_{\rm nd}$ is
non-degenerate and maps isomorphically onto $V/R$.
Every element of $\Sp(V/R)$ can then be represented as an element of
$\Sp(V_{\rm nd})$ and extended to an element of $\Sp^\#(V)$ by fixing $R$
pointwise.
This gives a section of $\pi$, so the sequence is split.
\end{proof}

We also record the corresponding generation criterion by non-radical
transvections, in the form needed later;
compare \cite[Theorem~2.7]{Brown_Humphries_1986a}.

\begin{lem}[Transitivity and the full degenerate symplectic group]\label{lem:non_radical_transvections_generate_sp_sharp}
Let $V\subseteq\Fn$ be a subspace, where the symplectic product on $V$ is the
restriction of $\sympempty$, and let $S\subseteq V\setminus\rad(V)$ be
non-empty.
Then:
\begin{enumerate}
    \item The non-radical transvections generate the full degenerate symplectic group, i.e.,
    \begin{equation}\label{eq:non_radical_transvections_generate_sp_sharp}
        \tvgroup{V\setminus\rad(V)}=\Sp^\#(V).
    \end{equation}
    \item The following are equivalent:
    \begin{nestedcaseenum}
        \item $\tvgroup{S}$ acts transitively on $V\setminus\rad(V)$;
        \item $\tvgroup{S}=\Sp^\#(V)$.
    \end{nestedcaseenum}
\end{enumerate}
\end{lem}
\begin{proof}
Set $R=\rad(V)$.
\emph{Generation by all non-radical transvections.}
We first prove Eq.~\eqref{eq:non_radical_transvections_generate_sp_sharp}.
For $v\in V$, the transvection $\tau_v$ preserves $V$ and fixes $R$ pointwise,
so it lies in $\Sp^\#(V)$, the subgroup fixing the radical pointwise from
Eq.~\eqref{eq:def:degenerate_symplectic_group}.
The quotient map from Lemma~\ref{lem:degenerate_symplectic_quotient_kernel}
sends the transvection $\tau_v$ with $v\notin R$ to the transvection with center
$v+R$ on the non-degenerate quotient $V/R$.
These quotient transvections generate $\Sp(V/R)$ by the standard
non-degenerate generation result for symplectic groups
\cite[Ch.~IX, \S 5, Ex.~11]{Bourbaki1959}.
It remains to generate the kernel $K$ of the quotient action.
By Lemma~\ref{lem:degenerate_symplectic_quotient_kernel}, this kernel is
identified with the additive group of linear maps from $V/R$ to $R$.
For $v\notin R$ and $r\in R$, the product
$\tau_{v+r}\tau_v$
lies in the kernel and acts as
$x\mapsto x+\symp{v}{x}r$.
Thus it realizes the rank-one map $x+R\mapsto\symp{v}{x}r$ from $V/R$ to $R$.
As $v$ varies modulo $R$, the functionals
$x+R\mapsto\symp{v}{x}$ exhaust $(V/R)^*$ by non-degeneracy of the induced
form; as $r$ varies in $R$, the corresponding rank-one maps span
$\operatorname{Hom}_{\F_2}(V/R,R)$.
Hence the non-radical transvections generate the quotient group and contain the
kernel.
By the exact sequence from Lemma~\ref{lem:degenerate_symplectic_quotient_kernel},
this proves Eq.~\eqref{eq:non_radical_transvections_generate_sp_sharp}.

\emph{Equivalence with transitivity.}
Assume now that $\tvgroup{S}$ acts transitively on $V\setminus R$.
Choose $s_0\in S$.
Then, for every $x\in V\setminus R$, there is $g\in\tvgroup{S}$ such that
$g(s_0)=x$.
Then
\[
    \tau_x=g\tau_{s_0} g^{-1}\in\tvgroup{S}.
\]
Thus $\tvgroup{S}$ contains all non-radical transvections, and so
$\tvgroup{S}=\Sp^\#(V)$ by
Eq.~\eqref{eq:non_radical_transvections_generate_sp_sharp}.

Conversely, assume $\tvgroup{S}=\Sp^\#(V)$.
We show that $\Sp^\#(V)$ acts transitively on $V\setminus R$.
Let $v,w\in V\setminus R$.
By Corollary~\ref{cor:symplectic-extension-f2}, any isometry between the
one-dimensional subspaces spanned by two non-zero vectors extends to an element
of $\Sp(V/R)$.
Thus $\Sp(V/R)$ is transitive on non-zero vectors.
By the surjectivity in Lemma~\ref{lem:degenerate_symplectic_quotient_kernel},
we can lift an element sending $v+R$ to $w+R$ to an element of $\Sp^\#(V)$.
After applying this element, the image of $v$ has the form $w+r$ for some
$r\in R$.
By Lemma~\ref{lem:degenerate_symplectic_quotient_kernel}, the kernel of the
quotient action consists of the maps
$x\mapsto x+\phi(x+R)$,
where $\phi$ is an arbitrary $\F_2$-linear map from $V/R$ to $R$.
Choose $\phi$ with $\phi(w+R)=r$.
The corresponding kernel element sends $w+r$ to $w$.
Therefore $v$ and $w$ lie in the same orbit of $\Sp^\#(V)$, which proves
transitivity.
\end{proof}

\subsection{Commutant-Symmetric Clifford Groups}\label{sec:commutant_symmetric_clifford_groups_lie_algebras}

We now translate the binary pointwise stabilizers from the previous subsection into finite Clifford subgroups.
A subspace $W\subseteq\Fn$ specifies the Pauli matrix algebra
\[
    \commalg=\Span[\C]{\inviso(W)}.
\]
Indeed, the vectors in $W$ label Pauli strings, and $\commalg$ is the complex span of those Pauli strings.
Here $\inviso=\isoempty^{-1}$ is the inverse of the map $\isoempty:\Fn\to\PP_n$ from Eq.~\eqref{eq:pauli:binary}.
Thus $\commalg$ is the Pauli symmetry algebra associated with $W$.
The corresponding finite Clifford subgroup is the subgroup of $\cl_n$ which centralizes this algebra, or equivalently is fixed by the adjoint action of these Pauli symmetries.
For this subgroup we use the fixed-point notation
\begin{equation}\label{eq:def:clifford_fixed_commutant}
    \cl_n^{\Ad_{\commalg}}
    :=
    \{U\in\cl_n\mid \Ad_C(U)=U
    \text{ for all } C\in\commalg\cap\PP_n\}.
\end{equation}
Equivalently, it is enough to impose the same adjoint fixed-point conditions for any set of Pauli strings whose generated matrix algebra is $\commalg$.
This subgroup is the same as the centralizer $\cent_{\cl_n}(\commalg)$, and also $\cl_n\cap\matalg$, where $\matalg=\commutant(\commalg)$ is the matrix algebra dual to $\commalg$.

The following lemma provides the precise binary image and Pauli kernel of this commutant-fixed subgroup.

\begin{lem}[Commutant-fixed Clifford subgroups]\label{lem:fixed_clifford_commutant_binary_image_kernel}
Let $W\subseteq\Fn$ and $\commalg=\Span[\C]{\inviso(W)}$.
\begin{enumerate}
\item\label{lem:fixed_clifford_commutant_binary_image_kernel:a}
The map $\SY$ from Eq.~\eqref{eq:proj:conj} induces an isomorphism
\begin{equation}
    \cl_n^{\Ad_{\commalg}}/
    (\cl_n^{\Ad_{\commalg}}\cap\Pgroup_n)
    \cong
    \Sp(2n,\F_2)_W.
\end{equation}
\item\label{lem:fixed_clifford_commutant_binary_image_kernel:b}
Equivalently, $\SY$ induces a short exact sequence
\begin{equation}\label{eq:quotient_fixed_clifford_commutant_to_binary}
    1\to
    \cl_n^{\Ad_{\commalg}}\cap\Pgroup_n
    \to
    \cl_n^{\Ad_{\commalg}}
    \xrightarrow{\SY}
    \Sp(2n,\F_2)_W
    \to 1.
\end{equation}
\item\label{lem:fixed_clifford_commutant_binary_image_kernel:c}
If a subgroup $H$ of $\cl_n$ satisfies
\begin{gather*}
    H\subseteq\cl_n^{\Ad_{\commalg}},
    \quad
    \SY(H)=\Sp(2n,\F_2)_W,\\
    H\cap\Pgroup_n=
    \cl_n^{\Ad_{\commalg}}\cap\Pgroup_n,
\end{gather*}
then $H=\cl_n^{\Ad_{\commalg}}$.
\item\label{lem:fixed_clifford_commutant_binary_image_kernel:d}
If a subgroup $H$ of $\cl_n$ satisfies
\begin{align*}
    &H\subseteq\cl_n^{\Ad_{\commalg}},
    \quad
    \SY(H)=\Sp(2n,\F_2)_W,\\
    &H\cap\Pgroup_n\simeq 
    \cl_n^{\Ad_{\commalg}}\cap\Pgroup_n,
\end{align*}
then the two subgroups agree after adjoining scalar phases,
\begin{equation*}
    H \simeq \cl_n^{\Ad_{\commalg}}.
\end{equation*}
\end{enumerate}
\end{lem}
\begin{proof}
Let $G=\cl_n^{\Ad_{\commalg}}$.
By Lemma~\ref{lem:finite_clifford_subgroups_over_binary_images}\ref{lem:finite_clifford_subgroups_over_binary_images:b}, the restriction of $\SY$ to $G$ has kernel $G\cap\Pgroup_n$.
It remains to identify its image.

First, let $U\in G$ and set $g=\SY(U)$.
For every $c\in W$, the Pauli string $\iso{c}$ lies in $\commalg\cap\PP_n$, and the condition $\Ad_{\iso{c}}(U)=U$ is equivalent to $U\iso{c}U^\dagger=\iso{c}$.
Thus $g c=c$, so $\SY(G)\subseteq\Sp(2n,\F_2)_W$.

Conversely, let $g\in\Sp(2n,\F_2)_W$.
By Lemma~\ref{lem:finite_clifford_subgroups_over_binary_images}\ref{lem:finite_clifford_subgroups_over_binary_images:a}, choose $U\in\cl_n$ with $\SY(U)=g$.
Since $g$ fixes $W$ pointwise, Eq.~\eqref{eq:eta} gives
\[
    U\iso{c}U^\dagger=(-1)^{\eta_U(c)}\iso{c}
    \qquad \text{for all } c\in W.
\]
For $c,d\in W$, Lemma~\ref{lem:sign:clifford}\ref{lem:sign:clifford:f} gives
\[
    \eta_U(c+d)+\eta_U(c)+\eta_U(d)
    =
    \sign{c}{d}+\sign{gc}{gd}=0,
\]
so $\eta_U|_W$ is linear.
By non-degeneracy of the symplectic form on $\Fn$, choose $p\in\Fn$ such that $\symp{p}{c}=\eta_U(c)$ for all $c\in W$.
Set $P=\iso{p}$.
Lemma~\ref{lem:sign:clifford}\ref{lem:sign:clifford:d} gives
\[
    \eta_{PU}(c)=\eta_U(c)+\symp{p}{gc}=0
    \qquad \text{for all } c\in W,
\]
because $gc=c$.
Thus $PU\in G$ and $\SY(PU)=g$.
Hence $\SY(G)=\Sp(2n,\F_2)_W$.

Items~\ref{lem:fixed_clifford_commutant_binary_image_kernel:a} and~\ref{lem:fixed_clifford_commutant_binary_image_kernel:b} follow from this image computation and Lemma~\ref{lem:finite_clifford_subgroups_over_binary_images}\ref{lem:finite_clifford_subgroups_over_binary_images:a},\ref{lem:finite_clifford_subgroups_over_binary_images:b}.
Items~\ref{lem:fixed_clifford_commutant_binary_image_kernel:c} and~\ref{lem:fixed_clifford_commutant_binary_image_kernel:d} are the equality and phase-adjoined equality criteria from Lemma~\ref{lem:finite_clifford_subgroups_over_binary_images}\ref{lem:finite_clifford_subgroups_over_binary_images:c},\ref{lem:finite_clifford_subgroups_over_binary_images:d}, applied with ambient subgroup $G$.
\end{proof}

This fixed-point notation will become especially useful when we shall later consider invariant bilinear forms as well.
There can be several subgroups of $\cl_n$ with binary image $\Sp(2n,\F_2)_W$, but the commutant condition singles out the subgroup in Eq.~\eqref{eq:def:clifford_fixed_commutant}.
For transvection groups, Eq.~\eqref{eq:lower_bound_paulis_in_transvection_groups} gives the basic lower bound on the Pauli kernel.
To prove equality with $\cl_n^{\Ad_{\commalg}}$, one must also verify the kernel condition in Lemma~\ref{lem:fixed_clifford_commutant_binary_image_kernel}\ref{lem:fixed_clifford_commutant_binary_image_kernel:c}; if only scalar phases are irrelevant, Lemma~\ref{lem:fixed_clifford_commutant_binary_image_kernel}\ref{lem:fixed_clifford_commutant_binary_image_kernel:d} is the corresponding phase-adjoined criterion.

These Clifford subgroups were studied in \cite{Mitsuhashi_Yoshioka_2023}, where a complete and unique circuit construction is given for Clifford groups symmetric under Pauli subgroups; see in particular \cite[Theorem~2]{Mitsuhashi_Yoshioka_2023}.
Specifically, they provided circuit decompositions via phase gates (which are transvections) and controlled gates.
Here we give the corresponding formulation in the commutant or orthogonal-complement language needed below, and later discuss these groups purely as transvection groups.

Finally, notice that the above discussion makes no assumption on the connectivity of the generating set nor rank, but simply on its symmetries.
Then, in the case $V=V^\perp = \rad(V)$, which corresponds to a commutant generated by a maximal set of commuting Pauli strings, we have that the symmetric Clifford group $\cl_n^{\Ad_\commalg}$ ($\commalg=(\C\id_1)^{\oplus 2^n}$), is the diagonal Clifford group (up to isomorphism), denoted as $\calD_n$.
Moreover, by Lemma~\ref{lem:symplectic_diagonal_kernel_transvection_generators}, given that $\DD{\rad(V)}$ is generated by transvections $h_i$ and $h_i+h_j$ from a basis of $\rad(V)$, we have that the diagonal Clifford group is also generated by the corresponding transvections $\tvu{\iso{h_i}}$ and $\tvu{\iso{h_i+h_j}}$, by the equality criterion in Proposition~\ref{prop:full_space_symplectic_symmetric_Clifford_groups}(e).
Then, the diagonal Clifford group also acquires a natural structure in terms of the additive group of binary symmetric matrices, which was also exploited to discuss diagonal gates in the more general Clifford hierarchy \cite{Rengaswamy_Calderbank_Pfister_2019,Chen_de_Silva_2024}.

In the summary below we specialize the preceding commutant notation to $W=V^\perp$, so that $V$ is the corresponding transvection generating subspace.
We summarize the above discussion in the following and specialize to the setting of transvection groups:
\begin{prop}\label{prop:full_space_symplectic_symmetric_Clifford_groups}
Let $V\subseteq\Fn$ be a subspace and let $\commalg=\Span[\C]{\inviso(V^\perp)}$.
\begin{enumerate}
\item The restriction of $\Sp(2n,\F_2)_{V^\perp}$ to $V$ gives a split short exact sequence
\begin{equation*}
    1
    \to
    \DD{\rad(V)}
    \to
    \Sp(2n,\F_2)_{V^\perp}
    \to
    \Sp^\#(V)
    \to
    1.
\end{equation*}
\item The kernel $\DD{\rad(V)}$ is isomorphic to the additive group of symmetric $r\times r$ matrices over $\F_2$, where $r=\dim\rad(V)$.
For any basis $\{h_i\}_{i=1}^r$ of $\rad(V)$, it is generated by the transvections with centers
\begin{equation*}
    \{h_i\}_{i=1}^r\cup\{h_i+h_j\}_{1\leq i<j\leq r}.
\end{equation*}
\item If $H\subseteq\Sp(2n,\F_2)_{V^\perp}$ maps onto $\Sp^\#(V)$ under the restriction map $\activemap\mapsto\activemap|_V$ and contains $\DD{\rad(V)}$, then $H=\Sp(2n,\F_2)_{V^\perp}$.
\item The map $\SY$ from Eq.~\eqref{eq:proj:conj} induces a short exact sequence
\begin{equation*}
    1\to
    \cl_n^{\Ad_{\commalg}}\cap\Pgroup_n
    \to
    \cl_n^{\Ad_{\commalg}}
    \xrightarrow{\SY}
    \Sp(2n,\F_2)_{V^\perp}
    \to 1.
\end{equation*}
\item If $H\subseteq\cl_n^{\Ad_{\commalg}}$ has binary image
$\SY(H)=\Sp(2n,\F_2)_{V^\perp}$ and Pauli kernel
\begin{equation*}
    H\cap\Pgroup_n
    =
    \cl_n^{\Ad_{\commalg}}\cap\Pgroup_n,
\end{equation*}
then $H=\cl_n^{\Ad_{\commalg}}$.
If instead the two Pauli kernels agree after adjoining $\ZZ(\Pgroup_n)$, then $H$ and $\cl_n^{\Ad_{\commalg}}$ agree after adjoining scalar phases.
\item Let $\vgens\subseteq\Fn$ be a binary generating set and $\tvgroup{\pgens}$ its transvection group, such that  $V=\Span{\vgens}$. 
If $\tvgroup{\pgens}$ contains $\DD{\rad(V)}$ and maps onto $\Sp^{\#}(V)$ under the restrict map to $V$, then $\tvgroup{\pgens}=\Sp(2n,\F_2)_{V^\perp}$.
\item Let $\pgens=\isolong{\vgens}\subseteq\PP_n$ be a Pauli generating set and $\cltvgroup{\pgens}$ its Clifford transvection group. 
If $V=\Span{\vgens}$ and $\cltvgroup{\pgens}$ has binary image $\Sp(2n,\F_2)_{V^\perp}$, $\cltvgroup{\pgens}$ and $\cl_n^{\Ad_\commalg}$ agree after adjoining scalar phases.
\end{enumerate}
\end{prop}
\begin{proof}
(a) is the short exact sequence from Lemma~\ref{lem:pointwise_symplectic_stabilizer_sequence} applied to the pointwise stabilizer of $V^\perp$.
(b) combines the symmetric-matrix description in Lemma~\ref{lem:symplectic_diagonal_kernel_structure} with the transvection generators from Lemma~\ref{lem:symplectic_diagonal_kernel_transvection_generators}.
For (c), use the restriction map from (a).
The ambient group is $G=\Sp(2n,\F_2)_{V^\perp}$, the quotient is $\Sp^\#(V)$, and the kernel is $\DD{\rad(V)}$.
If $H$ has the same quotient image and contains the full kernel, then Lemma~\ref{lem:maximality_of_quotient_group} gives $H=G$.
(d) is Lemma~\ref{lem:fixed_clifford_commutant_binary_image_kernel}\ref{lem:fixed_clifford_commutant_binary_image_kernel:b} with $W=V^\perp$.
(e) is Lemma~\ref{lem:fixed_clifford_commutant_binary_image_kernel}\ref{lem:fixed_clifford_commutant_binary_image_kernel:c},\ref{lem:fixed_clifford_commutant_binary_image_kernel:d}, again with $W=V^\perp$.
For (f), note that $\tvgroup{\vgens}$ fixes $V^\perp$ and is a simplectic group, hence $\tvgroup{\vgens}\subseteq \Sp(2n,\F_2)_{V^\perp}$. Hence it follows from (e).
For (g), by Eq.~\eqref{eq:lower_bound_paulis_in_transvection_groups}, all transvection groups contain the Pauli subgroup generated by all Paulis in the generating set.
Moreover, the transvection groups are also symmetric under their commutant $\commalg$, hence we have:
\begin{equation}
    \groupclosure{\im\pgens} \subseteq \cltvgroup{\pgens}\cap\Pgroup_n \subseteq \Pgroup_n^{\Ad_{\commalg}}
\end{equation}
Clearly, both the lower and upper bound descend to the same binary vector space $\Fn$. 
Hence, up to irrelevant phases $\{\pm 1,\pm\im\}$, $\cltvgroup{\pgens}\cap\Pgroup_n$ is precisely $\Pgroup_n^{\Ad_{\commalg}}$. 
Equivalently, after adjoining phases we have that $\groupclosure{\im\pgens}$ and $\Pgroup_n^{\Ad_{\commalg}} = \cl_n^{\Ad_\commalg}\cap\Pgroup_n$ coincide.
Hence, $\cltvgroup{\pgens}\cap\Pgroup_n$ also coincides $\cl_n^{\Ad_\commalg}\cap\Pgroup_n$ after adjoining phases.
Finally, the group $H = \cltvgroup{\pgens}$ satisfies (e), which means that $H$ and $\cl_n^{\Ad_\commalg}$ coincide after adjoining scalar phases.
\end{proof}

\subsection{Binary Orthogonal Stabilizers}\label{sec:binary_orthogonal_stabilizers}

We now pass to the case where the group or Lie algebra also conserves quadratic forms (in the $\F_2$-group case) or bilinear forms (in the Clifford case or $\R$-Pauli Lie algebra case).
Over the symplectic space $\Fn$, we first consider the group of transformations which are jointly orthogonal for multiple quadratic forms $\QQ_w$ in the notation of Definition~\ref{def:quadratic-form-associated-vector}:
\begin{equation}
    \lieO(\{\QQ_w\}_{w\in A}).
\end{equation}
We also use an orthogonal group which stabilizes a subspace $W$ pointwise:
\begin{equation}\label{eq:def:orthogonal_group_stabilizing_subspace}
    \lieO(\QQ_{w^*})_W := \Sp(2n,\F_2)_W\cap\lieO(\QQ_{w^*}).
\end{equation}
Here $\Sp(2n,\F_2)_W$ is the pointwise symplectic stabilizer of $W$ from Eq.~\eqref{eq:pointwise_symplectic_stabilizer}, i.e., the subgroup of $\Sp(2n,\F_2)$ which fixes every vector in $W$.
For a subspace $V\subseteq\Fn$ and a restricted quadratic form $\QQ^*=\QQ_{w^*}|_V$, we use $\Sp^\#(V)$ as in Lemma~\ref{lem:pointwise_symplectic_stabilizer_sequence}, namely for the subgroup of $\Sp(V)$ which fixes $\rad(V)$ pointwise, and write
\begin{equation}\label{eq:def:restricted_degenerate_orthogonal_group}
    \lieO^\#(\QQ^*) := \Sp^\#(V)\cap\lieO(\QQ^*).
\end{equation}
We record the elementary properties of these pointwise stabilizers.

\begin{lem}[Basic properties of pointwise orthogonal stabilizers]\label{lem:pointwise_orthogonal_stabilizer_properties}
Let $V\subseteq\Fn$ and let $\QQ^*=\QQ_{w^*}|_V$.
\begin{enumerate}
\item\label{lem:pointwise_orthogonal_stabilizer_properties:fixes-perp}
Every element $\activemap\in\lieO(\QQ_{w^*})_{V^\perp}$ fixes $V^\perp$ pointwise.
\item\label{lem:pointwise_orthogonal_stabilizer_properties:preserves-V}
Every element $\activemap\in\lieO(\QQ_{w^*})_{V^\perp}$ preserves $V=(V^\perp)^\perp$.
\item\label{lem:pointwise_orthogonal_stabilizer_properties:fixes-radical}
Every element $\activemap\in\lieO(\QQ_{w^*})_{V^\perp}$ fixes $\rad(V)=V\cap V^\perp$ pointwise.
\item\label{lem:pointwise_orthogonal_stabilizer_properties:preserves-restriction}
Every element $\activemap\in\lieO(\QQ_{w^*})_{V^\perp}$ preserves the restricted quadratic form $\QQ^*$ on $V$.
\item\label{lem:pointwise_orthogonal_stabilizer_properties:restriction-homomorphism}
The restriction map from $\lieO(\QQ_{w^*})_{V^\perp}$ to $\lieO^\#(\QQ^*)$ given by
$\activemap\mapsto \activemap|_V$
is a well-defined group homomorphism.
\end{enumerate}
\end{lem}
\begin{proof}
Item \ref{lem:pointwise_orthogonal_stabilizer_properties:fixes-perp} follows directly from the definition in Eq.~\eqref{eq:def:orthogonal_group_stabilizing_subspace}.
For \ref{lem:pointwise_orthogonal_stabilizer_properties:preserves-V}, let $v\in V=(V^\perp)^\perp$ and $u\in V^\perp$.
Then
$
    \symp{\activemap v}{u}
    =
    \symp{\activemap v}{\activemap u}
    =
    \symp{v}{u}
    =
    0
$,
so $\activemap v\in V$.
\ref{lem:pointwise_orthogonal_stabilizer_properties:fixes-radical} follows from \ref{lem:pointwise_orthogonal_stabilizer_properties:fixes-perp}, because $\rad(V)\subseteq V^\perp$.
\ref{lem:pointwise_orthogonal_stabilizer_properties:preserves-restriction} follows from the preservation of $\QQ_{w^*}$ in Eq.~\eqref{eq:def:orthogonal_group_stabilizing_subspace} together with \ref{lem:pointwise_orthogonal_stabilizer_properties:preserves-V}.
For \ref{lem:pointwise_orthogonal_stabilizer_properties:restriction-homomorphism}, \ref{lem:pointwise_orthogonal_stabilizer_properties:preserves-V} and~\ref{lem:pointwise_orthogonal_stabilizer_properties:fixes-radical} show that the restriction of $\activemap$ to $V$ lies in $\Sp^\#(V)$, in the sense of Lemma~\ref{lem:pointwise_symplectic_stabilizer_sequence}.
\ref{lem:pointwise_orthogonal_stabilizer_properties:preserves-restriction} shows that this restriction also preserves $\QQ^*$.
Thus the restriction lies in $\Sp^\#(V)\cap\lieO(\QQ^*)=\lieO^\#(\QQ^*)$ by Eq.~\eqref{eq:def:restricted_degenerate_orthogonal_group}.
It is a homomorphism because composition of maps restricts to composition on the invariant subspace $V$.
\end{proof}

The first useful point is that joint preservation of an affine space of quadratic forms is the same as preserving one reference quadratic form and fixing the associated subspace of the affine space.
This is the orthogonal analogue of the way a Pauli commutant is encoded by a pointwise stabilizer.

\begin{lem}[Joint stabilizer of affine quadratic forms]\label{lem:joint_orthogonal_stabilizer_affine_quadratic_forms}
Let $A\subseteq w^*+W\subseteq\Fn$ be a subset such that every element of $w^*+W$ can be written as $\sum_i c_i a_i$ with $a_i\in A$, $c_i\in\F_2$, and $\sum_i c_i=1$.
Then
\begin{equation}\label{eq:joint_orthogonal_stabilizer_affine_quadratic_forms}
    \lieO(\{\QQ_w\}_{w\in A})
    =
    \lieO(\QQ_{w^*})_W .
\end{equation}
\end{lem}
\begin{proof}
It is enough to understand how a fixed element $\activemap\in\Sp(2n,\F_2)$ acts on differences of quadratic forms.
For $u\in\Fn$, Definition~\ref{def:quadratic-form-associated-vector} gives
\[
    \QQ_{w+u}(v)=\QQ_w(v)+\symp{u}{v}.
\]
After pullback by $\activemap^{-1}$, this difference becomes
\[
    \symp{u}{\activemap^{-1}v}=\symp{\activemap u}{v}.
\]
Thus, once $\activemap$ preserves the reference form $\QQ_{w^*}$, it preserves $\QQ_{w^*+u}$ if and only if $\activemap u=u$.
Consequently, preserving every form in the affine space $w^*+W$ is equivalent to preserving $\QQ_{w^*}$ and fixing every vector in $W$.
It remains to pass from $A$ to all of $w^*+W$.
If $w=\sum_i c_i a_i$ with $a_i\in A$, $c_i\in\F_2$, and $\sum_i c_i=1$, then the corresponding quadratic form is obtained from the same affine combination of the forms $\QQ_{a_i}$.
Thus an element which preserves all $\QQ_{a_i}$ with $a_i\in A$ also preserves $\QQ_w$.
By the assumption on $A$, this gives preservation of all forms indexed by $w^*+W$.
This proves Eq.~\eqref{eq:joint_orthogonal_stabilizer_affine_quadratic_forms}.
\end{proof}

This formulation also makes visible the same exceptional collapse already encountered for transvection groups in Lemma~\ref{lem:exceptional_plus_affine_space_collapse}.
It is not caused by passing from the binary symplectic group to Clifford subgroups, but already occurs at the level of the binary joint stabilizer.
\begin{rem}[The exceptional one-pair collapse]\label{rem:exceptional_one_pair_collapse_orthogonal_stabilizers}
The joint orthogonal stabilizer with $\type(\QQ^*)=0$ and one non-degenerate symplectic pair ($m=1$) can coincide with the stabilizer described as a type $+$ case with one more radical vector.
For example, in the two-dimensional non-degenerate space with symplectic basis $e,f$, the group $\lieO^+(2,\F_2)\subseteq\Sp(2,\F_2)$ contains only the identity and the transvection $\tau_{e+f}$, which swaps $e$ and $f$.
Both elements fix $e+f$.
Thus the same group can also be viewed as the stabilizer of the type $0$ form with one-dimensional anisotropic radical generated by $e+f$, giving $\lieO^+(2,\F_2)\cong \lieO_0^{\#}(0,2,\F_2)$ in this notation.
\end{rem}

We next restrict the pointwise stabilizer to the subspace $V=W^\perp$.
This produces the orthogonal analogue of the symplectic restriction sequence from Lemma~\ref{lem:pointwise_symplectic_stabilizer_sequence}.

\begin{prop}[Restriction sequence for pointwise orthogonal stabilizers]\label{prop:pointwise_orthogonal_stabilizer_sequence}
Let $V\subseteq\Fn$, let $\QQ^*=\QQ_{w^*}|_V$, and set $W=V^\perp$.
Then restriction to $V$ gives a short exact sequence
\begin{equation}\label{eq:quotient_orthogonal_group_over_Fn}
1 \to \OD{\rad(V)}{\QQ_{w^*}} \to \lieO(\QQ_{w^*})_{V^\perp} \to \lieO^{\#}(\QQ^*) \to 1 ,
\end{equation}
where
\begin{equation}\label{eq:def:orthogonal_diagonal_group}
\OD{\rad(V)}{\QQ_{w^*}} := \lieO(\QQ_{w^*})\cap \DD{\rad(V)}
\end{equation}
is the \emph{orthogonal} diagonal group of transformations with reference invariant quadratic form $\QQ_{w^*}$.
\end{prop}
\begin{proof}
Lemma~\ref{lem:pointwise_orthogonal_stabilizer_properties}\ref{lem:pointwise_orthogonal_stabilizer_properties:restriction-homomorphism}
gives a well-defined homomorphism $\rho$ from $\lieO(\QQ_{w^*})_{V^\perp}$ to $\lieO^\#(\QQ^*)$ with
$\rho(\activemap)=\activemap|_V$.
We first identify its kernel.
An element of $\ker(\rho)$ fixes $V$ pointwise and, by definition of $\lieO(\QQ_{w^*})_{V^\perp}$, also fixes $V^\perp$ pointwise.
Thus it lies in the symplectic diagonal kernel $\DD{\rad(V)}$ from Lemma~\ref{lem:pointwise_symplectic_stabilizer_sequence}.
Since it also preserves $\QQ_{w^*}$, the kernel is precisely $\OD{\rad(V)}{\QQ_{w^*}}$.

It remains to see that $\rho$ is onto.
Let $\alpha\in\lieO^\#(\QQ^*)$.
On $V+V^\perp$, define a map by applying $\alpha$ to the $V$ part and the identity to the $V^\perp$ part.
This is well-defined because $\alpha$ fixes $\rad(V)=V\cap V^\perp$ pointwise.
It preserves the restriction of $\QQ_{w^*}$ to $V+V^\perp$: it preserves $\QQ^*$ on $V$, fixes $V^\perp$ pointwise, and the mixed symplectic products between $V$ and $V^\perp$ vanish.
By Witt extension over $\F_2$ in Proposition~\ref{prop:witt-extension-f2}, this isometry extends to an element of $\lieO(\QQ_{w^*})$ on $\Fn$.
The extension fixes $V^\perp$ pointwise, hence lies in $\lieO(\QQ_{w^*})_{V^\perp}$, and its restriction to $V$ is $\alpha$.
Thus $\rho$ is surjective.
\end{proof}

Up to an isomorphism, we introduce the notation $\lieO_\varepsilon^{\#}(2m,r,\F_2)$ for  $\lieO^{\#}(\QQ^*)$. Here, $\varepsilon\in\{0,\pm\}$ depends on the isomorphism class of $\QQ^*$, while $2m$ and $r$ are respectively the rank and dimension of the radical of $V$.
The kernel in Proposition~\ref{prop:pointwise_orthogonal_stabilizer_sequence} is analyzed explicitly in Section~\ref{sec:orthogonal_diagonal_kernels_maximality} by starting from the symplectic diagonal kernel and then imposing preservation of $\QQ_{w^*}$.
This gives the matrix condition in Lemma~\ref{lem:orthogonal_diagonal_kernel_candidate}; the corresponding transvection generators are recorded in Lemma~\ref{lem:orthogonal_diagonal_kernel_transvection_generators_candidate}, and both enter Proposition~\ref{prop:orthogonal_stabilizer_binary_maximality_candidate}.

Unlike the case of $\Sp^{\#}(V)$, we in general cannot directly quotient $\lieO_\varepsilon^{\#}(2m,r,\F_2)$ by the radical to obtain a non-degenerate orthogonal group.
Namely, if $\type(\QQ^*)=0$, then there is $u\in\rad(V)$ with $\QQ^*(u)=1$.
For such a vector, $\QQ^*(v+u)=\QQ^*(v)+1$, hence the restriction is not well-defined up to the entire radical.
We also record the remaining quotient behavior.
However, we can still consider the restriction from $V$ to $V/\radzero(V,\QQ^*)$, which results in the following short exact sequences:
\begin{gather}
    1 \to K \to \lieO_\pm^{\#}(2m,r,\F_2) \to \lieO_\pm(2m,\F_2) \to 1,
    \label{eq:orthogonal_pm_radical_quotient_sequence}\\
    1 \to K \to \lieO_0^{\#}(2m,r,\F_2) \to \lieO_0^{\#}(2m,1,\F_2) \to 1.
    \label{eq:orthogonal_zero_radical_quotient_sequence}
\end{gather}
with suitable kernels $K$.

The target of Eq.~\eqref{eq:orthogonal_zero_radical_quotient_sequence} still has a one-dimensional anisotropic radical.
The following lemma identifies the remaining quotient: its induced action on the non-degenerate quotient has image the full symplectic group.

\begin{lem}\label{lem:type_zero_anisotropic_radical_quotient}
Let $(V,\QQ)$ be a binary quadratic space with radical $R=\rad(V)$ such that $\dim(R)=1$ and $\QQ(R)=\F_2$.
Let $\lieO_0^\#(V)$ be the isometry group of $\QQ$ fixing $R$ pointwise.
Then the induced action on $V/R$ gives a split exact sequence
\begin{equation*}
    1\to \tilde{K}\to \lieO_0^\#(V)\to \Sp(V/R)\to 1 .
\end{equation*}
Equivalently, every symplectic automorphism of $V/R$ is induced by an element of $\lieO_0^\#(V)$.
\end{lem}

\begin{proof}
We first construct the quotient homomorphism and identify its kernel.
Every element of $\lieO_0^\#(V)$ preserves the alternating bilinear form associated with $\QQ$ and fixes $R$ pointwise, hence it induces a symplectic automorphism of the non-degenerate quotient $V/R$.
This gives a group homomorphism from $\lieO_0^\#(V)$ to $\Sp(V/R)$ with kernel $\tilde{K}$.

It remains to see that this homomorphism is surjective and split.
For this, we start with an arbitrary symplectic automorphism of the quotient and construct a lift which preserves the quadratic form.
Choose a complement $N$ to $R=\Span[\F_2]{h}$ such that $V=N\oplus R$ and $N$ is non-degenerate for the induced alternating form.
Since $\QQ(h)=1$, every $v\in V$ can be written as $v=x+\gamma h$ with $x\in N$ and
\begin{equation*}
    \QQ(x+\gamma h)=q_N(x)+\gamma,
\end{equation*}
where $q_N=\QQ|_N$.
The naive lift $x+\gamma h\mapsto gx+\gamma h$ need not preserve $\QQ$, since $g$ preserves the associated alternating form but not necessarily the quadratic form $q_N$.
Since $h\in R$ and $\QQ(h)=1$, changing the $h$-coordinate by $\delta\in\F_2$ changes the value of $\QQ$ by $\delta$.
We therefore correct the $h$-coordinate by the change in the quadratic value.
For any $g\in\Sp(N)$, define
\begin{equation*}
    \tilde{g}(x+\gamma h)
    =
    gx+\bigl(\gamma+q_N(x)+q_N(gx)\bigr)h .
\end{equation*}
Then $\tilde{g}$ fixes $h$, since $\tilde{g}(h)=\tilde{g}(0+1h)=h$.
It induces $g$ on $V/R\cong N$, and it preserves $\QQ$, because
\begin{align*}
    \QQ(\tilde{g}(x+\gamma h))
    &= q_N(gx)+\gamma+q_N(x)+q_N(gx)\\
    &= q_N(x)+\gamma
    = \QQ(x+\gamma h).
\end{align*}
Thus every element of $\Sp(V/R)$ has a lift in $\lieO_0^\#(V)$, and the displayed construction gives a splitting.
\end{proof}

\subsection{Orthogonal Diagonal Kernels and Maximality}\label{sec:orthogonal_diagonal_kernels_maximality}

We now isolate the restriction and kernel information needed for later maximality arguments in the case of binary orthogonal groups.

\begin{lem}[Restriction of orthogonal stabilizers]\label{lem:orthogonal_stabilizer_restriction_candidate}
Let $V\subseteq\Fn$ be a subspace, $\QQ_{w^*}$ be a quadratic form on $\Fn$, and $\QQ^*=\QQ_{w^*}|_V$.
The restriction map from $\lieO(\QQ_{w^*})_{V^\perp}$ to $V$ has image $\lieO^\#(\QQ^*)$ and kernel $\OD{\rad(V)}{\QQ_{w^*}}$ [see Eqs.~\eqref{eq:def:orthogonal_group_stabilizing_subspace} and~\eqref{eq:quotient_orthogonal_group_over_Fn}].
Equivalently, there exists a short exact sequence
\begin{equation*}
    1\to
    \OD{\rad(V)}{\QQ_{w^*}}
    \to
    \lieO(\QQ_{w^*})_{V^\perp}
    \to
    \lieO^\#(\QQ^*)
    \to 1 .
\end{equation*}
\end{lem}
\begin{proof}
We first check that restriction gives the claimed image.
By applying Lemma~\ref{lem:pointwise_orthogonal_stabilizer_properties}, we show that every element of $\lieO(\QQ_{w^*})_{V^\perp}$ preserves $V$, fixes $\rad(V)$ pointwise, and preserves the restricted quadratic form $\QQ^*$ on $V$.
Thus restriction to $V$ defines a homomorphism with image in $\lieO^\#(\QQ^*)$, as summarized in Lemma~\ref{lem:pointwise_orthogonal_stabilizer_properties}\ref{lem:pointwise_orthogonal_stabilizer_properties:restriction-homomorphism}.

The image is all of $\lieO^\#(\QQ^*)$ by the restriction sequence in Eq.~\eqref{eq:quotient_orthogonal_group_over_Fn}.
Equivalently, the exactness of that sequence states that every element of $\lieO^\#(\QQ^*)$ is obtained as the restriction of an element of $\lieO(\QQ_{w^*})_{V^\perp}$.

It remains to identify the kernel.
The kernel consists precisely of those elements of $\lieO(\QQ_{w^*})_{V^\perp}$ which act trivially on $V$.
By Lemma~\ref{lem:pointwise_symplectic_stabilizer_sequence}, the elements of $\Sp(2n,\F_2)_{V^\perp}$ which act trivially on $V$ are the elements of $\DD{\rad(V)}$.
Intersecting with the condition of preserving $\QQ_{w^*}$ yields the kernel
\[
    \OD{\rad(V)}{\QQ_{w^*}}
    =
    \lieO(\QQ_{w^*})\cap\DD{\rad(V)},
\]
as defined in Eq.~\eqref{eq:quotient_orthogonal_group_over_Fn}.
\end{proof}

The kernel in Lemma~\ref{lem:orthogonal_stabilizer_restriction_candidate} is the part that has to be generated
when one wants to upgrade a subgroup with the correct quotient action to the
full orthogonal stabilizer.
The following statement provides its explicit form.

\begin{lem}[Orthogonal diagonal kernel]\label{lem:orthogonal_diagonal_kernel_candidate}
In the setting of Lemma~\ref{lem:orthogonal_stabilizer_restriction_candidate}, let $r=\dim\rad(V)$.
The kernel $\OD{\rad(V)}{\QQ_{w^*}}$ is isomorphic to the following additive groups:
\begin{enumerate}
    \item if $\type(\QQ^*)\in\{+,-\}$, the $r\times r$ off-diagonal symmetric matrices over $\F_2$;
    \item if $\type(\QQ^*)=0$, the $r\times r$ symmetric matrices over $\F_2$ subject to the constraint $B_{ii}=B_{i1}$.
\end{enumerate}
\end{lem}
\begin{proof}
By Lemma~\ref{lem:orthogonal_stabilizer_restriction_candidate}, the kernel is
the subgroup of $\DD{\rad(V)}$ which also preserves $\QQ_{w^*}$.
Thus we may start from the explicit description of the symplectic kernel
$\DD{\rad(V)}$ in Lemma~\ref{lem:symplectic_diagonal_kernel_structure}.
Namely, its elements are the maps $v\mapsto v+\phi(v)$ with
$\Im(\phi)\subseteq\rad(V)$, and, after choosing a basis of $\rad(V)$ and dual
vectors as below, the coefficient matrix of $\phi$ is symmetric.
Conversely, every such symmetric coefficient matrix gives an element of
$\DD{\rad(V)}$.

It remains to impose preservation of $\QQ_{w^*}$.
Choose a basis $\{h_i\}_{i=1}^r$ of $\rad(V)$ and symplectically dual vectors
$\{\tilde h_i\}$, meaning $\symp{\tilde h_i}{h_j}=\delta_{ij}$, as in the calculation preceding
Eq.~\eqref{eq:quotient_orthogonal_group_over_Fn}, and write
\[
    \phi(\tilde h_i)=\sum_j B_{ij}h_j .
\]
The condition from Eq.~\eqref{eq:def:orthogonal_group_stabilizing_subspace}
that $v\mapsto v+\phi(v)$ preserve $\QQ_{w^*}$ is obtained by
expanding $\QQ_{w^*}(v+\phi(v))=\QQ_{w^*}(v)$ using the quadratic identity,
and gives
\[
    \QQ_{w^*}(\phi(v))=\symp{\phi(v)}{v}.
\]
For $v=\tilde h_i$, we use that $\phi(\tilde h_i)\in\rad(V)$, so $\QQ_{w^*}$
is additive on the span of the $h_j$, and that the chosen dual vectors satisfy
$\symp{h_j}{\tilde h_i}=\delta_{ij}$.
This results in
\begin{equation}\label{eq:orthogonal_diagonal_kernel_matrix_condition}
    \sum_j B_{ij}\QQ_{w^*}(h_j)=B_{ii}.
\end{equation}

If $\type(\QQ^*)\in\{+,-\}$, then
$\QQ_{w^*}$ vanishes on $\rad(V)$, so
Eq.~\eqref{eq:orthogonal_diagonal_kernel_matrix_condition} is simply
$B_{ii}=0$.
The kernel is therefore the additive group of off-diagonal symmetric matrices.
If $\type(\QQ^*)=0$, choose the radical basis so that
$\QQ_{w^*}(h_j)=\delta_{j1}$.
Then Eq.~\eqref{eq:orthogonal_diagonal_kernel_matrix_condition} is
$B_{ii}=B_{i1}=B_{1i}$, giving the second claimed
additive group.
\end{proof}

For later use with transvection groups, we also record explicit transvection
generators for the orthogonal diagonal kernel.

\begin{lem}[Transvection generators for the orthogonal diagonal kernel]\label{lem:orthogonal_diagonal_kernel_transvection_generators_candidate}
In the setting of Lemma~\ref{lem:orthogonal_diagonal_kernel_candidate}, the following hold:
\begin{enumerate}
    \item If $\type(\QQ^*)\in\{+,-\}$ and $z\in V$ is anisotropic, then the fourfold products
    \begin{equation*}
        \tau_{z+u+u'}\tau_{z+u}\tau_{z+u'}\tau_z,\qquad u,u'\in\rad(V),
    \end{equation*}
    generate $\OD{\rad(V)}{\QQ_{w^*}}$. 
    In particular, for $\{h_i\}_{i=1}^n$ any basis of $\rad(V)$, $\OD{\rad(V)}$ is generated by the transvections with centers $z+h_i$ and $z+h_i+h_j$ for $1\leq i<j\leq r$.
    \item If $\type(\QQ^*)\in\{+,-\}$, then no transvection $\tau_v$ with center $v$ in $\rad(V)$ lies in $\OD{\rad(V)}{\QQ_{w^*}}$.
    \item If $\type(\QQ^*)=0$, then the transvections $\tau_v$ with centers $v$ in $\radone(V,\QQ^*)$ generate $\OD{\rad(V)}{\QQ_{w^*}}$. 
    In particular, for $\{h_i\}_{i=1}^n$ a basis of $\rad(V)$ with $h_1\in\radone(V,\QQ^*)$ and $h_i\in\radzero(V,\QQ^*)$, $\OD{\rad(V)}$ is generated by the transvections with centers $h_1$, $h_1+h_i$ and $h_1+h_i+h_j$ for $1<i<j\leq r$.
\end{enumerate}
\end{lem}
\begin{proof}
We first use the same criterion in both cases.
A transvection $\tau_{\tilde{v}}$ preserves $\QQ_{w^*}$ precisely when its center $\tilde{v}$ is anisotropic for $\QQ_{w^*}$.
Thus, the transvections in the orthogonal diagonal kernel whose centers lie in $\rad(V)$ are exactly those with centers in $\radone(V,\QQ^*)$.

Assuming that $\type(\QQ^*)\in\{+,-\}$, Lemma~\ref{lem:isotropic:radical}(d),(g) gives
$\radzero(V,\QQ^*)=\rad(V)$ and $\radone(V,\QQ^*)=\emptyset$.
Therefore no transvection with center in $\rad(V)$ lies in $\OD{\rad(V)}{\QQ_{w^*}}$, proving (b).

It remains in the same case to generate the kernel without using transvections whose centers lie in $\rad(V)$.
Choose an anisotropic vector $z\in V$.
For $u,u'\in\rad(V)$, all four centers
$z$, $z+u$, $z+u'$ and $z+u+u'$ are anisotropic, and they are pairwise commuting because $u,u'\in\rad(V)$ and $z\in V$.
Hence each transvection in the following product preserves $\QQ_{w^*}$, and the product fixes $V$ after restriction:
\begin{align}
  &\tau_{z+u+u'}\tau_{z+u}\tau_{z+u'}\tau_z(x)
  =\tau_{z+u+u'}\tau_{z+u}\tau_{z+u'}\, (x+\symp{z}{x}z) \nonumber\\
  &=
  \begin{aligned}[t]
  &x+\symp{z}{x}z\\
  &+\symp{z+u'}{x}(z+u')+\symp{z+u}{x}(z+u)\\
  &+\symp{z+u+u'}{x}(z+u+u')
  \end{aligned}
  \nonumber\\
  &= x+\symp{u}{x}u' + \symp{u'}{x}u.
  \label{eq:four:transvections}
\end{align}
Taking $u=h_i$ and $u'=h_j$ gives the off-diagonal basis transformations of the additive group from Lemma~\ref{lem:orthogonal_diagonal_kernel_candidate}.
These transformations generate the whole kernel in the $\type(\QQ^*)\in\{+,-\}$ case, proving (a).

We now treat $\type(\QQ^*)=0$.
For $\type(\QQ^*)=0$, Lemma~\ref{lem:isotropic:radical}(e),(f),(h) says
that $\radzero(V,\QQ^*)$ has codimension one in $\rad(V)$ and that
$\radone(V,\QQ^*)$ is an affine coset over $\radzero(V,\QQ^*)$.
Choose $h_1\in\radone(V,\QQ^*)$ and
$h_i\in\radzero(V,\QQ^*)$ for $i>1$.
Then $\radone(V,\QQ^*)=h_1+\radzero(V,\QQ^*)$.
The generators of the additive group in Lemma~\ref{lem:orthogonal_diagonal_kernel_candidate} are obtained as follows.
First, the transvection $\tau_{h_1}$ gives the diagonal generator corresponding to $B_{11}$.
Second, Eq.~\eqref{eq:four:transvections}, applied with $z=h_1$, gives the off-diagonal generators with $1<i<j$.
Third, it remains to obtain the generators satisfying $B_{ii}=B_{i1}$.
For this, we use
\begin{align}
&\tau_{h_1+u}\tau_{h_1}(x) = \tau_{h_1+u}\, (x+\symp{h_1}{x}h_1)  \nonumber \\
&= 
\begin{aligned}[t]
&x+\symp{h_1}{x}h_1  \nonumber \\
&+\symp{h_1+u}{x+\symp{h_1}{x}h_1}(h_1+u)
\end{aligned}  \nonumber \\
&=x+\symp{h_1}{x}u+\symp{u}{x}h_1+\symp{u}{x}u,
\label{eq:two:transvections}
\end{align}
where both centers $h_1+u$ and $h_1$ lie in $\radone(V,\QQ^*)$, since
$h_1+\radzero(V,\QQ^*)=\radone(V,\QQ^*)$.
Taking $u=h_i$ in Eq.~\eqref{eq:two:transvections} gives the required
generators satisfying $B_{ii}=B_{i1}$.
Together these three families generate the whole kernel in the $\type(\QQ^*)=0$ case.
Each factor used has center in $\radone(V,\QQ^*)$, so the transvections with centers in $\radone(V,\QQ^*)$ generate $\OD{\rad(V)}{\QQ_{w^*}}$, proving (c).
\end{proof}

Combining the quotient action from Lemma~\ref{lem:orthogonal_stabilizer_restriction_candidate}
with the kernel description from Lemma~\ref{lem:orthogonal_diagonal_kernel_candidate}
and, in transvection applications, the generators from
Lemma~\ref{lem:orthogonal_diagonal_kernel_transvection_generators_candidate}
yields the following maximality criterion.

\begin{prop}[Binary maximality of the orthogonal stabilizer]\label{prop:orthogonal_stabilizer_binary_maximality_candidate}
Let $H\subseteq\Sp(2n,\F_2)$ be a subgroup which fixes $V^\perp$ pointwise and preserves $\QQ_{w^*}$.
If the restriction of $H$ to $V$ is $\lieO^\#(\QQ^*)$ and $H$ contains the orthogonal diagonal kernel $\OD{\rad(V)}{\QQ_{w^*}}$, then
\begin{equation*}
    H=\lieO(\QQ_{w^*})_{V^\perp}.
\end{equation*}
\end{prop}
\begin{proof}
Since $H$ fixes $V^\perp$ pointwise and preserves $\QQ_{w^*}$, we have
$H\subseteq\lieO(\QQ_{w^*})_{V^\perp}$ by Eq.~\eqref{eq:def:orthogonal_group_stabilizing_subspace}.
For the reverse inclusion, let
$\activemap\in\lieO(\QQ_{w^*})_{V^\perp}$.
By Lemma~\ref{lem:pointwise_orthogonal_stabilizer_properties}\ref{lem:pointwise_orthogonal_stabilizer_properties:restriction-homomorphism}, equivalently by Lemma~\ref{lem:orthogonal_stabilizer_restriction_candidate}, the
restriction of $\activemap$ to $V$ lies in $\lieO^\#(\QQ^*)$.
By assumption, the restriction of $H$ to $V$ is all of $\lieO^\#(\QQ^*)$,
so there is some $\activemap_H\in H$ with
$\activemap_H|_V=\activemap|_V$.
Then $\activemap_H^{-1}\activemap$ is trivial on $V$, hence belongs to the
kernel $\OD{\rad(V)}{\QQ_{w^*}}$ of the restriction map.
Since this kernel is contained in $H$ by assumption, we get
$\activemap_H^{-1}\activemap\in H$, and therefore
$\activemap\in H$.
Thus $\lieO(\QQ_{w^*})_{V^\perp}\subseteq H$, proving equality.
\end{proof}

\subsection{Clifford Subgroups of Isometries}\label{sec:clifford_fixed_subgroups_orthogonal_images}

The commutant-only case in Subsection~\ref{sec:commutant_symmetric_clifford_groups_lie_algebras} gives Clifford subgroups over pointwise symplectic stabilizers.
We now refine this by adding the fixed-point condition coming from a Pauli bilinear form, whose binary image is the pointwise orthogonal stabilizer from Subsection~\ref{sec:binary_orthogonal_stabilizers}.
We can now pass from the binary isometry groups to the corresponding subgroups of the finite Clifford group that respect the same bilinear symmetries and commutant symmetries.
For Pauli strings $B_i,C_j\in\PP_n$, we write
\begin{align}
    &\cl_n^{\{\Theta_{B_i}\},\{\Ad_{C_j}\}} :=\\
    &
    \{U\in\cl_n\mid
    \Theta_{B_i}(U)=U\ \forall i,\;
    \Ad_{C_j}(U)=U\ \forall j\}. \nonumber
\end{align}
In particular, if $\commalg$ is a Pauli matrix algebra, then
\begin{alignat}{3}\label{eq:def:clifford_fixed_bilinear_commutant}
    \cl_n^{\Theta_{B^*},\Ad_{\commalg}}
    :=
    \{U\in\cl_n\mid\;
    & \Theta_{B^*}(U)=U,\;
    \Ad_C(U)=U \nonumber \\
    &\text{for all } C\in\commalg\cap\PP_n\}
\end{alignat}
denotes the subgroup fixed by the single bilinear-form involution $\Theta_{B^*}$ and by the adjoint symmetries determined by $\commalg$.
For the purpose of classification, it is enough to impose these adjoint conditions for any set of Pauli strings whose generated matrix algebra is $\commalg$.

\begin{rem}[Binary constraints and the Clifford fixed points]\label{rem:binary_constraints_literal_clifford_fixed_points}
At the binary level, the corresponding bilinear-form constraints are encoded by
the set
\begin{equation*}
    B^*\commalg:=\{B^*C\mid C\in\commalg\cap\PP_n\}.
\end{equation*}
At the Clifford level, however, we keep the fixed subgroup
$\cl_n^{\Theta_{B^*},\Ad_{\commalg}}$ from
Eq.~\eqref{eq:def:clifford_fixed_bilinear_commutant} as the primary object.
Indeed, the involution $\Theta_{B^*}$ is sensitive to scalar phases; for example, $\Theta_{B^*}(\im I)=-\im I$.
Thus, when scalar phases are to be ignored, we compare groups after adjoining $\ZZ(\Pgroup_n)$ as in Lemma~\ref{lem:finite_clifford_subgroups_over_binary_images}\ref{lem:finite_clifford_subgroups_over_binary_images:d}.
\end{rem}

The following lemma provides the precise binary image and Pauli kernel of this fixed subgroup.

\begin{lem}[Clifford subgroups of isometries]\label{lem:fixed_clifford_bilinear_commutant_binary_image_kernel}
Let $V\subseteq\Fn$ and $\commalg=\Span[\C]{\inviso(V^\perp)}$, where $\inviso=\isoempty^{-1}$ as in Eq.~\eqref{eq:pauli:binary},
$w^*\in\Fn$, and $B^*=\iso{w^*}$.
\begin{enumerate}
\item The map $\SY$ from Eq.~\eqref{eq:proj:conj} induces an isomorphism
\begin{equation}
    \cl_n^{\Theta_{B^*},\Ad_{\commalg}}/
    (\cl_n^{\Theta_{B^*},\Ad_{\commalg}}\cap\Pgroup_n)
    \cong
    \lieO(\QQ_{w^*})_{V^\perp}.
\end{equation}
\item Equivalently, $\SY$ induces a short exact sequence
\begin{equation}\label{eq:quotient_fixed_clifford_bilinear_commutant_to_binary}
    1\to
    \cl_n^{\Theta_{B^*},\Ad_{\commalg}}\cap\Pgroup_n
    \to
    \cl_n^{\Theta_{B^*},\Ad_{\commalg}}
    \xrightarrow{\SY}
    \lieO(\QQ_{w^*})_{V^\perp}
    \to 1.
\end{equation}
\item If a subgroup $H$ of $\cl_n$ satisfies
\begin{gather*}
    H\subseteq\cl_n^{\Theta_{B^*},\Ad_{\commalg}},
    \quad
    \SY(H)=\lieO(\QQ_{w^*})_{V^\perp},\\
    H\cap\Pgroup_n=
    \cl_n^{\Theta_{B^*},\Ad_{\commalg}}\cap\Pgroup_n,
\end{gather*}
then $H=\cl_n^{\Theta_{B^*},\Ad_{\commalg}}$.
\item If a subgroup $H$ of $\cl_n$ satisfies
\begin{align*}
    &H\subseteq\cl_n^{\Theta_{B^*},\Ad_{\commalg}},
    \quad
    \SY(H)=\lieO(\QQ_{w^*})_{V^\perp},\\
    &H\cap\Pgroup_n \simeq \cl_n^{\Theta_{B^*},\Ad_{\commalg}}\cap\Pgroup_n,
\end{align*}
then the two subgroups agree after adjoining scalar phases, i.e.,
\begin{equation*}
    H \simeq \cl_n^{\Theta_{B^*},\Ad_{\commalg}}.
\end{equation*}
\end{enumerate}
\end{lem}
\begin{proof}
To apply Lemma~\ref{lem:finite_clifford_subgroups_over_binary_images}\ref{lem:finite_clifford_subgroups_over_binary_images:a},\ref{lem:finite_clifford_subgroups_over_binary_images:b}, it remains to identify the binary image of
\[
    G:=\cl_n^{\Theta_{B^*},\Ad_{\commalg}}
\]
under $\SY$.
Indeed, Lemma~\ref{lem:finite_clifford_subgroups_over_binary_images}\ref{lem:finite_clifford_subgroups_over_binary_images:b} says that the restriction of $\SY$ to any subgroup $G$ of $\cl_n$ has kernel $G\cap\Pgroup_n$.

First, let $U\in G$ and set $g=\SY(U)$.
The condition $\Theta_{B^*}(U)=U$ is equivalent to $U^TB^*U=B^*$.
By Lemma~\ref{lem:bilinear_pauli_and_binary_quadratic_forms}, this means that the corresponding binary quadratic form is unchanged, or $\QQ_{w^*}\circ g^{-1}=\QQ_{w^*}$.
Thus $g$ preserves $\QQ_{w^*}$.
Moreover, the conditions $\Ad_C(U)=UCU^\dagger=C$ for all $C\in\commalg\cap\PP_n$ imply that $g$ fixes $\inviso(C)$ for all such $C$, hence fixes $V^\perp$ pointwise.
Hence
\[
    \SY(G)\subseteq \lieO(\QQ_{w^*})_{V^\perp},
\]
with the latter group defined in Eq.~\eqref{eq:def:orthogonal_group_stabilizing_subspace}.

Second, we prove the reverse inclusion.
Let $g\in\lieO(\QQ_{w^*})_{V^\perp}$.
By Lemma~\ref{lem:finite_clifford_subgroups_over_binary_images}\ref{lem:finite_clifford_subgroups_over_binary_images:a}, choose $U\in\cl_n$ with $\SY(U)=g$.
Since $g$ fixes $V^\perp$ pointwise, conjugation by $U$ sends every Pauli string in $\commalg\cap\PP_n$ to itself up to a sign.
More explicitly, Eq.~\eqref{eq:eta} gives a function $\eta_U$ such that
\[
    U\iso{c}U^\dagger=(-1)^{\eta_U(c)}\iso{c}
    \qquad \text{for } c\in V^\perp.
\]
For $c,d\in V^\perp$, Lemma~\ref{lem:sign:clifford}\ref{lem:sign:clifford:f} gives
\[
    \eta_U(c+d)+\eta_U(c)+\eta_U(d)
    =
    \sign{c}{d}+\sign{\SY_Uc}{\SY_Ud}.
\]
Since $\SY_Uc=c$ and $\SY_Ud=d$, the right-hand side vanishes.
Thus the restriction of $\eta_U$ to $V^\perp$ is linear.
By non-degeneracy of the symplectic form on $\Fn$, choose $p\in\Fn$ such that $\symp{p}{c}=\eta_U(c)$ for all $c\in V^\perp$.
Set $P=\iso{p}$.
Lemma~\ref{lem:sign:clifford}\ref{lem:sign:clifford:d} says that left multiplication by $P$ changes the sign function by
\[
    \eta_{PU}(c)=\eta_U(c)+\symp{p}{\SY_Uc}.
\]
Since $\SY_Uc=c$ on $V^\perp$ and $\symp{p}{c}=\eta_U(c)$, we get $\eta_{PU}(c)=0$ for all $c\in V^\perp$.
Therefore
\[
    (PU)\iso{c}(PU)^\dagger=\iso{c}
    \qquad \text{for } c\in V^\perp.
\]
Thus $PU$ satisfies the required adjoint fixed-point conditions and still has binary image $g$.
It remains to check the $\Theta_{B^*}$ fixed-point condition.
Set $U_0=PU$.
Transporting the Pauli bilinear form $B^*$ by $U_0$ gives another Pauli bilinear form
\[
    B_{U_0}:=U_0^T B^* U_0 .
\]
By the correspondence from Lemma~\ref{lem:bilinear_pauli_and_binary_quadratic_forms}, applied after this Clifford transport, the corresponding binary quadratic form is $\QQ_{w^*}\circ g^{-1}$.
Since $g\in\lieO(\QQ_{w^*})_{V^\perp}$ preserves $\QQ_{w^*}$ by Eq.~\eqref{eq:def:orthogonal_group_stabilizing_subspace}, we have $\QQ_{w^*}\circ g^{-1}=\QQ_{w^*}$.
Hence, with our Pauli-string normalization, $B_{U_0}=\pm B^*$.
Equivalently,
\[
    U_0^TB^*U_0=\pm B^*,
\]
which implies $\Theta_{B^*}(U_0)=\pm U_0$.
If the sign is negative, replacing $PU$ by $\im PU$ makes the sign positive, because $\Theta_{B^*}(\im I)=-\im I$; this scalar phase again does not change the binary image.
Thus $g$ is the binary image of an element of $G$, so
\[
    \SY(G)=\lieO(\QQ_{w^*})_{V^\perp}.
\]
We have therefore identified both the image and the kernel of the restricted map
$\SY|_G$ from $G$ to $\Sp(2n,\F_2)$,
namely its image is $\lieO(\QQ_{w^*})_{V^\perp}$ and its kernel is $G\cap\Pgroup_n$.
Items (a) and (b) now follow from Lemma~\ref{lem:finite_clifford_subgroups_over_binary_images}\ref{lem:finite_clifford_subgroups_over_binary_images:a},\ref{lem:finite_clifford_subgroups_over_binary_images:b}.
Items (c) and (d) are the equality and phase-adjoined equality criteria from Lemma~\ref{lem:finite_clifford_subgroups_over_binary_images}\ref{lem:finite_clifford_subgroups_over_binary_images:c},\ref{lem:finite_clifford_subgroups_over_binary_images:d}, applied with ambient subgroup $G$.
\end{proof}

In the special case of a single Pauli bilinear form and trivial commutant, the fixed subgroup $\cl_n^{\Theta_B}$ has two familiar matrix realizations, depending on whether $B$ is symmetric or skew-symmetric:
\begin{equation*}
    \cl_n^{\Theta_B}
    \conjugated
    \begin{dcases}
        \cl_n\cap\lieO(2^n) & \text{if } B^T=B,\\
        \cl_n\cap\USp(2^n) & \text{if } B^T=-B.
    \end{dcases}
\end{equation*}
The symmetric case corresponds to the orthogonal or real Clifford group, which has received much attention in the literature \cite{Hashagen2018realrandomized,Nebe_Rains_Sloane_2001}.
If $B=\iso{w}$, then Lemma~\ref{lem:fixed_clifford_bilinear_commutant_binary_image_kernel} gives
\begin{equation*}
    \SY(\cl_n^{\Theta_B})=\lieO(\QQ_w)\cong\lieO_+(2n).
\end{equation*}
The skew-symmetric case corresponds to the symplectic Clifford group, a comparatively less studied counterpart, and for $B=\iso{w}$ one analogously has
\begin{equation*}
    \SY(\cl_n^{\Theta_B})=\lieO(\QQ_w)\cong\lieO_-(2n).
\end{equation*}
Notice that symplectic here refers to the alternating bilinear form over the state vectors in $\C^{2^n}$, not the alternating bilinear forms over the binary vectors in $\Fn$ (i.e. the symplectic structure of the Pauli strings).

The same exact-sequence viewpoint also gives the Clifford-level uniqueness statement.
Let $H$ be a subgroup of the Clifford group $\cl_n$ satisfying
\[
    H\subseteq \cl_n^{\Theta_{B^*},\Ad_{\commalg}},
    \qquad
    \SY(H)=\lieO(\QQ_{w^*})_{V^\perp}.
\]
If $H$ has the same Pauli kernel as $\cl_n^{\Theta_{B^*},\Ad_{\commalg}}$, then Lemma~\ref{lem:fixed_clifford_bilinear_commutant_binary_image_kernel}(c) gives $H=\cl_n^{\Theta_{B^*},\Ad_{\commalg}}$.
If instead the Pauli kernels agree only after adjoining scalar phases, then Lemma~\ref{lem:fixed_clifford_bilinear_commutant_binary_image_kernel}(d) gives equality after adjoining $\ZZ(\Pgroup_n)$.
Hence, we write explicitly the distinction between the equality of two subgroups and the equality after ignoring scalar phases.

The Pauli kernel in Lemma~\ref{lem:fixed_clifford_bilinear_commutant_binary_image_kernel} is the subgroup
\[
    \cl_n^{\Theta_{B^*},\Ad_{\commalg}}\cap\Pgroup_n.
\]
After adjoining scalar phases, this kernel has the same Pauli-string support as the commutant-symmetric Pauli subgroup:
\begin{align*}
    &\cl_n^{\Theta_{B^*},\Ad_{\commalg}}\cap\Pgroup_n \simeq \cl_n^{\Ad_{\commalg}}\cap\Pgroup_n.
\end{align*}
The role of $\Theta_{B^*}$ is to select scalar representatives inside this phase-adjoined Pauli subgroup; for example, $\Theta_{B^*}$ does not fix $\im I$, as discussed in Remark~\ref{rem:binary_constraints_literal_clifford_fixed_points}.

We can now summarize the previous discussion and specialize the statements to transvection groups which respect bilinear or quadratic forms, as in Proposition~\ref{prop:full_space_symplectic_symmetric_Clifford_groups}:
\begin{prop}\label{prop:full_space_isometry_symmetric_Clifford_groups}
Let $V\subseteq\Fn$ be a subspace, $B=\iso{w^*}\in\PP_n$ and let $\commalg=\Span[\C]{\inviso(V^\perp)}$, $\QQ^*=\QQ_{w^*}|_V$.
\begin{enumerate}
\item The restriction of $\lieO(\QQ_{w^*})_{V^\perp}$ to $V$ gives a split short exact sequence
\begin{equation*}
    1\to
    \OD{\rad(V)}{\QQ_{w^*}}
    \to
    \lieO(\QQ_{w^*})_{V^\perp}
    \to
    \lieO^\#(\QQ^*)
    \to 1 .
\end{equation*}
\item The kernel $\OD{\rad(V)}{\QQ_{w^*}}$ is isomorphic to the following additive groups:
\begin{nestedcaseenum}
    \item if $\type(\QQ^*)\in\{+,-\}$, the $r\times r$ off-diagonal symmetric matrices over $\F_2$; also, let $z\in V$ be anisotropic, for any basis $\{h_i\}_{i=1}^r$ of $\rad(V)$, it is generated by the transvections with centers
    \begin{equation*}
        \{z+h_i\}_{i=1}^r\cup\{z+h_i+h_j\}_{1\leq i<j\leq r}.
    \end{equation*}
    \item if $\type(\QQ^*)=0$, the $r\times r$ symmetric matrices over $\F_2$ subject to the constraint $B_{ii}=B_{i1}$; also, for $\{h_i\}_{i=1}^n$ a basis of $\rad(V)$ with $h_1\in\radone(V,\QQ^*)$ and $h_i\in\radzero(V,\QQ^*)$, it is generated by the transvections with centers
    \begin{equation*}
        \{h_1\}\cup\{h_1+h_i\}_{i=2}^r\cup\{h_1+h_i+h_j\}_{1<i<j\leq r}.
    \end{equation*}
\end{nestedcaseenum}
\item Let $H\subseteq\lieO(\QQ_{w^*})_{V^\perp}$. If $\OD{\rad(V)}{\QQ_{w^*}}$ is contained in $H$ and the restriction map
$\activemap\mapsto\activemap|_V$ maps $H$ onto $\lieO^\#(\QQ^*)$, then $H=\lieO(\QQ_{w^*})_{V^\perp}$.
\item The map $\SY$ from Eq.~\eqref{eq:proj:conj} induces a short exact sequence
\begin{equation*}
    1\to
    \cl_n^{\Theta_B,\Ad_\commalg}\cap\Pgroup_n
    \to
    \cl_n^{\Theta_B,\Ad_\commalg}
    \xrightarrow{\SY}
    \lieO(\QQ_{w^*})_{V^\perp}
    \to 1 .
\end{equation*}
\item If $H\subseteq\cl_n^{\Theta_B,\Ad_{\commalg}}$ has the binary image
$\SY(H)=\lieO(\QQ_{w^*})_{V^\perp}$ and the Pauli kernel
\begin{equation*}
    H\cap\Pgroup_n
    =
    \cl_n^{\Theta_B,\Ad_{\commalg}}\cap\Pgroup_n,
\end{equation*}
then $H=\cl_n^{\Theta_B,\Ad_{\commalg}}$.
If instead the two Pauli kernels agree after adjoining $\ZZ(\Pgroup_n)$, then $H$ and $\cl_n^{\Theta_B,\Ad_{\commalg}}$ agree after adjoining scalar phases.
\item Let $\vgens\subseteq\Fn$ be a binary generating set and $\tvgroup{\pgens}$ its transvection group, such that  $V=\Span{\vgens}$ and $w^*\in\quadratic(\vgens)$. 
If $\tvgroup{\pgens}$ contains $\OD{\rad(V)}{\QQ_{w^*}}$ and maps onto $\lieO^{\#}(\QQ^*)$ under the restriction map to $V$, then $\tvgroup{\pgens}=\lieO(\QQ_{w^*})_{V^\perp}$.
\item Let $\pgens=\isolong{\vgens}\subseteq\PP_n$ be a Pauli generating set and $\cltvgroup{\pgens}$ its Clifford transvection group, such that $V=\Span{\vgens}$ and $B\in\bilinear(\pgens)$. 
If $\cltvgroup{\pgens}$ has binary image $\lieO(\QQ_{w^*})_{V^\perp}$, $\cltvgroup{\pgens}$ and $\cl_n^{\Theta_B,\Ad_\commalg}$ agree after adjoining scalar phases.
\end{enumerate}
\end{prop}
\begin{proof}
(a) is the short exact sequence from Lemma~\ref{lem:orthogonal_stabilizer_restriction_candidate}.
(b) follows from Lemma~\ref{lem:orthogonal_diagonal_kernel_candidate} and Lemma~\ref{lem:orthogonal_diagonal_kernel_transvection_generators_candidate}.
(c) follows from Proposition~\ref{prop:orthogonal_stabilizer_binary_maximality_candidate}.
Lemma~\ref{lem:fixed_clifford_bilinear_commutant_binary_image_kernel} gives (d) and (e).
For (f), note that $\tvgroup{\vgens}$ fixes $V^\perp$, is a symplectic group and conserves $\QQ_{w^*}$, hence $\tvgroup{\vgens}\subseteq \lieO(\QQ_{w^*})_{V^\perp}$. Hence it follows from (e).
For (g), we can use the same argument as for Prop~\ref{prop:full_space_symplectic_symmetric_Clifford_groups}(g) to show that $\cltvgroup{\pgens}\cap\Pgroup_n$ coincides with $\cl_n^{\Theta_B,\Ad_\commalg}\cap\Pgroup_n$ after adjoining phases.
Finally, the group $H = \cltvgroup{\pgens}$ satisfies (e), which means that $\cltvgroup{\pgens}$ and $\cl_n^{\Theta_B\Ad_\commalg}$ coincide after adjoining scalar phases.
\end{proof} 

The preceding discussion gives a precise Clifford-level formulation of finite Clifford subgroups preserving invariant bilinear forms and commutant symmetries, including their binary images and Pauli kernels.
While the corresponding orthogonal and symplectic Clifford groups are natural finite counterparts of the usual compact symmetry groups, their structure under simultaneous invariant bilinear forms and Pauli linear symmetries has received less systematic attention.
We discuss related Clifford transvection groups further in Section~\ref{sec:classification_transvection_groups} and Section~\ref{sec:clifford_3_designs_groups}.

\subsection{Pauli Lie Algebras with Commutants and Isometries}\label{sec:pauli_lie_algebras_isometries}

The preceding subsections described finite Clifford subgroups through their binary stabilizers and Pauli kernels.
We now pass to the corresponding (compact) Pauli Lie algebras.
As a first step, we treat only the linear symmetry imposed by a Pauli commutant, before adding invariant bilinear forms below.
For Pauli strings $\{C_i\}_{i=1}^s\subseteq\PP_n$, we write
$\lieu(2^n)^{\{\Ad_{C_i}\}_{i=1}^s}$ for the joint fixed-point subspace.
Similarly, $\lieu(2^n)^{\Ad_{\commalg}}$ yields the joint fixed-point subspace under $\Ad_C$ for all $C\in\commalg\cap\PP_n$:
\begin{alignat}{3}
    \lieu(2^n)^{\{\Ad_{C_i}\}_{i=1}^s}
    &:=
    \{&&M\in\lieu(2^n)\;\text{with } C_iMC_i^{-1}=M \nonumber\\
    &&&\text{for all }i\in\{1,\ldots,s\}\},
    \label{eq:def:fixed_points_pauli_generators_written_out}\\
    \lieu(2^n)^{\Ad_{\commalg}}
    &:=
    \{&&M\in\lieu(2^n)\;\text{with } CMC^{-1}=M \nonumber\\
    &&&\text{for all }C\in\commalg\cap\PP_n\}.
    \label{eq:def:fixed_points_pauli_commutant_written_out}
\end{alignat}
The following lemma provides the equivalent fixed-point, commutant, and vector-based descriptions of the resulting Lie algebra.

\begin{lem}[Commutant-fixed Pauli Lie algebra]\label{lem:commutant_fixed_pauli_lie_algebra_equivalent_descriptions}
Let $V\subseteq\Fn$ be a binary subspace, let $\commalg=\Span[\C]{\inviso(V^\perp)}$ be the associated Pauli commutant algebra, and let $\matalg=\commutant(\commalg)$ be the Pauli matrix algebra dual to $\commalg$.
If $\{C_i\}_{i=1}^s$ is any set of Pauli strings generating $\commalg$ as a matrix algebra, then the following real Lie subalgebras of $\lieu(2^n)$ coincide:
\begin{align}
    &\lieu(2^n)^{\Ad_{\commalg}}
    =
    \lieu(2^n)^{\{\Ad_{C_i}\}_{i=1}^s} \nonumber\\
    &=
    \commutant(\commalg)\cap\lieu(2^n)
    =
    \matalg\cap\lieu(2^n) \nonumber\\
    &=
    \Span[\R]{\{\im P\in\im\PP_n \text{ with }
    \comm{P}{C}=0\text{ for all }C\in\commalg\}} \nonumber\\
    &=
    \Span[\R]{\im\isolong{V}}.
\label{eq:commutant_fixed_pauli_lie_algebra_equivalent_descriptions}
\end{align}
\end{lem}
\begin{proof}
The first equality follows because it is enough to impose the adjoint fixed-point conditions for Pauli generators of the matrix algebra $\commalg$.
For a Pauli string $C$, the condition $\Ad_C(M)=M$ is equivalent to $\comm{M}{C}=0$.
Thus the joint fixed points inside $\lieu(2^n)$ are precisely the skew-hermitian matrices in $\commutant(\commalg)$, which gives the second equality.
Since $\matalg=\commutant(\commalg)$, this also gives the third equality.
Taking the skew-hermitian Pauli basis elements $\im P$ of $\lieu(2^n)$ gives the fourth equality.
Finally, because $\commalg=\Span[\C]{\inviso(V^\perp)}$, the Pauli string $\iso{v}$ commutes with every Pauli string
$\iso{u}$ with $u\in V^\perp$ iff $v\in(V^\perp)^\perp=V$.
This gives the last equality.
\end{proof}

This motivates the following notation for the common Lie algebra in Lemma~\ref{lem:commutant_fixed_pauli_lie_algebra_equivalent_descriptions}.
\begin{defn}[Pauli Lie algebra of isometries for a commutant]\label{def:commutant_pauli_lie_algebra_isometries}
Let $V\subseteq\Fn$ be a binary subspace and $\commalg=\Span[\C]{\inviso(V^\perp)}$ is the associated Pauli commutant algebra.
We define the Pauli Lie algebra of isometries for $\commalg$ by
\begin{align}
    \lieiso(\commalg)&:=\commutant(\commalg)\cap\lieu(2^n).
\label{eq:lieiso}
\intertext{The equivalent fixed-point and vector-based descriptions are given in Lemma~\ref{lem:commutant_fixed_pauli_lie_algebra_equivalent_descriptions}.
Moreover,}
\lieiso^0(\commalg) &:=[\lieiso(\commalg),\lieiso(\commalg)]
\label{eq:lieiso:zero}
\end{align}
denotes the derived Lie algebra of $\lieiso(\commalg)$.
\end{defn}
We also write $\lieiso(\algclosure{\In}) = \lieu(2^n)$ to identify the full unitary Lie algebra, and $\lieiso^0(\algclosure{\In}) = \su(2^n)$ to identify the special unitary Lie algebra.

After fixing this notation, we can also record the representation-theoretic form of $\lieiso(\commalg)$.
These descriptions are not equalities inside the fixed ambient matrix algebra as the first is a
Lie-algebra isomorphism, while the second is a concrete block form up to conjugation.

\begin{lem}[Block form of the commutant-fixed Pauli Lie algebra]\label{lem:commutant_fixed_pauli_lie_algebra_block_form}
Let $V\subseteq\Fn$ be a binary subspace and $\commalg=\Span[\C]{\inviso(V^\perp)}$ is the associated Pauli commutant algebra, with $2\ell=\rank(\commalg)$, $r=\nullity(\commalg)$ and $m=n-\ell-r$.
Then, up to Lie algebra isomorphism,
\begin{align}
    \lieiso(\commalg) &\cong \lieu(2^m)^{\oplus 2^r}.
\intertext{Up to conjugation in the ambient matrix algebra,}
\label{eq:reductive_commutant_isometry_Lie_algebra_replacement}
        \lieiso(\commalg) &\conjugated \bigoplus_{\lambda=1}^{2^r} \lieu(2^m) \otimes \id_{2^\ell}.
\end{align}
Here each copy of $\lieu(2^m)$ acts in the standard representation on its block, while $\id_{2^\ell}$ is the multiplicity factor on which the corresponding copy acts trivially.
\end{lem}
\begin{proof}
By Lemma~\ref{lem:commutant_fixed_pauli_lie_algebra_equivalent_descriptions}, we have
$\lieiso(\commalg)=\matalg\cap\lieu(2^n)$ for the Pauli matrix algebra $\matalg=\commutant(\commalg)$ dual to $\commalg$.
The Pauli matrix algebra classification in Lemma~\ref{lem:Classification_Pauli_Matrix_Algebras}, applied to $\commalg$ and its dual algebra $\matalg$, gives a decomposition of $\matalg$ with $2^r$ isotypic blocks and multiplicity factor $2^\ell$ on each block.
Intersecting the corresponding simple matrix algebra on each block with $\lieu(2^n)$ gives one copy of $\lieu(2^m)$ per block.
This proves the isomorphism.
The same matrix-algebra decomposition is implemented by an invertible change of basis, and a Lie algebra and its generated matrix algebra have the same invariant subspace decomposition.
Therefore the same change of basis gives the concrete block form in Eq.~\eqref{eq:reductive_commutant_isometry_Lie_algebra_replacement}.
\end{proof}

The commutant symmetry also has a finite Clifford-group version:
\begin{equation}
    \cl_n^{\Ad_{\commalg}}=\cl_n\cap \lieU(2^n)^{\Ad_{\commalg}}.
\end{equation}
In Section~\ref{sec:classification_transvection_groups}, this group is studied as a Clifford transvection group and compared with the Pauli Lie group generated by $\lieiso(\commalg)$.

\begin{lem}[Derived commutant-fixed Pauli Lie algebra]\label{lem:commutant_fixed_pauli_lie_algebra_derived_part}
In the setting of Definition~\ref{def:commutant_pauli_lie_algebra_isometries}, write $m=n-\ell-r$.
The derived Lie algebra satisfies
\begin{align}
   \lieiso^0(\commalg)
   &=
   \Span[\R]{\{\im\iso{v}\text{ for } v\in V\setminus\rad(V)\}}.
   \label{eq:commutant_fixed_pauli_lie_algebra_derived_span}
\intertext{Moreover, up to Lie algebra isomorphism,}
   \lieiso^0(\commalg)
   &\cong \su(2^m)^{\oplus 2^r} \cong \lieiso(\commalg)/\ZZ(\lieiso(\commalg)),
\intertext{and, up to conjugation in the ambient matrix algebra,}
   \lieiso^0(\commalg)
   &\conjugated
   \bigoplus_{\lambda=1}^{2^r}\su(2^m)\otimes\id_{2^\ell}.
   \label{eq:semisimple_commutant_isometry_Lie_algebra_replacement}
\end{align}
\end{lem}

\begin{figure}
    \centering
    \includegraphics{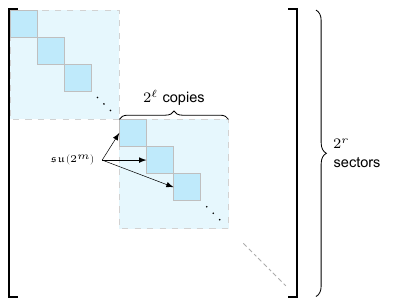}
    \caption{Schematic block decomposition of $\lieiso^0(\commalg)$ emphasizing the representation-theoretic structure in Eq.~\eqref{eq:semisimple_commutant_isometry_Lie_algebra_replacement}.
    In each of the $2^r$ isotypic sectors, one independent copy of $\su(2^m)$ acts simultaneously on each of the $2^m$ multiplicity copies of dimension $2^m$, reflecting the multiplicity tensor factor $\id_{2^\ell}$.
    Here $2\ell=\rank(\commalg)$, $r=\nullity(\commalg)$ and $m=n-\ell-r$.}
    \label{fig:block-su-representation}
\end{figure}

The block form in Eq.~\eqref{eq:semisimple_commutant_isometry_Lie_algebra_replacement} is illustrated schematically in Fig.~\ref{fig:block-su-representation}.
Notice that $\lieiso^0(\commalg)$ is not the joint fixed-point subalgebra of $\su(2^n)$ under $\Ad_{\commalg}$, which would still contain the non-identity central basis elements $\im\iso{v}$ with $v\in\rad(V)$.
\begin{proof}
By Lemma~\ref{lem:commutant_fixed_pauli_lie_algebra_equivalent_descriptions},
$\lieiso(\commalg)=\Span[\R]{\im\isolong{V}}$.
For $u,v\in V$, the bracket of $\im\iso{u}$ and $\im\iso{v}$ is nonzero exactly when $\symp{u}{v}=1$, and is then proportional to $\im\iso{u+v}$.
Thus the central basis elements in $\lieiso(\commalg)$ are precisely those $\im\iso{v}$ with $v\in\rad(V)$.
Taking the derived Lie algebra therefore removes exactly these central basis elements and gives Eq.~\eqref{eq:commutant_fixed_pauli_lie_algebra_derived_span}.
The same direct-sum separation of the central basis elements from the remaining basis elements gives the quotient isomorphism
$\lieiso^0(\commalg)\cong\lieiso(\commalg)/\ZZ(\lieiso(\commalg))$.

The block form of Lemma~\ref{lem:commutant_fixed_pauli_lie_algebra_block_form} identifies $\lieiso(\commalg)$ with a direct sum of $\lieu(2^m)$ blocks, each acting trivially on the tensor factor $\id_{2^\ell}$.
Taking derived Lie algebras replaces each factor $\lieu(2^m)$ by $\su(2^m)$, while this tensor factor $\id_{2^\ell}$ remains unchanged.
This yields both the isomorphism and the concrete block form.
\end{proof}

We now add a Pauli bilinear form as an additional invariant in the commutant-fixed setting.
Let $V\subseteq\Fn$, let $\commalg=\Span[\C]{\inviso(V^\perp)}$ be the associated Pauli commutant algebra, and let $B^*=\iso{w^*}$ be the Pauli bilinear form corresponding to the reference vector $w^*$.
Recall from Remark~\ref{rem:isometries_as_fixed_points_orientation} and Eq.~\eqref{eq:def:bilinear_isometry_involutions} that $\theta_B$ denotes the Lie-algebra involution associated with a non-degenerate bilinear form $B$ is given by
\begin{equation*}
    \theta_B(M)=-B^{-1}M^TB.
\end{equation*}
In the Pauli setting, the condition $\theta_{B^*}(M)=M$ imposes the bilinear-form symmetry, while the conditions $\Ad_C(M)=M$ for $C\in\commalg\cap\PP_n$ impose the commutant symmetry.
Using the fixed-point notation for Pauli commutants from Eq.~\eqref{eq:def:fixed_points_pauli_commutant_written_out} together with the compact bilinear-form convention from Remark~\ref{rem:pauli_compact_isometry_notation}, this simultaneous condition is written as
\begin{align}\label{eq:def:commutant_bilinear_fixed_points_written_out}
    &\lieu(2^n)^{\theta_{B^*},\Ad_{\commalg}}
    :=
    \{M\in\lieu(2^n)\;\text{with }\theta_{B^*}(M)=M \nonumber\\
    &\text{and }\Ad_C(M)=M 
    \text{ for all }C\in\commalg\cap\PP_n\}.
\end{align}
For Pauli bilinear forms and Pauli commutant operators, the maps $\theta_B$, $\theta_{B'}$ and $\Ad_C$ commute with one another, so the simultaneous fixed-point condition can be imposed in any order.
We now name this joint fixed-point Lie algebra together with its derived Lie algebra.

\begin{defn}[Pauli Lie algebra of isometries for a commutant and bilinear form]\label{def:commutant_bilinear_pauli_lie_algebra_isometries}
Let $V\subseteq\Fn$, let $\commalg=\Span[\C]{\inviso(V^\perp)}$, and let $B^*=\iso{w^*}$.
We define the Pauli Lie algebra of isometries for the pair $(\commalg,B^*)$ by
\begin{align}
    \lieiso(\commalg,B^*)
    &:=
    \lieu(2^n)^{\theta_{B^*},\Ad_{\commalg}},
    \label{eq:def:commutant_bilinear_pauli_lie_algebra_isometries}
\intertext{
where the fixed-point is detailed in Eq.~\eqref{eq:def:commutant_bilinear_fixed_points_written_out}.
We write}
    \lieiso^0(\commalg,B^*)
    &:=
    [\lieiso(\commalg,B^*),
    \lieiso(\commalg,B^*)]
    \label{eq:def:commutant_bilinear_pauli_lie_algebra_isometries_derived}
\end{align}
for its derived Lie algebra.
\end{defn}

The next lemma gives the corresponding description in terms of vectors in $\Fn$.
It is the bilinear-form analogue of Lemma~\ref{lem:commutant_fixed_pauli_lie_algebra_equivalent_descriptions}.

\begin{lem}[Equivalent descriptions with one invariant bilinear form]\label{lem:commutant_bilinear_pauli_lie_algebra_equivalent_descriptions}
In the setting of Definition~\ref{def:commutant_bilinear_pauli_lie_algebra_isometries}, set
\[
    B^*\commalg:=\{B^*C\text{ for } C\in\commalg\cap\PP_n\}.
\]
Then
\begin{align}
    \lieiso(\commalg,B^*)
    &=
    \lieiso(\commalg)\cap\lieu(2^n)^{\theta_{B^*}}
    =
    \lieu(2^n)^{\{\theta_B\}_{B\in B^*\commalg}}
    \nonumber\\
    &=
    \Span[\R]{\{\im\iso{v}\text{ for } v\in \QQ_{w^*}^{-1}(1)\cap V\}}.
    \label{eq:commutant_bilinear_pauli_lie_algebra_equivalent_descriptions}
\end{align}
Moreover, the vectors $v$ for which $\im\iso{v}$ spans $\lieiso(\commalg,B^*)$ are precisely the transvection centers 
\begin{equation}
    v \in \tvcenter{\lieO(\QQ_{w^*})_{V^\perp}}
    =
    \QQ_{w^*}^{-1}(1)\cap V
    \label{eq:commutant_bilinear_transvection_centers}
\end{equation}
for the pointwise orthogonal stabilizer from Eq.~\eqref{eq:def:orthogonal_group_stabilizing_subspace}.
\end{lem}
\begin{proof}
The first equality is just Definition~\ref{def:commutant_bilinear_pauli_lie_algebra_isometries} together with Definition~\ref{def:commutant_pauli_lie_algebra_isometries}.
For a vector $v\in\Fn$, Lemma~\ref{lem:commutant_fixed_pauli_lie_algebra_equivalent_descriptions} says that $\Ad_C(\im\iso{v})=\im\iso{v}$ for all $C\in\commalg\cap\PP_n$ is equivalent to $v\in V$.
Similarly, the condition $\theta_{B^*}(\im\iso{v})=\im\iso{v}$ is, in the notation of Definition~\ref{def:quadratic-form-associated-vector}, equivalent to $\QQ_{w^*}(v)=1$.
Thus the simultaneous fixed points have the basis description in the last line of Eq.~\eqref{eq:commutant_bilinear_pauli_lie_algebra_equivalent_descriptions}.

It remains to compare this with the fixed points for the translated family $B^*\commalg$.
If $C=\iso{u}$ with $u\in V^\perp$, then $B^*C$ corresponds, up to phase, to the vector $w^*+u$.
By Eq.~\eqref{eq:quadratic-form-associated-vector}, the condition $\QQ_{w^*+u}(v)=1$ is equivalent to
\[
    \QQ_{w^*}(v)+\symp{u}{v}=1.
\]
Requiring this for all $u\in V^\perp$ is therefore equivalent to $v\in(V^\perp)^\perp=V$ and $\QQ_{w^*}(v)=1$.
This proves the second equality in Eq.~\eqref{eq:commutant_bilinear_pauli_lie_algebra_equivalent_descriptions}.

Finally, considering the same vector $v$, membership in $\tvcenter{\lieO(\QQ_{w^*})_{V^\perp}}$ means precisely that the transvection $\tau_v$ lies in $\lieO(\QQ_{w^*})_{V^\perp}$.
This happens exactly when $\tau_v$ fixes $V^\perp$ pointwise and preserves $\QQ_{w^*}$.
The first condition is equivalent to $v\in V$, while the second is equivalent to $\QQ_{w^*}(v)=1$.
This proves Eq.~\eqref{eq:commutant_bilinear_transvection_centers}.
\end{proof}

\begin{figure}
    \centering
    \includegraphics{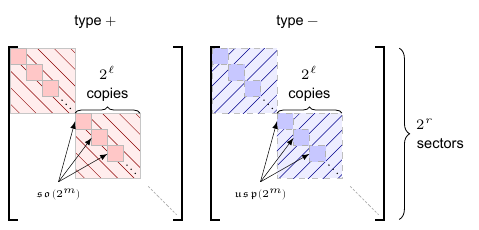}
    \caption{Block decomposition of $\lieiso^0(\commalg,B)$ for invariant bilinear forms of type $+$ (left) and type $-$ (right), following Fig.~\ref{fig:block-su-representation}.
    In each of the $2^r$ isotypic sectors, one independent copy of the simple Lie algebra $\so(2^m)$ for type $+$, or $\usp(2^m)$ for type $-$, acts simultaneously on each of the $2^\ell$ multiplicity copies of dimension $2^m$, reflecting the multiplicity tensor factor $\id_{2^\ell}$.
    Here $2\ell=\rank(\commalg)$, $r=\nullity(\commalg)$ and $m=n-\ell-r$.}
    \label{fig:block-so-representation}
\end{figure}

\begin{figure}
    \centering
    \includegraphics{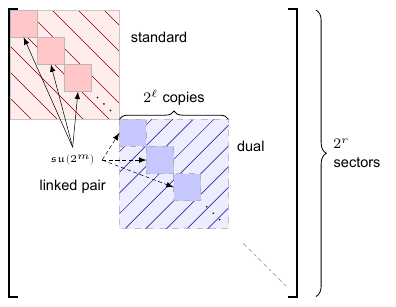}
    \caption{Block decomposition of the Pauli Lie algebra $\lieiso^0(\commalg,B) \cong \su(2^m)^{\oplus 2^{r-1}}$ for invariant bilinear forms of type $0$ in the nonabelian case,
    following Fig.~\ref{fig:block-su-representation}.
    The highlighted blocks show one linked pair of isotypic sectors.
    One independent copy of $\su(2^m)$ determines the action on both sectors in this pair, acting by the standard representation on one sector and by the dual representation on the other.
    The $2^r$ isotypic sectors are partitioned into $2^{r-1}$ linked pairs of this form.
    Each sector has the same $2^\ell$ multiplicity structure.
    Here $2\ell=\rank(\commalg)$, $r=\nullity(\commalg)$ and $m=n-\ell-r$.}
    \label{fig:block-type0-representation}
\end{figure}

We also describe the derived Lie algebra,
since this is the object classified below and illustrated in Figs.~\ref{fig:block-so-representation} and~\ref{fig:block-type0-representation}.
For the representative classes considered below, Lemma~\ref{lem:exceptional_plus_affine_space_unique} verifies the spanning hypothesis in Eq.~\eqref{eq:commutant_bilinear_anisotropic_vectors_span} below except in the exceptional type-$+$ case with $\ell=1$ from Lemma~\ref{lem:exceptional_plus_affine_space_collapse}.

\begin{lem}[Derived algebra with one invariant bilinear form]\label{lem:commutant_bilinear_pauli_lie_algebra_derived_part}
In the setting of Definition~\ref{def:commutant_bilinear_pauli_lie_algebra_isometries}, with $\QQ_{w^*}$ as in Definition~\ref{def:quadratic-form-associated-vector}, set
\begin{equation}
    A:=\QQ_{w^*}^{-1}(1)\cap V\;\text{ and }\; S:=\Span[\F_2]{A}.
    \label{eq:commutant_bilinear_anisotropic_span_def}
\end{equation}
Then the following hold:
\begin{enumerate}
\item In general,
\begin{align}
    \lieiso^0(\commalg,B^*)
    &=
    \spanempty_{\R}\{
    \begin{aligned}[t]
    &\im\iso{v}\text{ with } v\in A \text{ and }v\notin S^\perp\}
    \nonumber
    \end{aligned}\\
    &\cong
    \lieiso(\commalg,B^*)  /\ZZ(\lieiso(\commalg,B^*)).
    \label{eq:commutant_bilinear_pauli_lie_algebra_derived_general}
\end{align}
\item If, additionally,
\begin{equation}
    \Span[\F_2]{\QQ_{w^*}^{-1}(1)\cap V}=V.
    \label{eq:commutant_bilinear_anisotropic_vectors_span}
\end{equation}
then
\begin{align}
    \lieiso^0(\commalg,B^*)
    &=
    \spanempty_{\R}\{
    \begin{aligned}[t]
    &\im\iso{v}\text{ with } v\in\QQ_{w^*}^{-1}(1)\cap V \nonumber\\
    &\text{and }v\notin\rad(V)\}
    \nonumber
    \end{aligned}\\
    &\cong
    \lieiso(\commalg,B^*)  /\ZZ(\lieiso(\commalg,B^*)).
    \label{eq:commutant_bilinear_pauli_lie_algebra_derived_part}
\end{align}
\end{enumerate}
\end{lem}

\begin{proof}
By Lemma~\ref{lem:commutant_bilinear_pauli_lie_algebra_equivalent_descriptions}, $\lieiso(\commalg,B^*)$ is spanned by the elements $\im\iso{v}$ with $v\in A$.
For $u,v\in A$, the bracket of $\im\iso{u}$ and $\im\iso{v}$ is nonzero exactly when $\symp{u}{v}=1$, and is then proportional to $\im\iso{u+v}$.
Therefore the central basis elements in $\lieiso(\commalg,B^*)$ are exactly those $\im\iso{v}$ with $v\in A\cap S^\perp$.

It remains to see that every noncentral basis element $\im\iso{a}$ with $a\in A$ lies in the derived algebra.
Let $a\in A\setminus S^\perp$.
Since $S=\Span[\F_2]{A}$, there is some $b\in A$ with $\symp{a}{b}=1$.
Then $a+b\in A$, because
\[
    \QQ_{w^*}(a+b)=\QQ_{w^*}(a)+\QQ_{w^*}(b)+\symp{a}{b}=1+1+1=1,
\]
and the bracket of $\im\iso{b}$ with $\im\iso{a+b}$ is proportional to $\im\iso{a}$.
Hence the derived algebra is obtained from the basis in Eq.~\eqref{eq:commutant_bilinear_pauli_lie_algebra_equivalent_descriptions} by removing the central basis elements.
This gives the first equality in Eq.~\eqref{eq:commutant_bilinear_pauli_lie_algebra_derived_general}.
The quotient description follows from the same direct-sum separation of the center from the derived compact Pauli Lie algebra.

If Eq.~\eqref{eq:commutant_bilinear_anisotropic_vectors_span} holds, $S=V$.
Since $A\subseteq V$, the condition $v\in A\cap S^\perp$ is equivalent to $v\in A\cap V^\perp=A\cap\rad(V)$.
Substituting this into Eq.~\eqref{eq:commutant_bilinear_pauli_lie_algebra_derived_general} gives Eq.~\eqref{eq:commutant_bilinear_pauli_lie_algebra_derived_part}.
\end{proof}

When the commutant is trivial, $V=\Fn$ is non-degenerate and only the non-degenerate cases of 
type $+$ and $-$ occur.
They recover the standard compact orthogonal and unitary symplectic Lie algebras.
Here $\lieiso(I)$ and $\lieiso(Y_n)$ use the Pauli bilinear-form convention for compact isometry Lie algebras from Remark~\ref{rem:pauli_compact_isometry_notation} and Eq.~\eqref{eq:def:compact_bilinear_isometry_lie_algebra_fixed_points}, so that $\lieiso(B)=\lieu(2^n)^{\theta_B}$ for a Pauli bilinear form $B$:
\begin{align}
        \lieiso(I) &= \lieu(2^n)^{\theta_I} = \so(2^n),\\
        \lieiso(Y_n) &= \lieu(2^n)^{\theta_{Y_n}} = \usp(2^n).
\end{align}
With a nontrivial commutant, the restricted quadratic form $\QQ_{w^*}|_V$ may also have type $0$.
The resulting representation-theoretic block structures are summarized in Fig.~\ref{fig:block-so-representation} for the cases of types $\pm$ and in Fig.~\ref{fig:block-type0-representation} for the type-$0$ case.

We first separate the ingredients of the classification from the final lookup statement.
The first lemma provides canonical Pauli bases and the following proposition translates them into concrete block forms.

\begin{lem}[Canonical Pauli bases for commutant and bilinear-form isometries]\label{lem:canonical_pauli_bases_commutant_bilinear_isometries}
Let $W\subseteq\Fn$ label the Pauli commutant algebra $\commalg=\Span[\C]{\inviso(W)}$, set $V=W^\perp$, and let $B=\iso{w}$ be a Pauli bilinear form.
Write $\rank(W)=2\ell$, $\nullity(W)=r$, and $m=n-\ell-r$.
After a Pauli change of basis, preserving the invariants of $w+W$, the Pauli basis $\bas{\lieg}$ of $\lieg$, gives one of the following forms:
\begin{enumerate}
\item If $\type(w+W)=+$ and $m=0$, then
\[
    \bas{\lieiso(\commalg,B)}
    =
    \emptyset.
\]
\item If $\type(w+W)=+$ and $m\geq1$, then
\[
    \bas{\lieiso(\commalg,B)}
    =
    \PP_m^{\theta_I}\otimes\{I,Z\}^{\otimes r}\otimes I^{\otimes \ell}.
\]
\item If $\type(w+W)=-$ and $m\geq1$, then
\[
    \bas{\lieiso(\commalg,B)}
    =
    \PP_m^{\theta_{Y_m}}\otimes\{I,Z\}^{\otimes r}\otimes I^{\otimes \ell}.
\]
\item If $\type(w+W)=0$, then $r\geq1$ and
\begin{align}
    \bas{\lieiso(\commalg,B)}
    ={}
    &[\PP_m^{\theta_I}\otimes I\otimes\{I,Z\}^{\otimes(r-1)}\otimes I^{\otimes \ell}]
    \label{eq:pauli_lie_isometry_type_zero_basis}\\
    &{}\cup
    [\PP_m^{-\theta_I}\otimes Z\otimes\{I,Z\}^{\otimes(r-1)}\otimes I^{\otimes \ell}].
    \nonumber
\end{align}
\end{enumerate}
Here $\PP_m^{\theta_I}$ and $\PP_m^{-\theta_I}$ denote the Pauli strings on the non-degenerate factor, fixed and negated by $\theta_I$, respectively.
\end{lem}

\begin{proof}
By Lemma~\ref{lem:commutant_bilinear_pauli_lie_algebra_equivalent_descriptions}, the Pauli basis is obtained from the vectors in $\QQ_w^{-1}(1)\cap V$ via the Pauli-binary correspondence.
Choosing the canonical representative of the affine quadratic space $w+W$ gives the displayed cases: for type $+$ with $m=0$, the restricted form vanishes on $V=\rad(V)$, so there are no anisotropic vectors in $V$.
For type $+$ with $m\geq1$ and for type $-$, the cases are represented by the non-degenerate forms associated with $I$ and $Y_m$ on the non-degenerate factor, while the type-$0$ case is represented by adding one radical coordinate, which gives the two components in Eq.~\eqref{eq:pauli_lie_isometry_type_zero_basis}.
In this type-$0$ representative, the extra radical coordinate separates whether the Pauli string on the non-degenerate factor is fixed or negated by $\theta_I$, producing the two components in Eq.~\eqref{eq:pauli_lie_isometry_type_zero_basis}.
Equivalently, in Pauli terms, this is the representative with bilinear form $X_{m+1}$ and radical constraint $Z_{m+1}$.
The tensor factor $I^{\otimes \ell}$ is the multiplicity factor coming from the Pauli matrix algebra decomposition of $\commalg$.
\end{proof}

The canonical Pauli bases make the representation structure visible.
The next proposition gives the corresponding concrete block forms of the full compact Lie algebra before passing to the derived algebra.

\begin{prop}[Block forms for Pauli Lie algebras of isometries]\label{prop:pauli_lie_isometry_block_forms}
In the setting of Lemma~\ref{lem:canonical_pauli_bases_commutant_bilinear_isometries}, the full compact Pauli Lie algebra of isometries has the following block forms, up to conjugation in the ambient matrix algebra:
\begin{enumerate}
\item If $\type(w+W)=+$ and $m=0$, then the full compact Pauli Lie algebra of isometries is the zero Lie algebra, $\lieiso(\commalg,B)=\{0\}$.
\item If $\type(w+W)=+$ and $m\geq1$, then
\[
    \lieiso(\commalg,B)
    \conjugated
    \bigoplus_{\lambda=1}^{2^r}\so(2^m)\otimes\id_{2^\ell}.
\]
\item If $\type(w+W)=-$ and $m\geq1$, then
\[
    \lieiso(\commalg,B)
    \conjugated
    \bigoplus_{\lambda=1}^{2^r}\usp(2^{m})\otimes\id_{2^\ell}.
\]
\item If $\type(w+W)=0$, then
\[
    \lieiso(\commalg,B)
    \conjugated
    \bigoplus_{\lambda=1}^{2^{r-1}}\lieu(2^{m})\otimes\id_{2^\ell}.
\]
\end{enumerate}
For $\type(w+W)=0$, each copy acts on a pair of isotypic sectors indexed by $(+1,s_2,\ldots,s_r)$ and $(-1,s_2,\ldots,s_r)$, with the standard representation on one sector and the dual representation on the other.
Here $s_1$ is the distinguished radical coordinate on which the type-$0$ restricted quadratic form is non-zero, while $(s_2,\ldots,s_r)$ with $s_2,\ldots,s_r\in\{+,-\}$ labels the independent pairs.
This includes the abelian boundary case $\ell=0$, where the summands are copies of $\lieu(1)$.
For the $\pm$ types, the independent copies act sector by sector.
See Figs.~\ref{fig:block-so-representation} and~\ref{fig:block-type0-representation} for the corresponding block-representation pictures.
\end{prop}

\begin{proof}
Use the central projectors
\[
    \Pi_s=\prod_{j=1}^r\frac{1+s_jZ_{m+j}}{2}
    \quad \text{for } s=(s_1,\ldots,s_r)\in\{\pm1\}^r,
\]
for the isotypic sectors of the Pauli matrix algebra decomposition.
For type $+$ with $m=0$, Lemma~\ref{lem:canonical_pauli_bases_commutant_bilinear_isometries} gives the zero algebra.
For type $+$ with $m\geq1$ and for type $-$, Lemma~\ref{lem:canonical_pauli_bases_commutant_bilinear_isometries} separates independently over the projectors $\Pi_s$, giving one copy of $\so(2^m)$ or $\usp(2^m)$ in each sector.
For type $0$, we use the adapted radical basis in which the first radical coordinate is the one where the restricted quadratic form is non-zero.
For type $0$, Eq.~\eqref{eq:pauli_lie_isometry_type_zero_basis} pairs the sectors with labels $(+1,s_2,\ldots,s_r)$ and $(-1,s_2,\ldots,s_r)$.
The second component in Eq.~\eqref{eq:pauli_lie_isometry_type_zero_basis} changes sign between the two sectors, giving the dual action on the paired sector.
\end{proof}

Thus Proposition~\ref{prop:pauli_lie_isometry_block_forms} classifies the full compact Pauli Lie algebra of isometries.
The classification statement for the semisimple part is now obtained by taking derived Lie algebras of these block forms, with one separate exceptional case when the type-$+$ non-degenerate factor has dimension two.

\begin{thm}[Derived Pauli Lie algebras of isometries]\label{thm:derived_pauli_lie_algebras_isometries}
Let $W\subseteq\Fn$ label the Pauli commutant algebra $\commalg=\Span[\C]{\inviso(W)}$, set $V=W^\perp$, and let $B=\iso{w}$ be a Pauli bilinear form.
Write $\rank(W)=2\ell$, $\nullity(W)=r$, and $m=n-\ell-r$.
Then the possibilities for the derived Pauli Lie algebra $\lieiso^0(\commalg,B)$ are, up to Lie algebra isomorphism and with the block representations described in Prop.~\ref{prop:pauli_lie_isometry_block_forms}:
\begin{enumerate}
\item If $\type(w+W)=+$ and $m=0$, then the full compact Pauli Lie algebra of isometries is already zero, so
\[
    \lieiso^0(\commalg,B)=\{0\}.
\]
\item If $\type(w+W)=+$ and $m=1$, then the full compact Pauli Lie algebra of isometries is abelian, so its derived Lie algebra vanishes:
\[
    \lieiso^0(\commalg,B)=\{0\}.
\]
\item If $\type(w+W)=+$ and $m\geq2$, then
\[
    \lieiso^0(\commalg,B)
    \cong
    \so(2^m)^{\oplus 2^r}.
\]
\item If $\type(w+W)=-$ and $m\geq1$, then
\[
    \lieiso^0(\commalg,B)
    \cong
    \usp(2^m)^{\oplus 2^r}.
\]
\item If $\type(w+W)=0$ and $m\geq1$, then $r\geq1$ and
\[
    \lieiso^0(\commalg,B)
    \cong
    \su(2^m)^{\oplus 2^{r-1}}.
\]
\item If $\type(w+W)=0$ and $m=0$, then the full compact Pauli Lie algebra of isometries is abelian, so its derived Lie algebra vanishes:
\[
    \lieiso^0(\commalg,B)
    =
    \{0\}.
\]
\end{enumerate}
Restoring the multiplicity representation amounts to tensoring each block action with $\id_{2^\ell}$.
\end{thm}

\begin{proof}
The non-exceptional cases are precisely the cases where the anisotropic vectors of the restricted quadratic form span $V$, as in Eq.~\eqref{eq:commutant_bilinear_anisotropic_vectors_span}.
Thus Lemma~\ref{lem:commutant_bilinear_pauli_lie_algebra_derived_part}(b), together with the block forms in Proposition~\ref{prop:pauli_lie_isometry_block_forms}, gives the $+$ case with $m\geq2$, the $-$ case, and the type-$0$ case with $m\geq1$.
For type $+$ with $m=0$, the restricted form vanishes on $V=\rad(V)$, so $\QQ_w^{-1}(1)\cap V=\emptyset$ and $\lieiso(\commalg,B)=\{0\}$.
For the exceptional $+$ case with $m=1$, Lemma~\ref{lem:commutant_bilinear_pauli_lie_algebra_derived_part}(a) applies with the proper span $S=\Span[\F_2]{\QQ_w^{-1}(1)\cap V}$.
Equivalently, Proposition~\ref{prop:pauli_lie_isometry_block_forms} gives only copies of $\so(2)\cong\lieu(1)$, hence the algebra is abelian and its derived algebra is zero.
For type $0$ with $m=0$, the block-form proposition gives only copies of $\lieu(1)$, so the same abelian argument yields $\lieiso^0(\commalg,B)=\{0\}$.
\end{proof}

This also explains what remains visible after projecting to the irreducible sectors of the commutant algebra.
For $\type(w+W)=+$ and $\type(w+W)=-$, each sector still carries the corresponding invariant bilinear form, giving the orthogonal or unitary-symplectic block.
For $\type(w+W)=0$, the projection to a first sector of a linked pair gives only the standard representation of $\su(2^m)$, while the second sector carries the dual representation. The invariant bilinear form is therefore only visible before separating the linked pair.
This is the Lie algebra counterpart of Lemma~\ref{lem:quadratic_forms_under_projections_extensions}. After projecting over the full radical, the quadratic form is preserved precisely when the radical is isotropic, i.e., in the cases of type $+$ and $-$.

The same viewpoint accounts for the restrictions on the parameters.
For $\type(w+W)=0$, the radical must contain a vector on which the quadratic form is non-zero, so necessarily $r\geq1$; the semisimple type-$0$ contribution is non-zero only for $m\geq1$.
For $\type(w+W)=+$ with $m=0$, the restricted form vanishes on $V=\rad(V)$, so there are no anisotropic vectors in $V$ and the corresponding Lie algebra is zero.
The first nontrivial boundary case is $\type(w+W)=+$ with $m=1$: the block algebra is $\so(2)\cong\lieu(1)$, hence abelian, and its complex representation splits into two one-dimensional eigenspaces.
Over $\C$, this two-dimensional real block diagonalizes as $\C I\oplus(-\C I)$, so the real standard representation of $\so(2)$ is no longer irreducible after complexification; in this sense the block behaves like two one-dimensional commutant sectors.
Equivalently, the standard representation of $\so(2)=\lieu(2)^{\theta_I}$ becomes the block-diagonal representation of $\lieu(1)=\lieu(2)^{\theta_I,\theta_Y}$ after diagonalizing the block, which provides the Lie algebraic counterpart of the collapse of $\type(w+W)=+$ with $m=1$ to $\type(w+W)=0$ with $m=0$.

\section{Line Graphs of Multigraphs and the \texorpdfstring{$\calE_6$}{E6} condition}\label{sec:Line_Graphs_E6_Condition}

In the previous sections we have given the necessary details for the symmetry description of transvection groups and Pauli Lie algebras, which are sufficient to describe the generating sets $t$-equivalent to $\graphX_{2m-1,n_1}^1$, $\graphX_{2m-1,n_1}^2$ and $\graphX_{2m,n_1}^3$, as we shall see in more detail in Section~\ref{sec:classification:groups_lie_algebras}.
Therefore, it remains to deal with the cases $t$-equivalent to the blown-up Path Graphs $\graphP_{k,n_1}$.
Specifically, in this section we deal with \emph{invariant} properties of $\graphP_{k,n_1}$ under $t$-equivalence and highlight various criterions to identify the specific case we're considering.

\subsection{Line Graphs and Contractions}

\begin{figure}[t]
    \centering
    \includegraphics[width=\linewidth]{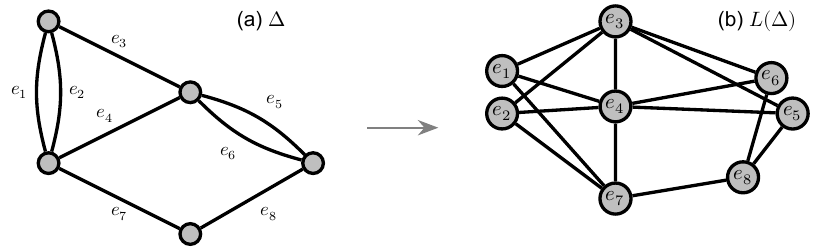}
    \caption{Example of multigraph $\Delta$ and its line-graph $L(\Delta)$.}
    \label{fig:line-graph-definition}
\end{figure}

We start by defining the line graph of a \emph{multigraph}:
\begin{defn}[Line Graphs of Multigraphs]\label{def:line_graphs}
We call a multigraph $\Delta$ a graph where edges between pairs of vertices can appear multiple times (see Fig.~\ref{fig:line-graph-definition}).
We define the line graph $L(\Delta)$ of a multigraph $\Delta$ as the simple graph whose vertices are the edges, which are connected if they share a single vertex in $\Delta$.
We say that $\Delta$ is a root graph of $L(\Delta)$.
\end{defn}

Notice that parallel edges are not connected in $L(\Delta)$ and in particular are \emph{twins}, since they share the same neighbors.
In fact, all twins in the line graph of a multigraph must correspond precisely to parallel edges.
Furthermore, if $\Delta$ has no parallel edges, then it coincides with the ordinary line graph of the simple graph $\Delta$.
As such, we also have that a graph is the line graph of a multigraph if and only if $\modout{\graphG}$ is the line graph of an ordinary graph \cite{Cuypers_2021_E6}.

\begin{rem}
We note that there are multiple ways of defining the line graph of a multigraph, which potentially depend on the specific application, see also \cite{beineke2021line,Bagga_2004}.
Indeed, in \cite{Cuypers_2021_E6} two classes of line graphs of a multigraph $\Delta$ are highlighted: the 1-line-graph $L_1(\Delta)$, as defined in \ref{def:line_graphs}, where edges of $\Delta$ are adjacent in $L_1(\Delta)$ iff they share precisely one vertex; the $\geq 1$-line-graph $L_{\geq 1}(\Delta)$, where instead edges of $\Delta$ are adjacent in $L_{\geq 1}(\Delta)$ iff they share at least one vertex.
However, for our purposes the first definition is more natural, hence we will consider the 1-line-graph definition as `canonical' and simply refer to it as line graph of a multigraph.
\end{rem}

\begin{figure}
    \centering
    \includegraphics{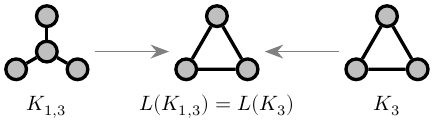}
    \caption{The only two non-isomorphic (simple) graphs which admit the same line graph are $K_{1,3}$ and $K_3$.}
    \label{fig:line-graph-claw}
\end{figure}

One can also consider also the uniqueness of a root graph of a line graph.
It is known \cite{Cuypers_2021_Whitney} that, up to a few exceptions, the root graph is unique:
\begin{thm}[{\cite[Theorem 1.2]{Cuypers_2021_Whitney}}]\label{thm:uniqueness_root_graph_line_graph}
Let $\Delta,\Delta'$ be two connected multigraphs such that $L(\Delta)$ and $L(\Delta')$ are isomorphic.
Then, $\Delta$ and $\Delta'$ are isomorphic, unless $\Delta$ or $\Delta'$ have four vertices.
Furthermore, there is a unique connected multigraph $\Delta''$ such that $\abs{\vertices(\Delta)}\neq 4$ and $L(\Delta'')$ is isomorphic to $L(\Delta)$ and $L(\Delta')$.
\end{thm}
The ambiguity on four vertices comes from the fact that, for ordinary line graphs, the claw $\graphK_{1,3}$ and cycle $\graphK_3$ have the same line graph $\graphK_3$ (see Fig.~\ref{fig:line-graph-claw}).

\begin{figure}
    \centering
    \includegraphics[width=0.9\linewidth]{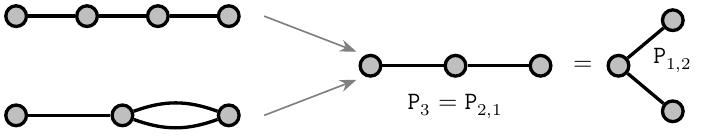}
    \caption{Examples of (multi)graphs which admit the same line graph.}
    \label{fig:line-graph-p3}
\end{figure}

Then, this naturally extends to line graphs by considering a natural multigraph extension, such that the cycle of length $3$, up to arbitrary twin vertices, has multiple non-isomorphic root graphs (see \cite[Proposition~2.5]{Cuypers_2021_Whitney} for a full characterization of the non-isomorphic multigraphs with isomorphic line graphs).
By choosing a root graph $\Delta$ without four vertices, the ambiguity is resolved and one can choose a unique root graph.
The natural root graph for this graph is then the cycle of length $3$ with a suitable number of parallel edges equal to the number of parallel twins.
In general, if $\graphG$ is a line graph, we refer to $R(\graphG)$ as the unique root graph of $\graphG$ specified in Theorem~\ref{thm:uniqueness_root_graph_line_graph}.
As another example for ambiguity, consider the path graph $\graphP_3$. This admits a root graph which is either $\graphP_4$ or the path multigraph on $3$ with two parallel edges at one end (see Fig.~\ref{fig:line-graph-p3}).
Indeed, for canonical $t$-representatives, the only cases where this ambiguity appears is precisely $\graphP_{2,n_1}$, which is reflected in the ambiguity $\graphP_{1,n_1+1} = \graphP_{2,n_1}$.
Namely, $\graphP_{2,n_1}$ admits as a root graph either a path multigraph on $4$ vertices with $n_1-1$ parallel edges on one end or a path multigraph on $3$ vertices with $n_1$ parallel edges on one end.
Then, $\graphP_{k,n_1}$ with $k\geq 3$ requires at least five vertices for the root graph, hence it is always uniquely defined.

\begin{figure}
    \centering
    \includegraphics[width=0.9\linewidth]{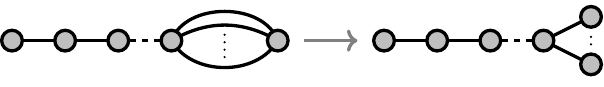}
    \caption{The blown-up path graph $\graphP_{k,n_1}$, $k\geq 1$, $n_1\geq 1$ (right) is the line graph of a multigraph (right), which consists of a path graph with $k+1$ vertices with $n_1$ parallel edges at one end.}
    \label{fig:line-graph-path}
\end{figure}

More generally, we find that $\graphP_{k,n_1}$ is a line graph whose root multigraph is the path graph of length $k+2$ with $n_1-1$ parallel edges at an extreme (see Fig.~\ref{fig:line-graph-path}).

\begin{figure}
    \centering
    \includegraphics[width=0.8\linewidth]{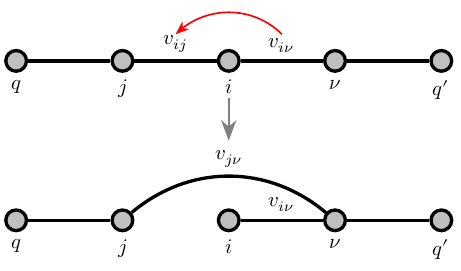}
    \caption{A contraction of a line graph may be seen at the level of the root graph as shifting one end of an edge to another.}
    \label{fig:line-graph-contraction}
\end{figure}

Indeed, we find that `line graphness' is precisely the invariant property which characterizes being $t$-equivalent to a blown-up path graph:
\begin{lem}
Valid contractions send line graphs into line graphs.
\end{lem}
\begin{proof}
Let $\graphG = R(\Delta)$ be a line graph of a multigraph.
We can identify each vertex of $\graphG$ with an edge from the root graph.
Let us consider first the case with no parallel edges, i.e. $\Delta$ is a simple graph, hence each edge is simply a pair of vertices $i$ and $j$ in $\Delta$.
Also, it is convenient to use the unique algebraically independent basis $\vgens$ associated to $\graphG$ to express whether two vertices in $\graphG$ ro edges in $\Delta$ are adjacent.
The following rules apply for the adjacency matrix over $\graphG$ for line graphs:
\begin{enumerate}
    \item $\symp{v_{\mu\nu}}{v_{ij}} = 0$ iff $\mu\nu$ and $ij$ are parallel or share no vertices
    \item $\symp{v_{\mu\nu}}{v_{ij}} = 1$ iff they share precisely one vertex, which we can choose as $i=\mu$ without loss of generality
\end{enumerate}
Notice that this is also a \emph{characterization} of line graphs, since for any choice of $\{\mu,\nu\}\subseteq[n]\times[n]$ the graph with adjacency matrix specified as above is a line graph.

If we now consider a valid contraction, we must be in case (b).
Hence, contraction of $v_{i\nu}$ onto $v_{ij}$ results in $v_{ij}+v_{i\nu}$, such that:
\begin{equation}
    \symp{v_{ij}+v_{i\nu}}{v_{\mu q}} = \symp{v_{ij}}{v_{\mu q}} + \symp{v_{i\nu}}{v_{\mu q}}
\end{equation}
This is equal to $1$ iff either $ij$ or $i\nu$ share precisely one vertex with $\mu q$, which excludes $i$ as common vertex.
Hence, $v_{ij}+v_{i\nu}$ is adjacent only to vectors of the form $v_{jq}$, $q\neq i$, or $v_{\nu q}$, $q\neq \nu$.
Then, we may call this vector $v_{j\nu}' = v_{ij}+v_{i\nu}$ and notice that the new adjacency matrix is still one of a line graph.
Specifically, the new root graph $\Delta'$ is the one where the edge $ij$ has been removed and the edge $j\nu$ has been added (see Fig.~\ref{fig:line-graph-contraction}).
Notice that an edge $j\nu$ may have already existed in $\Delta$, which case $\Delta'$ is a multigraph with two parallel edges between $j$ and $\nu$.

It is straightforward to generalize to the case of multiple parallel edges, where now each edge in $\Delta$ also has an additional index to distinguish it from its parallel edges, $(ij,d)$ where $d$ runs over the parallel edges at $i,j$.
Then, if we denote the corresponding vectors of $\graphG = L(\Delta)$ with $v_{(ij,d)}$, it is clear that the adjacency matrix respects the same rules independently of the additional index.

The result follows for general line graphs of multigraphs.
\end{proof}

We can also express this result with respect to the \emph{incidence matrix} of the root graph. We denote as $M(\Delta)$ and has size $\abs{\vertices(\Delta)}\times\abs{\edges(\Delta)} = \abs{\vertices(\Delta)}\times\abs{\vertices(\graphG)}$.
Let us denote vertices as $\vva\in\vertices(\Delta)$ and edges as
$\edgea\in\edges(\Delta) = \vertices(\graphG)$.
Succintly, the adjacency matrix of a line graph $\graphG=L(\Delta)$ has entries:
\begin{equation}
    A(\graphG)_{\edgea,\edgeb} = \begin{dcases}
        0 & \text{if }\edgea,\edgeb\text{ share }0\text{ or }2\text{ vertices}\\
        1 & \text{if }\edgea,\edgeb\text{ share }1\text{ vertex}
    \end{dcases}
\end{equation}

Let us define the incidence matrix of $\Delta$ as:
\begin{equation}\label{eq:def:incidence_matrix}
    M(\Delta)_{\vva,\edgea} = \begin{dcases}
        1 & \text{if }\edgea\text{ is adjacent to }\vva\\
        0 & \text{otherwise}
    \end{dcases}
\end{equation}
which results in the adjacency matrix (over $\F_2$):
\begin{align*}
        &(M^TM)_{\edgea,\edgeb}
        = \sum_{\vva} M_{\vva,\edgea}M_{\vva,\edgeb}\\
        &= \begin{dcases}
            0 & \text{if }\edgea,\edgeb\text{ share an even number of vertices}\\
            1 & \text{if }\edgea,\edgeb\text{ share an odd number of vertices}
        \end{dcases}\\
        &= A_{\edgea,\edgeb}
\end{align*}
Viceversa, a characterization of incidence matrices of (multi)graphs without self-loops is that it is a binary matrix whose columns sum to precisely $2$ (over the integers):

Over the integers we have instead:
\begin{align*}
        (M^TM)_{\edgea,\edgeb}
        &= \sum_{\vva} M_{\vva,\edgea}M_{\vva,\edgeb}\\
        &= \begin{dcases}
            0 & \text{if }\edgea,\edgeb\text{ share no vertices}\\
            1 & \text{if }\edgea,\edgeb\text{ share $1$ vertex}\\
            2 & \text{if }\edgea,\edgeb\text{ share $2$ vertices}\\
        \end{dcases}\\
        &= A_{\edgea,\edgeb} + 2\delta_{\edgea,\edgeb} + T_{\edgea,\edgeb}
\end{align*}
where $T_{\edgea,\edgeb}$ is $2$ if $\edgea,\edgeb$ are parallel edges, and
zero otherwise.
Hence $T$ is precisely the contribution coming from the multigraph setting.
In matrix form, we can write (over the integers):
\begin{equation}
    A(L(\Delta)) = M(\Delta)^T M(\Delta) - 2I - T
\end{equation}

We also use the following standard description of the cycle space in terms of
the incidence matrix. For ordinary graphs, we follow \cite{diestel2017} which proves that the kernel of
the incidence matrix over $\F_2$ is the cycle space
\cite[Prop.~1.9.6]{diestel2017}, and characterizes this cycle space as the
edge sets whose induced degrees are all even
\cite[Prop.~1.9.1]{diestel2017}. Moreover, \cite{diestel2017} extends the graph terminology to
multigraphs \cite[Sec.~1.10]{diestel2017}, where cycles of length two are
allowed and are given by pairs of parallel edges
\cite[Sec.~1.10]{diestel2017}. We need the following
version for multigraphs without loops:

\begin{lem}[cf.\ {\cite[Secs.~1.8--1.10]{diestel2017}}]\label{lem:cycle_space_incidence_matrix_multigraph}
Let $\Delta$ be a finite multigraph without loops, and let $M(\Delta)$ be its
incidence matrix over $\F_2$. Then $\ker(M(\Delta))$ is the cycle space of
$\Delta$.
\end{lem}
\begin{proof}
For a given edge set $F\subseteq\edges(\Delta)$, let
$x\in\F_2^{\edges(\Delta)}$ be its characteristic vector, with
$x_{\edgea}=1$ for $\edgea\in F$ and $x_{\edgea}=0$ otherwise.
Since every column of $M(\Delta)$ has a one exactly in the two rows corresponding to the
endpoints of that edge, the $\vva$th component of $M(\Delta)x$ is
\[
(M(\Delta)x)_{\vva}
=\sum_{\substack{\edgea\in\edges(\Delta)\\ \vva\in \edgea}}x_{\edgea}
\quad\text{in }\F_2 .
\]
Thus $(M(\Delta)x)_{\vva}$ is the parity of the number of edges of $F$ incident with
$\vva$. Hence $x\in\ker(M(\Delta))$ precisely when, at each vertex, an even number
of edges from $F$ is incident with that vertex.

If $F$ is the edge set of a cycle, then every vertex is incident with either
zero or two edges of $F$, including the case of a pair of parallel edges.
Hence the first paragraph shows that the edge
set of every cycle lies in $\ker(M(\Delta))$, and therefore the cycle space is
contained in $\ker(M(\Delta))$.

Conversely, let $x\in\ker(M(\Delta))$, and let $F=\{e\mid x_e=1\}$. By the first
paragraph, each vertex is incident with an even number of edges from $F$. For
ordinary graphs, \cite[Prop.~1.9.1]{diestel2017} identifies this condition with
belonging to the cycle space. For multigraphs without loops,
the same cycle-removal argument applies. The first step is to show that every
nonempty edge set with this even-incidence property contains the edge set of a
cycle. To see this, choose an edge in $F$ and walk through edges of $F$ without
reusing an edge. Whenever the walk enters a vertex, the number of edges of $F$
incident with that vertex is even, so unless all incident edges have already
been used there is another edge of $F$ through which the walk can leave. Since
the graph is finite, this process eventually repeats a vertex, and the segment
between the two occurrences is a cycle, possibly a length-two cycle formed by
parallel edges.

Remove the edge set of this cycle from $F$. At every vertex, the removed cycle
uses either zero or two incident edges, so the parity of the number of remaining
edges incident with that vertex is still even. Repeating the argument removes at
least one edge at each step and therefore terminates. When it terminates no
edges can remain: otherwise the remaining nonempty edge set would still have the
even-incidence property and would contain another cycle by the previous
paragraph. Thus all edges of $F$ have been removed as edge sets of cycles, so
$F$ is an $\F_2$-sum of edge sets of cycles. Hence $x$ lies in the cycle space,
proving the reverse inclusion.
\end{proof}

We now consider contractions on this line graph/root graph.
Let $P_{\sourceidx,\targetidx}$ be the edge-coordinate matrix on
$\F_2^{\edges(\Delta)}$ which performs the contraction of the vertex
$\sourceidx$ of $L(\Delta)$ onto the vertex $\targetidx$ of $L(\Delta)$,
equivalently of the corresponding edges
$\sourceidx,\targetidx\in\edges(\Delta)$.
Writing $\edgeunit{\edgea}$ for the edge-coordinate unit vector corresponding
to $\edgea\in\edges(\Delta)$, this means that
\begin{align}
    P_{\sourceidx,\targetidx}\edgeunit{\edgea}
    & = \edgeunit{\edgea} + \delta_{\edgea,\targetidx}\edgeunit{\sourceidx},\\
    (P_{\sourceidx,\targetidx})_{\edgea,\edgeb}
    &= \delta_{\edgea,\edgeb}
    + \delta_{\edgea,\sourceidx}\delta_{\targetidx,\edgeb}.
\end{align}
Then, consider the matrix (with matrix multiplication over $\F_2$):
\begin{equation}
    M' = MP_{\sourceidx,\targetidx}
\end{equation}
such that
$A(\graphG') = M'^TM'
= P_{\sourceidx,\targetidx}^TM(\Delta)^T M(\Delta)P_{\sourceidx,\targetidx}$
is the adjacency matrix of the new graph.
We now show that $M'$ is the incidence matrix of another multigraph, hence $\graphG'$ is still a line graph.
Indeed, we have (over $\F_2$):
\begin{align*}
        M'_{\vva,\edgea}
        &= (MP_{\sourceidx,\targetidx})_{\vva,\edgea}
        = \sum_{\edgeb} M_{\vva,\edgeb}
        (P_{\sourceidx,\targetidx})_{\edgeb,\edgea}\\
        &= \sum_{\edgeb} M_{\vva,\edgeb}
        (\delta_{\edgea,\edgeb}
        + \delta_{\edgeb,\sourceidx}\delta_{\targetidx,\edgea})
        = M_{\vva,\edgea}
        + \delta_{\edgea,\targetidx}M_{\vva,\sourceidx}
\end{align*}
which is $M_{\vva,\edgea}$ if $\edgea\neq \targetidx$ and
$M_{\vva,\targetidx}+M_{\vva,\sourceidx}$ otherwise.
If this is a valid contraction, $\sourceidx$ and $\targetidx$ are adjacent at
precisely one site, $M_{\vva,\targetidx}+M_{\vva,\sourceidx}$ will be non-zero
precisely at the two vertices $\vva$ which are adjacent to solely
$\targetidx$ and solely $\sourceidx$.
As such, we find that $M'$ is a $\{0,1\}$-matrix whose columns sum to $2$ (over the integers), which means that in fact $M(\Delta') = M(\Delta)P_{\sourceidx,\targetidx}$ is the incidence matrix of a multigraph and $\graphG'$ is the line graph of $\Delta'$.

We summarize this discussion in the following result:
\begin{cor}\label{cor:t_equivalence_contractions_on_multigraphs}
Let $\graphG = L(\Delta)$ be the line graph of a multigraph with incidence matrix $M(\Delta)$.
The graph $\graphG'$ obtained by performing the valid contraction of
$\sourceidx$ onto $\targetidx$ is the line graph of a multigraph $\Delta'$
whose incidence matrix is:
\begin{equation}
    M(\Delta') = M(\Delta)P_{\sourceidx,\targetidx}
\end{equation}
where multiplication is performed over $\F_2$.
Furthermore, if the edges $\sourceidx$ and $\targetidx$ are adjacent with
vertices $\{\vva,\vvb\}$ and $\{\vvb,\vvc\}$ respectively, then $\Delta'$ has
the same vertex set as
$\Delta$, and its edge multiset is obtained from the edge multiset of $\Delta$
by deleting the edge $\targetidx$ and adding one edge $\{\vva,\vvc\}$:
\begin{align*}
        \vertices(\Delta') &= \vertices(\Delta),\\
        \edges(\Delta') &= (\edges(\Delta)\setminus \{\targetidx\})
        \sqcup \{\{\vva,\vvc\}\}.
\end{align*}
\end{cor}

We will need the following tree-normalization step:
\begin{lem}\label{lem:root_tree_to_path_by_edge_slides}
Let $\Delta$ be a tree. Then $L(\Delta)$ is $t$-equivalent to the line graph of
a path on $\abs{\vertices(\Delta)}$ vertices.
\end{lem}
\begin{proof}
We use the edge-slide interpretation from
Corollary~\ref{cor:t_equivalence_contractions_on_multigraphs}. If
$s=\{a_{i-1},a_i\}$ and $t=\{a_i,b\}$ are adjacent edges of the root tree, then
contracting $s$ onto $t$ deletes $t$ and adds the edge $\{a_{i-1},b\}$. Thus a
branch edge attached at $a_i$ can be slid one step along an adjacent tree edge
toward $a_{i-1}$. This operation preserves the vertex set and the number of
edges, and the resulting graph is again a tree because the component attached
through $b$ remains connected to the rest of the graph at exactly one vertex.

Choose a longest path $\pathvar=(a_1,\ldots,a_h)$ in the current tree. If $\pathvar$
contains all vertices, the tree is already a path. Otherwise, since the graph is
a tree, each connected component of the induced subgraph on vertices outside
$\pathvar$ has exactly one edge to $\pathvar$, incident with a single vertex of $\pathvar$. Since
$\pathvar$ is longest, such an attachment vertex cannot be $a_1$ or $a_h$, so it is
some internal vertex $a_i$ with $1<i<h$.
Choose one attached component, let $\{a_i,b\}$ be the edge from $a_i$ to that
component, and let $c$ be a leaf of this component.

Repeatedly slide the attachment edge along the path
$\{a_{i-1}, a_i\},\{a_{i-2},a_{i-1}\},\ldots,\{a_1,a_2\}$ until the component is
attached at $a_1$. Because the tree property is preserved at each slide, the
path from $c$ to $a_h$ in the resulting tree contains all vertices
$a_1,\ldots,a_h$ and at least one additional vertex from the chosen
component. Thus the length of a longest path has strictly increased. Since the
number of vertices is finite, this process terminates, and at termination a
longest path contains all vertices. The resulting tree is therefore a path on
$\abs{\vertices(\Delta)}$ vertices.
\end{proof}

Now that it's clear how to use $t$-equivalence on multigraphs, we can provide explicit multigraph invariants which can distinguish between the different $t$-equivalence classes.
\begin{thm}\label{thm:t_equivalence_line_graphs_invariants}
Let $\graphG = L(\Delta)$ be the line graph of a connected multigraph $\Delta$ with incidence matrix $M(\Delta)$ and $n_\Delta$ vertices.
Let $\tilde q = \nullity(M(\Delta))$ be the dimension of $\F_2$-space of cycles over $\Delta$.
Then, $\graphG$ is $t$-equivalent to $\graphP_{n_\Delta-2,\tilde q+1}$.
\end{thm}
\begin{proof}
By Corollary~\ref{cor:t_equivalence_contractions_on_multigraphs},
contractions on the line graph can be read on the root multigraph as edge
slides; we use this rule throughout the proof. Choose a spanning tree
$T\subseteq\Delta$. Since $\Delta$ is connected,
the number of edges outside $T$ is
\[
\abs{\edges(\Delta)}-\abs{\edges(T)}
=\abs{\edges(\Delta)}-n_\Delta+1
=\tilde q.
\]
Let $e=\{u,v\}$ be one of these $\tilde q$ extra edges, and let
$(u=a_0,a_1,\ldots,a_s=v)$ be the unique path from $u$ to $v$ in $T$.
If $s=1$, then $e$ is already parallel to the tree edge $\{u,v\}$. If $s\geq2$,
repeatedly applying the edge-slide rule replaces $e=\{u,v\}$ by
$\{u,a_{s-1}\}$, then by $\{u,a_{s-2}\}$, and so on, until it becomes parallel
to the tree edge $\{u,a_1\}$. In both cases, this does not change the vertex
set, does not change the number of extra edges, and leaves the tree $T$ as a
connected spanning subgraph. Repeating this for all extra edges gives a
multigraph consisting of the same spanning tree $T$ together with $\tilde q$
additional edges parallel to tree edges.

By Lemma~\ref{lem:root_tree_to_path_by_edge_slides}, the tree $T$ can be
changed by edge slides into a path on $n_\Delta$ vertices. Whenever an edge
slide moves a tree edge, apply the same slide to each additional edge parallel
to that tree edge. Thus, after the tree has become a path, the $\tilde q$ additional
edges are still parallel to path edges. Finally, applying the edge-slide rule
along the path moves all $\tilde q$ additional parallel edges to one end of the path.
The resulting root multigraph is a path on $n_\Delta$ vertices with $\tilde q+1$
parallel edges at one end. Its line graph is
$\graphP_{n_\Delta-2,\tilde q+1}$, proving the statement.
\end{proof}

\begin{figure}
    \centering
    \includegraphics[width=0.9\linewidth]{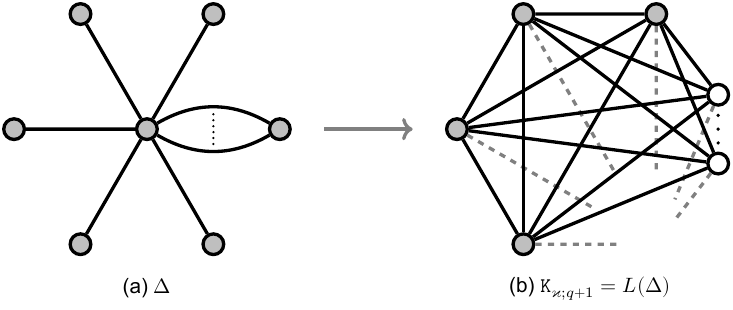}
    \caption{Alternative $t$-equivalent canonical representatives in the line graph case: (a) root graph is a star graph with $\tilde q+1$ parallel edges at one site (b) line graph is the complete graph with $\tilde q$ added twins at one vertex, denoted as $\graphK_{\varkappa;\tilde q+1}$.}
    \label{fig:line-graph-complete-graph}
\end{figure}

We can also provide an alternative construction, where we transport connections from each vertex in branch to a given \emph{central} vertex, thus producing a \emph{star} multigraph, with $\tilde q+1$ parallel edges at a pair of sites, see Fig.~\ref{fig:line-graph-complete-graph}(a).

The resulting line graph is then a complete graph over $n_\Delta-1$ vertices plus $\tilde q$ additional vertices which are twins to some reference vertex.
For $\varkappa\geq 1$ and $\tilde q\geq 0$, let $\graphK_{\varkappa;\tilde q+1}$ denote the graph
obtained from the complete graph $\graphK_{\varkappa+1}$ by adding $\tilde q$ twins of
one fixed vertex, see Fig.~\ref{fig:line-graph-complete-graph}(b).
Indeed, this is a representative graph also used in \cite{Janssen_1983}, which is
$\graphK_{n_\Delta-2;\tilde q+1}$ in this notation.

A particular interesting consequence of `playing' the $t$-equivalence game on the root graph, rather than the graph itself, is that loops can be turned into parallel edges.
This also show that regular line graphs may in fact be $t$-equivalent to line graphs of multigraphs.
Hence, we find that valid contractions do \emph{not} send regular line graphs into regular line graphs, which further motivates the necessity of multigraphs in the context of Lie algebras and transvection groups.

Theorem~\ref{thm:t_equivalence_line_graphs_invariants} produces two invariants, which uniquely determine the $t$-equivalence class:
\begin{enumerate}
    \item The number of vertices of the multigraph $\Delta$
    \item The number of independent cycles
\end{enumerate}
Notice that the theorem still applies even under the ambiguity in the $n_\Delta=4$ case, which does not distinguish between $\graphP_{2,n_1}$ and $\graphP_{1,n_1+1}$.
However, if we exclude this case then we can in fact say that a graph is $t$-equivalent to $\graphP_{k,n_1}$ if and only if it is the line graph of a multigraph with $k+2$ vertices and $n_1-1$ independent cycles.

Also notice that these two invariants are not independent, when given the size of the original graph.
This follows from an expression for the cyclomatic number $\tilde q = \abs{\edges(\Delta)} - n_\Delta + 1 = \abs{\vertices(\graphG)}-n_\Delta + 1$.
Moreover, we highlight that both of these may not be well defined in the presence of algebraic dependencies.
In particular, by looking at the representatives $\graphP_{n_\Delta-2,\tilde q+1}$, together with Proposition~\ref{prop:Limits_Lie_Algebraic_Dependencies_canonical}, it is clear that $n_\Delta$ is independent of minimality assumptions, whereas the number of cycles may in fact get smaller in the presence of algebraic dependencies.

Notice that the invariance of the number of cycles may also be shown algebraically.
One simply needs the fact that the kernel of the incidence matrix coincides with the $\F_2$-space of cycles over the multigraph $\Delta$, as shown in Lemma~\ref{lem:cycle_space_incidence_matrix_multigraph}.
Furthermore, using the fact that $P_{\sourceidx,\targetidx}$ is an involution,
a contraction is just a change of basis for the kernel:
\begin{equation}
    \ker(M') = \ker(MP_{\sourceidx,\targetidx})
    = P_{\sourceidx,\targetidx}\ker(M)
\end{equation}

\begin{figure}
    \centering
    \includegraphics[width=0.5\linewidth]{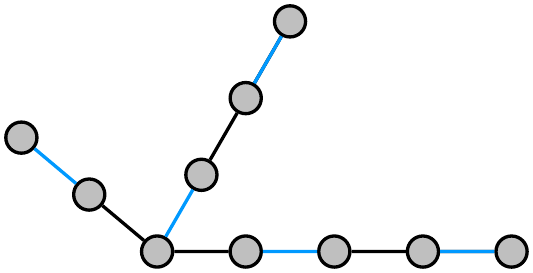}
    \caption{Example of T-join for a tree.}
    \label{fig:line-graph-T-join}
\end{figure}

\begin{figure}
    \centering
    \includegraphics[width=\linewidth]{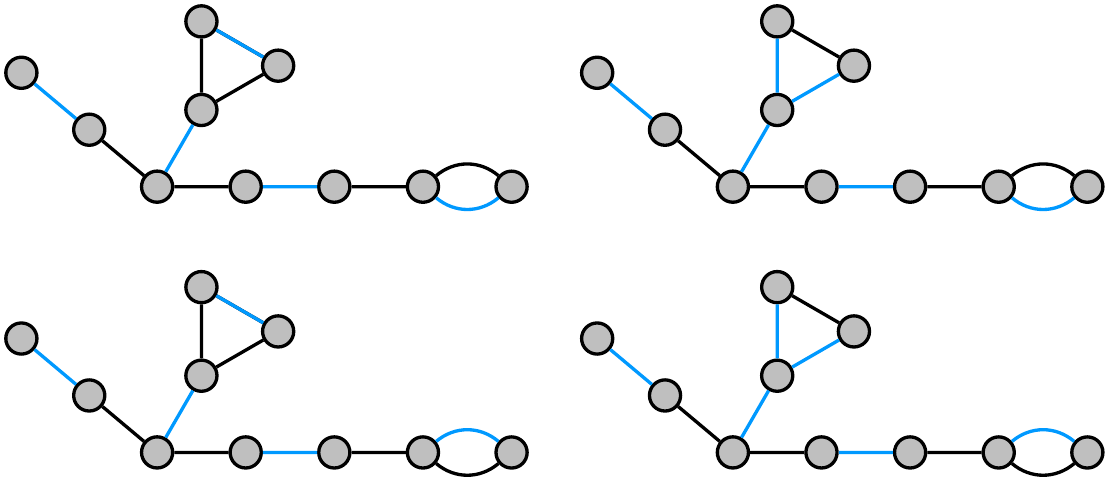}
    \caption{The T-joins for a multigraph with two independent cycles (including two parallel edges). Four possible T-joins are possible.}
    \label{fig:line-graph-T-join-w-cycles}
\end{figure}

We can also completely characterize the radical for line graphs of multigraphs, which also generalizes a result of \cite{Chapman_Flammia_2020} to the multigraph setting.
First, we define a \emph{T-join} for a (multi)graph:
\begin{defn}[T-join]\label{def:T-join}
Let $\Delta$ be a (multi)graph. Let $T\subseteq\vertices(\Delta)$ be a subset of vertices for $\Delta$. 
We say that a subset $E\subseteq\edges(\Delta)$ is a \emph{T-join} if every vertex in $T$ is adjacent to an odd number of edges in $E$.
Equivalently, one can say that the coloring defined on $E$ as
$\component{x}{\edgea}=1$ for $\edgea\in E$ is a T-join iff
\begin{equation}
    \sum_{\edgea\in\edges(\Delta)} M(\Delta)_{\vva,\edgea}
    \component{x}{\edgea} = 1,\,\forall \vva\in T
\end{equation}
If $T$ is not specified, we assume $T=\vertices(\Delta)$ and T-join for $\Delta$ is a T-join over all vertices. 
See also Figures~\ref{fig:line-graph-T-join} and \ref{fig:line-graph-T-join-w-cycles}.
\end{defn}

\begin{lem}\label{lem:radical_via_cycles_and_Tjoin}
Let $\graphG = L(\Delta)$ be the line graph of a connected multigraph with incidence matrix $M(\Delta)$ and $n_\Delta=\abs{\vertices(\Delta)}$.
As subspaces of $\F_2^{\edges(\Delta)}$,
\begin{enumerate}
    \item if $n_\Delta$ is odd, then $\ker(A(\graphG))=\ker(M(\Delta))$;
    \item if $n_\Delta$ is even, then
    \[
    \ker(A(\graphG))=\ker(M(\Delta))+\SpanS[\F_2]{\{x\}},
    \]
    for any $x\in\F_2^{\edges(\Delta)}$ with $M(\Delta)x=\mathbf 1$, where
    $\mathbf 1$ is the all-one vector in $\F_2^{\vertices(\Delta)}$.
\end{enumerate}
\end{lem}
\begin{proof}
Using $A(\graphG)=M(\Delta)^T M(\Delta)$, the kernel of $A(\graphG)$
consists of all edge-coordinate vectors $y$ such that $M(\Delta)y$ is in the
kernel of $M(\Delta)^T$.
For an edge $e=\{\vva,\vvb\}$, the condition $z\in\ker(M(\Delta)^T)$ is
\begin{equation}
    \sum_a (M(\Delta)^T)_{e,a}\component{z}{a}
    = \sum_a M(\Delta)_{a,e}\component{z}{a}
    = \component{z}{\vva}+\component{z}{\vvb}=0.
\end{equation}
which is either the all-zero or the all-one vector over $\F_2^{\vertices(\Delta)}$, since $\Delta$ is connected.
Then, the solutions resulting in the zero vector coincide with the kernel of $M(\Delta)$, hence the cycle space.

On the other hand, the ones vector is in the image of $M(\Delta)$ iff:
\begin{equation}
    \sum_e M(\Delta)_{a,e} \component{x}{e} = 1,\,\forall a\in\vertices(\Delta)
\end{equation}
i.e. there exists a coloring of the edges (hence, a coloring of vertices in
the line graph) such that each vertex is incident to an odd number of selected
edges.
One can construct such a coloring, i.e. a \emph{T-join}, if and
only if $n_\Delta$ is even. Any two edge-coordinate vectors $x$ with
$M(\Delta)x=\mathbf 1$ differ by an element of $\ker(M(\Delta))$, so the displayed
subspace in (b) is independent of the chosen $x$. This concludes the proof.
\end{proof}

Notice that a line graph is non-degenerate (invertible adjacency matrix) if and only if its root multigraph is a tree with an odd number of vertices.

We also highlight the necessity of using these tools to distinguish among the line graph cases.
Indeed, among the algebraically independent cases which are known to be line graphs, if one fixes rank and radical, there is still the ambiguity between $\graphP_{2m,n_1}$ and $\graphP_{2m-1,n_1+1}$.
If we also add information from the invariant quadratic forms (see Prop~\ref{prop:isomorphism_classes_quadratic_forms_arf_invariant_canonical}), it may still not be possible to distinguish the different cases.
For instance, this happens for $\vgens(\graphP_{4m+1,n_1})$ and $\vgens(\graphP_{4m+2,n_1})$.
This highlights the limitations of the symmetry tools to classify the line graph cases, even under the assumption of `line graphness', and the need for further tools.

When $\graphG=L(\Delta)$ and the vertex of $L(\Delta)$ corresponding to
an edge $\edgea\in\edges(\Delta)$ is labeled by $v_{\edgea}\in\vgens$, we use the linear
map $\edgetovec_{\Delta,\vgens}$ from $\F_2^{\edges(\Delta)}$ to $\SpanS[\F_2]{\vgens}$.
For $x\in\F_2^{\edges(\Delta)}$, let $\coloring_x\in\cspace(L(\Delta))$ be the
coloring corresponding to $x$ under the edge--vertex identification. Then
\begin{equation}\label{eq:edge_coordinate_to_label_map}
\edgetovec(x)=\edgetovec_{\Delta,\vgens}(x)
:=\coltovec_{\vgens}(\coloring_x)
=\sum_{\edgea\in\edges(\Delta)}x_{\edgea} v_{\edgea}.
\end{equation}
Thus an edge-coordinate vector first gives a coloring of $L(\Delta)$ and then
its realization in $\SpanS[\F_2]{\vgens}$. When $\Delta$ and $\vgens$ are clear,
we simply write $\edgetovec$.

Moreover, we can discuss the symplectic properties of vectors in $V=\Span{\vgens}$ directly on $\Delta$.
Indeed, while it is trivial that vertex colorings on $\graphG=L(\Delta)$ correspond to edge colorings on $\Delta$, we can further reduce the discussion from (arbitrary) edge colorings on $\Delta$ to \emph{vertex} colorings on $\Delta$:
\begin{lem}\label{lem:symplectic_on_root_multigraphs}
Let $\vgens=\{v_{\edgea}\}_{\edgea\in\edges(\Delta)}\subseteq\Fn$ be a set of binary
vectors with frustration graph $\graphG=L(\Delta)$, and let
$x,y\in\F_2^{\edges(\Delta)}$ be edge-coordinate vectors. Then:
\begin{enumerate}
    \item\label{lem:symplectic_on_root_multigraphs:a}
    $M(\Delta)x$ is the vertex coloring whose colored vertices are incident to an odd number of edges selected by $x$.
    \item\label{lem:symplectic_on_root_multigraphs:b}
    If $x$ is the characteristic vector of the edge set of a path
    $(a_1,\ldots,a_j)$ with edges $\edge{i}=\{a_i,a_{i+1}\}$, then
    $(M(\Delta)x)_a=\delta_{a,a_1}+\delta_{a,a_j}$.
    \item\label{lem:symplectic_on_root_multigraphs:c}
    If $\Delta$ is connected, then
    \[
    \Im(M(\Delta))=\bigg\{z\in\F_2^{\vertices(\Delta)}\text{ for }
    \sum_{a\in\vertices(\Delta)}z_a=0\bigg\},
    \]
    i.e. the vertex-coordinate vectors of even weight.
    \item\label{lem:symplectic_on_root_multigraphs:d}
    If $\Delta$ is a tree, then $M(\Delta)$ gives a bijection from
    edge-coordinate vectors in $\F_2^{\edges(\Delta)}$ to vertex-coordinate vectors of even weight.
    \item\label{lem:symplectic_on_root_multigraphs:e}
    $\symp{\edgetovec(x)}{\edgetovec(y)}=(M(\Delta)x)^T(M(\Delta)y)$, the number of simultaneous colored vertices for $x$ and $y$ in $\Delta$.
\end{enumerate}
\end{lem}
\begin{proof}
The vertex coloring $M(\Delta)x$ on $\Delta$ has components
\begin{equation}
    (M(\Delta)x)_{\vva}
    = \sum_{\edgea\in\edges(\Delta)} M(\Delta)_{\vva,\edgea}x_{\edgea}.
\end{equation}
This is non-zero precisely when the selected edge set has an odd number of
edges incident to $\vva$. This proves (a).

In particular, if $x$ is the characteristic vector of the edge set of a path, then all internal path vertices are incident to two selected edges and the two terminal vertices are incident to one selected edge. This proves (b).

Each column of $M(\Delta)$ has exactly two non-zero entries, so every
vector in $\Im(M(\Delta))$ has even coordinate sum. Conversely, if
$z\in\F_2^{\vertices(\Delta)}$ has even coordinate sum, pair up the vertices in
its support. Since $\Delta$ is connected, choose a path for each pair and let
$x$ be the sum over $\F_2$ of the characteristic vectors of these paths. By (b),
$M(\Delta)x=z$. This proves (c).

If $\Delta$ is a tree, then $\ker(M(\Delta))=0$; hence $M(\Delta)$ is
injective and therefore gives a bijection from edge-coordinate vectors to its
image, which is the even-weight vertex-coordinate space by (c). This proves (d).

For the line-graph convention used here, $A(L(\Delta))=M(\Delta)^TM(\Delta)$ over $\F_2$, so we can write
\begin{equation*}
    \symp{\edgetovec(x)}{\edgetovec(y)}
    = x^TA(L(\Delta))y
    = (M(\Delta)x)^T(M(\Delta)y)
\end{equation*}
which proves (e).
\end{proof}

\subsection{Checking for Line Graphs and the \texorpdfstring{$\calE_6$}{E6} Condition\label{sec:line:e6}}

Given criteria on how to distinguish the various line graph cases, it remains to provide suitable criteria to distinguish line graphs and all other graphs.
Furthermore, we will also state a possible way of constructing the root multigraph of a line graph, thus providing a computational way of computing the necessary invariants.

Recall from Definition~\ref{def:e6} that $\calE_6$ is the set of
six-vertex graphs $t$-equivalent to $\graphE_6$.
Then, we have the following:
\begin{thm}[\cite{Cuypers_2021_E6}]\label{thm:E6_free_characterization_line_graphs}
A connected graph $\graphG$ is a line graph of a multigraph if and only if it is $\calE_6$-free.
\end{thm}

Together with Proposition~\ref{prop:trees:e6}, this gives the following
equivalent ways of recognizing the blown-up path case:
\begin{thm}[Line Graph Conditions]\label{thm:equivalent_line_graph_conditions}
Let $\graphG$ be a connected simple graph with at least two vertices. The following are equivalent:
\begin{enumerate}
    \item $\graphG$ is $t$-equivalent to $\graphP_{k,n_1}$ for some $k,n_1$
    \item $\graphG$ is the line graph of a multigraph $\Delta$ with $k+2$ vertices and $n_1-1$ $\F_2$-independent cycles
    \item $\graphG$ is $\calE_6$-free
    \item $\graphG$ is $t$-equivalent to a graph obtained from a complete graph by adding twins of one vertex, i.e., $\graphK_{\varkappa;\tilde q+1}$ for some $\varkappa\geq 1$ and $\tilde q\geq 0$
    \item The quotient graph $\modout{\graphG}$ is the line graph of a regular graph
\end{enumerate}
\end{thm}
Condition (e) is particularly convenient, as it can reduce the search purely to regular line graphs.
For instance, one can use Beineke's nine forbidden subgraphs \cite{Beineke_1970} or the algorithm in \cite{Roussopoulos_1973}, which runs in time linear in the size of the graph and also returns the root graph (if this exists).
This also provides an explicit construction for the root multigraph.
Namely, given a root graph for $\modout{\graphG}$, it is clear that one only needs to add back the parallel edges corresponding to the twins in $\graphG$ to get a root multigraph for $\graphG$.
Indeed, this is the approach proposed in \cite{Cuypers_2021_Whitney} for computing the root multigraph of some line graph $\graphG$.
Notice that computationally checking for the $\calE_6$ condition may not be the most efficient option to determine whether a graph is a line graph, but it will turn out to be a simple condition for structured generating sets, as we discuss in the applications (e.g. Section~\ref{sec:qaoa}).
Then, we have a computational way of determining when a graph is $t$-equivalent to $\graphP_{k,n_1}$ \emph{and} for which values $k,n_1$ this happens.

Another interesting consequence of Thm~\ref{thm:equivalent_line_graph_conditions}(e) is that, for any generating set $\pgens\subseteq\PP_n$ which has an induced path of length at least $5$ with no algebraic dependencies, $\{v_i\}_{i=1}^5$, then there exist a single generator $v_6$ which makes it \emph{not} a line-graph.
Specifically, this is a solution of the linear equations $\symp{b_6}{v_i} = \delta_{i3}$ which always has a solution for $n\geq 3$ (given no algebraic dependencies, otherwise no solution exists).

Also, given that $\rank(A(\graphE_6))=6$, any graph with $\rank(A(\graphG))\leq 5$ is a line graph of a multigraph.
This also matches Eq.~\eqref{eq:small_rank_cases} with respect to $t$-equivalence of small rank graphs.

We also highlight a correction to \cite[Theorem~3.3]{Cuypers_2021_E6}, which states that $\modout{\graphG}$ is the line graph of a regular graph if and only if it is $t$-equivalent to the path graph $\graphP_n = \graphP_{n-1,1}$.
This is in fact not true.
Indeed, consider the line graph $\graphG$ of a cycle $\Delta=\graphC_n$ ($n\geq 5$), hence $\modout{\graphG} = \graphG$.
It is straightforward to show that $\graphC_n$ is in fact $t$-equivalent to $\graphP_{n-2,2}$, which is the line graph of a multigraph (namely, the path graph with a parallel edge at one end).
It is further clear that this can be generalized to the case of multiple loops in $\Delta$, which implies that one can obtain $\graphP_{k,n_1}$ with arbitrary $n_1$.
The statement still holds however if $\graphG$ has no cycles.

The correct characterization for graphs which are $t$-equivalent to $\graphP_n$ is instead given in \cite{Seven_2005}, which is also equivalent to one given by \cite{humphries_1985} and later corrected by \cite{Huang_2012} (with regards to transvection groups).
We collect equivalent criteria here:
\begin{thm}[Line Graph of Trees Conditions]\label{thm:equivalent_path_graph_conditions}
Let $\graphG$ be a simple graph over $n$ vertices. The following are equivalent:
\begin{enumerate}
    \item $\graphG$ is $t$-equivalent to $\graphP_n$
    \item $\graphG$ is the line graph of a tree with $n+1$ vertices
    \item $\graphG$ is $t$-equivalent to $\graphK_n$
    \item $\graphG$ is a claw-free block graph
    \item $\graphG$ does not contain a claw, diamond or cycles of size larger than $4$ as induces subgraphs
\end{enumerate}
\end{thm}
Hence, we can see that the property of $t$-equivalent to a \emph{path} graph is being the line graph of a \emph{tree}.
Indeed, this is shown implicitly in the proof of Theorem~\ref{thm:t_equivalence_line_graphs_invariants}.
This also bridges the gap regarding the properties of twin-free graphs discussed in Section~\ref{sec:Twins}.
Namely, even though being twin-free is not invariant under $t$-equivalence, being claw, diamond and $\graphC_{n\geq 4}$-free is invariant under $t$-equivalence.
Out of these forbidden subgraphs, twins only appears in claws and diamonds.

We also note that Thm~\ref{thm:equivalent_line_graph_conditions}(e) provides a clear connection between the classification of transvection groups and Lie algebras to the existence of \emph{free-fermionic} mappings, as described in \cite{Chapman_Flammia_2020}.
In \cite{Chapman_Flammia_2020}, it was described how to obtain mappings from sets of Pauli strings to quadratic Majoranas $\{\gamma_i\gamma_j\}$ under the assumption that the frustration graph is a regular line graph.
Hence, we see that the only gap between the existence of a free fermionic mapping and a set being $t$-equivalent to $\graphP_{k,n_1}$, or being a line graph of a multigraph, is purely related to the existence of \emph{twins}, which also briefly mentioned in \cite{Chapman_Flammia_2020} and also expanded upon in \cite{Ruh_Elman_2025}.
In both settings, the mapping is however treated up to symmetries, by performing suitable projectors which result in precisely quadratic Majoranas, which also potentially introduce algebraic dependencies.
Instead, since in our setting we are interested in precisely keeping track of all properties for a classification of the Lie algebra, we shall not perform such a projection.
On the other hand, since we are interested in Lie algebras, we shall be concerned with keeping tracks of phases due to commutation relations and products.
We will address the relation of these mappings to our problem in the next section and suitably generalize the mapping to any set of Pauli strings whose frustration graph is a line graph of a multigraph, independently of symmetries and algebraic dependencies.

\section{Orbit Radical and General Lie-Algebraic Dependencies}\label{sec:Lie_Algebraic_Dependencies_in_general}

In Section~\ref{sec:Line_Graphs_E6_Condition} we have provided criterions to distinguish line graphs from other graphs as well as how to identify the $t$-equivalent classes given the root multigraph.
Furthermore, in Section~\ref{sec:quadratic_bilinear_forms_for_pauli} we have considered criterions to distinguish the non-line graph cases using the invariant quadratic forms.
Hence, these also cover the problem of distinguishing algebraically independent sets under $t$-equivalence.

In this section we tackle two questions: how to determine (Lie-)algebraic dependencies \emph{for any generating set}, beyond the $t$-equivalent representatives (see Proposition~\ref{prop:Limits_Lie_Algebraic_Dependencies_canonical}); how to identify the cases which admit a minimal generating set with an algebraic dependency and distinguish them from the algebraically independent ones.
To do this we will introduce a tool which naturally addressed the problem of Lie-algebraic dependencies, as well as provide relevant properties in the study of the orbits.

\subsection{The Orbit Radical and its Properties}

To support our discussion on algebraic dependencies, we now introduce
the orbit radical.

\begin{defn}[Orbit radical \cite{Janssen_1983,Seven_2005}] \label{def:orbit:radical}
Given a set $\vgens \subset \Fn$ of binary vectors
and the generated transvection group $\tvgroup{\vgens}$ [see Eq.~\eqref{eq:tv:group}], the orbit radical
\begin{equation}\label{eq:orbit:radical}
    \radorbit(\vgens) :=  \rad(\vgens) \cap \{v_1+v_2 \text{ for } v_1,v_2\in\tvgroup{\vgens}\cdot\vgens \}
\end{equation}
is defined as the intersection of
the radical with all sums of two elements from the orbit $\tvgroup{\vgens}\cdot\vgens$.
\end{defn}

We derive several facts
about the orbit radical.

\begin{lem}[\cite{Janssen_1983,Seven_2005}] \label{lem:orbit:radical}
Given a set $\vgens \subset \Fn$ of binary vectors
and the transvection group $\tvgroup{\vgens}$ [see Eq.~\eqref{eq:tv:group}],
the orbit radical $\radorbit(\vgens)$
satisfies the following properties:
\begin{enumerate}
\item\label{lem:orbit:radical:a}
If $u \in \radorbit(\vgens)$, then $g u \in \radorbit(\vgens)$ for $g \in \tvgroup{\vgens}$.
\end{enumerate}
Additionally assuming that the frustration graph $\frustration{\vgens}$ is connected (see Lemma~\ref{lem:pauli:single_orbit_equivalent_to_connectedness}),
we obtain:
\begin{enumerate}[resume]
\item\label{lem:orbit:radical:b}
$\radorbit(\vgens) = \{ u \in \rad(\vgens) \text{ with }  v + u \in \tvgroup{\vgens} \cdot \vgens\}$
for any $v \in \tvgroup{\vgens} \cdot \vgens$.
\item\label{lem:orbit:radical:c}
If $v \in \tvgroup{\vgens} \cdot \vgens $, then $v + \radorbit(\vgens) \subseteq   \tvgroup{\vgens} \cdot \vgens$.
\item\label{lem:orbit:radical:d}
$\radorbit(\vgens)$ is a subspace of $\rad(\vgens)$.
\item\label{lem:orbit:radical:e}
If $\vgens'\subseteq\Fn$ is another set of binary vectors such that
$\tvgroup{\vgens'}\cdot\vgens' = \tvgroup{\vgens}\cdot\vgens$, then
$\radorbit(\vgens') = \radorbit(\vgens)$. In particular,
$\radorbit(\tvgroup{\vgens}\cdot\vgens)=\radorbit(\vgens)$.
\end{enumerate}
\end{lem}

\begin{proof}
In the following, fix $v \in \tvgroup{\vgens}\cdot\vgens$.

For (a), $v_1+v_2=u \in \rad(\vgens)$ for some
$v_1,v_2 \in \tvgroup{\vgens}\cdot\vgens$.
Since $g$ preserves the symplectic form, it also preserves the radical and $gu \in \rad(\vgens)$.
Also,
$gu=gv_1+gv_2$ then implies (a) as
$gv_1,gv_2 \in \tvgroup{\vgens}\cdot\vgens$.

For (b),
assume first that $u \in \radorbit(\vgens)$.
Then $u \in \rad(\vgens)$ and
$u=v_1+v_2$ for some $v_1,v_2 \in \tvgroup{\vgens}\cdot\vgens$.
Since $\frustration{\vgens}$ is connected, Lemma~\ref{lem:pauli:single_orbit_equivalent_to_connectedness}
implies that $\tvgroup{\vgens}\cdot\vgens$ is a single orbit.
Hence there exists $g \in \tvgroup{\vgens}$ such that $gv_1=v$.
By (a), $gu \in \radorbit(\vgens)$.
Moreover, since $u \in \rad(\vgens)$, every generating transvection fixes $u$, and hence $gu=u$.
Therefore
$u=gu=g(v_1+v_2)=v+gv_2$,
and hence
$v+u=gv_2 \in \tvgroup{\vgens}\cdot\vgens$.
This proves that
$\radorbit(\vgens) \subseteq \{ u \in \rad(\vgens) \text{ with } v+u \in \tvgroup{\vgens}\cdot\vgens\}$.
Conversely, let $u \in \rad(\vgens)$ satisfy
$v+u \in \tvgroup{\vgens}\cdot\vgens$.
Since also $v \in \tvgroup{\vgens}\cdot\vgens$, we may write
$u=v+(v+u)$
as a sum of two vectors from $\tvgroup{\vgens}\cdot\vgens$.
This implies
$u \in \radorbit(\vgens)$ and proves (b).

For (c), let $u \in \radorbit(\vgens) \subseteq \rad(\vgens)$.
Thus (b) implies $v+u \in \tvgroup{\vgens}\cdot\vgens$ and (c) holds.

For (d), by definition, $\radorbit(\vgens) \subseteq \rad(\vgens)$.
Also, $0 \in \radorbit(\vgens)$ since $0=v+v$ and $0 \in \rad(\vgens)$.
Let $u,u' \in \radorbit(\vgens)$.
Then $u+u' \in \rad(\vgens)$.
By (b), $v+u \in \tvgroup{\vgens}\cdot\vgens$.
Applying (b) again with $v+u$ in place of $v$, we get
$(v+u)+u' = v+(u+u') \in \tvgroup{\vgens}\cdot\vgens$.
Hence, by (b), $u+u' \in \radorbit(\vgens)$.
Therefore $\radorbit(\vgens)$ is a subspace of $\rad(\vgens)$ and (d) follows.

For (e), set $O:=\tvgroup{\vgens}\cdot\vgens
=\tvgroup{\vgens'}\cdot\vgens'$.
Since $\vgens\subseteq O\subseteq\SpanS[\F_2]{\vgens}$ and
$\vgens'\subseteq O\subseteq\SpanS[\F_2]{\vgens'}$, we have
$\SpanS[\F_2]{\vgens}=\SpanS[\F_2]{O}=\SpanS[\F_2]{\vgens'}$.
By Eq.~\eqref{eq:radical}, and using bilinearity to replace orthogonality to a
generating set by orthogonality to its span, the radical depends only on this
span. Hence $\rad(\vgens)=\rad(\vgens')$.
The definition of the orbit radical now gives
$$
\radorbit(\vgens)
=\rad(\vgens)\cap\{v_1+v_2\mid v_1,v_2\in O\}
=\radorbit(\vgens'),
$$
which proves the first assertion.
For the second assertion, take $\vgens'=O$.
The equality $\SpanS[\F_2]{O}=\SpanS[\F_2]{\vgens}$ gives
$\rad(O)=\rad(\vgens)$ by the same argument as above.
Moreover, $O$ is already the full $\tvgroup{\vgens}$-orbit of $\vgens$, hence
$\tvgroup{O}\cdot O=O$.
Therefore the same definition of the orbit radical yields
$\radorbit(O)=\radorbit(\vgens)$.
\end{proof}
In the Pauli formalism, we can identify a subspace of $\rad(\vgens)$ with a subalgebra of the center $\ZZ(\commalg)$ where $\commalg=\commutant(\pgens)$, $\pgens=\isolong{\vgens}$.
Then, we also write:
\begin{equation}\label{eq:def:ZZ_center_pauli_orbit}
    \ZZorbit(\pgens) = \Span[\C]{\isolong{\radorbit(\vgens)}}.
\end{equation}

We now recall from Eq.~\eqref{eq:isotropic:radical} the isotropic radical
$\radzero(V,\QQ) = \rad(V)\cap\QQ^{-1}(0)$ for a symplectic space
$V$ with a quadratic form $\QQ$. If a set $\vgens$ of binary vectors
is linearly independent, Corollary~\ref{cor:uniqueness_existence_quadratic_form_for_basis}
guarantees the existence of a unique standard quadratic form $\QQstandard{\vgens}$
with $\QQstandard{\vgens}(v)=1$ for $v \in \vgens$ (see Definition~\ref{def:standard:quadratic}).

\begin{defn}
Let $\vgens \subseteq \Fn$ be a linearly independent set of binary vectors
and $\QQstandard{\vgens}$ the associated standard quadratic form.
We define the isotropic radical of $\vgens$ as
\begin{align*}
\radst(\vgens) &:= \radzero(\Span[\F_2]{\vgens},\QQstandard{\vgens}) \\
&\phantom{:}= \rad(\vgens)\cap\QQstandard{\vgens}^{-1}(0).
\end{align*}
\end{defn}

\begin{lem}[\cite{Janssen_1983,Seven_2005}]\label{lem:orbit:radical:elementary}
Let $\vgens \subseteq \Fn$ be a linearly independent set of binary vectors, let
$V=\Span[\F_2]{\vgens}$ denote its span, and let $\frustration{\vgens}$ be its frustration graph.
Let $\QQstandard{\vgens}$ be the
standard quadratic form with $\QQstandard{\vgens}(v)=1$ for all $v\in \vgens$ (see Definition~\ref{def:standard:quadratic}).
Then:
\begin{enumerate}
\item $\radorbit(\vgens) \subseteq \radst(\vgens) \subseteq \rad(\vgens)$.
\end{enumerate}
If the frustration graph is connected, then:
\begin{enumerate}[resume]
\item $\radorbit(\vgens)$ is a subspace of $\radst(\vgens)$.
\end{enumerate}
\end{lem}

\begin{proof}
For (a),
$\radst(\vgens)\subseteq \rad(\vgens)$ follows from the definition.
Now let $u\in \radorbit(\vgens)$. By definition,
$u\in \rad(\vgens)$ and there exist $v_1,v_2\in \tvgroup{\vgens}\cdot\vgens$
with $u=v_1+v_2$. This implies $
0=\symp{u}{v_1}=\symp{v_2}{v_1}$. Corollary~\ref{cor:uniqueness_existence_quadratic_form_for_basis}
shows that  $\QQstandard{\vgens}$ is preserved by every
element $g\in \tvgroup{\vgens}$,
i.e., $\QQstandard{\vgens}(g u) = \QQstandard{\vgens}(u)$ for all $u \in V$.
Therefore
$\QQstandard{\vgens}(v_1)=\QQstandard{\vgens}(v_2)=1$
and
\begin{equation*}
\QQstandard{\vgens}(u)
=
\QQstandard{\vgens}(v_1{+}v_2)
=
\QQstandard{\vgens}(v_1)+\QQstandard{\vgens}(v_2)+\symp{v_1}{v_2} = 0.
\end{equation*}
Thus $u\in \radst(\vgens)$, proving (a).

For (b), Lemma~\ref{lem:orbit:radical}\ref{lem:orbit:radical:d} shows that $\radorbit(\vgens)$ is a subspace of $\rad(\vgens)$.
By (a), we have $\radorbit(\vgens)\subseteq \radst(\vgens)$.
From Lemma~\ref{lem:isotropic:radical}(b) we know that $\radst(\vgens)$ is a subspace of $\rad(\vgens)$ and hence
a vector space.
Thus $\radorbit(\vgens)$
is a subspace of $\radst(\vgens)$, proving (b).
\end{proof}

A critical step in our arguments relies on the invariance
of the various defined radicals under $t$-equivalence:

\begin{lem}\label{lem:invariance_under_t_equivalence}
Given a set $\vgens \subset \Fn$ of binary vectors
that is $t$-equivalent to a set $\vgens'$ of binary vectors, we have:
\begin{enumerate}
\item $\rad(\vgens)=\rad(\vgens')$.
\item $\radorbit(\vgens)=\radorbit(\vgens')$
\end{enumerate}
Also assuming that $\vgens$ is linearly independent,
we have:
\begin{enumerate}[resume]
\item $\radst(\vgens)=\radst(\vgens')$
\end{enumerate}
\end{lem}

\begin{proof}
(a) is Lemma~\ref{lem:properties:preserved:t:equivalence}(b).
Lemma~\ref{lem:properties:preserved:t:equivalence}(d) shows
$\tvgroup{\vgens}\cdot\vgens = \tvgroup{\vgens'}\cdot\vgens'$, which together with the
definition in Eq.~\eqref{eq:orbit:radical} verifies (b).
For (c), consider the standard quadratic form $\QQstandard{\vgens}$ of $\vgens$
with $\QQstandard{\vgens}(\vgens)=1$.
It suffices to verify (c) for a single contraction of $u$ onto $v$ in $\vgens$,
since $t$-equivalence is generated by such operations.
We compute
$\QQstandard{\vgens}(v+u)=\QQstandard{\vgens}(v)+\QQstandard{\vgens}(u)+\symp{v}{u}=1+1+1=1$.
Thus $\QQstandard{\vgens}(w)=1$ for all $w\in \vgens'$.
Since $\radst(\vgens)=\rad(\vgens)\cap \QQstandard{\vgens}^{-1}(0)$ and likewise for $\vgens'$, (a)
implies (c).
\end{proof}

Furthermore, we will also need to understand what happens in the presence of algebraic dependencies:
\begin{lem}\label{lem:orbit_radical_under_projection}
Let $\vgens\subseteq\Fn$ be a set of binary vectors, $V=\Span{\vgens}$ its span, and $\tilde{\vgens}$ the unique linearly independent extension with the same frustration graph and projection map $\projection$.
Then:
\begin{equation}
    \radorbit(\vgens) = \projection(\radorbit(\tilde{\vgens}))
\end{equation}
\end{lem}
\begin{proof}
By Lemma~\ref{lem:projection_extension_basic_properties}, $\projection(\rad(\tilde{\vgens})) = \rad(\vgens)$ and $\projection(\tvgroup{\tilde{\vgens}}\cdot\tilde{\vgens}) = \tvgroup{\vgens}\cdot\vgens$.
Furthermore, since $\projection$ is surjective, any sum of two vectors in $\tvgroup{\vgens}\cdot\vgens$ may be written as a sum the projection of the sum of two vectors in $\tvgroup{\tilde{\vgens}}\cdot\tilde{\vgens}$.
We apply the definition of the orbit radical to verify the statement.
\end{proof}

Depending on the canonical graph,
much more precise information is available:

\begin{lem}[\cite{Janssen_1983,Seven_2005}]\label{lem:orbit:radical:canonical_cases}
Let $\graphG$ be one of the canonical graph families listed in
Theorems~\ref{thm:classes:humphries} or~\ref{thm:classes:eisert}.
Let $\tilde{\vgens}(\graphG)$ be the associated unique linearly independent
extension.
Let $\wgens(\graphG)$ denote either $\tilde{\vgens}(\graphG)$ or, for
$\graphG\in\{\graphX_{2m,n_1}^3,\graphS_{n_2,n_1}^3,\graphP_{2m,n_1}\}$,
the linearly dependent generating set obtained from this linearly independent
extension by a projection
$\projection$ from  $\SpanS[\F_2]{\tilde{\vgens}(\graphG)}$
to $\SpanS[\F_2]{\wgens(\graphG)}$ whose kernel is generated by the unique
non-Lie-algebraic dependency from
Proposition~\ref{prop:Limits_Lie_Algebraic_Dependencies_canonical}.
Writing $b_1,\ldots,b_{n_1}$ for the vector labels of the length-one leaves, we have
$$
\radorbit(\wgens(\graphG))
=
\Span[\F_2]{b_j+b_{j+1}\mid 1\leq j<n_1}.
$$
In particular, $\nullorbit(\wgens(\graphG))=n_1-1$.
\end{lem}

\begin{proof}
Let $\tilde{\vgens}(\graphG)$ be the linearly independent extension and set,
as in Eq.~\eqref{eq:canonical-dependency-B},
\[
B=\Span[\F_2]{b_j+b_{j+1}\text{ for } 1\leq j<n_1}.
\]
We first prove the statement for $\tilde{\vgens}(\graphG)$.
For each $1\leq j<n_1$, Lemma~\ref{lem:Canonical_t_equivalent_Graph_Radical}
shows that $b_j+b_{j+1}\in\rad(\tilde{\vgens}(\graphG))$.
Since $b_j$ and $b_{j+1}$ are themselves generators, $b_j+b_{j+1}$ is a sum of
two elements of the transvection orbit of the generators.
Thus $B\subseteq\radorbit(\tilde{\vgens}(\graphG))$.

It remains to exclude all other possible radical elements.
If $\rad(\tilde{\vgens}(\graphG))=B$, this is immediate. Otherwise
Lemma~\ref{lem:canonical_quadratic_radical_values} identifies the possible
extra radical generator and its value under the standard quadratic form.
By Lemma~\ref{lem:orbit:radical:elementary}(a),
$\radorbit(\tilde{\vgens}(\graphG))\subseteq
\radst(\tilde{\vgens}(\graphG))$. Hence in the cases
$\graphX_{2m,n_1}^3$, $\graphS_{n_2,n_1}^3$, and
$\graphP_{2m,n_1}$ with $m$ even, the extra generator is anisotropic and
$\radst(\tilde{\vgens}(\graphG))=B$. Together with the already proved inclusion
$B\subseteq\radorbit(\tilde{\vgens}(\graphG))$, this gives
$B\subseteq\radorbit(\tilde{\vgens}(\graphG))
\subseteq\radst(\tilde{\vgens}(\graphG))=B$,
and hence $\radorbit(\tilde{\vgens}(\graphG))=B$.

For the remaining linearly independent case
$\graphG=\graphP_{2m,n_1}$ with $m$ odd, Lemma~\ref{lem:Canonical_t_equivalent_Graph_Radical}
leaves one additional radical generator, represented up to $B$ by the
alternating coloring on the spine.
Since $B\subseteq\radorbit(\tilde{\vgens}(\graphG))$ and
$\radorbit(\tilde{\vgens}(\graphG))$ is a subspace of $\rad(\tilde{\vgens}(\graphG))$ by
Lemma~\ref{lem:orbit:radical}\ref{lem:orbit:radical:d}, it suffices to show that this alternating
spine coloring is not contained in the orbit radical up to $B$.
Up to $B$, the length-one legs contribute only with their total parity, so this
reduces to the restricted transvection action on the spine path.
By Lemma~\ref{lem:local_path_orbit_interval}, each orbit element restricts to
an interval on this path.
Hence any sum of two orbit elements restricts to the symmetric difference of
two intervals, which has at most two connected components.
The alternating support $\sum_{j=1}^{m+1} a_{2j-1}$ has $m+1\geq 4$
components, because the even path family $\graphP_{2m,n_1}$ is considered
here only for $m\geq2$, and the present case has $m$ odd, hence $m\geq3$.
Thus it cannot lie in the orbit radical.
Thus also in this case
$\radorbit(\tilde{\vgens}(\graphG))=B$.

Finally consider the dependent cases in the statement.
Let $\projection$ be the projection from the statement.
By Lemma~\ref{lem:orbit_radical_under_projection},
$$
\radorbit(\wgens(\graphG))
=
\projection(\radorbit(\tilde{\vgens}(\graphG)))
=
\projection(B).
$$
By the definition of $\wgens(\graphG)$, the kernel $\ker(\projection)$ is
spanned by the unique non-Lie-algebraic dependency. By
Proposition~\ref{prop:Limits_Lie_Algebraic_Dependencies_canonical}, this
generator is not contained in $B$. Thus
$B\cap\ker(\projection)=\{0\}$, so $\projection$ is injective on $B$.
Hence $\projection(B)$ is spanned by the classes of
$b_j+b_{j+1}$ for $1\leq j<n_1$ and has dimension $n_1-1$, in all linearly dependent cases.
\end{proof}

Combining the canonical orbit-radical computation with the classification
Theorems~\ref{thm:classes:humphries}
and~\ref{thm:classes:eisert} gives the following general characterization.
\begin{cor}[\cite{Janssen_1983,Seven_2005}]\label{cor:orbit:radical:codim1}
Let $\vgens\subseteq\Fn$ be a set of vectors and $V=\Span[\F_2]{\vgens}$.
Assume that the frustration graph $\frustration{\vgens}$ is connected.
Then $\dim[\rad(\vgens)]-\dim[\radorbit(\vgens)] \le 1$.
Equivalently, $\radorbit(\vgens)$ has codimension at most one in $\rad(\vgens)$.

Moreover, if $\vgens$ is linearly independent, exactly one of the following cases holds:
\begin{enumerate}
\item $\radorbit(\vgens) = \radst(\vgens) \neq \rad(\vgens)$,
\item $\radorbit(\vgens) \neq \radst(\vgens) = \rad(\vgens)$,
\item $\radorbit(\vgens) = \radst(\vgens) = \rad(\vgens)$,
\end{enumerate}
If $\vgens$ is not necessarily linearly independent and admits an invariant quadratic form $\QQ$, we also have the equivalent cases with $\radzero(V,\QQ)$ in place of $\radst(\vgens)$.
\begin{enumerate}[resume]
\item $\radorbit(\vgens) = \radzero(V,\QQ) \neq \rad(\vgens)$,
\item $\radorbit(\vgens) \neq \radzero(V,\QQ) = \rad(\vgens)$,
\item $\radorbit(\vgens) = \radzero(V,\QQ) = \rad(\vgens)$.
\end{enumerate}
If $\vgens$ does not admit an invariant quadratic form (hence linearly dependent), we also have the additional case:
\begin{enumerate}[resume]
    \item $\radorbit(\vgens) = \rad(\vgens)$.
\end{enumerate}
\end{cor}
Notice that the cases (d-f) are generalizations of (a-c), though we highlight here the difference since there exist minimal generating sets which have invariant quadratic forms, but are not linearly independent, hence $\radst(\vgens)$ is not well-defined.
\begin{proof}
Theorems~\ref{thm:classes:humphries} and~\ref{thm:classes:eisert},
together with Lemma~\ref{lem:invariance_under_t_equivalence} and
Lemma~\ref{lem:orbit:radical}\ref{lem:orbit:radical:e} allow us to pass to the canonical minimal
generating sets. For these sets,
Lemma~\ref{lem:orbit:radical:canonical_cases} identifies the orbit radical with
the subspace $B$ generated by the pairs of length-one leaves, while
Lemma~\ref{lem:Canonical_t_equivalent_Graph_Radical} and
Lemma~\ref{lem:canonical_quadratic_radical_values} identify the full radical and,
when an invariant quadratic form exists, its isotropic radical. Hence
$\dim[\rad(\vgens)]-\dim[\radorbit(\vgens)]\leq 1$, and the linearly independent
cases are exactly (a)--(c).

If $\vgens$ admits an invariant quadratic form $\QQ$,
Lemma~\ref{lem:quadratic_forms_under_projections_extensions} identifies
$\radzero(V,\QQ)$ with the projection of the isotropic radical of the linearly
independent extension. Thus the same three alternatives hold with
$\radzero(V,\QQ)$ in place of $\radst(\vgens)$.

Finally, if $\vgens$ admits no invariant quadratic
form (hence must be linearly dipendent), the kernel of the projection is generated by the unique anisotropic extra
radical element of the linearly independent extension. Therefore the projection
removes precisely the radical generator outside $B$. By
Lemma~\ref{lem:orbit:radical:canonical_cases},
$\radorbit(\vgens)=\projection(B)$. On the other hand,
Lemma~\ref{lem:projection_extension_basic_properties}\ref{lem:projection_extension_basic_properties:a} gives
$\rad(\vgens)=\projection(\rad(\tilde{\vgens}))$. Since the kernel of
$\projection$ is generated by the unique extra radical element outside $B$,
we have $\projection(\rad(\tilde{\vgens}))=\projection(B)$. Hence
$\radorbit(\vgens)=\rad(\vgens)$.
\end{proof}

For linearly independent sets, $\radst(\vgens)$ is the isotropic radical
$\radzero(V,\QQstandard{\vgens})$. Thus the three alternatives in the first part
of Corollary~\ref{cor:orbit:radical:codim1} have the corresponding
linearly dependent versions with $\radzero(V,\QQ)$ in place of
$\radst(\vgens)$ whenever an invariant quadratic form exists. If no invariant
quadratic form exists, the corollary gives the additional alternative
$\radorbit(\vgens)=\rad(\vgens)$. With respect to these alternatives, the chosen
canonical cases fall into the following cases of Corollary~\ref{cor:orbit:radical:codim1}:
\begin{enumerate}[label=(\arabic*)]
    \item For $\vgens(\graphG)$ and $\graphG\in\{\graphX_{2m-1,n_1}^1,\graphX_{2m-1,n_1}^2,\graphP_{2m-1,n_1}\}$, case (c)
    \item For $\vgens(\graphG)$ and $\graphG\in\{\graphX_{2m,n_1}^3,\graphP_{2m,n_1}\text{ with }m\text{ even}\}$, case (a)
    \item For $\vgens(\graphG)$ and $\graphG=\graphP_{2m,n_1}$ with $m$ odd, case (b)
    \item For $\vgens^D(\graphP_{2m,n_1})$ with $m$ odd, case (e)
    \item For $\vgens^D(\graphG)$ and $\graphG\in\{\graphX_{2m,n_1}^3,\graphP_{2m,n_1}\text{ with }m\text{ even}\}$, case (g)
\end{enumerate}

For the non-line graph canonical graphs covered by
Corollary~\ref{cor:distinguishing_quasi_universal_cases}, the orbit radical is
therefore either the full radical or, when an invariant quadratic form exists,
the isotropic radical. Thus, in these cases, it can be computed directly from
the radical and the invariant quadratic form, without first choosing an explicit
canonical graph. For the line-graph cases, it is more convenient to use
the root multigraph, and we have the following characterization:
\begin{lem}[\cite{Janssen_1983,Seven_2005}]\label{lem:orbit:radical:line_graphs}
Let $\vgens$ be a linearly independent set whose frustration graph is
connected and is the line graph $\graphG=L(\Delta)$. Under the identification
of $\vgens$ with the edge set of $\Delta$, if $n_\Delta\neq4$, then
$\radorbit(\vgens)=\edgetovec(\ker(M(\Delta)))$.
\end{lem}
\begin{proof}
Let $\tilde q=\dim\ker(M(\Delta))$. By
Theorem~\ref{thm:t_equivalence_line_graphs_invariants}, the line graph
$L(\Delta)$ is $t$-equivalent to $\graphP_{n_\Delta-2,\tilde q+1}$. Under a valid
contraction, Corollary~\ref{cor:t_equivalence_contractions_on_multigraphs}
maps the root incidence matrix $M:=M(\Delta)$ to
$M'=MP_{\sourceidx,\targetidx}$, and hence
\[
\ker(M')=\ker(MP_{\sourceidx,\targetidx})
=P_{\sourceidx,\targetidx}\ker(M).
\]
Indeed, $P_{\sourceidx,\targetidx}^{-1}=P_{\sourceidx,\targetidx}$, and
$x\in\ker(MP_{\sourceidx,\targetidx})$ iff
$P_{\sourceidx,\targetidx}x\in\ker(M)$ iff
$x\in P_{\sourceidx,\targetidx}\ker(M)$.
Thus the cycle space of the root multigraph is transported by the same
basis change as the corresponding edge-coordinate vectors.

It therefore suffices to prove the claim for the canonical graph
$\graphP_{n_\Delta-2,\tilde q+1}$. Its root multigraph is a path on $n_\Delta$
vertices with $\tilde q+1$ parallel edges at one end. The kernel of its incidence
matrix is spanned by the $\tilde q$ differences of consecutive parallel edges. The
images under $\edgetovec$ of these differences are precisely the vectors
$b_j+b_{j+1}$ for consecutive length-one leaves.
By Lemma~\ref{lem:orbit:radical:canonical_cases}, these span
$\radorbit(\vgens)$. Hence
$\radorbit(\vgens)=\edgetovec(\ker(M(\Delta)))$.
\end{proof}
The preceding proof follows the canonical representatives. In
Subsection~\ref{sec:classification_free_orbit_radicals}, we give an independent
classification-free proof of Lemma~\ref{lem:orbit:radical:line_graphs}.
This also readily generalizes to the algebraically dependent case by performing a projection from its algebraically independent extension.
Finally, we also have a full characterization of the line graph cases, which complements Corollary~\ref{cor:distinguishing_quasi_universal_cases} for the non-line graph cases:
\begin{cor}\label{cor:distinguishing_line_graph_cases}
Let $\vgens\subseteq\Fn$ be a set of binary vectors with connected frustration graph which is the line graph of a multigraph $\graphG = L(\Delta)$.
Also, let $n_\Delta = \abs{\vertices(\Delta)}\neq 4$,
$\tilde{q}=\nullity(M(\Delta))$ and $r=\nullity(\vgens)$. Let
$\tilde{\vgens}$ be the unique linearly independent extension with projection
$\projection$,
$q = \nullorbit(\vgens)$ and
$\tilde{r} = \nullity(\tilde{\vgens})$. Then:
\begin{enumerate}
    \item If $r-q = \tilde{r}-\tilde{q}$, $\vgens$ has a minimal generating subset which is $t$-equivalent to $\vgens(\graphP_{n_\Delta-2,q+1})$.
    \item If $r-q = \tilde{r}-\tilde{q} - 1$, $n_\Delta$ is even and $\vgens$ has a minimal generating subset which is $t$-equivalent to $\vgens^D(\graphP_{n_\Delta-2,q+1})$.
\end{enumerate}
\end{cor}
\begin{proof}
Let $\tilde{\vgens}$ be the unique linearly independent extension of $\vgens$
with frustration graph $L(\Delta)$. Set $M:=M(\Delta)$ and
$\tilde{\edgetovec}:=\edgetovec_{\Delta,\tilde{\vgens}}$.
We use Corollary~\ref{cor:orbit:radical:codim1} and
Lemma~\ref{lem:orbit:radical:line_graphs} to separate the realized cycle-space
part of the radical from the possible additional radical vectors.
By Theorem~\ref{thm:t_equivalence_line_graphs_invariants}, the linearly
independent extension $\tilde{\vgens}$ is $t$-equivalent to a set with
frustration graph $\graphP_{n_\Delta-2,\tilde{q}+1}$.
By Lemma~\ref{lem:radical_via_cycles_and_Tjoin}, the realized radical of
the linearly independent extension is
\begin{align*}
\rad(\tilde{\vgens})
=
\begin{cases}
\tilde{\edgetovec}(\ker(M)), & n_\Delta \text{ odd},\\
\tilde{\edgetovec}(\ker(M))+ \SpanS[\F_2]{\{\tilde{\edgetovec}(x)\}}, & n_\Delta \text{ even},
\end{cases}
\end{align*}
where, in the even case, $x\in\F_2^{\edges(\Delta)}$ satisfies
$Mx=\mathbf 1$.
Since $\tilde q=\dim\ker(M)$ and
$\tilde r=\dim\rad(\tilde{\vgens})$, this gives
$\tilde r=\tilde q$ for odd $n_\Delta$ and
$\tilde r=\tilde q+1$ for even $n_\Delta$.
After projection, the surviving realized cycle space has dimension
$q=\dim(\projection(\tilde{\edgetovec}(\ker(M))))$.
In particular, by Lemma~\ref{lem:orbit:radical:line_graphs} and Lemma~\ref{lem:orbit_radical_under_projection}, this coincides with the orbit radical of $\vgens$, $\radorbit(\vgens) = \projection(\tilde{\edgetovec}(\ker(M)))$.
For a blown-up path
$\graphP_{k,n_1}$, the root multigraph has $n_1$ parallel edges at one end,
hence $n_1-1$ independent cycles. Therefore the surviving cycle dimension $q$
corresponds to $n_1=q+1$, and the minimal line-graph case has frustration graph
$\graphP_{n_\Delta-2,q+1}$.
If the minimal generating set is linearly independent, then $\vgens=\tilde{\vgens}$ and hence
$r=\tilde r$. Since also $q=\tilde q$ in this case, the preceding computation
gives $r=q$ for odd $n_\Delta$ and $r=q+1$ for even $n_\Delta$.

If a minimal generating set has a linear dependency, then the kernel of the
projection cannot contain a vector in the realized cycle space
$\tilde{\edgetovec}(\ker(M))$, since such vectors are Lie-algebraic dependencies by
Proposition~\ref{prop:Limits_Lie_Algebraic_Dependencies_canonical}. Thus the
only possible non-Lie-algebraic dependency is the extra realized T-join
vector $\tilde{\edgetovec}(x)$ from Lemma~\ref{lem:radical_via_cycles_and_Tjoin}, which exists
only when $n_\Delta$ is even. Projecting out this vector removes the
one-dimensional summand outside the realized cycle space, so the remaining radical has
dimension $r=q$.
Therefore, for a minimal generating set, the equality
$r-q=\tilde r-\tilde q$ characterizes the linearly independent case. Conversely,
if $r-q=\tilde r-\tilde q-1$, then a linear dependency must have been imposed;
by the preceding paragraph this can only remove $\tilde{\edgetovec}(x)$, so this case
can occur only when $n_\Delta$ is even.
This proves the theorem.
\end{proof}
In general, when the frustration graph is a line-graph over an even number of vertices, we say that a non-zero vector (or corresponding Pauli) in $\rad(\vgens)/\radorbit(\vgens)$ is a \emph{non-trivial} T-join symmetry for $\vgens$, and that a T-join is \emph{trivial} if $\rad(\vgens) = \radorbit(\vgens)$.
Given an edge coloring $x\in\F_2^{\edges(\Delta)}$, a T-join $\tilde{\edgetovec}(x)$ and $Mx = \mathbf{1}$ over the unique linearly independent extension, is a trivial T-join symmetry for $\vgens$ if and only if under projection we have
\begin{equation}
    \projection(\tilde{\edgetovec}(x)) \in\radorbit(\vgens).
\end{equation}
Of course, if $\projection(\tilde{\edgetovec}(x))$ is in $\radorbit(\vgens)$, so is $\projection(\tilde{\edgetovec}(x+\tilde{u}))$ for any cycle coloring $\tilde{u}\in\ker(M)$, i.e., a T-join is defined up to cycles. 
Since $\projection$ is surjective on the orbit radical, this means that we can choose $u$ such that $\projection(\tilde{\edgetovec}(x))+u = 0 \in\radorbit(\vgens)$, or equivalently $x+\tilde{u}$ is a T-join which leads to an algebraic dependency for $\vgens$.
Hence, without loss of generality, we say that a generating set $\vgens$ possesses a trivial T-join symmetry if it leads to an algebraic dependency (possibly up to cycles).
Hence, we have that a generating set whose frustration graph is a line-graph over $n_\Delta=2m+2$ vertices has a minimal generating set which is $t$-equivalent to $\vgens(\graphP_{2m,q+1})$ iff it has a non-trivial T-join symmetry, and it is $t$-equivalent to $\vgens^D(\graphP_{2m,q+1})$ iff any T-join over its root graph is trivial.
Also, we write the dimension of the orbit radical for $\pgens=\isolong{\vgens}$ as
\begin{equation}\label{eq:nullity_orbit_radical_paulis}
    \nullorbit(\pgens)=\nullorbit(\vgens):=\dim\radorbit(\vgens).
\end{equation}
This discussion and Corollary~\ref{cor:distinguishing_line_graph_cases}, motivates the following definition for generating sets whose frustration graphs are line graphs:
\begin{defn}\label{def:free-fermionic-types}
Let $\pgens=\isolong{\vgens}\subseteq\PP_n$ be a set of Pauli strings with connected frustration graph which is the line graph of a multigraph $\graphG = L(\Delta)$.
Also, let $n_\Delta = \abs{\vertices(\Delta)}\neq 4$, $\tilde{q}=\nullity(M(\Delta))$ and $r=\nullity(\vgens)$. Let $\tilde{\vgens}$ be the unique linearly independent extension with projection $\projection$, $q = \nullorbit(\vgens) = \dim(\projection(\edgetovec_{\Delta,\tilde{\vgens}}(\ker(M(\Delta)))))$ and $\tilde{r} = \nullity(\tilde{\vgens})$. Then:
\begin{enumerate}
    \item If $r-q = \tilde{r}-\tilde{q}=0$, we say that $\pgens$ ($\vgens$) is of \emph{odd} type
    \item If $r-q = \tilde{r}-\tilde{q}=1$, we say that $\pgens$ ($\vgens$) is of \emph{even} type
    \item If $r-q = \tilde{r}-\tilde{q} - 1=0$, we say that $\pgens$ ($\vgens$) is of \emph{exceptional} type
\end{enumerate}
\end{defn}
By Corollary~\ref{cor:distinguishing_line_graph_cases}, $\pgens$ is of a certain type if and only if it admits a minimal generating (sub)set which is $t$-equivalent to a canonical case of the same type.
Moreover: (a) $\vgens$ is of odd type iff $\Delta$ does not admit a T-join; (b) it is of even type iff $\Delta$ has a T-join $x$, $Mx = \mathbf 1$, with $\projection(\tilde \edgetovec(x))\not\in\radorbit(\vgens)$; (c) it is of exceptional type iff $\Delta$ has a T-join $x$, $Mx = \mathbf 1$, with $\projection(\tilde \edgetovec(x))\in\radorbit(\vgens)$.

\subsection{Classification-Free Proof of Lemma~\ref{lem:orbit:radical:line_graphs}\label{sec:classification_free_orbit_radicals}}

We aim at a classification-free proof of Lemma~\ref{lem:orbit:radical:line_graphs}.
To this
end, we state additional results for line graphs. When
the frustration graph $\frustration{\vgens}=L(\Delta)$ is a line graph of a
multigraph $\Delta$, each edge $\edgea\in\edges(\Delta)$ corresponds to a vertex of
$L(\Delta)$, and this vertex has a vector label from $\vgens$. We denote this
label by $v_{\edgea}$.
We write $\edgeunit{\edgea}\in\F_2^{\edges(\Delta)}$ for the edge-coordinate
vector supported on $\edgea$, and
$\mathbf{1}_{\vva}\in\F_2^{\vertices(\Delta)}$ for the vertex coloring
supported on the single vertex $\vva$.

\begin{lem}\label{lem:root_path_vectors_are_orbit_vectors}
Let $\vgens=\{v_1,\ldots,v_s\}\subseteq\Fn$ be a linearly independent set
whose connected frustration graph is the line graph $L(\Delta)$ of a
multigraph $\Delta$. For any path $\pathvar=(\vvc_1,\ldots, \vvc_j)$ in $\Delta$ with
$j\geq 2$, pairwise distinct vertices $\vvc_i$, and
edges $\edge{i}$ joining $\vvc_i$ and $\vvc_{i+1}$ for $1\leq i<j$, let
$x_{\pathvar}\in\F_2^{\edges(\Delta)}$ be the characteristic vector of
$\{\edge{1},\ldots,\edge{j-1}\}$. The path vector
\[
v_{\pathvar}:=\edgetovec(x_{\pathvar})=\sum_{i=1}^{j-1} v_{\edge{i}}
\]
belongs to the transvection orbit $\tvgroup{\vgens}\cdot\vgens$.
\end{lem}
\begin{proof}
For $1\leq i<j$, let $x_i$ be the characteristic vector of the path
$\{\edge{1},\ldots,\edge{i}\} \subseteq\{\edge{1},\ldots, \edge{j-1}\}$ and set
$w_i:=\edgetovec(x_i)=\sum_{o=1}^i v_{\edge{o}}$ as the corresponding path vector.
We prove by induction on $i$ that $w_i\in\tvgroup{\vgens}\cdot\vgens$.
The case $i=1$ is immediate, since $w_1=v_{\edge{1}}\in\vgens$.
Assume $w_i\in\tvgroup{\vgens}\cdot\vgens$ for some $i<j-1$.
By Lemma~\ref{lem:symplectic_on_root_multigraphs}\ref{lem:symplectic_on_root_multigraphs:b},
\[
M(\Delta)x_i=\mathbf{1}_{\vvc_1}+\mathbf{1}_{\vvc_{i+1}},
\quad
M(\Delta)\edgeunit{\edge{i+1}}
=\mathbf{1}_{\vvc_{i+1}}+\mathbf{1}_{\vvc_{i+2}}.
\]
Thus Lemma~\ref{lem:symplectic_on_root_multigraphs}\ref{lem:symplectic_on_root_multigraphs:e} gives
\[
\symp{w_i}{v_{\edge{i+1}}}
=\big(M(\Delta)x_i\big)^T\big(M(\Delta)\edgeunit{\edge{i+1}}\big)
=1.
\]
Hence the generating transvection $\tau_{v_{\edge{i+1}}}\in\tvgroup{\vgens}$
satisfies
\[
\tau_{v_{\edge{i+1}}}(w_i)=w_i+v_{\edge{i+1}}=w_{i+1}.
\]
Therefore $w_{i+1}\in\tvgroup{\vgens}\cdot\vgens$, completing the induction.
\end{proof}

The path vectors from Lemma~\ref{lem:root_path_vectors_are_orbit_vectors} allow
us to place the cycle space of the root multigraph inside the orbit radical.
For the following lemma, we use the edge-coordinate realization map
$\edgetovec_{\Delta,\vgens}$ from
Eq.~\eqref{eq:edge_coordinate_to_label_map} to regard the cycle space
$\ker(M(\Delta))$ as the subspace
$\edgetovec_{\Delta,\vgens}(\ker(M(\Delta)))\subseteq\SpanS[\F_2]{\vgens}$.
When $\Delta$ and $\vgens$ are clear, we write this map as $\edgetovec$.

\begin{lem}\label{lem:cycle_space_inside_orbit_radical}
Let $\vgens=\{v_1,\ldots,v_s\}\subseteq\Fn$ be a linearly independent set
whose connected frustration graph is the line graph $L(\Delta)$ of a
multigraph $\Delta$ without loops.
Then
\begin{enumerate}
    \item $\edgetovec(\ker(M(\Delta)))\subseteq\rad(\vgens)$.
    \item $\edgetovec(\ker(M(\Delta)))\subseteq\radorbit(\vgens)$.
\end{enumerate}
\end{lem}
\begin{proof}
We first prove (a). Let $x\in\ker(M(\Delta))$, and let
$\edgea\in\edges(\Delta)$. Since $v_{\edgea}=\edgetovec(\edgeunit{\edgea})$,
Lemma~\ref{lem:symplectic_on_root_multigraphs}\ref{lem:symplectic_on_root_multigraphs:e} gives
\[
\symp{\edgetovec(x)}{v_{\edgea}}
=\big(M(\Delta)x\big)^T\big(M(\Delta)\edgeunit{\edgea}\big)=0.
\]
Thus $\edgetovec(x)$ pairs trivially with every generator $v_{\edgea}$, and
therefore $\edgetovec(x)\in\rad(\vgens)$.

We now prove (b).
By Lemma~\ref{lem:cycle_space_incidence_matrix_multigraph}, every
$x\in\ker(M(\Delta))$ is an $\F_2$-sum of characteristic vectors of edge sets
of cycles.
Since $\radorbit(\vgens)$ is a subspace of $\SpanS[\F_2]{\vgens}$ by
Lemma~\ref{lem:orbit:radical}\ref{lem:orbit:radical:d}, it is enough to prove that, for every cycle
$C$ in $\Delta$, the realized cycle vector $\edgetovec(x_C)$ belongs to
$\radorbit(\vgens)$, where $x_C$ is the characteristic vector of $\edges(C)$.

If the cycle consists of two parallel edges $\edgea$ and $\edgeb$, then
$\edgetovec(x_C)=v_{\edgea}+v_{\edgeb}$. Since $x_C\in\ker(M(\Delta))$ by
Lemma~\ref{lem:cycle_space_incidence_matrix_multigraph}, this vector lies in
$\rad(\vgens)$ by (a). It is also a sum of two elements of
$\tvgroup{\vgens}\cdot\vgens$, since
$v_{\edgea},v_{\edgeb}\in\vgens$. Thus
$v_{\edgea}+v_{\edgeb}\in\radorbit(\vgens)$.

Otherwise, choose two distinct vertices on $C$. They split $C$ into two
paths $\pathvar_1$ and $\pathvar_2$ with pairwise distinct vertices in each path. Let
$x_{\pathvar_i}$ be the characteristic vector of $\edges(\pathvar_i)$ for $i\in\{1,2\}$.
Lemma~\ref{lem:root_path_vectors_are_orbit_vectors} gives
$\edgetovec(x_{\pathvar_1}),\edgetovec(x_{\pathvar_2})\in
\tvgroup{\vgens}\cdot\vgens$. Since the edge sets of $\pathvar_1$ and $\pathvar_2$ are
disjoint and their union is $\edges(C)$,
\[
\edgetovec(x_{\pathvar_1})+\edgetovec(x_{\pathvar_2})=\edgetovec(x_C).
\]
The vector $\edgetovec(x_C)$ lies in $\rad(\vgens)$ by (a), because
$x_C\in\ker(M(\Delta))$ by
Lemma~\ref{lem:cycle_space_incidence_matrix_multigraph}. Since the displayed
formula writes $\edgetovec(x_C)$ as a sum of two orbit vectors, it follows that
$\edgetovec(x_C)\in\radorbit(\vgens)$.
\end{proof}

The next missing ingredient is the converse control on orbit vectors: in a
line graph, vectors in a transvection orbit are path-like when viewed in the
root multigraph.

\begin{lem}\label{lem:line_graph_orbit_boundary}
Let $\vgens=\{v_1,\ldots,v_s\}\subseteq\Fn$ be a linearly independent set
whose connected frustration graph is the line graph $L(\Delta)$ of a
multigraph $\Delta$ without loops.  For
$u\in\tvgroup{\vgens}\cdot\vgens$, let
$x_u\in\F_2^{\edges(\Delta)}$ be the unique edge-coordinate vector with
$u=\edgetovec(x_u)$.
Then:
\begin{enumerate}[label=(\alph*)]
    \item There are vertices $\vva,\vvb\in\vertices(\Delta)$ such that
    \begin{equation}\label{eq:line_graph_orbit_boundary}
    M(\Delta)x_u=\mathbf{1}_{\vva}+\mathbf{1}_{\vvb}.
    \end{equation}
    Equivalently, the edge set encoded by $x_u$ has precisely the
    odd-incidence vertices $\vva$ and $\vvb$, with all other vertices with even incidence.
    \item If $\pathvar$ is any path in $\Delta$ from $\vva$ to $\vvb$, with edge-set
    characteristic vector $x_{\pathvar}$, then
    \begin{equation}\label{eq:line_graph_orbit_path_difference}
    x_u+x_{\pathvar}\in\ker(M(\Delta)).
    \end{equation}
    Equivalently, every orbit vector is represented modulo the cycle space
    $\ker(M(\Delta))$ by a path.
\end{enumerate}
\end{lem}
Applying $\edgetovec$ to Eq.~\eqref{eq:line_graph_orbit_path_difference},
Lemma~\ref{lem:cycle_space_inside_orbit_radical} implies that every orbit
vector $u\in\tvgroup{\vgens}\cdot\vgens$ differs from $\edgetovec(x_{\pathvar})$ for
some path $\pathvar$ in $\Delta$ by an element of $\radorbit(\vgens)$.  By
Lemma~\ref{lem:root_path_vectors_are_orbit_vectors}, the vector
$\edgetovec(x_{\pathvar})$ is itself an orbit vector.
\begin{proof}
Set $M:=M(\Delta)$.  By Lemma~\ref{lem:symplectic_on_root_multigraphs}\ref{lem:symplectic_on_root_multigraphs:a},
$Mx$ records the root vertices incident to an odd number of selected edges.

We first prove (a).  The property $\abs{\supp(Mx_u)}=2$ holds for every
generator $v_{\edgea}$, because $M\edgeunit{\edgea}$ is the characteristic
vector of the two endpoints of $\edgea$.  It remains to check that this
property is preserved by the generating transvections $\tau_{v_{\edgeb}}$ for
$\edgeb\in\edges(\Delta)$.
Let $\edgeb=\{\vva,\vvb\}$ be an edge of $\Delta$, and suppose
$\abs{\supp(Mx_u)}=2$.  By
Lemma~\ref{lem:symplectic_on_root_multigraphs}\ref{lem:symplectic_on_root_multigraphs:e},
\[
\symp{u}{v_{\edgeb}}=(Mx_u)^T(M\edgeunit{\edgeb}).
\]
If $\symp{u}{v_{\edgeb}}=0$, then $\tau_{v_{\edgeb}}(u)=u$, so the
edge-coordinate vector and its image under $M$ are unchanged.
Thus $\symp{u}{v_{\edgeb}}=1$ precisely when exactly one endpoint of $\edgeb$ lies in
$\supp(Mx_u)$.  Then
\[
\tau_{v_{\edgeb}}(u)=u+v_{\edgeb}=\edgetovec(x_u+\edgeunit{\edgeb}),
\]
where $\edgeunit{\edgeb}\in\F_2^{\edges(\Delta)}$ is the edge-coordinate vector
supported on the edge $\edgeb$.  Adding $\edgeunit{\edgeb}$ toggles the two
endpoint vertices of $\edgeb$ in $Mx_u$.  Since exactly one of these endpoint
vertices lies in
$\supp(Mx_u)$, the updated vector $M(x_u+\edgeunit{\edgeb})$ again has support of
size two.  Hence
the property is invariant under all generating
transvections, and thus holds for every vector in
$\tvgroup{\vgens}\cdot\vgens$, proving (a).

For (b), let $\pathvar$ be any path from the two vertices $\vva,\vvb$ in (a).
Lemma~\ref{lem:symplectic_on_root_multigraphs}\ref{lem:symplectic_on_root_multigraphs:b} says that the only
odd-incidence vertices of the edge set of $\pathvar$ are its two endpoints, so
\[
Mx_{\pathvar}=\mathbf{1}_{\vva}+\mathbf{1}_{\vvb}=Mx_u,
\]
where the last equality is Eq.~\eqref{eq:line_graph_orbit_boundary}.  Therefore
$M(x_u+x_{\pathvar})=0$, which is precisely
$x_u+x_{\pathvar}\in\ker(M)$.
\end{proof}

We can now combine the preceding lemmas to give a classification-free
proof of Lemma~\ref{lem:orbit:radical:line_graphs}.

\begin{proof}[Classification-free proof of Lemma~\ref{lem:orbit:radical:line_graphs}]
In this proof, set  $M:=M(\Delta)$. We apply
Lemma~\ref{lem:cycle_space_inside_orbit_radical}(b) to obtain the inclusion
$\edgetovec(\ker(M))\subseteq\radorbit(\vgens)$.
For the reverse inclusion, let $u\in\radorbit(\vgens)$ and write
$u=\edgetovec(x)$ with $x\in\F_2^{\edges(\Delta)}$.  This $x$ is unique because
$\vgens$ is linearly independent.  Since $u\in\rad(\vgens)$, it follows that
$x\in\ker(A(L(\Delta)))$.

If $n_\Delta$ is odd, Lemma~\ref{lem:radical_via_cycles_and_Tjoin} gives
$x\in\ker(M)$, and we are done.  Assume now that $n_\Delta$ is even.  If
$n_\Delta=2$, then connectedness of $L(\Delta)$ forces $L(\Delta)$ to have a
single vertex; hence $\radorbit(\vgens)=0=\edgetovec(\ker(M))$.  We may
therefore assume $n_\Delta\geq 6$, since the lemma excludes $n_\Delta=4$.

Suppose, for contradiction, that $x\notin\ker(M)$.  By
Lemma~\ref{lem:radical_via_cycles_and_Tjoin}, this implies
$Mx=\mathbf{1}$, the all-one vertex coloring on $\Delta$.
On the other hand, by the definition of $\radorbit(\vgens)$ there are
$y,z\in\tvgroup{\vgens}\cdot\vgens$ such that $u=y+z$.  Let
$x_y,x_z\in\F_2^{\edges(\Delta)}$ be their edge-coordinate vectors.  By
injectivity of $\edgetovec$, $x=x_y+x_z$.  By
Lemma~\ref{lem:line_graph_orbit_boundary}(a), each of $Mx_y$ and $Mx_z$ is
supported on exactly two root vertices.  Hence
\[
Mx=Mx_y+Mx_z
\]
has support of size at most four.  This contradicts $Mx=\mathbf{1}$, whose
support has size $n_\Delta\geq 6$.  Thus $x\in\ker(M)$, and therefore
$u=\edgetovec(x)\in\edgetovec(\ker(M))$.
\end{proof}

\subsection{Lie-Algebraic Dependencies for General Graphs}

Finally, given the tools describe in Sections~\ref{sec:quadratic_bilinear_forms_for_pauli} and \ref{sec:Line_Graphs_E6_Condition}, we are able to completely describe Lie-algebraic dependencies in a graph-independent way.

\begin{lem}\label{lem:Limits_Lie_Algebraic_Dependencies_general}
Let $\vgens$ be a generating set whose frustration graph
$\frustration{\vgens}$ belongs to one of the canonical graph families listed in
Theorems~\ref{thm:classes:humphries} or~\ref{thm:classes:eisert}.
Let $\tilde{\vgens}$ be its unique linearly independent extension with
projection map $\projection$ and set
$\tilde V=\SpanS[\F_2]{\tilde{\vgens}}$.
Given the correspondence between algebraic-dependency colorings
$\coloring$ of $\frustration{\vgens}$ and their realizations
$\tilde{v}\in\ker(\projection)\subseteq\tilde V$, we have:
\begin{enumerate}
    \item If $\tilde{v}\in \ker(\projection)\cap\radorbit(\tilde{\vgens})$, it also defines a Lie-algebraic dependency.
    \item If $\tilde{v}\in \ker(\projection)\setminus\radorbit(\tilde{\vgens})$, it does not define a Lie-algebraic dependency.
\end{enumerate}
In particular, if $\ker(\projection)\cap\radorbit(\tilde{\vgens}) = \{0\}$, the set is minimal.
\end{lem}
\begin{proof}
By Proposition~\ref{prop:Limits_Lie_Algebraic_Dependencies_canonical},
the Lie-algebraic dependencies for the listed canonical graph families are
precisely the algebraic dependencies whose extension vectors lie in the subspace
$B$ generated by the length-one leaf pairs.  By
Lemma~\ref{lem:orbit:radical:canonical_cases}, this subspace is
$\radorbit(\tilde{\vgens})$.  Hence the Lie-algebraic dependencies are exactly
those algebraic dependencies whose extension vectors lie in
$\ker(\projection)\cap\radorbit(\tilde{\vgens})$.
\end{proof}
Using Theorems~\ref{thm:classes:humphries}
and~\ref{thm:classes:eisert} and the invariance under $t$-equivalence, this
gives the corresponding criterion for arbitrary connected frustration graphs.
\begin{cor}\label{cor:Limits_Lie_Algebraic_Dependencies_general_classification}
Let $\vgens$ be a generating set with connected frustration graph, and let
$\tilde{\vgens}$ be its unique linearly independent extension with projection
map $\projection$. Then:
\begin{enumerate}
    \item If $\tilde{v}\in \ker(\projection)\cap\radorbit(\tilde{\vgens})$, then the corresponding algebraic dependency is Lie-algebraic.
    \item If $\tilde{v}\in \ker(\projection)\setminus\radorbit(\tilde{\vgens})$, then the corresponding algebraic dependency is not Lie-algebraic.
\end{enumerate}
In particular, if $\ker(\projection)\cap\radorbit(\tilde{\vgens})=\{0\}$,
then the corresponding generating set is minimal. Furthermore, a minimal
generating set has at most one algebraic dependency.
\end{cor}
\begin{proof}
By Theorems~\ref{thm:classes:humphries}
and~\ref{thm:classes:eisert}, the connected case can be represented by one of
the listed canonical graph families.  By
Lemma~\ref{lem:Contractions_conserve_Lie_algebraic_dependencies},
(Lie-)algebraic dependencies are conserved under $t$-equivalence, and by
Lemma~\ref{lem:invariance_under_t_equivalence}, the orbit radical is invariant
under $t$-equivalence.  The claim follows from
Lemma~\ref{lem:Limits_Lie_Algebraic_Dependencies_general}.
\end{proof}
Then, we can also compute all Lie-algebraic dependencies for both line graph (from the kernel of the incidence matrix) and non-line graph cases (from the isotropic radical of the extension).

We finally have all the necessary tools to distinguish the possible cases, in a way which avoid $t$-equivalence and minimality assumptions.
We shall discuss this in detail in Section~\ref{sec:classification:groups_lie_algebras}.

\ManuscriptPart{Majoranas and Free Fermions}{part:majorana_free_fermionic_structures}

Generating sets for which the frustration graph is a line graph are closely related to Majorana strings and free fermions \cite{Chapman_Flammia_2020,Ruh_Elman_2025}. In the following we explain this connection in detail, expand the free-fermionic mappings of \cite{Chapman_Flammia_2020} to faithfully represent symmetries and algebraic dependencies, and characterize the Lie algebras and transvection groups arising from such free-fermionic generating sets.

\section{Majorana Strings}\label{sec:Majorana_strings_definitions}

Let us now introduce the Majorana operators as an alternative generating set or basis for $\Fn$.
We refer to \cite{Bettaque_Swingle_2025} for details of the Majorana group, its even subgroup and their relation to the Pauli group.
We say that a set of Pauli strings $\{\gamma_i\}_{i=1}^{2n}\subseteq\PP_n$ realizes a set of \emph{Majorana operators} if they all mutually anti-commute, hence they generate, inside $\mat{2^n}{\C}$, the standard faithful representation of the complex Clifford algebra \cite{Bourbaki1959,Garling_2011}:
\begin{equation}\label{eq:def:majorana:operators}
    \acomm{\gamma_i}{\gamma_j} = 2\delta_{ij},\quad \text{for } i,j\in[2n].
\end{equation}
Equivalently, a set of Majorana operators is any set of $2n$ Pauli strings whose frustration graph is the complete graph $\graphK_{2n}$.
Notice that any such set must in fact be algebraically independent.
Indeed, by the coloring realization map from Eq.~\eqref{eq:coloring}, an
algebraic dependency would give a non-zero coloring whose colored vertices have
even adjacency to every vertex, but $\graphK_{2n}$ has no such coloring.
Hence, since the corresponding binary vectors form a basis of $\Fn$, the
operators $\{\gamma_i\}_{i=1}^{2n}$ generate all Pauli strings $\PP_n$ modulo
phases, and therefore generate the full matrix algebra $\matalg(2^n,\C)$.
In particular, this set is also unique up to isomorphism thanks to Lemma~\ref{lem:Bijection_Graphs_Linearly_Independent_Sets}, which justifies calling any such set \emph{Majorana operators}.
Using the Pauli-to-binary map $\inviso=\isoempty^{-1}$ from
Eq.~\eqref{eq:pauli:binary}, we denote the binary symplectic vector
corresponding to $\gamma_i$ by
\begin{equation}\label{eq:def:mu}
    \mu_i := \inviso(\gamma_i).
\end{equation}
We refer to $\{\mu_i\}_{i=1}^{2n}$ as a \emph{Majorana basis} of $\Fn$, which satisfies
$\frustration{\{\mu_i\}} = \graphK_{2n}$, or, equivalently,
\begin{equation}\label{eq:def:majorana_basis_pairing}
    \symp{\mu_i}{\mu_j}=1+\delta_{ij}
    \qquad\text{for all }i,j.
\end{equation}

As an explicit representation, one can use for instance the \emph{Jordan-Wigner} transformation:
\begin{equation}
    \gamma_{2i-1} = \Big(\prod_{j=1}^{i-1}Z_j\Big)\,Y_i,\quad \gamma_{2i} = \Big(\prod_{j=1}^{i-1}Z_j\Big)\,X_i.
\end{equation}
Via Eq.~\eqref{eq:def:mu}, we identify the Majorana operators
$\{\gamma_i\}_{i=1}^{2n}$ with their binary vectors
$\{\mu_i\}_{i=1}^{2n}$, where $\mu_i=\inviso(\gamma_i)$.
For coordinates in the chosen Majorana basis, we write $\majcomponent{v}{i}$ for the coefficient of $\mu_i$ in $v$.
For vectors $v = \sum_i\majcomponent{v}{i}\mu_i\in\Fn$,
we also define the \emph{Majorana length} (or \emph{weight})
    \begin{align}
    \majL(v) &:= \sum_{i=1}^{2n} \FtoN{\majcomponent{v}{i}} \in\N,
    \intertext{to be the number of non-zero coefficients in the Majorana basis expansion of $v$ and the \emph{Majorana strings}}
    \majiso{v} &:= \herm\Big(\prod_{i=1}^{2n}\gamma_i^{\majcomponent{v}{i}}\Big).
    \label{eq:def:majorana_strings}
    \end{align}
Here, the map $\herm$ from Eq.~\eqref{eq:def:herm_projection}
ensures that the Majorana string is also a Pauli string
$\majiso{v}\in\PP_n$ (given that a product of Paulis $\gamma_i\in\PP_n$ is in the Pauli group) and the map $\FtoNempty$
from $\F_2$ to $\N$ is the canonical inclusion with $\FtoN{0}=0$
and $\FtoN{1}=1$.
The product in Eq.~\eqref{eq:def:majorana_strings} is taken in increasing
Majorana index, i.e., only the factors with $\majcomponent{v}{i}=1$ are
included, ordered from smaller to larger $i$.
\begin{lem}\label{lem:majorana_string_equals_pauli_label}
Let $\{\gamma_i\}_{i=1}^{2n}\subseteq\PP_n$ be a choice of Majorana operators
as in Eq.~\eqref{eq:def:majorana:operators}, set
$\mu_i=\inviso(\gamma_i)$, and define $\majiso{v}$ as in
Eq.~\eqref{eq:def:majorana_strings}.
Then, for every $v\in\Fn$,
\begin{equation}
    \majiso{v}=\iso{v}.
\end{equation}
\end{lem}
\begin{proof}
Write $v=\sum_i\majcomponent{v}{i}\mu_i$ in the chosen Majorana basis.
We have $\gamma_i=\iso{\mu_i}$ by the definition of $\inviso$ on
$\PP_n$.
Repeatedly applying Eq.~\eqref{eq:Pauli:sign} gives
\[
    \prod_i\gamma_i^{\majcomponent{v}{i}}
    =
    \im^\varpi\isolong{\sum_i\majcomponent{v}{i}\mu_i}
    =
    \im^\varpi\iso{v}
\]
for some $\varpi\in\{0,1,2,3\}$.
Thus, by Eq.~\eqref{eq:def:herm_projection},
\[
    \majiso{v}
    =\herm\Big(\prod_i\gamma_i^{\majcomponent{v}{i}}\Big)
    =
    \herm(\im^\varpi\iso{v})
    =
    \iso{v}.\qedhere
\]
\end{proof}

Lemma~\ref{lem:majorana_string_equals_pauli_label} shows that the
phase-normalized Pauli string $\majiso{v}$ is independent of the chosen
Majorana basis.
What does depend on the chosen Majorana basis are the coordinates
$\majcomponent{v}{i}$ and the length $\majL(v)$ defined above, as well as the
parity and quadratic-form data introduced below.
Thus all coordinate-dependent objects in this section are understood with
respect to the chosen Majorana basis.

More generally, a \emph{Majorana product} is any finite product
$\gamma_{i_1}\gamma_{i_2}\cdots\gamma_{i_s}$ of Majorana operators, with the
scalar phase retained.
By reordering factors and cancelling repeated factors, every Majorana product
agrees up to scalar phase with a unique ordered product
$\prod_{i=1}^{2n}\gamma_i^{c_i}$, where $c_i\in\F_2$ and the product is taken
in increasing Majorana index.
Thus a Majorana product is an element of the Pauli group with its scalar phase
still present, whereas the Majorana string $\majiso{v}$ is the corresponding
phase-normalized Pauli string.
We write $A\simeq B$ when two Majorana products agree up to scalar
phase.
For sets of Majorana products $\calA$ and $\calB$, $\calA\simeq\calB$ if
\[
    \{\herm(a)\text{ for } a\in\calA\}=\{\herm(b) \text{ for } b\in\calB\}.
\]
In particular, different Majorana products with the same
phase-normalized representative are identified after applying $\herm$, so the
relation $\simeq$ compares the sets after this identification.

Let us define the level sets of the Majorana length, which takes values between $0$ and $2n$, i.e.,
\begin{subequations}\label{eq:Majorana_partitions_length}
    \begin{align}
        \MAJ_n &:= \{ \majiso{v} \text{ for } v\in\Fn\},\\
        \MAJ_{n,L} &:= \{ \majiso{v}\in\MAJ_n \text{ with } \majL(v) = L \},\\
        \majvec_{n,L} &:= \{ v\in\Fn \text{ with } \majL(v) = L \},
    \end{align}
\end{subequations}
which induce a natural grading on the Majorana strings and on $\Fn$ as disjoint unions of length level sets
\begin{equation*}
        \MAJ_n = \bigsqcup_{L=0}^{2n} \MAJ_{n,L} \;\text{ and }\;
        \Fn = \bigsqcup_{L=0}^{2n} \majvec_{n,L}.
\end{equation*}
We say that a Majorana string is \emph{even} if $\majL(v)=0\bmod2$ and \emph{odd} otherwise.
We denote with
\begin{equation}\label{eq:def:gamma}
\Gamma := \prod_{i=1}^{2n}\gamma_i
\end{equation}
the product of all Majorana operators and denote with $\majprod = \sum_{i=1}^{2n}\mu_i$ its binary representation.
Thus $\majiso{\majprod}\simeq \Gamma$ is its phase-normalized representative.
We will in particular be interested in the \emph{quadratic} Majorana strings, i.e., the strings $\majiso{\mu_i+\mu_j}$ with $\majL(\mu_i+\mu_j)=2$.
They are represented up to scalar phase by the products $\{\gamma_i\gamma_j\}_{i<j}$.
In particular these are generators for the even subalgebra $\{ \majiso{v} \mid \majL(v) = 0\bmod2\}$, which corresponds to elements
$P\in\PP_n$ with $P\Gamma =\Gamma P$.
Equivalently, $\{\mu_i+\mu_j\}_{i<j=1}^{2n}$ spans the corresponding subspace $\{ v\in\Fn \mid \majL(v) = 0\bmod2\} = \majprod^\perp$.

Given a Majorana basis, set $v=\sum_{i=1}^{2n}\majcomponent{v}{i}\mu_i$, then a convenient expression for the symplectic product is
\begin{align*}
        \symp{v}{w} &= \sum_{i,j}\majcomponent{v}{i}\majcomponent{w}{j} \symp{\mu_i}{\mu_j}
        = \sum_{i,j}\majcomponent{v}{i}\majcomponent{w}{j} (1+\delta_{ij}) \\
        &= \sum_i \majcomponent{v}{i}\majcomponent{w}{i} + \sum_{i,j}\majcomponent{v}{i}\majcomponent{w}{j}\\
        &= \sum_i \majcomponent{v}{i}\majcomponent{w}{i}
        + \Big(\sum_i\majcomponent{v}{i}\Big)
        \Big(\sum_j\majcomponent{w}{j}\Big).
\end{align*}
Now, if we regard $v,w$ as column vectors \emph{in the Majorana basis}, we have
\begin{equation}
    \symp{v}{w} = \majL(v)\majL(w) \bmod 2 + \sum_i\majcomponent{v}{i}\majcomponent{w}{i}.
\end{equation}
Hence, two Majorana strings $\majiso{v},\majiso{w}$ commute iff at least one is even and they overlap at an even number of sites, or they are both odd and they overlap at an odd number of sites.
In particular, if we choose $w=\majprod$, the even subspace satisfies $\symp{v}{\majprod} = 0$ iff $\sum_i\majcomponent{v}{i} = 0\bmod 2$ and the odd subspace satisfies $\symp{v}{\majprod} = 1$ iff $\sum_i\majcomponent{v}{i}=1\bmod 2$.
This implies $\symp{v}{\majprod} = \sum_i\symp{v}{\mu_i} = \sum_i\majcomponent{v}{i} = \majL(v)\bmod 2$.

Then, the symplectic product $\symp{v}{\mu_i}$ determines the components $\majcomponent{v}{i}$, i.e.,
\begin{subequations}\label{eq:majorana_components_via_symplectic_products}
\begin{align}
    \symp{v}{\mu_i} = \majL(v)\bmod 2 + \majcomponent{v}{i}
    \intertext{implies}
    \majcomponent{v}{i} = \symp{v}{\mu_i} + \majL(v)\bmod 2.
\end{align}
\end{subequations}
For later use, we define the reduced Majorana length over the first
$2m$ Majorana operators by
\begin{equation}
    \majL_m(v)
    := \sum_{i=1}^{2m}\FtoN{\majcomponent{v}{i}}\in\N.
\end{equation}
Consequently, Eq.~\eqref{eq:majorana_components_via_symplectic_products}
gives the equivalent expression
\begin{equation}
    \majL_m(v)
    =
    \sum_{i=1}^{2m}
    \FtoN{\symp{v}{\mu_i} + (\majL(v)\bmod 2)}.
\end{equation}
Here the addition inside $\FtoN{\cdot}$ takes place in $\F_2$, whereas the
outer sum takes place in $\N$.

Later, we shall also be interested in understanding the invariant quadratic forms with respect to the Majorana formalism.
Then, consider the unique invariant quadratic form for the Majorana basis $\{\gamma_i\}_{i=1}^{2n}$ and denote it $\QQ_\gamma$.
Using Corollary~\ref{cor:uniqueness_existence_quadratic_form_for_basis} and \ref{cor:quadratic_form_euler_characteristic}, we have
    \begin{align}\label{eq:reference_quadratic_form_majoranas}
        \QQ_\gamma(v) &= [\majL(v) + {\majL(v)(\majL(v)-1)}/{2}]\bmod 2 \nonumber \\
        &= {\majL(v)(\majL(v)+1)}/{2}\bmod 2 \nonumber \\
        &= \begin{dcases}
            0 & \text{if } \majL(v) \bmod 4 \in \{0,3\}, \\
            1 & \text{if } \majL(v) \bmod 4 \in \{1,2\}.
        \end{dcases}
    \end{align}
Since the Majorana generators have frustration graph $\graphK_{2n}$,
the induced graph on the support of $v$ is again complete, with
$\majL(v)$ vertices.
This gives the formula in Eq.~\eqref{eq:reference_quadratic_form_majoranas} for $\QQ_\gamma(v)$.
It is the specialization of the Arf-sum formula
Eq.~\eqref{eq:arf_invariant_binary_quadratic_form} to the invariant quadratic
form of the linearly independent generating set $\{\gamma_i\}_{i=1}^{2n}$.
By Lemma~\ref{thm:equivalent_path_graph_conditions}, this set is also
$t$-equivalent to a path graph over $2n$ vertices, $\graphP_{2n}$, so it is
covered by Proposition~\ref{prop:isomorphism_classes_quadratic_forms_arf_invariant_canonical}.
Then, by Proposition~\ref{prop:isomorphism_classes_quadratic_forms_arf_invariant_canonical}(g),
this invariant quadratic form over $\Fn$ has the following type depending on $n$:
\begin{equation*}
    \type(\QQ_\gamma) = \begin{cases}
    + & \text{if }\,m\bmod 4\in\{0,3\},\\
    - & \text{if }\,m\bmod 4\in\{1,2\}.
    \end{cases}
\end{equation*}
Moreover, $\QQ_\gamma$ is also a simple function of the Majorana length, and explicitly depends on its value modulo $4$.
This is analogous to $\QQ_0$, whose value corresponds to the number of $Y_j$ in the Pauli strings, by Lemma~\ref{lem:quadratic_forms_over_Fn_as_vectors}.

Then, for the analysis of Majorana generating sets, it becomes more convenient to work with $\QQ_\gamma$ as a reference invariant quadratic form.
Namely, using Lemma~\ref{lem:quadratic_forms_over_Fn_as_vectors}, we can obtain all other invariant quadratic forms by adding a linear term.
We write
\begin{equation*}
    \QQ_w^\gamma(v) = \QQ_\gamma(v) + \symp{w}{v}
\end{equation*}
for the quadratic form identified by the vector $w$ relative to the reference quadratic form $\QQ_\gamma$. Moreover, let
\begin{equation*}
\quadratic_\gamma(\vgens) = \{ w\in\Fn \text{ with } \QQ_w^\gamma(v) = 1 \text{ for all }v\in\vgens\}
\end{equation*}
denote the set of invariant quadratic forms.
Lemma~\ref{lem:affine_structure_invariant_quadratic_form_subspace} implies that
$\quadratic_\gamma(\vgens)$ is the affine space
$w^* + \vgens^\perp$.
Its type over $\Fn$ is given by
\begin{equation*}
    \type(\QQ_w^\gamma) = (-1)^{\Arf(\QQ_w^\gamma)} = (-1)^{\Arf(\QQ_\gamma)+\QQ_\gamma(w)}.
\end{equation*}

\section{Free-Fermionic Mappings for Line Graphs}\label{sec:Line_Graphs_Orbits_Lie_Algebras_Free_Fermions_Majoranas}

This section constructs the free-fermionic mappings which will be used in the
canonical orbit and Lie-algebra analysis below.
The objectives are:
\begin{enumerate}
    \item Introduce generalized \emph{free-fermionic mappings}, allowing symmetries and algebraic dependencies beyond the strictly quadratic mappings of \cite{Chapman_Flammia_2020} (Subsection~\ref{sec:Free-Fermionic_Mappings}).
    \item Construct such mappings for arbitrary line graphs of multigraphs (Subsection~\ref{sec:constructing_free_fermionic_mappings_line_graphs}).
    \item Choose adapted Majorana bases for these mappings, so that the later canonical descriptions can be written in Majorana coordinates (Subsection~\ref{sec:adapted_majorana_bases_free_fermionic_mappings}).
\end{enumerate}

\subsection{Free-Fermionic Mappings}\label{sec:Free-Fermionic_Mappings}

We highlight how the Majorana formalism naturally uses the line graph
property to construct free-fermionic mappings for arbitrary line graphs,
extending the strictly quadratic mappings for regular line graphs discussed in
\cite{Chapman_Flammia_2020}.
We then generalize this setup by keeping track of symmetries and algebraic
dependencies explicitly, which leads to the mixed, parity-extended, and
cycle-factor versions used in the construction below; see
Remark~\ref{rem:comparison_chapman_flammia_free_fermionic_mappings}
for the comparison with \cite{Chapman_Flammia_2020}.

Recall that $\binom{[m]}{2}$ denotes the set of two-element subsets of
$[m]$.
Now, for integers $m,q$, define the following sets in $\PP_n$, with $n\geq\lceil m/2\rceil + q$:
\begin{subequations}\label{eq:def_Majorana_bases}
    \begin{align}
        \MAJ_1^{(m)} &:= \{\majiso{\mu_i}\}_{i\in[m]} = \{\gamma_i\}_{i\in[m]},\\
        \MAJ_2^{(m)} &:= \{\majiso{\mu_i+\mu_j}\}_{\{i,j\}\in\binom{[m]}{2}}\nonumber\\
        &\phantom{:}\simeq \{\gamma_i\gamma_j\}_{\{i,j\}\in\binom{[m]}{2}},\\
        \MAJ_{1,2}^{(m)} &:= \MAJ_1^{(m)} \cup \MAJ_2^{(m)},\\
        \MAJ_{1,\Gamma}^{(m)} &:= \{\majiso{\mu_i},\majiso{\mu_i+\sum_{a=1}^m\mu_a}\}_{i\in[m]}\nonumber\\
        &\phantom{:}\simeq \{\gamma_i,\Gamma\gamma_i\}_{i\in[m]},\\
        \MAJ_{\gamma}^{(m)} &:= \MAJ_2^{(m)} \cup \MAJ_{1,\Gamma}^{(m)}
        \cup \{\majiso{\majprod}\} \nonumber\\
        &\phantom{:}\simeq \MAJ_2^{(m)} \cup \MAJ_{1,\Gamma}^{(m)} \cup \{\Gamma\},\\
        \Cyc^{(m,q)} &:= \{
        \majiso{\mu_{2\lceil m/2\rceil+2s-1}+\mu_{2\lceil m/2\rceil+2s}}
        \}_{s\in[q]} \nonumber\\
        &\phantom{:}\simeq \{ \gamma_{2\lceil m/2\rceil+2s-1}\gamma_{2\lceil m/2\rceil+2s}\}_{s\in[q]}.
    \end{align}
\end{subequations}
Let us unpack the notation in Eq.~\eqref{eq:def_Majorana_bases}.
We write
\[
    \Gamma= \Gamma_m:=\prod_{i=1}^{m}\gamma_i
    \;\text{ and }\;
    \majprod = \majprod_m:=\sum_{i=1}^{m}\mu_i,
\]
so that $\majiso{\majprod_m}\simeq \Gamma_m$.
When $m$ is fixed, we often suppress the subscript and write simply
$\Gamma$ and $\majprod$.
The set $\MAJ_1^{(m)}$ consists of the single Majorana strings on the first
$m$ Majorana operators, while $\MAJ_2^{(m)}$ consists of the corresponding
quadratic Majorana strings.
The set $\MAJ_{1,2}^{(m)}$ is the mixed set containing both of these.
The set $\MAJ_{1,\Gamma}^{(m)}$ is the parity-extended version of $\MAJ_1^{(m)}$, which
besides the strings $\gamma_i$ also contains the strings $\Gamma\gamma_i$,
up to scalar phase.
Since $\Gamma$ is the product of all $m$ Majorana operators, the product
$\Gamma\gamma_i$ removes the factor $\gamma_i$ and leaves the product of the
remaining $m-1$ Majorana operators, again up to scalar phase.
The set $\MAJ_\gamma^{(m)}$ is the largest Majorana set used below,
containing the quadratic strings, the parity-extended single strings, and the
total parity string $\Gamma$.

The set $\Cyc^{(m,q)}$ consists of certain quadratic Majorana strings.
Here $m$ fixes the first $m$ Majorana operators used by the Majorana
sets above, and $q$ specifies how many disjoint cycle-symmetry pairs
are placed after them.
More explicitly, the $s$th cycle-symmetry pair is
\begin{equation}\label{eq:cycle:symmetry:pair}
\{2\lceil m/2\rceil+2s-1,\,2\lceil m/2\rceil+2s\}.
\end{equation}
Thus the cycle factors use Majorana operators whose indices are disjoint
from the first $m$ Majorana operators and are grouped into adjacent
Jordan-Wigner pairs.
Thus we first pass to the next even label $2\lceil m/2\rceil$ and then take
the pairs in Eq.~\eqref{eq:cycle:symmetry:pair}.
Since the cycle-symmetry pairs are disjoint, the corresponding quadratic Majorana
strings mutually commute.
These commuting quadratic factors will later encode the independent cycle
symmetries arising from the line graph.
In a free-fermionic mapping, an image may be a product of one such cycle factor
and one Majorana string from $\MAJ_\gamma^{(m)}$. 

Let us introduce the product map
\begin{equation}\label{eq:def:free_fermionic_product_map}
    \prodmap\colon \MAJ_\gamma^{(m)}\times\Cyc^{(m,q)}\to\PP_n,
    \quad
    (M,C)\mapsto MC.
\end{equation}
For convenience, we usually denote the image of this map in the Pauli strings as a set operation [or see \eqref{eq:setprod_in_paulis}]:
\begin{align}
        \MAJ_\gamma^{(m)}\setprod\Cyc^{(m,q)} &:= \herm\bigl(\prodmap\bigl(\MAJ_\gamma^{(m)},\Cyc^{(m,q)}\bigr)\bigr)\\
        &\phantom{:}= \{ \herm(PC) | P\in \MAJ_\gamma^{(m)}, C\in \Cyc^{(m,q)}\}. \nonumber
\end{align}
The product map sends a Majorana factor and a cycle factor to their product
Pauli string.
The image is contained in the hermitian Pauli strings $\PP_n$. The cycle
factor uses auxiliary Majorana operators disjoint from those appearing in
$\MAJ_\gamma^{(m)}$, so the two factors commute.
We can now provide the suitable definition for free-fermionic mappings:
\begin{defn}[Free-Fermionic Mapping]\label{defn:free_fermionic_mappings}
Let $\pgens = \isolong{\vgens}\subseteq\PP_n$ be a set of Pauli strings.
We say that $\pgens$ admits a \emph{free-fermionic mapping} if there exist integers $m,q$ and an injective map $\inclusion\colon\pgens\to\PP_n$ such that
\[
    \inclusion(\pgens)\subseteq
    \MAJ_\gamma^{(m)}\setprod\Cyc^{(m,q)}
\]
and $\inclusion(\pgens)$ has the same frustration graph as $\pgens$.
\end{defn}
Thus $\inclusion$ assigns Pauli strings to the generators.
Equivalently, we also say that $\pgens$ has an $(m,q)$-free fermionic mapping.
Furthermore, we say that the mapping is
\begin{enumerate}
    \item symmetry-adapted if either $\ZZ(\inclusion(\pgens)) = \Cyc^{(m,q)}$, for any $m$, or, if $m$ is even,
    \begin{align*}
       \ZZ(\inclusion(\pgens))
        = 
        \{ \Gamma \} \setprod \Cyc^{(m,q)};
    \end{align*}
    \item quadratic if $\inclusion(\pgens)\subseteq
    \MAJ_2^{(m)}\setprod \Cyc^{(m,q)}$;
    \item \emph{strictly} quadratic if $\inclusion(\pgens) \subseteq \MAJ_2^{(m)}$;
    \item mixed if $\inclusion(\pgens)\subseteq
    \MAJ_{1,2}^{(m)}\setprod \Cyc^{(m,q)}$;
    \item parity-extended if
    \[
        \inclusion(\pgens)\cap
        (\{\majiso{\Gamma})\setprod\Cyc^{(m,q)}\}
        \neq\emptyset;
    \]
    \item faithful if $\pgens$ and $\inclusion(\pgens)$ have the same algebraic dependencies, i.e., the same colorings on the common frustration graph have vanishing realizations for $\pgens$ and for $\inclusion(\pgens)$, or
    $\coltovec_{\vgens}(\coloring)=0$ iff
    $\coltovec_{\inviso(\inclusion(\pgens))}(\coloring)=0$ with $\coltovec$ as in Eq.~\eqref{eq:coloring};
    \item algebraically independent if $\inclusion(\pgens)$ has no algebraic dependencies;
    \item minimal for some $m,q$ if there is no mapping with $m'\leq m$, $q'\leq q$, and $(m',q')\neq(m,q)$.
\end{enumerate}

A free fermionic mapping in the notation of \cite{Chapman_Flammia_2020} is specifically a strictly quadratic free-fermionic mapping with respect to our definitions. 
There is no guarantee that a strictly quadratic free-fermionic mapping in this
sense is faithful or minimal, which will instead be essential for our
applications.
The case of mixed mappings involve both quadratic and linear Majoranas, which provide \emph{parity-symmetry} breaking terms, and have also been considered in the context of efficient or exact solutions \cite{moussa2012generalizedunitarybogoliubovtransformation,Nishiyama_da_Providencia_2019,kaicher2026quantumsimulationgeneralspin12}.
We are not concerned here with keeping track of signs, since we are only interested in Lie algebras and groups, for which signs are irrelevant.
However, this becomes important for mappings of \emph{specific} Hamiltonians or circuits composed of such gates,
since $e^{\im \theta P} \neq e^{-\im \theta P}$ for arbitrary $\theta$.
We collect some immediate consequences of the definition:
\begin{lem}\label{lem:free_fermionic_mappings_basics}
Let $\pgens\subseteq\PP_n$ be a set with a $(m,q)$-free fermionic mapping $\inclusion$ with $\inclusion(\pgens)\subseteq\PP_{\lceil m/2\rceil + q}$.
Then
\begin{enumerate}
    \item $\pgens$ also admits a quadratic $(m{+}2,q)$-free fermionic mapping;
    \item $\frustration{\pgens}$ is a line graph of a multigraph; conversely,
    every Pauli generating set whose frustration graph is a line graph of a
    multigraph admits a free-fermionic mapping;
    \item If $\pgens'$ is obtained from $\pgens$ by a sequence of valid contractions, i.e.,
    $\pgens'\sim_t\pgens$ in the sense of Definition~\ref{defn:t:equivalence}, then there is a $(m,q)$-free fermionic mapping for $\pgens'$ such that $\inclusion(\pgens') \sim_t \inclusion(\pgens)$.
\end{enumerate}
\end{lem}
\begin{proof}

For (a), consider the following projection from $\MAJ_2^{(m+2)}$ to $\MAJ_{\gamma}^{(m)}$, up to the phase equivalence:
\begin{align*}
        \projection(\tilde{\gamma}_i\tilde{\gamma}_j) &\simeq \gamma_i\gamma_j \quad\text{for } 1\leq i<j\leq m, \\
        \projection(\tilde{\gamma}_i\tilde{\gamma}_{m+1}) &\simeq \gamma_i \quad\text{for } 1\leq i\leq m, \\
        \projection(\tilde{\gamma}_i\tilde{\gamma}_{m+2}) &\simeq \Gamma\gamma_i \quad\text{for } 1\leq i\leq m, \\
        \projection(\tilde{\gamma}_{m+1}\tilde{\gamma}_{m+2}) &\simeq \Gamma.
\end{align*}
We extend this to the presence of cycle factors by leaving them fixed:
for $C\in\Cyc^{(m,q)}$ and a Majorana string $M$ in the displayed domain,
we have $\projection(CM)\simeq C\projection(M)$.
This is a bijection between the sets and conserves the symplectic relations, or equivalently the frustration graph of $\MAJ_2^{(m+2)}$ and $\MAJ_{\gamma}^{(m)}$ coincide.
Then, $\projection(\inclusion(\pgens))$ is a quadratic $(m{+}2,q)$-free fermionic mapping for $\pgens$, which proves (a).

As the frustration graph is conserved by the quadratic mapping, it is sufficient to consider a mapping whose image is contained in
$\MAJ_2^{(m+2)}\setprod\Cyc^{(m+2,q)}$.
Equivalently, every mapped generator can be written as
$C_{\edgea}\gamma_i\gamma_j$ with
$C_{\edgea}\in\Cyc^{(m+2,q)}$ and $i,j\leq m+2$.
Ignoring the commuting cycle factor $C_{\edgea}$, the relevant quadratic part is indexed
by an edge $\edgea$ of a multigraph $\Delta$ on $m+2$ vertices, with endpoints
$a_{\edgea}$ and $b_{\edgea}$.
Clearly, two such operators anticommute if and only if the corresponding edges
share precisely one endpoint, which determines the line graph $L(\Delta)$ with
edge-labelled generators
$\{\gamma_{a_{\edgea}}\gamma_{b_{\edgea}}\}_{\edgea\in\edges(\Delta)}$.
This proves (b).

Again, regarding contractions, we can work over the quadratic mapping and perform the same contractions on $\pgens$ and on $\inclusion(\pgens)$.
A generator in this quadratic image has the form
$C_{\edgea}\gamma_i\gamma_j$, with
$C_{\edgea}\in\Cyc^{(m+2,q)}$ commuting with the Majorana factor.
Contracting two such generators replaces them by their commutator, which is
again, up to scalar phase, a product of a cycle factor in $\Cyc^{(m+2,q)}$ and
a quadratic Majorana string.
As such, any $t$-equivalent set to $\pgens'$ must also have a free fermionic mapping, which proves (c).
\end{proof}

\subsection{Constructing Free-Fermionic Mappings from Line Graphs}\label{sec:constructing_free_fermionic_mappings_line_graphs}

We now show how to construct free-fermionic mappings for arbitrary generating sets whose frustration graphs are line graphs.
Thus we fix a multigraph $\Delta$ whose line graph is the frustration graph, and write $n_\Delta$ for the number of vertices of $\Delta$.
We start by working with the algebraically independent case, which produces a symmetry-adapted mapping: always a quadratic one, and also a mixed one if $n_\Delta$ is odd.
First, we prove this for the algebraically independent sets, by working directly on the graphs. We generalize the approach by \cite{Chapman_Flammia_2020} by explicitly taking all symmetries into account:
\begin{lem}\label{lem:free_fermionic_mapping_alg_ind}
Let $\pgens=\isolong{\vgens}\subseteq\PP_n$ be a set of Pauli strings with connected frustration graph $\graphG$ which is the line graph of a multigraph $\Delta$ on $n_\Delta$ vertices and with $\tilde q$ independent cycles.

\begin{enumerate}
    \item For any $n_\Delta$, there is a symmetry-adapted quadratic $(n_\Delta,\tilde q)$-free fermionic mapping for $\pgens$, which is also algebraically independent.
    \item If $n_\Delta$ is odd for $\pgens$, there is also a symmetry-adapted mixed $(n_\Delta{-}1,\tilde q)$-free fermionic mapping.
\end{enumerate}
\end{lem}
\begin{proof}
\emph{Setup.}
Let $\Delta_T$ be a spanning tree for $\Delta$.
The construction only uses the endpoints of each edge of the multigraph.
Parallel edges are nevertheless kept as distinct elements of $\edges(\Delta)$.
They may have the same endpoints, but they correspond to different vertices in
the line graph $L(\Delta)$ and hence to different generators.
We first assign quadratic Majorana strings to the edges of $\Delta_T$ and use
one additional cycle factor for each edge in
$\edges(\Delta)\setminus\edges(\Delta_T)$.
This gives the quadratic mapping in (a); when $n_\Delta$ is odd, setting the
last Majorana to the identity gives the mixed mapping in (b).

For each edge $\edgea$ of the multigraph $\Delta$, including parallel copies as
distinct edges, write $a_{\edgea}$ and $b_{\edgea}$ for its two endpoints.
Since $\Delta_T$ is a spanning tree and $\Delta$ has $\tilde q$ independent
cycles, the edges outside $\Delta_T$ are ordered as
\begin{equation}\label{eq:free_fermionic_missing_edges}
    \edges(\Delta)\setminus\edges(\Delta_T)=\{\edge{1},\ldots,\edge{\tilde q}\}.
\end{equation}
For each $i$, the edge $\edge{i}$ closes one fundamental cycle when added to
$\Delta_T$.
We organize the proof as follows.
First we construct the quadratic mapping in (a), check that it has the same
frustration graph as $\pgens$, and then prove its algebraic independence by
examining cycle and T-join dependencies.
Second, when $n_\Delta$ is odd, we obtain the mixed mapping in (b) by removing
the last Majorana operator and checking that the commutation relations are
unchanged.
Finally, we identify the central product symmetries, which proves that the
mappings are symmetry-adapted.

\emph{Quadratic mapping.}
For (a), using the ordered missing edges from
Eq.~\eqref{eq:free_fermionic_missing_edges}, define
\begin{align}\label{eq:Free_Fermionic_Mapping_alg_ind_quadratic}
        \pgens_\gamma &=
        \begin{aligned}[t]
        &\{\gamma_{a_{\edgea}}\gamma_{b_{\edgea}}\}_{\edgea\in\edges(\Delta_T)}\\
        &\cup \{(\gamma_{m+2i-1}\gamma_{m+2i})\gamma_{a_{\edge{i}}}\gamma_{b_{\edge{i}}}\}_{i=1}^{\tilde q}
        \end{aligned}
\end{align}
as an algebraically independent set on $2n = 2(m + \tilde q)$ Majorana modes.
Here $m=\lceil n_\Delta/2\rceil$ and the $\tilde q$ phase Majorana pairs are
indexed by the edges $\edge{i}$ from
Eq.~\eqref{eq:free_fermionic_missing_edges}.
The frustration graph of $\pgens_\gamma$ is $L(\Delta)$.

\emph{Frustration graph.}
By assumption, $L(\Delta)$ is the frustration graph of $\pgens$.
On the other hand, the construction indexes the generators of $\pgens_\gamma$
by the same edges of $\Delta$ as the generators of $\pgens$.
For two edges $\edgea,\edgeb\in\edges(\Delta)$, the quadratic factors
$\gamma_{a_{\edgea}}\gamma_{b_{\edgea}}$ and
$\gamma_{a_{\edgeb}}\gamma_{b_{\edgeb}}$ anticommute exactly
when $\edgea$ and $\edgeb$ share one endpoint in $\Delta$.
The phase cycle factors use disjoint Majorana pairs and therefore do not
change these commutation relations.
Thus the same edge indexing identifies the frustration graph of
$\pgens_\gamma$ with $L(\Delta)$, and hence with the frustration graph of
$\pgens$.
This proves the frustration-graph claim in (a).
It remains, for (a), to show the algebraic independence of
$\pgens_\gamma$.

\emph{Algebraic independence.}
By Lemma~\ref{lem:radical_via_cycles_and_Tjoin}, the possible
edge-coordinate vectors of algebraic dependencies lie in
$\ker(A(L(\Delta)))$, where $A(L(\Delta))$ is the adjacency matrix of the
line graph.
For a connected multigraph $\Delta$, this kernel contains the cycle space
$\ker(M(\Delta))$; when $n_\Delta$ is even, it also contains the T-join
coloring from Definition~\ref{def:T-join}.
The corresponding fundamental cycles form a basis for the cycle space
$\ker(M(\Delta))$.
For the fundamental cycle associated with the missing edge $\edge{i}$, all
quadratic factors coming from the tree edges cancel in pairs, while the single
non-tree edge $\edge{i}$ contributes its cycle factor.
Up to scalar phase, the cycle product is
\[
    \gamma_{m+2i-1}\gamma_{m+2i}
    \simeq
    \majiso{\mu_{m+2i-1}+\mu_{m+2i}}\in\Cyc^{(m,\tilde q)}.
\]
Hence the cycle-space candidates reduce to the independent generators
of $\Cyc^{(m,\tilde q)}$, rather than to algebraic dependencies, as desired.

If $n_\Delta$ is even, a T-join $v_{\mathbf{1}}$ over all of $\Delta$ in the sense of
Definition~\ref{def:T-join} is specified by the unique T-join over
$\Delta_T$ (since it has no cycles).
Thus the strictly quadratic mapping already separates the T-join
vector from the identity.
Indeed, by definition of T-join, if $v_{\mathbf 1}$ denotes the vector realized by a T-join coloring
$x\in\F_2^{\edges(\Delta)}$, each vertex of $\Delta$ is incident to an odd number
of colored edges.
Multiplying the corresponding quadratic Majorana labels therefore leaves, up
to scalar phase, one copy of each vertex Majorana:
\[
    \isolong{v_{\mathbf 1}}
    \simeq
    \prod_{i=1}^{n_\Delta}\gamma_i
    =
    \Gamma.
\]
In particular, this T-join product is not an algebraic dependency (nor lies in the cycle space).
This proves the algebraic independence of the quadratic mapping, which
completes the proof of (a).

\emph{Odd mixed mapping.}
For (b), assume $n_\Delta=2m+1$ is odd.
Starting from the quadratic mapping constructed in (a), we obtain a
symmetry-adapted mixed mapping on $2n = 2(m + \tilde q)$ Majorana modes by replacing
every occurrence of the Majorana operator $\gamma_{n_\Delta}$ by the scalar
identity operator $I$.
Indeed, we only need to check pairs of edges for which at least one edge
is incident to the removed vertex $n_\Delta$.
An edge not incident to $n_\Delta$ has the same vector before and after the
restriction, namely $\mu_i+\mu_j$ with $i,j<n_\Delta$.
An edge incident to $n_\Delta$ has the vector $\mu_k+\mu_{n_\Delta}$ before the
restriction and becomes the linear vector $\mu_k$ after replacing
$\gamma_{n_\Delta}$ by $I$.
Since $\symp{\mu_i+\mu_j}{\mu_{n_\Delta}}=1+1=0$, we have
\[
    \symp{\mu_i+\mu_j}{\mu_k+\mu_{n_\Delta}}
    =
    \symp{\mu_i+\mu_j}{\mu_k}.
\]
For two edges incident to $n_\Delta$ with $i\neq j<n_\Delta$, we
similarly get
\[
\begin{aligned}
    &\symp{\mu_i+\mu_{n_\Delta}}{\mu_j+\mu_{n_\Delta}}\\
    &=
    \symp{\mu_i}{\mu_j}
    +\symp{\mu_i}{\mu_{n_\Delta}}
    +\symp{\mu_{n_\Delta}}{\mu_j}
    +\symp{\mu_{n_\Delta}}{\mu_{n_\Delta}}\\
    &=
    \symp{\mu_i}{\mu_j}+1+1+0
    =
    \symp{\mu_i}{\mu_j}.
\end{aligned}
\]
Thus replacing $\mu_k+\mu_{n_\Delta}$ by $\mu_k$ preserves the
symplectic products, hence the anti-commutation relations.
This gives the mixed mapping in (b); it remains only to check the
symmetry-adapted statement common to both constructions.

\emph{Symmetry-adaptedness.}
The same replacement does not change the product associated with any
cycle coloring. The endpoint Majoranas still cancel in pairs along the cycle,
and the same cycle factor remains.
Thus the product symmetries, and hence the absence or presence of algebraic
dependencies detected by them, are unchanged.
The center of the generated matrix algebra is generated by the commuting
cycle factors $\{\gamma_{m+2i-1}\gamma_{m+2i}\}_{i=1}^{\tilde q}$, for any $n_\Delta$ and
any mapping, together with
$\Gamma = \prod_{i=1}^{n_\Delta}\gamma_i$ if $n_\Delta$ is even.
These are precisely the symmetry factors included in the construction,
so the mappings are symmetry-adapted.
Also, the commutant for the $(n_\Delta,\tilde q)$-mapping for $n_\Delta$ odd contains $\gamma_{n_\Delta}$ as well.
This concludes the proof.
\end{proof}

The previous lemma treats the algebraically independent line-graph case.
We now pass to general line-graph generating sets by starting from the linearly
independent extension and then quotienting out the algebraic dependencies.
Cycle dependencies are absorbed into the cycle-symmetry factors, while the even
case may also involve a T-join vector.
For the linearly independent extension $\tilde{\vgens}$, the
edge-coordinate realization map $\edgetovec_{\Delta,\tilde{\vgens}}$ from
Eq.~\eqref{eq:edge_coordinate_to_label_map} sends the cycle space
$\ker(M(\Delta))$ of the incidence matrix $M(\Delta)$
(see Eq.~\eqref{eq:def:incidence_matrix} and
Lemma~\ref{lem:cycle_space_incidence_matrix_multigraph}) into
$\SpanS[\F_2]{\tilde{\vgens}}$.
After applying the projection $\projection$, the surviving cycle-symmetry space
has dimension
\begin{equation}\label{eq:def:surviving_cycle_symmetry_dimension}
q
:=\dim\qty(\projection(\edgetovec_{\Delta,\tilde{\vgens}}(\ker(M(\Delta)))))
=\dim\radorbit(\vgens),
\end{equation}
where the equality with $\dim\radorbit(\vgens)$ is the content of
Corollary~\ref{cor:distinguishing_line_graph_cases}.
When $n_\Delta$ is even, Lemma~\ref{lem:radical_via_cycles_and_Tjoin} supplies
a T-join vector $x\in\F_2^{\edges(\Delta)}$ with $M(\Delta)x=\mathbf 1$.
After projection, Corollary~\ref{cor:distinguishing_line_graph_cases} determines whether this T-join coloring
survives as a non-trivial T-join symmetry or becomes an algebraic dependency.
\begin{prop}\label{prop:free_fermionic_mapping_alg_dep}
Let $\pgens=\isolong{\vgens}\subseteq\PP_n$ be a set of Pauli strings with connected frustration graph $\graphG$ which is the line graph of a multigraph $\Delta$ on $n_\Delta$ vertices and with $\tilde{q}$ independent cycles.
Also, let $\tilde{\vgens}$ be the linearly independent extension with
projection $\projection$, and let $q$ be the surviving
cycle-symmetry dimension from
Eq.~\eqref{eq:def:surviving_cycle_symmetry_dimension}.
\begin{enumerate}
    \item If $n_\Delta$ is even, there is a symmetry-adapted faithful quadratic $(n_\Delta,q)$-free fermionic mapping for $\pgens$.
    \item If $n_\Delta$ is even and the T-join leads to an algebraic dependency, then there is a symmetry-adapted faithful parity-extended $(n_\Delta{-}2,q)$-free fermionic mapping.
    \item If $n_\Delta$ is odd, there is a symmetry-adapted faithful mixed $(n_\Delta{-}1,q)$-free fermionic mapping.
\end{enumerate}
\end{prop}
\begin{proof}
We organize the proof in four steps.
First, we start from the algebraically independent mapping of
Lemma~\ref{lem:free_fermionic_mapping_alg_ind} and project it to the given
generating set.
Second, we remove the algebraic dependencies coming from projected cycle
symmetries.
Third, we treat the two possible extra cases: the odd mixed mapping and the
even T-join dependency.
Finally, we identify the remaining central product symmetries, which verifies
that the resulting mappings are symmetry-adapted.

\emph{Projected mapping.}
Fix a spanning tree $\Delta_T$, write
$\{\edge{s}\}_{s=1}^{\tilde q}=\edges(\Delta)\setminus\edges(\Delta_T)$ for the
missing edges, and write $a_{\edgea}$ and $b_{\edgea}$ for the two endpoints
of an edge $\edgea\in\edges(\Delta)$.
We start with the algebraically independent mapping
$\tilde{\pgens}_\gamma$ provided by
Lemma~\ref{lem:free_fermionic_mapping_alg_ind}.
By the choice of linearly independent extension, there is a projection map
$\projection$ such that
$\pgens_\gamma=\projection(\tilde{\pgens}_\gamma)$ realizes the original
generating set $\pgens$.
Thus the frustration graph is already correct. We only need to rewrite the
projected cycle factors and to remove exactly those products which
become algebraic dependencies after the projection.

\emph{Cycle dependencies.}
In edge-coordinate space for the linearly independent extension,
the algebraic dependencies of the projected set
$\projection(\tilde{\pgens}_\gamma)$ are the elements of
\[
    \ker(\projection\circ\edgetovec_{\Delta,\tilde{\vgens}})
    \subseteq \ker(A(L(\Delta))).
\]
Here $A(L(\Delta))$ denotes the adjacency matrix of the line graph, in the
sense of the adjacency matrix convention following
Definition~\ref{defn:frustration}.
The inclusion holds because a projected dependency gives the zero vector and
therefore pairs trivially with every projected generator.
Let $R_C$ be the kernel of
$\projection\circ\edgetovec_{\Delta,\tilde{\vgens}}$ restricted to
$\ker(M(\Delta))$.
If $n_\Delta$ is odd, Lemma~\ref{lem:radical_via_cycles_and_Tjoin}
gives no T-join coloring, so the radical contribution is exactly the cycle
space $\ker(M(\Delta))$.
Thus the algebraic dependencies may only result from cycle symmetries, and
these are recorded by $R_C$.
If $n_\Delta$ is even, choose
$x\in\F_2^{\edges(\Delta)}$ with $M(\Delta)x=\mathbf 1$. Then
the remaining possible dependency is the T-join coloring: besides
$R_C$, there may be a one-dimensional summand $R_T$ represented by a T-join
vector modulo the cycle space.

Hence, the algebraic dependencies from $R_C$ are specified by a linear combination of cycle symmetries for $\tilde{\pgens}_\gamma$.
Choose a basis $\{\alpha^j\}_{j=1}^{\dim R_C}$ of $R_C$, written in the
fundamental-cycle basis as $\alpha^j=(\alpha_s^j)_{s=1}^{\tilde q}$.
For each such basis vector,
\begin{equation}
\projection\left[\prod_{s=1}^{\tilde{q}} (\tilde{\gamma}_{2s-1}\tilde{\gamma}_{2s})^{\alpha_s^j}\right] \simeq I
\end{equation}
for  $j\in\{1,\ldots,\dim R_C\}$.

We resolve the cycle dependencies by choosing an algebraically
independent subset of the projected cycle symmetries
$\{\projection(\tilde{\gamma}_{2s_i-1}\tilde{\gamma}_{2s_i})\}_{i=1}^{q}$
(which is empty for $q=0$), where $q$ is defined in
Eq.~\eqref{eq:def:surviving_cycle_symmetry_dimension} and
$\{s_i\}\subseteq[\tilde{q}]$, and then writing every other projected cycle
factor as a product of these chosen ones:
\begin{equation}
    \projection(\tilde{\gamma}_{2s-1}\tilde{\gamma}_{2s}) \simeq \prod_{i=1}^{q} \projection(\tilde{\gamma}_{2s_i-1}\tilde{\gamma}_{2s_i})^{\beta_i^s}
\end{equation}
for coefficients $\beta_i^s\in\F_2$ obtained by Gaussian elimination in
the projected cycle-symmetry space.
Equivalently, we choose a basis of the surviving projected cycle factors and
express every other projected cycle factor in that basis.

\emph{Quadratic mapping.}
After this reduction, the retained cycle factors are algebraically
independent.
If there is no additional T-join dependency, then the projected generators can
be written as
\begin{align*}
        &\pgens_\gamma =
        \{\gamma_{a_{\edgea}}\gamma_{b_{\edgea}}\}_{\edgea\in\edges(\Delta_T)}
        \cup \{ C_s \gamma_{a_{\edge{s}}}\gamma_{b_{\edge{s}}} \}_{s=1}^{\tilde{q}}\\
        &\text{with } C_s = \prod_{i=1}^{q} (\tilde{\gamma}_{2s_i-1}\tilde{\gamma}_{2s_i})^{\beta_i^s}.
\end{align*}
Here each $C_s$ lies in the subgroup generated by the $q$ retained cycle
factors, so $\pgens_\gamma$ is a faithful quadratic $(n_\Delta,q)$-free
fermionic mapping for $\pgens$.
This proves the faithful quadratic mapping, except that in the even case
with a T-join dependency we will also produce the parity-extended mapping
below.

\emph{Odd mixed mapping.}
If $n_\Delta$ is odd, then the same replacement as in
Lemma~\ref{lem:free_fermionic_mapping_alg_ind}, namely setting
$\gamma_{n_\Delta}=I$, preserves the commutation relations and algebraic dependencies, hence gives a
symmetry-adapted faithful mixed $(n_\Delta{-}1,q)$-mapping.

\emph{Even T-join dependency.}
It remains to handle the even case in which the T-join coloring becomes
an algebraic dependency.
Also, if $n_\Delta$ is even and $R_T$ is not empty, we have an algebraic dependency from a T-join, which takes the form
\begin{equation}
    \projection\left[\prod_{s=1}^{\tilde{q}} (\tilde{\gamma}_{2s-1}\tilde{\gamma}_{2s})^{\alpha_s^0}\prod_{i=1}^{n_\Delta}\gamma_i\right] \simeq I
\end{equation}
for suitable coefficients $\alpha_s^0\in\F_2$.

The following formal choice forces this T-join product to become the
identity while keeping the retained cycle factors as the independent symmetry
factors.
To resolve this additional algebraic dependency coming from the T-join, we can define the set
\begin{equation*}
        \pgens_\gamma =
        \{\gamma_{a_{\edgea}}\gamma_{b_{\edgea}}\}_{\edgea\in\edges(\Delta_T)}
        \cup \{ C_s \gamma_{a_{\edge{s}}}\gamma_{b_{\edge{s}}} \}_{s=1}^{\tilde{q}}
\end{equation*}
where the $C_s$ are defined as before and we write formally
\begin{align*}
        \gamma_{n_\Delta-1} &= I,\\
        \gamma_{n_\Delta} &\simeq \prod_{s=1}^{\tilde{q}} \prod_{i=1}^{q} (\tilde{\gamma}_{2s_i-1}\tilde{\gamma}_{2s_i})^{\alpha_s^0\beta_i^s} \prod_{i=1}^{n_\Delta-2}\gamma_i.
\end{align*}
Then, the new $\pgens_\gamma$ is a faithful parity-extended
$(n_\Delta{-}2,q)$-free fermionic mapping for $\pgens$.

\emph{Symmetry-adaptedness.}
We finish by checking
whether the mappings
constructed above are symmetry-adapted.
The surviving cycle-symmetry part of the commutant is generated, in all
cases, by the independent cycle factors
$\{\tilde{\gamma}_{2s_i-1}\tilde{\gamma}_{2s_i}\}_{i=1}^q$.
In the even case with $R_T=\emptyset$ for the $(n_\Delta,q)$-mapping, one
must also include the parity operator
$\Gamma=\prod_{i=1}^{n_\Delta}\gamma_i$ among the commutant generators.
In the odd construction, the unreduced quadratic model has the additional
central generators $\Gamma=\prod_{i=1}^{n_\Delta}\gamma_i$ and
$\gamma_{n_\Delta+1}$; after setting $\gamma_{n_\Delta}=I$, this gives the
corresponding symmetry data for the mixed mapping.
These are exactly the symmetry factors retained in the corresponding
mappings, so the constructed mappings are symmetry-adapted.
This concludes the proof.
\end{proof}
Combining Proposition~\ref{prop:free_fermionic_mapping_alg_dep}, which gives
the construction for line-graph generating sets, with
Lemma~\ref{lem:free_fermionic_mappings_basics}, which shows that
free-fermionic mappings have line-graph frustration graphs, we obtain the
following characterization of generating sets admitting such a mapping:
\begin{cor}\label{cor:free_fermionic_mappings_line_graphs}
Let $\vgens\subseteq\Fn$ be a set of binary vectors. Then, $\vgens$ admits a free-fermionic mapping if and only if $\frustration{\vgens}$ is the line graph of a multigraph.
\end{cor}

\begin{figure}
    \centering
    \includegraphics[width=\linewidth]{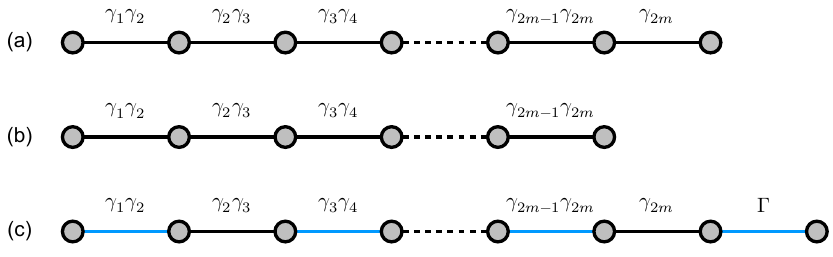}
    \caption{Representative free-fermionic mappings when the frustration graph $\graphG$ is a path graph, viewed in the line graph formalism on the root graph $\Delta$. Algebraic dependencies are highlighted as colorings. Lie algebras are obtained via Theorem~\ref{thm:Free_Fermionic_Lie_Algebras}. (a) $\graphG=\graphP_{2m} = L(\graphP_{2m+1})$, $\lie{\pgens} = \so(2m+1)$. (b) $\graphG=\graphP_{2m-1} = L(\graphP_{2m})$, $\lie{\pgens} = \so(2m)$ with a reducible representation. (c) $\graphG=\graphP_{2m+1}=L(\graphP_{2m+2})$, $\lie{\pgens} = \so(2m+2)$ with an irreducible representation and an algebraic dependency given by the odd vertices on $\graphG$ or odd edges on the root graph $\Delta=\graphP_{2m+2}$.}
    \label{fig:free-fermionic-paths}
\end{figure}

We now apply the construction to paths and cycles, which gives a
concrete comparison with the quadratic line-graph mappings of
\cite{Chapman_Flammia_2020}.
\begin{ex}[Paths]\label{ex:free_fermionic_mappings_paths}
Let $\vgens$ be such that $\frustration{\vgens} = \graphP_n$, $n\geq 4$, which is the line graph of $\Delta=\graphP_{n+1}$.
See also Fig.~\ref{fig:free-fermionic-paths}.

(a) If $n+1=2m+1$ is odd (or, equivalently, $n$ is even), no algebraic dependencies are possible, and the $(n,0)$-free fermionic mapping is of the form
\begin{equation}
    \{\gamma_i\gamma_{i+1}\}_{i=1}^{2m} \cup \{\gamma_{2m}\}.
\end{equation}
(b) If $n+1=2m$ is even (or, equivalently, $n$ is odd) and $\vgens$ has no algebraic dependencies, then the $(2m,0)$-free fermionic mapping is of the form
\begin{equation}
    \{\gamma_i\gamma_{i+1}\}_{i=1}^{2m-1}.
\end{equation}
(c) If $n+1=2m+2$ is even (or, equivalently, $n$ is odd) and there is an algebraic dependency, then the $(2m,0)$-free fermionic mapping is of the form
\begin{equation}
    \{\gamma_i\gamma_{i+1}\}_{i=1}^{2m-1} \cup \{ \gamma_{2m}, \Gamma\}
\end{equation}
with $\Gamma = \prod_{i=1}^{2m}\gamma_i$.
\end{ex}

\begin{figure*}
    \centering
    \includegraphics[width=0.8\linewidth]{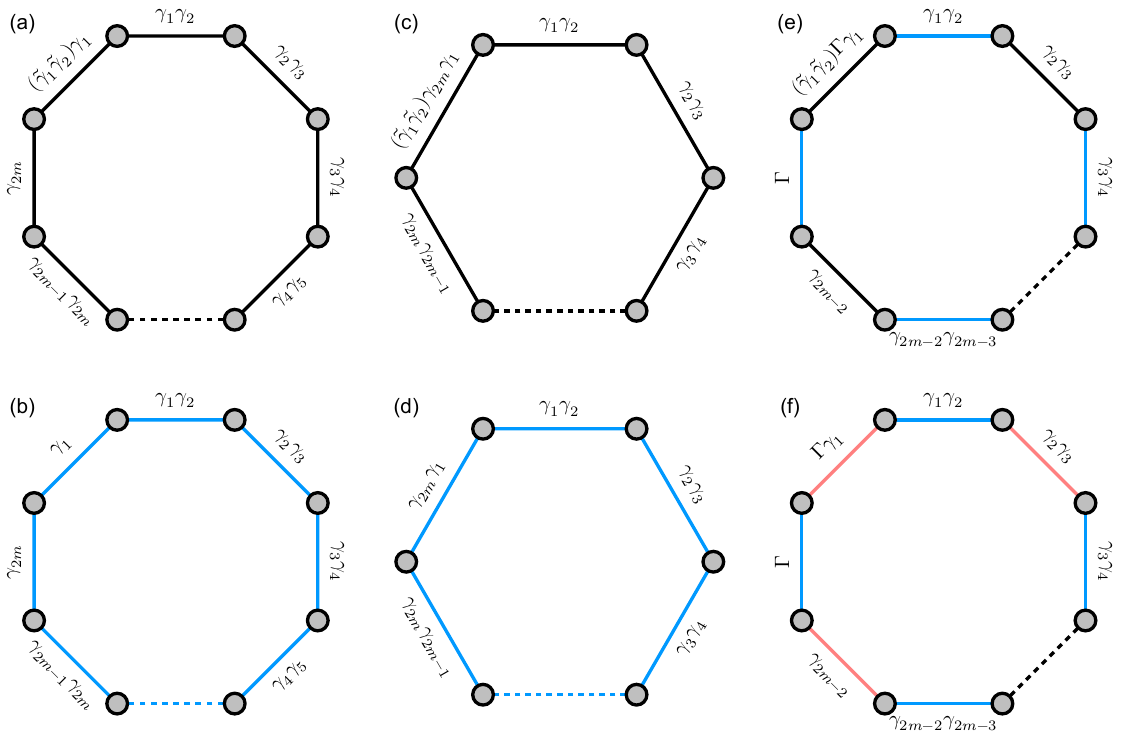}
    \caption{Representative free-fermionic mappings when the frustration graph $\graphG$ is a cycle graph, viewed in the line graph formalism on the root graph $\Delta$. Algebraic dependencies are highlighted as colorings, and we use different colors if multiple algebraic dependencies are present. Lie algebras are obtained via Theorem~\ref{thm:Free_Fermionic_Lie_Algebras}. (a) $\graphG=\graphC_{2m+1}$, $\lie{\pgens} = \so(2m+1)^{\oplus 2}$. (b) $\graphG=\graphC_{2m+1}$, $\lie{\pgens} = \so(2m+1)$ with the cycle algebraic dependency. (c) $\graphG=\graphC_{2m}$, $\lie{\pgens} = \so(2m)^{\oplus 2}$. (d) $\graphG=\graphC_{2m}$, $\lie{\pgens} = \so(2m)$ with the cycle algebraic dependency. (e) $\graphG=\graphC_{2m}$, $\lie{\pgens} = \so(2m+2)^{\oplus 2}$ with the alternating algebraic dependency. (f) $\graphG=\graphC_{2m}$, $\lie{\pgens} = \so(2m+2)$ with both cycle and alternating algebraic dependencies.}
    \label{fig:free-fermionic-cycles}
\end{figure*}

\begin{ex}[Cycles]\label{ex:free_fermionic_mappings_cycles}
Let $\vgens$ be such that $\frustration{\vgens} = \graphC_n$, $n\geq 4$, which is the line graph of itself $\Delta=\graphC_n$.
See also Fig.~\ref{fig:free-fermionic-cycles}.

(a) If $n=2m+1$ is odd and $\vgens$ has no algebraic dependencies, the $(n{-}1,1)$-free fermionic mapping is of the form
\begin{equation}
    \{\gamma_i\gamma_{i+1}\}_{i=1}^{2m-1} \cup \{\gamma_{2m}\}\cup \{(\tilde{\gamma}_1\tilde{\gamma}_2)\gamma_1\}.
\end{equation}
(b) If $n=2m+1$ is odd and the cycle symmetry is an algebraic dependency, the $(n{-}1,0)$-free fermionic mapping is of the form
\begin{equation}
    \{\gamma_i\gamma_{i+1}\}_{i=1}^{2m-1} \cup \{\gamma_{2m},\gamma_1\}.
\end{equation}

If $n=2m$ is even, one can check that $\ker(A(\graphC_{2m}))$ has dimension 2, spanned by a cycle symmetry (trivially) and a coloring which is over the even or odd vertices in $\graphG=\graphC_{2m}$ (or even and odd edges in its root graph $\Delta=\graphC_{2m}$).
Thus the even cycle has four cases, depending independently on whether
the cycle symmetry and the T-join coloring become algebraic dependencies.
In all cases, we can choose the spanning tree as the path from $1$ to $n$.

(c) We start with the algebraically independent case. It is a symmetry-adapted quadratic $(2m,1)$-free fermionic mapping with
\begin{equation}
    \{\gamma_i\gamma_{i+1}\}_{i=1}^{2m-1} \cup \{(\tilde{\gamma}_1\tilde{\gamma}_2)\gamma_{2m}\gamma_1\}.
\end{equation}
(d) Next, if the cycle symmetry is an algebraic dependency, we obtain a
$(2m,0)$-free fermionic mapping, i.e.,
\begin{equation}
    \{\gamma_i\gamma_{i+1}\}_{i=1}^{2m-1} \cup \{\gamma_{2m}\gamma_1\}.
\end{equation}
(e) If the T-join symmetry is an algebraic dependency while the cycle symmetry is not, then choosing the odd edges as T-join gives
\begin{equation}
    \{\gamma_i\gamma_{i+1}\}_{i=1}^{2m-3} \cup \{\gamma_{2m-2},\Gamma\} \cup \{(\tilde{\gamma}_1\tilde{\gamma}_2)\Gamma\gamma_1\}
\end{equation}
and choosing the even edges gives
\begin{equation}
    \{\gamma_i\gamma_{i+1}\}_{i=1}^{2m-3} \cup \{\gamma_{2m-2},(\tilde{\gamma}_1\tilde{\gamma}_2)\Gamma\} \cup \{\Gamma\gamma_1\}.
\end{equation}
(f) Finally, if both the cycle and alternating T-join symmetries are algebraic dependencies, then
\begin{equation}
    \{\gamma_i\gamma_{i+1}\}_{i=1}^{2m-3} \cup \{\gamma_{2m-2},\Gamma\} \cup \{\Gamma\gamma_1\}.
\end{equation}
\end{ex}

\begin{rem}[Comparison with Chapman and Flammia \cite{Chapman_Flammia_2020}]
\label{rem:comparison_chapman_flammia_free_fermionic_mappings}
The strictly quadratic mappings above recover the line-graph Majorana hopping
picture of \cite{Chapman_Flammia_2020}.
The mixed and parity-extended cases in
Example~\ref{ex:free_fermionic_mappings_paths}, together with the cycle and
T-join dependency cases in Example~\ref{ex:free_fermionic_mappings_cycles},
show why we keep cycle factors, T-join parity, and algebraic dependencies as
part of the mapping data.
First, cycle symmetries are retained as the explicit factors $\Cyc^{(m,q)}$,
rather than being absorbed by specializing the corresponding commuting
symmetries to scalar values.
Second, when $n_\Delta$ is even, the T-join or parity coloring is tracked
separately and may either survive as a symmetry or become an algebraic
dependency.
Third, allowing mixed and parity-extended mappings gives faithful
generator-level mappings in cases where a purely quadratic description would
not keep the T-join parity and algebraic-dependency data visible at the level
of generators.
Thus the present formulation treats multigraph or twin effects, cycle
symmetries, T-join parity, and algebraic dependencies uniformly.
This bookkeeping is used later in
Theorem~\ref{thm:Free_Fermionic_Lie_Algebras}, where the odd, even, and
exceptional free-fermionic cases are separated by precisely these cycle and
T-join or algebraic-dependency data.
\end{rem}

\subsection{Adapted Majorana Bases for Free-Fermionic Mappings}\label{sec:adapted_majorana_bases_free_fermionic_mappings}

We now show how to turn the faithful mappings produced by Proposition~\ref{prop:free_fermionic_mapping_alg_dep} into a free-fermionic mapping over all of $\PP_n\supseteq\pgens = \isolong{\vgens}$.
Namely, consider:
\begin{enumerate}
    \item The $(n_\Delta,q)$ mapping for the even type, which lies in $\PP_{n'}$ with $n'\geq n_\Delta/2+q$;
    \item The $(n_\Delta{-}1,q)$ mapping for the odd type (odd $n_\Delta$), which lies in $\PP_{n'}$ with $n'\geq (n_\Delta-1)/2+q$;
    \item The $(n_\Delta{-}2,q)$ mapping for the exceptional type, which lies in $\PP_{n'}$ with $n'\geq (n_\Delta-2)/2+q$.
\end{enumerate}
Since the mappings are faithful, they preserve both the commutation
relations and the algebraic dependencies of $\vgens$.
Thus the rank and radical dimension of $\vgens$ agree with those of the
corresponding free-fermionic model, yielding
\begin{enumerate}
\item $\rank(\vgens) = n_\Delta-2$, $\nullity(\vgens) = q+1$;
\item $\rank(\vgens) = n_\Delta-1$, $\nullity(\vgens) = q$;
\item $\rank(\vgens) = n_\Delta-2$, $\nullity(\vgens) = q$.
\end{enumerate}
Thus, in each of the three cases, the lower bound on the
free-fermionic ambient size $n'$ agrees with the lower bound on the original
Pauli ambient size $n$, namely
$n\geq \rank(\vgens)/2+\nullity(\vgens)$:
\begin{enumerate}
\item $n\geq (n_\Delta-2)/2 + q+1 = n_\Delta/2+q$;
\item $n\geq (n_\Delta-1)/2+q$;
\item $n\geq (n_\Delta-2)/2+q$.
\end{enumerate}

\begin{figure*}
    \centering
    \includegraphics[width=0.8\linewidth]{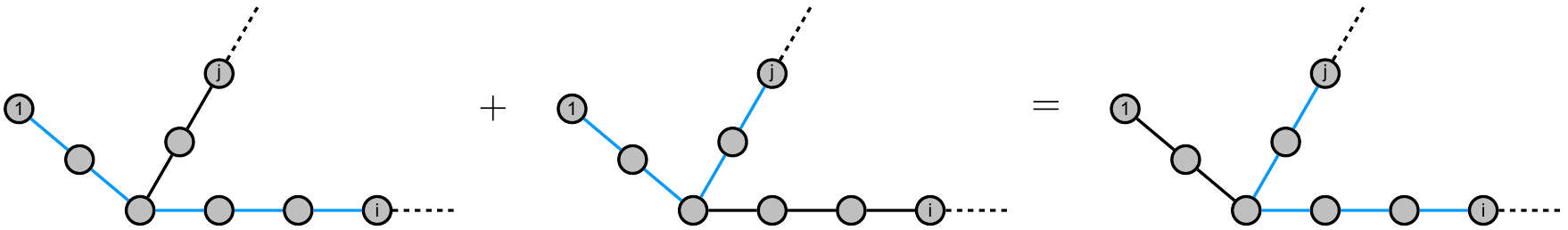}
    \caption{The sum of two path edge colorings on a multigraph which
    start at the same vertex $1$ cancels their common initial segment and gives
    the path edge coloring between the vertices $i$ and $j$.}
    \label{fig:sum-of-paths}
\end{figure*}

\begin{figure*}
    \centering
    \includegraphics{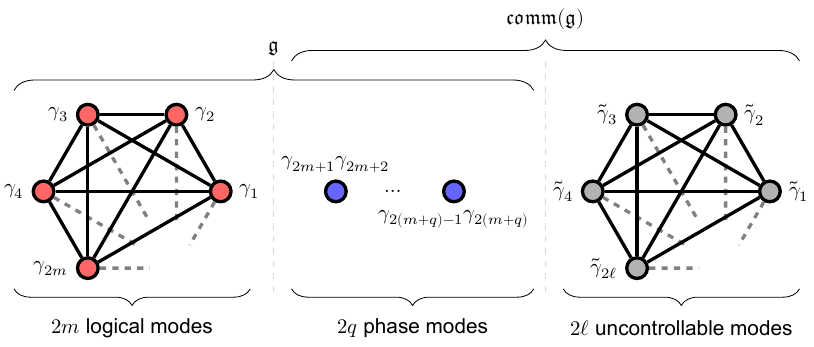}
    \caption{Canonical Majorana-mode partition for a line-graph Pauli Lie algebra
    $\lieg=\lie{\pgens}$ and its commutant $\commutant(\lieg)$.
    As indicated by the upper braces, $\lieg$ is supported on the logical and phase modes,
    while $\commutant(\lieg)$ is supported on the phase and uncontrollable modes.
    The lower braces show the corresponding mode counts: $2m$ logical modes,
    $2q$ phase modes, and $2\ell$ uncontrollable modes.
    }
    \label{fig:partition-Majorana}
\end{figure*}

Thus, for a line graph generating set in $\PP_n$, the faithful
free-fermionic mapping can be realized in the same $n$-qubit Pauli space.
This is what allows us to extend the mapped Majorana modes to a Majorana basis
on all of $\PP_n$.
We use the following terminology for the resulting adapted Majorana
basis, summarized later in Fig.~\ref{fig:partition-Majorana}.
The first $2m$ Majorana modes are called \emph{logical} modes because they
carry the free-fermionic image of the generators.
The next $2q$ modes are called \emph{phase} modes because their quadratic
pairs encode the surviving cycle symmetries.
The remaining $2\ell$ modes will be called \emph{uncontrollable} modes; they
represent $V^\perp/\rad(V)$ and do not enter the generating set itself.
We first record the existence statement in a form that only uses the free-fermionic mapping and the abstract extension of partial Majorana systems from Lemma~\ref{lem:partial_majorana_system_extension}.
\begin{prop}[Describing the adapted Majorana basis for a free-fermionic mapping]\label{prop:full_Fn_PPn_Majorana_basis_free_fermionic_mapping}
Let $\pgens=\isolong{\vgens}\subseteq\PP_n$ be a Pauli generating set whose connected frustration graph is a line graph, and let
$V=\Span[\F_2]{\vgens}$.
Assume that $\pgens$ has one of the faithful free-fermionic mappings in
Proposition~\ref{prop:free_fermionic_mapping_alg_dep}, written with $2m$ logical
Majorana modes and $q$ cycle symmetries, so that $n=\ell+m+q$.
Then there is a Majorana basis
$\{\mu_i\}_{i=1}^{2n}$ for $\Fn$ such that the first $2m$ modes are the logical
Majorana modes of the mapping and the next $2q$ modes are the phase modes.
Moreover:
\begin{enumerate}
    \item $\radorbit(\vgens)$ is spanned by
    $\{\mu_{2m+2s-1}+\mu_{2m+2s}\}_{s=1}^q$.
    \item In the even case, the additional
    radical vector can be chosen as
    $\majprod=\sum_{i=1}^{2m}\mu_i$.
    \item In the even case,
    $V^\perp/\rad(V)$ can be represented by
    $\{\mu_s\}_{s=2n-2\ell+1}^{2n}$.
    \item In the odd and exceptional cases, $V^\perp/\rad(V)$ can be represented by
    $\{\majprod+\mu_s\}_{s=2n-2\ell+1}^{2n}$.
\end{enumerate}
\end{prop}
\begin{proof}
By Proposition~\ref{prop:free_fermionic_mapping_alg_dep} and the rank and nullity count
above, the faithful free-fermionic mapping can be realized inside the same
$n$-qubit Pauli space.
By definition of a free-fermionic mapping, the mapped generators have
the same frustration graph as $\pgens$, hence the same commutation relations.
Since the mapping is faithful as introduced below
Definition~\ref{defn:free_fermionic_mappings}, the same colorings give
algebraic dependencies for the mapped generators and for $\pgens$.
Thus the logical Majorana modes and the $q$ phase-mode
pairs realize $V$ together with its orbit radical.
The phase-mode pairs give the displayed basis of $\radorbit(\vgens)$.
In the even case, the extra radical vector
is the total logical Majorana parity $\majprod$; in the exceptional case, this parity vector is projected
out, as in Proposition~\ref{prop:free_fermionic_mapping_alg_dep}.

It remains to complete the partial Majorana system to a basis of $\Fn$.
The quotient $V^\perp/\rad(V)$ has dimension $2\ell$.
After choosing representatives for a Majorana system on this quotient, the
partial Majorana system extends to a full Majorana basis of $\Fn$ by
Lemma~\ref{lem:partial_majorana_system_extension}.
In the even case these vectors may be chosen
as the final $2\ell$ Majorana modes.
In the odd and exceptional even cases, the free-fermionic image of the
generators contains linear Majoranas, so the final $2\ell$ Majorana modes themselves do not
represent $V^\perp/\rad(V)$; adding $\majprod$ gives the displayed
representatives instead.
\end{proof}

For later use it is helpful to keep track of how such a basis can be chosen from the graph data.
The next lemma gives an algorithmic refinement of Proposition~\ref{prop:full_Fn_PPn_Majorana_basis_free_fermionic_mapping}: after one initial linear choice, the spanning tree determines the logical Majorana modes, the fundamental cycles determine the phase modes, and the remaining modes are obtained by extending the resulting partial Majorana system.
\begin{lem}[Constructive adapted Majorana basis from a spanning tree]\label{lem:spanning_tree_adapted_majorana_basis_construction_teal}
In the setting of Proposition~\ref{prop:full_Fn_PPn_Majorana_basis_free_fermionic_mapping}, fix a spanning tree $\Delta_T$ of the root multigraph $\Delta$ used in the free-fermionic mapping.
Then the adapted Majorana basis in Proposition~\ref{prop:full_Fn_PPn_Majorana_basis_free_fermionic_mapping} can be chosen by the following finite procedure:
\begin{enumerate}
    \item choose an initial logical Majorana vector $\mu_1$ by solving
    the linear constraints imposed by the tree generators, the orbit
    radical and representatives of $V^\perp/\rad(V)$;
    \item define the remaining logical Majorana vectors by
    $\mu_i=v_{\pathvar_i}+\mu_1$, where $\pathvar_i$ is the unique path in $\Delta_T$ from the root
    vertex to $i$;
    \item for each fundamental cycle vector $u_s$, choose a phase-mode
    pair $\tilde{\mu}_{2s-1},\tilde{\mu}_{2s}$ with
    $\tilde{\mu}_{2s-1}+\tilde{\mu}_{2s}=u_s$ by solving the prescribed
    pairing constraints;
    \item choose representatives for $V^\perp/\rad(V)$ and, in the odd
    and exceptional even cases, use $\majprod+\mu_s$ as the corresponding
    representatives.
\end{enumerate}
The resulting basis has the four properties listed in Proposition~\ref{prop:full_Fn_PPn_Majorana_basis_free_fermionic_mapping}.
\end{lem}
\begin{proof}
\emph{Setup.}
We write the construction explicitly.
The common input is the root multigraph $\Delta$, a spanning tree $\Delta_T$,
and the Pauli vectors $v_{\edgea}\in\vgens$ attached to the edges
$\edgea\in\edges(\Delta)$.
Every choice below is made by solving a finite linear system over $\F_2$.
Thus the construction is algorithmic once the spanning tree, the fundamental cycle basis, and representative bases for the relevant quotients have been fixed.
The consistency of these systems follows from the rank and nullity count preceding Proposition~\ref{prop:full_Fn_PPn_Majorana_basis_free_fermionic_mapping}, together with the extension statement Lemma~\ref{lem:partial_majorana_system_extension} when the final quotient representatives are chosen.
We organize the proof as follows.
First, we construct the logical modes in the even case with non-trivial
T-join symmetry.
Second, we add the phase modes from the fundamental cycle symmetries.
Third, we choose the final representatives of $V^\perp/\rad(V)$.
Finally, we explain the modifications needed in the odd and exceptional even
cases.
Using the edge-coordinate realization map $\edgetovec_{\Delta,\vgens}$ from Eq.~\eqref{eq:edge_coordinate_to_label_map},
for a path $\pathvar$ in $\Delta_T$, let $x_{\pathvar}\in\F_2^{\edges(\Delta)}$ be
the edge indicator vector of $\pathvar$, i.e.,
$(x_{\pathvar})_{\edgea}=1$ exactly for the edges
$\edgea\in\pathvar$, and set
\[
    v_{\pathvar}:=\edgetovec_{\Delta,\vgens}(x_{\pathvar}).
\]
In the linear systems below, $\vecbas{V^\perp/\rad(V)}$ denotes any
chosen set of representatives in $V^\perp$ whose cosets under the quotient map
from $V^\perp$ to $V^\perp/\rad(V)$ form a basis of $V^\perp/\rad(V)$; a condition such as
$\symp{\vecbas{V^\perp/\rad(V)}}{\mu_1}=0$ means that $\mu_1$ has zero
symplectic product with each vector in this chosen set.

\emph{Logical modes in the even case.}
First suppose that $n_\Delta$ is even and the T-join coloring is a non-trivial symmetry, so that Proposition~\ref{prop:free_fermionic_mapping_alg_dep} gives the $(n_\Delta,q)$ mapping.
Choose a root vertex $1$.
The first unknown is $\mu_1\in\Fn$.
We choose it as any solution of
\[
    \begin{dcases}
        \symp{v_{\edgea}}{\mu_1}=M(\Delta)_{1,\edgea}, & \text{for } \edgea\in\edges(\Delta_T),\\
        \symp{\vecbas{\radorbit(\vgens)}}{\mu_1}=0,\\
        \symp{\vecbas{V^\perp/\rad(V)}}{\mu_1}=0.
    \end{dcases}
\]
The rank and nullity count preceding Proposition~\ref{prop:full_Fn_PPn_Majorana_basis_free_fermionic_mapping} shows that the set of solutions is an affine translate of $\rad(V)$.
Equivalently, the homogeneous solutions are exactly the allowed changes
of $\mu_1$ by vectors in $\rad(V)$.
In this construction, the relevant radical vectors are generated by the
cycle-symmetry vectors spanning $\radorbit(\vgens)$ and, in the even case under
consideration, by the T-join vector $\majprod$.
Thus changing $\mu_1$ by such a vector only changes the corresponding Majorana
operator by one of these symmetry factors, so the quadratic Majorana products
representing the tree generators are unchanged up to the symmetries encoded in
$\rad(V)$.

Once $\mu_1$ has been chosen, the tree determines all remaining logical
Majorana vectors.
For every vertex $i$ of $\Delta$, let $\pathvar_i$ be the unique path in $\Delta_T$ from $1$ to $i$ and define
\[
    \mu_i:=v_{\pathvar_i}+\mu_1 .
\]
Then $v_{\edgea}=\mu_i+\mu_j$ for every tree edge
$\edgea=\{i,j\}$.
Moreover, Lemma~\ref{lem:symplectic_on_root_multigraphs}\ref{lem:symplectic_on_root_multigraphs:b},\ref{lem:symplectic_on_root_multigraphs:e} identifies
$\symp{v_{\pathvar_i}}{v_{\pathvar_j}}$ with the dot product of the endpoint vectors of
$\pathvar_i$ and $\pathvar_j$.
Since all paths start at $1$, this dot product is $1+\delta_{ij}$, and therefore
\[
    \symp{\mu_i}{\mu_j}=1+\delta_{ij}
    \quad \text{for } i,j\in\vertices(\Delta).
\]
Thus the path construction gives the logical Majorana modes, and the quadratic products along the tree recover the tree generators.
This argument can also be interpreted as path-sum identity as illustrated in Fig.~\ref{fig:sum-of-paths}.
This proves the logical-mode part of the construction in the even case.

\emph{Phase modes.}
We now add the cycle-symmetry modes.
Let $u_1,\ldots,u_q$ be the basis of $\radorbit(\vgens)$ given by the fundamental cycles of $\Delta$ relative to $\Delta_T$.
For each $s\in\{1,\ldots,q\}$, solve for $\tilde{\mu}_{2s-1}\in\Fn$ and set
\[
    \tilde{\mu}_{2s}:=u_s+\tilde{\mu}_{2s-1}.
\]
The first pair $\{\tilde{\mu}_1, \tilde{\mu}_2\}$ is obtained from the system
\[
    \begin{dcases}
        \symp{\mu_i}{\tilde{\mu}_1}=1 & \text{for } i\in\vertices(\Delta),\\
        \symp{u_j}{\tilde{\mu}_1}=\delta_{j1} & \text{for } 1\leq j\leq q,\\
        \symp{\vecbas{V^\perp/\rad(V)}}{\tilde{\mu}_1}=0.
    \end{dcases}
\]
Inductively, after $\tilde{\mu}_1,\ldots,\tilde{\mu}_{2s-2}$ have been chosen, choose $\tilde{\mu}_{2s-1}$ from
\[
    \begin{dcases}
        \symp{\mu_i}{\tilde{\mu}_{2s-1}}=1 & \text{for } i\in\vertices(\Delta),\\
        \symp{\tilde{\mu}_j}{\tilde{\mu}_{2s-1}}=1 & \text{for } 1\leq j<2s-1,\\
        \symp{u_j}{\tilde{\mu}_{2s-1}}=\delta_{sj} & \text{for } s\leq j\leq q,\\
        \symp{\vecbas{V^\perp/\rad(V)}}{\tilde{\mu}_{2s-1}}=0.
    \end{dcases}
\]
The consistency of this system is the same inductive extension problem for the already constructed partial Majorana system.
The equations specify the pairings with all previously chosen Majorana vectors and force
$\tilde{\mu}_{2s-1}+\tilde{\mu}_{2s}=u_s$.
Consequently, after relabelling the $\tilde{\mu}$'s as the phase modes, we obtain
\[
    \radorbit(\vgens)
    =
    \SpanS[\F_2]{\{\mu_{2m+2s-1}+\mu_{2m+2s}\}_{s=1}^q}.
\]
Thus the phase-mode pairs realize precisely the orbit radical coming
from the surviving cycle symmetries.
In the even case, the remaining radical vector comes the T-join
\[
    \majprod=\sum_{i=1}^{2m}\mu_i .
\]
This proves the radical statements for the even case.

\emph{Uncontrollable modes.}
It remains to choose the final $2\ell$ modes, which will be the uncontrollable modes.
First compute a Majorana basis
$\bar{\eta}_1,\ldots,\bar{\eta}_{2\ell}$ of the non-degenerate quotient
$V^\perp/\rad(V)$ by applying the symplectic Gram--Schmidt procedure in the quotient.
Then choose representatives $\eta_a\in V^\perp$ of the classes $\bar{\eta}_a$.
We replace these representatives by vectors of the form
\[
    \mu_{2m+2q+a}=\eta_a+r_a\quad \text{for } r_a\in\rad(V),
\]
where the radical correction $r_a$ is chosen by solving the finite linear system
\[
    \begin{dcases}
        \symp{\eta_a+r_a}{\mu_b}=1 & \text{for } 1\leq b\leq 2m+2q,\\
        \symp{\eta_a+r_a}{\mu_{2m+2q+c}}=1+\delta_{ac} & \text{for } 1\leq c<a.
    \end{dcases}
\]
These equations say exactly that the new representatives keep their prescribed quotient classes and extend the already constructed partial Majorana system.
Lemma~\ref{lem:partial_majorana_system_extension} guarantees that the required radical corrections exist; algorithmically, this final step is again just linear algebra over $\F_2$.
In the present even case with non-trivial T-join symmetry, these final modes themselves represent $V^\perp/\rad(V)$.
This completes the constructive adapted Majorana basis in the even case
with non-trivial T-join symmetry.

\emph{Odd case.}
We now describe the two modifications needed in the other cases.
In the odd case, Proposition~\ref{prop:free_fermionic_mapping_alg_dep} uses the $(n_\Delta{-}1,q)$ mapping.
Choose an edge $\edgea\in\Delta_T$ incident to the vertex $n_\Delta$ with deleted Majorana $\mu_{n_\Delta}$ and define the first linear Majorana vector $\mu_i:=v_{\edgea}$ for the endpoint $i\in \edgea\setminus\{n_\Delta\}$ (such an $\edgea$ always exists since $\Delta$ is connected).
Use this endpoint as the root in the tree with the deleted vertex removed.
For every remaining vertex $j$, let $\pathvar_j$ be the unique path from this root to $j$ and set $\mu_j:=v_{\pathvar_j}+\mu_i$.
The same application of Lemma~\ref{lem:symplectic_on_root_multigraphs}\ref{lem:symplectic_on_root_multigraphs:b},\ref{lem:symplectic_on_root_multigraphs:e} verifies the required Majorana pairings and recovers the tree generators.
The cycle-symmetry systems above are unchanged.
Since the logical part now contains linear Majoranas, the final $2\ell$ Majorana modes cannot themselves represent $V^\perp/\rad(V)$.
Instead, if $\mu_s$ denotes one of the final modes, the representative in $V^\perp/\rad(V)$ is
\[
    \majprod+\mu_s .
\]
Thus the odd case differs only in the initial logical vector and in the
choice of representatives for $V^\perp/\rad(V)$; the phase-mode construction is
the same.

\emph{Exceptional even case.}
Finally consider the exceptional case, case where the T-join coloring is an algebraic dependency.
Here Proposition~\ref{prop:free_fermionic_mapping_alg_dep} uses the $(n_\Delta{-}2,q)$ mapping.
If there is an edge $\edgea\in\Delta_T$ with
$n_\Delta-1\in \edgea$ and $n_\Delta\notin \edgea$, choose
$\mu_i:=v_{\edgea}$ for the other endpoint
$i\in \edgea\setminus\{n_\Delta-1\}$ and continue with paths in the remaining tree.
If no such edge exists, then $n_\Delta-1$ is connected to $n_\Delta$ by an edge with vector $v_{n_\Delta-1,n_\Delta}$; choose an edge $\edgea\in\Delta_T$ incident to $n_\Delta$ but not to $n_\Delta-1$ and set
\[
    \mu_i:=v_{\edgea}+v_{n_\Delta-1,n_\Delta}
\]
for the other endpoint of $\edgea$.
After this initial choice, use the chosen endpoint as root and propagate along paths in the tree with the two dependent T-join vertices removed.
The same path construction produces the remaining logical modes, the same cycle-symmetry systems produce the phase modes, and the representatives of $V^\perp/\rad(V)$ are again $\majprod+\mu_s$ for the final modes $\mu_s$.
Thus the exceptional even case has the same structure as the odd case
after the appropriate initial logical vector has been chosen.
Combining the even, odd, and exceptional even constructions gives the adapted
Majorana basis with the four properties listed in
Proposition~\ref{prop:full_Fn_PPn_Majorana_basis_free_fermionic_mapping}.
\end{proof}

In the even case, with respect to Proposition~\ref{prop:full_Fn_PPn_Majorana_basis_free_fermionic_mapping}, we can choose a decomposition of $\Fn$ as $V\oplus V^\perp/\rad(V) \oplus W$ which generalizes Eq.~\eqref{eq:full_Pn_Pauli_decomposition_single_qubit_basis} to the Majorana setting.
Setting $n=\ell+m+q$ and $\majprod=\sum_{i=1}^{2m}\mu_i$, we obtain
\begin{subequations}\label{eq:full_Fn_decomposition_Majorana_even_case}
    \begin{align}
        \Fn &= V \oplus V^\perp/\rad(V) \oplus W\\
        \text{where } \vecbas{V/\rad(V)} &= \{\mu_i + \mu_{2m}\}_{i=1}^{2m-1}, \\
        \vecbas{\rad(V)} &= \{\majprod\}\cup\{\mu_{2j-1}+\mu_{2j}\}_{j=2m+1}^{2m+q},\\
        \vecbas{V^\perp/\rad(V)} &= \{\mu_s\}_{s=n-2\ell+1}^{2n},\\
        \vecbas{W} &= \{\mu_{2m}\}\cup\{\mu_{2j-1}\}_{j=2m+1}^{2m+q}.
    \end{align}
\end{subequations}
Correspondingly, over the Pauli strings $\PP_n$ we have the following representatives, up to phases:
\begin{subequations}\label{eq:full_Pn_decomposition_Majorana_even_case}
    \begin{align}
        \bas{\matalg/\ZZ(\matalg)}
        &\simeq \qty{\prod_{i=1}^{2m-1}(\gamma_i\gamma_{2m})^{\alpha_i} }_{\alpha\in\F_2^{2m-1}}, \\
        \bas{\ZZ(\matalg)}
        &\simeq \qty{\Gamma^{\delta_r}\prod_{s=1}^{q}(\gamma_{2m+2s-1}\gamma_{2m+2s})^{\delta_s}}_{\delta\in\F_2^r}, \\
        \bas{\commalg/\ZZ(\commalg)}
        &\simeq \qty{ \prod_{2n-s+1=1}^{2\ell}\gamma_s^{\beta_s} }_{\beta\in\F_2^{2\ell}}, \\
        \bas{\isolong{W}}
        &\simeq \qty{ \gamma_{2m}^{\delta_r}\prod_{s=1}^q\gamma_{2m+2s-1}^{\delta_s} }_{\delta\in\F_2^r}.
    \end{align}
\end{subequations}

In the odd and exceptional case, we can instead choose the following:
\begin{subequations}\label{eq:full_Fn_decomposition_Majorana_odd_exceptional_case}
    \begin{align}
        \Fn &= V \oplus \tilde{V} \oplus W\\
        \text{where } \vecbas{V/\rad(V)} &= \{\mu_i\}_{i=1}^{2m},\\
        \vecbas{\rad(V)} &= \{\mu_{2j-1}+\mu_{2j}\}_{j=2m+1}^{2m+q},\\
        \vecbas{V^\perp/\rad(V)} &= \{\majprod + \mu_s\}_{s=n-2\ell+1}^{2n},\\
        \vecbas{W} &= \{\mu_{2j-1}\}_{j=2m+1}^{2m+q},
    \end{align}
\end{subequations}
which corresponds to the following phase-normalized representatives in $\PP_n$:
\begin{subequations}\label{eq:full_Pn_decomposition_Majorana_odd_exceptional_case}
    \begin{align}
        \bas{\matalg/\ZZ(\matalg)}
        &\simeq \qty{ \prod_{i=1}^{2m}\gamma_i^{\alpha_i}}_{\alpha\in\F_2^{2m}}, \\
        \bas{\ZZ(\matalg)}
        &\simeq \qty{\prod_{s=1}^{q}(\gamma_{2m+2s-1}\gamma_{2m+2s})^{\delta_s}}_{\delta\in\F_2^q}, \\
        \bas{\commalg/\ZZ(\commalg)}
        &\simeq \qty{ \prod_{2n-s+1=1}^{2\ell}(\Gamma\gamma_s)^{\beta_s} }_{\beta\in\F_2^{2\ell}}, \\
        \bas{\isolong{W}}
        &\simeq \qty{ \prod_{s=1}^q\gamma_{2m+2s-1}^{\delta_s} }_{\delta\in\F_2^q}.
    \end{align}
\end{subequations}

Notice that the supplementary subspace $W$ in the Majorana decompositions above
is not chosen to be totally isotropic in the same way as in the symplectic-basis
decomposition Eq.~\eqref{eq:full_Fn_decomposition_symplectic_basis}.
Instead, it is chosen to be adapted to the Majorana basis; in particular, it may
have nonzero internal symplectic pairings, and in the even-dimensional cases it
can be non-degenerate.
In contrast, Eq.~\eqref{eq:full_Fn_decomposition_symplectic_basis} chooses
$W$ totally isotropic as $W=\rad(W)$.

This completes the construction of the adapted free-fermionic coordinates used
below.

In all cases, as in the Pauli picture, we divide the Majorana modes into three parts (see Fig.~\ref{fig:partition-Majorana}):
\begin{enumerate}
    \item The first $2m$ modes are the \emph{logical} Majorana modes
    \item The last $2\ell$ modes are the \emph{uncontrollable} Majorana modes
    \item The middle $2q$ modes are the \emph{phase} Majorana modes
\end{enumerate}
Notice that in the even case the generators described by logical Majorana modes do \emph{not} generate (via products) the full set of Majorana strings over the first $2m$ modes, but only the even Majorana strings.
On the other hand, the generators in the odd and exceptional case generate the logical $2m$ modes, much like in the Pauli case the generators produce the Pauli strings over the logical $m$ qubits.
This follows from the fact that we do not identify the parity symmetry for the even case as a phase symmetry,
given its special role for the orbit radical and orbits (as we will see in Section~\ref{sec:Orbits_Full_Space_Fn_Free_Fermionic_Case}).
Additionally, the phase pairs and linear uncontrollable modes generate the commutant only for the even case, whereas in the odd exceptional cases the commutant is generated by the phase pairs and $\Gamma\tilde{\gamma}_i$ (as shown in Proposition~\ref{prop:full_Fn_PPn_Majorana_basis_free_fermionic_mapping}).
Alternatively, one can view the uncontrollable modes as fermionic modes for another fermionic species, hence one can choose Majorana operators $\tilde{\gamma}_i$ as generators for $\commalg/\ZZ(\commalg)$ in both cases.

\section{Canonical Free-Fermionic Orbits, Lie Algebras, and Transvection Groups}\label{sec:canonical_free_fermionic_structures}

The adapted Majorana bases constructed above reduce line-graph generating sets
to canonical free-fermionic representatives.
We now use these representatives to describe the corresponding orbit
structure, Pauli Lie algebras, Majorana labels, and transvection groups.
The objectives of this section are:
\begin{enumerate}
    \item Starting from the canonical cases $\graphP_{k,n_1}$, describe the colorings which determine the orbit $\tvgroup{\vgens}\cdot\vgens$, and hence the Lie algebra basis, as well as all orbits in $V=\Span{\vgens}$ (Subsection~\ref{sec:Orbits_Canonical_Blown_Up_Path_graph}).
    \item Assign explicit \emph{Majorana} labels to classify the representations of the Lie algebras (Subsection~\ref{sec:Majorana_Strings_Free_Fermionic_Lie_Algebras})
    and their orbits (Subsection~\ref{sec:Orbits_Majorana_Formalism}) for arbitrary generating sets with line graphs as frustration graphs.
    \item Describe the generators and orbit structure of the transvection groups associated with line graphs, together with their relation to symmetric groups and transpositions (Subsection~\ref{sec:Transvection_Group_Line_Graph_Symmetric_Group}).
\end{enumerate}

\subsection{Orbits as Colorings in the Canonical Case}\label{sec:Orbits_Canonical_Blown_Up_Path_graph}

Given the ability to differentiate line graphs from other cases, it now remains to describe precisely the orbits, Lie algebras, and transvection groups.
We first describe the orbit $\tvgroup{\vgens}\cdot\vgens$.
Unlike the cases with invariant quadratic or bilinear forms, there is no clear symmetry criterion, and instead we rely mostly on the graph-theoretic formalism.
This also points to an important distinction between line graphs and other graphs: the latter may be described purely in terms of symmetry properties such as the orthogonal complement and invariant quadratic forms; the former admits no such description, and always relies on \emph{some} knowledge of the graph.
This is also reflected in the $\calE_6$-criterion for identifying free-fermionicness, which is based on the graph-theoretic formalism.

We will deal first with the minimal canonical cases and then, with the aid of the line graph formalism and free fermionic mapping, generalize to arbitrary generating sets with line graphs as frustration graphs.

Specifically, given Proposition~\ref{prop:Limits_Lie_Algebraic_Dependencies_canonical}, we now have only to deal with the following cases in the presence of minimal generating sets (see Corollary~\ref{cor:minimal_canonical_sets}):
\begin{enumerate}
    \item The algebraically independent case with $\graphP_{2m-1,n_1}$ and $m\geq 1$;
    \item The algebraically independent case with $\graphP_{2m,n_1}$ and $m\geq 2$;
    \item The algebraically \emph{dependent} case with $\graphP_{2m,n_1}$ and $m\geq 2$.
\end{enumerate}
Recall Definition~\ref{def:free-fermionic-types}, we refer to (a) as \emph{odd} type, (b) as \emph{even} type and (c) as \emph{exceptional} type.

By Proposition~\ref{prop:free_fermionic_mapping_alg_dep}, a generating set of even type admits a quadratic $(n_\Delta,q)$-free fermionic mapping, an odd type admits a mixed $(n_\Delta{-}1,q)$-mapping, and an exceptional type admits a parity-extended $(n_\Delta{-}2,q)$-mapping.
By Proposition~\ref{prop:full_Fn_PPn_Majorana_basis_free_fermionic_mapping}, if $\pgens\subseteq\PP_n$, there is also a Majorana basis for $\PP_n$ which respects this mapping.

As in Corollary~\ref{cor:minimal_canonical_sets}, we denote the sets as $\pgens(\graphG) = \isolong{\vgens(\graphG)}$ for the algebraically independent cases and $\pgens^D(\graphP_{2m,n_1}) = \isolong{\vgens(\graphP_{2m,n_1})}$ for the algebraically dependent one, with specified algebraic dependency.
As such, we follow a similar road as that in \cite{Janssen_1983} or \cite{Aguilar_Cichy_Eisert_Bittel_2024}.
In this section, we shall work purely with the generators, i.e., $\{a_i\}_{i=1}^{k+1}\cup\{b_j\}_{j=2}^{n_1}$ in the notation of Fig.~\ref{fig:humphries_classes} for $\vgens(\graphP_{k,n_1})$.
Here $k$ is the generic blown-up path parameter; in the canonical cases above it specializes to $k=2m-1$ for odd type and to $k=2m$ for even and exceptional type.
We will also occasionally use the redundant notation $b_1=a_{k+1}$.

\begin{figure}
    \centering
    \includegraphics[width=\linewidth]{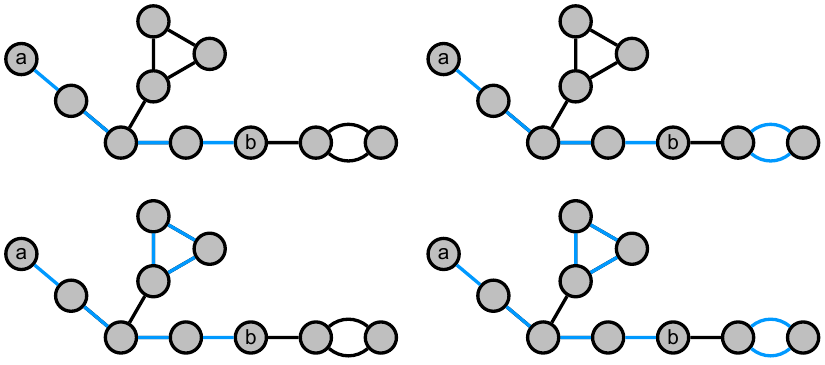}
    \caption{Valid colorings of a line graph $\graphG= L(\Delta)$ viewed as edge colorings on its root multigraph $\Delta$.
    Each valid edge coloring on $\Delta$ is a path coloring plus an
    arbitrary cycle coloring.}
    \label{fig:orbits-on-root-graph}
\end{figure}

We will separate the coloring on the path from the coloring on the
length-one legs.
In the canonical labels, recall that the orbit radical coincides with subspace spanned by an even number of length-one legs (Lemma~\ref{lem:orbit:radical:canonical_cases}):
\begin{equation}\label{eq:def:even_leg_coloring_subspace}
    \radorbit(\vgens)
    =
   \Big\{
    \sum_{s=1}^{n_1}\alpha_s b_s
    \text{ for }
    \alpha_s\in\F_2 \text{ with } \sum_{s=1}^{n_1}\alpha_s=0
    \Big\}.
\end{equation}
Thus two colorings on the length-one legs differ by an element of
$\radorbit(\vgens)$ precisely when their total parities agree.

In the exceptional and algebraically dependent case $\vgens^D(\graphP_{2m,n_1})$, we also have $a_1 + a_3 + \cdots + a_{2m+1} = 0$ as the unique algebraic dependency and we can fix $a_{2m+1} = \sum_{i=1}^m a_{2i-1}$ as linearly dependent on the rest where needed.
This suffices to find the orbits in $V=\Span{\vgens}$ and the Pauli Lie algebra elements as products of the generators, as well as the restricted transvection group $\tvgroup{\vgens}|_V$.

For the even and odd types we describe the orbit of the generators $\tvgroup{\vgens}\cdot\vgens$ or Pauli basis $\bas{\lieg}$ by the realizations of \emph{valid} colorings, which generalize Lemma~\ref{lem:local_path_orbit_interval} to take into account the legs of length one, and is a direct consequence of Lemmas~\ref{lem:root_path_vectors_are_orbit_vectors} and \ref{lem:line_graph_orbit_boundary} for arbitrary line graphs:
\begin{prop}[\cite{Janssen_1983,Aguilar_Cichy_Eisert_Bittel_2024}]\label{prop:valid_colorings_even_odd_cases_alg_ind}
Let $\vgens$ be linearly independent and assume that its connected
frustration graph is a line graph.
We obtain descriptions of the generator orbit:
\begin{enumerate}
    \item\label{prop:valid_colorings_even_odd_cases_alg_ind:line-graph}
    If $\frustration{\vgens}=\graphG=L(\Delta)$, then
    $\tvgroup{\vgens}\cdot\vgens$ consists of the realizations
    $\edgetovec(x)$ of edge colorings
    $x\in\F_2^{\edges(\Delta)}$ satisfying
    \[
        \abs{\supp(M(\Delta)x)} = 2.
    \]
    Equivalently, these are path colorings plus arbitrary cycle
    colorings (see Fig.~\ref{fig:orbits-on-root-graph}):
    \[
        x=x_{\pathvar} + x_C\qquad \text{for } x_C\in\ker(M(\Delta)).
    \]
    \item\label{prop:valid_colorings_even_odd_cases_alg_ind:blown-up-path}
    For $\vgens = \vgens(\graphP_{k,n_1})$, this specializes to the
    valid colorings for $\graphP_{k,n_1}$, namely interval colorings on the
    path, modulo the subspace
    $\radorbit(\vgens)$ from Eq.~\eqref{eq:def:even_leg_coloring_subspace}
    (see Fig.~\ref{fig:orbits-blown-up-path}).
    Using the canonical labels from Fig.~\ref{fig:humphries_classes},
\begin{align}\label{eq:orbits_blown_up_path_graph}
        \tvgroup{\vgens}\cdot\vgens &=
        \begin{aligned}[t]
        &\{ a_i + a_{i+1} + \cdots + a_j \}_{1\leq i\leq j\leq k+1} \\
        &+ \radorbit(\vgens).
        \end{aligned}
\end{align}
The number of elements in the orbit is
$$\abs{\tvgroup{\vgens}\cdot\vgens}=
\frac{(k+1)(k+2)}{2}\,2^{n_1-1}.$$
\end{enumerate}
\end{prop}

\begin{proof}
We first prove
\ref{prop:valid_colorings_even_odd_cases_alg_ind:line-graph}.
By Lemma~\ref{lem:line_graph_orbit_boundary}, the orbit $\tvgroup{\vgens}\cdot\vgens$ of a linearly independent generating set whose frustration graph $\graphG=L(\Delta)$ is a line graph contains edge colorings over the root multigraph which have precisely two vertices of odd degree.
Conversely, suppose that $x\in\F_2^{\edges(\Delta)}$ satisfies
$\abs{\supp(M(\Delta)x)}=2$, say
$M(\Delta)x=\mathbf 1_{\vva}+\mathbf 1_{\vvb}$.
Choose a path $\pathvar$ from $\vva$ to $\vvb$ in $\Delta$ and let
$x_{\pathvar}$ be its edge coloring.
Then $M(\Delta)(x+x_{\pathvar})=0$, so
$x=x_{\pathvar}+x_C$ with $x_C\in\ker(M(\Delta))$.
By Lemma~\ref{lem:root_path_vectors_are_orbit_vectors},
$\edgetovec(x_{\pathvar})$ lies in $\tvgroup{\vgens}\cdot\vgens$.
By Lemma~\ref{lem:cycle_space_inside_orbit_radical},
$\edgetovec(x_C)$ lies in $\radorbit(\vgens)$.
Hence $\edgetovec(x_{\pathvar}+x_C)$ also lies in
$\tvgroup{\vgens}\cdot\vgens$.
Thus the orbit consists exactly of the realizations of edge colorings
satisfying $\abs{\supp(M(\Delta)x)}=2$, equivalently of path colorings plus
cycle colorings.
This proves
\ref{prop:valid_colorings_even_odd_cases_alg_ind:line-graph}.

\begin{figure}
    \centering
    \includegraphics[width=\linewidth]{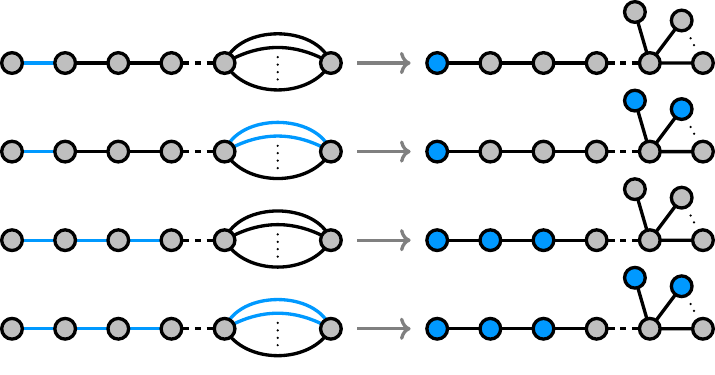}
    \caption{Valid colorings of the blown-up path graph $\graphP_{k,n_1} = L(\Delta)$ viewed as vertex colorings on $\graphP_{k,n_1}$ (right) and as edge colorings on its root multigraph $\Delta$ (left).
    As edge colorings on $\Delta$, each valid coloring is a path
    coloring plus a cycle coloring, where the cycle colorings come from the
    parallel edges.
    As vertex colorings on $\graphP_{k,n_1}$, this becomes an interval coloring
    on the path, modulo $\radorbit(\vgens)$.}
    \label{fig:orbits-blown-up-path}
\end{figure}

We now specialize to
\ref{prop:valid_colorings_even_odd_cases_alg_ind:blown-up-path}.
In the case $\graphP_{k,n_1}$ the root multigraph is a path graph over $k+2$ vertices with $n_1=q+1$ parallel edges at one end, where parallel edges provide $q$ independent cycles.
Path colorings in the root graph correspond to interval colorings over the
spine of the frustration graph.
The cycle space of this root multigraph is generated by differences of
parallel edges. Using canonical labels, these are the elements of
$\radorbit(\vgens)$ from Eq.~\eqref{eq:def:even_leg_coloring_subspace}.
Hence, the orbit is precisely that in Eq.~\eqref{eq:orbits_blown_up_path_graph}.
The path-coloring part is determined by choosing two distinct root
vertices in the path-like root graph, hence by $\binom{k+2}{2}$ choices.
Under the line-graph correspondence these choices give the intervals
$a_i+\cdots+a_j$ with $1\leq i\leq j\leq k+1$, where $i=j$ corresponds to a
single edge of the root path.
The cycle space has dimension $n_1-1$, so it contributes $2^{n_1-1}$ cycle
colorings. This gives the cardinality. This shows
\ref{prop:valid_colorings_even_odd_cases_alg_ind:blown-up-path}.
\end{proof}
As a corollary, we also get the orbit for the exceptional type by viewing the even type as an extension (see Lemma~\ref{lem:projection_extension_basic_properties}\ref{lem:projection_extension_basic_properties:e}) and then performing a projection such that $\projection(\sum_ia_{2i-1})=0$, or equivalently by setting $a_{2m+1} = \sum_{i=1}^m a_{2i-1}$.
We find that such a projection still acts injectively on the vectors in the orbits, so that the characterization as colorings remains:
\begin{prop}\label{prop:valid_colorings_orbits_line_graph_blown_up_path_graph_alg_dep_exceptional_case}
Let $\vgens = \vgens^D(\graphP_{2m,n_1})$ with $m\geq 2$
as in Corollary~\ref{cor:minimal_canonical_sets} and using the canonical
labels from Fig.~\ref{fig:humphries_classes}.
The orbit $\tvgroup{\vgens}\cdot\vgens$ consists of the realizations of colorings represented by an interval on the full spine, modulo the subspace $\radorbit(\vgens)$.
Explicitly:
\begin{align*}
        \tvgroup{\vgens}\cdot\vgens &=
         \begin{aligned}[t]
        &\{ a_i + a_{i+1} + \cdots + a_j \}_{1\leq i\leq j\leq 2m+1} \\
        &+ \radorbit(\vgens)
        \end{aligned}
\end{align*}
where the sum for $i=j$ is the single element $a_i$.
The number of elements in the orbit is
\begin{align*}
\abs{\tvgroup{\vgens}\cdot\vgens}
&=\frac{(2m+1)(2m+2)}{2}\,2^{n_1-1}\\
&=(2m+1)(m+1)\,2^{n_1-1}.
\end{align*}
\end{prop}
\begin{proof}
Denote with $\tilde{\vgens} = \vgens(\graphP_{2m,n_1})$ the unique linearly independent generating set with frustration graph $\graphP_{2m,n_1}$.
Then, $\tilde{\vgens}$ may be seen as an extension of $\vgens = \vgens^D(\graphP_{2m,n_1})$ with a projection map defined by $\projection(\tilde{\vgens})=\vgens$ and $\projection(\sum_{i=1}^{m+1}\tilde{a}_{2i-1}) = 0$.
The orbit for $\vgens$ is then the projection of the orbit of the extension $\projection(\tvgroup{\tilde{\vgens}}\cdot\tilde{\vgens})$ by Lemma~\ref{lem:projection_extension_basic_properties}\ref{lem:projection_extension_basic_properties:e}.

We show that $\projection$ is injective on this particular generator
orbit $\tvgroup{\tilde{\vgens}}\cdot\tilde{\vgens}$.
Namely, for any $\tilde{v},\tilde{v}'\in \tvgroup{\tilde{\vgens}}\cdot\tilde{\vgens}$, we have $\projection(\tilde{v})=\projection(\tilde{v}')$ if and only if $\tilde{v}+\tilde{v}' = \sum_{i=1}^{m+1}\tilde{a}_{2i-1}$ (i.e., the sum is in the kernel and non-zero).
By Proposition~\ref{prop:valid_colorings_even_odd_cases_alg_ind}, each of
$\tilde v$ and $\tilde v'$ has one connected path component, up to an
element of $\radorbit(\vgens)$.
Hence $\tilde v+\tilde v'$ has at most two connected components on the path
(the $\radorbit(\vgens)$ component is irrelevant for this comparison).
In contrast, the kernel vector
$\sum_{i=1}^{m+1}\tilde{a}_{2i-1}$ has $m+1$ singleton components on the path.
Since $m\geq2$, this gives $m+1\geq3$ path components, whereas
$\tilde v+\tilde v'$ has at most two. Hence the two colorings cannot agree.
Thus $\projection$ induces a bijection from
$\tvgroup{\tilde{\vgens}}\cdot\tilde{\vgens}$ to
$\tvgroup{\vgens}\cdot\vgens$.
Applying
Proposition~\ref{prop:valid_colorings_even_odd_cases_alg_ind}\ref{prop:valid_colorings_even_odd_cases_alg_ind:blown-up-path}
to the extension $\tilde{\vgens}=\vgens(\graphP_{2m,n_1})$ gives the
intervals $a_i+\cdots+a_j$ with $1\leq i\leq j\leq 2m+1$, together with the
elements of $\radorbit(\vgens)$.
The same bijection gives the displayed description and the stated
cardinality for the exceptional orbit.
\end{proof}
Notice that even though the orbits of the exceptional case are in bijection with those of the even case of the same size,
we still get a redundant description of the \emph{colorings}.
Namely, in the linearly dependent case, the same projected vector may
be represented either by a connected interval coloring or by the coloring
obtained from it after adding the vector
$\sum_{i=1}^{m+1} a_{2i-1}$ in the radical, together with an element of
$\radorbit(\vgens)$.
The second representative is generally disconnected; its precise number of
path components depends on the interval, but it represents the same vector in
$V$ because the kernel vector vanishes after projection.

Also, we leave the characterization of the Lie algebra generated by these elements for Subsection~\ref{sec:Majorana_Strings_Free_Fermionic_Lie_Algebras}, where we will give a Majorana labelling in all of $\Fn$ for these sets.
There is also a simple generalization of this result to \emph{any} orbit in $V$.
Compare \cite[Theorems~7.1 and~7.2]{Brown_Humphries_1986b} and \cite{Aguilar_Cichy_Eisert_Bittel_2024} for the pure vertex-coloring form of the statement, which we prove here through the line graph formalism.
\begin{lem}[Component criterion for blown-up paths]\label{lem:component_criterion_blown_up_paths}
Let $\vgens=\vgens(\graphP_{k,n_1})$
as in Corollary~\ref{cor:minimal_canonical_sets} and using the canonical
labels from Fig.~\ref{fig:humphries_classes}. Set $V=\Span{\vgens}$, and
write $\graphP_{k,n_1}=L(\Delta)$ for the root multigraph $\Delta$ from
Fig.~\ref{fig:line-graph-path}.
Let $v,v'\in V\setminus\rad(V)$. Then the following equivalent criteria hold:
\begin{enumerate}[label=(\alph*)]
    \item\emph{Edge-coloring form.}
    If $x,x'\in\F_2^{\edges(\Delta)}$ satisfy
    $v=\edgetovec_{\Delta,\vgens}(x)$ and
    $v'=\edgetovec_{\Delta,\vgens}(x')$, then $v$ and $v'$ lie in the same
    orbit under $\tvgroup{\vgens}$ if and only if the path-coloring parts of
    $x$ and $x'$ have the same number of connected components and their
    parallel-edge cycle parts differ by a cycle vector corresponding to an
    element of $\radorbit(\vgens)$ from
    Eq.~\eqref{eq:def:even_leg_coloring_subspace}.
    \item\emph{Vertex-coloring form.}
    If
    $\coloring,\coloring'\in\cspace(\graphP_{k,n_1})$ satisfy
    $v=\coltovec_{\vgens}(\coloring)$ and
    $v'=\coltovec_{\vgens}(\coloring')$, with $\coltovec_{\vgens}$ from
    Definition~\ref{def:coloring_to_vector_map}, then $v$ and $v'$ lie in the
    same orbit under $\tvgroup{\vgens}$ if and only if the interval parts of
    $\coloring$ and $\coloring'$ along the path $\graphP_{k,1}$ have the same number of connected
    components and their length-one leg colorings differ by an element of
    $\radorbit(\vgens)$ from Eq.~\eqref{eq:def:even_leg_coloring_subspace}.
\end{enumerate}
\end{lem}
\begin{proof}
We first prove (a).
The argument is the same boundary-coloring argument as in the proof of
Proposition~\ref{prop:valid_colorings_even_odd_cases_alg_ind}\ref{prop:valid_colorings_even_odd_cases_alg_ind:line-graph},
but with $2L$ boundary vertices instead of two.
Write the edge coloring $x$ as the sum of its path-coloring part and its
parallel-edge cycle part.
The cycle part lies in $\ker(M(\Delta))$, so it does not change the boundary
vertex coloring $M(\Delta)x$ on the root graph $\Delta$.
Thus $M(\Delta)x$ is the vertex coloring which is non-zero precisely on the endpoints of the connected components
of the path-coloring part (which holds by Lemma~\ref{lem:symplectic_on_root_multigraphs}\ref{lem:symplectic_on_root_multigraphs:a}).
If this path-coloring part has $L$ connected components, then
$M(\Delta)x$ has $2L$ colored vertices.
Conversely, since the underlying simple root graph is a path, the colored
vertices of $M(\Delta)x$ determine the $L$ path intervals (by Lemma~\ref{lem:symplectic_on_root_multigraphs}\ref{lem:symplectic_on_root_multigraphs:c}), up to adding cycle
colorings supported on the parallel edges.

We now need to compare this description with the transvection action.
For an edge $\edgea=\{\vva,\vvb\}$ of $\Delta$, the generator $v_{\edgea}$
changes an edge coloring $x$ by $x\mapsto x+\edgeunit{\edgea}$ exactly when
$\symp{\edgetovec_{\Delta,\vgens}(x)}{v_{\edgea}}=1$.
By Lemma~\ref{lem:symplectic_on_root_multigraphs}\ref{lem:symplectic_on_root_multigraphs:e}, this is equivalent to
saying that exactly one of $\vva$ and $\vvb$ is colored in $M(\Delta)x$, given that the symplectic product of edge colorings on $\Delta$ reduces to the usual dot product of vertex colorings on $\Delta$.
On the boundary coloring, the move therefore swaps a colored endpoint with an
adjacent uncolored endpoint.
These adjacent swaps generate all permutations of the vertices of the path-like
root graph, hence the endpoints of the connected components can be moved arbitrarily. 
Thus, the orbit of a non-radical edge coloring is determined by the
number of colored boundary vertices, equivalently by the number of connected
components of its path-coloring part.
The remaining freedom is exactly the cycle space generated by differences of
parallel edges.
By Lemma~\ref{lem:cycle_space_inside_orbit_radical}, the realized cycle
vectors lie in the orbit radical, and in the blown-up path case
Lemma~\ref{lem:orbit:radical:canonical_cases} identifies these cycle vectors with
the elements of $\radorbit(\vgens)$.
This proves (a).

We now prove (b) from (a).
Under the line-graph identification, an edge coloring
$x\in\F_2^{\edges(\Delta)}$ is the same data as a vertex coloring
$\coloring_x\in\cspace(L(\Delta))=\cspace(\graphP_{k,n_1})$, and
Eq.~\eqref{eq:edge_coordinate_to_label_map} gives
$\edgetovec_{\Delta,\vgens}(x)=\coltovec_{\vgens}(\coloring_x)$.
The path-coloring part of $x$ becomes the interval part of
$\coloring_x$ along $\graphP_{k,1}$, while the cycle colorings supported on the
parallel edges become the length-one leg colorings in
$\radorbit(\vgens)$.
Therefore the edge-coloring criterion in (a) is exactly the vertex-coloring
criterion in (b).
\end{proof}

We can state the following, which addresses the problem of orbit intersection and orbit classification for the odd and even types:
\begin{prop}[\cite{Brown_Humphries_1986b,Seven_2005}]\label{prop:orbits_subspace_even_odd_case_alg_ind}
Let $\vgens = \vgens(\graphP_{k,n_1})$
as in Corollary~\ref{cor:minimal_canonical_sets} and using the canonical
labels from Fig.~\ref{fig:humphries_classes}. Set $V=\Span{\vgens}$.
Let colorings be understood through the coloring-to-vector map
$\coltovec_{\vgens}$ from Definition~\ref{def:coloring_to_vector_map}.
Then:
\begin{enumerate}
    \item\label{prop:orbits_subspace_even_odd_case_alg_ind:criterion}
    \emph{General orbit criterion.}
    Two vectors $v,v'\in V$ are in the same orbit under
    $\tvgroup{\vgens}$ if and only if one of the following holds:
    \begin{nestedcaseenum}
        \item $v,v'\notin\rad(V)$, their unique colorings have the same
        number of components on the path, and their length-one leg colorings
        differ by an element of $\radorbit(\vgens)$;
        \item $v,v'\in\rad(V)$ and $v=v'$.
    \end{nestedcaseenum}

    \item\label{prop:orbits_subspace_even_odd_case_alg_ind:nontrivial-orbits}
    \emph{Non-trivial orbits.}
    For each
    $L\in\{1,2,\ldots,\lceil k/2\rceil\}$, the set of vectors whose
    coloring has $L$ connected components over the path, excluding the
    additional legs of length one, is a single orbit
\begin{align*}
    O_L = \{
    \begin{aligned}[t]
         & v\in V \text{ for } \frustration{\vgens,v}\text{ has } L\text{ components}\\
        &\text{in }\graphP_{k,1}\subseteq \graphP_{k,n_1}\}
    \end{aligned}
\end{align*}
and $\tvgroup{\vgens}\cdot O_L = O_L$.

    \item\label{prop:orbits_subspace_even_odd_case_alg_ind:trivial-orbits}
    \emph{Trivial orbits.}
    The remaining radical vectors are fixed pointwise by
    $\tvgroup{\vgens}$ and therefore give singleton orbits.
    If $k=2s-1$ is odd, there are $2^{n_1-1}$ such trivial orbits.
    If $k=2s$ is even, there are $2^{n_1}$ such trivial orbits.

    \item\label{prop:orbits_subspace_even_odd_case_alg_ind:exhaustion}
    \emph{Exhaustion.}
    If $k=2s-1$ is odd, the orbits in $V$ are exactly the
    $\lfloor k/2\rfloor+1 = s$ non-trivial orbits
    $\{O_L\}_{L=1}^{s}$ from
    \ref{prop:orbits_subspace_even_odd_case_alg_ind:nontrivial-orbits}
    together with the $2^{n_1-1}$ trivial orbits from
    \ref{prop:orbits_subspace_even_odd_case_alg_ind:trivial-orbits}.
    If $k=2s$ is even, the orbits in $V$ are exactly the
    $\lceil k/2\rceil = s$ non-trivial orbits
    $\{O_L\}_{L=1}^{s}$ together with the $2^{n_1}$ trivial orbits.
\end{enumerate}
\end{prop}
\begin{proof}
\emph{General orbit criterion.}
For non-radical vectors, the criterion is exactly
Lemma~\ref{lem:component_criterion_blown_up_paths}.
If $v\in\rad(V)$, then $\tau_g(v)=v$ for every generator $g\in\vgens$ because
$\symp{g}{v}=0$.
Hence every radical vector gives a singleton orbit, and a radical vector cannot
lie in the same orbit as a vector outside $\rad(V)$.
This proves
\ref{prop:orbits_subspace_even_odd_case_alg_ind:criterion}.

\emph{Non-trivial orbits.}
The non-trivial orbits are the orbits of the non-radical vectors.
By Lemma~\ref{lem:component_criterion_blown_up_paths}, they are classified by
the number $L$ of path components, modulo changes by elements of
$\radorbit(\vgens)$.
The largest possible non-radical component count is $\lceil k/2\rceil$.
Indeed, if $k=2s-1$, then the path has $2s$ vertices and the largest
component count is $s$.
If $k=2s$, then the path has $2s+1$ vertices; the coloring with $s+1$
singleton path components is the alternating radical vector from
Lemma~\ref{lem:Canonical_t_equivalent_Graph_Radical}, and hence belongs to
the trivial part.
Thus the non-trivial orbits are exactly the orbits $O_L$ with
$1\leq L\leq\lceil k/2\rceil$, proving
\ref{prop:orbits_subspace_even_odd_case_alg_ind:nontrivial-orbits}.

\emph{Trivial orbits.}
Lemma~\ref{lem:Canonical_t_equivalent_Graph_Radical} identifies the radical.
The length-one leaf pairs contribute a subspace of dimension $n_1-1$.
When $k=2s-1$, there is no further radical vector, so there are
$2^{n_1-1}$ singleton radical orbits.
When $k=2s$, the alternating path coloring gives one additional radical
vector, so there are $2^{n_1}$ singleton radical orbits.
This proves
\ref{prop:orbits_subspace_even_odd_case_alg_ind:trivial-orbits}.

\emph{Exhaustion.}
Every vector in $V$ either belongs to $\rad(V)$ or does not.
The radical vectors are precisely the trivial singleton orbits, while the
non-radical vectors are covered by the non-trivial orbit classification.
This proves
\ref{prop:orbits_subspace_even_odd_case_alg_ind:exhaustion}.
\end{proof}
The sets $O_L$ admit a concrete interval parametrization.
For $1\leq \mathsf{L}\leq \mathsf{R}\leq k+1$, write
\[
    e_{\mathsf{L},\mathsf{R}}:=a_{\mathsf{L}}+a_{\mathsf{L}+1}+\cdots+a_{\mathsf{R}}
\]
for the coloring supported on the interval from $\mathsf{L}$ to $\mathsf{R}$ along the path.
An element of $O_L$ is determined by $L$ mutually separated intervals,
\[
    1\leq \mathsf{L}_1\leq \mathsf{R}_1<\mathsf{L}_2-1\leq \mathsf{R}_2<\cdots<\mathsf{L}_L\leq \mathsf{R}_L\leq k+1,
\]
and by an element of $\radorbit(\vgens)$.
Thus it has the form
\begin{equation}
    v=\sum_{i=1}^L e_{\mathsf{L}_i,\mathsf{R}_i}+u
    \qquad\text{with }
    u\in \radorbit(\vgens).
\end{equation}
Consequently,
\[
    \abs{O_L}=2^{n_1-1}\binom{k+2}{2L},
\]
where the binomial coefficient counts the $2L$ interval endpoints on a path
with $k+1$ vertices.
The same count appears in the Majorana parametrization of the even case in
Eq.~\eqref{eq:majorana_even_orbit_interval_parametrization}.
Notice that this description explicitly relies on knowledge of the graph,
hence does not naturally generalize to arbitrary line graphs, even under the
assumption of minimality.

Again, through a projection from the even case $\graphP_{2m,n_1}$, we can also apply this to the exceptional case.
However, this time not all orbits in $V$ for $\vgens$ will admit a bijection to those in $\tilde{V}$ for $\tilde{\vgens}$, given that they have different sizes.
Specifically, we have $\rank(V) = \rank(\tilde{V}) = 2m$ and $\nullity(\tilde{\vgens}) = \nullity(\vgens)+1 = n_1$.
\begin{prop}\label{prop:orbits_subspace_exceptional_case_alg_dep}
Let $\vgens = \vgens^D(\graphP_{2m,n_1})$ with $m\geq2$
as in Corollary~\ref{cor:minimal_canonical_sets} and using the canonical
labels from Fig.~\ref{fig:humphries_classes}. Set
$V=\Span{\vgens}$.
Let colorings be understood through the coloring-to-vector map
$\coltovec_{\vgens}$ from Definition~\ref{def:coloring_to_vector_map}.
These orbits are obtained from the linearly independent case
$\tilde{\vgens}=\vgens(\graphP_{2m,n_1})$ by quotienting by the linear dependency
$\sum_{i=1}^{m+1}\tilde a_{2i-1}$.
Then:
\begin{enumerate}
    \item\label{prop:orbits_subspace_exceptional_case_alg_dep:criterion}
    \emph{General orbit criterion.}
    Two vectors $v,v'\in V$ are in the same orbit under
    $\tvgroup{\vgens}$ if and only if one of the following holds:
    \begin{nestedcaseenum}
        \item $v,v'\notin\rad(V)$, they admit colorings whose length-one leg
        colorings differ by an element of $\radorbit(\vgens)$ from
        Eq.~\eqref{eq:def:even_leg_coloring_subspace}, and whose path-component
        counts $L$ and $L'$ satisfy $L=L'$ or $L+L'=m+1$, the second
        possibility coming from adding the coloring for the linear dependence
        $a_1+a_3+\cdots+a_{2m+1}=0$;
        \item $v,v'\in\rad(V)$ and $v=v'$.
    \end{nestedcaseenum}

    \item\label{prop:orbits_subspace_exceptional_case_alg_dep:nontrivial-orbits}
    \emph{Non-trivial orbits.}
    For each
    $L\in\{1,\ldots,\lfloor(m+1)/2\rfloor\}$, the vectors admitting a coloring
    with $L$ or $m+1-L$ path components, where
    the two colorings differ by the coloring for the linear dependence
    $a_1+a_3+\cdots+a_{2m+1}=0$.
    We denote this orbit by $O_L$.
    Thus there are $\lfloor (m{+}1)/2\rfloor$ non-trivial orbits
    $\{O_L\}_{L=1}^{\lfloor (m+1)/2\rfloor}$.

    \item\label{prop:orbits_subspace_exceptional_case_alg_dep:trivial-orbits}
    \emph{Trivial orbits.}
    The remaining orbits are the singleton orbits of the radical
    vectors; equivalently, they are represented by colorings with no path
    component and leg coloring in $\radorbit(\vgens)$.
    There are $2^{n_1-1}$ such trivial orbits.

    \item\label{prop:orbits_subspace_exceptional_case_alg_dep:exhaustion}
    \emph{Exhaustion.}
    Every orbit of $\tvgroup{\vgens}$ in $V$ is either one of the
    non-trivial orbits from
    \ref{prop:orbits_subspace_exceptional_case_alg_dep:nontrivial-orbits}
    or one of the singleton radical orbits from
    \ref{prop:orbits_subspace_exceptional_case_alg_dep:trivial-orbits}.
    In total, there are
    \[
        \lfloor(m+1)/2\rfloor+2^{n_1-1}
    \]
    orbits in $V$.
\end{enumerate}
\end{prop}

\begin{proof}
\emph{Projection setup.}
Let $\tilde{\vgens}=\vgens(\graphP_{2m,n_1})$ be the linearly
independent extension of $\vgens$ from
Proposition~\ref{prop:valid_colorings_orbits_line_graph_blown_up_path_graph_alg_dep_exceptional_case}.
Thus the projection $\projection$ satisfies
$\projection(\tilde{\vgens})=\vgens$, and its kernel on
$\SpanS[\F_2]{\tilde{\vgens}}$ is generated by the kernel vector
\[
    u=\sum_{i=1}^{m+1}\tilde a_{2i-1}.
\]
By Lemma~\ref{lem:projection_extension_basic_properties}\ref{lem:projection_extension_basic_properties:e}, the
orbits for $\vgens$ are the projections of the orbits for $\tilde{\vgens}$.

\emph{Effect of the linear dependence on path components.}
We prove the connected component count which explains the exceptional
identification of $L$ and $m+1-L$.
Restrict a coloring $\coloring$ of $\graphP_{2m,n_1}$ to the of path vertices
$1,\ldots,2m+1$.
Set $\coloring(0)=\coloring(2m+2)=0$.
If this path coloring has $L$ connected components, then
\[
    \sum_{j=0}^{2m+1}\FtoN{\coloring(j)+\coloring(j{+}1)}=2L,
\]
where the additions inside $\FtoN{\cdot}$ take place in
$\F_2$, and the outer sum takes place in $\N$.
Indeed, the summand is equal to $1$ exactly when the coloring changes
between the adjacent path positions $j$ and $j+1$.
Thus the sum counts the boundary points of the support of the path coloring.
If the support has $L$ connected components, then it is a disjoint union of
$L$ maximal intervals, and each interval has one left and one right boundary.
Let $\coloring_u$ be the coloring which represents this kernel vector
$u$.
On the path vertices, $\coloring_u(j)=1$ exactly for odd $j$, and we
set $\coloring_u(0)=\coloring_u(2m+2)=0$.
Hence $\coloring_u(j)+\coloring_u(j+1)=1$ for all
$j=0,\ldots,2m+1$.
Therefore the translated coloring $\coloring' = \coloring+\coloring_u$ has
\[
    \sum_{j=0}^{2m+1}\FtoN{\coloring'(j)+\coloring'(j{+}1)}
    =
    (2m{+}2)-2L.
\]
Applying the same boundary-count argument to $\coloring'$ shows that
this sum is $2L'$, where $L'$ is the number of path components of
$\coloring'$.
Thus adding the coloring $\coloring_u$ sends a coloring with $L$ path
components to one with $L' = m+1-L$ path components.

\emph{General orbit criterion.}
We first prove
\ref{prop:orbits_subspace_exceptional_case_alg_dep:criterion}.
For the linearly independent extension $\tilde{\vgens}$,
Proposition~\ref{prop:orbits_subspace_even_odd_case_alg_ind}\ref{prop:orbits_subspace_even_odd_case_alg_ind:criterion}
shows that two non-radical vectors are in the same orbit precisely when their
path component counts agree and their length-one leg colorings differ by an
element of $\radorbit(\vgens)$.
After projection, colorings which differ by the coloring $\coloring_u$ in the kernel of the projection $\ker(\projection\circ \coltovec)$
represent the same vector, where $\coltovec$ is the coloring-to-vector map from Definition~\ref{def:coloring_to_vector_map}.
By the previous paragraph, this identifies the component counts $L$ and
$m+1-L$.
This proves the non-radical case in
\ref{prop:orbits_subspace_exceptional_case_alg_dep:criterion}.
If $v\in\rad(V)$, then $\tau_g(v)=v$ for every generator
$g\in\vgens$, because $\symp{g}{v}=0$.
Thus every radical vector is a singleton orbit.
This proves the radical case in
\ref{prop:orbits_subspace_exceptional_case_alg_dep:criterion}; it also excludes
an orbit containing both a radical vector and a vector outside $\rad(V)$.

\begin{figure}
    \centering
    \includegraphics[width=\linewidth]{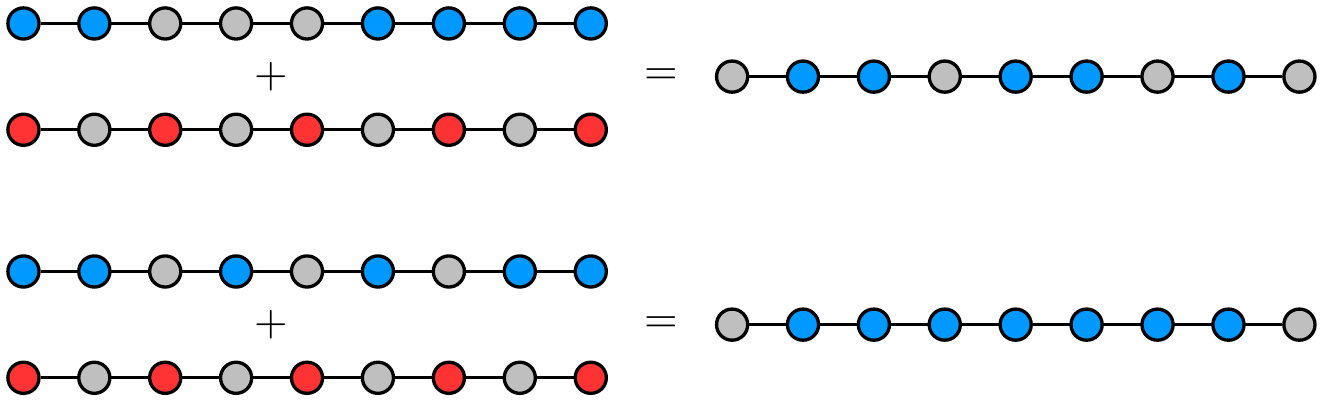}
    \caption{Over the exceptional, algebraically dependent generating set, adding the linear dependency $a_1+a_3+\cdots+a_{2m+1}=0$ identifies colorings whose path-component counts $L$ and $L'$ satisfy $L'=m+1-L$.
    In the displayed example $2m+1=9$, so $m+1=5$: the upper row has $(L,L')=(2,3)$, while the lower row has $(L,L')=(4,1)$.
    The blue colorings correspond to the same vector $\coltovec(\coloring) = \coltovec(\coloring')$, and the red coloring corresponds to a linear dependency $\coltovec(\coloring'')=0$ under the coloring-to-vector map from Definition~\ref{def:coloring_to_vector_map}.}
    \label{fig:projected-orbits-on-path}
\end{figure}

\emph{Non-trivial orbits.}
The non-trivial orbits are exactly the orbits of the vectors in
$V\setminus\rad(V)$.
By the general criterion, the component counts $L$ and $m+1-L$ describe the
same orbit.
Consequently, the non-trivial orbits can be represented without
redundancy by
\[
    L\in\{1,\ldots,\lfloor(m{+}1)/2\rfloor\},
\]
which gives exactly $\lfloor(m{+}1)/2\rfloor$ non-trivial orbits and
proves
\ref{prop:orbits_subspace_exceptional_case_alg_dep:nontrivial-orbits}.

\emph{Trivial orbits.}
The remaining vectors are represented by colorings with no path
component and leg coloring in $\radorbit(\vgens)$.
These are precisely the radical vectors in this parametrization, and each of
them is fixed by every transvection in $\tvgroup{\vgens}$.
There are $2^{n_1-1}$ elements of $\radorbit(\vgens)$.
This proves
\ref{prop:orbits_subspace_exceptional_case_alg_dep:trivial-orbits}.
\emph{Exhaustion.}
Every vector in $V$ either belongs to $\rad(V)$ or does not.
The vectors which belong to $\rad(V)$ give the trivial singleton orbits, while
the vectors which do not belong to $\rad(V)$ are covered by the non-trivial
orbit classification above.
This proves the exhaustion statement.
The total number of orbits is the sum of the
$\lfloor(m+1)/2\rfloor$ non-trivial orbits from
\ref{prop:orbits_subspace_exceptional_case_alg_dep:nontrivial-orbits}
and the $2^{n_1-1}$ trivial orbits from
\ref{prop:orbits_subspace_exceptional_case_alg_dep:trivial-orbits}.
This proves
\ref{prop:orbits_subspace_exceptional_case_alg_dep:exhaustion}.
\end{proof}
The parametrizations in
Proposition~\ref{prop:orbits_subspace_even_odd_case_alg_ind} and
Proposition~\ref{prop:orbits_subspace_exceptional_case_alg_dep} differ exactly by
the projection identified above.
In the exceptional dependent case, adding the linear dependence identifies
the number of connected components $L$ and $m+1-L$, so the non-trivial representatives may be restricted
to $1\leq L\leq\lfloor(m{+}1)/2\rfloor$.
This reflects the one-dimensional quotient from the linearly independent
extension to the dependent span.
Figure~\ref{fig:projected-orbits-on-path} illustrates this projection
mechanism.
The extension from orbits inside $V$ to orbits in all of $\Fn$ is
deferred to the full orbit classification in
Section~\ref{sec:classification:orbits}.

\subsection{The Lie Algebras and Free Fermionic Description}\label{sec:Majorana_Strings_Free_Fermionic_Lie_Algebras}

We now take advantage of the faithful free fermionic mappings to assign a full labelling inside of $\Fn$ or $\PP_n$ (since they are all equivalent up to change of basis), assuming the algebraic dependency is of the same form (i.e., by fixing all vectors but one).
This automatically provides a description of the Lie algebras and orbits which is independent of the specific generating set, but only depends on its free-fermionic type (as well as nullity and rank of the orthogonal complement).

By viewing each blown-up path graph as the line graph of a path graph
with parallel edges, Propositions~\ref{prop:free_fermionic_mapping_alg_dep} and
\ref{prop:full_Fn_PPn_Majorana_basis_free_fermionic_mapping} provide a
Majorana basis for $\PP_n$ or $\Fn$ in which the following representatives
describe the corresponding phase-normalized Pauli strings up to a phase:
\begin{subequations}\label{eq:majorana_string_basis_blown_up_path_graph_canonical}
    \begin{alignat}{3}
        \pgens(\graphP_{2m-2,n_1}) &\simeq\;
        &&\{ \gamma_i\gamma_{i+1} \}_{i=1}^{2m-1}\\
        &&&\cup \{ (\gamma_{2(m+i)-1}\gamma_{2(m+i)})\gamma_{2m-1}\gamma_{2m} \}_{i=1}^{n_1-1}, \nonumber\\
        \pgens(\graphP_{2m-1,n_1}) &\simeq
        &&\{ \gamma_i\gamma_{i+1} \}_{i=1}^{2m-1}\cup\{\gamma_{2m} \} \\
        &&&\cup\{ (\gamma_{2(m+i)-1}\gamma_{2(m+i)}) \gamma_{2m} \}_{i=1}^{n_1-1}, \nonumber\\
        \pgens^D(\graphP_{2m,n_1}) &\simeq
        &&\{ \gamma_i\gamma_{i+1} \}_{i=1}^{2m-1} \cup \{ \gamma_{2m}, \Gamma \} \\
        &&&\cup \{ (\gamma_{2(m+i)-1}\gamma_{2(m+i)}) \Gamma \}_{i=1}^{n_1-1}.\nonumber
    \end{alignat}
\end{subequations}
In each case the underlying spanning tree is a path graph, and each cycle may be seen as a pair of parallel edges on the last two vertices.

When $n_1=1$, there are no cycle factors, and the blown-up path
$\graphP_{k,1}$ is the ordinary path $\graphP_{k+1}$.
In this minimal case,
Eq.~\eqref{eq:majorana_string_basis_blown_up_path_graph_canonical} reduces,
up to scalar phase, to the following nearest-neighbour representatives inside
the Majorana sets from Eq.~\eqref{eq:def_Majorana_bases}:
\begin{align*}
        \pgens(\graphP_{2m-1}) &\simeq \{ \gamma_i\gamma_{i+1} \}_{i=1}^{2m-1},\\
        \pgens(\graphP_{2m}) &\simeq \{ \gamma_i\gamma_{i+1} \}_{i=1}^{2m-1}\cup\{\gamma_{2m} \},\\
        \pgens^D(\graphP_{2m+1}) &\simeq \{ \gamma_i\gamma_{i+1} \}_{i=1}^{2m-1} \cup \{ \gamma_{2m}, \Gamma \}.
\end{align*}
Thus the three families differ by which non-quadratic Majorana
operators are added to the nearest-neighbour quadratic chain:
the even case contains only nearest-neighbour quadratic products, the odd case
adds one linear Majorana operator, and the exceptional dependent case adds both
this linear operator and the parity product $\Gamma$.

For $n_1>1$, the extra generators in
Eq.~\eqref{eq:majorana_string_basis_blown_up_path_graph_canonical} are obtained
from the corresponding endpoint generator by multiplying it with one of the
commuting cycle factors $\gamma_{2(m+i)-1}\gamma_{2(m+i)}$.
These are the generators corresponding to the additional parallel edges in the
root multigraph.
The indexing in
Eq.~\eqref{eq:majorana_string_basis_blown_up_path_graph_canonical} is chosen
so that all three representatives are written using the same first $2m$
Majorana operators, even though their canonical graph labels are
$\graphP_{2m-2,n_1}$, $\graphP_{2m-1,n_1}$, and $\graphP_{2m,n_1}$.

Thus, at the level of displayed Majorana products, the independent
families appear as subfamilies of the exceptional list. This is only a
notational feature of the chosen representatives: the exceptional dependent
case is formally obtained from a suitable independent extension by projection.

\begin{lem}[Binary labels of the canonical Majorana representatives]\label{lem:binary_labels_canonical_majorana_representatives}
For the representatives in
Eq.~\eqref{eq:majorana_string_basis_blown_up_path_graph_canonical}, the
canonical generators have the following binary labels and interval labels, up
to the cycle factors:
\begin{enumerate}
    \item In the even case $\pgens(\graphP_{2m-2,n_1})$,
    \[
        a_i=\mu_i+\mu_{i+1}\qquad(1\leq i\leq 2m-1).
    \]
    Hence, for $1\leq i\leq j\leq 2m-1$,
    \[
        \isolong{a_i+\cdots+a_j}
        =
        \majiso{\mu_i+\mu_{j+1}}
        \simeq
        \gamma_i\gamma_{j+1}.
    \]
    \item In the odd case $\pgens(\graphP_{2m-1,n_1})$, in addition,
    \[
        a_{2m}=\mu_{2m}.
    \]
    Hence, for $1\leq i\leq 2m-1$,
    \[
        \isolong{a_i+\cdots+a_{2m}}
        =
        \majiso{\mu_i}
        =
        \gamma_i.
    \]
    \item In the exceptional case $\pgens^D(\graphP_{2m,n_1})$, in addition,
    \[
        a_{2m}=\mu_{2m},
        \qquad
        a_{2m+1}=\majprod,
        \qquad
        \majprod=\sum_{i=1}^{2m}\mu_i.
    \]
    Hence, for $1\leq i\leq 2m-1$,
    \[
        \isolong{a_i+\cdots+a_{2m+1}}
        =
        \majiso{\majprod+\mu_i}
        \simeq
        \Gamma\gamma_i.
    \]
\end{enumerate}
For $n_1>1$ and $1\leq i\leq n_1-1$, the $i$th additional parallel edge contributes the cycle factor
$\gamma_{2(m+i)-1}\gamma_{2(m+i)}$, whose binary label is
\[
    c_i:=\mu_{2(m+i)-1}+\mu_{2(m+i)}.
\]
Thus the additional generators have labels
$c_i+\mu_{2m-1}+\mu_{2m}$ in the even case, $c_i+\mu_{2m}$ in the odd case,
and $c_i+\majprod$ in the exceptional case.
\end{lem}
\begin{proof}
This is the binary-label translation of
Eq.~\eqref{eq:majorana_string_basis_blown_up_path_graph_canonical}.
By Lemma~\ref{lem:majorana_string_equals_pauli_label}, the phase-normalized
Majorana string $\majiso{v}$ has Pauli label $v$.
Hence $\gamma_i\gamma_{i+1}\simeq\majiso{\mu_i+\mu_{i+1}}$ gives
$a_i=\mu_i+\mu_{i+1}$, the factor $\gamma_{2m}=\majiso{\mu_{2m}}$ gives
$a_{2m}=\mu_{2m}$, and $\Gamma\simeq\majiso{\majprod}$ gives
$a_{2m+1}=\majprod$.
The interval formulas follow by summing the displayed labels over $\F_2$.
For example,
\[
    (\mu_i+\mu_{i+1})+\cdots+(\mu_j+\mu_{j+1})
    =
    \mu_i+\mu_{j+1},
\]
and the odd and exceptional endpoint formulas follow in the same way after
adding $a_{2m}=\mu_{2m}$ or $a_{2m+1}=\majprod$.
Multiplication by the cycle factor
$\gamma_{2(m+i)-1}\gamma_{2(m+i)}$ adds the corresponding binary label
$c_i=\mu_{2(m+i)-1}+\mu_{2(m+i)}$.
Applying this to the endpoint factors
$\gamma_{2m-1}\gamma_{2m}$, $\gamma_{2m}$, and $\Gamma$ gives respectively
the labels $c_i+\mu_{2m-1}+\mu_{2m}$, $c_i+\mu_{2m}$, and
$c_i+\majprod$.
\end{proof}

Notice that we refer to all these three cases as \emph{free-fermionic}, even though in the conventional setting only the even cases are considered as free-fermions.
We use this convention to highlight that all satisfy similar properties, especially regarding the existence of exact or computationally efficient solutions, which take advantage of small Lie algebras (at least when $n_1=1$).

Combining Propositions~\ref{prop:valid_colorings_even_odd_cases_alg_ind},
\ref{prop:valid_colorings_orbits_line_graph_blown_up_path_graph_alg_dep_exceptional_case},
and~\ref{lem:binary_labels_canonical_majorana_representatives}, we have the
following Pauli bases (see Eq.~\eqref{eq:def_Majorana_bases}):
\begin{align*}
        \baslong{\lie{\pgens(\graphP_{2m-2,n_1})}} &= \MAJ_2^{(2m)}\setprod\Cyc^{(2m,q)}\\
        \baslong{\lie{\pgens(\graphP_{2m-1,n_1})}} &= \MAJ_{1,2}^{(2m)}\setprod\Cyc^{(2m,q)}\\
        \baslong{\lie{\pgens^D(\graphP_{2m,n_1})}} &= \MAJ_{\gamma}^{(2m)}\setprod\Cyc^{(2m,q)}
\end{align*}

Finally, we arrive at the Lie algebra classification:
\begin{thm}[cf.\ \cite{Aguilar_Cichy_Eisert_Bittel_2024}]\label{thm:Free_Fermionic_Lie_Algebras}
Let $\pgens=\isolong{\vgens}\subseteq\PP_n$ be a generating set whose frustration graph is a line graph $\graphG=L(\Delta)$.
Let $r=\nullity(\vgens)$, let
$q=\nullorbit(\vgens)=\dim\radorbit(\vgens)$ as in
Eq.~\eqref{eq:nullity_orbit_radical_paulis}, and let
$2\ell=\rank(\commutant(\pgens))$.
Then, exactly one of the following holds:
\begin{enumerate}
    \item\label{thm:Free_Fermionic_Lie_Algebras:even} If $\pgens$ is of even type and $n_\Delta=2m$ is even, then
    \[
        \lie{\pgens}\conjugated
        \so(2m)^{\oplus 2^q}\otimes I^{\otimes \ell},
    \]
    where $\rank(\vgens)=2m-2$, $\nullity(\vgens)=q+1$, and $m\geq3$.
    \item\label{thm:Free_Fermionic_Lie_Algebras:odd} If $\pgens$ is of odd type and $n_\Delta=2m+1$ is odd, then
    \[
        \lie{\pgens}\conjugated
        \so(2m{+}1)^{\oplus 2^q}\otimes I^{\otimes \ell},
    \]
    where $\rank(\vgens)=2m$, $\nullity(\vgens)=q$, and $m\geq1$.
    \item\label{thm:Free_Fermionic_Lie_Algebras:exceptional} If $\pgens$ is of exceptional type and $n_\Delta=2m+2$ is even, then
    \[
        \lie{\pgens}\conjugated
        \so(2m{+}2)^{\oplus 2^q}\otimes I^{\otimes \ell},
    \]
    where $\rank(\vgens)=2m$, $\nullity(\vgens)=q$, and $m\geq2$.
\end{enumerate}
The ambient block representations are:
\begin{enumerate}[resume]
    \item
    In cases~\ref{thm:Free_Fermionic_Lie_Algebras:odd}
    and~\ref{thm:Free_Fermionic_Lie_Algebras:exceptional}, each copy of
    $\so(2m{+}1)$ or $\so(2m{+}2)$ acts on each $2^m$-dimensional irreducible
    block by a spinor representation, with a $2^{\ell}$ identity multiplicity
    factor. Refer to Fig.~\ref{fig:block-odd-exceptional}.
    \item
    In case~\ref{thm:Free_Fermionic_Lie_Algebras:even}, each copy of $\so(2m)$ acts on a linked pair of
    $2^{m-1}$-dimensional irreducible blocks by the two distinct spinor representations, again with
    a $2^{\ell}$ identity multiplicity factor. Refer to Fig.~\ref{fig:block-even-ff}.
\end{enumerate}
\end{thm}

\begin{proof}
\emph{Reduction to canonical representatives.}
By Corollary~\ref{cor:distinguishing_line_graph_cases}, the
line-graph case is distinguished by $\tilde r-\tilde q$, which determines
whether $n_\Delta$ is even or odd, and by $r-q$, which determines, for even
$n_\Delta$, whether we are in the even or exceptional case.
Thus it suffices to work with the canonical families
\[
    \pgens(\graphP_{2m-2,n_1}),\qquad
    \pgens(\graphP_{2m-1,n_1}),\qquad
    \pgens^D(\graphP_{2m,n_1}),
\]
with $n_1=q+1$, corresponding respectively to the even, odd, and
exceptional cases.

\emph{Removing cycle and commutant multiplicities.}
The proof proceeds as in
Lemma~\ref{lem:canonical_pauli_bases_commutant_bilinear_isometries} and
Proposition~\ref{prop:pauli_lie_isometry_block_forms}, by first identifying the
minimal block representation.
The $q=n_1-1$ cycle factors are commuting symmetries; fixing their eigenvalues
gives the $2^q$ direct-sum copies.
Equivalently, the commuting operators
$\im\gamma_{2(m+i)-1}\gamma_{2(m+i)}$ have simultaneous eigenspaces, with
projectors $(1\pm\im\gamma_{2(m+i)-1}\gamma_{2(m+i)})/2$.
The remaining $2\ell$ modes for the commutant give only the identity multiplicity
factor $I^{\otimes \ell}$.
Therefore it is enough to identify the simple Lie algebra and its irreducible
block representation for $n_1=1$ and $\ell=0$, and then restore these factors.

\begin{figure}
    \centering
    \includegraphics{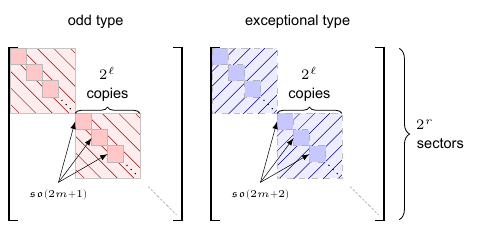}
    \caption{Block decomposition of the Pauli Lie algebra for free-fermionic generating sets of odd type (left) and exceptional type (right), following Fig.~\ref{fig:block-su-representation}.
    In each of the $2^r$ isotypic sectors, one independent copy of the simple Lie algebra $\so(2m{+}1)$ for odd type, or $\so(2m{+}2)$ for exceptional type, acts simultaneously on each of the $2^{\ell}$ multiplicity copies of dimension $2^m$, reflecting the multiplicity tensor factor $\id_{2^{\ell}}$.
    Here $2\ell=\rank(\commalg)$, $r=\nullity(\commalg)$ and $2m=\rank(\pgens)$.}
    \label{fig:block-odd-exceptional}
\end{figure}

\emph{Minimal Majorana bases.}
For $n_1=1$ and $\ell=0$, the relevant Pauli bases are, using the notation
from Eq.~\eqref{eq:def_Majorana_bases},
\begin{align*}
        \baslong{\lie{\pgens(\graphP_{2m-2,1})}} &= \MAJ_2^{(2m)},\\
        \baslong{\lie{\pgens(\graphP_{2m-1,1})}} &= \MAJ_{1,2}^{(2m)},\\
        \baslong{\lie{\pgens^D(\graphP_{2m,1})}} &= \MAJ_{\gamma}^{(2m)}.
\end{align*}

\emph{Odd and exceptional cases.}
As shown in
\cite[Appendix~B, Section~2]{Kazi_Larocca_Farinati_Coles_Cerezo_Zeier_2025},
the Lie algebra spanned by $\MAJ_{1,2}^{(2m)}$ is represented irreducibly as
$\so(2m{+}1)$ for $m\geq1$, and the Lie algebra spanned by
$\MAJ_{\gamma}^{(2m)}$ is represented irreducibly as $\so(2m{+}2)$ for
$m\geq2$.
Both representations have dimension $2^m$.
For $m\geq4$, these are the spinor representations of the corresponding
orthogonal Lie algebras.
In the $\so(2m{+}2)$ case there are two inequivalent spinor representations,
but each copy acts independently on one irreducible block, so this choice is
irrelevant up to outer automorphisms.
Restoring the $2^q$ cycle sectors and the $I^{\otimes \ell}$ multiplicity factor
proves the odd and exceptional cases, as well as the representation statement
for those cases.

\begin{figure}
    \centering
    \includegraphics{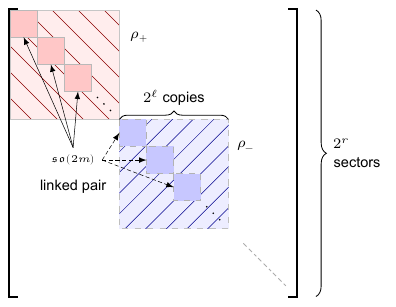}
    \caption{Block decomposition of the Pauli Lie algebra for free-fermionic generating sets of even type, following Fig.~\ref{fig:block-su-representation}.
    The highlighted blocks show one linked pair of isotypic sectors.
    One independent copy of $\so(2m)$ determines the action on both sectors in this pair, acting by a spinor representation $\rho_+$ on one sector and the other spinor representation $\rho_-$ on the other.
    The $2^r$ isotypic sectors are partitioned into $2^{r-1}$ linked pairs of this form.
    Each sector has the same $2^{\ell}$ multiplicity structure.
    Here $2\ell=\rank(\commalg)$, $r=\nullity(\commalg)$ and $2m-2=\rank(\pgens)$.}
    \label{fig:block-even-ff}
\end{figure}

\emph{Even case.}
By \cite[Proposition~19]{Kazi_Larocca_Farinati_Coles_Cerezo_Zeier_2025},
the Lie algebra spanned by $\MAJ_2^{(2m)}$ is represented as $\so(2m)$ for
$m\geq3$.
This representation is reducible: it splits into the two distinct spinor
representations of $\so(2m)$, each of dimension $2^{m-1}$.
Thus each copy of $\so(2m)$ acts on a linked pair of irreducible blocks by the
two spinor representations.
In the even case, the extra radical vector separates the $2^r=2^{q+1}$
isotypic sectors into $2^q$ linked pairs, and each linked pair carries one copy
of $\so(2m)$.
Restoring the $2^q$ cycle sectors and the $I^{\otimes \ell}$ multiplicity factor
proves the even case and its representation statement.
\end{proof}

For the exceptional type, the choice between the two half-spin
representations of $\so(2m{+}2)$ is immaterial for this classification:
each simple summand acts on a single irreducible block, and the two choices
are exchanged by an outer automorphism of that summand.
For the even type, one simple summand of $\so(2m)$ acts on a linked pair of
blocks, with the two half-spin representations appearing simultaneously.
Thus the distinction between the two half-spin modules is part of the block
representation in the even case.

\begin{rem}[Representation data beyond the abstract Lie algebra]\label{rem:free_fermionic_representation_data}
The corresponding statement in \cite{Aguilar_Cichy_Eisert_Bittel_2024}
identifies the free-fermionic Lie algebra type, but does not fully describe its
ambient representation.
The missing representation data comes from the commutant algebra.
When this commutant contains non-abelian matrix factors, these symmetries
appear as multiplicity spaces in the Pauli matrix algebra decomposition.
The generated algebra $\matalg=\algclosure{\pgens}$ acts trivially on those
multiplicity spaces, while the commutant
$\commalg=\commutant(\matalg)$ acts on them nontrivially.
Thus the same abstract simple summands may appear with nontrivial identity
multiplicity factors, and in the even case with linked pairs of spinor blocks.
Theorem~\ref{thm:Free_Fermionic_Lie_Algebras} provides this additional block
representation data.
\end{rem}

\subsection{Orbits in the Majorana Formalism}\label{sec:Orbits_Majorana_Formalism}

We now shift to the discussion of the orbits or commutator graph in the Majorana formalism, restricted to $V=\Span{\vgens}$ or $\algclosure{\pgens}$.
We shall discuss the full orbits over $\Fn$ or $\PP_n$ in Section~\ref{sec:Orbits_Full_Space_Fn_Free_Fermionic_Case}.

For the even case $\vgens(\graphP_{2m-2,n_1})$, Proposition~\ref{prop:orbits_subspace_even_odd_case_alg_ind} describes $O_L$ by $L$ connected interval components along the path, up to addition of cycle-symmetry vectors $u\in\radorbit(\vgens)$.
A single interval component from $\mathsf{L}$ to $\mathsf{R}$ has binary label
\begin{align*}
        e_{\mathsf{L},\mathsf{L}} &= \mu_{\mathsf{L}} + \mu_{\mathsf{L}+1}\\
        e_{\mathsf{L},\mathsf{R}} &= (\mu_{\mathsf{L}} + \mu_{\mathsf{L}+1}) + \cdots +(\mu_{\mathsf{R}} + \mu_{\mathsf{R}+1})
        = \mu_{\mathsf{L}} + \mu_{\mathsf{R}+1}
\end{align*}
Thus an element $x\in V$ lies in $O_L$ if and only if it is the sum of
$L$ such interval components and one cycle-symmetry vector, i.e., if it can be
written as $x=v+u$, with $u\in\radorbit(\vgens)$ and
\begin{align}\label{eq:majorana_even_orbit_interval_parametrization}
        v &= \sum_{i=1}^L e_{\mathsf{L}_i,\mathsf{R}_i}
        = \sum_{i=1}^L(\mu_{\mathsf{L}_i}+\mu_{\mathsf{R}_i+1}),\\
        \majiso{x} &\simeq
        \majiso{u}\,
        \gamma_{\mathsf{L}_1}\gamma_{\mathsf{R}_1+1}\cdots\gamma_{\mathsf{L}_L}\gamma_{\mathsf{R}_L+1}.
\end{align}
Equivalently, for $x\in V$, $x\in O_L$ iff there exists $u\in\radorbit(\vgens)$ such that
\[
    x+u\in\SpanS[\F_2]{\{\mu_i\}_{i=1}^{2m}}
    \text{ and }\majL_m(x+u)=2L.
\]
In the parametrization above, $x+u=v$.
The factor $u\in\radorbit(\vgens)$ provides the cycle-symmetry factors.
Conversely, given a Majorana string $\majiso{x}$ whose reduced length
over the first $2m$ modes is $2L$, the non-zero Majorana components in those
modes determine the extrema of the $L$ connected components along the path,
while the remaining cycle factors determine $u\in\radorbit(\vgens)$.
Thus $x\in O_L$.
It also shows that each non-trivial orbit has size $2^q\cdot \binom{2m}{2L}$, $1\leq L\leq m-1$, where $\abs{\radorbit(\vgens)}=2^q$.

Hence, we find a direct link between the number of connected components for a coloring in $\graphP_{k,n_1}$ and the length of its Majorana string in the labelling Eq.~\eqref{eq:majorana_string_basis_blown_up_path_graph_canonical}.

This is a well-known fact about quadratic Majoranas \cite{Diaz_GarciaMartin_Kazi_Larocca_Cerezo_2023,Sierant_Turkeshi_Tarabunga_2026,lastres2026geometryfreefermioncommutants}, which has been discussed in the literature, at least when $n_1=1$, and extends even over all of the Pauli strings.
We will nevertheless discuss this, as well as its generalization to the other free-fermionic cases, to connect to the transvection group and graph-theoretic approaches.
Then, the orbits $O_L$ for $\pgens=\pgens(\graphP_{2m-2,n_1})$ consists of those Majorana string of length $2L$,
up to multiplication by a Pauli string in the basis of the orbit-radical center
$\ZZorbit(\pgens)=\Span[\C]{\isolong{\radorbit(\vgens)}}$ from
Eq.~\eqref{eq:def:ZZ_center_pauli_orbit}.

In the odd case $\vgens(\graphP_{2m-1,n_1})$, the intervals not using
the linear endpoint are the same as in the even case.
The new possibility is that one interval ends at the linear generator, giving
\begin{equation}
    e_{2m-1,2m-1}=\mu_{2m-1},
    \qquad
    e_{\mathsf{L},2m-1}=\mu_{\mathsf{L}}.
\end{equation}
Thus an element $x\in V$ lies in $O_L$ if and only if, after adding a
cycle-symmetry vector $u\in\radorbit(\vgens)$, the vector $x+u$ is either an
even-case interval representative with $L$ components, or it has one component
ending at the linear generator.
In the latter case, writing the last component as
$e_{\mathsf{L}_L,\mathsf{R}_L}=e_{\mathsf{L}_L,2m-1}$ gives
\begin{align*}
        v &= \sum_{i=1}^{L} e_{\mathsf{L}_i,\mathsf{R}_i}
        = \sum_{i=1}^{L-1}(\mu_{\mathsf{L}_i}+\mu_{\mathsf{R}_i+1})+\mu_{\mathsf{L}_L},\\
        \majiso{x} &\simeq
        \majiso{u}\,
        \gamma_{\mathsf{L}_1}\gamma_{\mathsf{R}_1+1}
        \cdots
        \gamma_{\mathsf{L}_{L-1}}\gamma_{\mathsf{R}_{L-1}+1}\gamma_{\mathsf{L}_L}.
\end{align*}
Equivalently, for $x\in V$, $x\in O_L$ iff there exists $u\in\radorbit(\vgens)$ such that
\[
   x+u\in\SpanS[\F_2]{\{\mu_i\}_{i=1}^{2m}}
    \text{ and }\majL_m(x{+}u)\in\{2L{-}1,2L\}.
\]

In the exceptional case $\vgens^D(\graphP_{2m,n_1})$, there is one more
endpoint, corresponding to the parity product.
The additional interval components are
\begin{equation}
    e_{2m,2m}=\majprod,
    \qquad
    e_{\mathsf{L},2m}=\majprod+\mu_{\mathsf{L}}
    \qquad(\mathsf{L}\leq 2m).
\end{equation}
Thus, besides the even and odd representatives above, an element of
$O_L$ may have one component ending at this parity endpoint.
Writing this last component as $e_{\mathsf{L}_L,\mathsf{R}_L}=e_{\mathsf{L}_L,2m}$ gives
\begin{align*}
        v &= \sum_{i=1}^L e_{\mathsf{L}_i,\mathsf{R}_i}
        = \sum_{i=1}^{L-1}(\mu_{\mathsf{L}_i}+\mu_{\mathsf{R}_i+1})+\mu_{\mathsf{L}_L}+\majprod,\\
        \majiso{x} &\simeq
        \majiso{u}\,\Gamma
        \gamma_{\mathsf{L}_1}\gamma_{\mathsf{R}_1+1}
        \cdots
        \gamma_{\mathsf{L}_{L-1}}\gamma_{\mathsf{R}_{L-1}+1}\gamma_{\mathsf{L}_L}.
\end{align*}
Equivalently, for $x\in V$, $x\in O_L$ iff there exists
$u\in\radorbit(\vgens)$ such that
\begin{align*}
    &x+u\in\SpanS[\F_2]{\{\mu_i\}_{i=1}^{2m}}
    \text{ and}\\
    &\majL_m(x{+}u)\in
    \{2L{-}1,2L,2(m{-}L),2(m{-}L){+}1\}.
\end{align*}

Thus, after choosing a faithful free-fermionic mapping, the non-trivial
orbits inside $V=\Span{\vgens}$ can be described purely by the reduced Majorana
length and the cycle-symmetry factors, rather than by the original graph.
Let us also define the set of all Majorana strings of length $L$ over the first $2m$ modes as
\begin{equation}
    \MAJ_L^{(2m)} = \{ \majiso{v} \mid v\in\Span{\{\mu_i\}_{i=1}^{2m}},\, \majL_m(v) = L \}.
\end{equation}
We now summarize this discussion, using the language of Majorana strings:
\begin{prop}\label{prop:orbits_in_V_line_graphs_Majorana}
Let $\pgens=\isolong{\vgens}\subseteq\PP_n$ be a generating set with a frustration graph which is a line graph:
\begin{enumerate}
    \item If $\pgens$ is of even type with $(2m,q)$-free fermionic mapping,
    then the non-trivial orbits in $\algclosure{\pgens}$ are given by $O_L$ with $L\in\{1,\ldots,m{-}1\}$ and
    \begin{equation}
        \isolong{O_L} = \MAJ_{2L}^{(2m)}\setprod\Cyc^{(2m,q)}.
    \end{equation}
    \item If $\pgens$ is of odd type with $(2m,q)$-free fermionic mapping,
    then the non-trivial orbits in $\algclosure{\pgens}$ are given by $O_L$ with $L\in\{1,\ldots,m\}$ and 
    \begin{equation}
        \isolong{O_L} = (\MAJ_{2L-1}^{(2m)}\cup \MAJ_{2L}^{(2m)})\setprod\Cyc^{(2m,q)}.
    \end{equation}
    \item If $\pgens$ is of exceptional type with $(2m,q)$-free fermionic mapping,
    then the non-trivial orbits in $\algclosure{\pgens}$ are given by $O_L$ with $L\in\{1,\ldots,\lfloor (m{+}1)/2 \rfloor \}$ and
    \begin{alignat*}{3}
            \isolong{O_L} &=&&\; (\MAJ_{2L-1}^{(2m)}\cup \MAJ_{2L}^{(2m)}\cup \MAJ_{2(m-L)}^{(2m)}\\
            &&&\cup \MAJ_{2(m-L)+1}^{(2m)})\setprod\Cyc^{(2m,q)}.
    \end{alignat*}
\end{enumerate}
\end{prop}

\subsection{Transvections, Transpositions and the Symmetric Group}\label{sec:Transvection_Group_Line_Graph_Symmetric_Group}

We now discuss the relevant transvection groups, which will highlight properties shared among all generating sets whose frustration graphs are line graphs.
Namely, we shall find close relations between these groups and symmetric groups, which has also been used before in the treatment of Clifford subgroups related to matchgates or free-fermions \cite{Sierant_Turkeshi_Tarabunga_2026}.

We first define the transvection groups, parametrized in a way which highlights both the features related to the symmetric group over $d$ objects and the size of the radical, inside the space $\Fn$ with $n\geq m+r$:
\begin{subequations}\label{eq:def:binary_transvection_groups_free_fermionic}
\begin{align}
        \Sym(2m,r,n,\F_2) &= \tvgrouplong{\vgens(\graphP_{2m-2,r})} \\
        &\text{for } m\geq 3 \text{ and }r\geq 1, \nonumber\\
        \Sym(2m{+}1,r,n,\F_2) &= \tvgrouplong{\vgens(\graphP_{2m-1,r+1})} \\
        &\text{for } m\geq 1 \text{ and }r\geq 0, \nonumber\\
        \Sym^D(2m{+}2,r,n,\F_2) &= \tvgrouplong{\vgens^D(\graphP_{2m,r+1})} \\
        &\text{for } m\geq 2 \text{ and } r\geq 0.\nonumber
\end{align}
\end{subequations}
The lower bounds in the even and exceptional families exclude the ambiguous canonical path case
$\graphP_{2,n_1}$ as an even or exceptional representative, cf. the discussion
after Theorem~\ref{thm:uniqueness_root_graph_line_graph} and
Fig.~\ref{fig:line-graph-p3}.
The only retained boundary is the odd representative $\graphP_{1,n_1}$, which
is treated as part of the odd family.
Thus the displayed parameters determine the intended canonical representative.
Moreover, Corollary~\ref{cor:distinguishing_line_graph_cases} shows
that this notation depends only on the canonical line-graph invariants, not on
the particular generating set.
Concretely, these are the number $n_\Delta$ of vertices of the root multigraph,
the orbit-radical dimension $q=\dim\radorbit(\vgens)$, and the radical
dimension $r=\dim\rad(\vgens)$.
For even $n_\Delta$, the distinction between the even and exceptional cases is
whether the extra T-join radical vector lies outside the orbit radical
($r=q+1$) or collapses into it ($r=q$).

\begin{ex}[The low-dimensional odd boundary]\label{ex:low_dimensional_odd_boundary}
For $m=1$, the odd family in
Eq.~\eqref{eq:def:binary_transvection_groups_free_fermionic} gives
\[
    \graphP_{2m-1,r+1}=\graphP_{1,r+1}.
\]
By the canonical-path ambiguity, this can also be written as
$\graphP_{2,r}$, but in the present notation we treat it as the odd
representative.
For $r=0$, the ambient transvection group is
\[
    \Sym(3,0,n,\F_2)=\tvgrouplong{\vgens(\graphP_{1,1})}.
\]
Its induced action on $V=\Span[\F_2]{\vgens(\graphP_{1,1})}$ is
\[
    \Sym(3,0,n,\F_2)|_V\cong S_3\cong\Sp(2,\F_2).
\]
The corresponding Pauli Lie algebra is
\[
    \so(3)\cong\su(2),
\]
with Pauli basis $\MAJ_{1,2}^{(2)}$.
For $r>0$, the same $S_3$ target is extended by the blow-up kernel
appearing in Proposition~\ref{prop:free_fermionic_transvection_symmetric_groups}.
\end{ex}

When only the induced action on the span of the generators is relevant,
we suppress the ambient parameter $n$ from
Eq.~\eqref{eq:def:binary_transvection_groups_free_fermionic}.
For any ambient $n$ for which the preceding definitions make sense, define

\begin{subequations}
\begin{align}
    &\Sym(2m,r,\F_2) := \\
    &\Sym(2m,r,n,\F_2)|_{\Span{\vgens(\graphP_{2m-2,r})}},\nonumber\\
    &\Sym(2m{+}1,r,\F_2):= \label{eq:def:restricted_binary_transvection_groups_free_fermionic} \\
    &\Sym(2m{+}1,r,n,\F_2)|_{\Span{\vgens(\graphP_{2m-1,r+1})}},
    \nonumber \\
    &\Sym^D(2m{+}2,r,\F_2):= \\
    &\Sym^D(2m{+}2,r,n,\F_2)|_{\Span{\vgens^D(\graphP_{2m,r+1})}}.
    \nonumber
\end{align}
\end{subequations}
These equations describe the corresponding restricted transvection groups on
$V=\Span{\vgens}$.
The notation is chosen because the quotient groups appearing below are
symmetric groups, while the general groups are extensions of these symmetric
groups.
Specifically, the groups above are defined as matrix groups, hence as concrete
representations of symmetric groups over binary vector spaces.
To make the target symmetric-group actions explicit, we recall the
standard actions of permutations on a vector space.
Representation-theoretically, the same symmetric-group actions in
Proposition~\ref{prop:free_fermionic_transvection_symmetric_groups} below can be made
explicit using the permutation module and its even-weight
submodule \cite{Cameron_Hall_1991,Dickson_1908}.
For $d\geq 2$, write
\begin{equation}\label{eq:E:d}
    E_d:=\{x\in\F_2^d\mid \sum_{a=1}^d x_a=0\} \;\text{ and }\;
    \mathbf 1=(1,\ldots,1).
\end{equation}
The group $S_d$ acts on $\F_2^d$ by permuting coordinates and preserves
$E_d$.
Write $\basel_1,\ldots,\basel_d$ for the standard coordinate basis of $\F_2^d$.
For the even-weight submodule of the $\F_2$-permutation module, the symplectic transvections
which occur inside the symmetric-group action are exactly the transpositions \cite[p.~187]{Cameron_Hall_1991}.
In coordinates, the action of a transposition $(ij)$ on $x\in E_d$ is
\begin{align}\label{eq:cameron:hall:transvection}
    x \mapsto \tau_{\basel_i+\basel_j}(x) & =
    x+\symp{\basel_i+\basel_j}{x}(\basel_i+\basel_j) \nonumber \\
    &=
    x+(x_i+x_j)(\basel_i+\basel_j).
\end{align}
Here the restricted symplectic form on $E_d$ is the restriction of the
usual dot product, so $\symp{\basel_i+\basel_j}{x}=x_i+x_j$.
Thus, if $d$ is odd, the relevant representation is the action of $S_d$ on
$E_d$.
If $d$ is even, then $\mathbf 1\in E_d$ spans the radical of the restricted
symplectic form, and the corresponding non-degenerate representation is
the action of $S_d$ on $E_d/\SpanS[\F_2]{\{\mathbf 1\}}$.
As we now show in Proposition~\ref{prop:free_fermionic_transvection_symmetric_groups},
the odd case uses $E_{2m+1}$, the even case uses $E_{2m}$ with
non-degenerate action on
$E_{2m}/\SpanS[\F_2]{\{\mathbf 1\}}$, and the exceptional dependent case uses
the quotient
$E_{2m+2}/\SpanS[\F_2]{\{\mathbf 1\}}$.
With this choice of coordinates in mind, the relevant group isomorphisms,
orbit spanning statement (in the sense of Lemma~\ref{lem:one_generator_blowup_transvection_kernel_candidate}), and extension sequences are as follows.

\begin{prop}[Symmetric groups and transvection groups]\label{prop:free_fermionic_transvection_symmetric_groups}
With the notation from
Eq.~\eqref{eq:def:restricted_binary_transvection_groups_free_fermionic}, the
following hold.
\begin{enumerate}
    \item\label{prop:free_fermionic_transvection_symmetric_groups:isomorphisms}
    \emph{Target symmetric groups}
    \cite{Janssen_1983,humphries_1985,Brown_Humphries_1986b,Huang_2012,Cameron_Hall_1991,Dickson_1908}.
    \begin{align*}
        \Sym(2m,1,\F_2) & \cong S_{2m},\\
        \Sym(2m{+}1,0,\F_2) & \cong S_{2m+1},\\
        \Sym^D(2m{+}2,0,\F_2) & \cong S_{2m+2}.
    \end{align*}
    with representations described by the modules $E_{2m}$, $E_{2m+1}$ and $E_{2m+2}/\Span{\mathbf{1}}$, respectively.

    \item\label{prop:free_fermionic_transvection_symmetric_groups:exact-sequences}
    \emph{Exact sequences} \cite{Brown_Humphries_1986a,Brown_Humphries_1986b}.
    The projection-induced homomorphism to the quotient by the orbit radical
    gives short exact sequences of the form
    \begin{nestedcaseenum}
        \item The even case has
        \[
            \begin{aligned}
            1&\to \F_2^{(2m-2)(r-1)}
            \to \Sym(2m,r,\F_2)\\
            &\to \Sym(2m,1,\F_2)
            \to 1.
            \end{aligned}
        \]
        \item The odd case has
        \[
            \begin{aligned}
            1&\to \F_2^{2mr}
            \to \Sym(2m{+}1,r,\F_2)\\
            &\to \Sym(2m{+}1,0,\F_2)
            \to 1.
            \end{aligned}
        \]
        \item The exceptional case has
        \[
            \begin{aligned}
            1&\to \F_2^{2mr}
            \to \Sym^D(2m{+}2,r,\F_2)\\
            &\to \Sym^D(2m{+}2,0,\F_2)
            \to 1.
            \end{aligned}
        \]
    \end{nestedcaseenum}
\end{enumerate}
\end{prop}

\begin{proof}
\emph{Group isomorphisms.}
For item~\ref{prop:free_fermionic_transvection_symmetric_groups:isomorphisms},
we use the path-transvection argument from
\cite[Theorem~3.4]{Brown_Humphries_1986b}; see also
\cite{Janssen_1983,humphries_1985,Huang_2012,Cameron_Hall_1991,Dickson_1908}.
For the ordinary path representative $\graphP_{k,1}$, set
$d=k+2$, write the path generators as $a_1,\ldots,a_{d-1}$, and define
\[
    x_{ij}:=a_i+a_{i+1}+\cdots+a_{j-1}
    \qquad \text{for } 1\leq i<j\leq d.
\]
The interval sums $x_{ij}$ are the elements of the generator orbit from
Eq.~\eqref{eq:orbits_blown_up_path_graph} when $n_1=1$.
Using the choice of coordinates from Eq.~\eqref{eq:E:d}, identify $x_{ij}$ with
$\basel_i+\basel_j\in E_d$.
We can identify $a_i=x_{i,i+1}$ with $\basel_i+\basel_{i+1}$ (which defines a vector space isomorphism, since they are both $d-1=k+1$ linearly independent vectors).
By Eq.~\eqref{eq:cameron:hall:transvection}, the transvection by $a_i$ acts on
the interval vector $x_{pq}$ as the transvection $\tau_{\basel_i+\basel_{i+1}}$ acts on
$\basel_p+\basel_q$.
The symplectic product is $1$ exactly when the pair $\{p,q\}$ contains
precisely one of $i$ and $i+1$.
In that case the formula replaces that endpoint by the other one; otherwise
the pair is fixed.
Hence $\tau_{a_i}$ acts on the interval vectors as the simple transposition
$(i\,\,i{+}1)$ acts on the two-element subsets of $[d]$.
The path transvections $x_{i,i+1}$ therefore generate the permutation action of
$S_d$ on $E_d$.

This action is faithful for the target dimensions used here.
Indeed, if a permutation $\sigma\in S_d$ acts trivially on $E_d$, then it fixes
every vector $\basel_i+\basel_j$, equivalently every two-element subset $\{i,j\}$.
For $d\geq3$, the action of $S_d$ on the two-element subsets of $[d]$
is faithful as we explain now.
Suppose that $\sigma$ fixes every two-element subset.
If $\sigma(i)\neq i$, choose $j\notin\{i,\sigma(i)\}$, which is possible
because $d\geq3$.
Since $\sigma$ fixes the subset $\{i,j\}$, we must have
$\sigma(i)\in\{i,j\}$.
This contradicts both $\sigma(i)\neq i$ and $j\neq\sigma(i)$.
Hence $\sigma(i)=i$ for every $i$, so $\sigma$ is the identity.
Applying this calculation to the first two cases
gives the group isomorphisms $\Sym(2m,1,\F_2)\cong S_{2m}$ and
$\Sym(2m{+}1,0,\F_2)\cong S_{2m+1}$.

For $\vgens^D(\graphP_{2m,1})$, note first that
$\mathbf{1}=\sum_{i=1}^{2m+2}\basel_i=\sum_{i=1}^{m+1}(\basel_{2i-1}+\basel_{2i})$.
Under the identification above between $E_{2m+2}$ and
$\Span[\F_2]{\vgens(\graphP_{2m,1})}$, this vector corresponds to the linear
dependency $\sum_{i=1}^{m+1}a_{2i-1}=0$ for
$\vgens^D(\graphP_{2m,1})$.
Hence, we can identify $\Span[\F_2]{\vgens^D(\graphP_{2m,1})}$ with the quotient
$E_{2m+2}/\SpanS[\F_2]{\{\mathbf 1\}}$ described before the lemma.
The transvection action on this quotient is the natural action induced by the
projection, which is symplectic because $\mathbf{1}$ spans the radical of
$E_{2m+2}$.
This quotient action is still faithful for $m\geq2$.
If $\sigma\in S_{2m+2}$ acts trivially on
$E_{2m+2}/\SpanS[\F_2]{\{\mathbf 1\}}$, then for all $i<j$ we have
$\basel_{\sigma(i)}+\basel_{\sigma(j)}+\basel_i+\basel_j\in
\SpanS[\F_2]{\{\mathbf 1\}}$.
Here the weight of a vector in $\F_2^{2m+2}$ is the number of its nonzero
coordinates.
The left-hand side has weight $0$, $2$, or $4$, while $\mathbf 1$ has weight
$2m+2\geq6$.
Thus the left-hand side must be zero for every pair, so $\sigma$ fixes every
two-element subset and hence is the identity.
Therefore $\Sym^D(2m{+}2,0,\F_2)\cong S_{2m+2}$.

\begin{table}[t]
\caption{Input dimensions for the kernel formula in
Proposition~\ref{prop:free_fermionic_transvection_symmetric_groups}. Here
$W=\ker(\projection)$ is the kernel of the blow-up projection from $\tilde{V}$
to $V$. For each row, $\tvprojker$ denotes the kernel of the corresponding
induced transvection-group homomorphism $\tvproj$ in
\ref{prop:free_fermionic_transvection_symmetric_groups:exact-sequences}.}
\label{tab:free-fermionic-transvection-kernel-dimensions}
\footnotesize
\begin{tabular*}{\columnwidth}{@{\hspace{1mm}}l@{\extracolsep{\fill}}c@{\hspace{2mm}}c@{\hspace{2mm}}c@{\hspace{1mm}}}
\hline\hline
\\[-2.5mm]
case & $\dim(V/\rad(V))$ & $\dim(W)$ & $\dim \tvprojker$
\\[0.5mm] \hline
\\[-2.0mm]
$\Sym(2m,r,\F_2)$ & $2m-2$ & $r-1$ & $(2m-2)(r-1)$\\[1.5mm]
$\Sym(2m{+}1,r,\F_2)$ & $2m$ & $r$ & $2mr$\\[1.5mm]
$\Sym^D(2m{+}2,r,\F_2)$ & $2m$ & $r$ & $2mr$\\[1.5mm]
\hline\hline
\end{tabular*}
\end{table}

\emph{Blown-up projection.}
For \ref{prop:free_fermionic_transvection_symmetric_groups:exact-sequences},
we use
Lemma~\ref{lem:one_generator_blowup_transvection_kernel_candidate}, which is 
based on \cite[Proposition~6.6]{Brown_Humphries_1986a}.
In the canonical labels of Fig.~\ref{fig:humphries_classes}, the projection
from $\graphP_{k,n_1}$ to the corresponding target representative is
\[
    a_i\mapsto a_i,\qquad b_j\mapsto b_1=a_{k+1}.
\]
It collapses the false-twin class $\{b_1,\ldots,b_{n_1}\}$ and is
injective on the remaining generators, so
\[
    W=\ker(\projection)
    =
    \SpanS[\F_2]{\{b_j+b_1\}_{j=2}^{n_1}} = \radorbit(\vgens)
\]
and $\dim(W)=n_1-1$.

For the spanning hypothesis in
Lemma~\ref{lem:one_generator_blowup_transvection_kernel_candidate},
under the identifications in
\ref{prop:free_fermionic_transvection_symmetric_groups:isomorphisms}, the
collapsed generator $b_1$ is $x_{k+1,k+2} = a_{k+1}$.
The symmetric group acts transitively on the interval sum $x_{ij}$, and is generated by $\{x_{i,i+1}\}_{i=1}^{k+1}$ as a transvection group.
Thus, the orbit of $b_1+\rad(V)$ spans $V/\rad(V)$ for each of the three target
representatives in
\ref{prop:free_fermionic_transvection_symmetric_groups:isomorphisms}.

Hence, applying Lemma~\ref{lem:one_generator_blowup_transvection_kernel_candidate}, the kernel $\tvprojker$ of the induced homomorphism $\tvproj$ is
elementary abelian with
\[
    \dim \tvprojker=\dim(V/\rad(V))\dim(W).
\]

\emph{Dimension count.}
We now compute the quotient dimension $\dim(V/\rad(V))$ for each of the three
target representatives.
These computations use the canonical labels from Fig.~\ref{fig:humphries_classes}
and the radical basis in
Lemma~\ref{lem:Canonical_t_equivalent_Graph_Radical}.
For $\vgens(\graphP_{2m-2,1})$, there are $2m-1$ path generators.
Since $\graphP_{2m-2,1}=\graphP_{2(m-1),1}$, the additional spine radical
vector from
Lemma~\ref{lem:Canonical_t_equivalent_Graph_Radical}\ref{lem:Canonical_t_equivalent_Graph_Radical:path-extra}
is
$\sum_{j=1}^{m}a_{2j-1}$.
Hence
\[
    \dim(V/\rad(V))=(2m-1)-1=2m-2.
\]
For the set $\vgens(\graphP_{2m-1,1})$, there are $2m$ path generators and
Lemma~\ref{lem:Canonical_t_equivalent_Graph_Radical} gives no radical
generator when $n_1=1$.
Thus
\[
    \dim(V/\rad(V))=2m.
\]
For $\vgens^D(\graphP_{2m,1})$, the linearly independent extension has
$2m+1$ path generators and one spine radical vector
$\sum_{j=1}^{m+1}a_{2j-1}$.
The dependent representative is obtained by imposing exactly this vector as
the unique algebraic dependency, as recalled before
Proposition~\ref{prop:valid_colorings_even_odd_cases_alg_ind}.
Consequently the span has dimension $(2m+1)-1=2m$, and the remaining radical
is trivial in this target case because there are no leg radicals when $n_1=1$
and Lemma~\ref{lem:Canonical_t_equivalent_Graph_Radical} lists no other
radical generators.
Hence
\[
    \dim(V/\rad(V))=2m.
\]
Together with the already computed value $\dim(W)=n_1-1$, multiplying
$\dim(V/\rad(V))$ and $\dim(W)$ gives the exponents in
\ref{prop:free_fermionic_transvection_symmetric_groups:exact-sequences}.
The input dimensions are summarized in
Table~\ref{tab:free-fermionic-transvection-kernel-dimensions}.
The first two rows are the path blow-up calculation of
\cite[Theorem~4.3]{Brown_Humphries_1986b}, written in the present notation.
\end{proof}

The choice of coordinates used before the lemma also has a direct Majorana
interpretation.
In the Majorana realization, the same transvection formula becomes
especially concrete.
As used in the literature \cite{Sierant_Turkeshi_Tarabunga_2026}, the
transvection $\tau_{\mu_i+\mu_j}$ defined by a quadratic Majorana with
$i\neq j$ acts by swapping the two Majorana basis vectors $\mu_i$ and
$\mu_j$:
\begin{equation}\label{eq:majorana:transvection}
\begin{aligned}
    \tau_{\mu_i+\mu_j}\mu_o
    &=
    \mu_o+\symp{\mu_i+\mu_j}{\mu_o}(\mu_i+\mu_j)\\
    &=
    \begin{dcases}
        \mu_o & \text{if } o\neq i,j,\\
        \mu_i & \text{if } o = j,\\
        \mu_j & \text{if } o = i,
    \end{dcases}
\end{aligned}
\end{equation}
Thus these transvections $\tau_{\mu_i+\mu_j}$ realize the standard
coordinate-permutation action of transpositions on the Majorana basis.
Under the coordinate identification of $\basel_o$ with $\mu_o$, the center
$\basel_i+\basel_j$ corresponds to $\mu_i+\mu_j$.
Under the same identification, for $d=2n$, the even-weight subspace $E_d$
corresponds to the subspace
of vectors $v\in\Fn$ whose Majorana coordinates satisfy
$\sum_{o=1}^{2n}\majcomponent{v}{o}=0$.
This subspace is preserved by the transpositions, and restricting
Eq.~\eqref{eq:majorana:transvection} to it gives exactly the
formula on $E_d$ in Eq.~\eqref{eq:cameron:hall:transvection}, as in
\cite[p.~187]{Cameron_Hall_1991}.
This is the same symmetric-group mechanism as in
Proposition~\ref{prop:free_fermionic_transvection_symmetric_groups}, now written
directly in Majorana coordinates instead of canonical path-interval
coordinates.
This gives the following full Majorana-coordinate version for quadratic
Majorana generating sets whose root graph is connected.
\begin{lem}[Connected quadratic Majorana transvection groups]\label{lem:connected_quadratic_majorana_transvection_groups}
Let $n\geq 3$, let $\Delta$ be a connected simple root graph with vertex set
$[2n]$, and let
\[
    \vgens=\{\mu_i+\mu_j\mid \{i,j\}\in\edges(\Delta)\},
    \qquad
    \pgens=\isolong{\vgens}.
\]
Equivalently, $\pgens\simeq
\{\gamma_i\gamma_j\mid \{i,j\}\in\edges(\Delta)\}$ is the corresponding
quadratic Majorana generating set, up to scalar phases.
Thus, in the sense of Definition~\ref{def:line_graphs}, the frustration graph
of $\pgens$ is the line graph $L(\Delta)$.
The following holds:
\begin{enumerate}
    \item By Eq.~\eqref{eq:majorana:transvection}, the action of
    $\tvgroup{\vgens}$ on the Majorana basis $\{\mu_i\}_{i=1}^{2n}$ is the
    coordinate-permutation representation of $S_{2n}$.
    Its induced action on $V=\Span{\vgens}$ is
    $\Sym(2n,1,\F_2)\cong S_{2n}$.
    \item The Pauli basis of $\lie{\pgens}$ is
    \begin{align*}
        \baslong{\lie{\pgens}}
        &=
        \{\majiso{\mu_i+\mu_j}\mid 1\leq i<j\leq 2n\}\\
        &\simeq
        \{\gamma_i\gamma_j\mid 1\leq i<j\leq 2n\}\\
        &=
        \baslong{\lie{\pgens(\graphP_{2n-2,1})}}.
    \end{align*}
    In particular, $\lie{\pgens}=\so(2n)$.
    \item The center of the generated matrix algebra $\algclosure{\pgens}$ is spanned by $\baslong{\ZZ(\algclosure{\pgens})} = \{I,\Gamma\}$.
\end{enumerate}
\end{lem}
\begin{proof}
To prove (a), we use Eq.~\eqref{eq:majorana:transvection}: the
transvection with center $\mu_i+\mu_j$ acts as the transposition $(ij)$ on the
Majorana basis.
This is the Majorana-coordinate version of the transposition-generation
argument used in
Proposition~\ref{prop:free_fermionic_transvection_symmetric_groups}.
We also use the known fact \cite[Lem.~3.10.1, p.~52]{godsil2013algebraic} that a set of transpositions $\{S_{ij}\}_{ij}$ over $d$ objects generates $S_d$ if and only if $ij$ are edges for a connected graph.
Hence, the transvections with centers $\{\mu_i+\mu_j\}_{\{i,j\}\in\edges(\Delta)}$ generate $S_{2n}$ iff $\Delta$ is a connected graph.
This is precisely the restricted group $\Sym(2n,1,\F_2)$ from
Eq.~\eqref{eq:def:restricted_binary_transvection_groups_free_fermionic}, and
the action is the coordinate-permutation representation on
$\{\mu_i\}_{i=1}^{2n}$.

For (b), (a) gives all transpositions on the Majorana basis.
Since $\Delta$ is connected, for any two vertices $i$ and $j$ there is a path
from $i$ to $j$.
Repeated commutators of the quadratic Majoranas along this path generate
$\majiso{\mu_i+\mu_j}$ up to scalar phase.
Thus the Lie closure contains all quadratic Majorana strings
$\majiso{\mu_i+\mu_j}$ with $1\leq i<j\leq 2n$.
Equivalently, under the transposition action from (a), all quadratic Majoranas
lie in the same orbit as the path representative
$\pgens(\graphP_{2n-2,1})
\simeq \{\gamma_i\gamma_{i+1}\}_{i=1}^{2n-1}$.
This gives the displayed Pauli basis and hence $\lie{\pgens}=\so(2n)$.

Finally, (c) follows from the standard even Clifford algebra generated
by quadratic Majoranas.
All quadratic Majoranas commute with the parity operator $\Gamma$, and the
Pauli-string center of the full even Clifford algebra is exactly
$\Span[\C]{\{I,\Gamma\}}$.
\end{proof}
The even subspace $\majprod^\perp$ is the Majorana-coordinate version of
$E_{2n}$ above.
Notice that in the standard representation $S_{2n}$ is reducible, with an irreducible subspace given by the even subspace $\majprod^\perp$, since it only permutes components (in fact, it conserves the Majorana length itself, not just its parity).
It is however \emph{not} decomposable as a direct sum of even and odd
vectors.
Indeed, the odd vectors are not closed under addition, since the sum of two odd
vectors is even.
For any fixed odd vector $o$, they form instead the affine coset
$o+\majprod^\perp$ of the even subspace.
Finally, we also define the corresponding Clifford transvection groups:
\begin{subequations}\label{eq:free_fermionic_clifford_transvection_groups}
    \begin{align}
        \clSym(2m,r,n,\F_2) &= \cltvgrouplong{\pgens(\graphP_{2m-2,r})}\\
        &\text{for } m\geq 3,\, r\geq 1, \nonumber\displaybreak[2]\\
        \clSym(2m{+}1,r,n,\F_2) &= \cltvgrouplong{\pgens(\graphP_{2m-1,r+1})}\\
        &\text{for } m\geq 1,\, r\geq 0, \nonumber\displaybreak[2]\\
        \clSym^D(2m{+}2,r,n,\F_2) &= \cltvgrouplong{\pgens^D(\graphP_{2m,r+1})}\\
        &\text{for } m\geq 2,\, r\geq 0. \nonumber
    \end{align}
\end{subequations}
Applying the Clifford-to-binary map $\SY$ from Eq.~\eqref{eq:proj:conj}
to these Clifford transvection groups gives the corresponding binary
transvection groups, modulo the Pauli kernel as in
Lemma~\ref{lem:clifford_transvection_group_binary_image_kernel}.
Thus the associated binary action is the permutation-group action described
above.
This suggests a natural comparison with the Clifford elements inside the
corresponding Pauli Lie groups.
In the minimal even case, this is closely related to the Clifford intersection
of the matchgate group, which acts on Majorana operators by signed permutations
\cite{Sierant_Turkeshi_Tarabunga_2026}.
From the present binary viewpoint, the unsigned part of this signed-permutation
action is precisely the symmetric-group transvection action described above.
Thus one expects a close relationship between Clifford transvection groups and
Clifford intersections with the generated Lie groups, with the remaining point
being the comparison of the Pauli kernels, i.e., the sign or phase part
invisible to the binary symplectic action.

\ManuscriptPart{Classification}{part:final_classification}

\section{Classification of Transvection Groups and Pauli Lie Algebras}\label{sec:classification:groups_lie_algebras}

Finally, combining all our previous results, we state in this section the classification of transvection groups and Lie algebras for Pauli generating sets with connected frustration graph. 
Namely, in view of the distinction between line graphs and all other graphs, we give the following definition for generating sets:
\begin{defn}
A generating set $\pgens\subseteq\PP_n$ of Pauli strings or a set $\vgens\subseteq\Fn$ of binary vectors is said to be of \emph{free-fermionic type} if its frustration graph is the line graph of a multigraph, and it is of \emph{quasi-universal type} otherwise.
\end{defn}
Equivalently, we can view the generating sets of free-fermionic type as those which are $\calE_6$-free or $t$-equivalent to a blown-up path graph (by Theorem~\ref{thm:equivalent_line_graph_conditions}).
Hence, quasi-universal cases also possess the rank constraint $\rank(\vgens)\geq 6$, given that $\calE_6$ is a non-degenerate graph of rank $6$.
Under this definition, we have seven possible classes, with three of free-fermionic type and four of quasi-universal type, as in \cite{Janssen_1983}, which correspond to the minimal generating sets in Corollary~\ref{cor:minimal_canonical_sets}:
\begin{enumerate}
    \item Quasi-universal:
    \begin{enumerate}[label=(\arabic*)]
        \item No invariant quadratic forms
        \item Invariant quadratic forms of type $+$
        \item Invariant quadratic forms of type $-$
        \item Invariant quadratic forms of type $0$
    \end{enumerate}
    \item Free-fermionic:
    \begin{enumerate}[label=(\arabic*), resume]
        \item Odd type
        \item Even type
        \item Exceptional type
    \end{enumerate}
\end{enumerate}

We finally have all necessary tools to tackle the classification of transvection groups, Lie algebras and orbits through the use of objects which are manifestly invariant under $t$-equivalence or minimality conditions.
We briefly recall these, in the binary language:
\begin{enumerate}
    \item frustration graph $\frustration{\vgens}$ (Definition~\ref{defn:frustration})
    \item radical $\rad(\vgens)$ (see Eq.~\eqref{eq:radical}) and orthogonal complement  $\vgens^\perp$ (see Eq.~\eqref{eq:def:orthogonal_complement})
    \item unique linearly independent extension $\projection(\tilde{\vgens})=\vgens$ (see Lemmas~\ref{lem:Bijection_Graphs_Linearly_Independent_Sets} and \ref{lem:extensions}(b)) and the linear dependency space $\ker(\projection)$ (see Lemmas~\ref{lem:dependency}(b) and ~\ref{lem:projection_extension_basic_properties}(a))
    \item invariant quadratic forms $\quadratic(\vgens)$ (see Eq.~\eqref{eq:invariant_quadratic_forms_general_group_subsets}, Lemma~\ref{lem:affine_structure_invariant_quadratic_form_subspace}, and Lemma~\ref{cor:distinguishing_quasi_universal_cases})
    \item anisotropic radical $\radzero(V,\QQ)$
    and isotropic radical
    $\radone(V,\QQ)$ (see Definition~\ref{def:isotropic:radical})
    \item if $\frustration{\vgens}$ is a line graph of a multigraph $\Delta$ (see Definition~\ref{def:line_graphs}, Theorem~\ref{thm:uniqueness_root_graph_line_graph}, Theorem~\ref{thm:equivalent_line_graph_conditions}) and its incidence matrix $M(\Delta)$ (see Eq.~\eqref{eq:def:incidence_matrix})
    \begin{nestedcaseenum}
        \item space $\ker(M(\Delta))$ of cycles in $\Delta$ (see Lemma~\ref{lem:cycle_space_incidence_matrix_multigraph})
        \item number $n_\Delta$ of vertices (see Thms.~\ref{thm:equivalent_line_graph_conditions} and \ref{thm:equivalent_path_graph_conditions})
        \item free-fermionic mapping (see Def.~\ref{defn:free_fermionic_mappings}, Prop.~\ref{prop:full_Fn_PPn_Majorana_basis_free_fermionic_mapping})
    \end{nestedcaseenum}
    \item orbit radical $\radorbit(\vgens)$ (see Def.~\ref{def:orbit:radical}, Lemma~\ref{cor:distinguishing_line_graph_cases})
\end{enumerate}

\subsection{Free-Fermionic Case}

We start with the classification for the free-fermionic cases.
\begin{thm}[Free-Fermionic Algebraically Independent Sets]\label{thm:Classification_algebraically_independent_free_fermions_line_graphs_E6_free}
Let $\pgens = \isolong{\vgens}\subseteq\PP_n$ be a set of algebraically independent Pauli strings, with connected frustration graph $\frustration{\pgens} = \graphG$.
Also, $\rank(\vgens)=2m$, $r=\nullity(\vgens)$, $q = \nullorbit(\vgens)=\dim\radorbit(\vgens)$ (see Eq.~\eqref{eq:nullity_orbit_radical_paulis}) and $\delta=r-q+1$.
Then, the following are equivalent:
\begin{enumerate}[itemsep=0.05cm]
    \item $\graphG$ is the line graph of a multigraph $\Delta$,
    \item $\pgens$ admits a free-fermionic mapping,
    \item $\lie{\pgens} = \so(2m+\delta)^{\oplus 2^q}\otimes I^{\otimes(n-m-r)}$,
    \item $\tvgroup{\vgens} = \Sym(2m+\delta,r,n,\F_2)$,
\end{enumerate}
where the representations of the Lie algebra are described in Theorem~\ref{thm:Free_Fermionic_Lie_Algebras}\ref{thm:Free_Fermionic_Lie_Algebras:even} and~\ref{thm:Free_Fermionic_Lie_Algebras:odd}, depending on $\delta$.
\end{thm}

\begin{table*}[t]
\caption{Low-dimensional free-fermionic, connected Pauli Lie algebras and transvection groups, with $q=0$ (where the isomorphisms naturally extend to arbitrary $q$). We write $\lieiso(B)$ for $B$ a single invariant form and $\lieiso(\commalg,B)$ for an additional Pauli commutant $\commalg$. We choose $m$ such that $2m$ is the number of logical Majorana modes for the free-fermionic mapping $\pgens\subseteq\PP_m$ (given no non-abelian and cycle symmetries $q=\ell=0$), hence respects the parametrization in Corollary~\ref{cor:free_fermionic_arbitrary_generating_sets_classification}.}
\label{tab:exceptional_isomorphisms}
\centering
\footnotesize
\begin{tabular*}{\textwidth}{@{\hspace{1mm}}l@{\extracolsep{\fill}}C@{\hspace{2mm}}C@{\hspace{2mm}}C@{\hspace{2mm}}C@{\hspace{2mm}}l@{\hspace{2mm}}C@{\hspace{2mm}}C@{\hspace{1mm}}C@{\hspace{1mm}}}
\hline\hline
\\[-2.5mm]
        Type & \text{Canonical } \graphG & m & \lieg & \text{Exceptional } \lieg' \cong \lieg & Representation & \lieg \conjugated \lieiso^0 & \tvgroup{\vgens}|_V & \text{Exceptional } G\conjugated\tvgroup{\vgens}|_V\\[0.5mm]
\hline
\\[-2.0mm]
        odd & \graphP_2 & 1 & \so(3) & \su(2) \cong \usp(2) & irreducible & \lieiso^0(0) \conjugated \lieiso^0(\text{Y}) & \Sym(3,0,\F_2) & \Sp(2,\F_2) \conjugated \lieO^-(2,\F_2)\\[1mm]
        odd & \graphP_4 & 2 & \so(5) & \usp(4) & irreducible & \lieiso^0(\text{IY}) & \Sym(5,0,\F_2) & \lieO^-(4,\F_2)\\[1mm]
        even & \graphP_5 & 3 & \so(6) & \su(4) & reducible & \lieiso^0(\algclosure{\text{IIZ}}, \text{IIX}) & \Sym(6,1,\F_2) & \lieO_0^{\#}(4,1,\F_2)\\[1mm]
        exceptional & \graphP_5 & 2 & \so(6) & \su(4) & irreducible & \lieiso^0(0) & \Sym^D(6,0,\F_2) & \Sp(4,\F_2)\\[1mm]
        exceptional & \graphP_7 & 3 & \so(8) & - & irreducible & \lieiso^0(\text{III}) & \Sym^D(8,0,\F_2) & \lieO^+(6,\F_2)\\[1.0mm]
\hline\hline
\end{tabular*}
\end{table*}

\begin{proof}
Corollary~\ref{cor:free_fermionic_mappings_line_graphs} shows that (a) and (b) are equivalent.
Also, by Theorem~\ref{thm:Free_Fermionic_Lie_Algebras}, one can choose a basis for
$\so(2m+\delta)^{\oplus 2^q}\otimes I^{\otimes(n-m-r)}$ which is free-fermionic, hence (b) and (c) are equivalent.
By definition of $\Sym(2m+\delta,r,n,\F_2)$, $\vgens$ admits a minimal generating set which is $t$-equivalent to $\graphP_{2m,n_1}$ or $\graphP_{2m-1,n_1}$ (depending on $\delta$), hence it must be a line graph, or (d) and (a) are equivalent.
\end{proof}
For additional equivalent graph-theoretic conditions, see also Thm~\ref{thm:equivalent_line_graph_conditions} and \ref{thm:equivalent_path_graph_conditions}.
We can now state the classification for arbitrary generating sets of free-fermionic type, using Corollary~\ref{cor:distinguishing_line_graph_cases}, the Lie algebra classification Theorem~\ref{thm:Free_Fermionic_Lie_Algebras} and the definition of the transvection groups in the free-fermionic case Definition~\ref{eq:def:binary_transvection_groups_free_fermionic}:
\begin{cor}\label{cor:free_fermionic_arbitrary_generating_sets_classification}
Let $\pgens=\isolong{\vgens}\subseteq\PP_n$ be a generating set whose frustration graph is a line graph $\graphG=L(\Delta)$, $n_\Delta\neq 4$.
Let $2\ell=\rank(\vgens^\perp)$, $r=\nullity(\vgens)$ and $q=\nullorbit(\vgens)$.
Then, exactly one of the following holds:
\begin{enumerate}
    \item If $\pgens$ is of even type and $n_\Delta=2m$ even with $m\geq 3$, then
    \begin{nestedcaseenum}
        \item $\lie{\pgens} = \so(2m)^{\oplus 2^q}\otimes I^{\otimes \ell}$;
        \item $\tvgroup{\vgens} = \Sym(2m,r,n,\F_2)$;
        \item there exists a free-fermionic mapping such that $\baslong{\pgens} = \MAJ_2^{(2m)}\setprod\Cyc^{(2m,q)}$.
    \end{nestedcaseenum}
    \item If $\pgens$ is of odd type and $n_\Delta=2m+1$ odd with $m\geq 1$, then
    \begin{nestedcaseenum}
        \item $\lie{\pgens} =\so(2m{+}1)^{\oplus 2^q}\otimes I^{\otimes \ell}$;
        \item $\tvgroup{\vgens} = \Sym(2m{+}1,r,n,\F_2)$;
        \item there exists a free-fermionic mapping such that $\baslong{\pgens} = \MAJ_{1,2}^{(2m)}\setprod\Cyc^{(2m,q)}$.
    \end{nestedcaseenum}
    \item If $\pgens$ is of exceptional type and $n_\Delta=2m+2$ even with $m\geq 2$, then
    \begin{nestedcaseenum}
        \item $\lie{\pgens} =\so(2m{+}2)^{\oplus 2^q}\otimes I^{\otimes \ell}$;
        \item $\tvgroup{\vgens} = \Sym^D(2m{+}2,r,n,\F_2)$;
        \item there exists a free-fermionic mapping such that $\baslong{\pgens} = \MAJ_{\gamma}^{(2m)}\setprod\Cyc^{(2m,q)}$.
    \end{nestedcaseenum}
\end{enumerate}
\end{cor}

We close this subsection by discussing the low-rank cases and exceptional isomorphisms in terms of Lie algebras and binary transvection groups, mirroring the discussion in \cite{Janssen_1983}.
Specifically, we are interested in the free-fermionic cases which are also isometry Lie algebras or binary groups.
We can consider $\ell=0$ and $q=0$, since these only contribute multiple independent copies or degeneracies in the representation of the Lie algebra.
Similarly, for arbitrary $q$ and $\ell$, one recovers the corresponding transvection groups isomorphisms (here viewed as invertible change of basis, not just group isomorphisms).
See Table~\ref{tab:exceptional_isomorphisms} for a summary of these exceptional isomorphisms.
On the Lie algebra side, the exceptional isomorphisms are a known result of Lie-theory for semisimple Lie algebras \cite{hall2015}.
The corresponding statement for the transvection groups can be found in \cite{Janssen_1983}.
We comment here mostly on the Lie algebra side from the point of view of symmetries, though Table~\ref{tab:exceptional_isomorphisms} presents the isomorphisms also at the group level.

We briefly comment here on the Lie algebra side, highlighting not only the Lie algebra isomorphisms, but also why the specific \emph{representations} coincide, i.e., when free-fermionic Lie algebras can be viewed as (derived) Lie algebras of isometries.
The smallest case is $\lieg = \so(3)$ with canonical generating set $\vgens(\graphP_2)$.
Here, $\lieg$ is isomorphic to both $\su(2)$ and $\usp(2)$, where the free-fermionic representation of $\so(3)$ (of dimension $2$, as described in Theorem~\ref{thm:Free_Fermionic_Lie_Algebras}) coincides with the standard representations of $\su(2)$ and $\usp(2)$.
We find a double exceptional isomorphism, given that not only the derived Lie algebras of isometries $\lieiso^0(0) = \su(2)$ and $\lieiso(\text{Y})=\usp(2)$
are free-fermionic, but they also mutually coincide.
Indeed, one can check that all Pauli matrices conserve Y as an invariant bilinear form.
Equivalently, over two Majorana modes, all Majorana strings are linear or quadratic, hence conserve $\QQ_\gamma$, which is also of type $-$ for $m=1$.

The second smallest case is again of odd type, with $\lieg = \so(5)$.
This satisfies the exceptional isomorphism $\so(5)\cong \usp(4)$, hence lives in the standard $4$-dimensional representation of $\usp(4) = \lieiso(\text{IY})$, which is naturally defined on four Majorana modes.
From the binary point of view, this follows from the fact that the invariant quadratic form $\QQ_\gamma$ is $1$ for $\majL(v)\bmod 4\in\{1,2\}$
and here $\majL(v)\leq 4$, hence all anisotropic vectors for $\QQ_\gamma$ are either linear or quadratic, and $\QQ_\gamma$ is of type $-$ for $m=2$.

The only free-fermionic case of even type which admits an exceptional isomorphism appears for $m=3$, with $\so(6)\cong \su(4)$, in a reducible representation. 
From the Majorana perspective, the generators commute with $\majprod$ and conserve $\QQ_\gamma$ and $\QQ_{\majprod}^\gamma$, which are respectively of type $+$ and $-$ for $m=4$. Hence the set of invariant quadratic forms is of type $0$, which results in $\so(6) = \lieiso^0(\algclosure{\text{IIZ}}, \text{IIX}) \cong \su(4)$.
Over six modes, those Majorana strings which conserve $\QQ_\gamma$ and commute are those with $\majL(v)\in\{2,6\}$, where $\majL(v)=6$ iff $v=\majprod$.
Hence only the quadratic ones survive in the corresponding derived Lie algebra of isometries $\lieiso^0(\algclosure{\Gamma},B) \conjugated \so(6)$.

Finally, the remaining cases are both of exceptional type with $m=2$ and $m=3$.
For $m=2$, $\majL(v)\in \{1,2,3,4\}$, hence all non-zero Majorana strings are either the linear ones
with $\majL(v)=1$, the quadratic ones with $\majL(v)=2$, products of a linear string with the full parity product ($\majL(v)=3$) or the full parity
product with $\majL(v)=4$.  Thus $\so(6) \conjugated \lieiso^0(0) \cong \su(4)$.
For $m=3$, $\majL(v)\in[6]$, hence the Majorana strings in $\so(8)$ satisfy $\majL(v)\in\{1,2,5,6\}$. These
are precisely the Majorana strings that are anisotropic for $\QQ_\gamma$. Hence the type is $+$ for $m=3$ and $\so(8) \conjugated \lieiso^0(\text{I}) \cong \so(8)$.
The Lie algebras trivially coincide, however, from the representation-theoretic point of view, this is a consequence of the triality of $\so(8)$. The associated
three fundamental representations include the standard and the two spinor ones and they related by outer automorphisms.

\subsection{Quasi-Universal Case}

\begin{thm}[Quasi-Universal Algebraically Independent Sets]\label{thm:classification_quasi_universal_algebraically_independent_E6}
Let $\pgens = \isolong{\vgens}\subseteq\PP_n$ be a set of algebraically independent Pauli strings, with connected frustration graph $\frustration{\calB} = \graphG$ and $\rank(\algclosure{\pgens})\geq 6$. Also, $\commalg=\commutant(\pgens)$, $B=\iso{w}\in\bilinear(\pgens)\neq\emptyset$ and $\QQ$ the unique invariant quadratic form on $V=\Span{\vgens}$ (or $\QQ=\QQ_w|_V$).
Then, the following are equivalent:
\begin{enumerate}[itemsep=0.05cm]
    \item $\graphG$ is \emph{not} the line graph of a multigraph.
    \item $\tvgroup{\vgens}\cdot\vgens = \QQ^{-1}(1)\setminus\rad(\vgens)$.
    \item $\lie{\pgens} = \lieiso^0(\commalg,B)$.
    \item $\tvgroup{\vgens} = \lieO^{\#}(\QQ)$.
\end{enumerate}
\end{thm}
\begin{proof}
Let $\rank(V)=2m$ and $\nullity(V)=r$. We assume that $\vgens$ is quasi-universal (i.e., not $\calE_6$-free).
Then \cite[Theorem~3.5]{Janssen_1983} implies that the orbit of the generators consists of all anisotropic vectors which are not in the radical, i.e.,
$\tvgroup{\vgens}\cdot\vgens = \QQ^{-1}(1)\setminus\rad(V)$.
This is connected to the quasi-universal graphs being $t$-equivalent if and only if their invariant quadratic form is of the same type and rank (see Corollary~\ref{cor:distinguishing_quasi_universal_cases}).
This proves that (a) implies (b).

Assume now (b). We prove (a) by contradiction, hence assume that $\graphG$ is $\calE_6$-free.
Therefore, by Lemma~\ref{thm:equivalent_line_graph_conditions}, $\vgens$ is $t$-equivalent to an algebraically independent set $\vgens'$ whose frustration graph is $\graphP_{2m-1,n_1}$ or $\graphP_{2m,n_1}$ with $2m\geq 6$.
By Proposition~\ref{prop:valid_colorings_even_odd_cases_alg_ind}, the orbit $\tvgroup{\vgens'}\cdot\vgens'$ consists of the realizations of interval colorings on the path, modulo the orbit radical from Eq.~\eqref{eq:def:even_leg_coloring_subspace}.
On the other hand, if (b) holds, then the orbit consists of all realized
colorings whose vectors are anisotropic and not in the radical.
By Corollary~\ref{cor:quadratic_form_euler_characteristic}, anisotropic means
that the corresponding induced subgraph has Euler characteristic $1$ modulo
two. Since the graph is a blown-up path graph of diameter at least $6$, we can choose
the induced subgraph $\graphH$ supported on $\{a_1,a_3,a_5\}$.
This vector is not in the radical by
Lemma~\ref{lem:Canonical_t_equivalent_Graph_Radical}, but its induced subgraph
has Euler characteristic $1$ modulo two.
Hence it must lie in the orbit, contradicting the interval-coloring description
above.
We have now reached a contradiction, which proves that (b) implies (a).

By \cite[Theorem~3.8]{Janssen_1983}, the anisotropic transvections for $\QQ$ in $V\setminus\rad(V)$ generate $\lieO^{\#}(\QQ)$, hence (b) implies (d).
The reverse is also true, given that the restricted orthogonal group $\lieO^{\#}(\QQ)$ acts transitively on the level sets of $\QQ$ (excluding the radical),
hence in particular (d) implies (b).
Finally, by Corollary~\ref{cor:pauli:transvection} and by definition of $\lieiso^0(\commalg,B)$, (b) and (c) are equivalent.
This concludes the proof.
\end{proof}
Out of these cases, we highlight that one of smallest size, i.e.\ the one with $\abs{\pgens}=6=\rank(V)$, corresponds to the case with $\type(\QQ)=-$, hence $\lie{\pgens} = \usp(8)$ and $\tvgroup{\vgens}|_V = \lieO_-(6,\F_2)$.
Equivalently, $\pgens$ has a frustration graph that is contained in the set $\calE_6$.
Specifically, $\graphE_6$ is precisely the reference canonical graph for this class, since $\graphE_6 = \graphX_{5,1}^1 = \graphS_{2,1}^1$.
As a consequence, we also have that such Pauli Lie algebras (resp.\ transvection groups) are in fact subalgebras (resp.\ subgroups) of \emph{any} Pauli Lie algebra (resp.\ transvection group) generated by quasi-universal generating sets, independently of whether these are algebraically independent or not (since any graph in $\calE_6$ has trivial radical, hence no algebraic dependencies).

Given the characterization of the algebraically independent cases, it remains to understand the algebraically dependent ones.
For the quasi-universal types, we have the following characterization, independently of algebraic dependencies or minimality assumptions:
\begin{thm}[The Full Quasi-Universal Classification]\label{thm:Full_Classification_Theorem_w_E6_Quasi_Universal}
Let $\pgens = \isolong{\vgens}\subseteq\PP_n$ be a set of quasi-universal type. Also, let $V=\Span{\vgens}$ and $\commalg=\commutant(\pgens)$.
\begin{enumerate}
    \item If $\quadratic(\vgens)$ is empty, then $\tvgroup{\vgens}\cdot\vgens = V\setminus\rad(V)$, $\lie{\pgens} = \lieiso^0(\commalg)$ and $\tvgroup{\vgens}|_V = \Sp^{\#}(V)$.
    \item If $\exists w\in\quadratic(\vgens)$, then $\tvgroup{\vgens}\cdot\vgens = (\QQ_w^{-1}(1)\cap V)\setminus\rad(V)$, $\lie{\pgens} = \lieiso^0(\commalg,\iso{w})$ and $\tvgroup{\vgens}|_V = \lieO^{\#}(\QQ_w|_V)$.
\end{enumerate}
\end{thm}
\begin{proof}
By Corollary~\ref{cor:distinguishing_quasi_universal_cases}, we see that (lack of) quadratic forms correspond to algebraically independent (dependent) minimal generating sets.
If there are non-trivial quadratic forms, Theorem~\ref{thm:classification_quasi_universal_algebraically_independent_E6} applies.
If there are no invariant quadratic forms, it must of be of type $\graphX_{2m,n_1}^3$.
Then, there is an extension $\tilde{\vgens}$ with a unique invariant quadratic form $\tilde{\QQ}$ with $\type(\tilde{\QQ})=0$ over $\tilde{V}$, such that $\tvgroup{\tilde{\vgens}} = \tilde{\QQ}^{-1}(1)\setminus\rad(\tilde{V})$.
Also, algebraic dependencies for $\graphX_{2m,n_1}^3$ which are not Lie-algebraic dependencies are in the anisotropic radical by Lemma~\ref{lem:Limits_Lie_Algebraic_Dependencies_general}, or there is a vector $\tilde{a} = \sum_{i\in\vertices(\graphG)}\coloring(i)\tilde{v}_i\in\rad(\tilde{V})$ such that $\tilde{\QQ}(\tilde{a})=1$ and $\projection(\tilde{a})=0$.
Then, by Lemma~\ref{lem:projection_extension_basic_properties}\ref{lem:projection_extension_basic_properties:e}, the orbit for $\vgens$ is the projection $\projection(\tilde{\QQ}^{-1}(1)\setminus\rad(\tilde{V}))$.
For any $v\in V\setminus\rad(V)$, there is a $\tilde{v}$ such that $\projection(\tilde{v}) = v$ and $\tilde{v}\in\tilde{V}\setminus\rad(V)$.
If $\tilde{\QQ}(\tilde{v})=1$, then $v$ is in the orbit.
If $\tilde{\QQ}(\tilde{v})=0$, then we consider the vector $\tilde{v}+\tilde{a}$ such that
$\tilde{\QQ}(\tilde{v}+\tilde{a})=1$, $\tilde{v}+\tilde{a}\in \tilde{\QQ}^{-1}(1)\setminus\rad(\tilde{V})$ and $\projection(\tilde{v}+\tilde{a})=v$.
Again, $v$ is in the orbit.

Hence, $\tvgroup{\vgens}\cdot\vgens = V\setminus\rad(V)$, which implies $\lie{\pgens} = \lieiso^0(\commalg)$ by definition.
Finally, Lemma~\ref{lem:non_radical_transvections_generate_sp_sharp} implies
that the non-radical transvections generate the degenerate symplectic group,
or $\tvgroup{\vgens}|_V
    =
    \tvgroup{V\setminus\rad(V)}
    =
    \Sp^{\#}(V)$,
which concludes the proof.
\end{proof}

We also collect the following properties of sets of quasi-universal type
\begin{cor}\label{cor:quasi_universal_canonical_properties}
Let $\pgens = \iso{\vgens}\subseteq\PP_n$ be a generating set of quasi-universal type with $\graphG = \frustration{\pgens}$, $V=\Span{\vgens}$ and $\QQ$ being
the unique invariant quadratic form of $\graphG$.
The following holds:
\begin{enumerate}
    \item $\graphG$ is $t$-equivalent to one of $\graphX_{2m-1,n_1}^1, \graphX_{2m-1,n_1}^2$ or $\graphX_{2m,n_1}^3$ depending on the isomorphism class of $\QQ$ as in Proposition~\ref{prop:isomorphism_classes_quadratic_forms_arf_invariant_canonical}.
    \item $\graphG$ is $t$-equivalent to one of $\graphS_{n_2,n_1}^1, \graphS_{n_2,n_1}^2$ or $\graphS_{n_2,n_1}^3$ depending on the isomorphism class of $\QQ$ as in Proposition~\ref{prop:isomorphism_classes_quadratic_forms_arf_invariant_canonical}.
    \item $\graphG$ has an induced subgraph in $\calE_6$.
    \item There is a subgroup of $\tvgroup{\vgens}$, generated by transvections, which is isomorphic to $\lieO_-(6,\F_2)$.
    \item There is a subalgebra of $\lie{\pgens}$, spanned by Pauli strings, which is isomorphic to $\usp(8)$.
\end{enumerate}
\end{cor}

We also state a particularly useful criterion for \emph{strict} universality, up to linear symmetries, whenever the full transvection group cannot be generated by a set of free-fermionic type.
This follows immediately from Thm.~\ref{thm:Full_Classification_Theorem_w_E6_Quasi_Universal}:
\begin{cor}\label{cor:Symmetry_Breaking_su_Universality}
Let $\pgens=\isolong{\vgens}$ be a connected generating set with $\rank(\Span{\vgens})\geq 6$.
Then, $\lie{\pgens} = \lieiso^0(\commalg)$ if and only if $\pgens$ is of quasi-universal type and $\quadratic(\vgens)$ is empty.
\end{cor}
We can also phrase quasi-universality up to linear symmetries in terms of \emph{adding} an element to some generating set which conserves a quadratic form, while reformulating a result of \cite{Brown_Humphries_1986b} in the Pauli language:
\begin{cor}\label{cor:Symmetry_Breaking_su_Universality_with_basis}
Let $\pgens=\isolong{\vgens}$ be an algebraically independent generating set with $\rank(\Span{\vgens})\geq 6$ and let $\pgens' = \pgens\cup P$ with $P\in\algclosure{\pgens}$. Let $B\in\bilinear(\pgens)$ and $\commalg=\commutant(\pgens)$.
Then, $\lie{\pgens'} = \lieiso^0(\commalg)$ if and only if $\pgens'$ is of quasi-universal type (i.e., it has a frustration graph that connected and not a line graph) and $\theta_B(P)=-P$.
In particular, $\lie{\pgens'} = \su(2^n)$ if and only if $\pgens'$ is of quasi-universal, $\theta_B(P)=-P$ and $\commutant(\pgens)=\C I$.
\end{cor}
We can view this result as a symmetry-breaking requirement for universality, in the sense that, for a fixed commutant, a Pauli Lie algebra will be universal if and only if: (1) it breaks the free-fermionic property (by having an $\calE_6$ subgraph) and (2) it breaks all invariant quadratic forms.
Of course, this only works when $\Sp^{\#}(V)$ or $\lieiso^0(\commalg)$ cannot be generated by a set of free-fermionic type, hence the requirement that the rank be at least $6$, which excludes all such cases.

Notice that Theorem~\ref{thm:Full_Classification_Theorem_w_E6_Quasi_Universal} does not provide realizations of all Lie algebras and groups of isometries as Pauli Lie algebras and transvection groups, respectively, due to the rank constraint.
Hence, strictly, speaking, we cannot identify the quasi-universal cases with those cases with Lie algebras or binary groups of isometries.
We now discuss such cases.
This can be seen at the level of Table~\ref{tab:exceptional_isomorphisms}, which shows how low-rank cases are realized as free-fermionic types.
On the Lie algebra side, the missing cases are $\lieiso^0(\text{I})$, $\lieiso^0(\text{II})$ (i.e., the cases of type $+$ with $n\in\{1,2\}$), $\lieiso^0(\algclosure{\text{Z}},\text{X})$, $\lieiso^0(\algclosure{\text{IZ}},\text{IX})$ (i.e., the cases of type $0$ with $n\in\{1,2\}$).
On the binary group side, $\lieO_+(2,\F_2)$, $\lieO_+(4,\F_2)$, $\lieO_0^{\#}(0,1,\F_2)$ and $\lieO_0^{\#}(2,1,\F_2)$.
All other cases are either free-fermionic or quasi-universal.

For the Lie algebra which conserves an invariant bilinear form of $+$ type over $1$ qubit, there is a single skew-symmetric Pauli string over one qubit.
Thus $\lieiso(\text{I})= \so(2) \cong\lieu(1)$ and $\lieiso^0(\text{I}) \cong \{0\}$. 
Hence, this is excluded since we only consider connected generating sets of size at least two (where $\graphP_2$ is free-fermionic of odd-type).
Additionally, since $\lieiso(\text{I})$ is abelian, it also conserves the matrix algebra $\algclosure{\text{Y}}$, hence coincides with $\lieiso(\algclosure{Y},\text{I})$, which is of type $0$ over a single qubit.
Clearly, the addition of abelian symmetries $r\geq 1$ or non-abelian symmetries $\ell\geq 1$ still results in disconnected sets of Paulis
in the derived Lie algebra of isometries (i.e., it still remains abelian). Thus the corresponding derived Lie algebras are empty.
From the binary point of view, we have that $\lieO_+(2,\F_2) = \lieO_0(0,1,\F_2) \cong \Z_2$, and it consists of the identity and the transvection $\tau_{e+f}$.
  
Over two qubits, all skew-symmetric Pauli strings have a disconnected
frustration graph with components
$\{\text{IY,YX,YZ}\}$ and $\{\text{YI,XY,ZY}\}$.
Hence no connected generating set can generate
$\lieiso^0(\text{II})=\so(4)$ in its standard irreducible representation.
However, the Lie algebra
$\so(4)\cong\su(2)^{\oplus 2}\cong\so(3)^{\oplus 2}$
can still appear in a reducible representation, realized by a free-fermionic
set with a single cycle symmetry.
The corresponding canonical graph is $\graphP_{1,2}=\graphP_3$.
From the binary point of view, $\lieO_+(4,\F_2)$ instead acts transitively on all anisotropic vectors, hence the anisotropic transvections do not generate $\lieO_+(4,\F_2)$.
Indeed, they do not generate the swap, which conserves $B=\text{II}$ but is not obtained as a product of 2-qubit anisotropic transvections.

Finally, $\lieiso(\algclosure{\text{IZ}},\text{IX}) \cong \lieu(2)$ is spanned by the Paulis $\{\text{IZ, XZ, ZZ, YI}\}$, but $\lieiso^0(\algclosure{\text{IZ}},\text{IX}) \cong \su(2)$ is spanned by $\{\text{YI, XZ, ZZ}\}$, which has additional commutant elements beyond $\{\text{II, IZ}\}$, i.e., $\{\text{II, IZ, YY, YX}\}$.
This follows as the abelian factor $\lieu(1)$ acts differently in the two eigenspaces specified by IZ, whereas the non-abelian part $\su(2)$ acts the same in both.
From the point of view of representation theory, this exceptional behaviour follows from the fact that the standard representation of $\su(2)$ is dual to itself, whereas $\su(d)$ for any $d\geq 3$ has a dual representation which is distinct from the standard one.
Hence, a Lie algebra with invariant bilinear forms of type $0$ over two qubits can in fact be obtained as a free-fermionic case, though it requires non-abelian symmetries.

\subsection{Binary and Clifford Transvection Groups}\label{sec:classification_transvection_groups}

Given the knowledge of the restricted transvection groups $\tvgroup{\vgens}|_V$ for connected generating sets, we now also discuss the full transvection groups over $\Fn$ as well as the related Clifford transvection groups in $\matalg(2^n,\C)$.
In the quasi-universal case, we have:
\begin{enumerate}
    \item If $\vgens$ does not have invariant quadratic forms, then $\tvgroup{\vgens}\subseteq\Sp(2n,\F_2)_{V^\perp}$ with $\tvgroup{\vgens}|_V = \Sp^{\#}(V)$.
    \item If $\vgens$ has an invariant quadratic form $\QQ_{w^*}$, then $\tvgroup{\vgens}\subseteq\lieO(\QQ_{w^*})_{V^\perp}$ with $\tvgroup{\vgens}|_V = \lieO^{\#}(\QQ^*)$ and $\QQ^* = \QQ_{w^*}|_V$.
\end{enumerate}
Moreover, by Lemmas~\ref{lem:pointwise_symplectic_stabilizer_sequence}, \ref{lem:symplectic_diagonal_kernel_structure}, and~\ref{lem:symplectic_diagonal_kernel_transvection_generators}, Proposition~\ref{prop:full_space_symplectic_symmetric_Clifford_groups}, and \TAIL{Lemmas~\ref{lem:orthogonal_stabilizer_restriction_candidate}, \ref{lem:orthogonal_diagonal_kernel_candidate}, and~\ref{lem:orthogonal_diagonal_kernel_transvection_generators_candidate}}, the kernels of the restrictions of the symplectic and orthogonal groups to $V$ are as follows:
\begin{itemize}
    \item For the restriction of $\Sp(2n,\F_2)_{V^\perp}$ to $\Sp^{\#}(V)$, the kernel is given by $\DD{\rad(V)}$ and is generated by transvections in the radical.
    \item For the restriction of $\lieO(\QQ_{w^*})_{V^\perp}$ to $\lieO^{\#}(\QQ^*)$ and $\QQ^*$ being of type $\pm$, 
    the kernel is $\OD{\rad(V)}{\QQ_{w^*}}$ and it is generated by transvections of an anisotropic element plus linear combinations with the (isotropic) radical.
    \item For the restriction of $\lieO(\QQ_{w^*})_{V^\perp}$ to $\lieO^{\#}(\QQ^*)$ and $\QQ^*$ being of type $0$,
    the kernel is $\OD{\rad(V)}{\QQ_{w^*}}$ and it is generated by the transvections of the anisotropic radical.
\end{itemize}
In all cases, for proving that $\vgens$ generates the full group on $\Fn$, it suffices to prove that it generates the kernel. This follows from Lemma~\ref{lem:maximality_of_quotient_group}.
In particular, it follows from Eq.~\eqref{eq:four:transvections} that
all off-diagonal elements in any orthogonal case are generated by any anisotropic vector $z\in V$ with $\QQ_w(z)=1$.
We may always choose such an element in the orbit with
$z\in\tvgroup{\vgens}\cdot\vgens = (\QQ^*)^{-1}(1)\cap (V\setminus\rad(V))\neq\emptyset$, given that $\rank(V)\geq 6$ in the quasi-universal case.
In particular, this automatically proves that $\OD{\rad(V)}{\QQ_{w^*}}\subseteq \tvgroup{\vgens}$ for the $\pm$ types, hence that $\tvgroup{\vgens} = \lieO(\QQ_{w^*})_{V^\perp}$.

\begin{rem}\label{rem:pathological_for_kernel_quasi_universal}
Given Theorem~\ref{thm:Full_Classification_Theorem_w_E6_Quasi_Universal},
the restricted group $\tvgroup{\vgens}|_V$ is already known.
Thus, to identify the full group $\tvgroup{\vgens}\subseteq\Sp(2n,\F_2)$
(and hence $\cltvgroup{\pgens}\subseteq\cl_n$), it remains to determine
whether
$\tvgroup{\vgens}$ contains the kernel of the corresponding restriction map to
$V$.
Since the kernel descriptions above give explicit transvection generators, it
is enough to show that these kernel-generating transvections belong to
$\tvgroup{\vgens}$.
Since the generating set is connected and these transvections lie in the center, they must in fact be \emph{pathological} transvections (see Definition~\ref{def:pathological}), which are generated through arbitrary products of transvections, not through conjugation.
\end{rem}

We first treat the case without invariant quadratic form.
Let $u\in\rad(V)$ and let $v\in V$ commute with $u$.
Define
\[
    s_{v,u}:=\tau_{v+u}\tau_v .
\]
If both $v$ and $v+u$ lie in the orbit
$\tvgroup{\vgens}\cdot\vgens$, then $s_{v,u}\in\tvgroup{\vgens}$.
Moreover, if $v_1,v_2,u$ are pairwise commuting, then
\begin{equation}\label{eq:radical_transvection_from_orbit_pairs}
    \tau_u=s_{v_1+v_2,u}s_{v_2,u}s_{v_1,u}.
\end{equation}
Indeed, this follows by expanding the three factors using
$\tau_w(x)=x+\symp{w}{x}w$. The non-radical contributions cancel, and the
remaining term is $\symp{u}{x}u$.
By Corollary~\ref{cor:orbit:radical:codim1}, we have
$\tvgroup{\vgens}\cdot\vgens=V\setminus\rad(V)$ and
$\radorbit(\vgens)=\rad(V)$.
Since $\rank(V)\geq6$, we may choose, up to isomorphism, pairwise commuting
non-radical vectors $e_1,e_2,e_1+e_2\in V\setminus\rad(V)$.
For every $u\in\rad(V)$, the vectors
$e_1$, $e_2$, $e_1+e_2$ and their shifts by $u$ still lie in
$V\setminus\rad(V)$.
Thus Eq.~\eqref{eq:radical_transvection_from_orbit_pairs} shows that
$\tau_u\in\tvgroup{\vgens}$ for every $u\in\rad(V)$.
Hence $\DD{\rad(V)}\subseteq\tvgroup{\vgens}$, and therefore
$\tvgroup{\vgens}=\Sp(2n,\F_2)_{V^\perp}$.

\begin{table}[t]
\caption{A computational example for the case of $\type(\QQ^*)=0$.
The listed anisotropic vectors generate the pathological transvection with center $u=e_1\in\rad(\vgens)$.}
\label{tab:orthogonal-zero-pathological-transvection}
\centering
\begin{tabular}{@{\hspace{1mm}} l
@{\hspace{8mm}} l
@{\hspace{8mm}} l
@{\hspace{6mm}} l
@{\hspace{1mm}}}
\\[-3mm]
\hline\hline
\\[-3mm]
Paulis & Vectors & Basis of
&Center $u$
\\
$\PP_n$
& $\Fn$
& $\rad(\vgens)$
& of pathol.\ $\tau_u$
\\[0.5mm] \hline
\\[-2.5mm]
$\mathrm{IXYX}$
&
$(0 1 1 1 0 0 1 0)^T$
&
$(0 0 0 0 1 0 0 0)^T$
&
$(0 0 0 0 1 0 0 0)^T$
\\
$\mathrm{ZYYZ}$
&
$(0 1 1 0 1 1 1 1)^T$
&
&
\\
$\mathrm{IIXY}$
&
$(0 0 1 1 0 0 0 1)^T$
&
&
\\
$\mathrm{ZXZZ}$
&
$(0 1 0 0 1 0 1 1)^T$
&
&
\\
$\mathrm{ZZXI}$
&
$(0 0 1 0 1 1 0 0)^T$
&
\multicolumn{2}{l}{
$\tau_u =
\tau_{v_1} \tau_{v_2} \tau_{v_3} \tau_{v_4} \tau_{v_5} \tau_{v_6}\tau_{v_7}$}
\\
$\mathrm{IYZX}$
&
$(0 1 0 1 0 1 1 0)^T$
&
&
\\
$\mathrm{IZIY}$
&
$(0 0 0 1 0 1 0 1)^T$
&
&
\\[0.5mm]
\hline\hline
\end{tabular}
\end{table}

For the orthogonal case of $0$ type, it suffices to generate anisotropic transvections, which we accomplish by explicit construction.
Let us first consider the minimal case with $\rank(V)=6$, $\nullity(V)=1$ and $\rank(V^\perp)=0$, hence $n=4$.
Up to isomorphism, we can always choose $w^* = f_1$ or $B=X_1$ so that $\rad(V) = \Span{e_1}$.
One can check that Table~\ref{tab:orthogonal-zero-pathological-transvection} lists seven mutually commuting vectors in $(\QQ^*)^{-1}(1)\cap (V\setminus\rad(V))$ whose transvections multiply to $\tau_{e_1}$.
Hence, we can find a product of seven transvections whose centers are
anisotropic vectors in the orbit $\tvgroup{\vgens}\cdot\vgens$, and their
product produces the radical transvection $\tau_{e_1}$.
Here
$e_1\in\rad(V)$ is anisotropic for the quadratic form $\QQ^*$ of type $0$.

We now explain how this construction generates the kernel for arbitrary
$\rank(V)$, $\nullity(V)$, and $\rank(V^\perp)$.
Without loss of generality, we can relabel $w^*=f_{m+1}$ and $\rad(V) = \Span{e_i}_{i=m+1}^{m+r}$ as in Theorem~\ref{thm:classification_of_affine_subspaces_isomorphism_classes}.
The construction from Table~\ref{tab:orthogonal-zero-pathological-transvection} embeds into this normal form using, up to isomorphism, the first
three symplectic pairs $\{e_i,f_i\}_{i=1}^3$ together with the anisotropic radical
vector $e_{m+1}$.
The seven-vector product is supported only on this six-dimensional summand and
the radical vector $e_{m+1}$.
All additional basis vectors commute with these seven
vectors, so the product identity is unchanged.
Finally, we show how via a simple shift one can also generate transvections by centers in the entire anisotropic radical, for arbitrary radical dimension.
Explicitly, we can use the fact that pairwise commuting transvections act as an additive group (Lemma~\ref{lem:symplectic_diagonal_kernel_structure}(b)), such that for a commuting set of vectors $v_i$, the corresponding product of transvections acts as
\begin{equation}
    \prod_i\tau_{v_i}(x) = x + \sum_i \symp{v_i}{x}v_i.
\end{equation}
Hence, such a product is equal to $\tau_u$ iff
\begin{equation}\label{eq:pathological_transvection_generation}
    \sum_i \symp{v_i}{x}v_i = \symp{u}{x}u.
\end{equation}
For the seven vectors $v_1,\ldots,v_7$ listed in
Table~\ref{tab:orthogonal-zero-pathological-transvection}, we have
\[
    \sum_{i=1}^7 v_i=e_{m+1}
    \;\text{ and }\;
    \sum_{i=1}^7\symp{v_i}{x}v_i
    =
    \symp{e_{m+1}}{x}e_{m+1}.
\]
Thus their transvections multiply to the radical transvection
$\tau_{e_{m+1}}$.
To obtain a transvection with center $e_{m+1}+u$ for
$u\in\radzero(V,\QQ_{w^*})$, we shift each of the seven centers by $u$,
that is, we replace $v_i$ by $v_i+u$.
Explicitly, we show that Eq.~\eqref{eq:pathological_transvection_generation} holds also for all $u'=e_{m+1}+u$, by taking as generating vectors $v_i' = v_i + u$:
\begin{align}
    &\sum_{i=1}^7 \symp{v_i+u}{x}(v_i+u) \nonumber\\
    &= \sum_{i=1}^7 (\symp{v_i}{x}v_i + \symp{u}{x}v_i + \symp{v_i}{x}u + \symp{u}{x}u) \nonumber\\
    &= \symp{e_{m+1}}{x}e_{m+1} + \symp{u}{x}e_{m+1} + \symp{e_{m+1}}{x}u + \symp{u}{x}u \nonumber\\
    &= \symp{e_{m+1}+u}{x}(e_{m+1}+u).\label{eq:shift_anisotropic_product}
\end{align}
Here, $\QQ_{w^*}(v_i)=1$ if $v_i$ is anisotropic and not in the radical, so is $v_i+u$ for $u\in\radzero(V,\QQ_{w^*}) = \radorbit(\vgens)$.
Thus the transvection group contains all transvections with centers in
$e_{m+1}+\radzero(V,\QQ_{w^*})=\radone(V,\QQ_{w^*})$.
Consequently, by Lemma~\ref{lem:orthogonal_diagonal_kernel_transvection_generators_candidate}(c), the kernel $\OD{\rad(V)}{\QQ_{w^*}}$ is also in the transvection group.
We summarize this discussion in the following:
\begin{lem}\label{lem:transvection_kernels_quasi_universal}
Let $\vgens\subseteq\PP_n$ be of quasi-universal type.
Then, $\tvgroup{\vgens}$ contains the kernel under the restriction to $V=\Span{\vgens}$ of the respective symplectic or orthogonal stabilizer subgroup $\Sp(2n,\F_2)_{\vgens^\perp}$ or $\lieO(\QQ_{w^*})_{\vgens^\perp}$.
\begin{enumerate}
    \item If $\vgens$ has no invariant quadratic forms, the kernel $\DD{\rad(V)}$ is generated by transvections in the radical, which themselves are generated via products of transvections in $\tvgroup{\vgens}$ with
    \begin{equation}
        \tau_u = s_{v_1+v_2,u}s_{v_2,u}s_{v_1,u},
	    \end{equation}
	    for any distinct commuting $v_1,v_2,v_1+v_2\in\tvgroup{\vgens}\cdot\vgens = V\setminus\rad(V)$.
	    The condition $\rank(V)\geq6$ ensures that such a commuting triple exists;
	    up to isomorphism, we may take it to be $e_1,e_2,e_1+e_2$.
    \item If $\vgens$ has invariant quadratic forms of type $\pm$, the kernel
    $\OD{\rad(V)}{\QQ_{w^*}}$ is generated by the fourfold products
    \begin{equation}
        \tau_{z+u+u'}\tau_{z+u}\tau_{z+u'}\tau_z
    \end{equation}
    for any $z\in V\setminus\rad(V)$ with $\QQ^*(z)=1$ and
    $u,u'\in\rad(V)$, under the assumptions that $\rank(V)\geq6$ and
    $\radorbit(\vgens)=\radzero(V,\QQ_{w^*})=\rad(V)$.
    \item If the generating set $\vgens$ has invariant quadratic forms of type $0$, then the kernel $\OD{\rad(V)}{\QQ_{w^*}}$ is generated by the anisotropic transvections
    with centers in the anisotropic radical coset $\radone(V,\QQ_{w^*})$.
    A single anisotropic transvection can be generated as a product of seven transvections, which is specified, up to isomorphism, in Table~\ref{tab:orthogonal-zero-pathological-transvection}.
    All other transvections are obtained by shifting this product along isotropic radical elements as in Eq.~\eqref{eq:shift_anisotropic_product}.
\end{enumerate}
\end{lem}

Then, Lemma~\ref{lem:transvection_kernels_quasi_universal}, Proposition~\ref{prop:full_space_symplectic_symmetric_Clifford_groups} and Proposition~\ref{prop:full_space_isometry_symmetric_Clifford_groups} imply that the binary transvection groups $\tvgroup{\vgens}$ whose restricted $\tvgroup{\vgens}|_V$ transvection groups are respectively $\Sp^{\#}(V)$ and $\lieO(\QQ^*)$, are precisely $\Sp(2n,\F_2)_{V^\perp}$ and $\lieO(\QQ_{w^*})_{V^\perp}$.
Additionally, the Clifford transvection groups whose binary transvections groups are respectively the $\Sp(2n,\F_2)_{V^\perp}$ and $\lieO(\QQ_{w^*})_{V^\perp}$, are the corresponding Clifford groups $\cl_n^{\Ad_\commalg}$ and $\cl_n^{\Theta_{B^*},\Ad_\commalg}$ (after adjoining phases).
Following Remark~\ref{rem:pathological_for_kernel_quasi_universal} and Lemma~\ref{lem:transvection_kernels_quasi_universal}, we can now state:
\begin{prop}\label{prop:quasi_universal_transvection_groups}
Let $\pgens = \isolong{\vgens}\subseteq\PP_n$ be a set of quasi-universal type.
Also, let $V=\Span{\vgens}$ and $\commalg=\commutant(\pgens)$.
\begin{enumerate}
    \item If $\quadratic(\vgens)$ is empty, then $\tvgroup{\vgens} = \Sp(2n,\F_2)_{V^\perp}$ and
    $\cltvgroup{\pgens} \simeq\cl_n^{\Ad_\commalg}$.
    \item If exists $w\in\quadratic(\vgens)$ with $B=\iso{w}$, then $\tvgroup{\vgens} = \lieO(\QQ_w)_{V^\perp}$ and
    $\cltvgroup{\pgens}\simeq \cl_n^{\Theta_B,\Ad_\commalg}$.
\end{enumerate}
\end{prop}
Hence, up to phases, we have that for the quasi-universal case, the transvection groups are precisely the subgroups of the Clifford (symplectic) group which respect the same symmetries, namely, the commutant (orthogonal complement) and invariant bilinear (quadratic) forms.
See Sections~\ref{sec:commutant_symmetric_clifford_groups_lie_algebras}, \ref{sec:clifford_fixed_subgroups_orthogonal_images}, \ref{sec:orthogonal_diagonal_kernels_maximality}, and~\ref{sec:pauli_lie_algebras_isometries} for the definitions and more details on these groups.

For the free-fermionic cases, we have instead \emph{defined} the families
$\Sym(k,r,n,\F_2)$ and $\Sym^D(2m{+}2,r,n,\F_2)$ as transvection groups, for arbitrary $k\geq 1$, $k\neq 4$ and $m\geq 2$.
Together with Corollary~\ref{cor:distinguishing_line_graph_cases}, this
identifies $\tvgroup{\vgens}$ for every connected generating set whose
frustration graph is a line graph.
Combining this with Proposition~\ref{prop:quasi_universal_transvection_groups}
gives the full classification of the transvection groups $\tvgroup{\vgens}$ on
$\Fn$ for connected generating sets.

By definition of Clifford groups and Lie groups of isometries, we have
\begin{align*}
        \cl_n^{\Ad_{\commalg}} &= \cl_n \cap \lieISO(\commalg) = \cl_n \cap \lieU(2^n)^{\Ad_{\commalg}},\\
        \cl_n^{\Theta_B,\Ad_\commalg} &= \cl_n \cap \lieISO(B,\commalg) = \cl_n \cap \lieISO(B)^{\Ad_\commalg}.
\end{align*}
Also, by Eq.~\eqref{eq:clifford:transvection:inclusion}, transvection groups lie in the intersection of the respective Pauli Lie groups with the Clifford group.
Since the generators preserve the commutant $\commalg$ and, when present, the
bilinear-form involution $\Theta_B$, the generated Pauli Lie group is contained
in the corresponding compact isometry group:
\begin{align*}
     e^{\lie{\pgens}} &\subseteq \lieISO(\commalg),\\
     e^{\lie{\pgens}} &\subseteq \lieISO(\commalg,B).
\end{align*}
Now, let $\lieG = \lieISO(\commalg)$ or $\lieG = \lieISO(\commalg,B)$ depending on whether there is an invariant bilinear form or not.
Combining Proposition~\ref{prop:quasi_universal_transvection_groups} and the above, we find for $\pgens$ quasi-universal that
\begin{equation}
    \cl_n\cap\lieG \simeq \cltvgroup{\pgens}
\end{equation}
and
\begin{equation}
    \cltvgroup{\pgens} \subseteq \cl_n\cap e^{\lie{\pgens}} \subseteq \cl_n\cap \lieG.
\end{equation}
This implies that $\cltvgroup{\pgens}$ and $\cl_n \cap e^{\lie{\pgens}}$ in fact coincide after adjoining phases for arbitrary quasi-universal generating sets $\pgens$, i.e.,
\begin{equation}\label{eq:transvections_are_clifford_cap_lie_group}
    \cltvgroup{\pgens} \simeq \cl_n \cap e^{\lie{\pgens}}.
\end{equation}
We highlighted the corresponding free-fermionic statement at the end of Section~\ref{sec:Transvection_Group_Line_Graph_Symmetric_Group},
though we only prove this here for the quasi-universal case.

For $n\geq 2$, adjoining Pauli scalar phases does not change the complex
determinant, since $\det(i^\varpi I)=i^{\varpi 2^n}=1$.
Moreover, $e^{\lie{\pgens}}\subseteq \SU(2^n)$, because $\lie{\pgens}$ is
generated by traceless skew-Hermitian matrices.
Thus the equality up to phases in
Eq.~\eqref{eq:transvections_are_clifford_cap_lie_group} does not allow for
additional Clifford elements of determinant $-1$.
Since $e^{\lie{\pgens}}$ is connected, the subgroup
$\cl_n\cap e^{\lie{\pgens}}$ lies in the identity component of the relevant
compact isometry group $\lieG$.

\subsection{Full Classification and Algorithm}\label{sec:classification_and_algorithm_lie_algebras}

We can now formally state the classification theorem for Pauli Lie algebras of connected generating sets:
\begin{thm}[Classification of Connected Pauli Lie Algebras]\label{thm:full_classification_pauli_lie_algebras}
Let $\pgens=\isolong{\vgens}\subseteq\PP_n$ be a Pauli generating set with connected frustration graph $\graphG$.
Also, let $\commalg=\commutant(\pgens)$, $r=\nullity(\pgens)$, $q=\nullorbit(\pgens)=\dim\radorbit(\pgens)$
and $2\ell=\rank(\commalg)$.
Then, one of the following holds:
\begin{enumerate}
    \item Assume that $\graphG$ is \emph{not} the line graph of a multigraph.
    \begin{nestedcaseenum}
        \item If $\bilinear(\pgens)$ is trivial, then $\lie{\pgens} = \lieiso(\commalg)$.
        \item If there is some $B\in \bilinear_{\PP_n}(\pgens)$, then $\lie{\pgens} = \lieiso(\commalg,B)$.
    \end{nestedcaseenum}
    The Lie algebras and representations are described in Theorem~\ref{thm:derived_pauli_lie_algebras_isometries} and Proposition~\ref{prop:pauli_lie_isometry_block_forms}
    \item Assume that $\graphG$ is the line graph of a multigraph.
    \begin{nestedcaseenum}
        \item If $\pgens$ is of even type and $n_\Delta=2m$ with $m\geq 3$, then $\lieg = \so(2m)^{\oplus 2^q}\otimes I^{\otimes \ell}$.
        \item If $\pgens$ is of odd type and $n_\Delta=2m+1$ with $m\geq 1$, then $\lieg = \so(2m{+}1)^{\oplus 2^q}\otimes I^{\otimes \ell}$.
        \item If $\pgens$ is of exceptional type and $n_\Delta=2m+2$ with $m\geq 2$, then $\lieg = \so(2m{+}2)^{\oplus 2^q}\otimes I^{\otimes \ell}$.
    \end{nestedcaseenum}
    The representations are described in Theorem~\ref{thm:Free_Fermionic_Lie_Algebras}.
\end{enumerate}
\end{thm}
Similarly, for the binary transvection groups over the full space $\Fn$ we get:
\begin{thm}[Transvection Groups Classification]\label{thm:full_classification_transvection_groups}
Let $\pgens=\isolong{\vgens}\subseteq\PP_n$ be a Pauli generating set with connected frustration graph $\graphG$.
Also, let $\commalg=\commutant(\pgens)$, $r=\nullity(\pgens)$, $q=\nullorbit(\pgens)=\dim\radorbit(\pgens)$,
and $2\ell=\rank(\vgens^\perp)$.
Then, one of the following holds:
\begin{enumerate}
    \item Assume that $\graphG$ is \emph{not} the line graph of a multigraph.
    \begin{nestedcaseenum}
        \item If $\bilinear(\pgens)$ is trivial, then $\tvgroup{\vgens} = \Sp(2n,\F_2)_{V^\perp}$, $\cltvgroup{\pgens}$ and $\cl_n^{\Ad_\commalg}$ coincide after adjoining phases.
        \item If there is some $B=\iso{w}\in \bilinear(\pgens)$, then $\tvgroup{\vgens} = \lieO(\QQ_w)_{V^\perp}$, while
        $\cltvgroup{\pgens}$ and $\cl_n^{\Theta_B,\Ad_\commalg}$  coincide after adjoining phases.
    \end{nestedcaseenum}
    \item Assume that $\graphG$ is the line graph of a multigraph.
    \begin{nestedcaseenum}
        \item If $\pgens$ is of even type and $n_\Delta=2m$ with $m\geq 3$, then
        $\cltvgroup{\pgens} = \clSym(2m,r,n,\F_2)$ and $\tvgroup{\vgens} = \Sym(2m,r,n,\F_2)$.
        \item If $\pgens$ is of odd type and $n_\Delta=2m+1$ with $m\geq 1$, then
        $\cltvgroup{\pgens} = \clSym(2m{+}1,r,n,\F_2)$, and $\tvgroup{\vgens} = \Sym(2m{+}1,r,n,\F_2)$.
        \item If $\pgens$ is of exceptional type and $n_\Delta=2m+2$ with $m\geq 2$, then
        $\cltvgroup{\pgens} = \clSym^D(2m{+}2,r,n,\F_2)$, and $\tvgroup{\vgens} = \Sym^D(2m{+}2,r,n,\F_2)$.
    \end{nestedcaseenum}
\end{enumerate}
\end{thm}

Furthermore, Algorithm~\ref{alg:classification_lie_algebras} translates the
classification into a practical procedure for deciding the type of the Lie
algebra:
\begin{cor}\label{cor:algorithm_lie_algebras}
Let $\pgens = \isolong{\vgens}\subseteq\PP_n$ be a sequence (or set) of binary vectors with connected frustration graph $\graphG$.
There is an algorithm that can determine the Pauli Lie algebra type with a time complexity of $\BigO(\max\{\abs{\vertices(\graphG)}+\abs{\edges(\graphG)},2n\}^3)$.
\end{cor}
\begin{proof}
The algorithm is described in Algorithm~\ref{alg:classification_lie_algebras}.
We discuss here further details and its computational complexity.
Set $\abs{\graphG}:=\abs{\vertices(\graphG)}+\abs{\edges(\graphG)}$.
For the initial single-generator branch, the output is simply $\lieu(1)$ and
the Pauli Lie algebra is abelian.

The preprocessing step computes the frustration graph following
Definition~\ref{defn:frustration}, which has a time complexity of
$O(\abs{\pgens}^2)$.
Computing the orthogonal complement (see Eq.~\eqref{eq:def:orthogonal_complement}) and radical (see Eq.~\eqref{eq:radical})
requires solving linear equations over $\abs{\vgens}$ variables in at most a $2n$-dimensional system. Hence we obtain a time complexity of
$\BigO(\max\{2n,\abs{\vgens}\}^3)$.
In particular, these computations determine the rank of the orthogonal
complement $2\ell$, the radical dimension $r$, and consequently the remaining
non-degenerate rank $2m$ of $V$, with $m=n-\ell-r$ in the non-line-graph branch.

For the line-graph test, one can use
Theorem~\ref{thm:equivalent_line_graph_conditions} possibly combined with
Theorem~\ref{thm:equivalent_path_graph_conditions}, which reduces recognizing a
line graph of a multigraph to an ordinary line graph by removing twins.
Deciding line-graphness and computing the root multigraph then has time complexity of $O(\abs{\graphG})$.

In the odd line-graph branch, where $n_\Delta=2m+1\geq 3$, the Lie algebra is of odd free-fermionic type, and the output is the label $\so(2m{+}1)^{\oplus 2^q}\otimes I^{\otimes \ell}$ for the Pauli Lie algebra, with representation described in Theorem~\ref{thm:Free_Fermionic_Lie_Algebras}\ref{thm:Free_Fermionic_Lie_Algebras:odd}.
In this odd case, $q=r$.

In the even line-graph branch, where $n_\Delta$ is even, one also needs to
compute the dimension of the orbit radical.
This can be computed via Lemma~\ref{lem:orbit:radical:line_graphs} and Corollary~\ref{cor:distinguishing_line_graph_cases}.
Namely, one needs to compute the incidence matrix $M(\Delta)$ (see Eq.~\eqref{eq:def:incidence_matrix}) for $\Delta$ and its kernel, which takes at most $\BigO(\abs{\graphG}^2)$ time, given that $n_\Delta\leq\rank(A(\graphG)) \leq \abs{\graphG}$.
Then, one can project $\edgetovec(\ker(M(\Delta)))$ to the actual generators and compute the corresponding dimension, where $\edgetovec = \edgetovec_{\Delta,\tilde{\vgens}}$.
Explicitly, let $\Pi_\vgens$ be a matrix whose columns are the binary vectors in $\vgens\subseteq\Fn$, and $K(\Delta)$ a matrix whose columns are a basis for $\ker(M(\Delta))\subseteq\F_2^{\abs{\edges(\Delta)}}$. Assume that the ordering of the columns in $\Pi_\vgens$ and $M(\Delta)$ matches such that $\edgetovec$ is the identity map (i.e.\ we identify the linearly independent extension $\tilde{\vgens}$ and the edge colorings $\F_2^{\abs{\edges(\Delta)}}$). Then the columns of $R = \Pi_\vgens K(\Delta)$ span $\projection(\edgetovec(\ker(M(\Delta)))) = \radorbit(\vgens)$, but they are not necessarily linearly independent. We can find a basis for $\radorbit(\vgens)$ by Gaussian elimination on the columns of $R$, and $q$ is the matrix rank of $R$.
Again, this takes $\BigO(\max\{2n,\abs{\vgens}\}^3)$.
Then, if $r-q=1$, set $m=n_\Delta/2$.
The Lie algebra is of even type, and the output is $\so(2m)^{\oplus 2^q}\otimes I^{\otimes \ell}$, with representation described in Theorem~\ref{thm:Free_Fermionic_Lie_Algebras}\ref{thm:Free_Fermionic_Lie_Algebras:even}.
If instead $r-q=0$, set $m=(n_\Delta-2)/2$.
The Lie algebra is of exceptional type, and the output is $\so(2m{+}2)^{\oplus 2^q}\otimes I^{\otimes \ell}$, with representation described in Theorem~\ref{thm:Free_Fermionic_Lie_Algebras}\ref{thm:Free_Fermionic_Lie_Algebras:exceptional}.

\begin{algorithm}[t]
    \caption{Pseudocode for computing the Pauli Lie algebra.
        $\pgens=\isolong{\vgens}$ is the Pauli generating set.
        $\frustration{\vgens}$ is the frustration graph. $\vgens^\perp$ is the orthogonal complement and $\rad(\vgens)$ the radical, with symplectic bases $\vecbas{\vgens^\perp}$ and $\vecbas{\rad(\vgens)}$.
        If this is a line graph, its unique root graph is $\Delta$ on $n_\Delta\neq 4$ vertices. $\texttt{Solve}$ outputs False if there is no solution to a linear system of equations, otherwise it returns True and a solution.
        The solution $w^*\in\quadratic(\vgens)$ is an invariant quadratic form.
        $\texttt{IsomorphismClass}$ is defined in Algorithm~\ref{alg:quadratic_isomorphism_class} and outputs a string in $\{0,+,-\}$. }\label{alg:classification_lie_algebras}
    \SetKwInOut{Input}{input}
    \SetKwInOut{Output}{output}
    \SetKwFunction{FSolve}{Solve}
    \SetKwFunction{FClass}{IsomorphismClass}
    \SetKwProg{Fn}{Function}{}{end}

    \Input{A generating set $\pgens=\isolong{\vgens}$.}
    \Output{A name for the Pauli Lie algebra $\lie{\pgens}$ as a tuple.}
    \SetKwBlock{Beginn}{beginn}{ende}
    \Begin{
        \If{$\abs{\vgens} = 1$}{
            name $\gets \langle $`u', $0, 1, 2n{-}2\rangle$ \;
            \Return name
        }
        $\graphG \gets \frustration{\pgens}$\;
        $\vecbasempty_\perp \gets\vecbas{\vgens^\perp}, \vecbasempty_\rad \gets \vecbas{\rad(\vgens)}$\;
        $r \gets \nullity(\vgens)$\; $\ell \gets \rank(\vgens^\perp)/2$\; $m\gets n-\ell-r$\;
        \uIf{$\graphG$ is a line graph} {
            $\Delta \gets R(\graphG), n_\Delta \gets \abs{\vertices(\Delta)}$\;
            \uIf{$n_\Delta$ odd} {
                $m\gets (n_\Delta{-}1)/2$\;
                name $\gets \langle $`ff-odd', $m, r, \ell\rangle$\;
                \Return name
            }
            \Else{
                $M\gets M(\Delta)$\;
                $q\gets \dim(\projection(\edgetovec(\ker(M))))$\;
                \uIf{$r-q = 1$}{
                    $m\gets n_\Delta/2$\;
                    name $\gets \langle $`ff-even', $m, q, \ell\rangle$\;
                    \Return name
                }
                \ElseIf{$r-q = 0$}{
                    $m\gets (n_\Delta{-}2)/2$\;
                    name $\gets \langle $`ff-exc', $m, q, \ell\rangle$\;
                    \Return name
                }
            }
        }

        \Else{
            bool, $w^* \gets$ \FSolve{$\symp{x}{\vgens} = \QQ_0(\vgens)$}\;
            \uIf{not $\text{bool}$}{
                name $\gets \langle $`su', $m, r, \ell\rangle$\;
                \Return name
            }
            \Else{
                class $\gets $ \FClass{$w^*$, $\vecbasempty_\perp$, $\vecbasempty_\rad$}\;
                name $\gets \langle $`iso' + class, $m, r, \ell\rangle$\;
                \Return name
            }
        }
    }
\end{algorithm}

In the non-line-graph branch, computing a symplectic basis for $\vgens^\perp$
and finding a solution (if it exists) for an invariant quadratic form $w^*$
(see Eq.~\eqref{eq:invariant_quadratic_forms_general_group_subsets} and
Lemma~\ref{lem:affine_structure_invariant_quadratic_form_subspace}) involves
solving systems of linear equations over $\abs{\vgens}$ variables in a space of
$2n$ dimension plus Gaussian elimination, hence takes at most
$\BigO(\max\{2n,\abs{\vgens}\}^3)$-time.
If $w^*$ does not exist, output $\su(2^m)^{\oplus 2^r}\otimes I^{\otimes \ell}$.
If $w^*$ exists, one can use Lemma~\ref{lem:isomorphism_class_affine_subspaces_quadratic_forms} to compute the isomorphism class by evaluating $\QQ_{w^*}$ on the symplectic basis of $\vgens^\perp$.
This requires evaluating a quadratic form on at most $2\ell+r$ vectors, hence takes $\BigO(n^2)$-time.
Then, the output of the class is $\lieiso^0(\commalg,B)$, $B=\iso{w^*}$ and $\commalg=\Span[\C]{\isolong{\vgens^\perp}}$, with type specified in Theorem~\ref{thm:derived_pauli_lie_algebras_isometries} and representation specified in Proposition~\ref{prop:pauli_lie_isometry_block_forms}
(where only the knowledge of the type, $\rank(\vgens^\perp)$ and $\nullity(\vgens)$ is required).

No other cases exist, hence the algorithm always terminates and takes at most $\BigO(\max\{2n,\abs{\graphG}\}^3)$ time, which concludes the proof.
\end{proof}

Assuming that the size of the generating set grows polynomially with the number
of qubits, and typically only linearly, determining quasi-universality always
takes polynomial time and space.
Also, notice that, we have given different approaches for computing the symmetries, quadratic forms and deciding whether it is a line-graph, some of which may more or less efficient depending on the specific structure of the generators.
Even though this may prove more or less practical, it does not change the asymptotic runtime in general.

Moreover, despite the fact that minimal generating sets always have size $\BigO(n)$, it is not clear whether one can always guarantee a runtime of $\BigO(n^3)$ by cleverly choosing only a small subset of an initial generating set.
As $\abs{\Fn}=4^n$, the algorithm may run in exponential time if the generating set $\pgens$ has been chosen naively, i.e.\ $\abs{\pgens} = \BigO(2^n)$.
For instance, it may happen that a quasi-universal set of exponential size is free-fermionic up to a single vertex which provides an induced subgraph in $\calE_6$, which may be detected through a local search through the graph.

We also highlight here the differences of our algorithm with that of \cite{Cuypers_2026}.
The main steps are the same: (1) check if the frustration graph is a line graph, (2) if yes, compute the root multigraph, 
(3) if not, compute the isomorphism class of the quadratic space.
However the output itself is distinct, since for a connected frustration graph Algorithm~\ref{alg:classification_lie_algebras}
provides not only the Lie algebra up to isomorphism, but also a description of its concrete representation over $\C^{2^n}$ (together with the classification theorems).
The representation highlights the distinction between the even and exceptional free-fermionic types, as well as the quasi-universal type with no invariant quadratic forms and invariant quadratic forms of type $0$, which cannot be detected at the level of isomorphisms of Lie algebras.
Additionally, for the quasi-universal type, the computation of the radical and invariant quadratic form is performed not over $V$ itself but directly over $\Fn$, by using the reference $\QQ_0$, which also provides information about the orbits over $\Fn$.
Finally, the computation of the isomorphism class of the invariant quadratic form is performed over $V^\perp$ via Algorithm~\ref{alg:quadratic_isomorphism_class}.
This can potentially be more efficient, given a finite number of symmetries, compared to a computation over $V$ which typically grows with system size.

We can also compare our approach to that \cite{Aguilar_Cichy_Eisert_Bittel_2024}.
There, the approach requires as a first step finding the canonical $t$-equivalent representatives, which produces one of four possible classes. $\graphG\in\{\graphP_{\tilde{k},\tilde{n}_1},\graphS_{\tilde{n}_2,\tilde{n}_1}^1,\graphS_{\tilde{n}_2,\tilde{n}_1}^2,\graphS_{\tilde{n}_2,\tilde{n}_1}^3\}$, for some $\tilde{k}$, $\tilde{n}_2$ and $\tilde{n}_1$. 
Such an algorithm is not explicitly provided, though it is mentioned in \cite{Aguilar_Cichy_Eisert_Bittel_2024} that it can be implemented efficiently.
Given the new generating set, one systematically removes the Lie-algebraic dependencies by looking at algebraic dependencies in legs of length one (which here is stated via Proposition~\ref{prop:Limits_Lie_Algebraic_Dependencies_canonical}).
This provides a minimal generating set which is one of $\vgens(\graphG)$ with $\graphG\in\{\graphP_{k,n_1},\graphS_{n_2,n_1}^1,\graphS_{n_2,n_1}^2,\graphS_{n_2,n_1}^3\}$ or $\vgens^D(\graphG)$ with $\graphG\in\{\graphP_{2m,n_1}, \graphS_{n_2,n_1}^3\}$, for some suitable $k$, $m$, $n_1$ and $n_2$.
Finally, one checks whether a remaining algebraic dependence exists, which provides a Lie algebra out of six possible cases.
Hence, the approach in \cite{Aguilar_Cichy_Eisert_Bittel_2024} is almost purely graph-driven, with the exception of of the algebraic dependencies check, and requires transformation of the initial generating set.
On the other hand, we focus here on the invariant and symmetry properties of the generating set, such as the orthogonal complement (commutant), invariant quadratic (bilinear) forms and the orbit radical, as well as deciding if $\graphG$ is a line graph.
Out of these steps, none requires transformations of the generating set, and only the last two require a graph-theoretic approach.

Using alternative criterions for free-fermionicness and the isomorphism classes, one can choose alternative implementations of Algorithm~\ref{alg:quadratic_isomorphism_class} using different subroutines, possibly informed on the specific generating set.
However, given that all procedures are either linear algebra over a $2n$-dimensional binary vector space or some efficient combinatorial search, Algorithm~\ref{alg:quadratic_isomorphism_class} provides a simple and accessible implementation to identify the Lie algebra of a connected generating set, up to isomorphism and representation.
Finally, we highlight in the applications (e.g. Section~\ref{sec:example:2-local_paulis}) how we can use our criterions to describe practical examples \emph{exactly}, beyond an algorithmic computation.

\subsection{Disconnected Graphs}\label{sec:disconnected_graphs}

The classification of Pauli Lie algebras immediately generalizes to a generating set whose frustration graph admits multiple connected components.
Consider a generating set with two connected components
$\pgens = \pgens_1\sqcup\pgens_2$, $\pgens_i = \isolong{\vgens_i}$ for $i\in\{1,2\}$, hence $\symp{\vgens_1}{\vgens_2} = 0$.
Consequently the transvection subgroups $\tvgroup{\vgens_1},\tvgroup{\vgens_2}\subseteq\tvgroup{\vgens}$ mutually commute $g_1g_2 = g_2g_1,\,\forall g_1\in \tvgroup{\vgens_1},\,g_2\in\tvgroup{\vgens_2}$ (since their generators do), and also leave the other generating set invariant $\tvgroup{\vgens_1}\cdot\vgens_2 = \vgens_2$, and viceversa.
Hence, we must have
\begin{align*}
        \tvgroup{\vgens}\cdot\vgens_i &= \tvgroup{\vgens_i}\cdot\vgens_i\quad\text{with } i\in\{1,2\},\\
        \tvgroup{\vgens}\cdot\vgens_1&\cap \tvgroup{\vgens}\cdot\vgens_2 = \emptyset,\\
        \tvgroup{\vgens}\cdot\vgens &=\tvgroup{\vgens_1}\cdot\vgens_1\sqcup \tvgroup{\vgens_2}\cdot\vgens_2.
\end{align*}
Consequently, the Pauli Lie algebras generated by $\pgens_1$ and $\pgens_2$ must also commute, hence contribute to distinct summands, i.e.,
\begin{align*}
        \lie{\pgens} &\cong \lie{\pgens_1} \oplus \lie{\pgens_2},\\
        \comm{\lie{\pgens_1}}{\lie{\pgens_2}} &= 0,\\
        \lie{\pgens_1}\cap \lie{\pgens_2} &= \{0\}.
\end{align*}
We formalize this into the following:
\begin{lem}\label{lem:classification_lie_algebras_disconnected}
Let $\pgens = \isolong{\vgens}\subseteq\PP_n$ be a Pauli generating set (or sequence) whose frustration graph has connected components $\pgens = \bigsqcup_i\pgens_i$, $\pgens_i\cap\pgens_j = \emptyset$ for $i\neq j$. Then, $\lie{\pgens} \cong \bigoplus_i \lie{\pgens_i}$.
\end{lem}
Then, classification of the \emph{isomorphism classes} of Pauli Lie algebras gets reduced to iterating upon each connected component.
However, in order to understand the representation and the orbit structure, this is not sufficient.
Indeed, we cannot in general treat each component as independent, since
$\symp{\vgens_1}{\vgens_2}=0$ only implies
$V_1\cap V_2 \subseteq \rad(V_1\cup V_2)$.

A trivial case of disconnected generating set is the case of $\vgens$ consisting of a non-trivial connected component $\vgens_c$ plus isolated vertices $\{u_i\}_{i=1}^s$ which are pathological transvections (see Definition~\ref{def:pathological}) for the non-trivial connected component.
Then, clearly the addition of these vectors to the generating set does not change the transvection group $\tvgroup{\vgens}$ or the orbits $\orb(\vgens)$, but it does affect the orbits containing the generators, in the sense that $\tvgroup{\vgens}\cdot\vgens = (\tvgroup{\vgens_c}\cdot\vgens_c) \cup \{u_i\}$.
Hence, the corresponding Lie algebra is a reductive Lie algebra with the semisimple part coming from Theorem~\ref{thm:full_classification_pauli_lie_algebras}, plus an abelian part $\lieu(1)^{\oplus s}$, $\lieu(1) = \Span[\R]{\im\iso{u_i}}$, with each Pauli $\iso{u_i}$ projecting to $\pm\id$ in each irreducible block.
However, such additional terms can still affect the orbits and the representation, i.e.\ the action of $\lieg$ on the state (and operator) space.

\section{Classification of the Orbits}\label{sec:classification:orbits}

We now shift to our second objective, i.e. the classification and characterization of the orbits of transvection groups over the non-degenerate space $\Fn$.
We shall be using tools from throughout the manuscript.
Recall that the orbits of the transvection group $\tvgroup{\vgens}$ partition the binary space $\Fn$ into the sets $O\in\orb(\Fn)$.
Correspondingly these are also the orbits of the Clifford transvection $\cltvgroup{\pgens}$ acting on $\PP_n$, or the connected components of the commutator graph (see Lemma~\ref{lem:orbits_and_commutato_graph}).
Explicitly, we are interested in three tasks, for each of the seven cases of connected generating sets:
\begin{enumerate}
    \item Orbit intersection, i.e., when do $v,v'\in\Fn$ or Paulis $P,P'\in\PP_n$ belong in the same orbit
    \item Orbit labelling, i.e., identify a set of labels for all orbits orbits $O\in\orb(\Fn)$ (or correspondingly for the Pauli strings $\isolong{O}\in\orb(\PP_n)$)
    \item Orbit identification, i.e., given the list of labels for the orbits and a vector $v\in\Fn$ (Pauli $P\in\PP_n$), determine to which orbit $v$ ($P$) belongs, i.e. the label of the corresponding orbit
\end{enumerate}
We start by showing general properties of the actions of transvection groups on the full space $\Fn$ and how they relate to the action over $V$.
Hence, we will mostly be able to transfer knowledge of the orbits in $V$ to the orbits over $\Fn$.
Then, we shall discuss the quasi-universal and free-fermionic cases separately, given that the latter case is more easily discussed in the Majorana formalism.
The resulting labelling of the orbits will be made up of the values of commutation relations with a basis of the orthogonal complement $\vecbas{\vgens^\perp}$, the (an)isotropy of a vector with respect to a reference invariant quadratic form $\QQ_w$, and the Majorana length in the free-fermionic cases.
Finally, we show how the tasks (a-c) can be performed efficiently in Corollary~\ref{cor:algorithm_orbits}, given that they also reduce to linear algebra and graph-theoretic subroutines.

\subsection{General Properties}

By definition, the orthogonal complement $\vgens^\perp$ (commutant $\commutant(\pgens)$) consists of those elements which commute with the generators, hence are left invariant by their action:
$$ \tvgroup{\vgens}\cdot u = u,\;\forall u\in V^\perp $$
Hence, such elements provide singleton orbits, independently of the given generating set.
This motivates the following definition:
\begin{defn}
We say that an orbit $O\in\orb(\Fn)$ for a transvection group $\tvgroup{\vgens}$ is \emph{trivial} if it is a singleton orbit $\abs{O}=1$, i.e. $O=\{u\}$ for some $u\in \vgens^\perp$.
An orbit $O$ is \emph{non-trivial} if $\abs{O}\geq 2$.
The equivalent statement applies for the Pauli orbits $\iso{O}\in\orb(\PP_n)$ under $\cltvgroup{\pgens}$.
\end{defn}

We start with a some basic but fundamental results:
\begin{lem}\label{lem:orbits:basic}
Let $\pgens=\isolong{\vgens}\subseteq\Fn$ be any Pauli generating set such that $V = \Span[\F_2]{\vgens}$.
Let $O\in\orb(\Fn)$ be an orbit for $\tvgroup{\vgens}$ and $u\in\Fn$, then:
\begin{enumerate}
    \item $\tvgroup{\vgens}\cdot(u+V) = u+V$,
    \item if $u'\in (u+V)\cap V^\perp\neq\emptyset$, then $u'+O$ is also an orbit for $\tvgroup{\vgens}$,
    \item $(u+V)\cap V^\perp$ is a set of fixed points for $\tvgroup{\vgens}$.
\end{enumerate}
\end{lem}
\begin{proof}
We can write any element in $u+V$ as $u+v$, where $v\in V$.
Then for all $v'\in\vgens$:
\begin{align*}
        &\tau_{v'}(u+v) = \tau_{v'}u + \tau_{v'}v =\\
        &= u + (\symp{v'}{u}v' + v + \symp{v'}{v}v')
        = u + v''\in u + V,
\end{align*}
which proves (a). 
For (b), notice that $u+V=u'+V$. 
Since $u'\in V^\perp$, $\tvgroup{\vgens}\cdot (u'+v) = u' + \tvgroup{\cdot} v$ for any $v\in\Fn$. 
Hence, if $O$ is the orbit containing $v$, $u'+O$ is the orbit containing $u'+v$.
Regarding (c), simply note that $V^\perp$ is the set of fixed points, hence singleton orbits.
\end{proof}
In the Pauli language, Lemma~\ref{lem:orbits:basic}(a) implies that, for a given generating set $\pgens$ and Pauli $P\in\PP_n$, there are connected components of the commutator graph entirely in the set $P\cdot\algclosure{\pgens}$, given the Pauli basis $\baslong{\algclosure{\pgens}} = \isolong{V}$.
In particular, if we only \emph{shift} by some Pauli $P=C$ in the commutant, this affects the orbit structure in a trivial way from $\algclosure{\pgens}$ to $C\algclosure{\pgens}$, i.e. simply by multiplying all orbits in $\algclosure{\pgens}$ by $C$.
We refer to such orbits as \emph{parallel translates} of the original orbits in $\algclosure{\pgens}$ or $\Span{\vgens}$.
Notice that we can have parallel translates orbits of \emph{any} orbit $O$ simply by adding a $v\in V^\perp$.

Furthermore, we also have \emph{trivial} orbits in $(u+V)$ given by its intersection with $V^\perp$.

A fundamental step for the classification of orbits is the classification of affine subspaces with underlying subspace $V$, and in particular a simple way to check for a non trivial intersection $(u+V)\cap V^\perp$.
From now on we write $\bar{V}^\perp \cong V^\perp/\rad(V)$ for a non-degenerate subspace of $V^\perp$ which is isomorphic to $V^\perp/\rad(V)$.
We shall also use $\bar{V} \cong V/\rad(V)$ for a non-degenerate subspace of $V$ which is isomorphic to $V/\rad(V)$.
We can frame this in a way which is independent of the choice of basis:
\begin{lem}\label{lem:Canonical_Basis_Affine_Subspaces_Parametrization}
Let $V\subseteq\Fn$ be a subspace.
Also, consider a direct sum decomposition of $\Fn$ in $V$, we write $\bar{V}^\perp = V^\perp/\rad(V)$ and $W$ the joint supplementary subspace of $V$ and $V^\perp$ (see also \ref{eq:full_Fn_decomposition_symplectic_basis}). 
We have the following:
\begin{enumerate}
    \item For $u,u'\in\Fn$, $u+V= u'+V$ if and only if:
    \begin{equation}
        \symp{u}{v} = \symp{u'}{v},\,\forall v\in V^\perp.
    \end{equation}
    \item For any $u\in\Fn$, $\symp{u}{\rad(V)}=\F_2$ if and only if $(u+V)\cap V^\perp = \emptyset$, viceversa $u'\in (u+V)\cap V^\perp\neq\emptyset$ if and only if $\symp{u'}{\rad(V)}=0$.
    \item If $(u+V)\cap V^\perp\neq\emptyset$, then $(u+V)\cap V^\perp\neq\emptyset = u'+\rad(V)$.
    \item There is a bijection between affine subspaces over $V$ and vectors in $\tilde{w}\in W\oplus\bar{V}^\perp$, such that:
    \begin{gather}
        \tilde{w}+V = \tilde{w}'+V \iff \tilde{w} = \tilde{w}',\\
        \tilde{w} + V \cap (W\oplus\bar{V}^\perp) = \tilde{w}.
    \end{gather}
\end{enumerate}
\end{lem}
\begin{proof}
Consider (a) and write $u\in u'+V$ as $u = u'+v$. Then $u+u'=v\in V$ if and only if $\symp{u+u'}{V^\perp} = 0$, which proves (a).

For (b), assume that there is some $b\in\rad(V)$ such that $\symp{u}{b} = 1$. 
Then, we must have, for all $u+ v\in u+ V$:
\begin{equation}
    \symp{u+v}{b} = \symp{u}{b} = 1.
\end{equation}
The intersection $(u+V)\cap V^\perp$ consists of the vectors $u+v$, $v\in V$, which are also in $V^\perp$, hence $\symp{u+v}{V} = 0$ and in particular $\symp{u+v}{\rad(V)} = 0$, which is an absurd.
Therefore, $(u+V)\cap V^\perp = \emptyset$.
Viceversa, if $(u+V)\cap V^\perp = \emptyset$, then for each $u+v\in V$ we must have:
\begin{equation}
    \symp{u+v}{V} = \F_2 \then \symp{u+v}{\rad(V)} = \symp{u}{\rad(V)} = \F_2,
\end{equation}
which proves (b) (the other statement is its contrapositive).

Also, if $u'\in (u+V)\cap V^\perp \neq \emptyset$, then in particular $u'\in V^\perp$ and $\symp{u'}{\rad(V)} = 0$.
If $u'\in (u+V)\cap V^\perp$, then $u'+\rad(V)$ is also in the intersection, given that $\rad(V) = V\cap V^\perp$.
Also, for all $u'+v$ in $u+V$:
\begin{equation}
    \symp{u'+v}{V} = \symp{v}{V} = 0 \then v\in V\cap V^\perp = \rad(V),
\end{equation}
which implies $(u+V)\cap V^\perp = u'+\rad(V)$ and proves (c).

(d) follows from the fact that $V\oplus(W\oplus\bar{V}^\perp)$ is a direct sum decomposition of $\Fn$.
\end{proof}
Hence, we find that affine subspaces are completely determined by how they interface with $V^\perp$, or equivalently a basis of it (given the non-degenerate symplectic product on $\Fn$).
As a corollary, this means that there are precisely $2^{2\ell+r}$ distinct affine subspaces over $V$, given $\rank(V^\perp)=2\ell$ and $\nullity(V)=r$.
The elements $\tilde{w}\in\vgens^\perp$ are those with $\symp{\tilde{w}}{\rad(V)} = 0$, which means that the `problematic' elements will come from those $u\in V$ which do not commute with the radical.
Also, if $\symp{\tilde{w}}{\rad(V)}=0$, we also have that $\tilde{w}\in \bar{V}^\perp = V^\perp/\rad(V)$, i.e. the non-degenerate part of the commutant.

In the Pauli picture, if we consider the canonical bases as in Eq.~\eqref{eq:Canonical_Basis_Pauli_Matrix_Algebra} for $\matalg=\algclosure{\pgens} = \Span[\C]{\isolong{V}}$, we find that `affine' matrix algebras $P\cdot\matalg$ are of the following form (up to phases):
\begin{equation}
    \qty[I^{\otimes m}\otimes \prod_{j=1}^r X_j^{\delta_j} \otimes \prod_{i=1}^{\ell} Z_i^{\alpha_i}X_i^{\beta_i}]\cdot\matalg.
\end{equation}
Also, the choice of $\delta_j=0$ for all $j$ implies $P\in\commutant(\pgens)$, which correspond to the 'easy' shifts.
Indeed, this is to be expected, since in this picture $\pgens$ lives purely in the first $m+r$ qubits.
Thus, any added Pauli terms over the remaining $\ell$ qubits does not affect the way $\pgens$ acts on the first $m+r$ qubits.

In the Majorana picture, in order to describe $P\cdot\matalg$, we have to distinguish between the odd or exceptional case (Eq.~\eqref{eq:full_Pn_decomposition_Majorana_odd_exceptional_case}) and the even case (Eq.~\eqref{eq:full_Pn_decomposition_Majorana_even_case}).
For the former, the `affine' matrix algebras are of the form ($r=q$):
\begin{equation}
    \qty[\prod_{j-2m=1}^q\gamma_{2j-1}^{\delta_j}\prod_{2n-i+1=1}^{2\ell}\gamma_i^{\alpha_i}]\algclosure{\pgens}
\end{equation}
and for the even case, $r=q+1$:
\begin{equation}
    \qty[\gamma_{2m}^{\delta_r}\prod_{j-2m=1}^q\gamma_{2j-1}^{\delta_j}\prod_{2n-i+1=1}^{2\ell}\gamma_i^{\alpha_i}]\algclosure{\pgens}.
\end{equation}
Notice that here phases are irrelevant since the matrix algebras contain arbitrary complex linear combinations.

\subsection{Quasi-Universal Case}\label{sec:Orbits_Full_Space_Fn_Quasi_Universal_Case}

As one might expect from Theorem~\ref{thm:Full_Classification_Theorem_w_E6_Quasi_Universal}, the orbit structure of quasi-universal generating sets are quite straightforward in terms of symmetries and quadratic forms.
Without loss of generality, we shall consider only the non-trivial orbits, i.e. all orbits in $(\tilde{w}+V)\setminus V^\perp$.

Regarding the orbits inside of $V$, we have the following description:
\begin{lem}\label{lem:quasi_universal_orbits_inside_V}
Let $\pgens = \isolong{\vgens}\subseteq\PP_n$ be a set of quasi-universal type, with $V=\Span{\vgens}$.
\begin{enumerate}
    \item If $\quadratic(\vgens)$ is empty, the only non-trivial orbit is $V\setminus\rad(V)$.
    \item If there is some $w\in\quadratic(\vgens)$, there are two non-trivial orbits: $(\QQ_w^{-1}(1)\cap V)\setminus\rad(V)$ and $(\QQ_w^{-1}(0)\cap V)\setminus\rad(V)$.
\end{enumerate}
\end{lem}
\begin{proof}
(a) follows immediately from Theorem~\ref{thm:Full_Classification_Theorem_w_E6_Quasi_Universal}. For (b), we can invoke \cite[Theorem~10.1]{Brown_Humphries_1986b}.
\end{proof}

As an immediate corollary of (a), we find the orbits in the case $\quadratic(\vgens)=\emptyset$ over all of $\Fn$:

\begin{cor}\label{cor:orbits_full_Fn_strictly_universal_case}
Let $\pgens = \isolong{\vgens}\subseteq\PP_n$ be a set of quasi-universal type with trivial quadratic forms. Also, let $V=\Span{\vgens}$ and $\commalg=\commutant(\pgens)$ and $\tilde{w}$ as in Lemma~\ref{lem:Canonical_Basis_Affine_Subspaces_Parametrization}.
We write $2\ell = \rank(\vgens^\perp)$, $2m = \rank(\vgens)$ and $r=\nullity(\vgens)$.
The non-trivial orbits in $\tilde{w}+V$ are the following:
\begin{enumerate}
    \item If $\tilde{w}\in V^\perp$, then $(\tilde{w}+V)\setminus(\tilde{w}+\rad(V)) = \tilde{w}+(V\setminus\rad(V))$ is the only non-trivial orbit in $\tilde{w}+V$. There are $2^{2\ell}$ such orbits, of size $2^{2m}(2^r{-}1)$.
    \item If $\tilde{w}\not\in V^\perp$, then all of $\tilde{w}+V$ is an orbit. There are $2^{2\ell}(2^r{-}1)$ such orbits, of size $2^{2m+r}$.
\end{enumerate}
In total, there are $2^{2\ell+r}$ non-trivial orbits and $2^{2\ell+r}$ trivial orbits (coming from $V^\perp$).
\end{cor}
\begin{proof}
(a) follows from Lemma~\ref{lem:orbits:basic}(b).
Regarding (b), we simply need to show that there is an element $g\in\tvgroup{\vgens} = \tvgroup{V\setminus\rad(V)} = \Sp^{\#}(V)$ which connects $\tilde{w}+\rad(V)$ to $(\tilde{w}+V)\setminus(\tilde{w}+\rad(V))$, whenever $\symp{\tilde{w}}{\rad(V)}=\F_2$.
For any $\tilde{w}+u\in \tilde{w}+\rad(V)$, consider any $v\in \bar{V} \cong V/\rad(V)$ and let $u_1\in\rad(V)$ such that $\symp{\tilde{w}}{u_1}=1$, then:
\begin{align*}
        \tau_{v+u_1}(\tilde{w}+u) &= \tilde{w}+u + \symp{v+u_1}{\tilde{w}+u}(v+u_1) \\
        &= \tilde{w}+v+u+u_1
\end{align*}
which is in $(\tilde{w}+V)\setminus\rad(V)$.
This proves (b).
\end{proof}
In the Pauli picture, in the canonical form Eq.~\eqref{eq:Canonical_Basis_Pauli_Matrix_Algebra} for the commutant $\commalg$, we find that under $\lieiso^0(\commalg)$ the Pauli orbits are given by the following, up to phases:
\begin{enumerate}
    \item for fixed arbitrary $\alpha_i,\beta_i$ in $\F_2$
    $$\PP_m^0\otimes \{I,Z\}^{\otimes r} \otimes \prod_{i=1}^{\ell} Z_i^{\alpha_i}X_i^{\beta_i},$$
    \item for fixed $\delta_j\in\F_2$ not all zero and fixed arbitrary $\alpha_i,\beta_i$ in $\F_2$
    $$\PP_m\otimes \Big(\prod_{j=1}^rX_j^{\delta_j}\setprod\{I,Z\}^{\otimes r}\Big) \otimes \prod_{i=1}^{\ell} Z_i^{\alpha_i}X_i^{\beta_i},$$
    \item for fixed arbitrary $\delta_j$ and $\alpha_i,\beta_i$ in $\F_2$, the singleton orbits
    $$I^{\otimes m}\otimes \prod_{j=1}^r Z_j^{\delta_j}\otimes \prod_{i=1}^{\ell} Z_i^{\alpha_i}X_i^{\beta_i},$$
\end{enumerate}
where $\PP_m^0 = \PP_m\setminus\{I^{\otimes m}\}$ are the traceless Pauli strings.
Consider a Pauli string $P=P_1\otimes P_2\otimes P_3$ partitioned over the logical, phase and uncontrollable qubits.
We can describe the action of $\lieiso^0(C)$ as follows:
\begin{enumerate}
    \item The uncontrollable part is fixed;
    \item Over the logical qubits, one reach arbitrary Paulis, including the identity if there is a non commuting term over the phase qubits
    \item Over the phase qubits, only Pauli strings which have the same $X$-part as the initial $P_2$ can be reached.
\end{enumerate}
We can also give the Clifford invariant formulation, which still requires an explicit generating set of the commutant.
Let $\{C_i\}\cup\{A_j,B_j\}$ be a Pauli generating set of $\commalg= \commutant(\pgens)$ corresponding to a symplectic basis.
Then, the orbits are the joint fixed points under the adjoint operators $\Ad_P$ or their negations $-\Ad_P$, for $P$ corresponding to the Pauli generators of $\commalg$.
Namely, we have the following orbits:
\begin{enumerate}
    \item For arbitrary $\alpha_i,\beta_i\in\{\pm 1\}$
    $$\PP_n^{\{\Ad_{C_i}\}_{i=1}^r,\{\alpha_j\Ad_{A_j},\beta_j\Ad_{B_j}\}_{j=1}^{\ell}}\setminus\bas{\commalg},$$
    \item For $\delta_j\in\{\pm 1\}$ not all $1$ and arbitrary $\alpha_i,\beta_i\in\{\pm 1\}$
    $$\PP_n^{\{\delta_j\Ad_{C_i}\}_{i=1}^r,\{\alpha_j\Ad_{A_j},\beta_j\Ad_{B_j}\}_{j=1}^{\ell}},$$
    \item The elements of the Pauli basis $\bas{\commalg}$.
\end{enumerate}
Furthermore, in order to find the orbit in which a Pauli string $P$ belongs (orbit identification), we also simply need to check whether $P\in \commalg$ and the commutation relations with a generating set of $\commalg$.
Equivalently, for a given generating set of the commutant, the components of the commutator graph are completely determined by whether they commute or anti-commute with the given generators.
On the other hand, for the task of orbit \emph{intersection} for two vectors (Pauli strings), it suffices to check whether they have the same commutation relations with a given generating set of the orthogonal complement (commutant).
Equivalently, two Pauli strings are in the same orbit iff they coincide or the product given Pauli strings $P,P'$ is in the matrix algebra $PP'\in\algclosure{\pgens}$ and they are both not in the commutant.

In the case with non-trivial quadratic forms, we find instead the following description over arbitrary $\tilde{w}+V$:
\begin{lem}[{\cite[Theorem 6.2]{Shapiro_Shapiro_Vainshtein_1998},\cite[Theorem 2.14]{Seven_2005}}]\label{lem:orbits_full_Fn_quadratic_cases}
Let $\pgens = \isolong{\vgens}\subseteq\PP_n$ be a set of quasi-universal type with non-trivial quadratic forms $w\in\quadratic(\vgens)\neq\emptyset$. 
Also, let $V=\Span{\vgens}$ and $\tilde{w}\in W\oplus\bar{V}^\perp$ as in Lemma~\ref{lem:Canonical_Basis_Affine_Subspaces_Parametrization}.
We write $2\ell = \rank(\vgens^\perp)$, $2m = \rank(\vgens)$ and $r=\nullity(\vgens)$.

The non-trivial orbits in $\tilde{w}+V$ are the following:
\begin{enumerate}
    \item If $\tilde{w}\in V^\perp$, the non-trivial orbits are: $(\QQ_w^{-1}(1)\cap(\tilde{w}+(V\setminus\rad(V)))$ and $(\QQ_w^{-1}(0)\cap(\tilde{w}+(V\setminus\rad(V)))$. There are $2\cdot 2^{2\ell}$ such orbits, whose sizes depend on how $\QQ_w$ acts on $\tilde{w}+W$ as follows:
    \begin{nestedcaseenum}
        \item If $\type(\QQ_w|_{\tilde{w}+W})=+$ and $\QQ_w(\tilde{w})=\alpha$, these orbits respectively have size $2^r2^{m-1}(2^m-1)-(1-\alpha)2^r$ and $2^r2^{m-1}(2^m+1)-\alpha 2^r$,
        \item If $\type(\QQ_w|_{\tilde{w}+W})=-$, these orbits respectively have size $2^r2^{m-1}(2^m+1)-(1-\alpha)2^r$ and $2^r2^{m-1}(2^m-1)-\alpha 2^r$,
        \item If $\type(\QQ_w|_{\tilde{w}+W})=0$, these orbits are both of size $2^{2m+r-1}-2^{r-1}$.
    \end{nestedcaseenum}
    \item If $\tilde{w}\not\in V^\perp$, the non-trivial orbits are: $(\QQ_w^{-1}(1)\cap(\tilde{w}+V))$ and $(\QQ_w^{-1}(0)\cap(\tilde{w}+V))$. There are $2\cdot 2^{2\ell}(2^r-1)$ such orbits, whose sizes depend on how $\QQ_w$ acts on $\tilde{w}+W$ as follows:
    \begin{nestedcaseenum}
        \item If $\type(\QQ_w|_{\tilde{w}+W})=+$, these orbits respectively have size $2^r2^{m-1}(2^m-1)$ and $2^r2^{m-1}(2^m+1)$.
        \item If $\type(\QQ_w|_{\tilde{w}+W})=-$, these orbits respectively have size $2^r2^{m-1}(2^m+1)$ and $2^r2^{m-1}(2^m-1)$.
        \item If $\type(\QQ_w|_{\tilde{w}+W})=0$, these orbits are both of size $2^{2m+r-1}$.
    \end{nestedcaseenum}
\end{enumerate}
In total, there are $2\cdot 2^{2\ell+r}$ non-trivial orbits and $2^{2\ell+r}$ trivial orbits (coming from $V^\perp$).
\end{lem}
\begin{proof}
It suffices to adapt \cite[Theorem 6.2]{Shapiro_Shapiro_Vainshtein_1998}, which states that $\tvgroup{\vgens}$ has two non-trivial orbits in $(\tilde{w}+V)\setminus V^\perp$, which are intersections with the level sets of $\QQ_{w^*}$ with $w^*\in\quadratic(\vgens\cup\{\tilde{w}\})$: the anisotropic vectors $(\QQ_{w^*}^{-1}(1)\cap(\tilde{w}+V))\setminus V^\perp$ and isotropic vectors $(\QQ_{w^*}^{-1}(0)\cap(\tilde{w}+V))\setminus V^\perp$ in $\tilde{w}+V\setminus V^\perp$.
Notice that $\quadratic(\vgens\cup\{\tilde{w}\})$ is in fact non-trivial.
Since $\tilde{w}$ is either $0$ or not in $V$ and hence linearly independent from $\vgens$ (due to the direct sum decomposition).
Thus, if $\vgens$ has non-trivial invariant quadratic forms, so does $\vgens\cup\{\tilde{w}\}$.

Then, since $w^*$ and $w$ are both in $\quadratic(\vgens)$, there is a $u\in V^\perp$ such that $w^* = w+u$ by the affine structure of the quadratic forms, see Lemma~\ref{lem:affine_structure_invariant_quadratic_form_subspace}.
As such, the level sets for $\QQ_{w^*}$ in $\tilde{w}+V$ are also level sets for $\QQ_w$ but with a possibly different value:
\begin{align*}
        &\QQ_{w^*}(\tilde{w}+v) = \QQ_{w+u}(\tilde{w}+v) \\
        &= \QQ_w(\tilde{w}+v) + \symp{u}{\tilde{w}+v}
        = \QQ_w(\tilde{w}+v) + \symp{u}{\tilde{w}},
\end{align*}
hence $\QQ_{w^*}^{-1}(q)\cap(\tilde{w}+V) = \QQ_{w^*}^{-1}(q+\symp{u}{\tilde{w}})+(\tilde{w}+V)$.

As such, without loss of generality, for a fixed $w\in\quadratic(\vgens)$, the orbits in $\tilde{w}+V\setminus V^\perp$ are the corresponding level sets of $\QQ_w$, independently of $\tilde{w}\in W\oplus \bar{V}^\perp$.
Finally, the counting of the non-trivial orbits comes from counting the $\tilde{w}\in W\oplus\bar{V}^\perp$ which are in $V^\perp$ or not.
The trivial orbits are simply all the elements of $V^\perp$. This proves (a) and (b).
For the different orbit sizes for (a) and (b), see Lemma~\ref{lem:isotropic_sizes_over_affine_subspaces} and take the affine subspace as $\tilde{w} + V$, where $\QQ=\QQ_w$, $\rank(V)=2m$ and $\nullity(V)=r$.
This concludes the proof.
\end{proof}

In the Pauli picture, let $\{C_i\}\cup\{A_j,B_j\}$ be a Pauli generating set of $\commalg= \commutant(\pgens)$ corresponding to a symplectic basis.
Then, the orbits are the joint fixed points under the adjoint operators of the generators of $\commalg$, or their negations, as well as the (negative) $B$-transpose $\theta_B$, for any $B$:
\begin{enumerate}
    \item For arbitrary $\alpha_i,\beta_i\in\{\pm 1\}$
    \begin{align*}
        \PP_n^{\theta_B,\{\Ad_{C_i}\}_{i=1}^r,\{\alpha_j\Ad_{A_j},\beta_j\Ad_{B_j}\}_{j=1}^{\ell}}\setminus\bas{\commalg},\\
        \PP_n^{-\theta_B,\{\Ad_{C_i}\}_{i=1}^r,\{\alpha_j\Ad_{A_j},\beta_j\Ad_{B_j}\}_{j=1}^{\ell}}\setminus\bas{\commalg}.
    \end{align*}
    \item For $\delta_j\in\{\pm 1\}$ not all $1$ and arbitrary $\alpha_i,\beta_i\in\{\pm 1\}$
    \begin{align*}
        \PP_n^{\theta_B,\{\delta_j\Ad_{C_i}\}_{i=1}^r,\{\alpha_j\Ad_{A_j},\beta_j\Ad_{B_j}\}_{j=1}^{\ell}},\\  
        \PP_n^{-\theta_B,\{\delta_j\Ad_{C_i}\}_{i=1}^r,\{\alpha_j\Ad_{A_j},\beta_j\Ad_{B_j}\}_{j=1}^{\ell}}.
    \end{align*}
    \item The elements of the Pauli basis $\bas{\commalg}$.
\end{enumerate}
Now, for the task of orbit identification, compared to the case with no quadratic forms, one also needs to check whether $v$ is isotropic or anisotropic for a given reference quadratic form.
For orbit intersection of two vectors $v,v'$, one also needs to check whether $\QQ_w(v) = \QQ_w(v')$, independently of the type of invariant quadratic forms.
Notice that the canonical form of the orbits (with respect to Eq.~\eqref{eq:full_Pn_Pauli_decomposition_single_qubit_basis} and Corollary~\ref{cor:Canonical_Forms_Invariant_Bilinear_Forms}) has a more complex interplay between invariant quadratic forms and affine subspaces.
In particular, we can no longer easily factorize into the various qubits sites, unlike the quasi-universal case with no invariant quadratic forms.
For instance, one may have a Pauli string which is globally symmetric, but skew symmetric over the logical qubits, skew-symmetric over the phase qubits and symmetric over the uncontrollable qubits.
This also corresponds to a more subtle reachability of such architectures.
Namely, in the canonical form specified by Eq.~\eqref{eq:full_Pn_Pauli_decomposition_single_qubit_basis} and Corollary~\ref{cor:Canonical_Forms_Invariant_Bilinear_Forms}, consider a Pauli strings $P=P_1\otimes P_2\otimes P_3$ partitioned over the logical, phase and uncontrollable qubits. 
Also, take a canonical bilinear form $B$ in $\{I,Y_m\}$ (here $2m=\rank(\pgens)$ and $Y_m$ lives in the logical qubits).
If $P_3$ and $P_2$ are $B$-symmetric, then one has the usual reachability over the logical part, i.e. one can go from $P_1$ to $P_1'$ if and only if they are both $B$-symmetric (and possibly not the identity).
However, the combined reachability when taking into account both the logical and phase qubits is larger.
Namely, given that these three parts commute, globally the (an)isotropic of a Pauli string under $B$ or $\QQ_w$ ($w=\inviso(B)$), the following is an invariant (in the binary formalism):
\begin{equation}
    \QQ_w(v) = \QQ_w(v_1) + \QQ_w(v_2) + \QQ_w(v_3).
\end{equation}

\subsection{Free-Fermionic Case}\label{sec:Orbits_Full_Space_Fn_Free_Fermionic_Case}

We now deal with the free-fermionic or line-graph case, by taking advantage of the Majorana labels.
We start with the linearly independent case (i.e. even or odd types), by complimenting the classification over $V=\Span{\vgens}$ to all of $\Fn$:
\begin{thm}[\cite{Seven_2005}]\label{thm:orbits_in_Fn_alg_ind_even_odd_cases_free_fermionic}
Let $\vgens\subseteq\Fn$ be a free-fermionic generating set of either even or odd type. Let $w$ be a reference quadratic form in $\quadratic(\vgens)\neq\emptyset$ and $q=\nullorbit(\vgens)$. Also, consider some vector $u\in\Fn$.
Consider the (non-trivial) orbits in $u+V\setminus V^\perp$:
\begin{enumerate}
    \item If $\symp{u}{\radorbit(\vgens)} = 0$, then there is some $u'\in u+V$ such that $\frustration{\vgens\cup\{u'\}}$ is a line graph and the orbits for $\tvgroup{\vgens}$ over $u+V$ coincide with the intersection of $u+V$ and the orbits of $\tvgroup{\vgens\cup\{u'\}}$.
    \item If $\symp{u}{\radorbit(\vgens)} = \F_2$, the orbits are $\QQ_{w}^{-1}(1)\cap(u+V)$ and $\QQ_{w}^{-1}(0)\cap(u+V)$.
\end{enumerate}
\end{thm}
We have adjusted the original statement in \cite{Seven_2005} for Thm~\ref{thm:orbits_in_Fn_alg_ind_even_odd_cases_free_fermionic}(b) to reflect the use of invariant quadratic forms over $\Fn$, as done in Lemma~\ref{lem:orbits_full_Fn_quadratic_cases}.
Notice that the the orbits from case (a) are not necessarily parallel translates of orbits in $V$ in the even case, since $\rad(\vgens)/\radorbit(\vgens)$ is non trivial, hence one might still have $\symp{u}{\rad(\vgens)} = \F_2$.

However, since this $\radorbit(\vgens)$ has codimension $1$ in $\rad(\vgens)$, it suffices to check whether $u$ commutes or anti-commutes with one vector.
Namely, in the Majorana representation we can choose uniquely
$\tilde{w}\in W\oplus \bar{V}^{\perp}$ as
\begin{align*}
    &\majiso{\tilde{w}} \simeq
    \gamma_{2m}^{\delta_0}\,
    (\prod_{j=1}^{q}\gamma_{2m+2j-1}^{\delta_j})\, \majiso{v}
    \\
    &\text{with }
    \delta_0,\delta_j\in\F_2,\;
    \majiso{v}\in\commalg,\; v\in V^\perp/\rad(V).
\end{align*}
Here $\delta_0$ determines whether the extra odd-length factor
$\gamma_{2m}$ is present, while the coefficients $\delta_j$ result in the cycle-symmetry factors.
Out of these vectors, those with $\symp{u}{\radorbit(\vgens)} = 0$ are precisely
those with $\delta_j=0$ for all $j\in[q]$.
For such vectors $u$, we have $\symp{u}{\rad(\vgens)} = 0$ if $\delta_0=0$, and
$\symp{u}{\rad(\vgens)} = \F_2$ if $\delta_0=1$.
Furthermore, notice that all cases with $\symp{u}{\rad(\vgens)} = \F_2$ but $\symp{u}{\radorbit(\vgens)} = 0$ are parallel translates of each other.
Hence, it suffices to describe these only for $\majiso{v}=I$.

Then, we have two qualitatively distinct cases where $\symp{u}{\radorbit(\vgens)} = 0$, i.e.\ those with $\majiso{\tilde{w}} = I$ and those with $\majiso{\tilde{w}} = \gamma_{2m}$.
The former are simply the orbits in $V$ or $\algclosure{\pgens}$, i.e. the Majorana strings of even length, $\majL_m(v) = 0\bmod 2$ (up to cycle symmetries). 
These orbits are described in Proposition~\ref{prop:orbits_in_V_line_graphs_Majorana}(a), and consist of $\MAJ_{2L}^{(2m)}\setminus\Cyc^{(2m,q)}$.
The latter instead correspond to the orbits inside the space of Majorana strings of \emph{odd} length, $\majL_m(v) = 1\bmod 2$ (up to cycle symmetries).
Furthermore, it is a known fact \cite{Kaufman1949} that the purely quadratic Majoranas ($q=0$) conserve Majorana length, and in particular that the Clifford action produces orbits which depend solely on the Majorana length \cite{Sierant_Turkeshi_Tarabunga_2026}.
As such we immediately have that the orbits in $\mu_{2m}+V$ are those of the form:
\begin{equation}\label{eq:odd_orbits_even_type}
    \majiso{\tilde{w}}\MAJ_{2L-1}^{(2m)}\setprod\Cyc^{(2m,q)}.
\end{equation}

We can also use Theorem~\ref{thm:orbits_in_Fn_alg_ind_even_odd_cases_free_fermionic} to describe the orbits coming from $\tilde{w}+V$ with $\tilde{w}=\mu_{2m}$.
Since $\pgens$ is of even type and with free-fermionic mapping as in Proposition~\ref{prop:free_fermionic_mapping_alg_dep}(a), we have that $\pgens\cup\{\gamma_{2m}\}$ has free-fermionic mapping as in Proposition~\ref{prop:free_fermionic_mapping_alg_dep}(b)), $\pgens\cup\{\gamma_{2m}\}$ is a free fermionic generating set of odd type. 
Equivalently, choosing $\pgens$ as a minimal generating set with frustration $\graphP_{2m-2,n_1}$ as in Eq.~\eqref{eq:majorana_string_basis_blown_up_path_graph_canonical}, the set $\pgens\cup\{\gamma_1\}$ is also minimal with frustration graph $\graphP_{2m-1,n_1}$, and this set is $t$-equivalent to $\pgens\cup\{\gamma_{2m}\}$.
Hence, if we take $u' = \mu_{2m} = \inviso(\gamma_{2m})$, we can apply Theorem~\ref{thm:orbits_in_Fn_alg_ind_even_odd_cases_free_fermionic}(a) to $\vgens$ over the affine space $\mu_{2m}+V$, together with we can use the description of the orbits for free-fermionic generating sets of odd-type as in Proposition~\ref{prop:orbits_in_V_line_graphs_Majorana}(b).
The orbits of $\pgens\cup\{\gamma_{2m}\}$ over its matrix algebra (equivalently, $\Span{\vgens\cup\{\mu_{2m}\}}\supseteq \mu_{2m}+V$), as described in Proposition~\ref{prop:orbits_in_V_line_graphs_Majorana}(b), are then identified by the Majorana length, as $\isolong{O_L} = \{\MAJ_{2L-1}^{(2m)}\cup \MAJ_{2L}^{(2m)}\}\setminus\Cyc^{(2m,q)}$.
Hence, the intersection of these orbits with the the affine subspace $\tilde{w}+V$, i.e.\ all Majorana strings of odd length, are precisely those with only odd length $\majL_m(v)=2L-1$, together with the cycle symmetries, as in Eq.~\eqref{eq:odd_orbits_even_type}, as expected.

Given the dependence of the orbits on the invariant quadratic forms also for free-fermionic cases, we now describe their action as dependent on the Majorana length.
We first describe their action on their generated subspace.
In either case, it is sufficient to describe the action on a basis of $V/\radorbit(\vgens)$ (viewed as a subspace of $V$), given that $\radorbit(\vgens) \subseteq \radzero(V,\QQ)$ for an invariant quadratic form $\QQ$ over $V$. 
Thus, for $v\in V$, the values of $\QQ$ do not change upon adding $u\in\radorbit(\vgens)$, i.e.\ $\QQ(v+u) = \QQ(v)$.

In the odd case, it is clear that the invariant quadratic form, denoted as $\QQ_1$, is simply the restriction to the first $2m$ modes of $\QQ_\gamma$, or
\begin{equation}
    \QQ_1(v) = \begin{dcases}
        0 & \majL_m(v)\bmod 4 \in \{0,3\} \\
        1 & \majL_m(v)\bmod 4 \in \{1,2\}
    \end{dcases}.
\end{equation}
In the even case, we denote the invariant quadratic form over the even Majoranas as $\QQ_2$, such that $\QQ_2(\mu_i+\mu_j) = 1$.
We can also define $\QQ_2$ first on a basis, e.g. $\{\mu_i+\mu_{i+1}\}_{i=1}^{2m-1}$ and then extend it.
Notice that even Majorana string has a unique decomposition into $s$ pairs of non-overlapping even Majoranas $v = \sum_{j=1}^s \mu_{i_j} + \mu_{i_j'}$.
Hence, since the pairs $\mu_{i_j} + \mu_{i_j'}$ are anisotropic and commute for different $j$, the invariant quadratic form coincides precisely with $s$, $\QQ_2(v) = s\bmod 2$, where $\majL_m(v) = 2s$.
Equivalently, the isotropic vectors for $\QQ_2$ are the Majorana strings in $V$ with length $\majL_m(v) = 0\bmod 4$, and the anisotropic vectors are the Majorana strings with $\majL_m(v) = 2\bmod 4$.

In order to describe all of the orbits in $\Fn$, we now consider the full set of invariant quadratic form, using the as reference the natural invariant Majorana quadratic form $\QQ_\gamma$, defined in Eq.~\eqref{eq:reference_quadratic_form_majoranas}.
Then, the set of invariant quadratic form may be written in vector form as $\quadratic_\gamma(\vgens) = \{ w\in\Fn | \QQ_w^\gamma(\vgens) = 1\}$, such that $\QQ_w^\gamma$ restricts to $\QQ_1$ in the odd case and $\QQ_2$ in the even case.

We first consider $q=0$. 
In both the even and odd case with $(2m,0)$-free fermionic mappings, it is immediate to show that $\QQ_\gamma$ is an invariant quadratic form for both.
We can then take $w^* = 0$ and identify the invariant quadratic forms $\quadratic_\gamma(\vgens)$ directly with $\vgens^\perp$ in both cases.
In particular, when $\ell=0$ the only invariant quadratic form over $\Fn =V$ in the odd case is $\QQ_\gamma$, and for the even case both $\QQ_\gamma$ and $\QQ_{\majprod}^\gamma$ are invariant quadratic forms, which are respectively of type $(-1)^n$ and $+$ (given that $(-1)^{n+\QQ_\gamma(\majprod)} = +1$) for all $n$.
As a corollary, we find that $\QQ_\gamma$ restricted to the even Majorana strings coincides with $\QQ_2$.

When $q>0$, $\QQ_\gamma$ is no longer an invariant quadratic form, since under the presence of cycle symmetries the quadratic generators ($\calL_m(v)=2$) $(\tilde{\gamma}_{2s-1}\tilde{\gamma}_{2s})\gamma_i\gamma_j$ have length $\calL(v)=4$ and the linear ones ($\calL_m(v)=2$) $(\tilde{\gamma}_{2s-1}\tilde{\gamma}_{2s})\gamma_i$ have length $\calL(v)=3$ (see also Eq.~\eqref{eq:reference_quadratic_form_majoranas}).
Let us then solve for $w\in\quadratic_\gamma(\vgens) = \{ w\in\Fn | \QQ_w^\gamma(\vgens) = 1\}$ when $q>0$. 
In the odd case we can choose $\{\mu_i\}_{i=1}^{2m}\cup\{\mu_{2m}+\mu_{2(m+j)-1} + \mu_{2(m+j)}\}_{j=1}^q$ as generating set (which has frustration graph $\graphK_{2m-1;q+1} \sim_t \graphP_{2m-1,q+1}$).
Then, $w$ is a solution of the following system of linear equations:
\begin{equation*}
    \begin{dcases}
        \symp{w}{\mu_i} = 0 &\text{for } i\in[2m]\\
        \symp{w}{\mu_{2m}+\mu_{2(m+j)-1} + \mu_{2(m+j)}} = 1 &\text{for } j\in[q]
    \end{dcases}
\end{equation*}
which has rank $2m+q$ inside a space of dimension $2n = 2m+2q+2\ell$, hence has a space of solutions of dimension $q+2\ell$.
After some calculations, one finds a special solution which depends on the parity of $q$ as follows
\begin{equation}\label{eq:quadratic_forms_free_fermions_w_symmetries}
    w = \begin{dcases}
        \sum_{j=1}^q \mu_{2(m+j)-1} & q=0\bmod 2,\\
        \majprod + \sum_{j=1}^q \mu_{2(m+j)-1} & q=1\bmod 2
    \end{dcases}
\end{equation}
Notice that for arbitrary values of $q$, $w$ specified in Eq.~\eqref{eq:quadratic_forms_free_fermions_w_symmetries} is also an invariant quadratic for the even case, since the vectors $\mu_i+\mu_j$ commute with $\majprod$ and are isotropic for $\QQ_\gamma$.
However, the underlying subspaces for $\quadratic_\gamma(\vgens)$ differ depending on whether $\vgens$ is of even or odd type.
Consequently, also $\QQ = \QQ_w^\gamma|_V$ will differ depending on the given case, reducing to $\QQ_1$ in the odd case and $\QQ_2$ in the even case.

Finally, we summarize the discussion in the following.
\begin{lem}\label{lem:orbits_Fn_even_case}
Consider a generating set $\pgens$ of even type, with a $(2m,q)$ free fermionic mapping, $w$ taken as in Eq.~\eqref{eq:quadratic_forms_free_fermions_w_symmetries} and orthogonal complement of rank $2\ell$.
Let $w$ be as in Eq.~\eqref{eq:quadratic_forms_free_fermions_w_symmetries}.
The orbits over $\tilde{w}+V$ or $\majiso{\tilde{w}}\algclosure{\pgens}$ (with $\commalg = \commutant(\pgens)$) are the following:
\begin{enumerate}
    \item If $\majiso{\tilde{w}}\in\commalg$, there are $m-1$ non trivial orbits which are parallel translates of those in $\algclosure{\pgens}$
    \begin{equation}
        \majiso{\tilde{w}} \MAJ_{2L}^{(2m)}\setprod\Cyc^{(2m,q)}.
    \end{equation}
    There are $2^{2\ell}$ such distinct cases, by taking $\tilde{w}\in V^\perp/\rad(V)$, hence $\majiso{\tilde{w}}$ over the uncontrollable $2\ell$ Majorana modes.
    \item If $\majiso{\tilde{w}} \simeq \gamma_{2m}\cdot \majiso{v}$ with $\majiso{v}\in\commalg$, there are $m$ (non trivial) orbits which are parallel translates of those in $\gamma_{2m}\algclosure{\pgens}$
    \begin{equation}
        \majiso{\tilde{w}} \MAJ_{2L-1}^{(2m)}\setprod\Cyc^{(2m,q)}.
    \end{equation}
    There are $2^{2\ell}$ such distinct cases.
    \item If $\majiso{\tilde{w}} \simeq
    \gamma_{2m}^{\delta_0}
    (\prod_{j=1}^q\gamma_{2m+2j-1}^{\delta_j})\majiso{v}$ with
    $\{\delta_j\}_{j=1}^q$ not all zero and $\majiso{v}\in\commalg$, there are
    $2$ orbits in the corresponding binary affine subspace $\tilde{w}+V$,
    $(\QQ_w^\gamma)^{-1}(0)\cap(\tilde{w}+V)$ and
    $(\QQ_w^\gamma)^{-1}(1)\cap(\tilde{w}+V)$.
    There are $2^{2\ell+1}(2^q-1)$ such cases.
    \item The $2^{2\ell+q+1}$ trivial orbits coming from the orthogonal complement.
\end{enumerate}
In total, there are $2^{2\ell}(2m+1 + 6 (2^q-1) )$ orbits.
\end{lem}
\begin{lem}\label{lem:orbits_Fn_odd_case}
Consider a generating set $\pgens$ of odd type, again with a $(2m,q)$ free fermionic mapping and orthogonal complement of rank $2\ell$.
The orbits are as follows:
\begin{enumerate}
    \item If $\majiso{\tilde{w}}\in\commalg$, there are $m$ non trivial orbits which are parallel translates of those in $\algclosure{\pgens}$
    \begin{equation}
        \majiso{\tilde{w}} (\MAJ_{2L-1}^{(2m)}\cup\MAJ_{2L}^{(2m)})\setprod\Cyc^{(2m,q)}.
    \end{equation}
    There are $2^{2\ell}$ such cases.
    \item If $\majiso{\tilde{w}} \simeq
    (\prod_{j=1}^{q}\gamma_{2m+2j-1}^{\delta_j})\majiso{v}$ with
    $\{\delta_j\}_{j=1}^q$ not all zero and $\majiso{v}\in\commalg$, there are
    $2$ orbits: $(\QQ_w^\gamma)^{-1}(0)\cap(\tilde{w}+V)$ and
    $(\QQ_w^\gamma)^{-1}(1)\cap(\tilde{w}+V)$.
    There are $2^{2\ell}(2^q-1)$ such cases.
    \item The $2^{2\ell+q}$ trivial orbits coming from the orthogonal complement/commutant.
\end{enumerate}
In total, there are $2^{2\ell}(m+1 + 3(2^q-1))$ orbits.
\end{lem}
Notice that the orbits in some affine subspace $\tilde{w}+V$ for the even and odd cases do \emph{not} coincide, even when $\tilde{w}$ is taken for both, since the subspace $V=\Span{\vgens}$ generated by the odd case is larger than that generated by the even case, due to the presence of odd Majorana strings.
Indeed, under the parametrization above, we have that $\rank(\vgens)=2m-2$ and $\nullity(\vgens)=q+1$ in the even case, and $\rank(\vgens)=2m$ and $\nullity(\vgens)=q$ in the odd case.

We can now look at the exceptional case, which we can see as a generating set of odd type, together with $\Gamma = \prod_{i=1}^{2m}\gamma_i$ as additional generator and $n_\Delta=2m+2$ vertices for the root multigraph $\frustration{\pgens}$.

Notice that both the exceptional and odd case span the same subspace $V$, as well as the radical, orbit radical and orthogonal complement, under the usual free-fermionic mapping (see Eq.~\eqref{eq:majorana_string_basis_blown_up_path_graph_canonical}).
The resulting orbits in $V$ are described in \ref{prop:orbits_in_V_line_graphs_Majorana}(c), which also provide the orbits in $\tilde{w}+V$ when $\symp{\tilde{w}}{\rad(V)} = \symp{\tilde{w}}{\radorbit(V)}$, as parallel translates of those in $V$.
Consequently, it remains to deal with those orbits in $\tilde{w}+V$ such that $\symp{\tilde{w}}{\radorbit(\vgens)} = \F_2$, which for the odd case depende purely on the invariant quadratic form $\QQ_w^\gamma$.

We have that $\QQ_\gamma(\majprod) = m$, hence $\Gamma$ is isotropic for $m = 0\bmod 2$. Since $\Gamma$ commutes with itself and the non-overlapping Majorana strings (coming from the cycle symmetries), we also have $\QQ_w^\gamma(\majprod) = \QQ_\gamma(\majprod)= m$ for any $q\geq 0$, 
Hence the (an)isotropy of $\majprod$ with respect to $\QQ_w^\gamma$ dependes.
As such, the invariant quadratic forms in the odd and exceptional case with $(2m,q)$-free fermionic mapping coincide when $m$ is odd, and the exceptional case has no invariant quadratic forms for $m$ even (given that $\Gamma$ is already spanned by the generators of the odd case).
This is also expected from the classification of invariant quadratic forms of the canonical cases in Proposition~\ref{prop:isomorphism_classes_quadratic_forms_arf_invariant_canonical} and \ref{prop:quadratic_forms_arf_invariant_canonical_graph_t_representatives_alg_dep_cases}.

Then, for the orbits of the exceptional case with $\tilde{w}+V$ such that $\symp{\tilde{w}}{\radorbit(\vgens)} = \F_2$ we have the following:
\begin{enumerate}
    \item if $m$ is odd, they must be the same since the quadratic form is still conserved by $\Gamma$; 
    \item if $m$ is even, the orbit for the exceptional case is all of $V$, since $\Gamma$ connects isotropic and anisotropic vectors into a single orbit.
\end{enumerate}
Explicitly, to prove case (b), one needs to find a $u\in \tilde{w}+V$ which anti-commutes with $\majprod$, $\symp{u}{\majprod} = 1$. 
Then, since $\QQ_w^\gamma(\majprod) = m \bmod 2$ for $w$ as in Eq.~\eqref{eq:quadratic_forms_free_fermions_w_symmetries}, $\QQ_w^\gamma(\tau_{\majprod}(u)) = \QQ_w^\gamma(u) + (m+1)\bmod 2$, i.e. $\tau_{\majprod}$ changes the (an)isotropy of $u$ with respect to $\QQ_w^\gamma$ whenever $m$ is even.
Clearly, for any $\tilde{w}$, since $V$ contains linear Majoranas $\mu_i$, $i\in[2m]$, there exists both even and odd strings in $\tilde{w} + V$, hence a $u$ such that $\symp{u}{\majprod} = 1$. 
Hence, when $m$ is even, $\Gamma$ connects the two anisotropic and isotropic vectors (with respect to $\QQ_w^\gamma$) in $\tilde{w}+V$, which are single orbits for the odd case, resulting in a single orbit for the exceptional case.

If $\symp{\tilde{w}}{\radorbit(\vgens)} = 0$ instead we simply have parallel translates, hence the description comes from Proposition~\ref{prop:orbits_subspace_exceptional_case_alg_dep}.
We formalize this into the following:
\begin{prop}\label{prop:orbits_in_Fn_exceptional_case_free_fermionic}
Let $\vgens\subseteq\Fn$ be a free-fermionic generating set of exceptional type, with $\frustration{\vgens} = L(\Delta)$ and $n_\Delta = \abs{\vertices(\Delta)} = 2m+2$, hence $\rank(\vgens)=2m$. Let $q=\nullorbit(\vgens)$. 
If $m$ is even, let $w$ be a reference quadratic form in $\quadratic(\vgens)\neq\emptyset$. Also, consider some vector $u\in\Fn$.
\begin{enumerate}
    \item If $\symp{u}{\radorbit(\vgens)} = 0$, then there is some $u'\in u+V$ such that the orbits in $u+V$ are the parallel translates $u+O$ of the orbits $O$ in $V$.
    \item If $\symp{u}{\radorbit(\vgens)} = \F_2$,
    \begin{nestedcaseenum}
        \item If $m = 1\bmod 2$, there are two non-trivial orbits: $\QQ_w^{-1}(1)\cap(u+V)$ and $\QQ_w^{-1}(0)\cap(u+V)$.
        \item If $m = 0\bmod 2$, there is a single non-trivial orbit $u+V$.
    \end{nestedcaseenum}
\end{enumerate}
\end{prop}
Explicitly, we have the following description for the orbits a generating set of exceptional type with $(2m,q)$ free fermionic mapping:
\begin{lem}\label{lem:explicit_orbits_Fn_exceptional_case}
Consider a generating set $\pgens\subseteq\PP_n$ of exceptional type, with a $(2m,q)$ free fermionic mapping and orthogonal complement of rank $2\ell$.
The orbits are as follows:
\begin{enumerate}
    \item If $\majiso{\tilde{w}}\in\commalg$, there are $\lfloor (m+1)/2\rfloor$ non trivial orbits which are parallel translates of those in $\algclosure{\pgens}$:
    \begin{equation*}
            \majiso{\tilde{w}} (\MAJ_{2L-1}^{(2m)}\cup \MAJ_{2L}^{(2m)}\cup \MAJ_{2(m-L)-1}^{(2m)}
            \cup \MAJ_{2(m-L)}^{(2m)})\setprod\Cyc^{(2m,q)}
    \end{equation*}
    There are $2^{2\ell}$ such cases.
    \item If $\majiso{\tilde{w}} \simeq
    (\prod_{j=1}^{q}\gamma_{2m+2j-1}^{\delta_j})\majiso{v}$ with
    $\{\delta_j\}_{j=1}^q$ not all zero and $\majiso{v}\in\commalg$, there are
    $2$ orbits, $(\QQ_w^\gamma)^{-1}(0)\cap(\tilde{w}+V)$ and
    $(\QQ_w^\gamma)^{-1}(1)\cap(\tilde{w}+V)$.
    There are $2^{2\ell}(2^q-1)$ such cases.
    \item The $2^{2\ell+q}$ trivial orbits coming from the orthogonal complement/commutant.
\end{enumerate}
In total, there are $2^{2\ell}(\lfloor (m+1)/2\rfloor+1 + 2(2^q-1))$ orbits.
\end{lem}

This completes the classification of orbits over $\Fn$ or $\PP_n$ for all generating sets with connected frustration graph.

\subsection{The Algorithm}\label{sec:classification_and_algorithm_orbits}

As in the case of Pauli Lie algebras and transvection groups (Corollary~\ref{cor:algorithm_lie_algebras}), we now describe in the binary formalism the necessary computational steps required for three tasks regarding the orbits for a given Pauli generating set:
\begin{cor}\label{cor:algorithm_orbits}
Let $\pgens = \isolong{\vgens}\subseteq\PP_n$ be a sequence (or set) of binary vectors with a connected frustration graph $\graphG$.

If $\dim(\vgens^\perp)$ is a constant $\BigO(1)$, there is an algorithm which can return a list of labels the orbits for $\tvgroup{\vgens}$ over $\Fn$ in $\BigO(\max\{\abs{\vgens},2n\}^3)$-time.
Also, given such a list, it takes $\BigO(n^3)$-time to identify the specific orbit in which a Pauli $P$ belongs.

For any $\vgens$, it takes $\BigO(\max\{\abs{\vgens},2n\}^3)$-time to decide if two Pauli strings $P,P'\in\PP_n$ belong in the same orbit.
\end{cor}
See also Table~\ref{tab:orbit_tasks} for a summary of Corollary~\ref{cor:algorithm_orbits}.
We can highlight the main steps for classification, which take advantage of the results in the previous sections, namely: Corollary~\ref{cor:orbits_full_Fn_strictly_universal_case} and Lemmas~\ref{lem:orbits_full_Fn_quadratic_cases} for the quasi-universal cases, Theorem~\ref{thm:orbits_in_Fn_alg_ind_even_odd_cases_free_fermionic}, Lemmas \ref{lem:orbits_Fn_even_case} and~\ref{lem:orbits_Fn_odd_case}, and Proposition~\ref{prop:orbits_in_Fn_exceptional_case_free_fermionic} for the free-fermionic cases.
Using these results, we will need to compute symplectic products and quadratic forms, and solve systems of linear equations.

\begin{table}[t]
    \caption{The three basic tasks regarding the orbits for a given Pauli generating set $\pgens = \isolong{\vgens}$ with connected frustration graph $\graphG$ and their complexities.\\
    Orbit Intersection: check if $P,P'\in\PP_n$ are in the same orbit.
    Orbit Labelling: provide a list of labels for each orbit.
    Orbit Identification: given list of labels, find the orbit in which a given Pauli $P$ belongs.}
    \label{tab:orbit_tasks}
\centering
\setlength\extrarowheight{3pt}
\begin{tabular}{@{\hspace{1mm}} l
@{\hspace{4mm}} l
@{\hspace{1mm}}}
\hline\hline
\\[-5mm]
        Task & Complexity \\[0.5mm]
\hline
\\[-4mm]
        Orbit Intersection & $\BigO(\max\{\abs{\vgens},2n\}^3)$\\[1mm]
        Orbit Labelling & $\BigO(\max\{\abs{\vgens},2n\}^3)$ if $\dim(\vgens^\perp)=\BigO(1)$\\
                         & $\BigO(2^{\alpha n})$ if $\dim(\vgens^\perp)=\BigO(n)$ \\[1mm]
        Orbit Identification & $\BigO(n^2)$\\[1mm]
\hline\hline
\end{tabular}
\end{table}

\begin{proof}
Let us assume that one already knows the type of Lie algebra or transvection group via Corollary~\ref{cor:algorithm_lie_algebras}, so that we only need to evaluate the steps needed for the orbits.

Then, one needs to find a decomposition $\Fn = V\oplus W\oplus \bar{V}^\perp$ to obtain either a symplectic basis (as in Eq.~\eqref{eq:full_Fn_decomposition_symplectic_basis}) or a Majorana basis Eq.~\eqref{eq:full_Fn_decomposition_Majorana_odd_exceptional_case}, \eqref{eq:full_Fn_decomposition_Majorana_even_case}).
Both methods require solving systems of linear equations over $\BigO(\vgens)$ variables, hence take $\BigO(\max\{\abs{\vgens},2n\}^3)$-time.
Even if we assume a name for the Lie algebra or transvection group is given (via Algorithm~\ref{alg:classification_lie_algebras}), in order to get a suitable basis for $\Fn$, one still needs to solve systems of linear equations over $\vgens$ variables, so that the complexity does not depend purely on the dimension of the space $2n$.
In particular, this is needed for both orbits intersection and orbit labelling, since we require explicit (symplectic or Majorana) bases.
We discuss all cases individually.

In the quasi-universal cases without any invariant quadratic forms, the orbits are uniquely labelled by (at most) three terms:
\begin{enumerate}\label{enum:general_orbit_labels}
    \item\label{enum:general_orbit_labels:affine-representative}
    An $\F_2$-vector $\tilde{w}\in W\oplus\bar{V}^\perp$ of size $r+2\ell$, such that $u\in \tilde{w}+V$. Equivalently, for a given symplectic basis of $V^\perp$, the components of this vector are the symplectic products $\symp{e_i}{u}$, $\symp{f_i}{u}$ and $\symp{h_j}{u}$ for $i\in [\ell]$ and $j\in [r]$.
    \item\label{enum:general_orbit_labels:commutes-with-V}
    A boolean, regarding whether $\tilde{w}$ is in $\vgens^\perp$ or not, hence $\symp{u}{V} = 0$ or $\symp{u}{V}=\F_2$
    \item\label{enum:general_orbit_labels:trivial-or-nontrivial}
    If the above is zero, an additional field which says which determines whether $u$ belongs to a trivial ($u\in \vgens^\perp \iff \symp{u}{V}=0$) or non-trivial orbit in $u+V$ ($u\not\in\vgens^\perp \iff \symp{u}{V}=\F_2$).
\end{enumerate}
The labels \ref{enum:general_orbit_labels:affine-representative}--\ref{enum:general_orbit_labels:trivial-or-nontrivial} provide a list of size $2\cdot 2^{2\ell+r} = 2\cdot \dim(\vgens^\perp)$ whose elements have $\BigO(n)$ entries, hence outputting this list runs in $\BigO(n)$ time only if $2\ell+r$ is constant (does not grow with $n$).
If instead $2\ell+r = \BigO(n)$, then up to polynomial factors this step already takes exponential time $\BigO(2^n)$ due to the size of the list.
It takes $\BigO(n^2)$ to compute all symplectic products of a given vector with $V$ or $V^\perp$, which identify the orbit label for a given Pauli string $P=\iso{v}\in\PP_n$ or vector $v\in\Fn$.

For arbitrary $\dim(\vgens)$, the symplectic products are necessary to decide orbit intersection of two Pauli strings $P,P'$, it also takes $\BigO(n^2)$ time to decide orbit intersection.
Namely, for $u\neq u'\in\Fn$, one needs to check the values of $\symp{u}{\vecbas{V}}$ and $\symp{u'}{\vecbas{V}}$, as well as the symplectic products with the symplectic basis of $V^\perp$, $\symp{u}{\vecbas{V^\perp}}$ and $\symp{u}{\vecbas{V^\perp}}$.

For the quasi-universal case with invariant quadratic forms, one also consider an additional field to the ones above:
\begin{enumerate}[resume]
    \item\label{enum:general_orbit_labels:quadratic-form-value}
    The $\F_2$ value of the chosen invariant quadratic form $\QQ_w(u) \in\{0,1\}$.
\end{enumerate}
It takes $\BigO(n^2)$ time to evaluate the quadratic form, for a given Pauli string, which provides the orbit label.
This test also extends to orbit intersection.

Finally, in the free-fermionic case, we use the labels \ref{enum:general_orbit_labels:affine-representative}--\ref{enum:general_orbit_labels:trivial-or-nontrivial}, as well as additional ones depending on the Majorana length over the $2m$ logical modes.
In this case one needs to additionally write the vectors in the Majorana basis.
This can be done in $\BigO(n^2)$ time by evaluating symplectic products of the given vectors with the Majorana basis (see Eq.~\eqref{eq:majorana_components_via_symplectic_products}).
Given the components in this basis, it takes $\BigO(n)$ time to check the non-zero components, hence find the Majorana length.

In the even and odd case, or the exceptional case when $m$ is odd, we also use the label \ref{enum:general_orbit_labels:quadratic-form-value} given by value of the chosen invariant quadratic form.
The additional Majorana-length label is chosen as follows:
\begin{orbitlabelcaseenum}
    \item\label{enum:general_orbit_labels:majorana-length-even}
    In the even case, if $\symp{u}{\radorbit(\vgens)} = 0$, use the integer value of the Majorana length over the first $m$ modes,
    $\majL_m(u)\in\{1,\ldots, 2m{-}1\}$.
    \item\label{enum:general_orbit_labels:majorana-length-odd}
    In the odd case, use $\lceil \majL_m(u)/2\rceil\in\{1,\ldots,m\}$,
    the integer part of half the Majorana length over the first $m$ modes.
    \item\label{enum:general_orbit_labels:majorana-length-exceptional}
    In the exceptional case, use the integer part of the half Majorana length over the first $m$ modes, up to $2m$, given by
    \begin{gather*}
    \min\{\lceil (2m-\majL_m(u))/2\rceil, \lceil \majL_m(u)/2\rceil\}\\
    \in\{1,\ldots, \lfloor (m{+}1)/2\rfloor\}.
    \end{gather*}
\end{orbitlabelcaseenum}
Again, the list is of size at most $m\cdot 2^{2\ell+r}$, with $m\leq n$, hence for constant $\dim(\vgens^\perp)$, it takes at most $\BigO(n)$ time to output this list.
Also, in the exceptional case, one uses the value of the invariant quadratic form, field (d), only when it is defined, i.e. for $m=0\bmod 2$.

We can then also use this information to check for orbit intersection, independently of $\dim(\vgens^\perp)$.
This concludes the proof.
\end{proof}

\ManuscriptPart{Back Matter}{part:appendices}

\section*{Methodology}
Our work is a theory contribution, which neither presents
experimental results nor proposes any experiments.
Even though we present algorithms to address
a range of questions, none
of our presented results, examples, or applications rely on explicit
computations beyond
what would be possible by pen and pencil.

However, during the research leading to our results,
we used computational tools. In particular, computations were carried out using
the computer algebra system \textsc{Magma} \cite{Bosma_Cannon_Playoust_1997},
the programming language \textsc{Julia} \cite{Bezanson_Edelman_Karpinski_Shah_2017},
and the programming language \textsc{Python} \cite{Rossum_Boer_1991}.
Some of the corresponding programs were written with assistance from
large language models. Specifically, we have used OpenAI's \textsc{ChatGPT} and \textsc{Codex},
Google's \textsc{Gemini}, and Anthropic's \textsc{Claude Code}.

These large language models were also used as exploratory aids for identifying
relevant results.
They further assisted with exploratory computational checks.
They were also used to draft, reformulate, revise, polish, and cross-check
results and text at various stages of this work.

The conception, research direction, analysis of the literature, mathematical
development, and physical interpretation remain entirely those of the authors.
The same applies to the selection and presentation of the results, examples, and
applications, as well as to the final judgement throughout the work.
This work has been thoroughly checked by the authors.

\begin{acknowledgments}
We thank Gerard Aguilar, Simon Cichy, Jens Eisert and Lennart Bittel for useful discussions on \cite{Aguilar_Cichy_Eisert_Bittel_2024}.
R.G. and R.Z. thank Sahil Ugale for stimulating discussions on Pauli propagation \cite{Ugale_Master}.
R.Z. thanks Thomas Schulte-Herbrüggen, Zoltán Zimborás, and Michael Keyl for many discussions and insights
on symmetries of controlled quantum systems.

R.G. and R.Z. acknowledge 
funding by the Deutsche Forschungsgemeinschaft (DFG, German Research Foundation) under Germany's Excellence Strategy – Cluster of Excellence
Matter and Light for Quantum Computing (ML4Q) \href{https://gepris.dfg.de/gepris/projekt/390534769}{EXC 2004/1 – 390534769}
and by the Horizon Europe programme HORIZON-CL4-2022-QUANTUM-02-SGA via the project
\href{https://doi.org/10.3030/101113690}{101113690} (PASQuanS2.1). P.H. acknowledges funding from the Alexander von Humboldt Foundation and the Natural Sciences and Engineering Research Council of Canada (NSERC).
\end{acknowledgments}

\appendix

\section{Basic Notational Conventions \label{app:notation}}

We collect and define here basic notation so that
we can streamline the main text.
Throughout the manuscript, we sometimes use the shortcut
$$\{O_j\}_{j=1}^m:=\{O_j \text{ for } j \in \{1,\ldots,m\}   \}.$$
Also, for a function $f\colon X\to Y$ and $A\subseteq X$, we denote with $f^{-1}(A)$

Let $\F^d$ denote the vector space of dimension $d$ over the the
field $\F$.
Let $\mat{\F}{d}$ and $\F^{d \times d}$ each denote the set of $d \times d$ matrices
with entries from the field $\F$.
As this is usually clear from the context,
we do not distinguish
the zero vector $0 \in \F^{d}$ and the zero matrix
$0 \in \F^{d \times d}$
from the zero field element
$0 \in \F$.
We use $\id_d$ to describe the identity matrix. But we usually
use $\In$ for the identity matrix in a $n$-qubit system,
where $I$ is the $2\times 2$ identity matrix.

Let
$\matalg$ denote a set that acts on a set $\calX$ or one of its elements $x \in \calX$.
For example, both $\calA$ and $\calX$ can be sets of matrices
or sets of group elements where the action is given by multiplication.
We recall the notions
of the left and the right orbit
\begin{subequations}
\label{eq:orbit}
\begin{alignat}{2}
\calA \cdot x &= \calA x &&:=\{A \cdot x \,\text{ for all }\, A \in \calA\},
\label{eq:left:orbit} \\
 x \cdot \calA &=  x \calA &&:=\{ x \cdot A \,\text{ for all }\, A \in \calA\}.
\label{eq:right::orbit}
\end{alignat}
\end{subequations}
The set of all (left) orbits is then given by
\begin{equation}\label{eq:orbits}
\orb(\calX) = \orb_{\calA}(\calX) := \{  \calA \cdot x \,\text{ for }\, x \in \calX  \}.
\end{equation}
The stabilizer of a subset $\calY$ of $\calX$ is defined as
\begin{equation}
\stab_{\calA}(\calY) :=\{A \in \calA \,|\, A \cdot y \in \calY \,\text{ for all }\, y \in  \calY \}
\label{eq:stab:set}
\end{equation}
and we use the shortcut $\stab_{\calA}(x):=\stab_{\calA}(\{x\})$.
Moreover, the invariant elements of $\calX$ under $\calA$, or fixed points, are denoted by
\begin{equation}
\calX^{\calA}:= \{ x\in\calX \,|\, A \cdot x = x \,\text{ for all }\, A \in \calA\}.
\label{eq:invariant:elements}
\end{equation}
For a group $G$ and a subgroup $H$, the right coset is
\begin{equation}\label{eq:right:coset}
G/H:=\{gH \,\text{ for }\, g \in  G\},
\end{equation}
which is a quotient group if $H$ is a normal subgroup,
i.e., if $ghg^{-1} \in H$ for all $g\in G$ and $h\in H$.
We also recall the following standard consequence of the Second Isomorphism Theorem and the Correspondence Theorem.
If $H$ is a subgroup of $G$ and $N\subseteq G$ is normal, then the image of $H$ in $G/N$ is $NH/N\cong H/(H\cap N)$.
The following criterion records when the image and kernel determine a subgroup.
\begin{lem}[\cite{rotman2012introduction}]\label{lem:maximality_of_quotient_group}
Let $G$ be a group with normal subgroup $N$, let $\pi:G\to G/N$ be the quotient map, and let $H\subseteq H'$ be two subgroups of $G$.
If $\pi(H)=\pi(H')$ and $H\cap N = H'\cap N$, then $H = H'$.
\end{lem}
\begin{proof}
By the correspondence theorem from \cite[Theorem~2.28]{rotman2012introduction}, the equality $\pi(H)=\pi(H')$ implies that the subgroups $NH$ and $NH'$ coincide, since both contain $N$.

Now let $x\in H' \subseteq NH' = NH$. Then, there exists some $n\in N$ and $h\in H$ such that $x = nh$.
This implies $n = xh^{-1}\in H'$ since $h\in H\subseteq H'$ and thus $n\in N\cap H' = N\cap H$.
Going back, this implies $x = nh$ is in $H$, hence $H'\subseteq H$ and finally $H=H'$.
\end{proof}
In the context of normal subgroups, we also recall the definition of short exact sequence as a sequence of groups and group homorphisms. Let $A$, $B$ and $C$ be groups and $1$ the trivial group, then the short sequence:
\begin{equation}
    1 \to A \stackrel{f}{\to} B \stackrel{g}{\to} C \to 1
\end{equation}
is said to be exact if $A$ is homomorphic to $B$ via $f$ with kernel $1$ and $B$ is homomorphic via $g$ to $C$ with kernel $A$.
Hence, by viewing $A$ as embedded in $B$ via $f$, we have that $B/A \cong C$, hence $A$ is a normal subgroup of $B$ and the quotient of $B$ by $A$ results in $C$. 

For $G$ acting on $\calX$,
the stabilizer subgroup $G_x$ of an element $x\in \calX$ is
\begin{equation}
G_x := \stab_{G}(x) = \{ g \in G \,|\, g\cdot x=x\}.
\label{eq:stabilizer:subgroup}
\end{equation}
Let both $\calA$ and $\calB$ be now either sets of matrices
or sets of group elements. We recall the centralizer
\begin{gather}
\cent_{\calA}(\calB)  := \{ C \in \calA \, |\, C B = B C \,\text{ for all }\, B \in \calB\}
\label{eq:cent}
\intertext{and the center}
\ZZ(\calA) := \cent_{\calA}(\calA).
\label{eq:center}
\end{gather}
Given a set of matrices $ \calA \subseteq \Cd$, the commutant is given by
\begin{equation}
\commutant(\calA) = \cent_{\Cd}(\calA).
\label{eq:commutant}
\end{equation}

Throughout the manuscript, $r$ denotes the radical dimension or nullity parameter,
while $k$ is reserved for the generic blown-up path parameter in labels such as $\graphP_{k,n_1}$.

\section{Pathological Transvections\label{app:pathological}}

This appendix characterizes pathological transvections
(see Definition~\ref{def:pathological}) by proving Lemma~\ref{lem:pathological}.
This is done in a slightly more general framework and
we recall some additional notation.
The \emph{order} $\abs{g}$ of an element $g$ of a finite group
$G$ is the smallest positive integer such that $g^{\abs{g}} = \one$ is equal to the identity element $\one$ of $G$.
A subset $D$ of a finite group $G$ is called a \emph{set of $3$-transpositions of $G$}
if $D$ generates $G$,
$D = \{ g^{-1} d g \text{ for } d\in D, g\in G\}$,
$d^2 = \one$ for $d \in D$, and
$\abs{d_1 d_2} \in \{1,2,3\}$ for $d_j \in D$ \cite{Fischer1971,Aschbacher1997}. In this work, we limit us to finite groups and we explicitly allow
the identity element $\one$ of $G$ to be contained in $D$. In particular, the symplectic transvections in $\Sp(2n,\F_2)$ are a set of $3$-transpositions
of $\Sp(2n,\F_2)$ \cite{Fischer1971,Aschbacher1997}.

Given a subset $E \subseteq D$ of the set of $3$-transpositions,
it generates a subgroup $G_E$ of $G$. Clearly, $G_E \cap D$
is a set of $3$-transpositions of $G_E$ \cite{Fischer1971,Aschbacher1997}.
Following Definition~\ref{def:pathological}, a pathological $3$-transposition $d \in G_E \cap D$
fulfills the properties that $d \neq \one$, $d \notin E$, and $d$ is not conjugated
to an element of $E$, i.e., there are no $e \in E$ and $g \in G_E$ with
$d = g^{-1} e g$.
We now prove that each pathological element of $G_E$ is
contained in the center $\ZZ(G_E)$ of $G_E$. We now assume that $d \in G_E \cap D$
is pathological and $e \in E$. We consider the different cases $\abs{de} \in \{1,2,3\}$.

First we assume $\abs{de}=3$ which implies
$(de) d (de)^{-1} = deded = e$ by applying  $(de)^3=\one$ and $e^2 = \one$.
But now $d$ and $e$ are conjugated which is a contradiction to $d$ being pathological.
Second we assume $\abs{de}=1$. We conclude that $de = \one$ and $d = e$ which again
leads to a contradiction. Consequently, $\abs{de}=2$ and $(de)^2=\one$ for all $e\in E$.
This then implies that $de = (de)^{-1} = ed$. We infer that $d$ commutes with all generators
$e \in E$ and finally that  $d \in \ZZ(G_E)$.

We now prove that a $3$-transposition $d \in G_E \cap D$ with
$d \neq \one$ and  $d \notin E$ is pathological if and only if $d \in \ZZ(G_E)$.
We have just proved that the condition $d \in \ZZ(G_E)$ is necessary. But an element
$d \in \ZZ(G_E)$ with $d \neq \one$ and  $d \notin E$ can clearly not be conjugated to an element of $E$,
which proves that the condition $d \in \ZZ(G_E)$ is sufficient.

\section{Binary Quadratic Forms}\label{app:binary-quadratic-forms}

This appendix records the binary quadratic-form background used in the
main text. First, we recall the normal forms and type classification for
possibly degenerate binary quadratic forms. Second, we record the version of
Witt's extension theorem needed for binary quadratic spaces, together with its
one-dimensional transitivity consequence. We also fix the orthogonal-group
notation and the linear and affine extension consequences used in the
classification of affine spaces of quadratic forms.

\subsection{Normal Forms and Type Classification}

We collect here the background and proof for
Proposition~\ref{prop:Quadratic_Form_Isomorphism_Classes}. The classification
of quadratic forms in characteristic two is classical, but the standard
references emphasize different aspects. Early finite-field normal forms
already appear in \cite{Dickson_1901}. The algebraic normal-form
treatment in \cite{Bourbaki1959} and the Arf-invariant formulation,
including the role of the radical, in \cite{Browder_1972} give the forms
used here. For further context, we refer to
\cite{Lidl_Niederreiter_1997,Meyer_1972}.

\begin{lem}[Binary quadratic normal forms]\label{lem:binary_quadratic_normal_forms}
Let $\QQ$ be a quadratic form from a finite-dimensional binary vector
space $V$ to $\F_2$, let $B$ be its associated alternating bilinear form,
and let $R=\rad(V)$. A symplectic basis of $V$ is written as
$\{e_i,f_i\}_{i=1}^m\cup\{h_j\}_{j=1}^r$, where the $h_j$ span $R$
(and hence $\dim(V)=2m+r$ and $\dim(R)=r$).
Then:
\begin{enumerate}
    \item The restriction $\QQ|_R$ is linear. Thus either $\QQ|_R=0$, or
    $\QQ(R)=\F_2$.
    \item If $\QQ|_R\neq0$, then in a suitable symplectic basis,
    \begin{equation}\label{eq:binary_quadratic_type_zero_normal_form}
        \QQ(v)=
        \sum_{i=1}^m \component{v}{i}\component{v}{m+i}
        +\component{v}{2m+1}^2.
    \end{equation}
    Its isomorphism class is determined by $\dim(V)$ and $\dim(R)$.
    \item If $\QQ|_R=0$, then $\QQ$ is constant on cosets of $R$ and
    therefore defines an induced non-degenerate quadratic form on $V/R$.
    Its isomorphism class is determined by
    $\dim(V)$, $\dim(R)$, and the Arf invariant of this quotient form.
    In suitable symplectic bases, the two possible Arf values give the
    representatives
    \begin{equation}\label{eq:binary_quadratic_split_normal_form}
        \QQ(v)=\sum_{i=1}^m \component{v}{i}\component{v}{m+i}
    \end{equation}
    and
    \begin{equation}\label{eq:binary_quadratic_nonsplit_normal_form}
        \QQ(v)=
        \sum_{i=1}^m \component{v}{i}\component{v}{m+i}
        +\component{v}{m}^2+\component{v}{2m}^2.
    \end{equation}
\end{enumerate}
Moreover, the three representatives in
Eqs.~\eqref{eq:binary_quadratic_type_zero_normal_form}--\eqref{eq:binary_quadratic_nonsplit_normal_form}
satisfy $\abs{\QQ^{-1}(0)}=\abs{\QQ^{-1}(1)}$,
$\abs{\QQ^{-1}(0)}>\abs{\QQ^{-1}(1)}$, and
$\abs{\QQ^{-1}(0)}<\abs{\QQ^{-1}(1)}$, respectively.
\end{lem}

The assertions in Lemma~\ref{lem:binary_quadratic_normal_forms} are the
binary specialization of several standard formulations. The radical
linearity in (a) is the binary case of
Bourbaki~\cite[Ch.~IX, \S 6, Ex.~27(a)]{Bourbaki1959}; Browder also
separates the cases $\QQ|_R=0$ and $\QQ|_R\neq0$ in
\cite[Ch.~III, \S 1, Thm.~III.1.14]{Browder_1972}. The nonzero-radical
normal form and its uniqueness statement in (b) are the binary case of
Bourbaki~\cite[Ch.~IX, \S 6, Ex.~27(b)]{Bourbaki1959}, are stated in
Arf-invariant language by
\cite[Ch.~III, \S 1, Thm.~III.1.14]{Browder_1972}, and are compatible
with the classical normal forms over finite fields in
\cite[Ch.~VIII, \S 199, pp.~197--199]{Dickson_1901}. The quotient
classification in (c) corresponds to the two even-dimensional
non-degenerate cases in
Bourbaki~\cite[Ch.~IX, \S 6, Ex.~27(e)]{Bourbaki1959}: for the
$2m$-dimensional quotient forms represented in
Eqs.~\eqref{eq:binary_quadratic_split_normal_form}
and~\eqref{eq:binary_quadratic_nonsplit_normal_form}, the maximum
dimension of a totally singular subspace, that is, a subspace on which
the quadratic form vanishes identically, is respectively $m$ and $m-1$.
Browder gives the
two-dimensional representatives associated with a symplectic pair, their
reduction by Arf invariant, and the
radical extension in
\cite[Ch.~III, \S 1, Lem.~III.1.5, Prop.~III.1.7, Thm.~III.1.14]{Browder_1972}.
The cardinality comparisons between $\QQ^{-1}(0)$ and $\QQ^{-1}(1)$ agree
with Browder's majority criterion in the
non-degenerate case
\cite[Ch.~III, \S 1, Prop.~III.1.8]{Browder_1972}, and also follow by
directly counting the displayed normal forms.
For completeness, and to keep the binary case self-contained, we give a
direct proof.

\begin{proof}
By the standard symplectic Gram--Schmidt argument, equivalently the
Witt-decomposition argument used in Lemma~\ref{lem:symplectic_basis},
choose a symplectic basis
$\{e_i,f_i\}_{i=1}^m\cup\{h_j\}_{j=1}^r$ for $V$, where the $h_j$ span
$R$. Since $B$ vanishes on $R$, we have for all $u,u'\in R$ that
\begin{equation*}
    \QQ(u+u')=\QQ(u)+\QQ(u')+B(u,u')
    =\QQ(u)+\QQ(u').
\end{equation*}
Thus $\QQ|_R$ is linear. Since its codomain is $\F_2$, its image is
either $\{0\}$ or all of $\F_2$.

Assume first that $\QQ|_R\neq0$. Choose $h_1\in R$ with $\QQ(h_1)=1$,
and choose $h_2,\ldots,h_r$ as a basis of $\ker(\QQ|_R)$. Set
\begin{equation*}
    e_i'=e_i+\QQ(e_i)h_1,\qquad
    f_i'=f_i+\QQ(f_i)h_1.
\end{equation*}
The replacement of $e_i,f_i$ by $e_i',f_i'$ preserves all symplectic
products because $h_1\in R$. Moreover, each $e_i'$ and $f_i'$ is
isotropic for $\QQ$, since
$\QQ(x+\QQ(x)h_1)=\QQ(x)+\QQ(x)\QQ(h_1)=0$ for
$x\in\{e_i,f_i\}$. Thus, in the resulting symplectic basis, all
$e_i',f_i'$ and all $h_j$ with $j>1$ have $\QQ$-value zero, while
$\QQ(h_1)=1$. For
$v=\sum_i x_i e_i'+\sum_i y_i f_i'+\sum_j z_j h_j$, we repeatedly use
$\QQ(u+u')=\QQ(u)+\QQ(u')+B(u,u')$ and
$\QQ(a u)=a^2\QQ(u)$. This gives a contribution $z_1^2$ from the
values of the basis vectors. The only nonzero cross-terms come from
the symplectic pairs
$B(e_i',f_i')=1$, so they contribute $\sum_i x_i y_i$. Hence
\begin{equation*}
    \QQ(\sum_i x_i e_i'+\sum_i y_i f_i'+\sum_j z_j h_j)
    =\sum_i x_i y_i+z_1^2,
\end{equation*}
which is the normal form in
Eq.~\eqref{eq:binary_quadratic_type_zero_normal_form}.
This also shows that, in this case, the isomorphism class is fixed by
$m$ and $r$, equivalently by $\dim(V)=2m+r$ and $\dim(R)=r$.

Assume now that $\QQ|_R=0$. Then $\QQ(v+h)=\QQ(v)$ for all $h\in R$, so
$\QQ$ is constant on cosets of $R$ and defines an induced quadratic form
$\bar{\QQ}$ on $V/R$. The associated alternating bilinear form on $V/R$
is non-degenerate. Choose an
orthogonal symplectic basis of $V/R$. It remains
to classify the restrictions to the symplectic pairs. For a single
symplectic pair $e,f$, the identity
$\QQ(e+f)=\QQ(e)+\QQ(f)+1$ leaves two possibilities up to replacing one
of $e$ and $f$ by $e+f$. If $\QQ(e)$ and $\QQ(f)$ are mixed, this
replacement makes both values zero; if both values are one, then all
three nonzero vectors in $\SpanS[\F_2]{\{e,f\}}$ have value one. Thus
the restriction to $\SpanS[\F_2]{\{e,f\}}$ is isomorphic to exactly one
of the two pair forms
\begin{equation*}
    q_0(e)=q_0(f)=0,\qquad q_1(e)=q_1(f)=1,
\end{equation*}
with $q_0(e+f)=q_1(e+f)=1$. Hence any non-degenerate form is an
orthogonal direct sum of such pair forms. Two $q_1$ pairs cancel:
if $a,b$ and $c,d$ are symplectic pairs with $q(a)=q(b)=q(c)=q(d)=1$,
then
\begin{equation*}
    e_1=a+c,\, f_1=b+c,\,
    e_2=a+b+d,\, f_2=a+b+c+d
\end{equation*}
are two orthogonal symplectic pairs with
$q(e_1)=q(f_1)=q(e_2)=q(f_2)=0$. Thus $q_1\oplus q_1\cong q_0\oplus q_0$.
Consequently the non-degenerate form is isomorphic either to $m q_0$ or
to $q_1\oplus(m-1)q_0$, according to the parity of the number of
$q_1$ pairs. For this orthogonal symplectic basis, the Arf invariant is
$\sum_i\QQ(e_i)\QQ(f_i)$. Each $q_0$ pair contributes $0$ to this sum,
whereas each $q_1$ pair contributes $1$. Thus the Arf invariant is the
parity of the number of $q_1$ pairs. Pulling these two
normal forms back from $V/R$ gives the representatives
in Eqs.~\eqref{eq:binary_quadratic_split_normal_form}
and~\eqref{eq:binary_quadratic_nonsplit_normal_form}, and also
shows that the isomorphism class is determined by $\dim(V)$, $\dim(R)$,
and the Arf invariant.

It remains to compare the sets $\QQ^{-1}(0)$ and $\QQ^{-1}(1)$. For the
normal form in Eq.~\eqref{eq:binary_quadratic_type_zero_normal_form},
translation by $h_1$ changes the value of $\QQ$, since
$\QQ(v+h_1)=\QQ(v)+1$. Thus this translation gives a bijection between
$\QQ^{-1}(0)$ and $\QQ^{-1}(1)$, so the two sets have the same size. In
the case $\QQ|_R=0$, the radical variables do not change $\QQ$, so the
cardinalities are those of $\bar{\QQ}$ multiplied by $\abs{R}$. It is
therefore enough to count the non-degenerate pair forms. On the four
vectors $0$, $e$, $f$, $e+f$ of one symplectic pair, $q_0$ takes the values
$0$, $0$, $0$, $1$, while $q_1$ takes the values $0$, $1$, $1$, $1$. If
$\Phi=\Phi'\oplus\Phi''$ is an orthogonal direct
sum of two quadratic forms, then
\begin{align*}
    \abs{\Phi^{-1}(0)}-\abs{\Phi^{-1}(1)}
    &=
    (\abs{(\Phi')^{-1}(0)}-\abs{(\Phi')^{-1}(1)})\\
    &\quad\times
    (\abs{(\Phi'')^{-1}(0)}-\abs{(\Phi'')^{-1}(1)}),
\end{align*}
because zeros of $\Phi$ occur when $\Phi'$ and $\Phi''$ have equal
values, while ones occur when they have unequal values.
Hence $m q_0$ has more zeros than ones, while $q_1\oplus(m-1)q_0$ has
fewer zeros than ones.
\end{proof}

We can now recover Proposition~\ref{prop:Quadratic_Form_Isomorphism_Classes}
by translating Lemma~\ref{lem:binary_quadratic_normal_forms} into the
type notation and canonical representatives used in the main text.

\begin{proof}[Proof of Proposition~\ref{prop:Quadratic_Form_Isomorphism_Classes}]
Let $B=\genbilempty_\QQ$ and $R=\rad(V)$. Based on
Lemma~\ref{lem:symplectic_basis}, we may choose a symplectic basis
$\{e_i,f_i\}_{i=1}^m\cup\{h_j\}_{j=1}^r$ of $V$, where the $h_j$ span
$R$. By Lemma~\ref{lem:binary_quadratic_normal_forms}, either
$\QQ|_R=0$ or $\QQ(R)=\F_2$.

Assume first that $\QQ|_R\neq0$. Then
Eq.~\eqref{eq:binary_quadratic_type_zero_normal_form} gives the type $0$
representative.

Assume now that $\QQ|_R=0$. Then $\QQ$ is constant on cosets of $R$ and
defines an induced quadratic form $\bar{\QQ}$ on $V/R$. The associated
alternating bilinear form on $V/R$ is non-degenerate.
Lemma~\ref{lem:binary_quadratic_normal_forms}
gives the two classes in
Eqs.~\eqref{eq:binary_quadratic_split_normal_form}
and~\eqref{eq:binary_quadratic_nonsplit_normal_form}, distinguished by the
Arf invariant. Pulling these representatives back along the quotient map
$V\to V/R$ and adjoining the radical basis
gives the type $+$ and type $-$ representatives. In particular, in the
case $\QQ|_R=0$, the Arf invariant of $\QQ$ is the Arf invariant of the
induced non-degenerate form on $V/R$, and is therefore independent of
the chosen symplectic basis.

The cardinality comparison is the final assertion of
Lemma~\ref{lem:binary_quadratic_normal_forms}.
\end{proof}

\subsection{Witt Extension and Orthogonal Groups}\label{app:witt-extension-f2}

We first record the extension result needed to pass from isometries of subspaces
to ambient automorphisms, and then fix the orthogonal-group notation used in the
main text.
We only need the version of Witt's extension theorem
for binary quadratic spaces whose
associated alternating bilinear form is non-degenerate. In this setting,
the extension theorem for isometries between subspaces is given by
\cite[Ch.~IX, \S 4, no.~3, Thm.~1]{Bourbaki1959}:

\begin{prop}[Witt extension over $\F_2$]\label{prop:witt-extension-f2}
Let $\QQ$ be a quadratic form on a finite-dimensional binary vector space $V$,
and assume that its associated alternating bilinear form is non-degenerate. If
$U,U'\subseteq V$ are subspaces and $\phi$ is an isometry
from $U$ to $U'$, i.e.,
\[
    \QQ(\phi u)=\QQ(u)
    \qquad\text{for all }u\in U,
\]
then $\phi$ extends to an automorphism $\tilde{\phi}:V\to V$ satisfying
$\QQ(\tilde{\phi}v)=\QQ(v)$ for all $v\in V$.
\end{prop}

The alternating-form version follows from Proposition~\ref{prop:witt-extension-f2}
by choosing a quadratic refinement adapted to the given subspace isometry.

\begin{cor}[Symplectic extension over $\F_2$]\label{cor:symplectic-extension-f2}
Let $V$ be a finite-dimensional binary vector space with non-degenerate
alternating form $\genbilempty$.
If $U,U'\subseteq V$ are subspaces and $\phi\colon U\to U'$ is an isometry
for the restricted alternating form, i.e.,
\[
    \genbil{\phi u}{\phi v}=\genbil{u}{v}
    \qquad\text{for all }u,v\in U,
\]
then $\phi$ extends to an automorphism $\tilde{\phi}\in\Sp(V)$.
\end{cor}
\begin{proof}
We prove the corollary by reducing it to Proposition~\ref{prop:witt-extension-f2}.
Let $\QQ'$ be any quadratic form whose associated alternating form is
$\genbilempty$.

First, we modify $\QQ'$ by a linear form so that $\phi$ also preserves the
quadratic form on $U$.
Define
\[
    \lambda(u)=\QQ'(\phi u)+\QQ'(u)
\]
for $u\in U$.
This function is linear on $U$ since the quadratic identity for
$\QQ'$ and $u,v\in U$ gives
\[
    \lambda(u+v)+\lambda(u)+\lambda(v)
    =
    \genbil{\phi u}{\phi v}+\genbil{u}{v}
    =
    0,
\]
which follows as $\phi$ preserves $\genbilempty$.
We now turn this defect into a linear correction of $\QQ'$.
Let
\[
    S=\{u+\phi u\mid u\in U\}\subseteq V
\]
and define $\mu\colon S\to\F_2$ by $\mu(u+\phi u)=\lambda(u)$.
To see that $\mu$ is well-defined, suppose that the same element of $S$ is
written in two ways, say $u+\phi u=u'+\phi u'$ for $u,u'\in U$.
We must show that these two representatives give the same value, i.e.,
$\lambda(u)=\lambda(u')$.
Set $w=u+u'\in U$.
Then $w+\phi w=0$, so $\phi w=w$, and hence
\[
    \lambda(w)=\QQ'(\phi w)+\QQ'(w)=\QQ'(w)+\QQ'(w)=0.
\]
Since $\lambda$ is linear, $\lambda(w)=\lambda(u)+\lambda(u')=0$, and therefore
$\lambda(u)=\lambda(u')$.

Choose a linear extension $\tilde{\lambda}$ of $\mu$ to $V$ and set
\[
    \QQ(v)=\QQ'(v)+\tilde{\lambda}(v).
\]
The associated alternating form is unchanged:
\[
    \QQ(v+v')+\QQ(v)+\QQ(v')
    =
    \QQ'(v+v')+\QQ'(v)+\QQ'(v')
\]
for $v,v'\in V$, because the additional contribution
$\tilde{\lambda}(v+v')+\tilde{\lambda}(v)+\tilde{\lambda}(v')$ vanishes by
linearity as shown above.
Thus $\QQ$ also has associated alternating form $\genbilempty$.
Moreover, for $u\in U$, the correction gives
\begin{align*}
    \QQ(\phi u)+\QQ(u)
    &=
    \QQ'(\phi u)+\QQ'(u)
    +\tilde{\lambda}(\phi u)+\tilde{\lambda}(u)\\
    &=
    \lambda(u)+\tilde{\lambda}(\phi u+u)\\
    &=
    \lambda(u)+\mu(u+\phi u)
    =
    \lambda(u)+\lambda(u)
    =
    0.
\end{align*}
Thus $\phi$ is an isometry for the restricted quadratic forms.

Second, we apply Proposition~\ref{prop:witt-extension-f2} to this adapted
quadratic form $\QQ$.
By Proposition~\ref{prop:witt-extension-f2}, $\phi$ extends to an automorphism
$\tilde{\phi}$ preserving $\QQ$ on $V$.

Finally, because the associated alternating form of $\QQ$ is
$\genbilempty$, every automorphism preserving $\QQ$ also preserves
$\genbilempty$.
Therefore $\tilde{\phi}\in\Sp(V)$.
\end{proof}

For a quadratic form $\QQ$ on $V$,
we obtain from \cite[Ch.~IX, \S 6, no.~2]{Bourbaki1959} the definition
of the orthogonal group
\[
    \lieO(\QQ)=\{\activemap\in\GL(V) \mid \QQ(\activemap v)=\QQ(v)
    \text{ for all }v\in V\}.
\]
Thus a fixed quadratic form $\QQ$ determines its full orthogonal group as this
stabilizer. A different representative gives
a conjugate, and therefore isomorphic, group, i.e.,
\[
    \lieO(\QQ')=\activemap\lieO(\QQ)\activemap^{-1}
\]
if
$\QQ'=\QQ\circ\activemap^{-1}$.
We next apply Witt's extension theorem to the one-dimensional subspaces
spanned by two nonzero vectors of the same quadratic value.

\begin{cor}[Transitivity on vectors of fixed value]\label{cor:witt-transitivity-fixed-value}
Let $\QQ$ be a quadratic form on a finite-dimensional binary vector space $V$,
and assume that its associated alternating bilinear form is non-degenerate.
If $v,v'\in V\setminus\{0\}$ satisfy $\QQ(v)=\QQ(v')$, then there exists
$\activemap\in\lieO(\QQ)$ such that $\activemap v=v'$.
\end{cor}
\begin{proof}
The unique linear map from the vector space $\SpanS[\F_2]{\{v\}}$ to
$\SpanS[\F_2]{\{v'\}}$ sending $v$ to $v'$ is
an isometry of one-dimensional quadratic spaces. Proposition~\ref{prop:witt-extension-f2}
extends it to an element of $\lieO(\QQ)$.
\end{proof}

For a quadratic form $\QQ$ on $V$, let $\AO[\QQ]$ denote the affine
orthogonal group of $\QQ$, i.e., the group of affine maps
$\alpha$ from $V$ to $V$ of the form $\alpha(v)=b+\phi v$, with
$b\in V$ and $\phi\in\GL(V)$, satisfying
\[
    \QQ(\alpha(v))=\QQ(v)
    \qquad\text{for all }v\in V.
\]
The following direct-sum extension is the affine statement needed in the main
text.

\begin{cor}[Affine extension from a non-degenerate summand]\label{cor:affine-witt-extension-nondegenerate-summand}
Suppose that $S\subseteq V$ is non-degenerate and that $V=S\oplus S^\perp$ is
the corresponding orthogonal direct sum. Let $\alpha\in\AO[\QQ|_S]$ be given by
$\alpha(x)=b+\phi x$, with
$b\in S$ and $\phi\in\GL(S)$. Define the map $\tilde{\alpha}$ from $V$ to $V$ by
\[
    \tilde{\alpha}(x+y)=b+\phi x+y
    \qquad\text{for } x\in S,\ y\in S^\perp .
\]
Then $\tilde{\alpha}\in\AO[\QQ]$. Its linear part fixes $S^\perp$ pointwise.
\end{cor}
\begin{proof}
Let $v=x+y$ be the unique decomposition with $x\in S$ and $y\in S^\perp$.
Then $\tilde{\alpha}(v)=b+\phi x+y$. Using the orthogonal direct-sum
decomposition and the assumption $\alpha\in\AO[\QQ|_S]$, we obtain
\begin{align*}
    \QQ(\tilde{\alpha}(v))
    &=
    \QQ|_S(b+\phi x)+\QQ|_{S^\perp}(y)\\
    &=
    \QQ|_S(x)+\QQ|_{S^\perp}(y)
    =
    \QQ(v).
\end{align*}
Hence $\tilde{\alpha}\in\AO[\QQ]$. Its linear part is
$\phi\oplus\id_{S^\perp}$, which sends $x+y$ to $\phi x+y$. In particular,
it sends $y\in S^\perp$ to itself, and therefore fixes $S^\perp$ pointwise.
\end{proof}

\subsection{Majorana Basis Extensions}

We finish this appendix by recording a symplectic extension
used in the Majorana description of line-graph generating sets.
As in Eq.~\eqref{eq:def:majorana_basis_pairing}, a \emph{Majorana basis} of $\Fn$ means a basis
$\{\mu_i\}_{i=1}^{2n}$ satisfying
$\symp{\mu_i}{\mu_j}=1+\delta_{ij}$ for all $i$ and $j$.

\begin{lem}[Extending partial Majorana systems]\label{lem:partial_majorana_system_extension}
Let $\mu_1,\ldots,\mu_s\in\Fn$ be linearly independent vectors with
\[
    \symp{\mu_i}{\mu_j}=1+\delta_{ij}
    \qquad\text{for all }1\leq i,j\leq s.
\]
Then there are vectors $\mu_{s+1},\ldots,\mu_{2n}\in\Fn$ such that
$\{\mu_i\}_{i=1}^{2n}$ is a Majorana basis of $\Fn$.
\end{lem}
\begin{proof}
Let $g_{ij}=1+\delta_{ij}$ be the entries of a $2n\times 2n$ matrix $G$ over $\F_2$.
The corresponding alternating form is non-degenerate, since $2n$ is even.
Indeed, if $S=\sum_j x_j$, then $(Gx)_i=S+x_i$; hence $Gx=0$ implies
$x_i=S$ for all $i$.
If $S=1$, then $x$ is the all-one vector, but for this vector one has
$S=2n=0$ in $\F_2$, a contradiction.
Thus $S=0$ and hence $x=0$.
Thus it defines a non-degenerate alternating form on $\F_2^{2n}$.
By the symplectic-basis argument of Lemma~\ref{lem:symplectic_basis}, this alternating space is isometric to
$\Fn$ with the symplectic product $\sympempty$.
If $\tilde{\basel}_i$ denotes the standard basis of $\F_2^{2n}$ and
the map $\psi$ from $\F_2^{2n}$ to $\Fn$ is such an isometry, then
$\nu_i=\psi(\tilde{\basel}_i)$ satisfies
$\symp{\nu_i}{\nu_j}=g_{ij}=1+\delta_{ij}$ for all $i,j$.
Thus $\{\nu_i\}_{i=1}^{2n}$ is a Majorana basis in the sense above.
The assignment $\nu_i\mapsto\mu_i$ for $1\leq i\leq s$ defines an isometry from
$\SpanS[\F_2]{\{\nu_i\}_{i=1}^s}$ to
$\SpanS[\F_2]{\{\mu_i\}_{i=1}^s}$, since the two ordered families have the same
pairwise symplectic products, and hence induce the same restricted alternating
forms by bilinearity.
By the symplectic extension statement in
Corollary~\ref{cor:symplectic-extension-f2}, this isometry extends to an element
${\tilde{\phi}}\in\Sp(2n,\F_2)$.
For $i>s$, set $\mu_i={\tilde{\phi}}(\nu_i)$.
Then $\{\mu_i\}_{i=1}^{2n}$ is a basis and has the pairwise symplectic products
of a Majorana basis, and therefore is the desired extension.
\end{proof}

\section{Adjoint Commutant Basis for Pauli Lie Algebras and Transvection Groups}\label{app:sec:proof_adjoint_commutant}

In this section we prove Theorem~\ref{thm:Adjoint_Commutant_Paulis_Orbits}. 
For $A,B\in\mat{2^n}{\C}$, let
$\multmap{A}{B}\in\End_\C(\mat{2^n}{\C})$ denote the double-sided
multiplication map $M\mapsto AMB$, where $\End_\C$ denotes complex-linear
endomorphisms.
We first collect some basic facts about the adjoint representation, or conjugation action:

\begin{lem}\label{lem:adjoint_double_sided_multiplication_basis}\leavevmode
\begin{enumerate}
    \item For $U\in \lieU(2^n)$, we have
    \[
        \Ad_{\Ad_U}(\multmap{A}{B})
        =
        \multmap{\Ad_U(A)}{\Ad_U(B)}.
    \]
    \item All double-sided multiplication maps
    $\{\multmap{P}{P'}\}_{P,P'\in \PP_n}$ form together a basis of
    $\End_\C(\mat{2^n}{\C})$.
\end{enumerate}
\end{lem}

\begin{proof}
We first prove (a). For any $M\in\mat{2^n}{\C}$,
\begin{subequations}
    \begin{align*}
        &(\Ad_{\Ad_U}(\multmap{A}{B}))(M)
        = (\Ad_U\multmap{A}{B}\Ad_U^\dagger)(M)\\
        &= \Ad_U\big(\multmap{A}{B}(\Ad_U^\dagger(M))\big)
        = \Ad_U\big(\multmap{A}{B}(U^\dagger MU)\big)\\
        &= \Ad_U(AUMU^\dagger B) = U(AU^\dagger MU B)U^\dagger\\
        &= \Ad_U(A)\cdot M\cdot\Ad_U(B).
    \end{align*}
\end{subequations}

(b) follows from the isomorphism under vectorization, which sends
$\multmap{P}{P'}$ into $ P'^T\otimes P$.
Then $P'^T\otimes P=\pm P'\otimes P$ is an orthogonal basis of
$\mat{2^n}{\C}^{\otimes 2}$.
Through the vectorization isomorphism, the double-sided multiplication maps
$\multmap{P}{P'}$ therefore form a basis of $\End_\C(\mat{2^n}{\C})$.
\end{proof}

We now briefly recall the symplectic properties of Pauli strings and the Clifford group.
These are precisely defined in Sections~\ref{sec:pauli} and \ref{sec:clifford:transvections}, which also include additional details.
We can view the Pauli strings $P\in\PP_n = \{I,X,Y,Z\}^{\otimes n}$ as a binary symplectic space $\Fn$ under the isomorphism map $\isoempty$ from $\Fn$ to $\PP_n$
\begin{equation}
    P = \iso{v} = \prod_{i=1}^n \im^{\component{v}{i}\component{v}{i+n}} X_i^{\component{v}{i}}Z_i^{\component{v}{i+n}},
\end{equation}
and the symplectic product $\symp{v}{w}$ is $0$ whenever $\iso{v}$ and $\iso{w}$ commute, and $1$ if they anti-commute.
Also, given that the product of two Pauli strings is a phase $\im^\varpi$, $\varpi\in\{0,1,2,3\}$, times a Pauli string, we can define the sign function (see Eq.~\eqref{eq:Pauli:sign}):
\begin{equation}
    \iso{v}\iso{w} = \im^{\symp{v}{w}}(-1)^{\sign{v}{w}}\iso{v+w}.
\end{equation}
Recall from Eq.~\eqref{eq:eta} that a Clifford unitary acts by conjugation on
Pauli strings as follows:
\begin{equation}
    U \iso{w} U^{\dagger} = (-1)^{\eta_U(w)}\, \isolong{\SY_U w},
\end{equation}
where $\eta_U$ is a phase function, and $\SY_U\in\Sp(2n,\F_2)$ is a binary symplectic map on $\Fn$.

We now prove a useful lemma which applies to the adjoint commutant of arbitrary Clifford groups, or generating sets.
Recall that $\multmap{A}{B}$ denotes the double-sided multiplication map
$M\mapsto AMB$.
\begin{lem}\label{lem:General_Clifford_Adjoint_Commutant}
Consider a subset $X\subseteq\cl_n$ of the Clifford group. Then
$\Lambda = \sum_{v,v'}c_{v}^{v'}\multmap{\iso{v}}{\iso{v'}}\in\commutant_{\Ad}(X)$
if and only if, for all $U\in X$ and all pairs of vectors $v,v'$, we have
\begin{equation}\label{eq:identities_clifford_adjoint_commutant}
    c_{\SY_Uv}^{\SY_Uv'}
    =
    (-1)^{\eta_U(v)+\eta_U(v')} c_{v}^{v'}
\end{equation}
where $\Ad_U(\iso{v}) = (-1)^{\eta_U(v)}\isolong{\SY_Uv}$ for all $U\in X$.
The identity in Eq.~\eqref{eq:identities_clifford_adjoint_commutant} also holds for the entire subgroup generated by $X$, i.e.\ for all $U\in\groupclosure{X}\subseteq\cl_n$.
\end{lem}
\begin{proof}
We first prove the equivalence for one fixed Clifford unitary
$U\in X$ by expanding $\Lambda$ in the double-sided Pauli multiplication basis
and comparing coefficients.
Afterwards we show that the resulting coefficient identity is stable under
products, which gives the final statement.

\emph{Single Clifford unitary.}
Fix $U\in X$.
For this fixed $U$, the condition that $\Lambda$ commutes with $\Ad_U$ is
\begin{equation}
    \comm{\Lambda}{\Ad_U} = 0 \iff \Ad_{\Ad_U}(\Lambda) = \Lambda.
\end{equation}
By Lemma~\ref{lem:adjoint_double_sided_multiplication_basis}(a) and
Eq.~\eqref{eq:eta}, the action of $\Ad_{\Ad_U}$ on one double-sided
multiplication basis element is
\begin{align*}
    \Ad_{\Ad_U}(\multmap{\iso{v}}{\iso{v'}})
    &=
    \multmap{\Ad_U(\iso{v})}{\Ad_U(\iso{v'})}\\
    &=
    \multmap{
    (-1)^{\eta_U(v)}\isolong{\SY_Uv}}
    {(-1)^{\eta_U(v')}\isolong{\SY_Uv'}}\\
    &=
    (-1)^{\eta_U(v)+\eta_U(v')}
    \multmap{\isolong{\SY_Uv}}{\isolong{\SY_Uv'}}.
\end{align*}
Thus
\begin{equation*}
    \Ad_{\Ad_U}(\Lambda)
    =
    \sum_{v,v'}
    c_v^{v'}(-1)^{\eta_U(v)+\eta_U(v')}
    \multmap{\isolong{\SY_Uv}}{\isolong{\SY_Uv'}}.
\end{equation*}
Since $\SY_U$ is invertible, we may relabel this sum by
$a=\SY_Uv$ and $a'=\SY_Uv'$.
Then
\begin{equation*}
    \Ad_{\Ad_U}(\Lambda)
    =
    \sum_{a,a'}
    c_{\SY_U^{-1}a}^{\SY_U^{-1}a'}
    (-1)^{\eta_U(\SY_U^{-1}a)+\eta_U(\SY_U^{-1}a')}
    \multmap{\iso{a}}{\iso{a'}}.
\end{equation*}
In words, $\Ad_{\Ad_U}$ acts on the double-sided Pauli multiplication basis by
a signed permutation induced by $\SY_U$ on the two Pauli labels.
Thus commutation with $\Ad_U$ is equivalent to invariance of the coefficients
under this signed permutation.

\emph{Coefficient comparison.}
Substituting the relabelled expansion above and
\[
    \Lambda=\sum_{a,a'}c_a^{a'}\multmap{\iso{a}}{\iso{a'}},
\]
we see that both $\Ad_{\Ad_U}(\Lambda)$ and $\Lambda$ are expanded in the same
basis
$\{\multmap{\iso{a}}{\iso{a'}}\}_{a,a'}$.
By Lemma~\ref{lem:adjoint_double_sided_multiplication_basis}(b), this basis is
linearly independent.
Therefore
\[
    \Ad_{\Ad_U}(\Lambda)=\Lambda
\]
is equivalent to equality of the corresponding coefficients, i.e., for every
pair $a,a'$,
\[
    c_a^{a'}
    =
    c_{\SY_U^{-1}a}^{\SY_U^{-1}a'}
    (-1)^{\eta_U(\SY_U^{-1}a)+\eta_U(\SY_U^{-1}a')}.
\]
After setting $a=\SY_Uv$ and $a'=\SY_Uv'$, this condition becomes exactly
Eq.~\eqref{eq:identities_clifford_adjoint_commutant}.
Thus, for the fixed Clifford unitary $U$, the commutation relation
$\comm{\Lambda}{\Ad_U}=0$ is equivalent to
Eq.~\eqref{eq:identities_clifford_adjoint_commutant}.
Since $\Lambda\in\commutant_{\Ad}(X)$ means that
$\comm{\Lambda}{\Ad_U}=0$ for every $U\in X$, the coefficient
comparison above for single $U$ proves the equivalence after imposing
Eq.~\eqref{eq:identities_clifford_adjoint_commutant} for all $U\in X$.

\emph{Generated subgroup.}
It remains to show that the coefficient identity then holds for all
$U\in\groupclosure{X}$.
It suffices to check that it is preserved under products.
Assume it holds for $U$ and $U'$.
Then, using
Lemma~\ref{lem:sign:clifford}\ref{lem:sign:clifford:e}, we find
\begin{align*}
        c_{\SY_{UU'}v}^{\SY_{UU'}v'} &= c_{\SY_U\SY_{U'}v}^{\SY_U\SY_{U'}v'}\\
        &= (-1)^{\eta_U(\SY_{U'}v)+\eta_U(\SY_{U'}v')} c_{\SY_{U'}v}^{\SY_{U'}v'}\\
        &= (-1)^{\eta_U(\SY_{U'}v)+\eta_U(\SY_{U'}v') + \eta_{U'}(v) + \eta_{U'}(v')} c_v^{v'}\\
        &= (-1)^{\eta_{UU'}(v)+\eta_{UU'}(v')} c_v^{v'}.
\end{align*}
In words, the coefficient identity for $U'$ first moves the labels by
$\SY_{U'}$, and the coefficient identity for $U$ then moves the resulting
labels by $\SY_U$.
The two sign contributions combine into the phase function of the product
$UU'$ by Lemma~\ref{lem:sign:clifford}\ref{lem:sign:clifford:h}.
Hence, the identity holds for products of generators, and therefore for the
entire generated group.
\end{proof}

The previous lemma has a particularly simple consequence for Clifford
transvections, which is the only case needed below.
\begin{cor}\label{cor:clifford_adjoint_commutant_transvection_parity}
Let $U=\tvu{\iso{w}}$ be a Clifford transvection, and let
\[
    \Lambda=\sum_{v,v'}c_v^{v'}\multmap{\iso{v}}{\iso{v'}}
    \in\commutant_{\Ad}(U).
\]
If $c_v^{v'}\neq 0$, then
\begin{equation}\label{eq:clifford_adjoint_commutant_transvection_parity}
    \symp{w}{v}=\symp{w}{v'}.
\end{equation}
\end{cor}
\begin{proof}
For a Clifford transvection, the binary action is $\SY_U=\tau_w$, and
$\tau_w$ is an involution.
By Lemma~\ref{lem:General_Clifford_Adjoint_Commutant}, the coefficient identity
for the pair $(v,v')$ gives
\[
    c_{\tau_wv}^{\tau_wv'}
    =
    (-1)^{\eta_U(v)+\eta_U(v')}c_v^{v'}.
\]
Applying the same identity to $(\tau_wv,\tau_wv')$ gives
\[
    c_v^{v'}
    =
    (-1)^{\eta_U(\tau_wv)+\eta_U(\tau_wv')}c_{\tau_wv}^{\tau_wv'}.
\]
Substituting the first equation into the second and using $c_v^{v'}\neq0$ gives
\begin{equation}\label{eq:adjoint_commutant_transvection_sign_cycle}
    \eta_U(v)+\eta_U(v')
    +\eta_U(\tau_wv)+\eta_U(\tau_wv')=0.
\end{equation}
We claim that
\begin{equation}\label{eq:eta_transvection_center_shift}
    \eta_U(\tau_wx)+\eta_U(x)=\symp{w}{x}.
\end{equation}
Indeed, Lemma~\ref{lem:sign:clifford}\ref{lem:sign:clifford:h}, applied to the product $U^2$, gives
\[
    \eta_{U^2}(x)=\eta_U(x)+\eta_U(\tau_wx),
\]
because $\SY_U=\tau_w$.
On the other hand, $U^2=\im\iso{w}$, and scalar phases do not affect
conjugation.
Hence $\eta_{U^2}(x)=\eta_{\iso{w}}(x)$.
Lemma~\ref{lem:sign:clifford}\ref{lem:sign:clifford:c}, applied with $U=\In$ and Pauli factor
$\iso{w}$, together with Lemma~\ref{lem:sign:clifford}\ref{lem:sign:clifford:a}, gives
\[
    \eta_{\iso{w}}(x)
    =
    \eta_{\In}(x)+\symp{w}{x}
    =
    \symp{w}{x}.
\]
This proves Eq.~\eqref{eq:eta_transvection_center_shift}.
Substituting this formula into
Eq.~\eqref{eq:adjoint_commutant_transvection_sign_cycle}
gives $\symp{w}{v}+\symp{w}{v'}=0$.
\end{proof}

This gives the following condition for double-sided Pauli
multiplication maps in the adjoint commutant.
\begin{lem}\label{lem:adjoint_commutant_support_condition}
Let
\[
    \Lambda=\sum_{v,v'}c_v^{v'}\multmap{\iso{v}}{\iso{v'}}
    \in \commutant_{\Ad}(\{\tvu{G}\}_{G\in\pgens}).
\]
If $c_v^{v'}\neq0$, then $v+v'\in\vgens^\perp$.
Equivalently, $\Lambda$ can be written as
\[
    \Lambda
    =
    \sum_{u\in\vgens^\perp}\sum_{v\in\Fn}
    c_v^{u+v}\multmap{\iso{v}}{\iso{u+v}}.
\]
\end{lem}
\begin{proof}
Apply
Corollary~\ref{cor:clifford_adjoint_commutant_transvection_parity} to every
generator $G=\iso{w}\in\pgens$.
If $c_v^{v'}\neq0$, then $\symp{w}{v}=\symp{w}{v'}$ for all such $w$, or
equivalently $\symp{w}{v+v'}=0$ for all $w\in\vgens$.
Thus $v+v'\in\vgens^\perp$.
Writing $u=v+v'$ and collecting the terms with the same $u$ gives the claimed
expansion.
\end{proof}

The remaining condition is the covariance along Pauli orbits.  We prove
it separately.
\begin{lem}\label{lem:adjoint_commutant_orbit_covariance}
Let $C\in\cent_{\PP_n}(\pgens)$.
For each orbit $O\in\orb(\PP_n)$ of the Clifford transvection group
$\cltvgroup{\pgens}$, define
\begin{equation}\label{eq:adjoint_commutant_orbit_sum_basis_element}
    \Lambda_{O,C}:=\sum_{P\in O}\multmap{P}{CP}.
\end{equation}
Then $\Lambda_{O,C}$ is fixed by
$\Ad_{\Ad_U}$ for every $U\in\cltvgroup{\pgens}$.
Moreover, every element of
\begin{equation}\label{eq:adjoint_commutant_fixed_C_span}
    \SpanL{\{\multmap{P}{CP}\mid P\in\PP_n\}}
\end{equation}
that is fixed by all $\Ad_{\Ad_U}$ is a linear combination of the
$\Lambda_{O,C}$.
\end{lem}
\begin{proof}
Since $C\in\cent_{\PP_n}(\pgens)$, each generating Clifford transvection
with center in $\pgens$ fixes $C$.
Hence every element of the generated group fixes $C$, so $\Ad_U(C)=C$ for every
$U\in\cltvgroup{\pgens}$.
If $\Ad_U(P)=sP'$ with $s\in\{\pm1\}$, then
\[
    \Ad_{\Ad_U}(\multmap{P}{CP})
    =
    \multmap{sP'}{CsP'}
    =
    \multmap{P'}{CP'}.
\]
Thus $\cltvgroup{\pgens}$ acts on the basis elements
$\multmap{P}{CP}$ by the permutation action on the Pauli label $P$, with no
remaining sign.
The fixed vectors of a permutation representation are precisely the sums over
its orbits.
Therefore the fixed elements are exactly the linear combinations of the orbit
sums $\Lambda_{O,C}$.
\end{proof}

We now use this support condition to compute the adjoint commutant basis
in Theorem~\ref{thm:Adjoint_Commutant_Paulis_Orbits}.
\begin{proof}[Proof of Theorem~\ref{thm:Adjoint_Commutant_Paulis_Orbits}]
We now prove the theorem in four steps.
First, we replace the Lie-algebra adjoint commutant by the adjoint commutant
of the corresponding Clifford transvection group.
Second, we use Lemma~\ref{lem:adjoint_commutant_support_condition} to identify
the possible double-sided Pauli multiplication maps on which $\Lambda$ is supported, and which lie in the subspace defined in Eq.~\eqref{eq:adjoint_commutant_fixed_C_span}.
Third, we apply Lemma~\ref{lem:adjoint_commutant_orbit_covariance} to impose
the remaining covariance condition along each Pauli orbit.
Finally, we verify that the resulting orbit sums form a basis.

\emph{Reduction to Clifford transvections.}
By Lemma~\ref{lem:adjoint_commutant_lie_and_clifford}, the adjoint commutant of the Lie algebra $\lie{\pgens}$ and the adjoint commutant of the Clifford transvection group $\cltvgroup{\pgens}$ coincide:
\begin{equation}
    \commutant_{\ad}(\pgens) = \commutant_{\Ad}(\{\tvu{G}\}_{G\in\pgens})
\end{equation}
We now compute a basis for this commutant in the Clifford formalism, to highlight an alternative proof to that of \cite{Diaz_GarciaMartin_Kazi_Larocca_Cerezo_2023}.

\emph{Support.}
By Lemma~\ref{lem:adjoint_commutant_support_condition}, every element of this
adjoint commutant has the form
\[
    \Lambda
    =
    \sum_{u\in\vgens^\perp}\sum_{v\in\Fn}
    c_v^{u+v}\multmap{\iso{v}}{\iso{u+v}}.
\]
Equivalently, writing $C=\iso{u}$ and $P=\iso{v}$, the only possible
double-sided multiplication maps in $\Lambda$ are $\multmap{P}{CP}$.
Indeed, by Eq.~\eqref{eq:Pauli:sign},
\[
    CP=\iso{u}\iso{v}
    =
    \lambda(u,v)\iso{u+v}
    \qquad \text{for }
    \lambda(u,v)\in\{\pm1,\pm\im\}.
\]
Since the double-sided multiplication map is linear in its second argument,
\[
    \multmap{\iso{v}}{CP}
    =
    \lambda(u,v)\multmap{\iso{v}}{\isolong{u+v}}.
\]
Thus replacing $\multmap{\iso{v}}{\isolong{u+v}}$ by
$\multmap{P}{CP}$ only changes the complex coefficient of that basis element.
This rescaling does not change the span because the coefficients in the
expansion are arbitrary complex numbers.
The condition $u\in\vgens^\perp$ means that $\iso{u}$ commutes with every
generator in $\pgens$, so $C=\iso{u}\in\cent_{\PP_n}(\pgens)$.

\emph{Orbit covariance.}
For each fixed $C\in\cent_{\PP_n}(\pgens)$, the maps $\multmap{P}{CP}$ are
permuted by the Clifford transvection group according to the orbit of $P$.
By Lemma~\ref{lem:adjoint_commutant_orbit_covariance}, the elements fixed by
this permutation action are precisely the linear combinations of the orbit sums
$\Lambda_{O,C}$ from
Eq.~\eqref{eq:adjoint_commutant_orbit_sum_basis_element}, with
$O\in\orb(\PP_n)$.

\emph{Basis.}
The support step shows that every adjoint-commutant element is a sum of terms
inside the subspaces from
Eq.~\eqref{eq:adjoint_commutant_fixed_C_span}, with
$C\in\cent_{\PP_n}(\pgens)$.
For each such fixed $C$, the covariance step identifies the part fixed by the
Clifford transvection group as the span of the orbit sums $\Lambda_{O,C}$.
Hence the orbit sums in
Eq.~\eqref{eq:adjoint_commutant_orbit_sum_basis_element} span the whole adjoint
commutant.
They are linearly independent because distinct pairs $(O,C)$ have disjoint
supports in the double-sided Pauli multiplication basis
$\{\multmap{P}{P'}\}_{P,P'\in\PP_n}$.
Therefore they form a basis, which proves the theorem.
\end{proof}

\section{Operator Invariant Subspaces}\label{sec:app:operator_invariant_subspaces}

In this section we prove
Theorem~\ref{thm:symmetry_adapated_orbit_subspaces}.
The proof has two independent ingredients.
First, the commuting Pauli symmetries in the commutant split the operator
space into joint left and right eigenspaces.
Second, the Clifford transvection group splits the Pauli basis into orbit
spans.
Thus the first goal is to prove the two decompositions
\begin{align}
    \matalg(2^n,\C)
    &=
    \bigoplus_{\textbf{s},\textbf{s}'\in\{\pm1\}^{\times(\ell+r)}}
    \invV_{\textbf{s},\textbf{s}'},
    \label{eq:app:operator_symmetry_block_decomposition}\\
    \matalg(2^n,\C)
    &=
    \bigoplus_{O\in\orb(\PP_n)}\invV_O .
    \label{eq:app:operator_orbit_block_decomposition}
\end{align}
The invariant subspaces in the theorem are then obtained as the intersection
\begin{equation}\label{eq:app:operator_common_refinement}
    \invV_{\textbf{s},\textbf{s}'}^O
    =
    \invV_{\textbf{s},\textbf{s}'}\cap\invV_O .
\end{equation}
The remaining work is to identify the symmetry blocks explicitly and to prove
the dimension bound
\begin{equation}\label{eq:app:operator_refined_block_dimension_bound}
    \dim \invV_{\textbf{s},\textbf{s}'}^O
    \leq
    \min\{2^{2m},\abs{O}\}.
\end{equation}

We use the notation of the theorem throughout:
$\rank(\vgens^\perp)=2\ell$, $\nullity(\vgens^\perp)=r$, and
$C_i=\iso{u_i}$ are algebraically independent mutually commuting Pauli
symmetries, indexed by $1\leq i\leq \ell+r$.
\[
    W=\SpanS[\F_2]{\{u_i\}_{i=1}^{\ell+r}}.
\]
Since the $C_i$ commute, the restriction of the symplectic form to $W$ is zero,
so $W=\rad(W)$.

We first isolate the decomposition coming from the commuting Pauli
symmetries.
\begin{lem}[Block decomposition]\label{lem:operator_invariant_symmetry_blocks}
Let $\{C_i\}_{i=1}^{\ell+r}$ with $C_i=\iso{u_i}$ be algebraically independent
mutually commuting Pauli strings, and set
$W=\SpanS[\F_2]{\{u_i\}_{i=1}^{\ell+r}}$.
For $\textbf{s},\textbf{s}'\in\{\pm1\}^{\times(\ell+r)}$, let
$\invV_{\textbf{s},\textbf{s}'}$ be the joint eigenspace of left multiplication
by $C_i$ with eigenvalue $s_i$ and right multiplication by $C_i$ with
eigenvalue $s_i'$.
Then Eq.~\eqref{eq:app:operator_symmetry_block_decomposition} holds.
Moreover, if $w_{\textbf{s},\textbf{s}'}$ is any vector satisfying
\[
    \symp{w_{\textbf{s},\textbf{s}'}}{u_i}
    =
    [(1{-}s_is_i')/{2}]\bmod 2
\]
for all $i$, then
\begin{align}
        \invV_{\textbf{s},\textbf{s}'} =
        \spanempty\Big(\Big\{
        &\prod_{i=1}^{\ell+r}(1{+}s_i\iso{u_i})
        \isolong{w_{\textbf{s},\textbf{s}'}{+}v} \label{eq:app:symmetry_subspaces} \\
        & \text{for representatives } v\in W^\perp/W\Big\}\Big). \nonumber
\end{align}
Each symmetry block has dimension
$2^{2n-2(\ell+r)}$.
\end{lem}
\begin{proof}
Let $L_{C_i}$ and $R_{C_i}$ denote left and right multiplication by $C_i$.
The operators $L_{C_i}$ and $R_{C_j}$ commute for all $i,j$ because the $C_i$
commute with each other.
They are also involutions.
Therefore the simultaneous projectors
\begin{equation}
    \Pi_{\textbf{s},\textbf{s}'}
    =
    \prod_{i=1}^{\ell+r}
    \frac{1+s_iL_{C_i}}{2}
    \frac{1+s_i'R_{C_i}}{2}
\end{equation}
give the direct-sum decomposition in
Eq.~\eqref{eq:app:operator_symmetry_block_decomposition}.
Expanding this projector, every non-identity term is a left and/or right
multiplication by a non-identity Pauli string and therefore has trace zero.
Indeed, under vectorization, left and right multiplication by $A$ and $B$ is
represented by $B^T\otimes A$, whose trace is $\Tr(B^T)\Tr(A)$.
The algebraic independence of the $C_i$ ensures that no non-empty product of
the $C_i$ is the identity.
Thus
\[
    \Tr(\Pi_{\textbf{s},\textbf{s}'})
    =
    4^n/4^{\ell+r}
    =
    2^{2n-2(\ell+r)}.
\]

It remains to identify a basis of the image of
$\Pi_{\textbf{s},\textbf{s}'}$.
The displayed condition on $w_{\textbf{s},\textbf{s}'}$ says precisely that
$P=\isolong{w_{\textbf{s},\textbf{s}'}+v}$ satisfies
$(-1)^{\symp{u_i}{w_{\textbf{s},\textbf{s}'}+v}}=s_is_i'$ for every $i$
whenever $v\in W^\perp$.
Equivalently, $C_iP=(s_is_i')PC_i$.
Therefore
\begin{align*}
    C_i(1+s_iC_i)P &= s_i(1+s_iC_i)P,\\
    (1+s_iC_i)PC_i &= s_i'(1+s_iC_i)P .
\end{align*}
Since the $C_i$ commute with one another, the same eigenvalue calculation
applies to the full product over $i$.
Hence the projected Pauli strings
\begin{equation}\label{eq:app:projected_basis_operator_subspaces}
    \prod_{i=1}^{\ell+r}(1{+}s_i\iso{u_i})
    \isolong{w_{\textbf{s},\textbf{s}'}{+}v}
    \qquad \text{for } v\in W^\perp/W
\end{equation}
lie in $\invV_{\textbf{s},\textbf{s}'}$.
Choosing representatives for $W^\perp/W$ avoids double counting the same
projected vector, and distinct representatives give linearly independent
vectors because their Pauli supports are disjoint.

Finally, since $\dim W=\ell+r$ and $W\subseteq W^\perp$, we have
\[
    \dim(W^\perp/W)=2n-2(\ell+r).
\]
Thus Eq.~\eqref{eq:app:projected_basis_operator_subspaces} gives
$2^{2n-2(\ell+r)}$ linearly independent vectors.
This equals the dimension of the image of
$\Pi_{\textbf{s},\textbf{s}'}$, so these vectors form a basis of
$\invV_{\textbf{s},\textbf{s}'}$.
\end{proof}

We next separate the decomposition coming from the Clifford-transvection
orbits.
\begin{lem}[Orbit decomposition]\label{lem:operator_invariant_orbit_blocks}
Let $\orb(\PP_n)$ be the set of Pauli orbits under
$\cltvgroup{\pgens}$, and set
\begin{equation}\label{eq:app:orbit_subspaces}
    \invV_O=\SpanL{\{\iso{v}\mid v\in O\}}
    \qquad \text{for } O\in\orb(\PP_n).
\end{equation}
Then Eq.~\eqref{eq:app:operator_orbit_block_decomposition} holds.
Moreover, each $\invV_O$ is invariant under $\cltvgroup{\pgens}$.
\end{lem}
\begin{proof}
The Pauli strings form an orthogonal basis of $\matalg(2^n,\C)$, and distinct
orbits are disjoint.
This gives the direct-sum decomposition in
Eq.~\eqref{eq:app:operator_orbit_block_decomposition}.
Since $\cltvgroup{\pgens}$ sends each Pauli string in an orbit to another
Pauli string in the same orbit, up to sign, each span $\invV_O$ is invariant.
\end{proof}

\begin{proof}[Proof of Theorem~\ref{thm:symmetry_adapated_orbit_subspaces}]
We now prove the theorem in four steps.
First, we show the invariance under the Pauli Lie algebra and the Clifford
transvection group.
Second, we use the commuting Pauli symmetries to obtain the block
decomposition.
Third, we use Pauli orbits to obtain the orbit decomposition.
Finally, we combine the two decompositions and derive the dimension bound.

\emph{Invariance under generators.}
We use two elementary facts.
First, if $G=\iso{v}\in\pgens$ and $P=\iso{w}$ is a Pauli string, then
Eq.~\eqref{eq:Pauli:comm} shows that $\comm{G}{P}$ is either zero or
proportional to $\iso{v+w}$.
In the non-zero case, this is precisely the Pauli label obtained from $P$ by
the Clifford transvection $\tvu{G}$, whose binary action is $\tau_v$ by
Lemma~\ref{lem:clifford_transvection_binary_action}(b).
Thus every orbit span $\invV_O$ is invariant under the adjoint action of the
Pauli Lie algebra $\lie{\pgens}$, and hence under the connected Pauli Lie group.
Second, if $C_i\in\commutant(\pgens)$, then left and right multiplication by
$C_i$ commute with $\ad_G$ for every $G\in\pgens$ and with
$\Ad_{\tvu{G}}$.
Consequently, any common eigenspace of these left and right multiplication maps is
invariant under both the Pauli Lie algebra and the Clifford transvection group.

\emph{Block decomposition.}
Lemma~\ref{lem:operator_invariant_symmetry_blocks} gives the direct-sum
decomposition into the joint left and right eigenspaces
$\invV_{\textbf{s},\textbf{s}'}$ and the explicit basis in
Eq.~\eqref{eq:app:symmetry_subspaces}.
By the preceding paragraph, these joint eigenspaces are invariant under the
Pauli Lie algebra, the connected Pauli Lie group, and the Clifford
transvection group.

\emph{Orbit blocks.}
Lemma~\ref{lem:operator_invariant_orbit_blocks} gives the orbit decomposition
and shows invariance under the Clifford transvection group.
The same orbit spans are invariant under the Pauli Lie algebra by the
commutator calculation in the first step.

\emph{Intersection.}
The intersection of two invariant subspaces is invariant.
Hence Eq.~\eqref{eq:app:operator_common_refinement} defines an invariant
subspace for every $O$ and every pair of block labels
$\textbf{s},\textbf{s}'$.
After discarding zero and repeated intersections, these subspaces give a refinement of the orbit decomposition and the symmetry-block
decomposition.

Finally, Lemma~\ref{lem:operator_invariant_symmetry_blocks} shows that the
symmetry blocks have dimension $2^{2n-2(\ell+r)}$.
Under the notation of Theorem~\ref{thm:symmetry_adapated_orbit_subspaces}, the
assumption $\dim(\algclosure{\pgens})=2^{2m+r}$ identifies this dimension with
$2^{2m}$.
Since
$\invV_{\textbf{s},\textbf{s}'}^O
\subseteq \invV_{\textbf{s},\textbf{s}'}$
and
$\invV_{\textbf{s},\textbf{s}'}^O\subseteq\invV_O$,
we obtain Eq.~\eqref{eq:app:operator_refined_block_dimension_bound}.
This proves the theorem.
\end{proof}

\bibliography{large_bibliography}

\end{document}